# NEW INFORMATION TECHNOLOGIES, SIMULATION AND AUTOMATION

MONOGRAPH
Scientific publication (issue)
**Editor-in-Chief**
**Sergii Kotlyk**

Velychko V., Voinova S., Granyak V., Gurskiy O. Zavertailo K., Ivanova L., Kotlyk D., Kotlyk S., Kudriashova A., Kunup T., Malakhov K., Pikh I., Punchenko N., Senkivskyy V., Sergeeva O. Sokolova O., Fedosov S., Khoshaba O.,Tsyra O., Chaplinskyy Y.

# NEW INFORMATION TECHNOLOGIES, SIMULATION AND AUTOMATION

MONOGRAPH
Scientific publication (issue)

**Editor-in-Chief**
**Sergii Kotlyk**

UDC 004.01/08
H73

Recommended by the Academic Council of
The Odesa National University of Technology
(Protocol № 11 dated 05.04.2022)

**Reviewers:**

**Olexandr Romanyuk,** DSc, Prof., Vinnytsia National Technical University
**Valery Plotnikov,** DSc, Prof., Odesa National Academy of Food Technologies
**Olexandr Shpinkovski,** PhD, Docent, Odesa Polytechnic State University

**Editor-in-Chief**
**Sergii Kotlyk**

PhD, Docent, Odesa National University of Technology

**Team of Authors:**

Velychko V., Voinova S., Granyak V., Gurskiy O., Zavertailo K., Ivanova L., Kotlyk D., Kotlyk S., Kudriashova A., Kunup T., Malakhov K., Pikh I., Punchenko N., Senkivskyy V., Sergeeva O., Sokolova O., Fedosov S., Khoshaba O., Tsyra O., Chaplinskyy Y.

**H73** **New information technologies, simulation and automation:** Monograph / Velychko V., Voinova S., Granyak V., et al; Editor-in-Chief Kotlyk S. Iowa State University Digital Press.

The monograph summarizes and analyzes the current state of development of computer and mathematical simulation/modeling, the automation of management processes, the use of information technologies in education, the design of information systems and software complexes, the development of computer telecommunication networks and technologies — most areas that are united by the term Industry 4.0.

The monograph will be useful both for experts and employees of companies engaged in the field of IT and automation, as well as for educators, masters, students and postgraduates of higher educational institutions, and everyone interested in issues related to Industry 4.0.

# Preface

The fourth industrial revolution (Industry 4.0) envisages a new approach to production based on the mass introduction of information technologies into the industry, large-scale automation of business processes, and the spread of artificial intelligence. The benefits of the Fourth Industrial Revolution are obvious: increased productivity, more significant safety for employees due to the reduction of jobs in hazardous working conditions, increased competitiveness, fundamentally new products, and much more. However, it also has shortcomings that can negatively affect society's development, thus, studying the evolution of Industry 4.0 directions is a necessary condition for the practical application of modern science.

The monograph summarizes and analyzes the current state of development of computer and mathematical simulation/modeling, the automation of management processes, the use of information technologies in education, the design of information systems and software complexes, the development of computer telecommunication networks and technologies — most areas that are united by the term Industry 4.0.

The monograph was compiled based on the results of the XIV International Scientific and Practical Conference "Information Technologies and Automation - 2021", which took place in October 2021 at The Odesa National University of Technology (the former Odesa National Academy of Food Technologies).

The range of problems presented in the monograph is extremely wide — the application of information technologies for the design of post-printing processes and new food products, the development of decision-making theory, mathematical simulation/modeling in ferroelectric polymers and polarized films, the development of logic control algorithms, the automation of maintenance of powerful electric machines and shipping mooring systems, applications of information technologies in education, digital health and distribution between computing complexes.

The presented monograph is a significant help to experts, educators, students, graduate students who are trying to learn about the current state of science in the field of Industry 4.0. This information can be used to solve a wide range of problems in the specified sections that arise both in the educational process and in research and scientific plans.

# НОВІ ІНФОРМАЦІЙНІ ТЕХНОЛОГІЇ, МОДЕЛЮВАННЯ ТА АВТОМАТИЗАЦІЯ

Монографія

За загальною редакцією
**С. В. Котлика**

Колектив авторів:

**В. Ю. Величко, С. О. Воінова, В. Ф. Граняк, О. О. Гурський, К. С. Завертайло, Л. В. Іванова, Д. О. Котлик, С. В. Котлик, А. В. Кудряшова, Т. В. Кунуп, К. С. Малахов, І. В. Піх, Н. О. Пунченко, В. М. Сеньківський, О. Є. Сергєєва, О. П. Соколова, С. Н. Федосов, О. М. Хошаба, О. В. Цира, Ю. П. Чаплінський**

Рецензенти:

**О. Н. Романюк**, д. т. н., професор, зав. кафедри програмного забезпечення Вінницького національного технічного університету;

**В. М. Плотніков**, д. т. н., професор, зав. кафедри інформаційних технологій та кібербезпеки Одеської національної академії харчових технологій;

**О. А. Шпинковський**, к. т. н., доцент кафедри інформаційних систем Державного університету «Одеська політехніка»

Рекомендовано до друкування рішенням вченої ради Одеської національної академії харчових технологій *(протокол № 11 від 5 квітня 2022 р.)*

У монографії узагальнено і проаналізовано рівень сучасного стану розвитку комп'ютерного та математичного моделювання, автоматизації процесів управління, застосування інформаційних технологій в освіті, проектування інформаційних систем і програмних комплексів, розвитку комп'ютерних телекомунікаційних мереж та технологій — більшості напрямків, які об'єднуються терміном Індустрія 4.0.

Монографія буде корисною як для фахівців і працівників фірм, зайнятих в галузі ІТ і автоматизації, так і для викладачів, магістрів, студентів і аспірантів вищих навчальних закладів, і всіх, хто цікавиться питаннями, пов'язаними з Індустрією 4.0.

# Зміст

*Розділ IV*
## ПРОЕКТУВАННЯ ІНФОРМАЦІЙНИХ СИСТЕМ
## І ПРОГРАМНИХ КОМПЛЕКСІВ

*Розділ V*
## КОМП'ЮТЕРНІ ТЕЛЕКОМУНІКАЦІЙНІ МЕРЕЖІ
## ТА ТЕХНОЛОГІЇ

# Передмова

Четверта промислова революція (Індустрія 4.0) передбачає новий підхід до виробництва, що базується на масовому впровадженні інформаційних технологій у промисловість, масштабній автоматизації бізнес-процесів та поширенні штучного інтелекту.

Переваги Четвертої промислової революції очевидні: підвищення продуктивності, велика безпека працівників за рахунок скорочення робочих місць у небезпечних умовах праці, підвищення конкурентоспроможності, принципово нові продукти та багато іншого. Однак вона має й недоліки, які можуть негативно впливати на розвиток суспільства, тому вивчення розвитку напрямів Індустрії 4.0 — необхідна умова практичного застосування сучасної науки.

У колективній монографії представлені результати практичних і теоретичних досліджень в області комп'ютерного та математичного моделювання, автоматизації процесів управління, застосування інформаційних технологій в освіті, проектування інформаційних систем і програмних комплексів, розвитку комп'ютерних телекомунікаційних мереж та технологій — більшості напрямків, які об'єднуються терміном Індустрія 4.0.

Монографія складена за підсумками проведення XIV Міжнародної науково-практичної конференції «Інформаційні технології та автоматизація — 2021», яка відбулася в жовтні 2021 року в Одеському національному технологічному університеті (колишня Одеська національна академія харчових технологій).

Спектр представлених у монографії проблем надзвичайно широкий — застосування інформаційних технологій для проектування післядрукарських процесів і нових харчових продуктів, розвиток теорії прийняття рішень, математичне моделювання в сегнетоелектричних полімерах і поляризованих плівках, розробка алгоритмів логічного управління, автоматизація обслуговування потужних електричних машин і систем швартування судноплавства, застосування інформаційних технологій в освіті та розподіл між обчислювальними комплексами.

Представлена монографія являє собою істотну підмогу фахівцям, викладачам, студентам, аспірантам, які намагаються дізнатися про сучасний стан науки в галузі Індустрія 4.0. Ця інформація може бути використана для вирішення широкого кола проблем в зазначених розділах, що виникають як в навчальному процесі, так і в дослідницькому і науковому планах.

## КОНТЕКСТНО-ОНТОЛОГІЧНА СИСТЕМНА ОПТИМІЗАЦІЯ ПРОБЛЕМНО-ОРІЄНТОВАНОЇ ПІДТРИМКИ ПРИЙНЯТТЯ РІШЕНЬ

*Чаплінський Ю. П.*

*Показана актуальність використання знання-орієнтованих підходів до прийняття рішень, що базується на використанні системної оптимізації, онтологій та контексту. Описана технологія підтримки прийняття рішень для розв'язання управлінських задач, яка ґрунтується на методології системної оптимізації. Досліджуються різні ситуації при прийнятті рішень в рамках запропонованої технології. Процес прийняття рішень на основі системної оптимізації розглядається через модель певного контексту. Визначені контекстна онтологія та її складові, які дозволяють розпізнати, зрозуміти, представити та підтримати розв'язання задачі прийняття рішень. Представлені онтологія шарів та онтологія аспектів.*

*The actuality of the usage of knowledge-oriented decision-making approach based on the use of system optimization, ontologies and context is shown. The decision — making technology for solving management problems, which is based on the methodology of system optimization, is described. Various situations of decision making within proposed technology are researched. The decision-making process based on system optimization through the model of some context is considered. Contextual ontology and its components, which allow to recognize, understand, present and support the solution of the decision-making problem are identified. The ontology of layers and the aspects ontology are presented.*

Сьогодні комплексна та системна підтримка прийняття рішень є домінуючим динамічним діловим середовищем. Це визначається характерними рисами сучасного прийняття рішень: інтеграція наукових знань, зростання кількості міждисциплінарних проблем, комплексність проблем та необхідність їх вивчення у єдності технічних, економічних, соціальних, психологічних, управлінських та інших аспектів; ускладнення аналізованих проблем та об'єктів; динамічність ситуацій прийняття рішень; дефіцитність ресурсів; підвищення рівня стандар-тизації та автоматизації елементів виробничих та управлінських процесів; глобалізація конкуренції, виробництва, кооперації, стандартизації тощо; підвищення ролі людського фактора в управлінні та ін.

При цьому прийняття рішень відбудеться як через горизонтальні перехресні вузли, так і через вертикальні перехресні ієрархічні зв'язки, при цьому можливо отримання раніше недоступної інформації, що в подальшому дає змогу розвивати нові знання та розуміння. При цьому неможливо із загального процесу прийняття рішення виділити будь-які окремі задачі, оскільки вони об'єднані в одну загальну задачу. Це означає, що діяльність як окремих людей, так і підприємств все більшою мірою залежить від наявних у них знань як одного з найцінніших ресурсів і можливості їхнього ефективного використання.

Управління знаннями сьогодні розглядається як потужна конкурентна перевага на підприємстві, орієнтованому на постійні зміни ділових процесів. Під представленням знань розуміється їх структуризація з метою формалізації процесів рішення задач у певній проблемній області.

Такий розгляд прийняття рішень визначає перехід від вузькодисциплінарного прийняття рішень до взаємодіючої множини предметних областей, що об'єднує різні аспекти розгляду: представлення, зміст, інтерпретацію та використання; постійної зміни середовища прийняття рішень, постійного накопичення нових знань, використання активних знань.

Слід також зазначити, що знання в таких складних предметних областях дуже швидко змінюються або застарівають, з'являються нові задачі та нові методи розв'язання. При цьому необхідно розглядати мультидисциплінарні сфери, що пов'язані з відповідною прикладною проблемою, їх взаємодію та інтеграцію.

Результати розв'язання задач прийняття рішень є результатом поєднання та інтеграції знань, розуміння та ідей розв'язання множин взаємопов'язаних задач з різних предметних областей, кожна з яких має свої специфічні передумови. Вони характеризуються різноманітністю, багатовимірністю, багаторівневістю. Реалізацію ефективного прийняття рішень будемо розглядати на основі методології систем підтримки прийняття рішень, основою якої є між-/мульти-/трансдисциплінарність, контекст та онтологія, як засоби розуміння та представлення предметних областей і процесів прийняття рішень та інтеграції методів системного, процесного та ситуаційного аналізу.

В рамках такого прийняття рішень людині, що приймає рішення (ЛПРу), необхідно врахувати множину властивостей, що визначаються та використовуються одночасно. Це вимагає розгляду процесів, структур, ресурсів, навколишнього середовища, а також взаємодії між акторами процесу прийняття рішень. Для цього необхідно використовувати тільки ті особливості дійсності, які є найважливішими для ситуації чи проблеми. При цьому необхідно сконцентрувати увагу на деяких конкретних характеристиках, які визначаються через точки зору (аспекти розгляду). Це надає можливість використовувати аспекти або точки зору для того, щоб формулювати, розв'язувати та керувати складними та взаємозв'язними ситуаціями, які можуть базуватися не тільки на знаннях окремої предметної області, а на деякій сукупності проблемних областей. При цьому необхідно розуміти відношення між елементами середовища прийняття рішень.

Тому є актуальною є задача розв'язання проблемних ситуацій з використанням відповідних інтелектуальних засобів, що розроблені на принципах інженерії знань для сукупності певних проблемних областей. При цьому необхідно використовувати та враховувати когнітивні знання («знаю, що»); прикладні знання застосування («знаю, як»); системне розуміння («знаю, чому»); особисту мотивацію («хочу знати, чому»). Для цього всі знання, що використовуються, розглядаються в розрізі знань, що описують контент, та знань, що описують контекст.

З іншого боку, прийняття рішень у системах управління описується взаємозалежними задачами. При чому, як правило, такі задачі виявляються несумісними через їхню структуру, що склалася, та обмежуючі фактори, так званими «вузькими місцями», до яких відносять вимоги до функціонування системи, обсяги фінансування; наявність достатніх людських ресурсів, виробничі можливості підприємств, нормативні чи фактичні годинні етапи життєвого циклу виробництва продукції тощо. При чому прийняття рішень в таких задачах вимагає врахування таких особливостей, системності, альтернативності, неспільності (суперечності), багатокритеріальності, врахування думок аналітиків та експертів. Застосування традиційних методів для розв'язання таких задач у класичній постановці, тобто знаходження розв'язання в незмінній протягом рішення моделі, вимагає внесення всіх варіацій параметрів (нових технологій, додаткових ресурсів) до початкової постановки, а це веде до надмірної розмірності задачі, і, отже, складнощів розв'язання задачі і неможливості отримання рішення за прийнятний час і прийнятної точності.

Таким особливостям задач прийняття рішень задовольняє технологія системної оптимізації, яка була запропонована В. М. Глушковим [1]. Суть якої полягає в цілеспрямованій зміні моделей прийняття рішень для досягнення спільності й у виборі найбільш прийнятного рішення поставленої задачі, що формулюються як задачі багатокритеріального лінійного програмування та для різних видів припустимих варіацій параметрів.

Створення різних засобів підтримки прийняття рішень — це безперервний процес формування, уточнення вимог та розв'язання. При цьому необхідно враховувати, що функціонування систем відбувається в умовах інформаційної та реалізаційної неоднорідності, розподіленості та автономності інформаційних ресурсів системи. Інформаційна неоднорідність ресурсів полягає в різноманітності їхніх прикладних контекстів. Реалізаційна неоднорідність джерел проявляється у використанні різноманітних комп'ютерних платформ, засобів управління базами даних, моделей даних і знань і таке інше. Таким чином, потрібна підтримка розвитку систем та підсистем до складніших, інтегрованих систем, що базуються на інтероперабельній взаємодії компонентів. Реалізація такої підтримки прийняття рішень базується на підтримці повного циклу прийняття рішень для того, щоб пройти від формулювання проблеми, визначення відповідних моделей та алгоритмів розв'язання до використання розв'язувача, вимагає застосування знань для прийняття рішень для конкретної задачі з врахуванням навичок та досвіду користувача.

Метою роботи є представлення онтологокерованої підтримки прийняття управлінських рішень на основі методів та алгоритмів системної оптимізації й онтологічних методів представлення та обробки знань з урахуванням контекстів розв'язання задач прийняття рішень.

Для врахування цих особливостей і властивостей прикладних систем управління та багатьох інших вимог, що виникають в процесі функціонування різних систем управління, потрібна побудова єдиної технології прийняття рішень, що дозволяє виробляти найбільш прийнятні рішення. При цьому вибір того або іншого рішення не повинен порушувати системність розгляду і цілісність процесу прийняття рішень. Слід зазначити, що сучасне розв'язання задач вимагає використання інформації різного походження.

Це визначає необхідність зрозуміти складність проблеми, взяти до уваги різноманіття оточуючого світу та науковий розгляд про-блеми, поєднати абстрактне і конкретне знання, розвивати знання та діяльність в напрямку досягнення результатів. При цьому необхідно враховувати, що використання інформації та знань у процесі прийняття рішень, як правило, відбувається в контексті складної структури процесу прийняття рішень, який часто формується за допомогою ряду чинників. Такі системи з точки зору прийняття рішень включають:

• наявність складного змістовного об'єкта (системи), з яким пов'язана загальна проблема (задача) прийняття рішення;

• розбиття даної системи на взаємозв'язані підсистеми, з відповідною декомпозицією загального завдання на підзадачі;

• наявність спільної мети при розподілі функцій по підсистемах;

• фізична або віртуальна відособленість кожної з підсистем; можливість відносно самостійного вибору своїх станів;

• наявність засобів обміну станами між підсистемами, а також засобів узгодження, подолання протиріч і синхронізації процедур розв'язання підзадач.

В роботі [2] пропонується розглядати прийняття рішень в рамках трьох етапів: аналіз, розробка та вибір. Аналіз (опис системи, розуміння поведінки системи, оцінка поточної ситуації, формулювання цілей) включає в себе пошук середовища для умов виклику прийняття рішення. Розробка (формулювання моделі, генерація альтернатив) належить до створення, розробки та аналізу можливих варіантів дій, в той час як вибір (оцінка впливу альтернатив, оцінка та прийняття рішення, пояснення: візалізація та спілкування) включає в себе вибір напрямку дій з наявних. Основною частиною підтримки прийняття рішень є збір, оцінка, організація та перетворення цієї інформації в форми, що придатні для аналізу.

Для успішної розробки та впровадження систем підтримки прийняття рішень (СППР) необхідно [3; 4]: участь кінцевих користувачів в розробці СППР; проектування СППР для потреб кінцевих користувачів, а не потреб, як їх розуміє розробник; гнучкість, адаптивність та оновлюваність системи; простий інтерфейс, який вимагає обмеженого часу для навчання користування системою; візуальне відображення результатів; врахування факторів, що стосуються якості системи, якості інформації та представлення інформації.

Область прийняття рішень будемо розглядати як багаторівневу структуру, що включає область проблем, область моделей, область методу та область реалізацій. Область прийняття рішень можна деком-позувати на елементарні об'єкти, кожен з яких описується сукупністю атрибутів. В рамках такого розгляду необхідно визначити поняття та конструкції, за якими можуть бути визначені природа, структура та представлення процесу формування та прийняття рішень та відповідних складових областей, що описують такий процес.

В роботі під прийняттям рішень будемо розуміти інтерактивний процес нагромадження, обробки, використання та поширення знань, що дає можливість обміну інформацією, знаннями і досвідом та підвищення рівня інформованості, можливість отримання та вироблення усвідомленого вибору між альтернативними рішеннями, можливість підвищення фахового рівня. Метою такого процесу є розв'язання проблем: надання знанням доступності та корисності, так, що відповідні достовірна інформація та знання, що впливають на прийняття рішень та розуміння проблеми, будуть донесені відповідному користувачу в відповідному форматі в відповідний час. Таким чином, підтримка прийняття рішень, що реалізована як певна система прийняття рішень, повинна відповідати SMART-критеріям, тобто рішення мають бути конкретними, вимірними, погодженими, реалістичними, чітко прив'язаними до часу та простору.

Будемо розуміти під підтримкою прийняття рішень інтелектуальну комп'ютерну технологію посилення можливостей ЛПР у процесі спостереження за станом предметної області, діагностики проблемних ситуацій і цілей дій, планування дій і генерацію способів їх реалізації, формування раціональних варіантів рішень з використанням експертних знань і методів моделювання та оптимізації. При цьому прийняття рішень реалізується на основі інформаційних моделей даних; на основі логічних моделей; на основі формальних моделей; на основі типових рішень або прецедентів. Під основними етапами прийняття рішень будемо розглядати:

• моніторинг і збір достовірних даних про процеси функціонування системи;

• розпізнавання, прогнозування розвитку й оцінка штатних і критичних ситуацій, що мають місце у діяльності системи;

• постановку цілей і пошук альтернативних дій з їх досягнення в умовах ситуацій, що складаються в підсистемах підприємства і в системі в цілому;

• адекватну оцінку можливих способів дій, аналіз наслідків і вибір найбільш ефективних з них з комплексним аналізом всього спектру характеристик альтернативних рішень;

• організацію виконання рішень, що включає оцінку і вибір напрямів робіт з реалізації рішень, конкретних заходів і термінів, розподіл ресурсів для реалізації рішень;

• контроль виконання рішень на основі оцінки і порівняння станів і результатів (проміжних у зіставленні з бажаними кінцевими) діяльності, оцінку якості рішень, що приймалися, і правильності організації їх вироблення.

Прийняття рішень можна представити у вигляді багаторівневої системи, що складається з сукупності завдань, що знаходяться на різних рівнях ієрархії та відповідають за певну функцію чи діяльність та пов'язані з відповідною логічною структурою. Прийняття рішень в такій системі будемо розглядати як через горизонтальні перехресні вузли (перетину кордону), так і через вертикальні перехресні ієрархічні зв'язки (перетин ієрархічних рівнів), при цьому можливо отримання раніше недоступної інформації, що в подальшому дає змогу розвивати нові знання та розуміння. Кожна задача, що відповідає конкретному напрямку(ам) діяльності, може мати підзадачі. Задача та підзадачі описуються відповідними формалізованими завданнями, які описуються комплексами взаємопов'язаних моделей. Формалізовані моделі реалізуються певними методами, алгоритмами. Сам процес будемо розглядати як систему, яка складається з деякого набору підсистем (етапів) та їх елементів (процедур, дій, операцій), які взаємодіють між собою, кількість та склад яких можуть змінюватись у залежності від умов та розв'язуваних завдань. При цьому інтеграція рішень, що приймаються, в рамках підсистем досягається за рахунок прийняття узгоджених рішень у завданнях, а інтеграція управління усією системою в цілому буде отримана шляхом узгодження дій між пов'язаними підсистемами, що належать одному або різним рівням.

Визначимо, що між різними підсистемами, функціональними задачами (підзадачами), моделями можливі різні види взаємодії. Така взаємодія може реалізовуватися через відношення прямого підпорядкування, інформаційного обміну, функціонального підпорядкування, функціонального узгодження і координації. Відношення прямого підпорядкування і функціонального підпорядкування є базою для опису побудови системи управління за організаційною та функціональною ознаками. Ці відношення можуть бути задані при визначенні ієрархії функціональних задач, що розв'язуються окремими підсистемами, та пріоритетів їхньої взаємодії. Відношення ін-формаційного обміну визначається при описі взаємодій окремої підсистеми (задачі) з іншими підсистемами (задачами) у рамках цілісної системи. Це може бути задано при визначенні для даної підсистеми деякої нормативно-довідкової інформації, яку використовує у процесі реалізації своїх функцій дана підсистема або задача. Відношення функціонального узгодження і координації визначається при описі функціональних задач підсистеми. Це відношення задається вхідними і вихідними параметрами функціональних задач підсистеми, а також ресурсами підсистеми. Відношення функціонального узгодження і координації реалізується в процесі розв'язання конкретних взаємозв'язаних задач, що виникають всередині системи управління. Тип зв'язків між окремими підсистемами визначається в процесі опису організаційно-функціональної структури системи управління.

При реалізації прийняття рішень будемо розрізняти три стратегії прийняття рішень. Це — створення, інтеграція та адаптація. Створення означає «абсолютно нову проблему», або «на порожньому місці» концепцію розв'язання проблеми в ситуації, коли не існує відповідної моделі, методу та/або алгоритму розв'язання, що могло би використовуватися як основа для прийняття рішень. Це важливо, якщо деяка частина з процесу прийняття рішень має бути спроектована без підтримки існуючого. Інтеграція означає концепцію розв'язання проблеми, згідно з якою побудовано процес розв'язання проблеми, збираючи компоненти з існуючого. Чим більше компонентів багатократного використання, з яких складений процес прийняття рішень, тим легше процес інтеграції. Адаптація означає концепцію процесу прийняття рішень, згідно з якою побудовано розв'язання проблеми, знижуючись або змінюючи деяку частину(и) існуючого, або розширюючи існуюче деякою новою частиною(ами).

Розв'язання задач на основі системної оптимізації можна представити в вигляді послідовності людино-машинних процедур, що включають формування моделі початкової задачі в термінах предметної області, переведення сформованої моделі в область задач, наприклад, математичного програмування, та розв'язання задачі математичного програмування в багатокритеріальній постановці.

При цьому рішення задачі складається з перевірки здійснимості вимог за якістю функціонування системи (директивні вимоги) в області власних можливостей системи, і в разі їх нездійсненності — знаходження «вузьких місць», вироблення заходів, спрямованих на усунення нездійсненності директивних вимог, і в виборі найбільш прийнятного рішення. Таким чином, видно, що системна оптимізація дає можливість представлення рішення досить складних задач у вигляді послідовності рішення простіших задач.

При реалізації системної оптимізації враховується те, що обмін інформацією про рішення, які приймаються, здійснюються між задачами (етапами) з чітко вираженими зв'язками та існує пріоритет в прийнятті рішень між задачами (етапами) як з точки зору правил їх взаємодії, так і часу їх виконання.

1) Аналіз виконання управлінських рішень, отримані з задачі (етапу), що має більший пріоритет за взаємодією чи за часом розв'язання, або заданих власними цілями з функціонування даної задачі (етапу), в рамках існуючих можливостей задачі (етапу), що розв'язується.

У разі неможливості реалізації цих рішень задачею (етапом) потрібно виконати такі кроки:

• підготувати пропозиції щодо бажаної та можливої зміни отриманих рішень;

• ініціювати процес взаємодії з відповідними задачами (етапами).

2) Пошук управлінських рішень з урахуванням власних можливостей завдання (етапу) й існуючих взаємних зв'язків з іншими задачами (етапами).

3) Визначення завдань наступного завдання (етапу) для реалізації отриманих рішень.

В ході процесу взаємодії при пошуки узгоджених рішень структура взаємодії, тобто множина взаємопов'язаних задач (етапів) і відповідних осіб, які приймають рішення, не може бути задана заздалегідь. Ця структура генерується безпосередньо під час пошуку рішення.

У відповідності з цим інструментарій повинен забезпечити:

• передачу і прийом управлінських рішень або генерацію власного напрямку(ів) рішення поставленого завдання (такі рішення будемо називати директивними);

• модельне представлення власних можливостей і інтересів даного завдання (етапу), директивних рішень і взаємних зв'язків з іншими задачами (етапами);

• ініціювання процесу взаємодії з відповідними особами, які приймають рішення;

• формування модельного представлення задач пошуку узгоджених рішень з урахуванням можливостей даного завдання (етапу);

• методи і алгоритми визначення розв'язку сформованих задач.

Такий розгляд дозволяє запропонувати підхід до реалізації взаємодії між підсистемами і відповідно певними задачами (та надалі його використати при розгляді певних моделей, формалізованих задач тощо), що базується на понятті відношення пріоритету взаємодії. Це відношення на множині взаємопов'язаних локальних задач визначає характер впливу відповідних підсистем і задач один на одного. Таке відношення може бути розглянуте як відношення нестрогого порядку $R$, визначене таким чином: якщо $K$ — множина взаємопов'язаних задач, то $iRj$ ($i, j \in K$) означає, що задача $i$ має пріоритет у прийнятті рішення по відношенню до задачі $j$, тобто її рішення є обов'язковими (директивними) для задачі $j$ і входять в її модель як деякі параметри. Це відношення визначає відношення пріоритету взаємодії. Оскільки $R$ є відношенням нестрогого порядку, то воно розбиває множину взаємопов'язаних задач на класи еквівалентності, які розглянемо в подальшому. Опишемо реалізацію відношень. Для цієї мети ми введемо такі позначення (без врахування структури системи): $I$ — множина підсистем системи, $I_l$ — множина задач для $l$-ї структурної одиниці (підсистеми);

Припустимо, що задачі прийняття рішень в структурній одиниці структуровані, а задача вибору рішення у всій системі в цілому має бути сформульована через інтеграцію розподілених підсистем і відповідно через інтеграцію задач, що реалізуються в підсистемах.

Для формалізації специфічних проблем інтеграції функціональних задач введемо позначення: $x^p$ — вектор, що визначає вибір дії в $p$-й функціональній задачі; $s^p = \{s^p_j, j \in J_p\}$ — вектор, що визначає вплив (відношення функціонального узгодження і координації) інших функціональних задач, які описують множину $J_p, J_p \in I$, на $p$-ту функціональну задачу; $z^p = \{z^p_j, j \in Z_p\}$ — вектор, що визначає відношення інформаційного обміну інших функціональних задач, які описують множину $Z_p, Z_p \in I$, з $p$-ю функціональною задачею; $u^p = \{u^p_{ji}, j \in J^p, i \in I_l\}$ — вектор директивного впливу (підпорядкування) через пріоритет взаємодії на $p$-ту функціональну задачу, який може визначати дію як інших функціональних задач $l$-ї підсистеми, так і інших функціональних задач підсистем, що мають директивні взаємозв'язки з цією задачею, що визначають множину $J^p$, $J^p \in I$.

При цьому визначимо вектори: $s_p = \{s^j_p, j \in J^v_p\}$, $z_p = \{z^j_p, j \in Z^z_p\}$, $u_p = \{u^j_{pi}, j \in J^p_u, i \in I_l\}$, які описують відповідні вектори та множини, що визначають вплив, інформаційний обмін та директивний вплив даної задачі на інші задачі.

В рамках СППР взаємодія між множиною підсистем багаторівневої системи та множинами супутніх задач в цих підсистемах, які відповідають за різні напрямки діяльності системи, реалізується через принцип системності.

Принцип незалежності в функціонуванні СППР підсистем базується на тому, що кожна підсистема за своєю функціональною задачею може робити свій вибір власної дії, яка описується вектором $x^p$, у відповідності зі своєю власною моделлю вибору. Така дія дасть змогу розв'язання задачі самоуправління з виконання підсистемою своєї функціональної діяльності. Проте принцип цілісності вимагає побудови такої моделі задачі вибору рішень, область припустимих рішень якої враховувала б вплив підсистем та функціональних задач.

З цією метою введемо такі типи припустимих областей задач прийняття рішень в СППР: $D_0(z^p)$ — область, що визначає область вибору припустимих рішень (дій) на підставі власних можливостей відповідної підсистеми в $p$-й функціональній задачі при виборі власних рішень $x^p$ з урахуванням вектора $z^p$; $D_0(s^p)$ — область припустимих рішень $x^p$ $p$-ї функціональної задачі, що визначається вектором $s^p$; $D_0(u^p)$ — область припустимих рішень $x^p$, що описує директивну область, утворену вектором $u^p$.

Як відомо, більшість задач прийняття рішення розв'язується при врахуванні деякої множини $J = \overline{1, M}$ характеристик оцінки рішення $x^{p^*}$, яке може визначитися в кількісній або якісній шкалі за допомогою множини критеріальних функцій $f = \{f_i(x^{p^*}), i \in J\}$, при цьому критерії можуть носити як кількісний, так і якісний характер.

Перш за все припустимо, що $D_0(z^p) \bigcap D_0(s^p) \bigcap D_0(u^p) \neq \varnothing$. В даному випадку ми зможемо розв'язати задачу та знайти $x^{p^*}$ у врахуванням множини критеріальних функцій $f$ та припустимих областей.

Якщо $D_0(z^p) \bigcap D_0(s^p) \bigcap D_0(u^p) = \varnothing$ та $D_0(z^p) \bigcap D_0(u^p) = \varnothing$, то область $D_0(u^p)$ не має припустимих рішень з областю $D_0(z^p)$ з врахуванням впливу інших задач, і нам необхідно змінювати область $D_0(z^p)$ за рахунок можливостей задачі з метою одержання сумісності з областю $D_0(u^p)$ або у разі неможливості досягнення сумісності формувати обмеження на вектор $u^p$ з метою інформування більш пріоритетної задачі про неможливість розв'язання задачі.

Якщо $D_0(z^p) \bigcap D_0(s^p) \bigcap D_0(u^p) = \varnothing$ і $D_0(z^p) \bigcap D_0(u^p) \neq \varnothing$, то область $D_0(s^p)$ не має припустимих рішень з областю $D_0(z^p)$ з урахуванням директивного впливу, і нам необхідно формувати вектор $s^p$ для множини $J_p, J_p \in I$ з метою реалізації процедури узгодження рішень взаємопов'язаних задач.

Зазначимо, що модель задачі, метод та алгоритм розв'язання задач можуть не тільки бути з області математичного програмування, а і описуватися в області інформаційних та логічних моделей.

У випадку реалізації стратегії створення формалізований опис деякої локальної задачі, що формулюється як задача математичного програмування та розв'язується в системі підтримки прийняття рішень. Оскільки будь-яка математична модель задачі прийняття рішення включає декілька критеріїв оптимальності і системи обмежень, що описує множину припустимих альтернатив, то всі види впливу на цю модель можуть бути зведені до впливу на критерій і впливу на обмеження. Розглянемо останній з них. В цьому випадку множина обмежень локальної задачі включатиме обмеження, що описують зв'язки з іншими задачами, і обмеження, що описують локальну область припустимих рішень.

Розглянемо модель локальної задачі прийняття рішення у багаторівневій організаційній системі, яка має загальний вигляд:

$M_i = \{C^{xi}, X_0, X(u^{i-1}), X(u^i), U(x), U(x^{i+1})\}$, де $i$ — індекс задачі, що розглядається ($i \in I = \overline{1, M}$), $C^{xi} = \{C_j^{xi}(x) \rightarrow extr, j \in J^i = \overline{1, N}\}$ — множина оцінок вибору рішення задачі $M_i$;

$X_0^i$ — область можливих рішень, що визначається локальними обмеженнями задачі $M_i$ (область $D_0(z^p)$);

$X(u^{i-1})$ — область бажаних рішень, яка визначається обмеженнями, які називають директивними (область $D_0(u^p)$);

$X(u^i)$ — область рішень, яка визначається з урахуванням компромісних зв'язків із завданнями, які володіють однаковими з даним завданням пріоритетами взаємодії (область $D_0(s^p)$);

$U(x)$ — область змінних $u$, яка залежить від рішення $x^*$ даної задачі (вектор $s_p$);

$U(x^{i+1})$ — область змінних, що характеризують вплив даної задачі на пов'язані з нею завдання з меншим пріоритетом взаємодії (вектор $u_p$).

Наявність у завданнях прийняття рішення локальних цілей та пріоритетів взаємодії призводить до різних ситуацій взаємодії між від-повідними завданнями. Ці ситуації визначаються взаємним розташуванням областей відносно одна одної.

Таким чином, процес прийняття рішень може складатися з послідовності етапів, кожен з яких включає такі елементи:

1. визначення рішень локальних задач з урахуванням результатів, отриманих на попередніх етапах;

2. узгодження рішень пов'язаних локальних задач.

Перший етап полягає в аналізі моделей локальних задач. Якщо припустимих рішень в локальній задачі не існує, то виникає необхідність у цілеспрямованій зміні області $X_0^i$ для виконання директивних вимог, що визначаються областю $X(u^{i-1})$, де $u^{i-1}$ отримано при розв'язанні більш пріоритетних задач. Така задача корекції $X_0^i$ інтерпретується як задача системної оптимізації [5].

Таким чином рішення локальної задачі $y = (x, u^{i-1}, u^i)$ (локальне припустиме рішення) буде знайдено безпосередньо або буде отримано в результаті розв'язання задачі системної оптимізації, тобто $X_0 \bigcap X(u^{i-1}) \neq \varnothing$.

Оскільки рішення $y$ визначено без врахування області зв'язків $X(u^i)$, то значення параметра $u$ визначені незалежно в кожній із пов'язаних задач і можуть не збігатися. Тоді узгодження рішень полягає у знаходженні таких локально допустимих (оптимальних, компромісних) рішень, для яких значення параметрів зв'язку рівні. Можливі підходи до реалізації алгоритмів узгодження рішень по параметрах зв'язку наведено в [6].

У разі відсутності таких узгоджених рішень необхідне корегування моделей пов'язаних задач для досягнення сумісності в просторі параметрів $u$, яка може бути зведена до задачі системної оптимізації. Основною проблемою при цьому є вибір напрямку і величини корегування областей $X_0^i$, $X(u^{i-1})$. Отримане рішення $y$ визначить значення параметра $u^{i+1}$, що характеризує вплив даної задачі на пов'язані з нею задачі з меншим пріоритетом.

Розглянемо формулювання локальної задачі як задачу лінійного математичного програмування, що розв'язується в системі підтримки прийняття рішень. В цьому випадку множина $C^{xi}$ задається через деяку множина критеріальних функцій $f^i = \sum_{j=1}^{n} c_{ij} * x_j \rightarrow extr, i \in I$; область директивних вимог $D^g$ щодо функціонування системи управління визначається множиною $D_0^g = \{x : x_j = x_j^{*(g)}, j = \overline{1,n}\}$, або областю $P^g = \{x : x_j^b \le x_j \le x_j^u, j = \overline{1,n}\}$, або областю $D_2^g = \{x : \sum_{j=1}^{n} a_{ij} * x_j \le u_i^0, i \in Q^g\}$; область припустимих рішень описується множиною $D^0 = \{x : \sum_{j=1}^{n} b_{ij}^0 * x_j \le b_i^0, i \in Q, x_j \ge 0, j = \overline{1,n}\}$.

Згідно з методологією системної оптимізації, необхідність у розв'занні задачі системної оптимізації виникає у разі неспільності області директивних вимог $D^g$ та області припустимих рішень $D^0$. Основна мета алгоритмів системної оптимізації полягає в побудові нової області припустимих рішень відповідно до первинної області $D^0$ і додаткової обмеженої області варіацій $\Delta b_{ij}, \Delta b_i, i \in Q, j = \overline{1,n}$ параметрів $b_{ij}, b_i$, що будується в процесі рішення з урахуванням того, узгоджуються директивні вимоги й інтереси даної системи управління $D^g$ чи ні, в якій існуватимуть рішення із значеннями по всіх критеріальних функціях більшими або рівними значеннями критеріїв, що задаються вимогами людини, що приймає рішення (ЛПР), в області директивних вимог $D^g$. Ці алгоритми носять ітераційний характер.

Збіжність процедур в рамках системної оптимізації реалізується через ітераційне відсікання неприпустимих варіантів рішення, при цьому гарантується, що припустимі варіанти не будуть відсікатися. Що в підсумку дає нам можливість або отримати рішення поставленої задачі, або зробити висновок про неможливість розв'язання поставленої задачі.

Перевірка виконання вимог $D^g$ в області $D^0$ проводиться або прямою підстановкою $x = x^{*(g)}$ в систему обмежень області $D^0$ при $D^g = D_0^g$ або на основі якого-небудь методу лінійного програмування для областей та у разі порожнього перетину $D^g$ і $D^0$ можливі різні випадки взаєморозташування області $D^g$ та області $D^0$ щодо критеріальних функцій, які в загальному випадку можуть бути такими:

1. Всі точки директивної області $D^g$ мають кращі значення за всіма критеріями в порівнянні зі значеннями, що досягаються у відповідних їм за перевагою точках області $D^0$, тобто повне узгодження.

2. Для будь-якої точки з директивної області $D^g$ в області припустимих рішень $D^0$ існує точка з кращими значеннями за всіма критеріями одночасно, тобто директивні вимоги не узгоджуються з цілями даної системи, що задані набором критеріальних функцій.

3. Тільки частину точок директивної області $D^g$ дає поліпшення значень за всіма критеріями одночасно, тобто вимоги лише частково узгоджуються з цілями даної системи.

У разі виконання другого варіанту ЛПРу потрібно перевизначити область директивних вимог $D^g$, оскільки в області $D^g$ не будуть виконуватися вимоги до значень критеріальних функцій.

З урахуванням інших варіантів розташування, а також з урахуванням вигляду директивних областей можна виділити обмеження області, які перешкоджають спільності й узгодженості рішень з $D^g$ і області припустимих рішень $D^0$ і які називають суттєвими обмеженнями. Множину індексів суттєвих обмежень позначимо як $Q_0$.

У разі директивної області вигляду $D_0^g$ суттєвими обмеженнями будуть співвідношення, що порушуються при підстановці $x = x^{*g}$.

При виділенні суттєвих обмежень можна розглянути випадки з урахуванням заданої множини критеріїв і без урахування цільових функцій.

При врахуванні критеріїв після з'ясування реалізованого варіанту узгодження виділяється множина точок, яку будемо називати областю захоплення і яка апроксимує область $D^g$ при реалізації першого варіанту або будується область захоплення $X^g \subseteq D^g$ (яка містить точки, що мають кращі значення за всіма критеріями одночасно в порівнянні з розв'язками області $D^0$) при реалізації третього варіанту.

Якщо ж вирішується задача системної оптимізації без урахування множини критеріальних функцій, то визначення суттєвих обмежень залежить від вибраної ЛПРом області захоплення $X^g$, $X^g \subseteq D^g$, яка може описуватися паралелепіпедом або деякою областю або точкою $x^{*'}, x^{*''} \in D^g$. При цьому задачу системної оптимізації будемо розв'язувати відносно деякої точки $x^{*g}, x^{*g} \in X^g$, яка є вершиною відповідного багатокутника. Отримана область захоплення $X^g$ визначить відповідні множину точок і множину індексів суттєвих обмежень, щодо яких будемо розв'язувати задачу системної оптимізації.

Для приналежності заданої області захоплення $X^g$ змінній моделі будується система обмежень, що описує область $P$.

Для визначення можливості зміни параметрів моделі задачі для досягнення вимог з $D^0$ і отже можливості рішення самої задачі сис-темної оптимізації, так і сформованої локальної задачі, вихідної побудуємо перетин області $P$ варіації параметрів обмежень множини $Q_0$ і області $P_0$ припустимих варіацій цих параметрів.

Якщо $P \cap P_0 \neq \varnothing$, то область зміни параметрів моделі буде обмежена і це дозволить вирішити задачу побудови нової моделі $D_1$, в якій виконуються вимоги з області $D^0$.

В цьому випадку необхідно або змінювати чи перевизначити ЛПРом свої вимоги або обмеження $P_0$.

Для корекції області припустимих варіацій $P_0$ можна зокрема побудувати паралелепіпед, який вписано в область $P$, на основі якого ЛПР зможе задати або відкорегувати обмеження області $P_0$, так що $P \cap P_0 \neq \varnothing$.

Задача вибору варіацій параметрів $\Delta b_{ij}, \Delta b_i, i \in Q, j = \overline{1,n}$ при непорожньому перетині областей $P$ і $P_0$ зводиться до задачі оптимізації, в якій як критерії вибрані витрати, що пов'язані зі змінами параметрів моделі $C(\Delta B, \Delta b)$.

Якщо функцію витрат побудувати неможливо, то задача вибору формується як багатокритеріальна задача, в якій кожен параметр виступає як окремий критерій і залежно від фізичної суті може максимізуватися або мінімізуватися.

Таким чином, нова область допустимих рішень згідно з умовами побудови в новій області $D_1$, забезпечується здійсненість вимог задачі і існують розв'язки зі значеннями критеріїв не гірше бажаних.

Умови розв'язання задачі системної оптимізації дозволяють алгоритму сходитися до відповідного ефективного рішення задачі. Цей процес ітераційний. Збіжність процедур у рамках системної оптимізації реалізується через ітераційне відсікання недопустимих варіантів рішення, при цьому гарантується, що допустимі варіанти не будуть відсікатися. Це в підсумку дає нам можливість отримання рішення поставленого завдання або дає можливість зробити висновок про неможливість вибору варіанту інвестиційного проекту або неможливість реалізувати даний проект.

У разі реалізації стратегії адаптація та інтеграції можна використати апарат прецедентів, який допомагає визначити рішення для поточної ситуації на основі прецедентів, які вже мали місце у минулому при розв'язанні подібних задач. В загальному випадку прецедент може включати такі компоненти: опис задачі (проблемної ситуації); рішення задачі (діагноз із проблемної ситуації і рекомендації ЛПР), результат (або прогноз) застосування рішення, результат використання знайденого рішення.

До основних переваг технології прецедентів можна віднести можливість безпосередньо використовувати досвід, накопичений системою, без інтенсивного залучення експертів в тій чи іншій предметній області, а також можливість виключення отримання помилкового рішення. Істотними недоліками цього підходу є зниження продуктивності системи при великій кількості прецедентів у базі прецедентів і неможливість отримання рішення задач, для яких немає прецедентів у бібліотеці прецедентів (БП) системи.

В цьому випадку процес підтримки прийняття рішень формально представляється як: $СППР = <KB\{Rule, Case\}, M, S(M), Dec>$, де $KB\{Rule, Case\}$ — база знань, що містить множину правил $Rule$ та множину $Case$ прецедентів. $Case = (x_1, x_2, ..., x_n, Sol)$, де $x_1, x_2, ..., x_n$ — параметри ситуації, яка описує даний прецедент, $N$ — кількість параметрів для опису прецедента, а $X_1, X_2, ..., X_n$ — області допустимих значень відповідних параметрів, $Sol$ — рішення (діагноз, рекомендації ЛПР). $Rule = \{R_i\}$, де $R_i$ — $i$-те правило, $i = \overline{1, ..., I}$. Правила $R_i \in Rule$ визначаються в такій формі: $<A, a_1, U_1, ..., a_n, U_n; P_1, ..., P_m; b, U_b, S, S'>$, де $a_i \in A$ є передумови проблемної ситуації (ПС); $U_i \in U$ — необхідні оцінки міри упевненості в передумовах; $P \in P^V$ є предикати, $V \geq 1, m \geq 0$; $b \in B$ — укладення с оцінкою міри упевненості $U_b$; $S$ — початкова проблемна ситуація (ПС); $S'$ — проблемна ситуація, що виникає в результаті прийнятого рішення. $M = \{M_1, ..., M_N\}$ — множина моделей, що реалізують функції процесу прийняття рішень; $S(M)$ — модуль, що реалізує функцію вибору необхідної моделі (моделей) для даної задачі; $Dec$ — модуль формування рішень на основі бази знань. Це дозволяє реалізувати званий цикл прецедентів або цикл навчання за прецедентами, що включає такі етапи:

• витягання найбільш відповідного (подібного) прецедента (прецедентів) для проблемної ситуації, що склалася, з бібліотеки прецедентів (БП);

• повторне використання витягнутого прецедента для спроби розв'язання поточної проблеми;

• перегляд та адаптація в разі потреби отриманого рішення відповідно до поточної проблеми;

• збереження знову прийнятого рішення як частини нового прецеденту.

Пошук рішення на основі прецедентів полягає в визначенні міри схожості поточної ситуації з ситуаціями прецедентів з БП. При цьому враховуються ваги параметрів для ситуацій з БП, задані експертом. Міра схожості залежить від близькості поточної ситуації до ситуації прецедента і визначається за допомогою алгоритму пошуку найближчого сусіда за допомогою простого зіставлення поточної ситуації з ситуацією прецедента (кожен параметр для опису ситуацій з БП розглядається як одна з координат). В результаті визначається відстань $D$ між поточною ситуацією і ситуацією прецедента і максимальна відстань $D_{max}$ на основі меж діапазонів параметрів для ситуацій прецедентів. Потім обчислюється значення міри схожості $sim = 1 - D / D_{max}$. Для цього можна використати метод найближчого сусіда (найближчих сусідів) (К найближчих сусідів).

Необхідно враховувати, що міркування на основі прецедентів може не привести до необхідного рішення проблемної ситуації, що виникла, наприклад, у разі відсутності подібної (аналогічної) ситуації у БП. Ця проблема може бути розв'язана через можливість поповнення БП безпосередньо в процесі міркування (висновку). Це може включати: визначення існуючої практики, яка є прийнятою та реалізованою; використання часткової стратегії автоматизації узгодження розрізненої інформації з декількох джерел інформації на основі математичних моделей та експертних систем; використання стратегічної мети усунення невизначеності, неповноти інформації та врахування суб'єктивної експертної інформації від декількох джерел інформації.

Як показує вищепредставлене, процес прийняття рішень складається з декількох етапів. На кожному етапі розв'язуються свої задачі. Задача приймає вхідні дані і виробляє певний результат. Вхідними для задач є ситуації, кожна з яких представляє собою множину пов'язаних відношеннями об'єктів предметної області. Виділяється клас проблемних ситуацій, тобто ситуацій, в яких значення атрибутів деяких об'єктів виходять за область нормальних значень або критично близько підходять до її кордонів. Результатом рішення задачі може бути повідомлення, ситуація або задача. Повідомлення — це остаточний результат рішення задачі, який користувач приймає до відома. Ситуація — це результат, який може бути підданий подальшому аналізу. Ситуація може являти собою наслідки прийнятих рішень або ж початкові рішення, які повинні привести до бажаних результатів. Якщо в якості розв'язання отримано кілька ситуацій — альтернатив, то може бути згенерована нова задача, яка буде оцінювати отримані альтернативи і вибирати з них найбільш прийнятні. Для розв'язання задач використовуються різні методи підтримки прийняття рішень. Деякі з них можуть мати комп'ютерну реалізацію, тобто можуть бути реалізовані в деякому програмному модулі, який, у свою чергу, інтерпретується тим чи іншим розв'язувачем. Інші методи можуть не мати програмної підтримки. В цьому випадку використовується текстовий (можливо, формалізований) опис методу, а в якості розв'язувача, інтерпретуючого такий модуль, виступає людина — учасник процесу прийняття рішень. Учасниками можуть бути ЛПРи, власники проблеми, різні активні групи, експерти та фахівці з прийняття рішень.

Таким чином, представлення знань про розв'язання задачі за допомогою технології системної оптимізації необхідно описати:

• моделі, що описують вихідне завдання та виникають у процесі реалізації технології системної оптимізації;

• методи та алгоритми розв'язання сформованих моделей;

• процес розв'язання задачі за допомогою технології системної оптимізації. Цей процес реалізується через певні етапи [5].

Таким чином, для реалізації системної оптимізації необхідно описати та використовувати підходи та засоби для:

• формування рішень з урахуванням даних. Тут розглядається область деталізованих даних, тобто пошук інформації з використанням засобів СУБД як в окремих базах даних, так у загальному сховищі даних, область агрегованих показників, тобто збір у сховище даних відповідної інформації, її узагальнення та агрегація, гіперкубічне подання та багатовимірний аналіз (оперативна аналітична обробка даних (OLAP)), сфера закономірностей, тобто пошук функціональних та логічних закономірностей у накопиченій інформації, побудова моделей та правил, які пояснюють знайдені аномалії та/або прогнозують розвиток деяких процесів (інтелектуальна обробка даних (Data Mining));

• формування рішень на основі логічних моделей та правил (прийняття рішень на основі продукційних моделей, семантичних мереж тощо);

• формування рішень на основі математичних моделей (оптимізація через використання аналітичних формул, оптимізація через алгоритми, оптимізація вибору з багатьох альтернатив тощо);

• формування рішень на основі типових рішень або прецедентів (типові рішення та моделі, прецеденти проблемних ситуацій).

При цьому необхідно розглядати різні аспекти прийняття рішень. Такими можуть бути, наприклад, поведінковий аспект (описує ситуації прийняття рішень та порядок, в якому розглядаються завдання та в якому виконуються відповідні дії), організаційний аспект (описує структуру середовища прийняття рішень, ресурси та засоби та визначає організаційну структуру, в якій розв'язання задачі виконується або буде виконуватися, і відносини між елементами структури), інформаційний аспект (описує інформацію, яка використовується при прийнятті рішень, як вона представляється та як вона може застосовуватись).

В рамках такого представлення прийняття рішень необхідно ідентифікувати модель предметної області; визначити взаємодії між задачами та відповідними моделями на підставі відносин пріоритетів взаємодії; визначити усі види впливу даної задачі (моделі) на інші задачі (моделі); визначити всі можливі випадки активізації даної задачі (моделі) як для локальних ситуацій прийняття рішення, так і для розподіленого прийняття рішень; визначити можливі схеми реалізації даної задачі (моделі) в відповідній предметній області; визначити множини даних, на яких реалізується дана задача (модель) і які описують результат розв'язання задачі на вибраній моделі.

При цьому прийняття рішень будемо описувати через три виміри (світи) розуміння процесу прийняття рішень: світ 1: реальний світ (прикладний світ), світ 2: формальний світ (формальні моделі, методи, алгоритми тощо) та світ 3: світ програмного забезпечення (програмні засоби, платформи тощо).

При реалізації прийняття рішень в розрізі моделей реалізується ефект тріади: за допомогою сприйняття та концептуалізації побудувати модель прикладної області (модель представляється з точки зору опису (об'єкти, процеси, відношення, властивості та характеристики) та з точки зору діяльності (визначення процесів, побудова концептуальної моделі) і за допомогою знаків або мови, зробити формалізацію відносин (вплив, регулювання, управління) та створити формалізовану модель, наприклад, символьну модель (модель представляється з точки зору опису, як математична модель, та з точки зору діяльності через визначення структури моделі, оцінку параметрів, достовірні властивості та характеристики). Зв'язок між формальною моделлю та моделлю програмного забезпечення (модель обчислювань, програмні модулі та визначення програмної концепції, узгодження програмних модулів) визначає методи та алгоритми, які необхідні для розв'язання формальної системи.

Основою для використання знань та реалізації процесу прийняття рішень за допомогою системної оптимізації, представлення відповідного інтегрованого середовища прийняття рішень, взаємодії між складовими частинами середовища, опису предметних областей та розв'язання задач в такому середовищі є онтологія, як засіб явного розуміння та представлення областей та процесів прийняття рішень.

Під онтологією [7] будемо розуміти систему, що описує структуру певної проблемної області або множини проблемних областей та складається з множини класів понять, пов'язаних відношеннями, їх визначень та аксіом, що задають обмеження на інтерпретацію цих понять в рамках даної проблемної області або їх множини.

Така онтологія [8] базується на взаємопов'язаній множині онтологій, що представляє собою багаторівневу асоціативну структуру, що включає метаонтологію або онтологію верхнього рівня, базову онтологію, контекстну онтологію, множина онтологій представлення процесу прийняття рішень, що включає представлення задач та їх розв'язання на рівні проблемної області, онтологій предметно-формального та формального представлення та реалізацій цього процесу, онтологію реалізацій, що включає опис програмного забезпечення для підтримки прийняття рішень, онтологію представлення користувача та взаємодії з ним, модель машини виведення, що асоціюється з побудованою онтологічною моделлю. Мета онтології полягає у забезпеченні інтегрованої концептуальної основи для того, щоб вона була визначена, зрозуміла, структурована та представляла явища при прийнятті рішень за допомогою систем підтримки прийняття рішень. Метаонтологія розглядається як засіб інтеграції різних складових реалізації процесу підтримки прийняття рішень та найбільш загального його опису. Сутностями метаонтології є такі поняття, як об'єкт, атрибут, значення, відношення і т. п., наприклад, описувати метаінформацію на основі моделі Захмана. Мета базової онтології полягає у забезпеченні ключових понять та конструкцій для того, щоб визначити, зрозуміти, структурувати та представити основні принципи області прийняття рішень, в рамках якої функціонує СППР. Контекстна онтологія реалізує контекстну систему, що допомагає розпізнати, зрозуміти та представити прийняття рішень через контексти та в межах контекстів. Множина онтологій представлення процесу прийняття рішень розглядається як компонента бази знань при роботі з конкретною проблемною областю та є, у свою чергу, шаблоном для побудови динамічної ком-поненти бази знань, що змінюється при переході від дослідження однієї конкретної задачі до іншої. Онтологія реалізацій, що включає опис програмного забезпечення для підтримки прийняття рішень: функціональний, поведінковий, організаційний та інформаційний. При цьому опис ґрунтується на функціональних (що робить програмне забезпечення) та нефункціональних вимогах (обмеження використання). Онтологія представлення користувача та взаємодії з ним реалізує формування моделі сценарію та компонентів діалогу (автоматично або автоматизовано).

Реалізація онтологічного представлення перш за все базується на визначенні та взаємодії понять та термінів для опису предметних областей та розв'язання певних задач у відповідних предметних областях. Будемо розглядати проблемну область прийняття рішень як множину предметних областей та задач, що розв'язуються в них. Таке онтологічне представлення складається з консолідованого представлення певних проблемних областей, через які прийняття рішень може бути представлене та визначене на основі вибраної точки зору (стан проблеми або проблемної області, поведінка проблеми або проблемної області та розв'язання проблеми).

Поняття та терміни, що стосується проблемної області, включають такі поняття: об'єкт, задача (проблема), модель (формулювання проблеми), методологія (сценарій, метод, алгоритм), система характеристик (властивостей), що їх описують, значення, відношення. Об'єкт — термін або поняття (сутність), що визначається семантичним представленням та з яким пов'язані відповідні властивості, реалізується певний зв'язок з іншим терміном(ами), з задачами та моделями, що ініціювали присутність цього терміна. Задача — кожен екземпляр цього класу визначає задачу для конкретного об'єкта, має ідентифікатор, вказує на об'єкт та термін або властивість та значення властивості, що ініціюють цю задачу тощо. Модель — кожен екземпляр цього класу визначає опис об'єкта на певній мові, зокрема формалізованій, що складений з метою вивчення його властивостей. До такого опису вносяться, наприклад, чинники, що впливають на вибір моделі, такі як період часу, змінні рішення, критерії оцінки, числові параметри та відношення, включаючи математичні. Моделі інтегруються в класи моделей. Є кілька класів моделей для прийняття рішень, які, у свою чергу, можуть бути розв'язані декількома альтернативними методами. Кожен клас моделі краще підходить для представлення певних видів процесів прийняття рішень. Властивість — певна ознака, що характеризує термін, має властивості, аналогічні класу термінів. Окрім цього він вказує на екземпляри класу значень, які визначають його в задачах. Значення — визначають дані, що використовуються при пошуку в семантичному представленні, вказують на задачу та на властивість, рішенням якої вони є. Відношення — визначає зв'язок між двома термінами та вказує на терміни або властивості та об'єкти знань. Інші терміни визначають ті поняття, що пов'язані з системою характеристик (структура, обмеження, середовище, контекст, рівень узагальнення тощо), проблемою (предмет проблеми, проблема верхнього рівня, проблема нижнього рівня, методи (сценарії) рішення, складність проблеми, атомна проблема, складена проблема, опис проблеми, проблемна тема, контекст проблеми, власність проблеми, відповідальність за проблему, оцінка проблеми, проблемна область, вплив на проблеми, вплив з проблем, ініціювання, час, взаємодія, актор тощо), моделлю (мета, обмеження, контекст, проблемна область, проблема, методологія, об'єкти, вхідні параметри, вихідні параметри, інші параметри, умови, тригери (який випадок запускає), передумови (що на початку), післяумови (що в кінці) та пов'язані знанням проблемної області (область знання, функціональні знання, структурне знання, знання обробки тощо). Поняття з прийняття рішень, включаючи системну оптимізацію, пов'язані між собою відношеннями класифікації, узагальнення, агрегації та групування, асоціативними відношеннями, визначення яких здійснюється через представлення проблемної та предметних областей.

В рамках такого підходу прийняття рішень базується на представленні багаторівневої системи управління та розглядається через один або декілька взаємопов'язаних контекстів (модель певного контексту), в яких хтось (актор) щось робить (дія) з деяких причин (цілі) для когось (об'єкт) за допомоги деякого (об'єкта), десь (місце знаходження) та іноді (час). Представлення контексту складається зі змісту, що базується на онтологіях, які охоплюють певну частину моделі контексту.

Для опису контексту необхідно знайти поняття та конструкції, які визначають природу, структуру та представлення процесу формування та прийняття рішень і відповідних складових областей, які описують такий процес. Контекст повинен бути описаний стандартизованим способом. Представлення знань процесу прийняття рішень має підтримувати операції, що необхідні для представлення контексту та управління ним.

Контекст будемо розглядати як концептуальну або інтелектуальну конструкцію, яка складається з понять в межах відповідних контекстних областей та допомагає нам зрозуміти, проаналізувати та використовувати природу, значення та ефекти через елементарні сутності у відповідному середовищі або обставинах. Також контекст представляє ціле, що визначається через певні сутності, які є важливими для даного розгляду.

Це дозволяє розглядати контекст як будь-яку інформацію, яка може бути використана для опису ситуації, в якій щось існує чи відбувається та яка може допомогти пояснити ситуацію та визначити напрямок її розв'язання. Ця ситуація залежить від знань, світогляду, практики та обставин, які можуть бути використані для побудови «нескінченної і частково відомої сукупності припущень» [8], які визначають інтегральне розв'язання проблем та які забезпечують умови для створення, підтримки та застосування знань.

При цьому, по-перше, контекст є невід'ємною властивістю випадків взаємодії, а не є стабільним об'єктивним набором функцій, які зовнішньо характеризують діяльність. Контекст залишається критично важливим для розуміння, контекстуалізації та нерозуміння форм діяльності та інформації, але саме в контексті природи необхідно постійно домовлятися та переглядати його. По-друге, ці контекстні властивості беруть на себе їх значення або релевантність через їх зв'язок з формами практики, тобто займаються діями навколо артефактів та інформації, яка робить ці артефакти значущими та актуальними для людей. Тоді сенс технології не може бути відірваний від способів, яким люди мають його використовувати.

Такі моделі контексту мають давати змогу розв'язувати проблеми, що характеризуються контекстно-залежними властивостями:
- неоднорідність та мобільність;
- відносини та залежності;
- своєчасність;
- недосконалість;
- міркування;
- відповідність формалізму прийняття рішень;
- ефективне контекстне забезпечення.

Розгляд використання контексту в проблемних областях допомагає виявити всі семантичні відношення, надати всю необхідну інформацію та правильні інтерпретації для прийняття рішень, оскільки використання інформації в процесі прийняття рішень, як правило, відбувається в контексті складної структури процесу прийняття рішень, який часто формується за допомогою ряду чинників.

Як показано в [10], контекстна система допомагає розпізнати, зрозуміти та подати відповідні елементи прийняття рішень як контексти та в рамках контекстів. Контекстом є будь-яка інформація, яка може бути використана або характеризує відповідну проблемну область.

Контекст є важливим фактором у процесі прийняття рішень, допомагає визначити, яка інформація необхідна для підтримки прийняття рішень та представляється множиною взаємопов'язаних компонентів [11].

В [12] визначено, що контекст можна розглядати як представлення проблеми, беручи до уваги такі властивості контексту [12; 13]:

• контекст — це форма інформації, тобто контекст розглядається як те, що може бути відомо, представлено та закодовано;

• контекст є вичерпним, тобто вважається можливим сказати, що заздалегідь визначається як контекст для конкретного використання;

• контекст є стабільним, тобто, коли контекст може відрізнятися від застосування до програми, він не відрізняється від екземпляра до примірника взаємодії з додатком;

• контекст та діяльність є розділеними, тобто контекст використовується для опису особливостей середовища, в межах якого здійснюється діяльність, але елементи діяльності не належать до самого контексту та не розглядаються як контексти.

Основним недоліком багатьох існуючих систем, що базуються на контексті, є неможливість реалізації динамічного опису контексту. Існуючі контекстні моделі є «статичними» або обмеженими.

Для того, щоб додати до властивостей контексту динамізму, будемо розглядати контекст, який можна охарактеризувати як: виникає через простір і час; причинно-наслідковий процес прийняття рішень; визначений, але не обов'язково передбачуваний; семантична інтерпретація відносин між актором, завданням або діяльністю та середовищем, в яких вони знаходяться; обмежувальні критерії, за допомогою яких можна моделювати цілеспрямовану діяльність.

Обмеження контексту також зменшує складність часу обчислення потенційних рішень для діяльності. Ми використовуємо такі обмежувальні критерії: наявність даних або відсутність; повнота; набори включення / виключення; часові межі; просторові межі; область діяльності.

Контекст розглядається як динамічні відношення між актором, цілеспрямованим завданням та оточенням, в яких вони знаходяться. Такий розгляд дозволяє контекстним зв'язкам виникати, змінюватися або зникати через час і простір та охоплювати складність просторово-часової динаміки. Ми застосовуємо таке представлення контексту, оскільки воно дозволяє нам моделювати контекстну динаміку таким чином, що виникає в процесі розв'язання задачі, а не тільки вибирається на етапі формулювання проблеми та процесу розв'язання задачі. Модель контексту передбачає суб'єктивний погляд на проблемні рішення ситуації. Таким чином, ми моделюємо контекст з практичної точки зору та представляємо структуру контексту, яка успадковується від традиційних моделей контексту.

Такий погляд використовує позицію щодо властивостей контексту. По-перше, замість того, щоб розглядати контекст як інформацію, він стверджує, що контекстуальність є реляційною властивістю, яка визначається між об'єктами, діями, задачами, середовищем і т. д. Тобто щось є або не є контекстом; Навпаки, вона може або не може бути контекстуально актуальною для розв'язання певної задачі. По-друге, можна стверджувати, що множина контекстних функцій визначається динамічно. По-третє, можна стверджувати, що контекст є особливим для кожного випадку проблеми, задачі, діяльності, дії тощо. Контекст — це властивість, що пов'язана з певними налаштуваннями, окремими випадками проблем, задач, середовищ, дій та особами, що беруть участь у процесі прийняття рішень. По-четверте, замість того, щоб розглядати контекст та контент як два відокремленими об'єкта, можна стверджувати, що контекст виникає в результаті діяльності. Іншими словами, контекст необхідно розглядати не тільки як проблему представлення, а й як проблему взаємодії.

Оскільки контекст розглядається як множина динамічних відношень між актором, цілеспрямованою діяльністю, ресурсами, можливостями, часом, розташуванням та середовищем, в яких вони знаходяться або використовуються. Це дозволяє контекстним зв'язкам виникати, змінюватися або зникати через час і простір та охоплювати складність просторово-часової динаміки. Ми застосовуємо таке представлення контексту, оскільки воно дозволяє нам моделювати контекстну динаміку таким чином, що виникає в процесі розв'язання задачі, а не тільки вибирається на етапі формулювання проблеми та процесу розв'язання задачі.

Для опису контексту необхідно з'ясувати поняття та конструкції, які визначають природу, структуру та представлення процесу формування та прийняття рішень і відповідних складових областей, які описують такий процес. Контекст має бути описаний стандартизованим способом. Представлення знань процесу прийняття рішень повинне підтримувати операції, що необхідні для представлення контексту та управління ним.

Для нагромадження, інтерпретації, подання та управління контекстом пропонується загальна модель управління контекстом. Контекстна онтологія складається з трьох основних компонентів: контексту семантики (онтологія), даних екземплярів контексту та контексту, що пов'язаний з правилами. Онтологія представляє семантики, концепти і відношення в рамках контексту. Така онтологія утворюється в результаті злиття онтології, що описує абстрактні, конкретні контексти та контексти реалізації. Правила є аксіомами виведення, які використовують контекстно-орієнтовані системи для отримання рішення та міркування щодо дій, які необхідно виконати. Ці правила мають два джерела; правила, які явно визначені, та правила, які неявно отримані самою системою.

Складність в реалізації прийняття рішень полягає в необхідності синтезу різних точок зору зацікавлених сторін на проблему, управління великою кількістю інформації, що стосується завдання, та розуміння рішень, які визначили такий розгляд задачі прийняття рішень та самого процесу прийняття рішень. Крім того, знання, пов'язані з проблемою, розподіляються серед різних зацікавлених сторонім як власників проблеми та ЛПРами.

Це визначає необхідність розгляду процесів прийняття рішень на основі представлення багаторівневої системи управління та прийняття рішення в ній через модель певного контексту [14].

Контекстна онтологія $O_{cntxt}$ з урахуванням результатів [10] включає компонентні онтології: онтологія контексту, онтологія шарів і онтологія точок зору допомагають розпізнати, зрозуміти та представити відповідні явища як контексти та в межах контекстів. Загальна мета контекстної онтології полягає в тому, щоб визначити поняття та конструкції, які допомагають нам зрозуміти природу, цілі та значення окремих сутностей. Таким чином, замість того, щоб розглядати проблемну область як базову структуру сутностей, онтологія контексту визначає розгляд сутності в контексті від спеціальних ролей або значень. Така онтологія представляється у вигляді взаємопов'язаної множини онтологій, що є асоціативною структурою такого вигляду (рис. 1):

$$O_{cntxt} = \left\langle O_{ctx}, O_{layer}, O_{aspect} \right\rangle,$$

де $O_{ctx}$ — онтологія контексту; $O_{layer}$ — онтологія шарів; $O_{aspect}$ — онтологія аспектів (точок зору).

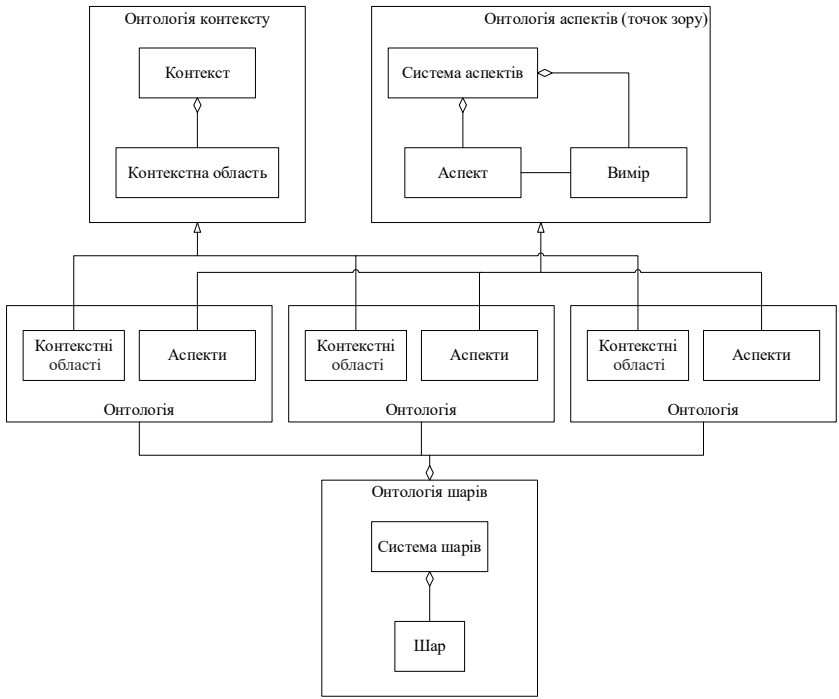

Рис. 1. Складові контекстної онтології у взаємозв'язках

Онтологія $O_{ctx}$ визначає такі контекстні області: область мети/результату, область актора, область процесу/дії, область об'єкта, область середовища, область можливостей, область засобів, область представлення, область розташування та область часу. Контекст визначається як конструкція, яка складається з понять в межах десяти контекстних областей. Кожна контекстна область визначається відповідними поняттями та конструкціями. Онтологія контексту $O_{ctx}$ містить деталізовані поняття та конструкції контекстних областей та міжобласних відношень.

Онтологія шарів $O_{layer}$ підтримує структуру прийняття рішень та описує відношення на загальному рівні складових прийняття рішень та їх реалізацію на відповідних рівнях: проблема, модель, метод та реалізація в рамках системи результатів, системи об'єктів, системи використання та системи управління. Онтологія шарів $O_{layer}$ забезпечує поняття та конструкції для розуміння та структуризації прийняття рішень, особливо через поняття СППР, системи об'єктів та системи використання та також служить концептуальною основою для того, щоб структурувати прийняття рішень за чотирма шарами (реалізація, метод, модель та проблема).

Онтологія аспектів (точок зору) $O_{aspect}$ отримується з онтології шарів та онтології контексту. Онтологія $O_{aspect}$ підтримує множину визначених аспектів розгляду для конкретного представлення процесу прийняття рішень в СППР та структурування сприйняття складових прийняття рішень, зокрема з системної, концептуальної, функціональної, інформаційної та реалізаційної точок зору.

Ієрархічна організація, формальний характер, стандартизованість, підтримка ефективної аргументації, підтримка різних рівнів абстракції та взаємодій є одними з головних особливостей контекстної онтології.

При цьому контекст розглядається на абстрактному, конкретному та реалізаційному рівнях [14]. Абстрактний контекст є онтологічною моделлю багаторівневої системи, побудованої на підставі інтеграції знань прикладних проблемних областей, що розглядаються при функціонуванні системи та релевантні конкретній задачі. Конкретний контекст є представленням певного абстрактного контексту для конкретної задачі в відповідності з існуючими даними та визначеними вимогами до процесу прийняття рішень. Контекст реалізації є представленням кожного конкретного контексту в рамках існуючих реалізацій, зокрема програмного забезпечення.

На підставі виявлених властивостей контексту та задач, що виникають при використанні контексту, сформульовані вимоги до управління контекстом. Контекст повинен бути описаний стандартизованими способами, що забезпечують незалежність способу представлення від платформи, модель представлення знань повинна підтримувати операції, необхідні для представлення контексту та управління ним. Контекст має надавати релевантну, реальну та доступну інформацію для розв'язання конкретної задачі або для розуміння поточної ситуації, що включається в контекст, інформація повинна містити безпосередньо одержувані дані, історію отримання цих даних і знання, які на даний момент відомі взаємодіючим об'єктам.

Будемо розглядати контекст як конструкцію, що складається з понять в межах відповідних контекстних областей та описується онтологією контексту через таку структуру контекстних областей [14]:

$$O_{ctx} = \left\langle \begin{matrix} O_{ctx}^{AR}, O_{ctx}^{A}, O_{ctx}^{PA}, O_{ctx}^{O}, O_{ctx}^{E}, \\ O_{ctx}^{F}, O_{ctx}^{Fclt}, O_{ctx}^{R}, O_{ctx}^{Plc}, O_{ctx}^{T} \end{matrix} \right\rangle.$$

На загальному рівні $O_{ctx}$ описується контекстними областями:
$O_{ctx}^{AR}$ — мета/результат, $O_{ctx}^{A}$ — актор, $O_{ctx}^{PA}$ — процес/дія, $O_{ctx}^{O}$ — об'єкт, $O_{ctx}^{E}$ — середовище, $O_{ctx}^{F}$ — можливості, $O_{ctx}^{Fclt}$ — засоби, $O_{ctx}^{R}$ — представлення, $O_{ctx}^{Plc}$ — розташування, $O_{ctx}^{T}$ — час.

Для показу контекстних областей будемо використовувати класи об'єктів, відношень та атрибутів, що дає можливість представляти їх як семантичні аспекти, де семантика умов та відношень між ними визначені явним чином (роблячи кожен аспект формальною онтологією). Використання таких категорій дозволяє зробити формалізацію таких аспектів в логіці опису (дескрипційна логіка) (DL).

Контекстні поняття взаємозв'язані через контекстні відношення, включаючи внутрішньообласні, міжобласні та міжконтекстні відношення, тобто такі відношення включають не тільки відношення між компонентами однієї області, а й відношення між іншими контекстами. Такі поняття та конструкції необхідні для того, щоб визначити, зрозуміти, структурувати та представити сутності як контексти та/або в межах контекстів, щоб зрозуміти природу, цілі та значення відповідних сутностей задач та процесу прийняття рішень.

Для розв'язання будь-якої задачі необхідно описати процес прийняття рішень, який ґрунтується на онтології контекстів за спеціалізацією. Це будемо реалізовувати через онтологію шарів $O_{layer}^{w}$.

Онтологія шарів допомагає розпізнати, зрозуміти та представити структуру прийняття рішень на основі контекстів. Онтологія шарів описує відношення складових прийняття рішень на загальному рівні та їх реалізацію на відповідних рівнях: проблема, модель, метод та реалізація в рамках системи реалізації, системи об'єктів, системи використання та системи управління.

Ми визначаємо онтологію шарів, яка забезпечує поняття та конструкції, щоб визначити, зрозуміти, структурувати та представити статичні та динамічні особливості представлення процесу прийняття рішень в розрізі чотирьох шарів.

$O_{layer} = \left\langle O_{layer}^w, O_{layer}^s \right\rangle$ — онтологія шарів. $O_{layer}^w$ містить поняття та конструкції, які пов'язані з процесом прийняття рішень в цілому. $O_{layer}^s$ представляє процес прийняття рішень структуровано та пов'язано з визначеною системою шарів $O_{layer}^w$.

Ми будемо розрізняти $O_{layer}^w = \left\langle Prblm_{layer}^w, Mdl_{layer}^w, Mth_{layer}^w, Rlztn_{layer}^w \right\rangle$ як систему з чотирьох шарів: проблема, модель, метод та реалізація. Розглянемо деякі з них. В будь-якому контексті, що охоплює розв'язання задач, людина використовує конструкції, які можуть допомогти визначити, проаналізувати, розробити та реалізувати розв'язання проблеми.

Поняття та конструкції $O_{layer}^w$ отримуються з базової онтології [15], онтології контексту [14], онтології шару $O_{layer}^s$ та онтології точок зору та взаємодіють з онтологіями предметно-формального та формального представлення, онтологією реалізацій, що включає опис програмного забезпечення для підтримки прийняття рішень, онтологією представлення користувача та взаємодії з ним.

Шар $Prblm_{layer}^w = \left\langle Level, View\mathrm{Pr}, Rel^{prb} \right\rangle$ забезпечує поняття та конструкції для того, щоб визначити, зрозуміти, структурувати і представити сутності з точки зору проблеми в межах множини проблем СППР, що можуть бути розв'язані в рамках системи прийняття рішень. Контекст проблеми (задачі) пов'язаний з контекстами моделі. Поняття та конструкції $Prblm_{layer}^w$ отримуються з базової онтології [15], онтології контексту [14], онтології шару $O_{layer}^s$ та онтології точок зору та взаємодіють з онтологіями предметно-формального та формального представлення. Контекст проблеми пов'язаний з контекстами моделі.

Іншим шаром є моделі, що описує онтологія $Mdl_{layer}^w$:

$Mdl_{layer}^w = \left\langle Level, ViewP, Rel^{mdl} \right\rangle$, яка забезпечує поняття та конструкції для того, щоб визначити, зрозуміти, структурувати і представити сутності з точки зору моделей в межах моделі системи. Будемо виділяти такі основні точки зору до поняття моделі: $Mdl_{layer}^w$ розглядається з системної, концептуальної та реалізаційної точок зору. Визначен-ня моделі має завжди висувати на перший план аспекти з цих трьох точок зору. Модель будемо розглядати як сутність, що використовується, щоб допомогти або дозволити розуміння, комунікацію, аналіз, розробку та/або виконання деякої іншої сутності, до якої звертається модель. Модель будемо представляти в одній з трьох форм, а саме як концептуальна конструкція (опис), як вираз на певній мові або як реалізація. Контекст моделі пов'язаний з контекстами методу та контекстами проблеми.

Наступним шаром є методи, які описуються через онтологію $Mth_{layer}^{w}$. Знання методів $Mth_{layer}^{w}$ складається з чотирьох компонентів: знання процесу виконання методу, знання проблемної області, знання технологій реалізації та знання поведінки користувача. Знання процесу виконання означає знання, які стосуються виконання методу. Знання проблемної області означає знання, яке стосується реалізації прийняття рішень, її системи використання та її системи об'єктів. В кожній проблемній області є власні специфіки, які необхідні, щоб знати та виконати метод. Знання технології означає знання, яке стосується пошуку, використання та налаштування апаратного та програмного забезпечення для прийняття рішень у визначеній задачі. Знання поведінки користувача означає знання, що визначають особливості проблем людини та її поведінки, а також соціальні та організаційні аспекти, які мають бути взяті до уваги в розробці та в організації роботи методу. Метод забезпечує явне знання в формі принципів, процедур і т. д. Методи можна розділити на технології, сценарії, механізми та алгоритми. Технологія включає мову представлення та процедуру. Контекст методу пов'язаний з контекстами моделі та контекстами реалізації. Цільові контексти методу влючають контексти, для яких було визначено модель.

Механізми та алгоритми використовуються процедурою прийняття рішень, що описуються через шар реалізації $Rlztn_{layer}^{w}$: $Rlztn_{layer}^{w} = \langle Level, ViewP, Rel^{rlztn} \rangle$. Вони можуть або бути загальними, тобто застосованими до усіх мов, що можуть бути використаними при прийнятті рішень, визначеними, тобто застосованими тільки до особливих мов, або гібридними, тобто з певними частинами, що є загальними та визначеними частинами, що є визначеними або пристосовуваними. Контекст реалізації пов'язаний з контекстами методу.

Щоб включити інформацію і знання та їх використати в різних формах та на різних шарах, будемо розрізняти чотири види систем, які тісно пов'язані з прийняттям рішень:

1. ті, які описують інформацію та знання,

2. ті, які пов'язані зі накопиченням, зберіганням, обробкою та застосуванням інформації,

3. ті, які використовують інформацію,

4. ті, які управляють та можливо змінюються на основі результатів прийняття рішень та використання інформації.

Таким чином будемо розглядати систему $O_{layer}^s$, що інтегрує систему об'єктів $So_{layer}$, систему реалізацій $Sr_{layer}$, систему використання $Su_{layer}$ та систему управління $Sm_{layer}$:

$O_{layer}^s = \langle Sr_{layer}, So_{layer}, Su_{layer}, Sm_{layer} \rangle$ — контекстні знання, що пов'язані з прийняттям рішень: система реалізацій $Sr_{layer}$, система об'єктів $So_{layer}$, система використання $Su_{layer}$, система управління $Sm_{layer}$.

$Sr_{layer}$ можна визначити як систему, що описує акторів, інформацію та дані, засоби та розташування і яка збирає, зберігає, обробляє та поширює інформацію про результати, що представляється системою об'єктів, для того, щоб реалізувати та/або поліпшити дії системи використання. Структурно $Sr_{layer}$ складається з акторів, дій, інформації та даних, засобів (включаючи програмне забезпечення) та розташування, визначає відповідні результати на рівнях прийняття рішень, які реалізують розв'язання задачі. $Sr_{layer}$ існує для надання інформації, що відповідає критеріям релевантності, своєчасності тощо, для того, щоб задовольнити потреби користувачів у результатах процесу прийняття рішень. $Sr_{layer}$ — функціональна одиниця, яка збирає, зберігає, обробляє та поширює інформацію про результати прийняття рішень, що представляються системою об'єктів $So_{layer}$, та для того, щоб реалізувати та/або покращити дії системи використання $Su_{layer}$. Користувачі системи реалізацій $Sr_{layer}$ є акторами, що бажають підвищити рівень знань про систему об'єктів $So_{layer}$ за допомогою системи реалізацій. Це також вносить зміни в здатність користувачів виконувати завдання, які стосуються системи управління $Sm_{layer}$.

$So_{layer}$ представляє систему, що збирає, зберігає, обробляє та поширює інформацію для та внаслідок інтересів системи використання $Su_{layer}$. Межа $So_{layer}$ повністю визначається інтересами системи використання $Su_{layer}$. $So_{layer}$ є частиною дійсності, яку розглядають як проблемну область для прийняття рішень.

$Su_{layer}$ можна представити як систему, яка використовує послуги, забезпечені системою реалізації $Sr_{layer}$, в процесі прийняття рішень для того, щоб планувати та виконати зміни (тобто зміни стану (переходи) за допомогою системи управління $Sm_{layer}$. Актори в системі використання $Su_{layer}$ — користувачі, програмні компоненти системи реалізацій $Sr_{layer}$. В рамках $Su_{layer}$ ми можемо розрізнити два види користувачів: кінцеві користувачі, які збільшують своє знання, взаємодіючи безпосередньо з СППР; непрямі користувачі, які збільшують своє знання, отримуючи результати СППР через користувачів СППР. $Su_{layer}$ можна класифікувати за різними критеріями, наприклад, розглядати на стратегічному, тактичному або оперативному рівнях. Результати $Su_{layer}$ можуть стосуватися людини-актора, програми-актора. $Su_{layer}$ можна визначити як систему, яка використовує послуги, що реалізуються системою реалізації $Sr_{layer}$, для прийняття рішень або інших дій, щоб планувати та виконати зміни (тобто зміни стану) в системі управління $Sm_{layer}$. $Sm_{layer}$ — система, яка використовує систему використання $Su_{layer}$.

Між системою реалізацій $Sr_{layer}$, системою об'єктів $So_{layer}$, системою використання $Su_{layer}$ та системою управління $Sm_{layer}$ існують певні відношення. Інформаційні об'єкти системи реалізацій представляють сутності системи об'єктів. Також інформаційні об'єкти системи використання представляють сутності системи об'єктів. Відношення між системи об'єктів і іншими системами залежить від того, чи ці системи перетинаються чи не перетинаються. Ми можемо визначити чотири різних випадки щодо того, в якій частина система об'єктів є частиною інших систем. В першому випадку система об'єктів повністю не перетинається з іншими системами. Це означає, наприклад, що інформація збирається з абсолютно різних сутностей у порівнянні з тими, які знаходяться під впливом системи використання. Це, звичайно, дуже рідкісна ситуація. В другому випадку система об'єктів збігається з системою управління. В третьому випадку система об'єктів перетинається з системою використання. В цьому випадку система реалізацій використовується, наприклад, для планування, контролю або виконання роботи в системі використання. Нарешті система об'єктів може перетинатися з системою реалізацій.

Як показано вище, третьою складовою контекстної онтології $O_{ctx}$ є онтологія аспектів (точок зору) $O_{aspect}$.

Поняття аспекту або точки зору не має чіткого визначення, тому будемо використовувати аспект (точку зору) як певний спосіб розгляду або оцінювання. Використання аспекту призводить до обмеженої або визначеної концепції певних сутностей та їх властивостей в реальності. Щоб отримати та пов'язати ці представлення, визначається певна структура.

Онтологія аспектів (точок зору)

$$O_{aspect} = \left\langle \begin{matrix} VofP_{aspect}^{Sys}, VofP_{aspect}^{Cncpt}, VofP_{aspect}^{Man}, VofP_{aspect}^{Inf}, \\ VofP_{aspect}^{Rlz}, Dim_{aspect}, Rel_{aspect} \end{matrix} \right\rangle,$$

де $VofP_{aspect}^{Sys}$ — системна точка зору, що відбиває склад взаємодіючих у процесах об'єктів проблемної області та відбиває взаємодію у процесах прийняття рішень;

$VofP_{aspect}^{Cncpt}$ — концептуальна точка зору, що відбиває зміст об'єктів проблемної області та їх взаємодію в процесах прийняття рішень;

$VofP_{aspect}^{Man}$ — точка зору управління, що відбиває події та правила, які виникають, використовуються та впливають на виконання процесів прийняття рішень;

$VofP_{aspect}^{Inf}$ — інформаційна точка зору, що відбиває взаємозв'язок функцій (дій) щодо перетворення об'єктів у процесах прийняття рішень;

$VofP_{aspect}^{Rlz}$ — реалізаційна точка зору, яка описує засоби реалізації елементів СППР та прийняття рішень;

$Dim_{aspect}$ — виміри розгляду точок зору;

$Rel_{aspect}$ — відношення точок зору.

В цьому випадку контекст є як результатом інтеграції релевантних сформованих вимог до розв'язання задачі частин декількох онтологій. Під інтеграцією розуміється інтеграція декількох частин вихідних онтологій, результатом якої є уніфікована онтологія або контекст, в якому однаково представлені знання з інтегрованих частин, і повністю підтримується логічний висновок, що заданий в цих частинах, та які можуть бути отримані з множини однорідних або різнорідних джерел.

Системна точка зору базується на системі об'єктів та системі використання, яка складається з пов'язаних точок зору розгляду кожного з шарів та пов'язаних контекстних областей. Тут визначають, наприклад, для проблеми, за яких умов може виникнути, які можуть існувати впливи на проблему щодо започаткування або використання, як може бути використана, які існують або були реалізовані контексти моделі, методу та реалізації тощо. Для шару метод визначає, які контексти моделі мають бути враховані стосовно методу, які є цілі, можливі актори та обмеження використання методу з урахуванням як контекстів проблеми та моделі, так і контексту реалізації тощо.

Концептуальна точка зору розглядає прийняття рішень через семантичний зміст інформаційних об'єктів, який означає, що точка зору адресується контексту системи об'єктів. Ця точка зору зосереджується на концептуальному змісті відповідних об'єктів у врахуванням структурної та динамічної складових.

З управлінської точки зору система розглядається як система управління з відповідними подіями та правилами функціонування такої системи. Для цього визначають акторів (людина, програмна система), як вони можуть взаємодіяти, де вони знаходяться, як можуть бути сформульовані відповідні процедури та алгоритми в рамках відповідних шарів тощо.

З інформаційної точки зору розглядається система, що базується на системі об'єктів, яка вважається функціональною структурою інформаційної обробки мети, дій та об'єктів, незалежно від будь-яких особливостей представлення, реалізації та використання, тобто визначається, яка інформація обробляється і чому, які дії та правила обміну та обробки тощо.

З реалізаційної точки зору розглядається система, що базується на системі реалізації, яка пов'язана з конкретним організаційним, управлінським та технологічним контекстами, тобто визначаються актори, що виконують дії в процесі реалізації прийняття рішень, як вони взаємодіють і де вони розташовані, де та як зберігаються необхідні дані, які засоби використовуються та коли, яке апаратне та програмне забезпечення використовуються, і як вони пов'язані між собою тощо

Відношення між точками зору можуть бути побудовані на декількох критеріях. Оскільки неможливо знайти щось, що покривало би всі необхідні особливості прийняття рішень та забезпечувало необхідні поняття та конструкції. При цьому вибір критеріїв визначає, що точки зору повинні підтримати структурований розгляд багатогранних особливостей процесу прийняття рішень. Критерії мають дозволяти розглянути кожну точку зору з урахування процесу прийняття рішень, що дозволяє визначити, що відповідає точці зору та що має бути проігноровано.

Представлена система аспектів (точок зору) розглядається з трьох вимірів: вимір розкладання, концептуальний вимір, незалежність реалізації — вимір залежності.

Використання точок зору, таких як системна, інформаційна, управлінська, надає можливість руху вздовж вимірювання розкладання, тобто від «чорного ящика» до системи, яка складена з цілей, дій та об'єктів. У цьому процесі переважно застосовані принципи розкладання та спеціалізації.

Системна, управлінська та реалізаційна точки зору дають змогу проаналізувати зміни вздовж виміру розкладання, з одного боку, та вздовж незалежності реалізації — вимірювання залежності, з іншого боку.

Таким чином, може бути отримано певне ієрархічне представлення системи точок зору, що будується на критерії залежності реалізації, оскільки кожний шар визначає більш конкретні поняття, і відношення, що розгорнуті на нижчих рівнях абстракції. Отже об'єкти, що реалізуються через певні інформаційні об'єкти, дії, що розділені на дії людини та дії програмної системи, і тимчасові конкретні специфікації. У цьому процесі визначаються ієрархія мети, засобів, структури розкладання дії, структура розкладання об'єкта.

Такий розгляд контексту в рамках задач прийняття рішень дозволяє, не впливаючи безпосередньо на процес прийняття рішень, обмежити його лише значущими для даного контексту правилами / процедурами. Це дозволяє: 1) логічно виводити новий контекст з наявних; 2) повторно використовувати контекст за допомогою застосування контекстів вищих рівнів абстракції, їх інтеграції та конкретизації для відповідних умов і завдань; 3) отримувати контекст більш високого рівня абстракції з відповідного розглянутого контексту; 4) розбивати контекст на складові відповідні логічно пов'язані внутрішньо узгоджені контексти.

Реалізація інтегрованого погляду на прийняття рішень через систему аспектів (точок зору) надає можливість використання інформації, яка міститься в декількох контекстах та визначає контекст, який може бути використаний, наприклад, прикладною програмою для розв'язання певних завдань, підвищити достовірність контекстної інформації. Аспекти або точки зору дозволяють використовувати тільки ту частину даних, інформації або знань, яка є релевантною для задачі, що розв'язується. Також вони дозволяють копіювати фрагменти контексту, повторно використовувати їх для інших цілей тощо.

Використання системної оптимізації, яке базується на використанні знань у вигляді онтології та контексту, дає можливість внести до організації процесу прийняття рішень ряд важливих властивостей, перш за все дає можливість перейти до безперервного аналізу ситуацій та планування дій, забезпечує проведення корекції процесу прийняття рішень без порушення технологічної цілісності та взаємозв'язків, допускає багатоваріантність рішень та можливість їх отримання за різними критеріями і моделями, будує взаємопов'язану систему підготовки та вибору рішень, як для даної проблеми, так і для взаємодії з іншими комплексами проблем і завдань, дозволяє приймати рішення з урахуванням наслідків їх реалізації. При цьому в рамках таких технологій вдасться врахувати взаємозалежність рішень, негативні наслідки реалізації, обмеження поведінки, інформаційні обмеження, час та середовище, що постійно змінюється, визначеність, ризик, невизначеність тощо.

Результати роботи використано в рамках науково-дослідної роботи «Розробити типові онтологокеровані процедури системної оптимізації для розв'язання прикладних задач».

# ТЕОРЕТИЧНІ ОСНОВИ ІНФОРМАЦІЙНОЇ ТЕХНОЛОГІЇ ПРОГНОСТИЧНОГО ОЦІНЮВАННЯ ЯКОСТІ ПРОЄКТУВАННЯ ПІСЛЯДРУКАРСЬКИХ ПРОЦЕСІВ

*Сеньківський В. М., Піх І. В., Кудряшова А. В.*

*Розроблення інформаційної технології прогностичного оцінювання якості проєктування післядрукарського опрацювання книжкових видань на первинному етапі передбачає опис предметної області, зокрема означення особливостей реалізації, послідовності післядрукарських процесів та перебігу їх проєктування. На основі теоретичного обґрунтування та експертних висловлювань виокремлено ключові фактори впливу та здійснено функціональне моделювання визначених процедур. Наведено контекстні діаграми, де основними функціями систем є реалізація та проєктування післядрукарських процесів. Описано процес декомпозиції кожної з них. Для формалізації наведених знань та отримання можливості встановлення оптимального розв'язку основної та побічних задач застосовано розроблені онтології проєктування післядрукарських процесів.*

*Внаслідок аналізу предметної області відбувається синтезування моделі пріоритетного впливу факторів на якість проєктування після-*

друкарських процесів. Зокрема описано особливості побудови семантичної мережі та опису зв'язків між факторами за допомогою логіки предикатів, встановлено пріоритетності впливу факторів на досліджувані процеси методами математичного моделювання ієрархій та ранжування факторів.

Наступним етапом є оптимізація моделі пріоритетного впливу факторів на якість проєктування післядрукарських процесів, що полягає у покращенні вхідних даних, та здійснюється за методом аналізу ієрархій, який передбачає розв'язання ряду задач: побудову матриці попарних порівнянь; обчислення компонент та нормалізацію значень головного власного вектора матриці; перевірку результатів оптимізації за критерієм максимального значення головного власного вектора, нормативних значень індексу узгодженості та відношення узгодженості; синтез оптимізованої моделі.

Для встановлення оптимальної альтернативи реалізації проєктування післядрукарських процесів обрано два методи: багатофакторний вибір альтернатив на основі лінійного згортання критеріїв та на основі нечіткого відношення переваги.

Отримання конкретного кількісного показника якості реалізації досліджуваного процесу реалізовано методами та засобами нечіткої логіки.

На основі виокремлених етапів побудовано структурно-функціональну модель та IDEF0-моделі інформаційної технології.

The development of the information technology for prognostic assessment of the design quality of post-press processing of book editions at the initial stage involves the description of the subject area, including the characteristics of implementation, sequence of post-press processes and the course of their design. Key factors of influence have been identified and functional modelling of certain procedures has been carried out on the basis of theoretical substantiation and expert statements. Contextual diagrams are presented, where the main functions of the systems are the implementation and design of post-press processes. The process of decomposition of each of them is described. To formalize the knowledge and get the opportunity to establish the optimal solution of the main and secondary problems, it is suggested to develop an ontology of design of post-press processes.

As a result of the analysis of the subject area, a model of the priority influence of factors on the design quality of post-press processes is synthesized. In particular, the paper describes the features of constructing a semantic network and describing the relationships between factors using predicate logic, prioritizing factors by mathematical modelling of hierarchies and factors ranking.

The next step is to optimize the model of priority influence of factors on the design quality of post-press processes, which consists in improving the input data and is carried out by the method of hierarchy analysis, which involves solving a number of problems: the construction of a matrix of pairwise comparisons; the calculation of components and the normalization of the values of the main eigenvector; checking the results of optimization by the criterion of the maximum value of the main eigen-

*vector, the normative values of the consistency index and the consistency ratio; the synthesis of an optimized model.*

*To establish the optimal alternative for the implementation of design of post-press processes, two methods have been chosen: a multifactor selection of alternatives based on linear convergence of criteria and on the basis of fuzzy benefit ratio.*

*Obtaining a specific quantitative indicator of the quality of implementation of the studied process is presented by methods and means of fuzzy logic.*

*The structural-functional model and IDEF0-models are constructed on the basis of the selected stages of the information technology.*

**Постановка проблеми.** Завершальний етап технології виготовлення книжкової продукції, до якого відносяться брошурувально-палітурні процеси, часто помилково ототожнюють із набором механічних, циклічно повторюваних дій, позбавляючи їх високоінтелектуальної інформаційної складової. Такий підхід призводить до підвищення ймовірності часткового чи повного відбракування тиражу. Типовою хибою є також невідповідність виготовленої продукції її функціональним та експлуатаційним характеристикам. Так, для прикладу, стосовно видання, яке повинно служити десятиліття, застосовують клейове скріплення органічного походження, непридатне для забезпечення прийнятих вимог, та обирають невідповідний оздоблювальний матеріал [79].

Останніми роками використовується моделювання вказаних процесів за допомогою комп'ютерної техніки та спеціального програмного забезпечення [5; 59]. Важливо враховувати той факт, що обладнання для виконання окремих операцій брошурувально-палітурних процесів та матеріали, що використовуються для різних видів продукції, є індивідуальними [7; 9; 17; 19; 24; 35]. Активно застосовується принцип вертикального проєктування, при якому розрізняють процедури аналізу і синтезу. У результаті синтезу створюються описи об'єктів, які відображають їхню структуру і параметри [11; 13]. Вибір технології та післядрукарського устаткування залежить від виду друкованої продукції, її призначення, обсягів виробництва, економічних та фінансових показників діяльності друкарень [1; 23; 31; 37; 74; 75]. Суттєвою проблемою є дотримання стандартів на виготовлення видань, метрологічні характеристики, що стосуються якості у поліграфії, моделювання бізнес-процесів [40; 45; 77], що є важливими чинниками планування та ефективного функціонування поліграфічних підприємств [79].

Слід зазначити, що автоматизація з використанням комп'ютеризованих технологій не приносить очікуваних результатів, адже застосовані процедури не пов'язуються при цьому в єдину, нероздільну систему. За таких умов доцільним та необхідним є пооопераційний інформаційний супровід, наслідком якого стане прогностичне оцінювання якості майбутньої продукції. Подібний підхід при наявності умов невизначеності вимагає виокремлення ключових факторів впливу на якість проєктування післядрукарських процесів, встановлення міри важливості кожного з них та пріоритетності впливу на досліджуваний процес; формування, розрахунку та багатокритеріального оцінювання альтернативних варіантів реалізації післядрукарських процесів на основі лінійного згортання критеріїв та нечітких відношень переваги і визначення оптимального з них; обчислення інтегрального показника рівня якості проєктування післядрукарських процесів; розроблення інформаційної технології прогностичного оцінювання вказаних процедур, що слугуватиме методологічною основою для отримання продукції належної якості [79].

**Основним завданням дослідження** є розроблення інформаційної технології прогностичного оцінювання якості проєктування післядрукарських процесів, що передбачає виконання таких підпорядкованих завдань:

— проаналізувати етапи проєктування та реалізації післядрукарських процесів, дослідити основні операції та функції, розробити онтологію;

— виокремити фактори впливу на якість проєктування післядрукарських процесів та сформувати семантичні мережі зв'язків між ними;

— синтезувати та оптимізувати моделі пріоритетного впливу факторів на якість проєктування післядрукарських процесів;

— визначити оптимальні альтернативні варіанти реалізації за методами багатофакторного вибору альтернатив на основі лінійного згортання критеріїв та нечіткого відношення переваги;

— побудувати функції належності лінгвістичних змінних і розрахувати їх значення із використанням нечітких логічних рівнянь;

— визначити інтегральний показник якості проєктування післядрукарських процесів шляхом дефазифікації нечітких множин за принципом центра ваги;

— розробити структурно-функціональну модель інформаційної технології прогностичного оцінювання якості проєктування післядрукарських процесів;

— розробити IDEF0-моделі інформаційної технології прогностичного оцінювання якості проєктування післядрукарських процесів.

**Виклад суті дослідження.** Незважаючи на багатовікову історію зусиль вчених та практиків, скерованих на формування узагальнюючих методів, способів та засобів апріорного оцінювання якості процесів, пов'язаних з виробництвом різнорідної продукції, проблема на сьогодні до кінця не вирішена. Спробуємо пояснити причини такого стану.

Найчастіше поняття «якість» співвідносять з виробом, що на перший погляд є логічним, оскільки споживача цікавить насамперед добротність саме готової продукції [12; 47]. І тут постає дилема — що вважати якістю і як її трактувати. Адже користувачі у переважній більшості не знайомі зі стандартами, тому кожний по-своєму оцінює товар, не кажучи вже про те, що його не цікавлять деталі технології виготовлення. У цьому випадку якість стає категорією суб'єктивною. З іншого боку, наявність і строге дотримання нормативів і стандартів якості, згідно з якими оцінюється продукція і розробляються вимоги до технології, машин та режимів реалізації процесів та окремих процедур, стають об'єктивною передумовою отримання якісних результатів.

Загальною парадигмою як підсумок до сказаного слугує той факт, що якість виступає в ролі апостеріорної категорії, отриманої для характеристики виходу виробничого процесу у статичному режимі. У цьому випадку якість — це сукупність властивостей продукції, які обумовлюють задоволення потреб користувача у відповідності з її призначенням [47].

Основна увага пропонованого дослідження буде звернена на динаміку формування якості книжкових видань, тобто механізм апріорного встановлення прогнозованого показника критерію ефективності, як оцінки якості видавничо-поліграфічних етапів засобами сучасних інформаційних технологій. Враховуючи сказане, можна стверджувати факт існування упорядкованих взаємозв'язків між рівнями технологічного процесу випуску книжкових видань. Структурування ходу дослідження встановить послідовність дій, а також підтвердить ефективність та більш строго обґрунтує логіку застосування інформаційної концепції до прогнозування якості поліграфічної продукції.

З огляду на сказане, на початку ери становлення електронних інформаційних засобів могло здатися, що друковану продукцію чекає повний занепад, про що в останні роки йшли серйозні дискусії. Однак дані про темпи та обсяги випуску друкованих видань (особливо книжкових) в Україні свідчать про те, що скептики традиційної (па-перової) поліграфії не врахували багатьох суттєвих чинників. Так, у публікації [4] вказано на фактори, що свідчать про відродження українського книгарства.

На початок третього тисячоліття в Україні склався добрий гурт професійно підготовлених і рішуче налаштованих на працю видавців і друкарів різних форм власності, які вміло продовжують кращі традиції своїх попередників на книговидавничому полі.

Незважаючи на економічні негаразди, в українському суспільстві існує стабільно високий попит на добротну українську книгу: художню, навчальну, наукову, пізнавальну, довідкову тощо.

За прикладом країн Західної Європи Україна все більше починає відчувати вплив щорічного ярмаркового буму в книжковій справі. Створено групи приватних видавництв в обласних центрах, які за короткий час своєї діяльності змогли серйозно заявити про себе на загальнодержавному рівні.

Технологія виготовлення друкованої продукції є складовою частиною інформаційних технологій, адже від оперативності та досконалості друкарських процесів та охоплення ними усіх сфер суспільної діяльності залежать обсяги та швидкість розповсюдження інформації — основи існування та поступу людства. Інформаційні видавничо-поліграфічні технології належать до одного з видів сучасних технологій, пов'язаних з виготовленням як паперових, так і електронних носіїв даних і знань. Як різновид соціальних інформаційних технологій, вони породжені суспільною необхідністю удосконалення процесу виготовлення твердих та електронних носіїв інформації. Ця технологія виникла не через появу комп'ютерної техніки як такої, а через суспільне усвідомлення можливості організувати видавничий процес більш ефективно, оперативно включитися в загальнолюдську інформаційну систему, стати її активним джерелом і споживачем у реальній інформаційній ситуації. І ця технологія активно впроваджується у видавничий процес, є ефективною, найбільш автоматизованою технологією виготовлення книги, журналу, газети чи іншої друкованої та «електронної» продукції [3; 36; 48; 65; 66].

Останніми роками поліграфічні корпорації створили та успішно використовують концептуально нову інформаційну технологію організації та функціонування видавничо-поліграфічного комплексу, названу терміном «робочий потік» (Workflow) [8; 14; 39; 49; 71]. Він забезпечує послідовність реалізації конкретних операцій, пов'язаних з даними, відображеними форматами файлів PDF, CIP3, CIP4, про-грамним та апаратним забезпеченням, а також взаємодію апаратного і програмного забезпечення. Мета полягала в тому, щоб об'єднати технічно й організаційно потоки даних Workflow і перекинути міст між клієнтами, друкарнями і брошурувальними підрозділами. Ці потоки використовуються для обробки цифрової інформації на всіх етапах поліграфічного виробництва; вони забезпечують інтеграцію систем CtP (Computer to Plate) з цифровим Workflow, а також з системами кольоропроби. Сюди входять процеси прийому даних, виробництво, коректура, управління кольорами, поділ на кольори, спуск полос і їх виведення.

На сучасному етапі спостерігається тенденція зменшення тиражів. Це привело до появи друку на вимогу PoD (Print on Demand). Кожна система PoD має своє призначення і свої можливості. Спільним є те, що друкарські комунікації здійснюються за допомогою цифрових способів друку, а також післядрукарських технологій, орієнтованих на нього.

Аналіз теперішнього стану технологій друкарства та системних і програмних засобів їх реалізації свідчить про відсутність на даний час універсального механізму апріорного оцінювання ефективності реалізації етапів, стадій чи окремих операцій видавничо-поліграфічного циклу саме на інформаційному рівні, що унеможливлює апріорне досягнення очікуваної якості за допомогою автоматизованих систем, орієнтованих на експертно-прогностичне вирішення вказаної проблеми.

Підставою для формування структурно-функціональної моделі інформаційної технології прогностичного оцінювання якості проєктування післядрукарських процесів є виокремлення та систематизація основних етапів інформаційної технології [10; 32; 46; 67].

**Етап 1. Аналіз предметної області**
*1.1. Узагальнений опис операцій та технологій післядрукарського опрацювання книжкової продукції*

Післядрукарські процеси — це сукупність послідовних дій, направлених на перетворення віддрукованих аркушів та інших конструкційних елементів у готову книгу [35; 61].

Післядрукарське опрацювання книжкових видань можна розділити на два великі блоки: брошурувальні та палітурні процеси. До брошурувальних процесів належать: виготовлення зошитів, комплектування та скріплення блоків. Можливе також з'єднання блоків з обкладинками та обрізування з трьох сторін. До палітурних процесів належать: опрацювання книжкових блоків, виготовлення та оздоблення палітурок, з'єднання книжкових блоків з палітурками, кінцеве опрацювання книг [6; 35]. Розглянемо згадані операції детальніше.

Виготовлення зошитів полягає в зіштовхуванні, розрізуванні аркушів на частини, фальцюванні, пресуванні та приклеюванні додаткових елементів. Зіштовхування виконується для покращення точності підрізання та розрізання аркушів. Ця операція полягає у вирівнюванні країв аркушів за горизонтальним та вертикальним краями пачки. Розрізування — це поділ друкарських чи палітурних аркушів на частини. Фальцювання полягає у згинанні аркушів у визначеному порядку з фіксацією згинів з метою одержання зошитів бажаного формату та конструкції. Фальцювання класифікується за такими ознаками: число згинів (однозгинне, двозгинне, трьохзгинне, чотирьохзгинне, багатозгинне), взаємне розміщення згинів (паралельне, перпендикулярне, комбіноване), розміщення згинів на аркуші (симетричне, зміщене), число полос (одинарне, двійником, четверником), число аркушів (один, два і більше, більше чотирьох згинів). Пресування і упаковування зошитів здійснюється для зручності транспортування та зберігання перед наступними операціями. В якості додаткових елементів можуть бути приєднані форзаци, ілюстрації, частини аркушів.

Комплектування блока — це операція, спрямована на розміщення аркушів або сфальцьованих аркушів у правильній послідовності в межах книжкового блоку. Є два основні способи комплектування: вкладанням (для видань обсягом до 64—80 сторінок; аркуші вставляються один в один) та підбиранням (для видань обсягом понад 80 сторінок; сфальцьовані аркуші накладають один на одного). Скріплення зошитів скомплектованого блоку може здійснюватися шиттям дротом, шиттям нитками або за допомогою незшивних способів. Якщо покривним матеріалом є обкладинка, то здійснюється з'єднання книжкового блоку з обкладинкою та обрізування з трьох сторін [6; 9; 35].

Якщо покривним матеріалом є палітурка, то подальше опрацювання книжкових блоків може полягати у пресуванні, заклеюванні корінця, сушінні блоку, обтискуванні корінця, обрізуванні блоку з трьох сторін, зафарбовуванні обрізів, зміні форми корінця, кашируванні, приклеюванні лясе, наклеюванні каптала, наклеюванні смужки паперу, гільзи тощо. Виготовлення палітурки загально може складатися з трьох основних операцій: розкрій покривного матеріалу, збір та з'єднання деталей, круглення корінця. При потребі покривний матеріал оздоблюється. Вставлення блоку в палітурку виконується одним з чотирьох способів: звичайне вставлення, на гільзу, «глухе», в кишені. Після вставлення блоків у палітурки здійснюється пресування книг, штрихування, обгортання книг суперобкладинкою. Завершальною операцією є пакування книжкової продукції [6; 9; 35; 76].

Таким чином, виготовлення книжкових видань в обкладинці передбачає виконання лише брошурувальних процесів, а в палітурці — брошурувальних і палітурних. Загалом брошурувально-палітурне виробництво характеризується неабиякою варіативністю операцій, що пов'язано зі значною кількістю елементів виробів, різновидами напівфабрикатів, тривалим технологічним ланцюжком, різноманітністю матеріалів. Разом з використанням класичних методів опрацювання спостерігається постійне вдосконалення напрямків автоматизації післядрукарських процесів та бажання прогностичного оцінювання результатів діяльності. Проєктування досліджуваних процесів є ключовим етапом для досягнення успішної реалізації необхідних операцій і забезпечення якості книжкового видання [35; 76].

### 1.2. Функціональне моделювання післядрукарського опрацювання книжкової продукції

З огляду на системний характер проєктування післядрукарських процесів, його доцільно розглядати та досліджувати як певну систему. При вивченні системи використовуються системний підхід та системний аналіз. Системний підхід полягає у дослідженні об'єкта як системи, виявленні та дослідженні сукупності відношень і зв'язків у ньому. Основними принципами системного підходу є принцип взаємозв'язку, принцип багатоплановості, принцип багатомірності, принцип ієрархічності, принцип різнопорядковості, принцип динамічності. Системний аналіз являє собою сукупність методів та алгоритмів, спрямованих на вирішення проблеми. Основною ідеєю системного аналізу є перетворення складної проблеми у чітку послідовність знань, розв'язок яких є уже відомим або до яких можна застосувати відомі методи вирішення. Процедура системного аналізу складається з двох частин: аналізу та синтезу. Аналіз полягає у розкладанні основної проблеми на підпроблеми та застосуванні оптимальних методів їх вирішення. Об'єднання окремих розв'язків підпроблем в один загальний розв'язок проблеми називається синтезом. Фактично аналіз та синтез є дзеркальними процедурами. Перш ніж застосовувати системний аналіз, необхідно чітко сформулювати проблему, яка потребує вирішення, та визначити межі системи в яких буде вирішуватися дана проблема.

Розрізняють два підходи системного аналізу. Перший полягає у застосуванні математичних прийомів, зокрема теорії оптимізації та дослідження операцій. При цьому ставиться математична задача, метою якої є знайдення оптимального проєкту системи чи/та найкращого режиму її функціонування. Основою другого підходу є логіка системного аналізу, яка використовується у тих випадках, коли застосування математичного підходу є неефективним.

Логіка системного аналізу випливає зі специфіки задач, для вирішення яких він застосовується, та реалізованого підходу. Системний аналіз застосовують для погано структурованих задач. При наявності невизначеностей у системному аналізі, метою якого є вплив на вибір способу дії, присутні такі елементи, як проблема та проблематика, цілі, засоби для досягнення цілей, альтернативи, ресурси, які потрібні для кожної альтернативи, моделі, критерії вибору оптимальної альтернативи.

Етап формулювання проблеми полягає у її розширенні до проблематики, тобто виявленні системи пов'язаних із нею проблем, без врахування яких ключова проблема не може бути розв'язана. Наступним за важливістю є етап виявлення цілей, на якому визначається, що потрібно зробити для розв'язання проблеми. Цілі є антиподом проблеми. На наступних етапах визначається, яким чином потрібно розв'язати проблему [59].

Підвищенню ефективності та удосконаленню післядрукарського опрацювання книжкових видань сприяє використання сучасних методів системного аналізу, які реалізуються за допомогою комп'ютерної техніки та спеціальних програмних продуктів — CASE-технологій. CASE-засоби підтримують процеси аналізу і формулювання вимог до різноманітних складних систем, процеси створення і супроводження інформаційних систем тощо. Одним з напрямів CASE-технологій є SADT-технології, спрямовані на створення, аналіз та подальше використання моделей складних систем [59]. На базі SADT-методології розроблена методологія IDEF0-моделі, що передбачає побудову контекстних діаграм деревовидної структури, створених за принципом декомпозиції. Контекстну діаграму позначають як А-0, а діаграму декомпозиції першого рівня — А0. Стрілками типу вхід (те, що опрацьовується) є множина значень $I = \{I_1, ..., I_n\}$, стрілками типу керування (процедури та стратегії управління) — множина $C = \{C_1, ..., C_n\}$, стрілками типу вихід (результат) — множина $O = \{O_1, ..., O_n\}$, а стрілками типу механізми (необхідні ресурси) — множина $M = \{M_1, ..., M_n\}$ [40; 77].

Основною функцією є реалізація післядрукарських процесів. Зв'язок системи із навколишнім середовищем ілюструється такими граничними стрілками: $I_1$ — віддруковані аркуші, $I_2$ — покривний матеріал, $I_2$ — інші матеріали, $C_1$ — нормативно-технічна та технологічна документація, $C_2$ — проєкт, $C_3$ — альтернативи реалізації, $O_1$ — рівень якості післядрукарських процесів, $O_2$ — готові видання, $M_1$ — брошурувально-палітурне устаткування, інші знаряддя праці, $M_2$ — особовий склад працівників, експерти з предметної області, зацікавлені особи.

Проаналізуємо інформаційне навантаження компонент множин граничних стрілок IDEF0 моделі реалізації післядрукарських процесів:

Граничні стрілки типу «Вхід» (Input):

— $I_1$ (віддруковані аркуші). Результатом додрукарського опрацювання авторських оригіналів та друкування накладу є віддруковані паперові аркуші, які надходять на дільницю післядрукарського опрацювання.

— $I_2$ (покривний матеріал). Аркуші покривного матеріалу, які слугують для виготовлення обкладинок чи палітурок (залежно від характеристик).

— $I_3$ (інші матеріали). Матеріали для скріплення, оздоблення тощо.

Граничні стрілки типу «Контроль» (Control):

— $C_1$ (нормативно-технічна та технологічна документація). До нормативно-технічної документації належать технічні вимоги та законодавчі положення, зокрема: закони, стандарти, технічні умови, кодекси усталеної практики та ін.

— $C_2$ (проєкт). Визначає перебіг усіх технологічних дій, направлених на реалізацію післядрукарських процесів.

— $C_3$ (альтернативи реалізації). Парето-оптимальні альтернативи, визначені оцінюванням нечітких відношень на заданій множині альтернатив.

Граничні стрілки типу «Вихід» (Output):

— $O_1$ (рівень якості післядрукарських процесів). Результат, отриманий внаслідок реалізації післядрукарського опрацювання книжкових видань.

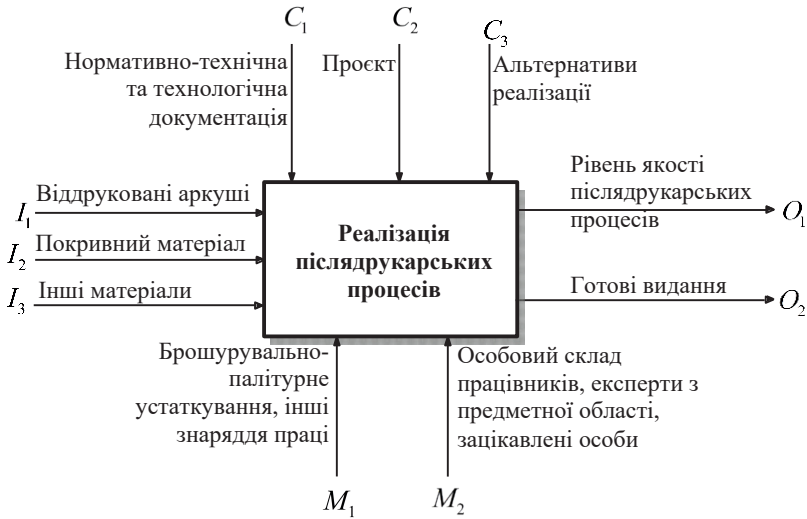

Рис. 1. Контекстна діаграма А-0 моделі IDEF0 реалізації
післядрукарських процесів

— $O_2$ (готові видання). Книжкові видання в обкладинках чи палі-турках, готові до розповсюдження.

Граничні стрілки типу «Механізми» (Mechanism):

— $M_1$ (брошурувально-палітурне устаткування, інші знаряддя праці). Устаткування, необхідне для реалізації післядрукарських про-цесів. Можливе також використання спеціалізованих знарядь праці при виконанні деяких операцій вручну.

— $M_2$ (особовий склад працівників, експерти з предметної облас-ті, зацікавлені особи). Реалізація післядрукарських процесів передба-чає участь працівників брошурувально-палітурної дільниці, кількість та кваліфікація яких залежать від обсягу та рівня автоматизації вироб-ництва. Можливе залучення профільних експертів, зокрема науков-ців та інших зацікавлених осіб.

Процес функціональної декомпозиції контекстної діаграми, на-веденої на рис. 1, полягає у її розділенні на функції нижчого порядку та встановленні напрямів граничних стрілок, що сприяє деталізації діяльності в межах досліджуваного процесу [77].

Діаграма першого рівня декомпозиції А0 моделі IDEF0 реалізації післядрукарських процесів містить такі функціональні блоки:

— РБП (реалізація брошурувальних процесів).

— РПП (реалізація палітурних процесів).

Діаграма другого рівня декомпозиції А1 моделі IDEF0:

— ВЗ (виготовлення зошитів).

— КБ (комплектування блоків).

— СБ (скріплення блоків).

— ВО (виготовлення обкладинок).

— ПБО (покриття блоків обкладинками).

— КОВО (кінцеве опрацювання видань в обкладинках).

Діаграма другого рівня декомпозиції А2 моделі IDEF0:

— ОБ (опрацювання блоків).

— ВОП (виготовлення та оздоблення палітурок).

— ЗБП (з'єднання блоків з палітурками).

— КОК (кінцеве опрацювання книг).

Для відображення ієрархічної залежності функцій доцільно використовувати діаграму дерева вузлів.

### 1.3. Аналіз факторів впливу на якість проєктування післядрукарських процесів. Розроблення онтології

Проєктування післядрукарських процесів є ключовим етапом для досягнення успішної реалізації необхідних операцій і забезпечення якості книжкового видання.

Важливим моментом при дослідженні післядрукарських процесів вважатимемо наявність технологічних характеристик чи параметрів, від яких залежить результативність проходження видання у загальному циклі його виготовлення. Узагальнюючи подібні чинники, вводимо поняття факторів, що стають основними елементами моделей визначення пріоритетності впливу факторів на хід реалізації та прогностичного оцінювання якості післядрукарських процесів. Встановлення пріоритетності компонент сформованої множини слугуватиме раціоналізації післядрукарських процесів та сприятиме отриманню готової продукції очікуваної якості.

Проєктування післядрукарських процесів слугує усвідомленому та впорядкованому виконанню запланованих технологічних дій, направлених на перетворення віддрукованих аркушів та інших конструкційних елементів у завершене книжкове видання високої якості. Відсутність етапу проєктування унеможливлює отримання прогнозованого результату.

Нехай $R = \{R_1, R_2, R_3, R_4, R_5, R_6, R_7, R_8\}$ — множина факторів проєктування післядрукарських процесів, де $R_1$ — показники видання;

$R_2$ — конструкційні особливості; $R_3$ — умови експлуатації; $R_4$ — тип виробництва; $R_5$ — матеріали; $R_6$ — тип обладнання; $R_7$ — технологічні та економічні розрахунки; $R_8$ — схема технологічного процесу.

Розглянемо детальніше кожен фактор досліджуваного процесу.

*Показники видання.* До основних показників книжкового видання відносяться: вид і тип видання, формат видання та його обсяг.

Вид видання — це сукупність видань, що об'єднані за однією чи кількома типологічними ознаками. До таких типологічних ознак належать: знакова природа інформації, спосіб виготовлення, періодичність, матеріальна конструкція, склад основного тексту, мовна ознака, ступінь аналітико-синтетичного перероблення інформації, цільове призначення, характер інформації, структура, повторюваність випуску, обсяг, формат. За знаковою природою інформації видання поділяються на текстові, нотні, картографічні (атлас, мапа, карта), образотворчі (альбоми, образотворчі картки, образотворчі плакати, художні репродукції, естампи, наочні посібники), видання брайлівським шрифтом. За способом виготовлення розрізняють друковані та електронні видання. За періодичністю: неперіодичні, серіальні, періодичні, продовжувані. За матеріальною конструкцією: блочне видання (кодексне видання), книжкове видання (книга-перекрутка, алігат), журнальне видання (журнал-перекрутка, алігат), аркушеве видання (плакат, буклет, газетне видання, карткове видання), комплектне видання, комбіноване видання, книжка-іграшка. За складом основного тексту бувають такі види: моновидання, полівидання (збірник, альманах, антологія), вибрані твори, зібрання творів (академічне видання). Види видань за мовною ознакою: оригінальне, перекладне, одномовне, багатомовне (видання з паралельним текстом), паралельне видання. За ступенем аналітико-синтетичного перероблення інформації: інформаційне видання (бібліографічне, реферативне (експрес-інформація, інформаційний листок), оглядове), дайджест. За цільовим призначенням: офіційне, суспільно-політичне, наукове, науково-популярне, популярне, виробничо-практичне, навчальне, літературно-художнє, релігійне, довідкове, рекламне, видання для дозвілля. За характером інформації: офіційне (нормативно-правове видання, нормативне видання (стандарт, технічні умови), нормативно-інструктивне (інструкція), наукове (монографія, автореферат дисертації, препринт, тези доповідей, тези повідомлень, матеріали конференції, матеріали з'їзду, матеріали симпозіуму, збірник наукових праць), виробничо-практичне (прак-тичний порадник, практичний посібник, методичні рекомендації, методичні настанови, методичний посібник, пам'ятка, паспорт (на вибір), навчальні (навчальна програма, підручник, навчальний посібник (навчально-методичний посібник, навчальний наочний посібник, хрестоматія, практикум, робочий зошит), довідкові і рекламні (енциклопедія (енциклопедичний словник), мовний словник, лінгвістичний словник, довідник, каталог, путівник, прейскурант, проспект, афіша). За структурою: однотомне видання, багатотомне видання, серія (підсерія), серійне видання, додаток. За повторюваністю випуску: перше, повторне (перевидання (видання без змін, доповнене видання, перероблене видання, виправлене видання), передрук (репринтове видання, факсимільне видання), нове. За обсягом видання поділяються на такі види: книга, брошура, листівка (аркушівка). За форматом видання бувають: мініатюрне, малоформатне, портативне, фоліант. Окремо виділяють види періодичних та продовжуваних видань [1; 59].

Також книжкові видання бувають звичайного, покращеного та сувенірно-подарункового (рекламного) типу. До звичайного типу належать видання, що видаються великими накладами, без або з невеликою кількістю ілюстрацій в одну чи дві фарби: вибрані твори, зібрання творів, окремі видання, масові серії, масові рекламно-інформаційні видання. Видання покращеного типу призначені для довготривалого використання, містять середню кількість ілюстрацій в три або чотири фарби: вибрані твори, зібрання творів, збірники, окремі видання, навчальні видання, покращені серії. Для видань сувенірно-подарункового типу характерні невеликі наклади, високоякісні матеріали, велика кількість багатоколірних ілюстрацій, нестандартні формати, великі поля, додаткове оздоблення та пакування, підвищена ціна. До них належать: видання виготовлені за індивідуальним замовленням, сувенірні, подарункові, ювілейні, факсимільні видання [35; 59].

Формат видання вказує на розмір готового книжкового блоку в міліметрах або друкарського аркуша в сантиметрах і частку аркуша. Обирається видавництвом, за погодженням з друкарнею. Залежить здебільшого від виду та типу видання. При поліграфічному відтворенні книжкових видань формат умовно позначають розміром аркуша паперу в сантиметрах та часткою аркуша (долею) та вказують у випускних даних [2; 51; 59].

Формат у міліметрах для видання в обкладинці визначають за розмірами книги після обрізки з трьох боків. Формат у міліметрах для видання в палітурці визначають за розмірами книжкового блока, обрізаного з трьох сторін. При цьому максимальні відхилення у форматах не можуть перевищувати 1 мм по ширині та 1 мм по висоті.

Формати аркушів та відповідні формати книжкових видань регламентуються ДСТУ 4489:2005. Основні формати книжкових видань наведені у таблицях 1—3. При цьому на довжину аркуша для рулонних машин вказує машинний напрям паперу [2].

Також за форматом і часткою аркуші бувають великі, середні, маленькі та мініатюрні. Загалом використовують 22 формати книжкових видань. Окремо виділяють формат друкування, який не завжди співпадає з форматом видання.

Подальший вибір технологічного процесу виготовлення видання та необхідного устаткування за ідеальних умов залежить від обраного формату [51]. В тих випадках, коли про альтернативний вибір друкарні не йдеться, а можливості обраної є обмеженими, залежність є оберненою, адже формат обирається з огляду не лише на тематичні та експлуатаційні вимоги видання, а й варіативність наявного устаткування та економічна доцільність того чи іншого варіанту виготовлення.

Також виділяють формат сторінки складання (формат набору), тобто ширину і довжину сторінки без полів. Вимірюється в друкарських одиницях. Бувають текстові, ілюстраційні і змішані сторінки складання.

Таблиця 1

**Формати книжкових видань для 1/8 частки аркуша**

| Формат аркуша паперу, мм | Позначка формату аркуша* | Формат книжкових видань | | |
|---|---|---|---|---|
| | | Сфальцьова-ного аркуша | видання | |
| | | | оптимальний | мінімальний |
| 840 М×1080 | | 270×420 | 265×410 | 262×408 |
| 700 М×1080 | | 270×350 | 265×340 | 262×338 |
| 700 М×1000 | В1 | 250×350 | 245×340 | 242×338 |
| 700 М×900 | | 225×350 | 220×340 | 217×338 |
| 610 М×860 | RA1 | 215×305 | 210×297 | 208×295 |
| 600 М×900 | | 225×300 | 220×290 | 205×275 |
| 600 М×840 | А1 | 210×300 | 205×290 | 202×288 |
| 500 М×700 | В2 | 175×250 | 169×239 | 165×235 |
| 460 М×640 | SRA2 | 160×225 | 148×210 | 145×208 |

*Згідно з ISO 216 та ISO 217.

М — розмір вздовж машинного напряму паперу.

**Формати книжкових видань для 1/16 частки аркуша**

| Формат арку-ша паперу, мм | Позначка формату аркуша* | Формат книжкових видань | | |
|---|---|---|---|---|
| | | Сфальцьова-ного аркуша | видання | |
| | | | оптимальний | мінімальний |
| 890×1240 М | | 222×310 | 210×297 | 208×295 |
| 860×1220 М | RA0 | 215×305 | 210×297 | 208×295 |
| 840×1080 М | | 210×270 | 205×260 | 192×255 |
| 750×900 М | | 185×225 | 182×216 | 179×213 |
| 700×1000 М | В1 | 175×250 | 169×239 | 165×235 |
| 700×900 М | | 175×225 | 170×215 | 167×213 |
| 600×900 М | | 150×225 | 145×215 | 142×213 |
| 600×840 М | А1 | 150×210 | 145×200 | 142×198 |

*Згідно з ISO 216 та ISO 217.

М — розмір вздовж машинного напряму паперу.

**Формати книжкових видань для 1/32 частки аркуша**

| Формат арку-ша паперу, мм | Позначка формату аркуша* | Формат книжкових видань | | |
|---|---|---|---|---|
| | | Сфальцьова-ного аркуша | видання | |
| | | | оптимальний | мінімальний |
| 1000 М×1400 | В0 | 175×250 | 169×239 | 165×235 |
| 890 М×1240 | | 155×222 | 148×210 | 145×208 |
| 880 М×1120 | | 140×220 | 136×210 | 133×208 |
| 860 М×1220 | RA0 | 152×215 | 148×210 | 145×208 |
| 840 М×1080 | | 135×210 | 130×200 | 127×188 |
| 750 М×900 | | 112×187 | 107×177 | 104×175 |
| 700 М×1080 | | 135×175 | 130×165 | 127×163 |
| 700 М×1000 | В1 | 125×175 | 120×165 | 117×163 |
| 700 М×900 | | 112×175 | 107×165 | 104×163 |
| 600 М×900 | | 112×150 | 107×140 | 104×138 |
| 600 М×840 | | 105×150 | 100×140 | 97×138 |

*Згідно з ISO 216 та ISO 217.

М — розмір вздовж машинного напряму паперу.

Поля сторінки — це незаповнені ділянки навколо сторінки складання, розміри яких визначаються різницею формату сторінки та формату сторінки складання. Кожна сторінка має чотири поля: верхнє (головкове), нижнє (хвостове), зовнішнє (переднє) і внутрішнє (корінцеве). Оптимальні розміри полів завжди є пропорційними одне до одного.

Рекомендовані розміри сторінки складання та полів для певних форматів залежать від варіантів оформлення видань (перший — найбільш економний, з невеликими полями; другий — звичайний, з середніми полями; третій — покращений, з великими полями) [2; 59].

<div align="right">Таблиця 4</div>

**Рекомендовані розміри полів**

| Формат паперу, см та частка аркуша | Формат сторінки складання, кв. | Розміри полів до обрізування (корінцеве, верхнє, зовнішнє, нижнє), мм |
|---|---|---|
| Перший варіант оформлення | | |
| 60×84/32 | . | 9, 13, 15, 20 |
| 60×90/32 | . | 9, 13, 18, 20 |
| 70×90/32 | . | 9, 13, 18, 23 |
| 75×90/32 | . | 9, 13, 18, 21 |
| 70×100/32 | . | 9, 13, 21, 23 |
| 70×108/32 | $6 \times 7\frac{3}{4}$ | 9, 13, 18, 23 |
| 84×108/32 | $6 \times 9\frac{3}{4}$ | 9, 13, 18, 23 |
| 60×84/16 | $6\frac{3}{4} \times 9\frac{3}{4}$ | 11, 16, 17, 19 |
| 60×90/16 | $6\frac{3}{4} \times 10\frac{1}{2}$ | 11, 16, 17, 20 |
| 70×90/16 | $8 \times 10\frac{1}{4}$ | 11, 16, 20, 25 |
| 75×90/16 | $8\frac{3}{4} \times 10\frac{1}{4}$ | 11, 16, 19, 25 |
| 70×100/16 | $8 \times 11\frac{1}{2}$ | 11, 16, 20, 27 |
| 70×108/16 | $8 \times 12\frac{1}{2}$ | 11, 16, 20, 29 |
| 84×108/16 | $9\frac{3}{4} \times 12\frac{1}{2}$ | 11, 16, 23, 29 |
| 60×84/8 | $9\frac{3}{4} \times 14$ | 13, 18, 21, 30 |
| 60×90/8 | $10\frac{1}{2} \times 14\frac{1}{4}$ | 13, 18, 23, 26 |
| 70×100/8 | $12 \times 17$ | 13, 18, 21, 26 |
| 70×108/8 | $13 \times 17$ | 13, 18, 23, 26 |
| 84×108/8 | $13 \times 20\frac{3}{4}$ | 13, 18, 23, 29 |
| Другий варіант оформлення | | |
| 60×84/32 | $4\frac{1}{4} \times 6\frac{1}{4}$ | 11, 16, 18, 22 |

| Формат паперу, см та частка аркуша | Формат сторінки складання, кв. | Розміри полів до обрізування (корінцеве, верхнє, зовнішнє, нижнє), мм |
|---|---|---|
| 60×90/32 | $4\frac{1}{2} \times 6\frac{1}{4}$ | 11, 16, 20, 22 |
| 70×90/32 | $4\frac{1}{2} \times 7\frac{1}{2}$ | 11, 16, 20, 24 |
| 75×90/32 | $4\frac{1}{2} \times 8\frac{1}{4}$ | 11, 16, 20, 22 |
| 70×100/32 | $5 \times 7\frac{1}{2}$ | 11, 16, 24, 24 |
| 70×108/32 | $5\frac{3}{4} \times 7\frac{1}{2}$ | 11, 16, 21, 24 |
| 84×108/32 | $5\frac{3}{4} \times 9\frac{1}{2}$ | 11, 16, 21, 23 |
| 60×84/16 | $6\frac{1}{2} \times 9\frac{1}{2}$ | 13, 18, 20, 21 |
| 60×90/16 | $6\frac{1}{2} \times 10\frac{1}{4}$ | 13, 18, 20, 23 |
| 70×90/16 | $7\frac{3}{4} \times 10$ | 13, 18, 22, 27 |
| 75×90/16 | $8\frac{1}{2} \times 10$ | 13, 18, 21, 27 |
| 70×100/16 | $7\frac{3}{4} \times 11\frac{1}{4}$ | 13, 18, 22, 30 |
| 70×108/16 | $7\frac{3}{4} \times 12\frac{1}{4}$ | 13, 18, 22, 31 |
| 84×108/16 | $9\frac{1}{2} \times 12\frac{1}{4}$ | 13, 18, 26, 31 |
| 60×84/8 | $9\frac{1}{2} \times 13\frac{3}{4}$ | 16, 20, 23, 33 |
| 60×90/8 | $10\frac{1}{4} \times 14$ | 16, 20, 24, 28 |
| 70×100/8 | $11\frac{3}{4} \times 16\frac{3}{4}$ | 16, 20, 23, 29 |
| 70×108/8 | $12\frac{3}{4} \times 16\frac{3}{4}$ | 16, 20, 25, 29 |
| 84×108/8 | $12\frac{3}{4} \times 20\frac{1}{2}$ | 16, 20, 25, 31 |
| Третій варіант оформлення | | |
| 60×84/32 | $4 \times 6$ | 13, 18, 20, 24 |
| 60×90/32 | $4\frac{1}{4} \times 6$ | 13, 18, 23, 24 |
| 70×90/32 | $4\frac{1}{4} \times 7\frac{1}{4}$ | 13, 18, 20, 27 |
| 75×90/32 | $4\frac{1}{4} \times 8$ | 13, 18, 20, 25 |

| Формат паперу, см та частка аркуша | Формат сторінки складання, кв. | Розміри полів до обрізування (корінцеве, верхнє, зовнішнє, нижнє), мм |
|---|---|---|
| 70×100/32 | $4\frac{3}{4} \times 7\frac{1}{4}$ | 13, 18, 26, 27 |
| 70×108/32 | $5\frac{1}{2} \times 7\frac{1}{4}$ | 13, 18, 23, 27 |
| 84×108/32 | $5\frac{1}{2} \times 9\frac{1}{4}$ | 13, 18, 23, 26 |
| 60×84/16 | $6\frac{1}{4} \times 9\frac{1}{4}$ | 16, 20, 22, 24 |
| 60×90/16 | $6\frac{1}{4} \times 10$ | 16, 20, 22, 25 |
| 70×90/16 | $7\frac{1}{2} \times 9\frac{3}{4}$ | 16, 20, 24, 29 |
| 75×90/16 | $8\frac{1}{4} \times 9\frac{3}{4}$ | 16, 20, 23, 30 |
| 70×100/16 | $7\frac{1}{2} \times 11$ | 16, 20, 24, 32 |
| 70×108/16 | $7\frac{1}{2} \times 12$ | 16, 20, 24, 24 |
| 84×108/16 | $9\frac{1}{4} \times 12$ | 16, 20, 27, 24 |
| 60×84/8 | $9\frac{1}{4} \times 13\frac{1}{2}$ | 18, 22, 26, 35 |
| 60×90/8 | $10\frac{1}{4} \times 13\frac{3}{4}$ | 18, 22, 27, 31 |
| 70×100/8 | $11\frac{1}{2} \times 16\frac{1}{2}$ | 18, 22, 25, 31 |
| 70×108/8 | $12\frac{1}{2} \times 16\frac{1}{2}$ | 18, 22, 27, 31 |
| 84×108/8 | $12\frac{1}{2} \times 20\frac{1}{4}$ | 18, 22, 27, 34 |

Обсяг — це кількість сторінок або аркушів в одному примірнику видання. Розрізняють такі види аркушів: паперовий, фізичний, умовний, авторський, обліково-видавничий. Паперовий аркуш — облікова одиниця виміру кількості паперу, необхідного для друкування видання. Один паперовий аркуш дорівнює двом фізичним аркушам. Фізичний друкарський аркуш — це фізичний обсяг видання, що дорівнює площі аркушу визначеного формату (84×108 см, 84×90 см, 70×100 см та ін.), задрукованого з однієї сторони. Кількість фізичних аркушів у два рази більша за кількість паперових аркушів. Умовний друкарський аркуш — облікова одиниця обсягу видання, площею 60×90 см, при-значена для порівняння обсягу видань різних форматів. Вираження фізичних друкарських аркушів в умовних друкарських аркушах і навпаки здійснюється за допомогою коефіцієнта переведення:

$$K_{np} = \frac{S_{\phi a}}{S_{ya}},\qquad(1)$$

де $S_{\phi a}$ — площа фізичного аркуша, $S_{ya}$ — площа умовного аркуша.

Тоді обсяг видання в умовних друкарських аркушах визначається за формулою:

$$O_{ya} = O_{\phi a} \cdot K_{np},\qquad(2)$$

де $O_{\phi a}$ — обсяг в фізичних аркушах.

Авторський аркуш — облікова одиниця обсягу твору. Один авторський аркуш містить 40000 знаків прозового тексту (в тому числі пробіли, розділові знаки, цифри тощо) або 700 рядків віршованого тексту, або 3000 см$^2$ ілюстрацій.

Обліково-видавничий аркуш призначений для вимірювання обсягу авторського твору з урахуванням матеріалів, доданих видавництвом. Визначається так само, як і авторський аркуш, але враховує об'єкти, що не є наслідком авторської праці (видавничу анотацію, зміст, титульні елементи, вихідні та випускні дані, передмову, дані на обкладинці, палітурці, суперобкладинці, колонцифру тощо).

Обсяг видання в аркушах слугує для обліку робіт, уніфікації видань, підрахунку витрат та здійснення технологічних розрахунків. Кількість умовних та обліково-видавничих аркушів зазначається у випускних відомостях книги. Ще однією обліковою одиницею обсягу видання є кількість сторінок, що зазначається в бібліографічному описі. Від обсягу видання залежать вид та тип покрівельного матеріалу, спосіб фальцювання та комплектування, технологія скріплення книжкового блоку та ін. [35; 51; 59].

*Конструкційні особливості.* Враховується спосіб комплектування, спосіб скріплення книжкового блоку, вид і тип покрівельного матеріалу, наявність та вид додаткових елементів.

Комплектування книжкових блоків відбувається двома способами: вкладанням та підбиранням. Комплектування вкладанням полягає у вкладанні сфальцьованих аркушів один в один, використовується для невеликих за обсягом видань. Накладання сфальцьованих зошитів або аркушів один на один називається комплектуванням видань підбиранням (для видань обсягом понад 80 сторінок) [35].

Способи скріплення поділяються на швейні, безшвейні і комбіновані. При швейному скріпленні використовують дріт або нитки, при безшвейному — клей чи механічні способи. Також виділяють позошитне та поблочне скріплення. При позошитному скріпленні блок повинен бути скомплектованим підбиранням. Тоді кожен зошит один за одним прошивається через фальц і скріплюється (швейне скріплення). При поблочному скріпленні блок комплектується вкладанням або підбиранням і скріплюється за один робочий цикл (швейне, незшивне клейове чи комбіноване скріплення). Поблочний спосіб скріплення є економічнішим, особливо для видань великого об'єму.

Скріплення дротом найчастіше використовується для видань із малим чи середнім терміном використання, зазвичай для брошур і книжок у м'якій обкладинці. Забезпечує високу продуктивність, міцність та низьку собівартість. Є чотири способи шиття дротом: ушивкою, вшиттям, врознім, зустрічними скобами.

При шитті ушивкою дротяні скоби проходять через згин корінця і загинаються всередину книги. Для запобігання корозії використовується дріт із покриттям або іноді латунний дріт. Шиття ушивкою застосовується до блоків, скомплектованих вкладанням із накинутою зверху обкладинкою.

Шиття вшиттям дроту використовується для видань, скомплектованих підбиранням. Прошивання здійснюється дротяними скобами на відстані 4—5 мм від краю корінця. Кінці скоби загинаються паралельно до спинки скоби на корінцевому полі останньої сторінки. Для закриття спинки та ніжок скоби приклеюють обкладинку.

Шиття дротом врознім використовується для видань у палітурках. При цьому скоби загинаються поверх блока на корінець. Шиття врознім може бути як поблочним (для видань, скомплектованих вкладанням), так і позошитним.

Шиття дротом зустрічними скобами використовується для блоків товщиною більше 15 мм, скомплектованих підбиранням. Зазвичай застосовується для виготовлення відкритих календарів. Міцність забезпечується невеликою відстанню між ніжками скоб (не менше 5 мм).

Шиття блоків нитками є одним з найбільш розповсюджених та надійних способів скріплення, дозволяє обробляти зошити на поопераційному обладнанні та потокових лініях. Є чотири способи шиття нитками: впрострочку, вшиттям, позошитне, позошитне на марлі.

Шиття нитками впрострочку використовується для видань невеликого обсягу, скомплектованих вкладанням. Прошивання здійснюється неперервним швом вздовж усього згину.

Шиття вшиттям ниток застосовується до блоків, скомплектованих підбиранням. Прошивання здійснюється вздовж усього корінця з відступом від краю 4—5 мм.

Позошитне шиття полягає у послідовному прошиванні корінцевих згинів. Відбувається не лише зшиття кожного аркуша, а й зошитів між собою. Допускається зшиття блоків на корінцевому матеріалі, наприклад, на марлі (для книг у палітурках), та без нього (для книг в обкладинках та палітурках).

Незшивне скріплення здійснюється за допомогою клейових плівок чи різних механічних пристроїв.

Незшивне клейове скріплення реалізується різними клеями та у різний спосіб і поділяється так: з повним зрізанням, частковим зрізанням і без зрізання корінцевих фальців.

Незшивне клейове скріплення з повним зрізанням корінцевих фальців при застосуванні полівінілацетатної дисперсії (холодне скріплення) здійснюється шляхом утворення клейової плівки внаслідок випаровування води з клею та його часткового вбирання папером. Обладнання, що використовується для незшивного клейового скріплення з повним зрізанням корінцевих фальців при застосуванні термоклеїв (гаряче скріплення), повинно мати підігрівач бачка з клеєм, а при застосуванні полівінілацетатної дисперсії (холодне скріплення) — стіл з підігрівом чи сушильну секцію. Загалом більшість операцій при холодному та гарячому скріпленнях є однаковими чи подібними, тож часто використовують універсальне обладнання, яке можна переналаштовувати.

Незшивне клейове скріплення з частковим руйнуванням корінцевих фальців характеризується вищою міцністю порівняно з попереднім способом, за рахунок збереження частини фальців. При цьому клей склеює зошити між собою та проникає у прорізи, склеюючи аркуші. Блоки комплектуються підбиранням, а обсяг зошитів становить 8 чи 16 сторінок. Можливі такі способи: скріплення блоків за допомогою шнурів чи ниток у прорізах на корінці; скріплення блоків з прорізами, канавками поперек корінця; «флекстабіль», скріплення блоків з вирубуванням окремих зон; скріплення блоків з зошитів, перфорованих за корінцем.

Незшивне клейове скріплення без руйнування корінцевих фальців поділяється на такі способи: скріплення однозгинних зошитів,

скріплення дво- і тризгинних зошитів у корінцевих згинах вузькими смужками рідкого холодного клею, скріплення зошитів за корінцевими фальцами з використанням термоклею, незшивне клейове скріплення з попереднім нанесенням на середину корінцевих полів смужок холодного поліамідного клею.

Механічні незшивні способи скріплення призначені для з'єднання блоків за допомогою механічних замків. При цьому блоки складаються з окремих аркушів. Застосовуються для виготовлення дитячих видань, рекламних проспектів, каталогів, альбомів та ін.

Швейно-клейове скріплення (комбіноване) здійснюється за допомогою термониток. При цьому блоки повинні бути скомплектовані підбиранням. Шиття відбувається при фальцюванні [9; 35; 75].

Виділяють чотири типи обкладинок та п'ять типів палітурок.

Обкладинка — зовнішній покрівельний матеріал, що скріплюється з книжковим блоком без застосування форзаців. Розрізняють чотири типи обкладинок.

Тип 1 — проста обкладинка для покриття блока наопашки. Застосовується для видань обсягом до 64 сторінок. Обкладинка складається з одного аркушу, який накидається на зошит і скріплюється з ним дротяними скобами чи нитками. Блок при цьому зазвичай комплектується вкладанням. Найчастіше виготовляється з паперу, який, для збільшення довговічності, може бути покритий прозорим полімерним шаром з однієї або двох сторін.

Тип 2 — проста обкладинка для звичайного покриття блока. Скріплюється з книжковим блоком шляхом приклеювання по корінцю. Містить подвійне бігування. На відміну від обкладинки типу 1 може бути покрита прозорим полімерним шаром тільки з зовнішнього боку (щоб була змога приклеїти її по корінцю). Блок комплектується підбиранням. При відкриванні основне навантаження припадає на біги, тому може виникати відривання обкладинки від книжкового блока.

Тип 3 — проста обкладинка для покриття блока врозпуск. Скріплюється з книжковим блоком шляхом приклеювання. При цьому клей наноситься не лише на корінець, а й на бокові сторони блоку (на кілька міліметрів корінцевого поля першої та останньої сторінки). Блок комплектується підбиранням. Такий тип обкладинки найбільш поширений, адже є довговічнішим за тип 2. Це пов'язано з наявністю чотирьох бігувань.

Тип 4 — складена обкладинка з обкантованим корінцем. Матеріалом боковин може бути папір або палітурний картон, а як обкантовку використовують палітурний матеріал. Характеризується значно вищою міцністю та складністю виготовлення порівняно з обкладинками типів 1, 2 та 3. Зазвичай застосовується для видань великого обсягу.

Тобто, за конструкцією обкладинки типи 1, 2, 3 складаються з однієї деталі, а обкладинка типу 4 з боковин обкладинки та обкантовки [13; 35].

Палітурка — цупкий, захисний зовнішній покрівельний елемент книги, який скріплюється з книжковим блоком за допомогою форзаців. Розрізняють п'ять типів палітурок.

Тип 5 — складена. Складається з картонних боковин, корінця, відставу, вкритих різними покрівельними матеріалами. Для видань товщиною до 10 мм можна виготовляти без розставів. Використовується для дитячої, художньої, наукової літератури, підручників для середньої школи, невеликих за обсягом довідників. Характеризується невисокою собівартістю, достатньою міцністю та великими можливостями оформлення. Може задруковуватися з подальшим лакуванням чи припресовуванням плівки, що підвищує довговічність та стійкість до стирання.

Тип 6 — палітурка з однієї деталі. Складається з одного суцільного матеріалу. Зазвичай використовується для довідникових видань малого формату та для паперово-білових виробів.

Тип 7 — суцільнокрита. Складається з картонних боковин та відставу, вкритих суцільним покрівельним матеріалом. Для видань товщиною до 10 мм можна виготовляти без розставів. Характеризується невеликою собівартістю та більшою міцністю, порівняно з палітурками типів 5, 8 та 9. Завдяки своїм характеристикам отримала широке застосування, зокрема для покриття підручників для закладів вищої та професійно-технічної освіти, передплатних, науково-популярних, наукових видань.

Тип 8 — палітурка з накладними боковинками і накладним корінцем. Складається з картонних боковинок, відставу, накладних боковинок, обкладених покрівельним матеріалом з усіх сторін, та накладного корінця. Характеризується невисокою міцністю, середньою собівартістю, привабливим виглядом. Використовуються при виготовленні наукових, науково-довідкових, науково-популярних видань.

Тип 9 — палітурка з накладними боковинками і обкантованим корінцем. Складається з картонних боковинок, накладних боковинок та обкантовувального матеріалу. Зазвичай слугує для покриття під-ручників для початкової та середньої школи, довідкових та науково-популярних видань [35].

Видання також може мати додаткові елементи: форзаци (для скріплення книжкових блоків з палітурками та оформлення видань), ілюстрації (для оформлення видань). Форзаци поділяються за такими характеристиками: характером оформлення (прості незадруковані, виготовлені з кольорового паперу та незадруковані, фонові, декоративно-орнаментні, тематичні), кількістю задрукованих сторін (односторонні, двосторонні), фарбовістю (однофарбові, двофарбові, багатофарбові), способом приєднання (приклейні, пришивні, прошивні), конструкцією (суцільнопаперові, обкантовані, прикантовані, накидні, складені). Додаткові ілюстративні елементи класифікуються залежно від місця розміщення в зошиті та способу приєднання до нього: приклейки (прості, складнофальцьовані, з окантуванням, в рамку, на стержень, на паспарту (на стержні, з плюром), з відігнутим фальцем, з бігуванням), вклейки (в роз'єм зошитів, з розрізуванням фальців зошитів, прості, складнофальцьовані), накидки, вкладки, окремий зошит. Додатковими елементами також можуть бути частини аркушів — зошити з іншою кількістю сторінок, ніж основні. Обсяг додаткових зошитів повинен бути кратним чотирьом [35].

*Умови експлуатації.* Цей фактор містить дві основні складові: термін та інтенсивність експлуатації книжкових видань.

Термін служби — це календарний час експлуатації видання чи його довговічність, які залежать від конструкційних особливостей, інформаційної цінності, місця його використання та вікової категорії читачів [35].

За віковою категорією видання поділяються на ті, що призначені для дорослих читачів та для дітей. Крім того дітей-читачів поділяють за швидкістю читання (досвідчені та читачі-початківці) та за віком (дошкільнята (до 6 років включно), читачі молодшого шкільного віку (від 7 до 10 років), читачі середнього шкільного віку (від 11 до 14 років) та читачі старшого шкільного віку (від 15 до 17 років). Звісно, що зазвичай видання для дітей молодшого шкільного віку будуть менш довговічними, ніж для старшого [70].

Виділяють малий (до 2 років), середній (до 5—10 років) та великий (до 20 років і більше) термін служби видань [35].

За терміном використання також розрізняють видання для тривалого, разового та разового тривалого користування. Видання для тривалого користування можуть неодноразово перечитуватися од-ним чи кількома читачами впродовж великого проміжку часу. Наприклад, мистецькі видання, літературна класика, вузькопрофільні видання. Видання для разового використання актуальні недовготривалий період, зазвичай використовуються лише один раз та втрачають своє функціональне призначення. До таких видань належать програми святкових заходів, концертів, конференцій тощо. Видання для разового тривалого користування призначені для читання одним користувачем один раз протягом відносно тривалого часу, після чого можуть бути використані іншим читачем за тим самим принципом. Сюди відносяться навчально-методичні матеріали, зокрема методичні вказівки, робочі програми тощо [70].

Інтенсивність експлуатації визначається числом подвійних перегинів елементів книги. Чим більша кількість перегинів, тим вищою є інтенсивність експлуатації. Розрізняють малу та велику інтенсивність. При чому вона не залежить від терміну служби, адже, наприклад, при малому терміні служби інтенсивність експлуатації може бути як малою, так і великою [35].

*Тип виробництва.* Тип виробництва — багатоскладова характеристика організаційного та технічного рівнів виробництва, яка поширюється на обсяг виробництва, номенклатуру продукції, характер завантаження робочих місць, випуск однотипної продукції, собівартість продукції та кваліфікацію робітників. Іншими словами, це рівень постійного завантаження робочих місць однотипною роботою. Розрізняють такі типи організації виробничого процесу: одиничне, серійне, масове та змішане.

Одиничне виробництво характеризується високою собівартістю продукції, тривалим терміном виробництва, великою часткою ручної праці та відсутністю закріплених операцій за робочими місцями. Впроваджується при наявності великої кількості номенклатур при невеликих тиражах, зазвичай для випуску книг на замовлення (Book on Demand).

При серійному виробництві виготовляється обмежений асортимент продукції, тобто робота проводиться з певними партіями. Характеризується значною механізацією праці, паралельно-послідовним переміщенням предметів праці, закріпленням періодично повторюваних операцій за визначеними робочими місцями, великою номенклатурою, однак меншою, ніж при одиничному виробництві. Собівартість книжкової продукції також нижча, ніж при одиничному виробництві. Буває дрібно-, середньо- та великосерійне виробництво.

Масове виробництво характеризується виготовленням книжкової продукції великими накладами на вузькоспеціалізованих робочих місцях. Характерною є висока механізація, автоматизація виробничого процесу та значно нижча собівартість продукції. Можливе використання потокових ліній [38].

*Матеріали*. Основними матеріалами, характеристики яких враховуються при проєктуванні післядрукарських процесів, є вид і параметри паперу, на якому друкується наклад; палітурні матеріали; матеріали для скріплення книжкових блоків; оздоблювальні матеріали та ін.

Загальними для всіх видів паперу є такі вимоги:

— достатня механічна міцність;

— незасміченість;

— однорідність товщини, щільності та структури в межах однієї партії та всередині кожного аркуша;

— вологість 6—8 %;

— чітка прямокутна форма аркушів (допустимі відхилення косини не більше 2 мм).

Будову та структуру паперу характеризують такі параметри, як товщина, маса квадратного метра, щільність, пористість.

Товщина є основною характеристикою, що впливає на механічні та оптичні властивості паперу. Папір товщиною від 0,03 до 0,25 мм використовують для друкування (зазвичай 0,07—0,1 мм). Матеріал із більшою товщиною, але до 3 мм називається картоном. Товщина паперу визначає масивність видання та його економічні показники. Для прикладу, від товщини корінця книги залежать витрати палітурних матеріалів. Чим тонший папір, тим компактніший книжковий блок.

Ще одним важливим показником характеристики паперу є маса квадратного метра, яка пропорційна середній товщині паперу. Для друкування використовують папір масою від 30 до 250 г/м². Матеріал масою більше 250 г/м² називають картоном. Сорти паперу однакової маси можуть мати різну товщину та щільність.

На щільність паперу впливає кількість наповнювача, ступінь розмелу волокон, каландрування паперу тощо. Для друкування використовують папір щільністю від 0,5 до 1,35 г/м². Щільність паперу пов'язана з його пористістю.

Пористість — це ступінь присутності порожнин у міжволокнистому просторі паперу. Чим більша пористість, тим вища вбирна здатність паперу та, відповідно, швидкість всотування фарби. Однак при значній пористості зменшується контрастність друкарських відбитків.

Неоднорідність структури пов'язана з технологічними особливостями виготовлення паперу і спостерігається між поперечними і поздовжніми волокнами. Поперечному напрямку притаманні менша цупкість, більше розширення структури при зволоженні (ніж видовження структури при зволоженні у поздовжньому напрямку). Визначення напрямку паперу здійснюється за допомогою дослідження на надрив, де при перпендикулярному розриві аркуша в поперечному напрямку виникає рваний надрив, а в поздовжньому — гладкий. Також різниться лицевий та зворотній бік паперу. Лицевий бік гладкіший, адже при виготовленні зворотній бік контактує з сіткою. Ці особливості важливі не лише при друкуванні, а й при брошурувально-палітурних процесах. Наприклад, фальцювання краще відбувається вздовж напрямку відливу. Для уникнення поперечних складок і хвилеподібності поверхні блоку напрям волокон для книжкового видання повинен бути паралельним корінцю блоку.

Важливим показником характеристики поверхні паперу є гладкість. Чим вища гладкість, тим краща якість віддрукованих зображень, тому для друкування високоякісних ілюстраційних видань використовують гладкий крейдяний папір.

До механічних властивостей паперу належать міцність та деформація. Міцність — це здатність паперу чинити опір руйнуванню під дією механічних сил. Властивості міцності та деформації залежать від складу та структури: наявності наповнювача, поверхневої проклейки, ступеня розробки рослинних волокнистих напівфабрикатів та каландрування, вологості паперу. При дослідженні міцності паперу послуговуються такими характеристиками: міцність на розрив і видовження, міцність на згин, міцність на надрив, міцність поверхні до стирання. Деформація паперу виникає під дією навантаження. Розрізняють зворотну (зникає при відсутності тиску) та незворотну (залишається після припинення навантаження) деформацію. Кожна з них використовується для певних цілей, наприклад, при тисненні на палітурці потрібно, аби рельєф залишався, а не зникав з часом, а при високому друку — навпаки. За характером деформація поділяється на пружну, еластичну та пластичну. Пружність — це властивість, що дозволяє паперу змінювати свою форму під час дії механічних сил, а після припинення цієї дії миттєво відновлювати початкову форму. Еластичність дозволяє поступово відновлювати форму після припинення дії механічних сил. Пластична деформація є незворотною, адже папір не може відновити свою початкову форму після усунення напруги.

До оптичних властивостей паперу належать білизна, глянець, прозорість, світлопроникність. Білизна — здатність паперу рівномірно відбивати світло, характеризується коефіцієнтом відбивання (відношення кількості відбитого світла поверхнею паперу до кількості світла, що падає на цю поверхню). В цілому білизна паперу коливається від 60 до 98 %, а оптимальною для читання вважається від 70 до 80 %. Глянець — частково дзеркальне відбивання світла від поверхні. За цим показником папір буває глянцевим (глянець може доходити до 75—80 %) і матовим (до 30 %). Прозорість — один з випадків світлопроникності, який визначає здатність паперу пропускати крізь себе світло без розсіювання. Зазвичай це негативне явище, яке призводить до видимості надрукованого на зворотному боці паперу.

Друкарський папір поділяється на групи за певними ознаками:

— призначенням: для офсетного, високого, глибокого способів друку;

— форматом: аркушевий, рулонний;

— видом друкарської продукції: книжково-журнальний, газетний, картографічний та ін.;

— волокнистим складом;

— масою метра квадратного тощо.

Класифікація паперу у різних країнах відрізняється [19].

Картон у поліграфії використовується для виготовлення палітурок (палітурний картон), суцільнокартонних обкладинок (кольоровий пресшпан), упаковки різного виду (крейдяний хром-ерзац коробковий). Також використовують гофрований картон. Поверхня палітурного картону повинна бути гладкою, рівною, нежолобленою, без плям і складок. Є чотири марки палітурного картону: А (для ручного та механічного виготовлення палітурок, для виготовлення палітурок з приклеєним ззовні покривним матеріалом), Б (для виготовлення футлярів книг, палітурок до малоформатних видань, палітурок з приклеєним ззовні покривним матеріалом), В (для виготовлення суцільнокартонних палітурок типу 6 без поверхневої проклейки), Г (для виготовлення палітурок з приклеєним ззовні покривним матеріалом) [19].

Форзацний папір використовується для виготовлення форзаців. Щільність форзацного паперу становить 900 кг/м³, а ступінь проклейки 1 мм. Якщо вказані показники вищі спостерігається ускладнення фальцювання, погіршення сприйняття клею та скручування. Небажаною також є підвищена пористість, що призводить до надмірного намокання при склеюванні. Неоднорідність щільності спричиняє жо-лоблення, утворення пухирців та зморшок. Недостатня міцність унеможливлює задруковування. При намащуванні клеєм деформація і нахил до скручування повинні бути мінімальними. В цілому форзацний папір повинен бути достатньо міцним на згин та розрив, адже забезпечує довговічність видання. Форзацний папір поділяється на дві марки: А (для незадрукованих форзаців) та О (для багатофарбових форзаців).

Для виготовлення обкладинок та обклейки палітурок використовують обкладинковий папір трьох марок: А (глазурований), Б (матовий), В (містить волокна деревної маси та поступається міцністю маркам А та Б). Такий папір не повинен змінювати розмір при зволоженні і повинен мати високу міцність на розрив і згин.

Папір для відставу має масу 210 г/м². Повинен бути пружним, щільним, цупким, не ламким.

Для склеювання корінця книжкового блоку використовують міцний, непроклеєний та неглазурований папір, який добре сприймає клей.

Покривні матеріали повинні відповідати таким вимогам:

— мати високу міцність на надрив, розрив та стирання;

— мати достатню щільність, аби глибоко не всмоктувати клеї та фарби;

— витримувати багаторазові згини впродовж тривалого часу;

— бути водо- та світлостійкими;

— сприймати друкарські фарби та тиснення фольгою;

— мати естетичний вигляд тощо.

Основними властивостями палітурного матеріалу є колір, яскравість, художньо-технічні елементи, що визначаються призначенням та змістом видання. За видом основа покривних палітурних матеріалів може бути ткана, паперова та неткана [19; 35].

Зазвичай як тканеву основу покривних палітурних матеріалів використовують міткаль — міцну тканину простого полотняного переплетення, що слугує основою для виготовлення палітурного коленкору й ледерину. Також можуть використовувати дук (сильно апретована, товста бавовняна тканина, з рідким полотняним переплетенням; призначена для виготовлення суцільнотканинних оправ високохудожніх видань), рогожку (міцна, груба бавовняна тканина, з рідким полотняним переплетенням, зафарбована у природні кольори волокон; призначена для виготовлення суцільнотканинних оправ високохудожніх видань) та шифон (міцна і тонка шовкова тканина; використовується для приклейки форзаца та як стержні для вкле-йок). Коленкор — це бавовняна тканина полотняного переплетення, що просочена розчином з крохмального клею, каоліну та барвника. Коленкор не використовують для видань із тривалим терміном користування, адже швидко брудниться, а від надмірної вологості може пліснявіти та загнивати. Ледерин — це палітурний матеріал з нітроцелюлозним покриттям, який виготовляється на основі міткалі з просоченням крохмально-каоліновим розчином і нітроцелюлозним покриттям на лицевому боці. Йому властива міцність, водостійкість, світло- і термостійкість, стійкість до згинання. Ззовні нагадує шкіру. Має підвищену жорсткість, тож вимагає використання дуже липкого клею. З часом спостерігається старіння цього матеріалу, що призводить до підвищення жорсткості, крихкості та руйнування згинів.

До покривних матеріалів на паперовій основі належать: ледерин на папері, папвініл, тевін та ін. Ледерин на папері — це міцний ізоляційний папір, що виготовляється з волокон небіленої сульфатної хвойної целюлози, покритий шаром нітроцелюлозної сульфатної плівки. Значно дешевший, ніж ледерин на тканевій основі. Зазвичай застосовується для виготовлення палітурок малоформатних видань. Папвініл — це матеріал з полівінілхлоридним покриттям, що має високу водостійкість, міцність до стирання і згинання, однак з часом може розтріскуватися. Зовні подібний до шкіри. Тевін — покривний матеріал з вініловим покриттям, має широку колірну гаму та витримує понад 2000 подвійних згинів.

До покривних матеріалів на нетканій основі належать: неткор, сінтоніт, сканвініл, ламінар, маленіт. Неткор покритий крохмально-каоліновим покриттям, а його нетканева основа складається зі склеєних між собою лавсанових і віскозних волокон. Використовується для виготовлення палітурок масових видань. Сінтоніт покритий нітроцелюлозою, зовні подібний до ледерину, використовується для виготовлення палітурок об'ємних видань. Сканвініл покритий поліхлорвінілом, за властивостями нагадує папвініл, використовується для оформлення палітурок цінних видань. Ламінар — дубльований палітурний матеріал, який складається з несклеєних нетканих та паперових полотен, використовується для виготовлення палітурок художніх та наукових видань. Основа маленіту виготовляється з нетканого нітроцелюлозного полотна з відходів низькосортної бавовняної пряжі.

Виділяють також покривні матеріали без основи — пластмасова плівка товщиною від 0,2 до 0,45 мм, яку використовують для виготовлення паперо-білових товарів, а не книжкових видань.

Для виготовлення оправ ювілейних та подарункових альбомів можуть застосовувати шкіру. Це ефектний, однак дорогий матеріал. Найчастіше використовують гладку козячу шкіру товщиною від 0,4 до 1 мм — сап'ян. Окрім звичайного, буває ще левантський сап'ян — зі шкіри гірського козла. Сап'ян з тисненням на лицевому боці називають шагреневою шкірою. Також застосовують товсту, м'яку телячу шкіру (опойок) та шкіру жирового дублення, що отримують зі шкір лосів, оленів, овець та диких кіз (замшу) [19].

Для скріплення книжкових блоків використовують дріт, нитки, термонитки, марлю, каптал, матеріал для обклейки корінців, клеї тощо.

Для скріплення ушивкою застосовують дріт поліграфічний або стальний низьковуглецевий загального призначення, іноді — латунний дріт. Для видань, скомплектованих вкладанням і масою паперу основного тексту до 80 г/м$^2$ діаметр дроту варіюється в межах від 0,4 до 0,7 мм залежно від товщини блоку. Якщо маса паперу більша 80 г/м$^2$, то діаметр дроту від 0,45 до 0,7 мм. Маса 1000 м дроту залежить від діаметру дроту і обирається в межах від 5,9 до 39,45 кг. При шитті дротом в рознім можуть використовуватися бавовняна поліграфічна марля НШ та дротяні скоби кількістю від 2 до 4 шт, залежно від висоти видання. Під час шиття вшиттям дроту використовують від 2 до 3 скоб (залежно від висоти видання) з товщиною дроту від 0,4 до 0,85 мм (залежно від товщини корінця). Шиття дротом зустрічними скобами передбачає розміщення скоб на відстані не менше ніж 5 мм одна від одної.

Стальний дріт, що використовується у поліграфії, повинен мати однакову товщину, гладку блискучу поверхню, бути м'яким та гнучким. Для уникнення корозії дріт можуть покривати тонким шаром міді, олова, цинку або лаку.

Для зшивання зошитів також застосовують бавовняні нитки, синтетичні та термонитки. Бавовняні нитки складаються з шести скручених між собою ниток, просочених крохмальними речовинами. Вони мають стабільні властивості, практично не плутаються, не обриваються, не розрізають папір під час зшивання. Синтетичні нитки удвічі міцніші за бавовняні, хоча й значно тонші та економічніші, однак дорожчі та можуть різати папір при зшиванні, плутатися та розтягуватися. Майже не обриваються та не торочаться. Виготовлені з поліамідних полімерів. Бавовняні та синтетичні нитки іноді поєднують. Термонитки застосовують для скріплення блоків у корінцевих фальцах. Вони виготовлені з віскозного шовку, покритого поліпропіленом. Використання термониток уможливлює автоматизацію брошурувально-палітурних процесів [19; 35].

Поліграфічна марля — це бавовняна тканина з рідким полотняним переплетенням. Виробляється двох марок: НШ та БО. Марля НШ застосовується для шиття ниткошвейними машинами. Є добре апретованою та просоченою клеєм, що забезпечує достатню жорсткість. Марля БО характеризується меншою жорсткістю та використовується для наклеювання корінця у блокооброблних агрегатах. Попри невисоку міцність вона значно зміцнює корінець. Замінником марлі БО може бути мікрокрепірований папір [19].

Для наклеювання видань на корінець у блокооброблних агрегатах та для окантовки корінця при безшвейному клейовому скріпленні чи скріпленні термонитками можуть використовувати мікрокрепірований папір.

В обклеювально-каптальних машинах і агрегатах для обклейки корінця використовують папір з сульфатної целюлози масою від 60 до 80 г/м².

Для видань обсягом понад 10 аркушів використовують каптал, який являє собою стрічку шириною від 13 до 15 см з потовщеним краєм в 1,5—2 мм і виготовляється тканням різнокольорових шовкових, напівшовкових та бавовняних ниток [19; 35].

При виготовленні поліграфічної продукції також застосовують клеї, які повинні легко і рівномірно розмащуватися, добре змочувати матеріал, мати високу швидкість скріплювання, бути світлими, щоб не залишати плям, не вступати в хімічні реакції з матеріалами, що скріплюються, не пліснявіти, не старіти та ін. Усі палітурні клеї поділяються на групи: водяної дисперсії (латексний, ПВАД та ін.), водяних клейових розчинів (кістковий, крохмальний, декстриновий), термопластичних полімерів (термоклеї), у вигляді розчинів у органічних розчинниках, термореактивні клеї. До клеїв рослинного походження належать крохмальний та декстриновий клеї. Клеями тваринного походження є кістковий, казеїновий. Синтетичні клеї: полівінілацетатний, епоксидний, латексний на основі бутадієнстирольного каучуку, карбоксиметилцелюлозний, термоклей, клеї у вигляді розчинів у органічних розчинниках. В цілому вибір клею залежить від характеру поліграфічного матеріалу та умов склеювання. Для прикладу, при склеюванні пористого паперу доцільно застосовувати в'язкий клей, а для приклеювання пружного покривного матеріалу — більш липкий.

Для заклеювання корінця можна використати палітурні клеї з хорошою еластичністю та високою міцністю клейової плівки [19].

У подарункових чи мистецьких виданнях часто використовують ляссе — закладку у вигляді шовкової стрічки шириною від 3 до 8 мм, зазвичай червоного кольору. Також ляссе може бути виготовленою з товстого паперу чи пластмаси, мати орнамент чи медальйон.

Для оздоблення палітурних матеріалів здійснюють тиснення палітурною фольгою, лакування, припресовування плівки.

Палітурна фольга використовується для нанесення кольорового чи металевого зображення шляхом тиснення. Фольга може бути на паперовій або лавсановій основі, є багатошаровим матеріалом. Буває кольорова, бронзова, «ювілейна» та голографічна. Кольорова фольга буває різних кольорів та відтінків, що уможливлює втілення найрізноманітніших ідей оформлення поліграфічної продукції. Її поверхня буває матовою та глянцевою. Відбитки є стійкими до впливу зовнішніх факторів. Тиснення бронзовою фольгою візуально нагадує тиснення золотом, однак з часом тьмяніє, тож рідко використовується. «Ювілейній» фользі притаманний хороший блиск, який не тьмяніє з часом. Вона міцно тримається на палітурному матеріалі. Завдяки розсіювання відбитого світла голографічна фольга створює ефект об'ємності зображення. За іншими властивостями нагадує «ювілейну». При виборі фольги слід враховувати її сумісність з іншими. Фольга, виготовлена на водяних розчинах, не буде друкуватися по фользі, що виготовлена на спиртових розчинах. На якість відбитків впливають: питомий тиск; температура штампа; швидкість та час тиснення; вид, характер та вологість покривних матеріалів; характер та площа друкарських елементів штампа; відповідність адгезійного шару фольги поверхні друкарського матеріалу; вид і товщина матеріалу декеля [19; 34; 35].

Лакування служить для додаткового оздоблення поліграфічної продукції, захисту від стирання, підвищення міцності та довговічності. За призначенням лаки поділяються на: ґрунтувальні (створюють адгезійний шар для подальшого нанесення іншого лаку чи фарби), матові, глянцеві, підвищеної стійкості до стирання, для термозварювання за допомогою ультразвуку, для термозварювання за допомогою мікрохвильових пристроїв, для полегшення або ускладнення руху задруковуваного пакувального матеріалу, для каландрування, спеціального призначення. Лакове покриття може наноситися як на сухий, так і на мокрий відбиток на лакувальних машинах або в лакувальних секціях. Для цього застосовують такі типи лаків: друкарські (на масляній основі), дисперсійні, лаки ультрафіолетового закріплення, лаки на основі летких розчинників. Друкарські лаки містять смоли, льняну оліфу, алкіди, сикативи та ін. і закріплюються вибірковим всмоктуванням та окислювальною полімеризацією ненасичених сполук. Головними компонентами дисперсійних лаків є полімери на основі стирола-акрилата. Закріплення відбувається через всмоктування і випаровування води, у зв'язку з цим окремі полімерні частинки зближуються і, внаслідок зростання капілярного тиску, мікрочастинки з'єднуються в однорідну плівку. Лаки ультрафіолетового закріплення поділяються на такі види: радикального і катіонного закріплення (за вийнятком лаків, що накладаються офсетним способом зі зволоженням — засобами радикальної полімеризації). Лаки на основі летючих розчинників закріплюються шляхом випаровування спирту. Основними недоліками цих лаків є надмірна липкість та високий рівень забруднення навколишнього середовища.

Припресовування плівки підвищує вологостійкість матеріалу, міцність, довговічність, надає блиску та естетичного вигляду. Для оздоблення покривних матеріалів застосовують синтетичні полімерні плівки, які повинні бути міцними, безбарвними, прозорими, еластичними, не призводити до скручування та жолоблення відбитку, якнайменше деформуватися в процесі старіння, мати рівномірну товщину. Виокремлюють три способи припресовування плівки: клейовий, безклейовий, спосіб перенесення. Клейовий спосіб полягає у нанесенні на плівку тонкого клейового шару, який висихає під дією інфрачервоних променів. Потім плівка розігрівається разом з відбитком і припресовується до нього. Таким способом оздоблюють видання в обкладинках типів 1, 2, 3, 4 та у палітурках типів 6, 7, 8, 9, а також суперобкладинки. При цьому використовують ацетилцелюлозні, поліпропіленові, поліетилентерефталатні плівки та клеї (розчин полімерів у летких органічних розчинниках), латекси (водяні дисперсії полімерів). Вид і склад клею обирається залежно від плівки та паперу. Припресовуванням плівки безклейовим способом називається процес з'єднання поліграфічної продукції з термопластичними полімерами чи плівками (поліетилтерефталатними, поліамідними, целофановими та ін.) із нанесеним заздалегідь клейовим шаром. Плівка нагрівається, підплавляється і припресовується до поверхні матеріалу. Такий спосіб використовується для оздоблення видань у палітурках типу 5, 7, 8, 9. За способом перенесення на відбиток наноситься прозора плівка (поліетилентерефталатна) на основі. Згодом основа відділяється та може бути використана повторно. Цей спосіб використовують для оздоблення обкладинок та суперобкладинок [19; 35].

*Тип обладнання*. На основі ключових характеристик видання та схеми технологічного процесу здійснюється вибір обладнання.

Для кожної операції може бути обране специфічне обладнання. Також обладнання може бути універсальним, з можливістю зміни налаштувань.

Для розрізування і підрізування паперових аркушів чи палітурних матеріалів використовують одноножеві паперорізальні машини. При відсутності спеціалізованого устаткування вони можуть також використовуватися для обрізування книжкових блоків з трьох сторін. В цілому поділяються на три категорії: малі (ширина стопи до 70 см), середні (до 90 см), великі (більше 90 см). Також існують два способи різання: марзанний (ніж у кінці руху врізається у пластмасову деталь, розташовану нижче стопи, — марзан) і безмарзанний (для розрізування використовується ніж та контрніж). Одноножеві різальні машини характеризуються такими параметрами: довжина різу, мінімальна та максимальна відстань від площини подавача до лінії різу, ширина переднього стола, зусилля тиску притискача, швидкість роботи, мінімальна та максимальна швидкість подавача, пружність головного приводу, відстань від підлоги до поверхні стола, маса машини. В якості допоміжних пристроїв паперорізальних машин можуть бути: «повітряна подушка» (пневматична система для полегшення ручного пересування або повороту стопи), гідравлічні стопопідйомники, пристрій для заміни ножа, система вилучення обрізків тощо. Також використовують велику кількість пристроїв для механізації допоміжних операцій з підготовки стопи.

Фальцювання аркушів можливе вручну, однак найчастіше цю операцію виконують механізовано. Фальцювальні машини поділяються на чотири групи: ножові, касетні, комбіновані, спеціальні. У касетних машинах фальцювання відбувається за допомогою касет з упором і рухомих валиків. Ножове фальцювання складається з чотирьох етапів: попереднього рівняння, бічного рівняння, утворення петлі за допомогою ножа, обтискання валиками місця згину. Ножові фальцапарати використовуються зазвичай в комбінованих фальцювальних машинах. Комбіновані машини мають ножові та касетні фальцапарати: перший згин утворюється в касетній фальцсекції, а всі решта — в ножевих. Спеціальні фальцмашини використовують для фальцюван-ня стосу з 10—15 аркушів, при цьому весь папір згинається за один удар. В цілому фальцмашини складаються з самонакладу, привода, контрольно-блокувальної системи, пневматичної системи.

Пресування та пакування зошитів може здійснюватися за допомогою пакувально-обтисних пресів, які можна поділити на такі групи: пакувально-обтисні преси для обтискування та обв'язування сфальцьованих аркушів, блокообтисні преси для обтискування книжкових блоків та корінців, палітурно-обтисні преси для обтискування книг.

Для автоматизації приклеювання форзаців, ілюстрацій, дробових частин зошита та інших додаткових елементів використовують приклеювальні автомати. Для обкантування зошита з приклеєними форзацами — обкантовувальні автомати. Також автомати можуть бути комбінованими: спочатку виконують приклеювання, а потім обкантовування.

Для комплектування блоків вкладанням використовуються вкладально-швейні або вкладально-швейно-різальні автомати. Комплектування блоків підбиранням здійснюється на аркушепідбиральних машинах, які повинні забезпечувати послідовність, комплектність зошитів і хороше зіштовхування [35; 75].

Для шиття блоків дротом використовують дротошвейні машини: операційні, вкладально-швейні, підбирально-швейні, а також дротошвейні секції вкладально-швейно-різальних агрегатів.

Для шиття блоків нитками застосовують ниткошвейні машини, які можуть бути автоматичними (усі операції виконуються без участі обслуговуючого персоналу) та напівавтоматичними (зазвичай механізовані усі операції крім подачі та розкриття зошитів), універсальними (можуть виконувати брошурне і палітурне шиття різними видами стібків на корінцевому матеріалі і без нього, призначені для видань різних форматів) та спеціалізованими (розраховані на шиття простим брошурним стібком без марлі видань обмеженого формату).

Залежно від технології, машини безшвейного скріплення можуть включати такі основні вузли та пристрої: пристрій введення блока в затискачі транспортера, вирівнювальний пристрій, фрезерна секція, торшонувальна секція, клейовий апарат для нанесення клею на корінець книжкового блока, клейовий апарат для нанесення смужки клею на бокові зошити блока, сушильний пристрій, охолоджувальний пристрій, секція подачі і приклеювання обкладинки, обкантувальна секція, обтискуючий пристрій, пристрій виведення блока з машини, транспортувальний пристрій [75].

Фальцювання і шиття термонитками відбувається на фальцювальному автоматі, який можна підключати до касетних і комбінованих фальцмашин.

Заклеювання корінця книжкового блоку може проводитися на різному обладнанні, наприклад, на блокозаклеювальному верстаті неперервної дії [35; 76]. Сушіння корінців рекомендовано проводити в спеціалізованих сушильних пристроях, конвекційним способом (повітря кімнатної температури або нагріте, що подається вентиляторами), радіаційно-конвекційним способом (теплоносієм є повітря та електромагнітні хвилі інфрачервоного і видимого діапазонів), опроміненням. Після сушіння здійснюється обтискування корінців у пресах з гідравлічним приводом пресувальної колодки [35].

Обрізування книжкових блоків з трьох сторін зазвичай відбувається на спеціальних різальних машинах, які можна поділити на такі види: одноножеві з поворотним столом, триножеві однопозиційні, для поштучного обрізування блоків.

Оздоблення корінця здійснюється на потокових лініях зі спеціальними секціями або на поопераційному обладнанні. Верстат для зафарбовування обрізів зазвичай складається з самонакладу, секції зволоження обрізів, секції зафарбовування обрізів, сушильного пристрою, приймального транспортеру. Золочення обрізів проводиться тисненням спеціальної фольги на обладнанні, у якому обріз блока шліфують, полірують, при потребі ґрунтують і припресовують фольгу за допомогою нагрітого валика з термостійкої гуми.

Закруглювання корінця блока здійснюється на закруглювальних машинах, в секціях блокообробних агрегатів потокових ліній, на заокруглювально-відгинальних автоматах. Після закруглювання корінця блок передається на відгинання фальців [35; 76].

Приклеювання лясе на великих підприємствах проводиться на спеціалізованих автоматах, які можна підключати до потокової лінії, призначеної для оброблення видань покращеного типу. Приклеювання капталів, паперової смужки і корінцевого матеріалу може відбуватися вручну, на напівавтоматах, на блокообробних агрегатах і автоматах [35].

В цілому для заклеювання та сушіння, кругління корінців, відгинання фальців, приклеювання до корінця зміцнювальних елементів можуть використовувати найрізноманітніше устаткування: операційні машини, що призначені для виконання лише однієї операції (заклеювальний верстат, блокообтискний прес, закруглювальний верстат тощо), машини для двох операцій (заклеювально-сушильні, закруглювально-каширувальні тощо), агрегати для трьох і більше операцій, потокові лінії [76].

Для розрізування задрукованої та незадрукованої аркушевої та рулонної продукції, паперу, картону, палітурних та обкантувальних матеріалів, марлі, полімерної плівки застосовують заготовельно-розкрійне устаткування: аркушерізальні, картонорізальні, картонорозкрійні, бобінорізальні, тканинорозкрійні машини.

Палітурки можуть виготовляти вручну, напівмеханізованим і механізованим способом на палітуркоробних машинах. Сучасні палітуркоробні машини є повністю автоматизованими і потребують втручання оператора лише для нагляду за процесом, переналагодження машини на новий тираж палітурок, подання заготовок або заміни бобін відставу, прийняття готової продукції. Більшість машин призначені для виготовлення суцільнокритих палітурок. Палітуркоробні машини класифікуються за напрямком технологічного процесу (з вертикальним, горизонтальним, комбінованим і карусельним ходом технологічного процесу), за характером руху напівфабрикату палітурки (з періодичним і неперервним рухом), за швидкістю (середньошвидкісні, швидкісні, високошвидкісні) [35; 76].

Преси для тиснення на палітурках зазвичай будуються за тигельним принципом: тиск створюється двома пресувальними плитами, одна з яких нерухома, а інша має зворотно-поступальний рух. Преси класифікуються за конструкцією (з горизонтальною і вертикальною площиною тиснення), призначенням (для опрацювання великих чи малих тиражів), ступенем механізації (автоматичні, напівавтоматичні), принципом будови (тигельні, плоскодрукарські, ротаційні), технологічним призначенням (легкого і важкого типу). Усі преси, як правило, складаються з механізму тиснення, станини, фольгоподавального механізму, пристрою підігріву штампу, пристрою розміщення палітурки відносно штампу, пристрою регулювання глибини тиснення, приводу.

Вставляння блока в палітурку відбувається за принципом вертикального переміщення блока знизу вверх. Книговставні машини складаються з таких механізмів та пристроїв: поштучної подачі блоків, розкриття блока посередині і базування за товщиною і форматом, вертикального конвеєра з крилами для транспортування блоків, клейових апаратів, самонакладу з пристроєм кругління корінця, базування палітурки перед вставлянням, суміщення блока з палітуркою та їх обтиснення, знімання книги з крила і виведення на приймальний пристрій. Такі машини класіфікують за швидкістю (тихохідні, напівавтоматичні і автоматичні, високошвидкісні), ступенем агрегатування (операційні і агрегатовані), конструкцією (з одним і кількома крилами) [76].

Після вставляння блоків у палітурки здійснюється пресування і сушіння книг, штрихування, обгортання суперобкладинкою, комплектування стосів з книг, упаковування книжкової продукції. Пресування книг здійснюється на палітурнообтискних пресах. Штрихувальне устаткування поділяється на операційне (використовується невеликими і середніми підприємствами) і комбіноване. Зазначеного поширення набули комбіновані пресувально-штрихувальні машини неперервної і періодичної дії, які працюють разом з книговставними машинами. Обгортання книг суперобкладинкою здійснюється вручну з використанням покривної машини або автоматизовано на спеціальних машинах. Пакування книг буває ручним, механізованим (із застосуванням комплектувальних, пакувальних і обв'язувальних машин) і автоматизованим.

У палітурному виробництві також застосовують потокові технологічні лінії, які мають такі переваги: розташування устаткування послідовно виконанню операцій, синхронізація операцій, оперативна передача напівфабрикатів за допомогою транспортно-передавальних пристроїв, періодичність запуску напівфабрикатів на потік, виконання операцій над ними і виведення з потоку. Потокові лінії класифікуються за однорідністю продукції (сталого та змінного потоку), продуктивністю (синхронного та несинхронного потоку), неперевністю руху напівфабрикатів (неперервного та перервного потоку), ступенем автоматизації та механізації (механізовані, комплексно-механізовані, автоматизовані потокові лінії).

Для лакування можуть використовувати спеціалізовані машини для лакування всієї поверхні, системи зволоження офсетних машин для лакування всієї поверхні або окремих ділянок, самостійні лакувальні секції друкарських машин. Устаткування для припресовування плівки до аркушевих матеріалів називають ламінаторами [35; 76].

*Технологічні та економічні розрахунки.* За відповідними формулами визначаються необхідна кількість матеріалів, розмір деталей, термін експлуатації та ін. Відповідно до економічних розрахунків здійснюється вибір оптимального варіанту виготовлення видання [23, 35].

*Схема технологічного процесу* відображає взаємопов'язану послідовність виконання технологічних операцій. Технологічна схема брошурувально-палітурних процесів обирається залежно від технічних характеристик видання (обсягу, формату, особливостей конструкції, призначення видання), накладу, очікуваної собівартості та технічного оснащення поліграфічного підприємства. Розрізняють типові та індивідуальні технологічні схеми. Можлива також адаптація типових схем, враховуючи конкретні умови проєктування та виготовлення поліграфічної продукції [35; 56].Основною метою розроблення будь-якої інформаційної технології є технологізація певного соціально значимого процесу, тобто цілеспрямований вплив на його перебіг із використанням комп'ютерно-обчислювальної техніки. Вихідними даними при цьому є певна недостатньо систематизована інформація. Високий рівень поділу процесів на етапи, системна повнота, регулярність та однозначність сприяють його раціоналізації, завершеності, стандартизації й уніфікації, а, отже, плануванню й прогнозуванню.

Для формалізації наведених знань доцільно розробити онтологію проєктування післядрукарських процесів. Структура онтології безпосередньо впливає на здатність встановлювати оптимальний розв'язок основної чи побічних задач. Ітеративний підхід до створення полягає у поетапному навчанні (наповненні) онтології. Можливе постійне додавання нових класів та зв'язків між ними. У процесі збільшення моделі виникає необхідність оптимізації шляхом видалення застарілих класів.

Існують декілька основних типів онтологій:

— метаонтології: для опису загальних понять, що не належать до предметної області;

— онтологія предметної області: формальний опис та визначення термінологічної бази предметної області;

— онтологія конкретної задачі: визначення термінологічної бази поставленої задачі;

— мережеві онтології: для опису результатів дії об'єктів предметної області чи задачі.

Основними принципами побудови онтологій є:

— «формальна онтологія», запропонована Гуаріно, яка містить теорії частин, цілісності, рівності, залежності, узагальнень та передбачає такі принципи побудови: потреба у розумінні всієї предметної області, чіткість ідентифікації, класифікація структури, встановлення ролей;

— скелетна методологія побудови онтології вручну, запропонована Усолдом та Грунінгером, яка передбачає: встановлення мети та меж, побудову онтології, оцінювання, документування, визначення принципів керування попередніми етапами;

— Ontological Design Patterns (ODPs): для визначення структур, термінів, семантики.

В загальному побудова моделі складається з кількох етапів:

— нагромадження знань про предметну область;

— декомпозиція: розділення досліджуваного процесу на окремі елементи, які стануть основою моделі;

— ідентифікація елементів;

— класифікація: визначення класів та елементів, що до них належать (ієрархія класів);

— опис властивостей;

— присвоєння значень властивостей;

— створення зв'язків;

— розширення та конкретизація онтології;

— перевірка;

— впровадження онтології [33].

Опис предметної області здійснюється за допомогою класів — основних структурних одиниць онтологічної моделі, які можуть містити інші класи та/або екземпляри. Класи — це загальні поняття, колекції, набори об'єктів. Екземпляри виступають суб'єктами. Зв'язок між екземпляром і класом, до якого він належить, задається предикатом rdf: type. Класи в онтології організовуються у таксономію (ієрархічну класифікацію). Так, для прикладу, Обкладинка та Палітурка є підкласами класу Вид_покрівельного_матеріалу, який є підкласом Конструкційні_ особливості: Вид_покрівельного_матеріалу SubClassOf Конструкційні_особливості, Конструкційні_особливості SubClassOf Проєктування_післядрукарських_процесів. Властивості-відношення визначають існуючі зв'язки між екземплярами. Наприклад: Значний_обсяг Визначає Комплектування_підбиранням. Властивості-дані визначають конкретні характеристики екземплярів певних класів. Наприклад, для екземпляра Малий_обсяг класу Обсяг — Кількість_сторінок 64.

Не менш важливими для опису є глосарій та тезаурус. Оформлені належним чином онтологічні словники сприяють полегшенню подальшого процесу створення онтології [73].

Для запису завжди істинних тверджень використовують аксіоми. Редактор Protégé 5 дозволяє використання таких аксіом: аксіоми класів (SubClassOf, EquivalentClasses, DisjointClasses та ін.), аксіоми властивостей об'єкта (SubObjectPropertyOf, EquivalentObjectProperties, InverseObjectProperties, FunctionalObjectProperty та ін.), аксіоми властивостей даних (SubObjectPropertyOf, EquivalentDataProperties, DisjointDataProperties та ін.), індивідуальні аксіоми (ClassAssertion, ObjectPropertyAssertion, DataPropertyAssertion, NegativeObjectPropertyAssertion та ін.), аксіоми анотації (AnnotationAssertion, SubAnnotationPropertyOf, AnnotationPropertyDomain, AnnotationPropertyRange) [78].

### 1.4. Функціональне моделювання проєктування післядрукарських процесів

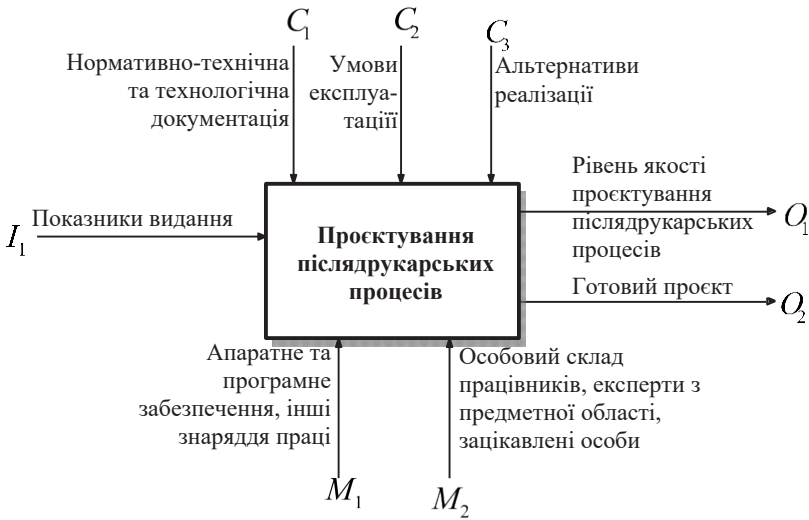

Рис. 2. Контекстна діаграма А-0 моделі IDEF0 проєктування післядрукарських процесів

Використаємо методологію IDEF0 [40; 77] для функціонального моделювання проєктування післядрукарських процесів. Контекстна діаграма зображена на рис. 2. При цьому основною функцією системи є проєктування післядрукарських процесів, а зв'язок системи із навколишнім середовищем зображується граничними стрілками: $I_1$ — показники видання, $C_1$ — нормативно-технічна та технологічна документація, $C_2$ — умови експлуатації, $C_3$ — альтернативи реалізації, $O_1$ — рівень якості проєктування післядрукарських процесів, $O_2$ — готовий проєкт, $M_1$ — апаратне та програмне забезпечення, інші знаряддя праці, $M_2$ — особовий склад працівників, експерти з предметної області, зацікавлені особи.

Проаналізуємо інформаційне навантаження компонент множин граничних стрілок IDEF0 моделі:

Граничні стрілки типу «Вхід» (Input):

— $I_1$ (показники видання). Ключовими показниками книжкових видань є вид, тип, формат та обсяг.

Граничні стрілки типу «Контроль» (Control):

— $C_1$ (нормативно-технічна та технологічна документація). До нормативно-технічної та технологічної документації належать: технічні вимоги та законодавчі положення, зокрема: закони, стандарти, технічні умови, кодекси усталеної практики та ін.

— $C_2$ (умови експлуатації). Умови експлуатації включають термін та інтенсивність експлуатації готового видання.

— $C_3$ (альтернативи реалізації). Парето-оптимальні альтернативи, визначені оцінюванням нечітких відношень на множині альтернатив.

Граничні стрілки типу «Вихід» (Output):

— $O_1$ (рівень якості проєктування післядрукарських процесів). Результатом діяльності, спрямованої на створення проєкту, є відповідний рівень якості.

— $O_2$ (готовий проєкт). Визначає перебіг усіх технологічних дій, направлених на реалізацію післядрукарських процесів.

Граничні стрілки типу «Механізми» (Mechanism):

— $M_1$ (апаратне та програмне забезпечення, інші знаряддя праці). Процес проєктування передбачає використання сучасних технічних та програмних засобів, в тому числі специфічного, вузькопрофільного програмного забезпечення.

— $M_2$ (особовий склад працівників, експерти з предметної області, зацікавлені особи). Проєктування післядрукарських процесів передбачає участь висококваліфікованих працівників, обізнаних із тонкощами реалізації досліджуваних процесів, задля оцінювання вихідних даних та прогнозування результату. Можливе залучення експертів, зокрема науковців та зацікавлених осіб (замовників, маркетологів та ін.).

Діаграма першого рівня декомпозиції А0 моделі IDEF0 містить такі блоки:

— ВКВ (визначення конструкції видання);

— ВВВ (визначення вимог до готового видання);

— ВПО (визначення послідовності технологічних операцій);

— ВРО (визначення режимів опрацювання).

Діаграма другого рівня декомпозиції A1 моделі IDEF0:

— ВСК (вибір способу комплектування);

— ВСС (вибір способу скріплення);

— ВПМ (вибір покривного матеріалу);

— ПДЕ (проєктування додаткових елементів).

Діаграма третього рівня декомпозиції A2 моделі IDEF0:

— ВВКБ (визначення вимог до книжкового блоку);

— ВВДЕ (визначення вимог до додаткових елементів);

— ВВПМ (визначення вимог до покривного матеріалу);

— ВВО (визначення вимог до оздоблення);

— ВВПБО (визначення вимог до покриття блоку обкладинкою);

— ВВВБП (визначення вимог до вставлення блоків у палітурку);

— ВВП (визначення вимог до пакування).

Діаграма третього рівня декомпозиції A3 моделі IDEF0:

— ВПОБП (визначення послідовності операцій брошурувальних процесів);

— ВПОПП (визначення послідовності операцій палітурних процесів).

Діаграма третього рівня декомпозиції A4 моделі IDEF0:

— ВРОБП (визначення режимів опрацювання брошурувальних процесів);

— ВРОПП (визначення режимів опрацювання палітурних процесів).

**Етап 2. Синтез моделей факторів проєктування післядрукарських процесів**

*2.1. Розроблення семантичної мережі взаємозв'язків між факторами проєктування післядрукарських процесів*

На основі експертних суджень формується деяка множина факторів $R=\{R_1, R_2, ..., R_n\}$, що вміщає найбільш суттєві фактори. Для виокремлення характерних чинників процесу залучаються представники наукової спільноти та фахівці-практики. Використання експертного оцінювання дозволяє одержати кількісну оцінку ступеня важливості кожного з факторів, що формують множину значень чинників впливу на якість виконання процесу.

Зв'язки між визначеними факторами, необхідні для формування підґрунтя подальшого опису предметної області, кількісного оціню-вання їх вагових значень та, відповідно, встановлення домінантності, визначаються та візуалізуються на основі теорії графів та семантичних мереж. Вузли семантичної мережі відображатимуть семантику понять, тобто факторів, які згодом будуть представлені у вигляді абстрактних лінгвістичних змінних. Дуги відтворюють функціональні (семантичні) відносини чи зв'язки між ними. Поєднання мовознавства (семантика лінгвістичних змінних) та математики (мережі як варіант графа) забезпечує, з одного боку, використання звичайної мови для опису бази знань досліджуваного процесу, з іншого — уможливлює застосування формальних методів та нечіткої логіки для дослідження, кінцевою метою якого є прогностичне оцінювання. Вузлами семантичної мережі стають елементи множини $R$, а дугами — функціональні зв'язки з певними смисловими навантаженнями $(R_i, R_j)$.

Модель семантичної мережі створює базу для подальшого конструктивного опису предметної області, є наочною та інтуїтивно зрозумілою, адже є аналогом сучасних уявлень про фізіологічні механізми пам'яті людини [59].

### *2.2. Формалізація з'язків між факторами за допомогою предикатних формул*

Логіка предикатів є частиною математичної логіки, її формальна мова представлена термами та взаємовідносинами між ними — предикатами. До термів, як словотвірних елементів, відносять такі конструкції мови предикатів: константи (конкретні реальні об'єкти), змінні (узагальнені можливі об'єкти, у нашому випадку фактори), функції (послідовність констант чи змінних, обмежених круглими дужками), функтори (оператори перед функцією, що повертають певне значення після впливу на об'єкт). Предикатом називають логічну функцію, яка приймає значення «істина», якщо відношення між її аргументами мають смисл, або «фальш» у противному випадку.

Таким чином використання логіки предикатів полягає у виведенні усіх зв'язків між факторами, враховуючи структуру семантичної мережі [59; 61; 72].

Означимо впливи кожного фактора проєктування післядрукарських процесів: $R_1 - R_2$ — визначає; $R_1 - R_3$ — визначає; $R_1 - R_4$ — визначає; $R_1 - R_5$ — обумовлює; $R_1 - R_6$ — обумовлює; $R_1 - R_7$ — формує; $R_1 - R_8$ — обумовлює; $R_2 - R_5$ — визначає; $R_2 - R_6$ — впливає на вибір; $R_2 - R_7$ — формує; $R_2 - R_8$ — обумовлює; $R_3 - R_5$ — впливає на вибір; $R_3 - R_8$ — обумовлює; $R_4 - R_6$ — визначає; $R_4 - R_7$ — формує;

$R_5 - R_6$ — впливає на вибір; $R_5 - R_7$ — формує; $R_6 - R_7$ — формує; $R_8 - R_5$ — обумовлює; $R_8 - R_6$ — визначає; $R_8 - R_7$ — формує.

Для формалізації опису відносин між термами семантичних мереж використано предикатні формули, що включають такі конструкції: $\wedge$ — логічне «і»; $\leftarrow$ — «якщо»; $\forall$ — квантор спільності (для всіх); $\exists$ — квантор існування (існує принаймні одне) [59].

($\forall\ R_i$) [ $\exists$ ( $R_1$, показники видання) $\leftarrow$ визначає ( $R_1, R_2$ ) $\wedge$ визначає ( $R_1, R_3$ ) $\wedge$ визначає ( $R_1, R_4$ ) $\wedge$ обумовлює ( $R_1, R_5$ ) $\wedge$ обумовлює ( $R_1, R_6$ ) $\wedge$ формує ( $R_1, R_7$ ) $\wedge$ обумовлює ( $R_1, R_8$ )];

($\forall\ R_i$) [ $\exists$ ( $R_2$, конструкційні особливості) $\leftarrow$ визначає ( $R_2, R_5$ ) $\wedge$ впливає на вибір ( $R_2, R_6$ ) $\wedge$ формує ( $R_2, R_7$ ) $\wedge$ обумовлює ( $R_2, R_8$ ) $\wedge$ визначається ( $R_2, R_1$ ) $\wedge$ обирається залежно від ( $R_2, R_3$ )];

($\forall\ R_i$) [ $\exists$ ( $R_3$, умови експлуатації) $\leftarrow$ впливає на вибір ( $R_3, R_2$ ) $\wedge$ впливає на вибір ( $R_3, R_5$ ) $\wedge$ обумовлює ( $R_3, R_8$ )];

($\forall\ R_i$) [ $\exists$ ( $R_4$, тип виробництва) $\leftarrow$ визначає ( $R_4, R_6$ ) $\wedge$ формує ( $R_4, R_7$ ) $\wedge$ визначається ( $R_4, R_1$ )];

($\forall\ R_i$) [ $\exists$ ( $R_5$, матеріали) $\leftarrow$ впливає на вибір ( $R_5, R_6$ ) $\wedge$ формує ( $R_5, R_7$ ) $\wedge$ обумовлюється ( $R_5, R_1$ ) $\wedge$ визначається ( $R_5, R_2$ ) $\wedge$ обирається залежно від ( $R_5, R_3$ ) $\wedge$ обумовлюється ( $R_5, R_8$ )];

($\forall\ R_i$) [ $\exists$ ( $R_6$, тип обладнання) $\leftarrow$ формує ( $R_6, R_7$ ) $\wedge$ обумовлюється ( $R_6, R_1$ ) $\wedge$ обирається залежно від ( $R_6, R_2$ ) $\wedge$ визначається ( $R_6, R_4$ ) $\wedge$ обирається залежно від ( $R_6, R_5$ ) $\wedge$ визначається ( $R_6, R_8$ )];

($\forall\ R_i$) [ $\exists$ ( $R_7$, технологічні та економічні розрахунки) $\leftarrow$ формується ( $R_7, R_1$ ) $\wedge$ формується ( $R_7, R_2$ ) $\wedge$ формується ( $R_7, R_4$ ) $\wedge$ формується ( $R_7, R_5$ ) $\wedge$ формується ( $R_7, R_6$ ) $\wedge$ формується ( $R_7, R_8$ )];

($\forall\ R_i$) [ $\exists$ ( $R_8$, схема технологічного процесу) $\leftarrow$ обумовлює ( $R_8, R_5$ ) $\wedge$ визначає ( $R_8, R_6$ ) $\wedge$ формує ( $R_8, R_7$ ) $\wedge$ обумовлюється ( $R_8, R_1$ ) $\wedge$ обумовлюється ( $R_8, R_2$ ) $\wedge$ обумовлюється ( $R_8, R_3$ )] [61].

### *2.3. Побудова моделі пріоритетного впливу факторів на якість проєктування післядрукарських процесів за методом математичного моделювання ієрархій*

Окрім виокремлення необхідних лінгвістичних змінних сучасні умови виробництва вимагають чіткого розуміння їх пріоритетності. Саме завдяки ієрархічному впорядкуванню досліджуваних факторів можна сформувати цілісну картину необхідних технічних та інтелектуальних компонент.

Для встановлення рівнів пріоритетності факторів на основі семантичної мережі використовуємо метод математичного моделювання ієрархій. Для початку будується матриця досяжності $A$, бінарні елементи якої визначаються за таким правилом:

$$R_{ij} = \begin{cases} 1, \textit{ якщо з вершини і можна попасти у вершину j} \\ 0, \textit{в іншому випадку} \end{cases}. \qquad (3)$$

Досяжність вершини $R_j$ $(j=1, 2, ..., n)$ відносно вершини $R_i$ $(i=1, 2, ..., n)$ обумовлюється наявністю зв'язку певного типу (прямого чи опосередкованого). Позначимо підмножину досяжних вершин $K(R_i)$. При цьому вершина $R_i$, для якої можлива зворотня досяжність з вершини $R_j$, буде її попередницею. Сукупність вершин попередниць формує підмножину $P(R_i)$. Перетин вершин сформованих підмножин $H(R_i)=K(R_i)\cap P(R_i)$, за умови $P(R_i)=H(R_i)$, визначає домінантність дії факторів, що ототожнюються з цими вершинами та встановлюється шляхом аналізу так званих ітераційних таблиць. Внаслідок виконання означених операцій над елементами семантичної мережі отримуємо багаторівневу модель, що відображає домінантність дії факторів на аналізований технологічний процес [58].

На основі синтезованої семантичної мережі будується матриця досяжності за принципом (3). Для зручності відображення матрицю доцільно поміщати у таблицю, додавши позначення факторів.

Для подальшого встановлення пріоритетності факторів за матрицею досяжності будуються ітераційні таблиці, що міститимуть чотири колонки, де $i$ — порядковий номер фактора у множині. При цьому для формування стовпця $K(R_i)$ ітераційних таблиць використовуємо дані, наведені у рядках матриці досяжності, а для формування стовпця $P(R_i)$ — дані, наведені у стовпцях цієї матриці. У стовпці $K(R_i)\cap P(R_i)$ подамо спільні для $K(R_i)$ та $P(R_i)$ фактори [58]. На основі отриманих даних синтезується модель пріоритетного впливу факторів на якість проєктування післядрукарських процесів.

Таким чином, найвищий пріоритет належить фактору $R_1$ (показники видання), що є логічним з технологічної точки зору, адже вид і тип видання, формат видання та його обсяг справді є визначальними при створенні проєкту реалізації післядрукарських процесів. На другому рівні знаходяться фактори $R_3$ (умови експлуатації) та $R_4$ (тип виробництва). Третім за пріоритетністю є фактор $R_2$ (конструкційні особливості), четвертим — $R_8$ (схема технологічного процесу), п'ятим — $R_5$ (матеріали), шостим — $R_6$ (тип обладнання). Найнижчий рівень пріоритетності свідчить про підрядний характер фактора $R_7$ (технологічні та економічні розрахунки).

### 2.4. Побудова моделі пріоритетного впливу факторів на якість проєктування післядрукарських процесів за методом ранжування

Уточнення чи підтвердження пріоритетності факторів проєктування післядрукарських процесів здійснюється шляхом встановлення їх рангів за методом ранжування, який полягає у синтезуванні деревовидних моделей на основі аналізу взаємозв'язків між виокремленими факторами. Слід зазначити, що згадані зв'язки поділяються на два типи: впливи та залежності, які передбачають прямі та опосередковані дії. Така методика дозволяє наблизити візуалізацію до реальних умов перебігу досліджуваного процесу. При цьому кожному типу присвоюються відповідні числові показники, що уможливлює подальше математичне оцінювання.

У дослідженні доцільно враховувати такі означення і твердження.

Означення 1. Будь-який технологічний процес поліграфічного виробництва містить деяку множину факторів, які здійснюють визначальний вплив на якість його реалізації, відповідно й на якість друкованої продукції.

З огляду на те, що кожен процес у поліграфії містить певну множину факторів, що впливають на його якість, нехай $D = \{d_1, d_2, ..., d_m\}$ буде довільною множиною технологічних процесів, а $R = \left\{ r_{1_m}, r_{2_m}, ..., r_{n_m} \right\}$ — множиною факторів, що впливають на якість конкретного процесу, де $n_m$ — це кількість факторів $m$-го технологічного процесу. При цьому:

$$C(S_k) = \bigcup_{j=1}^{n} \omega(S_{jk}), \quad (k = 1, 2, ..., m), \qquad (4)$$

де $C(S_k)$ — значення функції якості $m$-го процесу; $\omega(S_{jk})$ — ваговий показник додаткової якості, принесеної $j$-м фактором у $k$-й технологічний процес. Тоді подамо означення таким чином:

$$(\exists p)(\forall r) C(r_k); \quad d \in D; \quad r \in R. \qquad (5)$$

Означення 2. Ранг та пріоритет фактора визначається ваговим коефіцієнтом. Серед будь-якої множини факторів можна виокремити хоча б один пріоритетний.

Тобто для множини ваг факторів $W = \left\{ w_{1_m}, w_{2_m}, ..., w_{n_m} \right\}$, якщо $P(w) = max\left\{ w_{1_m}, w_{2_m}, ..., w_{n_m} \right\}$, матимемо:

$$(\exists p)(\forall w)P(w); \ d \in D; \ w \in W. \tag{6}$$

Твердження 1. Існування зв'язків між факторами є передумовою для їх формалізованого відображення у вигляді графа.

Твердження 2. Облік та аналіз впливів та залежностей між факторами у вихідній графічній моделі, побудованій на основі експертних суджень, дозволяє визначити початкові ранги факторів.

Твердження 3. При порівнянні факторів у межах вихідного графа синтезована багаторівнева модель показує лише переваги між ними.

Твердження 4. Виявлення кінцевих вагових значень, які визначають ранг та ступінь впливу факторів на $m$-й технологічний процес поліграфічного виробництва, можливе шляхом створення та обробки матриці попарних порівнянь і обчислення нормалізованих компонент головного власного вектора матриці.

Означення 3. Множина факторів, упорядкованих за спаданням їх нормалізованих вагових значень, не містить абсолютно ідентичних за ступенем впливу на технологічний процес.

Якщо $A(w) = w_j > w_{j+1}$ для $(j = 1, 2, ... n - 1)$, то вірним буде наступний запис:

$$(\forall w)A(w); \ w \in W. \tag{7}$$

Згідно з твердженнями 1—4, синтез моделі пріоритетного впливу факторів на $m$-й технологічний процес поліграфічного виробництва здійснюється шляхом виокремлення характерних для аналізованого процесу факторів, створення, аналіз та обробку вихідної графічної моделі, у якій на основі експертних суджень встановлено зв'язки між факторами.

За основу методу ранжування взято числові показники, які стосуються кількостей впливів і залежностей між факторами та відповідних їм вагових коефіцієнтів. При цьому розрізняємо прямі дії, назвавши їх впливами 1-го порядку, та непрямі — 2-го порядку. Залежності також розрізнятимемо, встановивши для них аналогічно 1-й і 2-й порядки важливості.

Для розрахунку сумарних вагових значень прямого та опосередкованого впливів факторів та їх інтегральної залежності від інших факторів введемо відповідні позначення. Нехай $k_{ij}$ — кількість впли-вів чи залежностей для $j$-го фактора $j = 1,...,n$); $w_i$ — вага $i$-го типу. Ідентифікуємо числові значення індексів наступним чином: $i = 1$ для впливів 1-го порядку, $i = 2$ для впливів 2-го порядку, $i = 3$ для залежностей 1-го порядку, $i = 4$ для залежностей 2-го порядку. Вважатимемо, що для впливів обох типів ваги будуть додатними, тобто $w_1 > 0$, $w_2 = w_1 / 2$, відповідно для залежностей — від'ємними, а саме: $w_3 < 0$, $w_4 = w_3 / 2$. Нехай $R_{ij}$ — інтегральні вагові значення факторів за сумами ваг усіх типів зв'язків. Тоді формула для розрахунків матиме вид:

$$R_{ij} = \sum_{i=1}^{4} \sum_{j=1}^{n} q_{ij} w_i, \qquad (8)$$

де $n$ — номер фактора досліджуваного процесу.

Відповідно до початкових умов $w_3 < 0$ і $w_4 < 0$, отже $R_{3j} < 0$ і $R_{4j} < 0$. Щоб привести вагові значення «до початку координат», тобто отримати додаткові величини, слід перемістити гістограму інтегрального графічного відображення усіх типів зв'язків вверх за таким співвідношенням:

$$\Delta_j = max\left|R_{3j}\right| + max\left|R_{4j}\right|, \left(j = 1, 2,...,n\right). \qquad (9)$$

На основі заданих умов отримаємо формулу підсумкових вагових значень факторів:

$$R_{Fj} = \sum_{i=1}^{4} \sum_{j=1}^{9} \left(k_{ij} w_i + \Delta_j\right), \qquad (10)$$

де $\Delta_j = max\left|R_{3j}\right| + max\left|R_{4j}\right|, \left(j = 1, 2,...,n\right)$ [30; 58].

Величини $R_{Fj}$ служать підставою для ранжування ваг, тобто встановлення рівнів факторів якості реалізації технологічного процесу. За результатами ранжування здійснюється синтез графічної моделі за отриманими ваговими значеннями, що відображають пріоритетність впливу факторів на процес.

Для реалізації методу стосовно кожного з факторів проєктування післядрукарських процесів на основі розробленої семантичної мережі будуються ієрархічні дерева зв'язків з іншими факторами, враховуючи прямі та непрямі впливи і прямі та опосередковані залежності [30; 58; 59].

За допомогою методу ранжування встановлюються вагові значення факторів та уточнюється їх пріоритетність. Таким чином фактори $R_3$ (умови експлуатації) та $R_4$ (тип виробництва), що знаходилися на одному рівні, отримали відповідно другий та четвертий рівні пріори-тетності. При цьому фактор $R_2$ (конструкційні особливості) змістився на третю позицію у моделі. Пріоритетність інших факторів лише підтвердилася.

**Етап 3. Оптимізація моделі пріоритетного впливу факторів на якість проєктування післядрукарських процесів**

*3.1. Формування матриці попарних порівнянь факторів відповідно до шкали відносної важливості об'єктів за Сааті*

Оптимізація вагових значень факторів і синтез моделі здійснюються за методами багатокритеріальної оптимізації та попарних порівнянь. Первинне визначення вагових значень факторів технологічних процесів на основі методу ранжування передбачає отримання укрупнених результатів, що потребують подальшого експертного опрацювання. Метод аналізу ієрархій, реалізований на основі шкали відносної важливості об'єктів за Сааті, дозволяє встановити уточнені (оптимізовані) вагові значення та передбачає побудову матриці попарних порівнянь, обчислення компонент її головного власного вектора та їх нормалізацію, а також перевірку результатів за ключовими критеріями. Унаслідок оптимізації здійснюється деталізація перебігу досліджуваного процесу, що позитивно впливає на його подальшу реалізацію.

З цією метою будуємо квадратну обернено-симетричну матрицю попарних порівнянь (МПП), порядок якої визначається числом аналізованих факторів. Алгоритм її організації такий. Порівнюються умовні міри впливу кожного із факторів першого стовпця матриці досяжності та кожний із факторів верхнього рядка матриці. Додатковими умовами при порівнянні служать отримані при ранжуванні вагові значення факторів. На перетині рядка і кожного зі стовпців МПП заносимо числове значення переваги фактора, використовуючи шкалу відносної важливості об'єктів (табл. 5). Так, для двох факторів (напр., $r_i$ і $r_j$), які порівнюються між собою, в залежності від їх важливості та міри впливу на проєктування післядрукарських процесів матимемо пропоновані у таблиці значення відповідного елемента матриці попарних порівнянь у позиції ($r_i$, $r_j$). Зрозуміло, що при такому алгоритмі діагональні елементи МПП рівні одиниці.

Нижня частина матриці попарних порівнянь заповнюється оберненими значеннями. Так, у позицію ($r_i$, $r_j$) заносимо відповідно 1, 1/3, 1/5, 1/7, 1/9. При незначних відмінностях між вагами критеріїв використовують парні числа 2, 4, 6, 8 та їх обернені значення [58; 59; 68].

**Шкала відносної важливості об'єктів**

| Оцінка важливості | Критерії порівняння | Пояснення щодо вибору критерію |
|---|---|---|
| 1 | Об'єкти рівноцінні | Відсутність переваги $r_i$ над $r_j$ |
| 3 | Один об'єкт дещо переважає інший | Існує підстава наявності слабкої переваги $r_i$ над $r_j$ |
| 5 | Один об'єкт переважає інший | Існує підстава наявності суттєвої переваги $r_i$ над $r_j$ |
| 7 | Один об'єкт значно переважає інший | Існує підстава присутності явної переваги $r_i$ над $r_j$ |
| 9 | Один об'єкт абсолютно переважає інший | Абсолютна перевага $r_i$ над $r_j$ не викликає сумніву |
| 2, 4, 6, 8 | Компромісні проміжні значення | Допоміжні порівняльні оцінки |

### 3.2. Визначення компонент головного власного вектора матриці попарних порівнянь

Головний власний вектор $R(r_1, r_2, ..., r_n)$ МПП визначається як середнє геометричне компонент кожного рядка матриці:

$$R_i = \sqrt[n]{a_{i1} \cdot a_{i2} \cdot a_{in}} \quad i = \overline{1, n}, \tag{11}$$

де $n$ — кількість використаних факторів.

Для одержання головного власного вектора (тобто вектора пріоритетів) матриці попарних порівнянь використаємо метод, запропонований Сааті [55]. Розрахунки за вказаним методом з використанням ідей теорії імітаційного моделювання здійснюються за допомогою програми «Імітаційне моделювання в системному аналізі методом бінарних порівнянь» [57], розробленої на кафедрі комп'ютерних наук та інформаційних технологій Української академії друкарства. Після завантаження програми отримуємо інтерфейс у вигляді діалогового вікна. Опція «Введіть число критеріїв» обумовлює кількість факторів, далі — кнопка «задати». «Введіть назви критеріїв» — вводимо цифрові номери факторів, кнопка «застосувати». Заповнюємо таблицю вікна «Задання експертних оцінок переваг критеріїв» елементами матриці попарних порівнянь, після кнопка «застосувати». Результати опрацювання — у вікні «Вивід проміжних результатів», стовпець якого *En* відтворює компоненти нормалізованого вектора $R$, що ідентифікують розраховані вагові значення факторів досліджуваного процесу.

### 3.3. Визначення компонент нормалізованого вектора матриці попарних порівнянь

Нормалізуємо значення компонент головного власного вектора $R_n$ МПП [58; 59; 68], встановивши попередній результат розв'язання задачі:

$$R_{in} = \frac{\sqrt[n]{a_{i1} \cdot a_{i2} \cdot a_{in}} \quad i = \overline{1, n}}{\sum_{i=1}^{n} \sqrt[n]{a_{i1} \cdot a_{i2} \cdot a_{in}}}. \tag{12}$$

Для зручнішого подання вагових значень факторів множимо оптимізовані компоненти вектора $R_n$ на довільний коефіцієнт $k$. Нехай $k = 500$.

Оцінка узгодженості вагових значень факторів обчислюється шляхом множення матриці попарних порівнянь справа на вектор $R_n$. В результаті обчислення одержимо нормалізований вектор $R_{n1}$.

Компоненти власного вектора $R_{n2}$ матриці попарних порівнянь отримаємо, поділивши компоненти вектора $R_{n1}$ на відповідні компоненти вектора $R_n$.

### 3.4. Аналіз результатів оптимізації за максимальним значенням головного власного вектора матриці попарних порівнянь, індексом узгодженості та відношенням узгодженості

Максимальне власне значення $\lambda_{\max}$ додатної обернено-симетричної матриці $A$ визначається як середнє арифметичне компонент вектора $R_{n2}$.

Оцінка одержаного рішення визначається індексом узгодженості $IU$, який вираховується за формулою:

$$IU = \frac{\lambda_{\max} - n}{n - 1}. \tag{13}$$

Отримані значення порівнюють з еталонними значеннями показника узгодженості — випадковим індексом $RI$. Результати можна вважати задовільними, якщо отримане шляхом обрахунків значення індекса узгодженості $IU$ не перевищує 10 % еталонного значення випадкового індексу $RI$, обраного з урахуванням кількості аналізо-ваних факторів. Отже для підтвердження адекватності розв'язку поставленої задачі повинна виконуватися нерівність $IU < 0,1 \times RI$.

Нижче наведена таблиця величин випадкового індекса для матриць різного порядку, в якій порядок матриці відповідає кількості аналізованих об'єктів (факторів) і вказується у першому рядку, а еталонне значення показника узгодженості для кожного порядку вказується у другому рядку.

**Значення випадкового індекса для матриць різного порядку**

| Кількість об'єктів | 3 | 4 | 5 | 6 | 7 | 8 | 9 | 10 | 11 | 12 | 13 | 14 |
|---|---|---|---|---|---|---|---|---|---|---|---|---|
| Еталонне значення індекса | 0,58 | 0,90 | 1,12 | 1,24 | 1,32 | 1,41 | 1,45 | 1,49 | 1,51 | 1,54 | 1,56 | 1,57 |

Додатково результати оцінюють відношенням узгодженості, величину якого отримують із виразу: $RU = IU/RI$. Результати попарних порівнянь можна вважати задовільними, якщо $RU \leq 0,1$. Це свідчитиме про достатній рівень збіжності процесу та належну узгодженість експертних суджень стосовно попарних порівнянь факторів, відображених у відповідній матриці.

При незадовільних значеннях індекса узгодженості та відношення узгодженості треба переглянути вихідний граф зв'язків між факторами, уточнити значення величин відповідних їм попарних порівнянь, тобто розв'язати деяку обернену задачу, достовірність розв'язку якої перевіряється за наведеними вище критеріями.

Цікавими в цьому контексті можуть бути такі міркування. Якщо максимальне значення власного вектора матриці попарних порівнянь і величина відношення узгодженості не виходять за межі допустимих значень, то їх можна вважати критеріями оптимізації одержаної ієрархічної моделі впливу факторів на ефективність проєктування післядрукарських процесів. За цими значеннями встановлюється адекватність ієрархічної моделі реальній ситуації та її узгодженість з експертними оцінками важливості факторів [68].

Відповідно $\lambda_{max} = 8,483$, $IU = 0,069$. Еталонне значення індекса $RI$ для матриці 8-го порядку становить 1,41, що не перевищує 10 % індекса узгодженості $IU$. Отже нерівність $IU < 0,1 \times RI$ є вірною і підтверджує адекватність розв'язку задачі. $RU = 0,049$, тож результати попарних порівнянь можна вважати коректними.

### 3.5. Візуалізація співвідношень компонент вихідного та нормалізованого векторів. Побудова оптимізованої моделі пріоритетного впливу факторів на якість проєктування післядрукарських процесів

Для одержання вагових значень факторів на основі отриманих моделей їм присвоюється градація умовних числових позначень, відповідно до рівня домінантності факторів, починаючи відлік з найнижчого. Нехай вага найнижчого рівня буде рівною 20 умовним одиницям, а вага кожного наступного збільшуватиметься на 20 умовних одиниць відносно попереднього фактора. Отримані числові значення факторів подаються у вигляді компонент вихідного вектора $R_0$ згідно з порядком їх розміщення у матриці.

На підставі отриманих вагових значень, представлених векторами $R_n$ та $R_0$, будуються гістограма і порівняльний графік.

Порівняльне графічне відображення дає підставу стверджувати, що компоненти векторів, розрахованих за методом ранжування факторів та отриманих у результаті застосування методу попарних порівнянь, незважаючи на деяку різницю у вагових значеннях, по суті відтворюють останній порядок і суть слідування факторів. Вказане уможливлює використання вагових компонент нормалізованого вектора як основи для синтезування оптимізованої моделі пріоритетного впливу виокремлених факторів на якість проєктування післядрукарських процесів.

Оптимізація дозволила уточнити значення ваг факторів досліджуваного процесу та деталізувати міру впливу кожного з них. Оптимізаційні результати підтвердили достовірність проведених досліджень, не змінивши порядок пріоритетів факторів [68].

Після детального аналізу та порівняння вихідного та нормалізованого векторів синтезується оптимізована модель пріоритетного впливу факторів на процес. Вона служить підставою для проєктування альтернативних та розрахунку оптимальних варіантів реалізації технологічного процесу, його етапів чи окремих операцій, фактори яких упорядковані за ваговими коефіцієнтами важливості.

**Етап 4. Визначення оптимальних альтернатив реалізації проєктування післядрукарських процесів**

*4.1. Багатофакторний вибір альтернативи на основі лінійного згортання критеріїв*

Залежно від глибини пізнання проблеми розділяють на три класи: добре структуровані, неструктуровані та погано структуровані. У добре структурованих проблемах існуючі залежності добре з'ясовані, тому можуть бути виражені у символах і числах та в підсумку давати числові оцінки. Неструктуровані проблеми містять лише описи ресурсів, характеристик та ознак, причому кількісні залежності між ними є невідомими. Погано структуровані проблеми, у свою чергу, містять як якісні, так і кількісні елементи, а якісні невизначені та маловідомі сторони проблеми мають домінуючу тенденцію.

Визначення класу сформульованої проблеми дозволяє обрати методику її вирішення. Так, добре структуровані проблеми вирішуються шляхом використання методології дослідження операцій, неструктуровані — за допомогою евристичного методу, а погано структуровані — використовуючи системний аналіз.

З огляду на вищенаведені факти можна зробити висновок, що проблема встановлення оптимальної альтернативи проєктування післядрукарського опрацювання книжкових видань є погано структурованою. Таким чином, визначення альтернативних варіантів вирішення проблеми реалізовується за допомогою системного аналізу.

Генерування множини альтернатив уможливлює подальший вибір оптимальної альтернативи із цієї множини. Сформована множина альтернатив відображає можливі способи досягнення поставлених цілей. Вибір оптимальної альтернативи здійснюється із врахуванням обмежень та критерію оптимальності [59].

У ході дослідження були встановлені вагові значення факторів аналізованих технологічних процесів та побудовані моделі їх пріоритетного впливу. Отримана інформація є основою для планування стратегії реалізації проєктування післядрукарських процесів, яка полягає у виборі оптимального альтернативного варіанту. Це дасть нам уявлення про необхідну затрату трудомісткості та міру важливості факторів.

Розв'язок поставленого завдання здійснюється за допомогою багатокритеріальної (в нашому випадку в ролі критерію виступає фактор, отже багатофакторної) оптимізації. При цьому є достатнім викорис-тання лише домінуючих факторів, що обумовлено принципом Парето [22], суть якого полягає у використанні взаємно недомінованих факторів, які утворюють множину Парето $P(D)$, де $D \subset R^n$ — множина допустимих розв'язків. Фактори із помітно нижчими ваговими значеннями просто відкидаються.

Синтезована раніше семантична мережа є підставою для побудови матриці попарних порівнянь, опрацювання якої приводить до отримання умовних вагових значень, що визначають числові пріоритети факторів — міри важливості їх для технологічного процесу. Далі — розрахунок та визначення оптимального (серед альтернативних) варіанту реалізації проєктування післядрукарських процесів.

Багатокритеріальна оптимізація функцій $r(x)=\left(r_1(x),...,r_n(x)\right)$ на множині $B$ полягає у виокремленні максимального значення функцій корисності $r_i(x) \to \max\limits_{x \in B}, \quad i = \overline{1,n}$. Відповідно за методом лінійного згортання критеріїв об'єднання часткових цільових функціоналів $r_1,...r_n$ здійснюється за формулою [25; 59]:

$$R(w,x) = \sum_{i=1}^{n} w_i r_i(x) \to \max\limits_{x \in D}; w \notin W, \qquad (14)$$

$$W = \left\{ w = \left(w_1,...,w_n\right)^T; w_j > 0; \sum_{i=1}^{n} w_i = 1 \right\},$$

де $w_i$ — ваги факторів множини Парето.

Для факторів незалежних за корисністю та перевагою існує така функція корисності [25]:

$$U(x) = \sum_{i=1}^{n} w_i u_i(y_i), \qquad (15)$$

де $U(x)$ — багатокритеріальна функція корисності $(0 \le U(x) \le 1)$ певної альтернативи $x$; $w_i$ — встановлене вагове значення $i$-го критерію, причому $0 < w_i < 1$, $\sum_{i=1}^{n} w_i = 1$; $u_i(y_i)$ — функція корисності $i$-го критерію $(0 \le u_i(y_i) \le 1)$; $y_i$ — значення альтернативи $x$ за $i$-м критерієм [18].

Для реалізації сформованої задачі виконуються такі дії:

1. Формується множина Парето, взявши до уваги лише фактори з найвищою пріоритетністю: $R_1$ — показники видання (188 у. о.); $R_3$ — умови експлуатації (105 у. о.); $R_2$ — конструкційні особливості (74 у. о.); $R_4$ — тип виробництва (49,5 у. о.). Фактори з суттєво нижчою пріоритетністю відкидаються [25; 54; 59].

2. Задаються три альтернативні варіанти реалізації досліджуваного процесу, які позначаються як $A_1, A_2, A_3$. Формується таблиця оцінювання альтернатив на основі міри важливості кожного виокремленого фактора. Міра важливості кожного аналізованого фактора для заданих альтернативних варіантів виражається у відсотках. Існує великий перелік можливих комбінацій (табл. 7). Реальні значення залежать від конкретного виробничого завдання. При цьому загальна сума усіх альтернатив одного фактора не повинна перевищувати 100 %.

Таблиця 7

**Комбінації значень факторів**

| Комбінації значень факторів у відсотках | | | |
|---|---|---|---|
| 10—10—80 | 20—10—70 | 30—10—60 | 40—10—50 |
| 10—20—70 | 20—20—60 | 30—20—50 | 40—20—40 |
| 10—30—60 | 20—30—50 | 30—30—40 | 40—30—30 |
| 10—40—50 | 20—40—40 | 30—40—30 | 40—40—20 |
| 10—50—40 | 20—50—30 | 30—50—20 | 40—50—10 |
| 10—60—30 | 20—60—20 | 30—60—10 | 50—10—40 |
| 10—70—20 | 20—70—10 | 60—10—30 | 50—20—30 |
| 10—80—10 | 70—10—20 | 60—20—20 | 50—30—20 |
| 80—10—10 | 70—20—10 | 60—30—10 | 50—40—10 |

3. Створюється матриця попарних порівнянь вагових значень факторів, оцінених за шкалою відносної важливості об'єктів за Сааті.

4. Здійснюється нормалізація головного власного вектора матриці попарних порівнянь у програмі «Імітаційне моделювання в системному аналізі методом бінарних порівнянь» [57]. Унаслідок нормалізації отримуються оптимізовані вагові значення факторів. Встановлюються критерії нормалізації.

Перевірка правильності розв'язку задачі здійснюється шляхом виконання нерівностей $IU < 0,1 \times RI$ та $RU \leq 0,1$, де $RI$ — випадковий індекс для матриці 4-го порядку (табл. 6), $RI = 0,9$.

4. Визначаються функції корисності кожної запроєктованої альтернативи за факторами множини Парето.

5. Визначаються багатокритеріальні оцінки корисності для трьох запроєктованих альтернатив.

Підставивиши у формулу (15) значення $R_2$: $n = 4; u_i(y_i) = u_{ij}$ — корисність $j$-ї альтернативи $(j = 1, 2, 3)$ за $i$-м фактором $(i = 1, ..., 4)$, отримаємо наступне:

$$U_j = \sum_{i=1}^{4} w_i u_{ij}; j = 1, 2, 3, \tag{16}$$

де $U_j$ — багатофакторна оцінка корисності $j$-ї альтернативи.

На основі формули (16) формуються такі відношення:

$$U_1 = w_1 \times u_{11} + w_2 \times u_{21} + w_3 \times u_{31} + w_4 \times u_{41};$$
$$U_2 = w_1 \times u_{12} + w_2 \times u_{22} + w_3 \times u_{32} + w_4 \times u_{42};$$
$$U_3 = w_1 \times u_{13} + w_2 \times u_{23} + w_3 \times u_{33} + w_4 \times u_{43} \ [25]. \tag{17}$$

При цьому $U_1 = 0,264$; $U_2 = 0,322$; $U_3 = 0,414$. Найкраща альтернатива реалізації проєктування післядрукарських процесів обирається за максимальним значенням $U_j$, $(i = 1, 2, 3)$. Відповідно альтернатива $A_3$ є оптимальною для досліджуваного процесу, а визначальним є фактор «Показники видання» *($R_1$)*.

### 4.2. Багатофакторний вибір альтернативи на основі нечіткого відношення переваги

Прийняття управлінських рішень щодо альтернативних варіантів реалізації технологічних процесів може ускладнюватися відсутністю інформації про їхню пріоритетність та неможливістю кількісного оцінювання переваг. Натомість, можливо здійснити попарне порівняння альтернатив на відрізку [0;1] та представити дані у числовому вигляді. Оцінювання здійснюється на основі багатокритеріальної оптимізації, де в ролі критеріїв виступають фактори технологічного процесу. За принципом Парето [22], як і в методі встановлення оптимальної альтернативи на основі лінійного згортання критеріїв, достатнім вважається вибір лише домінуючих факторів із найвищими ваговими показниками, які формують множину Парето. Відповідно при нечіткому відношенні переваги на множині альтернатив прийняття рішень буде здійснюватися за Парето-оптимальними альтернативами.

Введення чіткого відношення нестрогої переваги $R_i$ на множині альтернатив $X = \{x_1, ..., x_n\}$ дозволяє висловити одне з наведених тверджень для будь-якої пари альтернатив $(x, y)$: $x$ не гірша $y$, тоб-то $x \geq y$, $(x,y) \in R$; $y$ не гірша $x$, що записується як $y \geq x$, $(y,x) \in R$, $x$ та $y$ неможливо порівняти між собою, $(x,y) \notin R$, $(y,x) \notin R$. Такий підхід уможливлює звуження класу раціонального вибору.

Якщо існує строга перевага $(x,y) \in R_z$, альтернатива $x$ переважає $y$, тобто $x > y$. За умови чітких функцій корисності $r_j$ множини $X$ альтернатива $x$, що має вищу оцінку $r_j(x)$ за фактором $j$, є кращою, ніж альтернатива $y$, оцінка якої $r_j(y)$. Подане твердження описується чітким відношенням переваги $R_j$ множини $X$:

$$R_j = \left\{ (x,y) : r_j(x) \geq r_j(y), x,y \in X \right\}. \tag{18}$$

Визначимо якість проєктування післядрукарських процесів шляхом оцінювання нечітких відношень переваги $R_i$ на множині альтернатив $X = \{x_1, x_2, x_3\}$: $R_1$ (Показники видання) — $x_1 = x_2, x_2 < x_3$; $R_2$ (Умови експлуатації) — $x_1 < x_3$, $x_2 > x_3$; $R_3$ (Конструкційні особливості) — $x_1 > x_2, x_2 = x_3$; $R_4$ (Тип виробництва) — $x_1 > x_2, x_2 = x_3$.

Для виокремлення Парето-оптимальної альтернативи необхідно обрати альтернативу $x_0 \in X$ із найвищою оцінкою корисності на множині усіх факторів:

$$r_j(x_0) \geq r_j(y), \forall j = 1,m; \ \forall y \in X. \tag{19}$$

Згортка усіх критеріїв сформованої множини Парето в єдиний скалярний здійснюється за способом перетину [22; 59; 79].

Позначимо $Q_1 = \bigcap_{j=1}^{m} R_j$. Таким чином, множина альтернатив $X = \{x_1, ..., x_n\}$ із відношенням переваги $Q_1$ є відповідною до множини альтернатив з функціями корисності $r_j(x)$. Визначення недомінованих альтернатив за нечітким відношенням переваги $Q_1$ полягає у заміні кількох відношень $R_j(j = 1,m)$ на перетин між ними. Вважатимемо, що $\mu_j(x,y)$ є функцією належності чіткого відношення переваги $r_j$. Сформована умова матиме вид:

$$\mu_j(x,y) = \begin{cases} 1, \textit{якщо } r_j(x) \geq r_j(y), \textit{тобто } (x,y) \in R_j \\ 0, \textit{якщо } (x,y) \notin R \end{cases}. \tag{20}$$

Відповідно функція належності згортки $Q_1$ запишеться таким чином:

$$\mu_{Q_1}(x,y) = \min\left\{ \mu_1(x,y), \mu_2(x,y), ..., \mu_n(x,y) \right\}. \tag{21}$$

Згортка критеріїв із врахуванням вагових значень факторів технологічного процесу $v_j$ та відповідних функцій корисності матиме вигляд:

$$Q(x) = \min_j v_j r_j(x).$$  (22)

Згортка вихідних відношень $Q_2$ також формується ваговими значеннями аналізованих факторів $v_j$ і відповідними функціями корисності:

$$Q_2 = \sum_{j=1}^{m} v_j r_j(x), \ \textit{де} \sum_{j=1}^{m} v_j = 1, \ v_j \geq 0.$$  (23)

Їй відповідає така функція належності [58]:

$$\mu_{Q_2}(x,y) = \sum_{j=1}^{m} v_j \mu_j(x,y).$$  (24)

У результаті обчислень отримаємо:

$$\mu_{Q_2}^{\textit{нд}}(x_i) = [0,4; 0,74; 1,26].$$

За перетином множин $Q_1^{\textit{нд}}$ та $Q_2^{\textit{нд}}$ максимальне значення функції належності $\mu_Q^{\textit{нд}}(x_i)$ належить $x_3$, тобто оптимальним вважається третій варіант.

### 4.3. Перевірка результатів

Внаслідок проведення багатофакторного вибору оптимальної альтернативи на основі лінійного згортання критеріїв визначено багатофакторні оцінки корисності: $U_1 = 0,264$; $U_2 = 0,322$; $U_3 = 0,414$. Максимальною оцінкою корисності є $U_3$, отже оптимальною є третя альтернатива.

У результаті багатофакторного вибору альтернативи на основі нечіткого відношення переваги одержано такі значення функції належності: $\mu_Q^{\textit{нд}}(x_i)$: $x_1 = 0,4$; $x_2 = 0,74$; $x_3 = 1,26$. Таким чином, можна стверджувати, що варіант $x_3$ є оптимальною альтернативою проєктування післядрукарських процесів.

Результати, отримані внаслідок встановлення оптимальної альтернативи за методом лінійного згортання критеріїв та за методом на основі нечіткого відношення переваги, є тотожними, що свідчить про достовірність проведеного дослідження [79].

**Етап 5. Визначення інтегрального показника якості проєктування післядрукарських процесів**

Остаточно моделювання системи прогностичного оцінювання якості проєктування післядрукарських процесів на базі нечіткої логіки зводиться до розв'язання таких завдань [15; 16; 44; 64]:

— встановлення універсальної терм-множини значень та відповідних їй лінгвістичних термів виокремлених факторів (лінгвістичних змінних);

— побудова багаторівневої моделі логічного виведення, структура якої відтворює ієрархію факторів та лінгвістичних термів, що впливають на якість реалізації процесу. Компонента найвищого рівня визначає вихідний прогнозований показник якості досліджуваного процесу у вигляді нечіткої множини;

— побудова та опрацювання матриць попарних порівнянь для множини лінгвістичних термів відносно квантів поділу інтервалів значень універсальної множини та отримання для кожної з лінгвістичних змінних функцій належності;

— нормування значень функцій належності та співвіднесення їх із квантами поділу універсальної множини;

— побудова суміщених графіків за нормованими значеннями функцій належності для лінгвістичних змінних і відповідних їм лінгвістичних термів;

— розроблення нечіткої бази знань (або матриці знань) з використанням нечітких логічних висловлювань типу «якщо ⟨умова⟩, тоді ⟨висновок (або дія)⟩», що відтворює алгоритм формування якості проєктування післядрукарських процесів в залежності від рівня якості лінгвістичних термів;

— побудова нечітких логічних рівнянь на підставі матриці знань та функцій належності, які визначають зв'язок між функціями належності вхідних та вихідних даних;

— побудова аналітичного виразу для формалізованої ідентифікації прогнозованого результату у вигляді нечіткої множини, отриманої на підставі багаторівневої моделі логічного виведення та нечіткої бази знань;

— дефазифікація нечіткої множини, суть якої полягає у розрахунку числового показника прогнозованої якості за методом центра мас або центра ваги плоскої фігури, обмеженої графіком функції належності і віссю абсцис.

При дефазифікації нечіткої множини використовуються значення функцій належності лінгвістичних змінних, область існування яких визначена універсальною множиною.

### *5.1. Фазифікація нечіткої множини*

Проєктування післядрукарських процесів — це необхідна складова забезпечення якості готової книжкової продукції, яка включає ряд послідовних операцій, спрямованих на досягнення поставленої мети. Однак фактори, які здійснюють неопосередкований вплив на реалізацію досліджуваного процесу, не завжди містять кількісну складову. Натомість, значно інформативнішими стають певні лінгвістичні характеристики. Виникає необхідність заміни понять чіткої множини поняттями нечіткої множини. Саме тому для забезпечення точності моделювання доцільно використовувати методи та засоби нечіткої логіки.

Суттєвим елементом та перевагою нечіткої логіки є можливість фазифікації, тобто заміни компонент чіткої множини відповідними їм поняттями нечіткої множини. Відомо, що суть її полягає у зіставленні терм-множини значень аналізованих факторів відповідника нечіткого формату змінних величин — функцій належності. Тобто змінні, які не можуть бути чітко вираженими за допомогою кількісних значень та які зручно описувати словами чи словосполученнями, вважаються лінгвістичними змінними, такими як: «Показники видання», «Конструкційні особливості» чи ін. При цьому значення кожної лінгвістичної змінної формуються у певну сукупність — терм-множину, компоненти якої називаються термами. Так, терм-множина лінгвістичної змінної «Показники видання» складається з термів «просте», «ускладнене», «складне». Лінгвістичною вважається змінна, значення якої виражено засобами звичайної мови — словами або словосполученнями. При цьому множину можливих значень лінгвістичної змінної прийнято називати терм-множиною, а довільний її елемент — термом. Так, для лінгвістичної змінної «Тип обладнання» термами будуть лінгвістичні оцінки «ручне», «механічне», «автоматизоване», що утворюватимуть терм-множину значень.

Фазифікація забезпечує доволі високий рівень відповідності моделі реальному об'єкту і служить, як буде показано пізніше, основою для подальшого моделювання прогностичного оцінювання проєктування післядрукарських процесів.

У роботах основоположника нечіткої логіки Заде [20; 21] вводиться поняття універсальної множини $D$, як такої, що стосується всієї проблемної області. Тоді нечітку підмножину $M$ множини $D$ визначають через шкалу $D$ і функцію належності $\mu_M(d)$ [20], тобто

$$M = \left\{ \left( \mu_M(d), d \right), d \in D \right\}, \tag{25}$$

де $\left( 0 \leq \mu_M(d) \leq 1 \right)$.

Функція належності встановлює міру приналежності кожного елемента нечіткої множини універсальній множині, тобто $M \in D$. За умови дискретності і скінченності базової шкали (тобто поділеної на кванти чи проміжки) нечітка множина

$$M = \left( \mu_M(d_1) / d_1, \mu_M(d_2) / d_2, ..., \mu_M(d_n) / d_n \right) = \sum_{i=1}^{n} \mu_M(d_i) / d_i, \tag{26}$$

або спрощено: $M = \sum_{i=1}^{n} \mu_i / d_i$. Запис означає «прикріплення» функції належності $\mu_M(d_i)$ до елемента $d_i$.

Остаточно функції належності виступають ідентифікатором вхідних значень лінгвістичних змінних у нечіткому форматі, тобто множині значень змінної $d$ ставляться у відповідність функції належності $\mu(d)$.

Вважатимемо, що реалізація визначеного технологічного процесу буде певною функцією $G$, у якості аргументів якої будуть виокремлені фактори $r_{1_m}, r_{2_m}, ..., r_{n_m}$, тоді:

$$G = F\left( r_{1_m}, r_{2_m}, ..., r_{n_m} \right), \tag{27}$$

де $n_m$ — кількість факторів $m$-го технологічного процесу.

Отже досліджуваний процес є процедурою з множиною початкових змінних $r_i \left( i = \overline{1,n} \right)$ та кінцевою змінною $G$.

Існування кількісних значень змінних уможливлює задання проміжку, що виражатиметься граничними значеннями цих змінних [13; 14]: $\left| \underline{r_i}, \overline{r_i} \right|, i = \overline{1,n}; \left| \underline{G}, \overline{G} \right|$. Зважаючи на те, що виокремлені фактори є якісними змінними, постає потреба формування множини та меж задання значень: $D = \left\{ d^{(1)}, d^{(2)}, ..., d^{(j)} \right\}$, де $d^{(k)}, k = \overline{1,j}$ — множина кількісних чи якісних умовних одиниць, потужність якої визначає індекс $j$. Тоді результуюча змінна $G$ із граничними межами також може подаватися в умовних одиницях множиною $G = \left\{ g^{(1)}, g^{(2)}, ..., g^{(j)} \right\}$. Такі універсальні множини забезпечують виконання залежності (27). В той же час доцільно оцінювати лінгвістичні змінні засобами природної мови (наприклад: «просте», «ускладнене», «складне»), формуючи лінгвістичні терм-множини.

З огляду на наведені твердження побудова багаторівневої моделі нечіткого логічного виведення встановлення інтегрального показ-ника якості досліджуваного технологічного процесу передбачає формування часткових показників якості; виокремлення універсальної множини значень та терм-множини кожної лінгвістичної змінної; безпосередню візуалізацію ієрархічної залежності.

Вважатимемо процес проектування післядрукарських процесів функцією $G = F\left(R_1, R_2, R_3, R_4, R_5, R_6, R_7, R_8,\right)$ з такими аргументами: $R_1$ — показники видання; $R_2$ — конструкційні особливості; $R_3$ — умови експлуатації; $R_4$ — тип виробництва; $R_5$ — матеріали; $R_6$ — тип обладнання; $R_7$ — технологічні та економічні розрахунки; $R_8$ — схема технологічного процесу [27; 30]. Інтегральний показник якості проєктування післядрукарських процесів визначатиметься за принципом ієрархізації структури процесу. Відповідно залежність якості проєктування видання може бути виражена через якість часткових показників:

$$G = F_G\left(M, O, P\right). \tag{28}$$

Аргумент $M$ визначає якість формування видання:

$$M = F_M\left(m_1, m_2, m_3\right), \tag{29}$$

де $m_1$ — лінгвістична змінна «показники видання», $m_2$ — лінгвістична змінна «конструкційні особливості», $m_3$ — лінгвістична змінна «умови експлуатації».

Аргумент $O$ визначає якість організації виробництва:

$$O = F_O\left(o_1, o_2, o_3\right), \tag{30}$$

де $o_1$ — лінгвістична змінна «тип виробництва»; $o_2$ — лінгвістична змінна «матеріали»; $o_3$ — лінгвістична змінна «тип обладнання».

Аргумент $P$ визначає якість опрацювання видання:

$$P = F_P\left(p_1, p_2\right), \tag{31}$$

де $p_1$ — лінгвістична змінна «редагування»; $p_2$ — лінгвістична змінна «коректура».

Сформуємо таблицю, вказавши лінгвістичну суть кожної змінної, універсальні множини значень та відповідні лінгвістичні терми.

Значення умов експлуатації сформовано на основі груп довговічності користування книжковим виданням: 1-ша група — нетривалий термін служби (до двох років) з малою інтенсивністю користування; та 2-га група — нетривалий термін служби (до двох років) з великою інтенсивністю користування; 3-тя група — середній термін користування (від 2 до 10 років) з малою інтенсивністю користування; 4-та група — середній термін користування (від 2 до 10 років) з великою інтенсивністю користування; 5-та та 6-та групи — тривалий термін (від 10 років і більше) з малою чи високою інтенсивністю користування [27; 35; 70].

**Терм-множини значень лінгвістичних змінних**

| Змін-на | Лінгвістична суть змінної | Універсальна множина зна-чень (множина $D$) | Лінгвістичні терми (множина $L$) |
|---|---|---|---|
| $m_1$ | Показники видання | (1—5) у. о. | Просте видання, ускладнене видання, складне видання |
| $m_2$ | Конструкційні особли-вості | (1—5) у. о. | Проста конструкція, ускладнена конструкція, складна конструкція |
| $m_3$ | Умови експлуатації (гру-пи довговічності корис-тування) | (1—5) кате-горія | Нормальні умови, робочі умови, граничні умови |
| $o_1$ | Тип виробництва | (1—5) у. о. | Одиничне виробництво, серійне виробництво, масове виробництво |
| $o_2$ | Матеріали (складність опрацюван-ня) | (1—5) у. о. | Низька складність, середня складність, ви-сока складність |
| $o_3$ | Тип обладнання | (1—5) у. о. | Ручне, механічне, авто-матизоване |
| $p_1$ | Технологічні та економіч-ні розрахунки (ефективність виробни-цтва) | (10—90) % | Низька ефективність, середня ефективність, висока ефективність |
| $p_2$ | Схема технологічного процесу | (1—5) у. о. | Проста, ускладнена, складна |

Для візуалізації залежності якості проєктування післядрукарських процесів від значення лінгвістичних термів виокремлених факторів синтезується багаторівнева модель нечіткого логічного виведення [27; 59]. Використання багаторівневої моделі нечіткого логічного ви-ведення сприяє послідовному встановленню прогнозу якості реалі-зації проєктування післядрукарських процесів шляхом накопичення знань від найнижчого до найвищого її рівнів. Ця модель включає підпорядковані моделі: модель якості формування видання, модель якості організації виробництва, модель якості планування.

Рівень якості досліджуваного процесу позначено лінгвістичним термом $G$. При цьому універсальна множина $D$ ділиться на частини (кванти). У точках поділу задаються означені нами лінгвістичні змінні та ранги $r_g(d_i)$, що ідентифікують лінгвістичні терми. Отже, вихідною базою даних буде множина $D = \{d_1, d_2, \ldots d_n\}$ і ранги $r_g(d_i)$, що встановлюють пріоритетність лінгвістичних термів у діапазонах $d_i$ $(i = 1, \ldots, n)$. З урахуванням сказаного лінгвістичний терм «рівень якості технологічного процесу» $G$ подається у вигляді деякої нечіткої множини, елементи якої утворюють сукупності пар [15; 28; 41]:

$$G = \left\{ \frac{\mu_g(d_1)}{d_1}, \frac{\mu_g(d_2)}{d_2}, \ldots, \frac{\mu_g(d_n)}{d_n} \right\}, \tag{32}$$

де $G \subset D$; $\mu_g(d_i)$ — міра належності елемента $d_i \in D$ до множини $G$.

Міри або функції належності $\mu_g(d_i)$ є базовими складовими логічних рівнянь, розв'язання яких забезпечує числове значення функції належності лінгвістичного терму $G$. Для функцій належності виконується умова нормування: $\mu_1 + \mu_2 + \ldots + \mu_n = 1$.

При цьому розподіл мір (функцій) належності відповідає таким умовам:

$$\frac{\mu_1}{r_1} = \frac{\mu_2}{r_2} = \ldots = \frac{\mu_n}{r_n}, \tag{33}$$

де $\mu_i = \mu_g(d_i)$; $r_i = r_g(d_i)$ для всіх $i = 1, \ldots, n$.

Числові значення функцій належності, що слугують для встановлення рангів факторів проєктування післядрукарських процесів, отримуються із співвідношень [28; 41; 43]:

$$\left.\begin{array}{l} \mu_1 = \left( 1 + \dfrac{r_2}{r_1} + \dfrac{r_3}{r_1} + \ldots + \dfrac{r_n}{r_1} \right)^{-1} \\[3mm] \mu_2 = \left( \dfrac{r_1}{r_2} + 1 + \dfrac{r_3}{r_2} + \ldots + \dfrac{r_n}{r_2} \right)^{-1} \\[3mm] \ldots\ldots\ldots\ldots\ldots\ldots\ldots\ldots\ldots\ldots\ldots\ldots \\[2mm] \mu_n = \left( \dfrac{r_1}{r_n} + \dfrac{r_2}{r_n} + \dfrac{r_3}{r_n} + \ldots + 1 \right)^{-1} \end{array}\right\}. \tag{34}$$

На основі наведеного теоретичного обгрунтування формуються ключові задачі:

$$\left.\begin{array}{l} G_F = F\left(m_a, o_b, p_c\right) \rightarrow \max, \quad a = \overline{1,3}; \, b = \overline{1,3}; \, c = \overline{1,2} \\ m_a > 0, \, o_b > 0, \, p_c > 0 \\ \mu_g\left(y_i\right) \rightarrow \max, \, y_i \in Y, \, G_F \subset Y, \, i = \overline{1,3} \end{array}\right\}. \tag{35}$$

Для графічного відображення лінгвістичних термів діапазон значень лінгвістичних змінних ділиться на чотири частини, у результаті чого виникає п'ять точок $\left(d_1, d_2, d_3, d_4, d_5\right)$ [52, 53].

При відомих, або отриманих на основі матриць попарних порівнянь, рангах для кожного з лінгвістичних термів розраховуються функції належності $\mu_i$ у результаті опрацювання матриці

$$A = \begin{bmatrix} 1 & \dfrac{r_2}{r_1} & \dfrac{r_3}{r_1} & \dfrac{r_4}{r_1} & \dfrac{r_5}{r_1} \\ \dfrac{r_1}{r_2} & 1 & \dfrac{r_3}{r_2} & \dfrac{r_4}{r_2} & \dfrac{r_5}{r_2} \\ \dots & \dots & \dots & \dots & \dots \\ \dfrac{r_1}{r_5} & \dfrac{r_2}{r_5} & \dfrac{r_3}{r_5} & \dfrac{r_4}{r_5} & 1 \end{bmatrix}. \tag{36}$$

Отримання підсумкового результату полягає у досягненні максимального значення функції, що характеризує рівень якості процесу при максимальних значеннях функцій належності термів оцінювання факторів — лінгвістичних змінних.

Внаслідок побудови матриць попарних порівнянь для кожного терму аналізованих лінгвістичних змінних проєктування післядрукарських процесів формується певне кількісне уявлення про взаємовідношення рангів у точках універсальної множини. Тобто використання шкали відносної важливості об'єктів за Сааті та методології створення квадратних обернено-симетричних матриць значно полегшує сприйняття якісних ознак. Однак порівняння відбувається в межах рангів одного терму, що не дає уявлення про взаємозв'язок термів лінгвістичної змінної. Опрацювання функцій належності, на основі матриць попарних порівнянь, уможливлює перетворення експертної думки у кількісні показники [28; 29; 63].

Будуємо матриці попарних порівнянь для лінгвістичної змінної «показники видання» з терм-множиною значень $L(m_1) = $ <просте, ускладнене, складне> та універсальною множиною значень $D(m_1) = [1;2;3;4;5]$ у. о., що характеризують кількісні ознаки.

$$S_{просте}(m_1) = \begin{bmatrix} 1 & 7/9 & 5/9 & 3/9 & 1/9 \\ 9/7 & 1 & 5/7 & 3/7 & 1/7 \\ 9/5 & 7/5 & 1 & 3/5 & 1/5 \\ 9/3 & 7/3 & 5/3 & 1 & 1/3 \\ 9 & 7 & 5 & 3 & 1 \end{bmatrix};$$

$$S_{ускладнене}(m_1) = \begin{bmatrix} 1 & 5 & 8 & 3 & 1 \\ 1/5 & 1 & 8/5 & 3/5 & 1/5 \\ 1/8 & 5/8 & 1 & 3/8 & 1/8 \\ 1/3 & 5/3 & 8/3 & 1 & 1/3 \\ 1 & 5 & 8 & 3 & 1 \end{bmatrix};$$

$$S_{складне}(m_1) = \begin{bmatrix} 1 & 5 & 7 & 8 & 9 \\ 1/5 & 1 & 7/5 & 8/5 & 9/5 \\ 1/7 & 5/7 & 1 & 8/7 & 9/7 \\ 1/8 & 5/8 & 7/8 & 1 & 9/8 \\ 1/9 & 5/9 & 7/9 & 8/9 & 1 \end{bmatrix}.$$

Внаслідок обчислення матриць значення функцій належності для термів «просте», «ускладнене» та «складне» лінгвістичної змінної $m_1$ «показники видання» будуть наступними:

$$\mu_{просте}(y_1) = 0,36 \; ; \; \mu_{просте}(y_2) = 0,28 \; ; \; \mu_{просте}(y_3) = 0,2 \; ;$$
$$\mu_{просте}(y_4) = 0,12 \; ; \; \mu_{просте}(y_5) = 0,04 \; ;$$
$$\mu_{ускладнене}(y_1) = 0,055 \; ; \; \mu_{ускладнене}(y_2) = 0,277 \; ; \; \mu_{ускладнене}(y_3) = 0,444 \; ;$$
$$\mu_{ускладнене}(y_4) = 0,166 \; ; \; \mu_{ускладнене}(y_5) = 0,055 \; ;$$
$$\mu_{складне}(y_1) = 0,033 \; ; \; \mu_{складне}(y_2) = 0,166 \; ; \; \mu_{складне}(y_3) = 0,233 \; ;$$
$$\mu_{складне}(y_4) = 0,266 \; ; \; \mu_{складне}(y_5) = 0,3 \; .$$

Пронормовані відносно одиниці значення функцій належностей (коефіцієнт нормування $k_e = 1/\max \mu_e(y_i)$, $(i=1,2,3)$; $\mu_{e_n}(y_i) = k_e \times \mu_e(y_i)$) матимуть вид:

$$\mu_{\text{просте}_n}(y_1) = 1\,;\ \mu_{\text{просте}_n}(y_2) = 0{,}778\,;\ \mu_{\text{просте}_n}(y_3) = 0{,}556\,;$$

$$\mu_{\text{просте}_n}(y_4) = 0{,}333\,;\ \mu_{\text{просте}_n}(y_5) = 0{,}111\,;$$

$$\mu_{\text{ускладнене}_n}(y_1) = 0{,}124\,;\ \mu_{\text{ускладнене}_n}(y_2) = 0{,}624\,;\ \mu_{\text{ускладнене}_n}(y_3) = 1\,;$$

$$\mu_{\text{ускладнене}_n}(y_4) = 0{,}374\,;\ \mu_{\text{ускладнене}_n}(y_5) = 0{,}124\,;$$

$$\mu_{\text{складне}_n}(y_1) = 0{,}11\,;\ \mu_{\text{складне}_n}(y_2) = 0{,}553\,;\ \mu_{\text{складне}_n}(y_3) = 0{,}777\,;$$

$$\mu_{\text{складне}_n}(y_4) = 0{,}887\,;\ \mu_{\text{складне}_n}(y_5) = 1\,.$$

Утворимо нечіткі множини за формулою (32):

$$\text{просте видання} = \left\{ \frac{1}{1}; \frac{0{,}778}{2}; \frac{0{,}556}{3}; \frac{0{,}333}{4}; \frac{0{,}111}{5} \right\}\ \text{у. о.};$$

$$\text{ускладнене видання} = \left\{ \frac{0{,}124}{1}; \frac{0{,}624}{2}; \frac{1}{3}; \frac{0{,}374}{4}; \frac{0{,}124}{5} \right\}\ \text{у. о.};$$

$$\text{складне видання} = \left\{ \frac{0{,}11}{1}; \frac{0{,}553}{2}; \frac{0{,}777}{3}; \frac{0{,}887}{4}; \frac{1}{5} \right\}\ \text{у. о.}$$

За нечіткими множинами побудуємо графік функцій належності термів «просте», «ускладнена», «складне». При цьому по осі абсцис відобразимо універсальну множину значень, а по осі ординат — нормовані значення функцій належності термів лінгвістичної змінної «показники видання» [29; 69].

Опускаючи подібні викладки для решти лінгвістичних змінних, перейдемо до наступної компоненти нечіткої логіки.

База нечітких знань може бути представлена у вигляді матриці знань, яка пов'язує вхідні змінні (фактори $m$-го технологічного процесу) з вихідною змінною (результатом реалізації $m$-го технологічного процесу). Для побудови матриці знань використовується система висловлювань «якщо — і — тоді», «якщо — тоді — інакше», «якщо — або — тоді — інакше». На основі матриці знань створюється система нечітких логічних рівнянь, яка дозволяє отримати числові значення функцій належності та інтегрального прогнозу якості $m$-го технологічного процесу [26; 42].

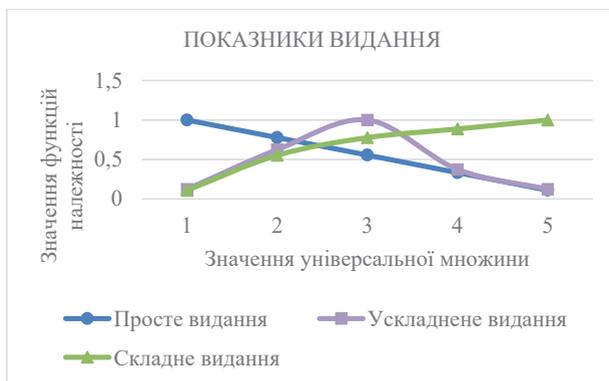

Рис. 3. Візуалізація функцій належності лінгвістичної
змінної «показники видання»

Наведемо комбінації отримання результату для двох значень
функцій належності $\mu_1$ та $\mu_2$:

$$\mu_1 \vee \mu_2 = \max(\mu_1, \mu_2) = \begin{cases} \mu_1, \textit{якщо } \mu_1 \geq \mu_2, \\ \mu_2, \textit{якщо } \mu_1 < \mu_2, \end{cases} \tag{37}$$

$$\mu_1 \wedge \mu_2 = \min(\mu_1, \mu_2) = \begin{cases} \mu_1, \textit{якщо } \mu_1 \leq \mu_2, \\ \mu_2, \textit{якщо } \mu_1 > \mu_2, \end{cases} \tag{38}$$

де операція $\vee$ у нечітких логічних рівняннях вказує на отримання
максимального значення, а операція $\wedge$ — мінімального значення.

Нехай для лінгвістичних змінних $M$ (якість формування видан-
ня), $O$ (якість організації виробництва) та $P$ (якість опрацювання
видання) термами будуть «низька», «середня», «висока». Відпо-
відно інтегральний показник G (якість проєктування післядру-
карських процесів) описуватиметься такими ж термами [27; 60;
64]. Тоді нечітка база знань для відношення $G = F_G(M, O, P)$ мати-
ме вид:

ЯКЩО ($M$ = низька) І ($M$ = середня) І ($M$ = висока);

І ($O$ = низька) І ($O$ = середня) І ($O$ = висока);

І ($P$ = низька) І ($P$ = середня) І ($P$ = висока);

ТОДІ ($G$ = низька) І ($G$ = середня) І ($G$ = висока).

Сформовані умови відображаються у матриці знань.

**Матриця знань для лінгвістичної змінної G**

| Якість вихідних даних видання $M$ | Якість опрацювання видання $O$ | Якість оформлення видання $P$ | Якість проєктування післядрукарських процесів $G$ |
|---|---|---|---|
| низька | низька | низька | низька |
| низька | середня | низька | |
| середня | низька | середня | середня |
| висока | середня | середня | |
| висока | висока | висока | висока |
| висока | середня | висока | |

Нечіткі логічні рівняння для термів «низька», «середня», «висока» інтегрального показника $G$ матимуть вид:

$$\mu_{\text{низька}}(G) = \mu_{\text{низька}}(M) \wedge \mu_{\text{низька}}(O) \wedge \mu_{\text{низька}}(P) \vee$$
$$\vee \mu_{\text{низька}}(M) \wedge \mu_{\text{середня}}(O) \wedge \mu_{\text{низька}}(P);$$
$$\mu_{\text{середня}}(G) = \mu_{\text{середня}}(M) \wedge \mu_{\text{низька}}(O) \wedge \mu_{\text{середня}}(P) \vee$$
$$\vee \mu_{\text{висока}}(M) \wedge \mu_{\text{середня}}(O) \wedge \mu_{\text{середня}}(P);$$
$$\mu_{\text{висока}}(G) = \mu_{\text{висока}}(M) \wedge \mu_{\text{висока}}(O) \wedge \mu_{\text{висока}}(P) \vee$$
$$\vee \mu_{\text{висока}}(M) \wedge \mu_{\text{середня}}(O) \wedge \mu_{\text{висока}}(P).$$

Зважаючи на те, що залежності якості формування видання, організації виробництва та опрацювання видання також можуть бути виражені через якість часткових показників $M = F_M(m_1, m_2, m_3)$, $O = F_O(o_1, o_2, o_3)$, $P = F_P(p_1, p_2)$ і на основі експертних суджень щодо множин $L(m_1, m_2, m_3)$, $L(o_1, o_2, o_3)$, $L(p_1, p_2)$ формуються нечіткі бази знань, матриці знань і нечіткі логічні рівняння лінгвістичних змінних проєктування післядрукарських процесів [59].

Логічні висловлювання стосовно таких лінгвістичних змінних:

— «якість формування видання»;
— «якість організації виробництва»;
— «якість опрацювання видання».

ЯКЩО ($m_1$) = (просте, ускладнене, складне),
I ($m_2$) = (проста, ускладнена, складна);
I ($m_3$) = (нормальні, робочі, граничні);
ТОДІ ($M$) = (низька, середня, висока);

ЯКЩО ($o_1$) = (одиничне, серійне, масове),
I ($o_2$) = (низька, середня, висока);
I ($o_3$) = (ручне, механічне, автоматизоване);
ТОДІ ($O$) = (низька, середня, висока);

ЯКЩО ($p_1$) = (низька, середня, висока),
I ($p_2$) = (проста, ускладнена, складна);
ТОДІ ($P$) = (низька, середня, висока).

Далі будуються матриці знань для аналізованих лінгвістичних змінних. Для зручності вони відображаються у табличній формі [50; 60; 64].

Таблиця 10

**Матриця знань для лінгвістичної змінної $M$ (якість формування видання)**

| Показники видання $m_1$ | Конструкційні особливості (складність конструкції) $m_2$ | Умови експлуатації $m_3$ | Якість вихідних даних видання $M$ |
|---|---|---|---|
| складне | складна | граничні | низька |
| складне | складна | робочі | |
| складне | ускладнена | робочі | середня |
| ускладнене | ускладнена | нормальні | |
| ускладнене | проста | нормальні | висока |
| просте | проста | нормальні | |

Таблиця 11

**Матриця знань для лінгвістичної змінної $O$ (якість організації виробництва)**

| Тип виробництва $o_1$ | Матеріали (складність опрацювання) $o_2$ | Тип обладнання $o_3$ | Якість організації виробництва $O$ |
|---|---|---|---|
| одиничне | висока | ручне | низька |
| серійне | висока | ручне | |
| серійне | середня | механічне | середня |
| серійне | середня | автоматизоване | |
| серійне | низька | автоматизоване | висока |
| масове | низька | автоматизоване | |

**Матриця знань для лінгвістичної змінної $P$ (якість опрацювання видання)**

| Технологічні та економічні розрахунки (ефективність виробництва) $p_1$ | Схема технологічного процесу $p_2$ | Якість опрацювання видання $P$ |
|---|---|---|
| низька | складна | низька |
| низька | ускладнена | |
| середня | проста | середня |
| середня | ускладнена | |
| висока | проста | висока |
| висока | ускладнена | |

Нечіткі логічні рівняння для визначених термів матимуть вид:
— для лінгвістичної змінної «якість формування видання»

$$\mu_{\text{низька}}(M) = \mu_{\text{складне}}(m_1) \wedge \mu_{\text{складна}}(m_2) \wedge \mu_{\text{граничні}}(m_3) \vee$$
$$\vee \mu_{\text{складне}}(m_1) \wedge \mu_{\text{складна}}(m_2) \wedge \mu_{\text{робочі}}(m_3),$$

$$\mu_{\text{середня}}(M) = \mu_{\text{складне}}(m_1) \wedge \mu_{\text{ускладнена}}(m_2) \wedge \mu_{\text{робочі}}(m_3) \vee$$
$$\vee \mu_{\text{ускладнене}}(m_1) \wedge \mu_{\text{ускладнена}}(m_2) \wedge \mu_{\text{нормальні}}(m_3),$$

$$\mu_{\text{висока}}(M) = \mu_{\text{ускладнене}}(m_1) \wedge \mu_{\text{проста}}(m_2) \wedge \mu_{\text{нормальні}}(m_3) \vee$$
$$\vee \mu_{\text{просте}}(m_1) \wedge \mu_{\text{проста}}(m_2) \wedge \mu_{\text{нормальні}}(m_3);$$

— для лінгвістичної змінної «якість організації виробництва»

$$\mu_{\text{низька}}(O) = \mu_{\text{одиничне}}(o_1) \wedge \mu_{\text{висока}}(o_2) \wedge \mu_{\text{ручне}}(o_3) \vee$$
$$\vee \mu_{\text{серійне}}(o_1) \wedge \mu_{\text{висока}}(o_2) \wedge \mu_{\text{ручне}}(o_3),$$

$$\mu_{\text{середня}}(O) = \mu_{\text{серійне}}(o_1) \wedge \mu_{\text{середня}}(o_2) \wedge \mu_{\text{механічне}}(o_3) \vee$$
$$\vee \mu_{\text{серійне}}(o_1) \wedge \mu_{\text{середня}}(o_2) \wedge \mu_{\text{автоматизоване}}(o_3),$$

$$\mu_{\text{висока}}(O) = \mu_{\text{серійне}}(o_1) \wedge \mu_{\text{низька}}(o_2) \wedge \mu_{\text{автоматизоване}}(o_3) \vee$$
$$\vee \mu_{\text{масове}}(o_1) \wedge \mu_{\text{низька}}(o_2) \wedge \mu_{\text{автоматизоване}}(o_3);$$

— для лінгвістичної змінної «якість опрацювання видання»

$$\mu_{\text{низька}}(P) = \mu_{\text{низька}}(p_1) \wedge \mu_{\text{складна}}(p_2) \vee \mu_{\text{низька}}(p_1) \wedge \mu_{\text{ускладнена}}(p_2),$$
$$\mu_{\text{середня}}(P) = \mu_{\text{середня}}(p_1) \wedge \mu_{\text{проста}}(p_2) \vee \mu_{\text{середня}}(p_1) \wedge \mu_{\text{ускладнена}}(p_2),$$
$$\mu_{\text{висока}}(P) = \mu_{\text{висока}}(p_1) \wedge \mu_{\text{проста}}(p_2) \vee \mu_{\text{висока}}(p_1) \wedge \mu_{\text{ускладнена}}(p_2).$$

### 5.2. Дефазифікація нечіткої множини

Для встановлення інтегрального показника якості проєктування післядрукарських процесів здійснюється процес дефазифікації, враховуючи розподілення якості за частковими показниками $G = F_G(M, O, P)$.

Слід зазначити, що дефазифікація є одним з ключових процесів нечіткої логіки, який полягає у перетворенні значень нечіткої множини у кількісний показник. Дефазифікація передбачає наявність сформованих нечітких баз знань та нечітких логічних рівнянь досліджуваного технологічного процесу для подальшого формування таблиць на основі терм-множин з пронормованими значеннями функцій належності у визначених точках поділу універсальної множини значень виокремлених лінгвістичних змінних та підставлення значень термів у нечіткі логічні рівняння. Для здійснення числових розрахунків у дослідженні обрано метод центру ваги, згідно з яким кількісне значення початкової змінної рівне абсцисі центру ваги площі, що обмежена графіком кривої функції належності аналізованої змінної [22; 62; 69; 72].

Відповідно до наведених тверджень формуються таблиці значень функцій належності для кожної лінгвістичної змінної за точками поділу універсальної множини значень та терм-множинами (табл. 6—13) [22; 26; 62; 72].

Як приклад наведемо таблицю значень терм-множини $D(m_1)$ лінгвістичної змінної «Показники видання»:

Таблиця 13

**Функції належності терм-множини $D(m_1)$ (показники видання)**

| $y_i$, умовні одиниці | 1 | 2 | 3 | 4 | 5 |
|---|---|---|---|---|---|
| $\mu_{просте}(y_i)$ | 1 | 0,778 | 0,556 | 0,333 | 0,111 |
| $\mu_{ускладнене}(y_i)$ | 0,124 | 0,624 | 1 | 0,374 | 0,124 |
| $\mu_{складне}(y_i)$ | 0,11 | 0,553 | 0,777 | 0,887 | 1 |

Наведемо нечіткі логічні рівняння для термів «низька», «середня», «висока» найвищого рівня $G$:

$$\mu_{низька}(G) = 0,667 \wedge 0,443 \wedge 0,334 \vee 0,667 \wedge 1 \wedge 0,334 = 0,334,$$
$$\mu_{середня}(G) = 0,777 \wedge 0,443 \wedge 1 \vee 0,333 \wedge 1 \wedge 1 = 0,443,$$
$$\mu_{висока}(G) = 0,333 \wedge 0,375 \wedge 0,666 \vee 0,333 \wedge 1 \wedge 0,666 = 0,333.$$

### 5.3. Визначення числового значення інтегрального показника якості

На основі отриманих даних виконується дефазифікація нечіткої множини за формулою [22; 26; 62; 72]:

$$G = \frac{\sum_{i=1}^{m}\left[\underline{G}+(i-1)\frac{\overline{G}-\underline{G}}{m-1}\right]\mu_i(G)}{\sum_{i=1}^{m}\mu_i(G)}, \tag{39}$$

де $\underline{G}$ — найменше значення показника якості; $\overline{G}$ — найбільше значення показника якості; $m$ — кількість нечітких термів [22; 26; 62; 72].

Приймаються умовні межі для змінної $G$: $\underline{G} = 1\%$, $\overline{G} = 100\%$. Обчислення виконується за трьома точками поділу: 1 %, 50 %, 100 %. У результаті обчислення встановлюється числове значення інтегрального показника якості проєктування післядрукарських процесів:

$$G_{прогноз.} = \frac{1 \cdot 0{,}334 + 50 \cdot 0{,}443 + 100 \cdot 0{,}333}{0{,}334 + 0{,}443 + 0{,}333} = 50{,}256\,\%.$$

Опираючись на сформовану послідовність етапів дослідження, наведемо синтезовану структурно-функціональну модель інформаційної технології прогностичного оцінювання якості проєктування післядрукарських процесів (рис. 4).

Таким чином, розроблена структурно-функціональна модель складається з п'яти основних етапів, кожен з яких розділений на відповідні підетапи, що визначають окрему дію з отримання, моделювання, аналізу та синтезу інформації, задля визначення якості досліджуваного процесу. Така деталізація, внаслідок впровадження розробленої інформаційної технології в реальні виробничі умови уможливлює обдумане, прогнозоване формування проєкту реалізації післядрукарських процесів, підвищення економічної ефективності, доцільності операцій, спрощення післядрукарського опрацювання книжкової продукції.

Використаємо методологію IDEF0 для функціонального моделювання інформаційної технології проєктування післядрукарських процесів. Побудуємо контекстну діаграму А-0, діаграму першого рівня декомпозиції А0, діаграми другого рівня декомпозиції А1, А2, А3, А4, А5, діаграми третього рівня декомпозиції А23, А24, А41, А42, А51, А52 [40; 77].

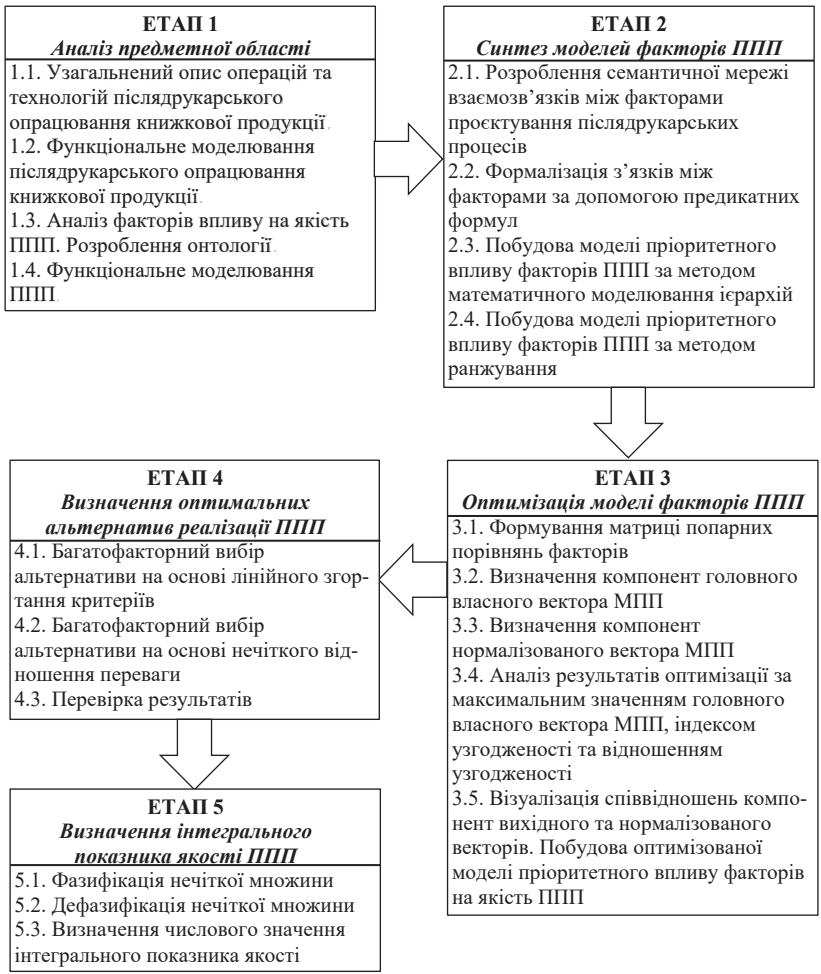

**ЕТАП 1**
*Аналіз предметної області*
1.1. Узагальнений опис операцій та технологій післядрукарського опрацювання книжкової продукції
1.2. Функціональне моделювання післядрукарського опрацювання книжкової продукції.
1.3. Аналіз факторів впливу на якість ППП. Розроблення онтології.
1.4. Функціональне моделювання ППП.

**ЕТАП 2**
*Синтез моделей факторів ППП*
2.1. Розроблення семантичної мережі взаємозв'язків між факторами проєктування післядрукарських процесів
2.2. Формалізація з'язків між факторами за допомогою предикатних формул
2.3. Побудова моделі пріоритетного впливу факторів ППП за методом математичного моделювання ієрархій
2.4. Побудова моделі пріоритетного впливу факторів ППП за методом ранжування

**ЕТАП 4**
*Визначення оптимальних альтернатив реалізації ППП*
4.1. Багатофакторний вибір альтернативи на основі лінійного згортання критеріїв
4.2. Багатофакторний вибір альтернативи на основі нечіткого відношення переваги
4.3. Перевірка результатів

**ЕТАП 3**
*Оптимизація моделі факторів ППП*
3.1. Формування матриці попарних порівнянь факторів
3.2. Визначення компонент головного власного вектора МПП
3.3. Визначення компонент нормалізованого вектора МПП
3.4. Аналіз результатів оптимізації за максимальним значенням головного власного вектора МПП, індексом узгодженості та відношенням узгодженості
3.5. Візуалізація співвідношень компонент вихідного та нормалізованого векторів. Побудова оптимізованої моделі пріоритетного впливу факторів на якість ППП

**ЕТАП 5**
*Визначення інтегрального показника якості ППП*
5.1. Фазифікація нечіткої множини
5.2. Дефазифікація нечіткої множини
5.3. Визначення числового значення інтегрального показника якості

Рис. 4. Структурно-функціональна модель інформаційної технології прогностичного оцінювання якості проєктування післядрукарських процесів

Контекстна діаграма зображена на рис. 5. При цьому основною функцією системи є інформаційна технологія прогностичного оцінювання якості проєктування післядрукарських процесів, а зв'язок системи із навколишнім середовищем зображується граничними стрілками: $I_1$ — потреба у розробленні інформаційної технології, $I_2$ — погано структурована задача, $C_1$ — нормативно-технічна та технологічна документація, $C_2$ — теорії, методи, методики, принципи, $O_1$ — оптимальна альтернатива реалізації проєктування післядрукарських процесів, $O_2$ — інтегральний показник якості проєктування післядрукарських процесів, $M_1$ — апаратне та програмне забезпечення, $M_2$ — дослідники, експерти з предметної області, інші зацікавлені особи.

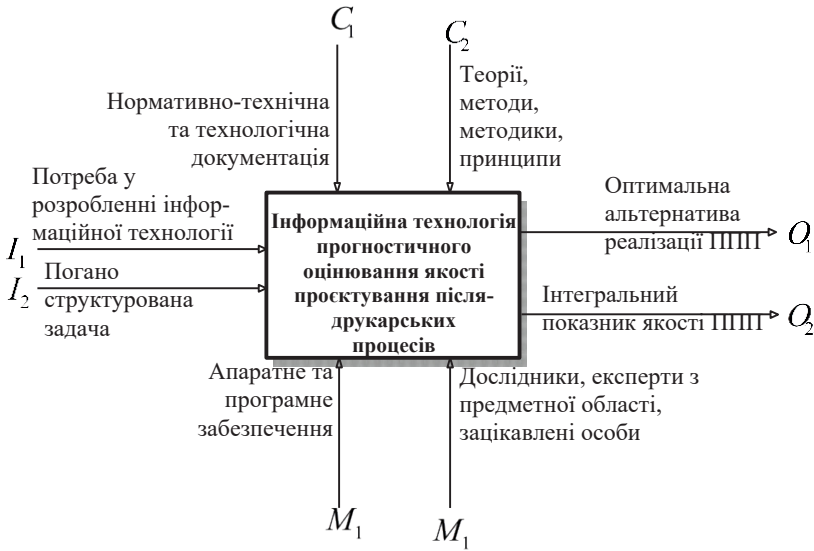

Рис. 5. Контекстна діаграма А-0 моделі IDEF0 інформаційної технології прогностичного оцінювання якості проєктування післядрукарських процесів

Опишемо кожну граничну стрілку, враховуючи поділ за типами.

Граничні стрілки типу «Вхід» (Input):

— $I_1$ (потреба у розробленні інформаційної технології). Потреба в організації інформаційних процесів з використанням засобів обчислювальної техніки, що пришвидшує опрацювання даних, пошук інформації та спрощує доступ до неї;

— $I_2$ (погано структурована задача). За глибиною пізнання розрізняють три класи проблем: добре структуровані, неструктуровані та погано структуровані. Останні характеризуються наявністю як якісних, так і кількісних показників, з явною перевагою маловідомих, недостатньо досліджених якісних характеристик проблеми. До погано структурованих проблем відносяться великомасштабні задачі, яким притаманні значна кількість альтернатив, залежність від сучасних технологій, невизначеність щодо тривалості виконання, кількості фінансових ресурсів та матеріалів, наявність ризиків. Проєктування післядрукарських процесів належить до погано структурованих проблем.

Граничні стрілки типу «Контроль» (Control):

— $C_1$ (нормативно-технічна та технологічна документація). До нормативно-технічної документації належать технічні вимоги та законодавчі положення, зокрема: закони, стандарти, технічні умови, кодекси усталеної практики та ін.;

— $C_2$ (теорії методи, методики, принципи). Кожен інформаційний процес виконується на основі загальновідомих чи новітніх теорій, методів, методик та принципів. Так, наприклад, виокремлення та формалізація зв'язків між факторами проєктування післядрукарських процесів здійснюється за методами системного та матричного аналізу, теорією графів та семантичних мереж, логікою предикатів; створення моделі пріоритетного впливу факторів відбувається за теорією ієрархічних багаторівневих систем і т. д. [58; 59].

Граничні стрілки типу «Вихід» (Output):

— $O_1$ (оптимальна альтернатива реалізації проєктування післядрукарських процесів). Виконання будь-якого технологічного завдання передбачає наявність можливих альтернатив. Важливим етапом є вибір найкращої альтернативи реалізації серед множини існуючих [25; 79];

— $O_2$ (інтегральний показник якості проєктування післядрукарських процесів). Основною метою виконання будь-якого процесу є отримання якісного результату. При цьому прогнозування якості уможливлює досягнення мети. Встановлення кількісного показника якості проєктування післядрукарських процесів здійснюється за допомогою методів та засобів нечіткої логіки [26; 42; 62].

Граничні стрілки типу «Механізми» (Mechanism):

— $M_1$ (апаратне та програмне забезпечення). Пошук, опрацювання, зберігання та передавання інформації передбачає використання сучасних технічних та програмних засобів;

— $M_2$ (дослідники, експерти з предметної області, зацікавлені особи). Реалізація інформаційної технології прогностичного оцінювання якості проєктування післядрукарських процесів передбачає проведення ряду досліджень із залученням фахових науковців, формуванням експертних висновків, апробацією та консультуванням із зацікавленими особами.

Діаграма першого рівня декомпозиції A0 моделі IDEF0 утворена шляхом декомпозиції контекстної діаграми та містить такі функціональні блоки:

— АПО (аналіз предметної області). Здійснюється для означення теоретичної складової досліджуваного процесу та виокремлення послідовності операцій шляхом функціонального моделювання. При цьому розглядається не лише створення проєкту, а й сам процес післядрукарського опрацювання книжкової продукції, адже важливо розуміти особливості об'єкта проєктування;

— СМФ ППП (синтез моделей факторів проєктування післядрукарських процесів). Полягає у розробленні семантичної мережі та формалізації зв'язків між факторами, використовуючи елементи логіки предикатів, а також у визначенні рівнів домінантності факторів та побудові моделі пріоритетного впливу. Пріоритетність факторів визначається за методом математичного моделювання ієрархій і уточнюється за методом ранжування [58, 59];

— ОМФ ППП (оптимізація моделі факторів проєктування післядрукарських процесів). Передбачає встановлення оптимізованих вагових значень факторів досліджуваного процесу та побудову оптимізованої моделі. Цей етап дозволяє деталізувати пріоритетність факторів і, за наявності, уникнути розміщення кількох факторів на однаковому рівні [68];

— ВОАР ППП (встановлення оптимальних альтернатив реалізації проєктування післядрукарських процесів). Здійснюється проєктування та дослідження можливих альтернатив реалізації аналізованого процесу та обирається оптимальна. При цьому опрацювання здійснюється за двома методами, а результати порівнюються. Збіжність отриманих результатів свідчить про адекватність розв'язку задачі [25; 61];

— ВІПЯ ППП (встановлення інтегрального показника якості проєктування післядрукарських процесів). На основі методів та засобів нечіткої логіки встановлюються прогнозовані числові параметри досліджуваного процесу та, відповідно до заданих умов, визначається інтегральний показник якості [26; 42; 62].

Наступним етапом є побудова діаграм другого рівня декомпозиції, тобто декомпозиція діаграми першого рівня. Діаграма A1 складається з чотирьох функціональних блоків:

— ООТ ПОКП (опис операцій та технологій післядрукарського опрацювання книжкової продукції). Описуються можливі операції та технології післядрукарського опрацювання, умови їх вибору та реа-лізації. Такий опис уможливлює подальше моделювання та проєктування;

— ФМ ПОКП (функціональне моделювання післядрукарського опрацювання книжкової продукції). Функціональне моделювання здійснюється за методологією IDEF0. Полягає у формуванні контекстної діаграми, де основною функцією системи є післядрукарське опрацювання книжкової продукції, декомпозиції контекстної діаграми (створенні діаграми першого рівня декомпозиції) та декомпозиції діаграми першого рівня (створенні діаграм другого рівня декомпозиції). Формується деревовидна ієрархічна модель післядрукарського опрацювання книжкової продукції, яка ілюструє відношення між батьківськими та дочірніми вузлами моделі IDEF0 [40; 77];

— АФВ та РО ППП (аналіз факторів впливу та розроблення онтології проєктування післядрукарських процесів). Означується інформаційна складова факторів впливу на якість досліджуваного процесу. Здійснюється деталізований опис виокремлених факторів: показники видання, конструкційні особливості, умови експлуатації, тип виробництва, матеріали, тип обладнання, технологічні та економічні розрахунки, схема технологічного процесу [30]. Розробляється онтологія [33];

— ФМ ППП (функціональне моделювання проєктування післядрукарських процесів). Будується контекстна діаграма, діаграма першого рівня декомпозиції та діаграми другого рівня декомпозиції. Основною функцією системи є проєктування післядрукарських процесів. Також формується деревовидна ієрархічна модель, вершиною якої є контекстна діаграма [40; 77].

Діаграма А2 містить такі функціональні блоки:

— РСМ (розроблення семантичної мережі). Модель семантичної мережі є основою для подальшого дослідження. За структурою це орієнтований граф, сукупність вузлів якого відповідає множині факторів, а дуги — зв'язкам між ними [61];

— ФЗФ (формалізація зв'язків між факторами). Формалізований опис зв'язків між факторами здійснюється з використанням елементів логіки предикатів [59; 61];

— ПМПВФ за МММІ (побудова моделі пріоритетного впливу факторів за методом математичного моделювання ієрархій). Метод математичного моделювання ієрархій полягає у встановленні рівнів пріоритетності факторів шляхом побудови матриці досяжності та ітераційних таблиць [58];

— ПМПВФ за МР (побудова моделі пріоритетного впливу факторів за методом ранжування). Метод ранжування передбачає побудову ієрархічних дерев, що ілюструють зв'язки між факторами і встановлення пріоритетності факторів за ваговими значеннями [30].

Для третього та четвертого функціональних блоків діаграми А2 доцільно продовжити декомпозицію. Діаграма А23 містить такі функціональні блоки:

— СМД (створення матриці досяжності). Матриця досяжності для зручності відображення даних будується у вигляді таблиці. Наявність прямого чи опосередкованого впливу позначається одиницею, а відсутність — нулем;

— СІТ (створення ітераційних таблиць). Ітераційні таблиці містять чотири колонки: порядковий номер фактора у множині; порядкові номери факторів, на які впливає визначений фактор; порядкові номери факторів, від яких залежить визначений фактор; спільні порядкові номери впливаючих та залежних факторів. Кожна ітерація полягає у викресленні рядка (рядків) ітераційної таблиці, у якому співпали дані у третьому та четвертому стовпцях. Фактор, що відповідає першому викресленому рядку, має найвищий рівень пріоритетності, а фактор, що відповідає останньому викресленому рядку, — найнижчий рівень пріоритетності;

— МПФ (моделювання пріоритетності факторів). Дані, отримані внаслідок ітерації, використовуються для побудови моделі пріоритетного впливу факторів [58].

Діаграма А24:

— СІД (створення ієрархічних дерев). Для кожного фактора множини будуються ієрархічні дерева, що ілюструють прямі та опосередковані впливи визначеного фактора на інші фактори та ієрархічні дерева, що ілюструють прямі та опосередковані залежності;

— ВВЗФ (встановлення вагових значень факторів). За ієрархічними деревами визначається кількість прямих та опосередкованих впливів і залежностей. Приймаються умовні вагові коефіцієнти. Обчислюються інтегральні вагові величини факторів за сумами ваг усіх типів зв'язків;

— ВРФ (встановлення рангів факторів). Згідно з інтегральними ваговими значеннями визначаються ранги факторів. При цьому фактору із найменшим інтегральним значенням належить найнижчий ранг — перший. Кільком факторам можуть бути присвоєні однакові ранги;

— ВРПФ (встановлення рівня пріоритетності факторів). Рівень пріоритетності встановлюється за рангом фактора. Фактору з найвищим рангом належить найвищий рівень пріоритетності — перший. Кілька факторів можуть бути однаковими за пріоритетністю;

— МПФ (моделювання пріоритетності факторів). На основі встановлених рівнів пріоритетності факторів, отриманих внаслідок ітераційних процесів та уточнених шляхом ранжування, синтезується ієрархічна модель пріоритетного впливу факторів. При цьому найвищий рівень моделі відповідає фактору з найбільшим пріоритетом, а найнижчий — з найменшим пріоритетом серед виокремленої множини [30; 59].

Розглянемо функціональні блоки діаграми А3:

— ФМППФ (формування матриці попарних порівнянь факторів). На основі шкали відносної важливості об'єктів за Сааті формується матриця попарних порівнянь факторів. За критеріями порівняння обирається необхідна оцінка корисності від 1 до 9. Якщо об'єкти рівноцінні, оцінка корисності 1. Якщо один об'єкт абсолютно переважає інший — 9. Для опису проміжних відношень слугують інші оцінки в межах вказаної шкали [25; 68];

— ВКГВВ (визначення компонент головного власного вектора). Головний власний вектор визначається як середнє геометричне елементів кожного рядка матриці попарних порівнянь;

— ВКНВ (визначення компонент нормалізованого вектора). Компоненти нормалізованого вектора визначають числові пріоритети факторів та дозволяють уточнити їх вагові значення;

— АРО (аналіз результатів оптимізації). Здійснюється перевірка отриманих результатів за нормативними значеннями індекса узгодженості, відношення узгодженості та максимальним значенням головного власного вектора матриці попарних порівнянь;

— ВСКВ (візуалізація співвідношень компонент векторів). Будуються гістограма та порівняльний графік вагових значень компонент вихідного та нормалізованого векторів. Вихідний вектор формується на основі моделі пріоритетного впливу факторів за присвоєними ваговими значеннями. Для зручності візуалізації компоненти нормалізованого вектора адаптуються за довільним коефіцієнтом;

— ПОМПВФ (побудова оптимізованої моделі пріоритетного впливу факторів). За перевіреними результатами оптимізації синтезується оптимізована модель пріоритетного впливу факторів на якість проєктування післядрукарських процесів. Найвищий рівень моделі від-повідає фактору з найбільшим пріоритетом, а найнижчий — з найменшим. Отримана модель є основою подальшого прогностичного оцінювання [68].

Діаграма А4 містить такі блоки:

— БВА ЛЗК (багатофакторний вибір альтернатив на основі лінійного згортання критеріїв). Встановлення оптимального варіанту реалізації проєктування післядрукарських процесів за методом лінійного згортання критеріїв полягає у лінійному об'єднанні часткових цільових функціоналів в один, а задача багатокритеріальної (багатофакторної) оптимізації — у знаходженні максимального значення функцій корисності [25; 59];

— БВА НВП (багатофакторний вибір альтернатив на основі нечіткого відношення переваги). Багатофакторний вибір альтернатив на основі нечіткого відношення переваги полягає у встановленні попарних переваг між запроєктованими альтернативами факторів проєктування післядрукарських процесів та їх кількісному представленні;

— ПР (перевірка результатів). Порівнюється, чи однаковими є встановлені оптимальні альтернативи за двома вищеописаними методами. Тотожність результатів свідчить про адекватність розв'язку задачі [59; 79].

Для першого та другого функціональних блоків діаграми А4 доцільно продовжити декомпозицію. Діаграма А41 містить:

— ФМП (формування множини Парето). Множина Парето включає тільки фактори з суттєво вищою пріоритетністю, фактори з низькою пріоритетністю відкидаються;

— ПА (проєктування альтернатив). Проєктується необхідна кількість альтернативних варіантів реалізації проєктування післядрукарських процесів;

— ОА (оцінювання альтернатив). Здійснюється відсоткове вираження міри впливу факторів множини Парето для кожної альтернативи;

— СМППФ (створення матриці попарних порівнянь факторів). Згідно із ваговими даними факторів множини Парето та за шкалою відносної важливості об'єктів формується матриця попарних порівнянь;

— НГВВ (нормалізація головного власного вектора). Внаслідок нормалізації головного власного вектора матриці попарних порівнянь факторів множини Парето встановлюються вагові значення, необхідні для подальших обчислень;

— ВК та БОК (визначення корисності та багатокритеріальних оцінок корисності). Формуються матриці попарних порівнянь альтернативних

варіантів реалізації щодо кожного фактора множини Парето, за якими визначаються корисності альтернатив. Багатокритеріальні оцінки корисності кожної альтернативи обчислюються як суми добутків вагових значень факторів та корисності відповідних альтернатив [25];

— ВОА (вибір оптимальної альтернативи). Оптимальна альтернатива реалізації проєктування післядрукарських процесів обирається за максимальним значенням багатокритеріальної оцінки корисності [59].

Діаграма А42:

— ОНВП на МА (оцінювання нечітких відношень переваги на множині альтернатив). Формуються відношення нестрогої переваги між альтернативами кожного фактора множини Парето;

— ФМВФ (формування матриць відношень для факторів). На основі відношень переваги формуються матриці відношень. Причому наявність переваги позначається одиницею, а непорівнюваність альтернатив між собою — нулем;

— ПЗВ (побудова згорток відношень). Формується згортка відношень за усіма факторами множини Парето, де одиницею позначається наявність переваги між альтернативами для усіх факторів, а нулем — непорівнюваність альтернатив хоча б за одним фактором. Інший тип згортки відношень визначається за ваговими значеннями факторів та відповідних функцій корисності;

— ВПНА (визначення підмножин недомінованих альтернатив). Визначаються на основі згорток відношень та за відповідними формулами;

— ВФНСМ (визначення функцій належності спільної множини). Визначається спільна множина недомінованих альтернатив та функції належності;

— ВОА (вибір оптимальної альтернативи). Оптимальною є альтернатива із максимальним значенням функції належності [25; 59].

Діаграма А5 містить три функціональні блоки, а саме:

— Ф (фазифікація). Процес фазифікації полягає у зіставленні множини значень її функцій належності;

— Д (дефазифікація). Процес дефазифікації є зворотним до фазифікації. Дефазифікація нечіткої множини здійснюється за принципом центра ваги;

— ВЧЗІПЯ (визначення числового значення інтегрального показника якості). Визначається для можливості прогностичного оцінювання якості проєктування післядрукарських процесів за певних визначених умов. Виражається у відсотках [27—29; 60; 62].

Подальшій декомпозиції підлягають перший та другий функціональні блоки діаграми А5. Відповідно діаграма А51 містить такі блоки:

— ФЧПЯ (формування часткових показників якості). Для встановлення інтегрального показника якості проєктування післядрукарських процесів доцільно сформувати часткові показники якості лінгвістичних змінних та згрупувати її за спільними ознаками та призначенням;

— ВУМЗ та ТМАЗ (виокремлення універсальної множини значень та терм-множини аналізованих змінних). Для кожної лінгвістичної змінної формується універсальна множина значень зі встановленими межами та одиницями вимірювання і терм-множина, яка словесно описує градацію універсальної множини;

— ПБМНЛВ (побудова багаторівневої моделі нечіткого логічного виводу). Багаторівнева модель нечіткого логічного виводу будується для ієрархічного представлення залежності між якістю проєктування післядрукарських процесів та значеннями лінгвістичних термів виокремлених факторів [27];

— ОФНЛЗ (опрацювання функцій належності лінгвістичних змінних). За відносними оцінками рангів лінгвістичних термів створюються квадратні обернені симетричні матриці, внаслідок обчислення яких встановлюються числові значення функцій належності у п'яти точках поділу універсальної множини. Отримані нечіткі множини візуалізуються за допомогою графіків [28; 29];

— ФБЗ (формування баз знань). Формується нечітка база знань для інтегрального показника якості та для кожної лінгвістичної змінної, враховуючи ієрархію, наведену у багаторівневій моделі нечіткого логічного виводу;

— ФМЗ (формування матриць знань). За сформованими базами знань синтезуються матриці знань для якості проєктування післядрукарських процесів та для кожного часткового показника;

— ФНЛР (формування нечітких логічних рівнянь). За сформованими матрицями знань будуються нечіткі логічні рівняння для кожного терму часткових показників якості та для термів лінгвістичної змінної «якість проєктування післядрукарських процесів» [50; 60; 64].

Діаграма А52 включає:

— ФТЗФНЛЗ (формування таблиць значень функцій належності лінгвістичних змінних). За терм-множинами з пронормованими значеннями функцій належності у п'яти точках поділу універсальної множини створюються таблиці значень для кожної лінгвістичної змінної;

— ФНЛР (формування нечітких логічних рівнянь). Здійснюється підстановка значень з таблиць значень у нечіткі логічні рівняння для термів «низька», «середня», «висока» кожної лінгвістичної змінної. Відповідно проводяться обчислення підсумкових значень функцій належності [22; 26; 62; 72].

Для відображення ієрархічної залежності функцій використаємо діаграму дерева вузлів, у якій верхній рівень відповідає контекстній діаграмі (батьківському елементу), а нижні — декомпозиції потоків (дочірнім елементам). При цьому взаємозв'язки між функціональними блоками не відображаються, лише ієрархічна впорядкованість. Такий підхід уможливлює цілісний аналіз ієрархії функціональних блоків моделі IDEF0 [77].

**Висновки.** У дослідженні наведено розв'язання актуального науково-прикладного завдання розроблення інформаційної технології прогностичного оцінювання якості проєктування післядрукарських процесів на основі дослідження домінантності виокремлених факторів і застосування нечіткої логіки для отримання інтегрального показника якості. Виокремлено основні етапи інформаційної технології.

Наведено загальну характеристику реалізації та проєктування післядрукарського опрацювання книжкових видань. Виокремлено та описано фактори впливу на якість проєктування післядрукарських процесів. Здійснено моделювання функцій реалізації та проєктування досліджуваного процесу. Створено IDEF0-моделі, які складаються з сукупності ієрархічно впорядкованих та взаємопов'язаних діаграм: контекстної діаграми, декомпозиції контекстної діаграми (діаграми першого рівня декомпозиції) та декомпозиції функціональних блоків діаграми першого та другого рівнів. Описано основні підходи до створення онтології та основні типи онтологій.

Подано методологію побудови семантичної мережі, що відтворює зв'язки між факторами впливу на якість аналізованого процесу. За допомогою логіки предикатів здійснено формалізоване відображення зв'язків між ними.

Наведено особливості синтезу та оптимізації моделі пріоритетного впливу факторів на якість проєктування післядрукарських процесів. Критерії оптимізації становлять: власне значення матриці $\lambda_{max} = 8,483$, індекс узгодженості $IU = 0,069$, відношення узгодженості $RU = 0,049$. Критерії оптимізації знаходяться в допустимих межах.

Визначено оптимальні альтернативні варіанти реалізації проєктування післядрукарських процесів за методами багатофакторного ви-бору альтернатив на основі лінійного згортання критеріїв та на основі нечіткого відношення переваги. При цьому за методом багатофакторного вибору альтернатив на основі лінійного згортання критеріїв максимальне значення отримала оцінка корисності $U_3 = 0,414$ альтернативи $A_3$. За методом багатофакторного вибору альтернатив на основі нечіткого відношення переваги максимальне значення отримала функція належності $\mu_Q^{нд}(x_3) = [0,4; 0,74; 1,26]$, тобто оптимальним вважається третій варіант. Порівняно результати пошуку оптимальних альтернатив та встановлено тотожність варіантів.

Отримано значення функцій належності лінгвістичних змінних аналізованого технологічного процесу шляхом обчислення матриць попарних порівнянь для кожної лінгвістичної змінної та відповідної їй терм-множини значень. Терми лінгвістичних змінних представлено нечіткими множинами.

Описано отримання значення оцінки якості проєктування післядрукарських процесів шляхом дефазифікації нечітких множин за принципом центра ваги. Інтегральний показник якості за обраних умов становить $G_{прогноз.} = 50,256\,\%$ при максимальних значеннях $100\,\%$.

Розроблено структурно-функціональну модель інформаційної технології прогностичного оцінювання якості проєктування післядрукарських процесів, що враховує етапи дослідження та уможливлює апріорне забезпечення якості друкованої продукції. Створено IDEF0-моделі інформаційної технології: побудовано контекстну діаграму A-0, діаграму першого рівня декомпозиції A0, діаграми другого рівня декомпозиції A1, A2, A3, A4, A5, діаграми третього рівня декомпозиції A23, A24, A41, A42, A51, A52 та діаграму дерева вузлів.

# THERMALLY STIMULATED PROCESSES
# AND PYROELECTRICITY IN FERROELECTRIC POLYMERS

*Sergeeva A. E.*

*Стаття присвячена експериментальному дослідженню тонких плівок полівініліденфториду (ПВДФ), його сополімеру з тетрафторетиленом П(ВДФ-ТФЄ), а також композитів на основі ПВДФ та неорганічних керамічних матеріалів титанату барію BaTiO₃ та титанату цирконату свинцю (ЦТС). Всі ці матеріали є сегнетоелектриками. Вивчено їхню поведінку при різних температурних впливах, зокрема струми термостимульованої поляризації (ТСП) та деполяризації (ТСД), а також піроелектричний ефект у цих матеріалах. Встановлено, що термічний вплив є важливим при формуванні сегнетоелектричної поляризації та забезпечення її стабільності.*

*Встановлено важливу роль об'ємного заряду в сегнетоелектричних полімерах на величину та стабільність залишкової поляризації. Запропоновано методи поділу гомозаряду та гетерозаряду у полімерних плівках.*

*Отримані результати мають як наукове, так і практичне значення, оскільки сегнетоелектричні полімери широко використовуються для виготовлення різних сенсорів і датчиків.*

*This article is devoted to the experimental study of polyvinylidene fluoride (PVDF) thin films, its copolymer with tetrafluoroethylene P(VDF-TFE), as well as composites based on PVDF and inorganic ceramic materials of barium titanate BaTiO₃ and lead zirconate titanate (PZT). All of these materials are ferroelectrics. Their behavior was studied under different temperature influences, in particular, currents of thermally stimulated polarization (TSP) and depolarization (TSD), as well as the pyroelectric effect in these materials.*

*It was found that thermal action affects the formation of ferroelectric polarization and its stability. The important role of the space charge in ferroelectric polymers on the magnitude and stability of polarization has been established. Methods for separating homocharge and heterocharge in polymer films were proposed.*

*The results obtained are of both scientific and practical importance, since ferroelectric polymers are widely used in the manufacture of various kinds of sensors and transducers.*

## 1. Introduction

To measure the thermal relaxation of the residual polarization, the measurement of the thermally stimulated depolarization currents (TSD) is the most appropriate. Despite numerous experimental studies, there is still no theory of the TSD current method for the case of poled ferroelectric polymers like polyvinylidene fluoride (PVDF). Therefore, the description and interpretation of measured current peaks are usually qualitative and hypothetical.

To detect the nature of the TSD current peaks, their connection with polarization and pyroelectricity must be taken into account, as well as processes occurring in the amorphous and crystalline phases. Such an attempt was made [1; 2] by considering individual contributions to TSD currents of pyroelectric processes, polarization in the amorphous phase, "charge-induced interphase polarization" (called the Maxwell-Wagner effect) and the ferroelectric polarization in the crystalline phase. In qualitative description of the predicted processes, in addition to compensating charges, "injected surplus charges" were taken into account, that is, space charges. Often it is considered as a self-evident that there is polarization in PVDF, compensating charges, and space charges. While there is no doubt about the existence of polarization, the existence of the space charge in addition to compensating charges is questionable, since PVDF has a sufficiently high specific conductivity about $g = 10^{-11}$ Sm/m at room temperature [3; 4] and the dielectric permittivity $\varepsilon = 10-20$.

Therefore, any surplus charges will be neutralized with the Maxwell relaxation time of approximately

$$\tau = \frac{\varepsilon_0 \varepsilon}{g} \approx 13s \; . \tag{1}$$

This indicates that in the short-circuited sample, after about 40 seconds, any electric field caused by charges (other than compensating charges and polarization charges of the ferroelectric crystals) will disappear.

Homogeneously polarized two-phase materials, like PVDF, consist of ferroelectric crystallites scattered in the amorphous matrix. Compensation of the depolarizing field in the ferroelectric crystallites is possible only due to the charges localized at two sides of the crystallites. Full compensation by the electrode charges, as in the case of single crystals or 100 % crystalline materials, is not possible. In PVDF and other biphasic ferroelectric polymers, the accumulation of charge at the boundaries between the crystalline and amorphous phases only occurs to a small extent due to the difference in dielectric constant, and is the most likely due to presence of the ferroelectric polarization. von Seggern and Fedosov proposed a model of a layered structure with alternating ferroelectric and non-ferroelectric layers for the description of initial poling [3, 4], switching of polarization [5], short circuiting and the back switching [6] in PVDF.

The theory of the TSD method [7] was developed only for electrets with dipole and/or space charge polarization. It is assumed that polarization or space charge is initially thermally frozen, and then it defrosts under linear heating. For application the TSD method to ferroelectric polymers, the theoretical basis has not yet been constructed, since polarization in PVDF is not thermally frozen, but has the ferroelectric nature and can occur in high fields without heating during poling. In addition to the irreversible relaxation currents, in the TSD experiments on ferroelectric polymers, there are reversible pyroelectric currents, and the separation of these components is not an easy task. However, the TSD method, due to its informative and versatile nature, has been widely used for studying the polymer ferroelectric as well.

The TSD current peak at $50-60$ °C has been observed by many researchers in the non-ferroelectric $\alpha$-PVDF and the ferroelectric $\beta$-PVDF poled in a wide range of temperatures, electric fields and times. The only common feature of all PVDF samples on which these experiments were carried out was the presence of the amorphous phase. Thus, this TSD current peak with high probability can be associated with processes occurring in the amorphous phase of PVDF.

In many cases, it was assumed that the TSD current peak near $50-60$ °C is associated with the so-called $\alpha_c$-process [8]. However, data on the nature of the $\alpha_c$-process are still controversial. Moreover, an interrelation between structural transformations and TSD currents should be established, since the TSD currents are the result of electrical, but not always structural processes. Making assumptions about the nature of the TSD currents peaks, it is necessary to take into account not only structural transformations, but also fundamental electrical principles and laws. Otherwise, a qualitative explanation without the corresponding formulas and equations may be false. Two high-temperature peaks, in addition to the usual $\alpha$- and $\beta$-processes in PVDF, were found by the TSD method. One of the processes was attributed to charges trapped on the boundary between the nonpolar crystalline $\alpha$-phase and the amorphous region. The other was assumed to be related to charges on polar $\beta$-crystallites. The authors believe that both peaks are due to the positive homocharge injected from the anode, although it is quite probable that they are related to the poling temperature.

Eliasson [9], studying the influence of the polarizing field, temperature, and electrode material on TSD currents, believed that the formation of a bulk charge was influenced by injection, although her results did not match with the position and magnitude of peaks in the data of Ieda and others [10]. The picture becomes even more complicated if we compare the data on the TSD obtained by different researchers. The variety of relaxation processes determined by the TSD method is due to the lack of clear methods for identifying peaks (sometimes they take for relaxation peaks the smallest bends at the TSD current curve). Neagu and others [11] believe that the high influence on the TSD currents has the material of electrodes and the uncontrolled composition of the atmosphere, in particular presence of oxygen, nitrogen and water vapors.

It follows from the theory of the TSD currents [12] that in a one-component homogeneously polarized dipolar sample, the measured current is equal to the displacement current dP/dt, so that the integral of the TSD current is equal to the value of the initial polarization. As shown by von Seggern and Fedosov [5], the integral of the TSD current is smaller than the residual polarization in the case of a two-phase system.

Possibilities of thermally stimulated research methods are extremely wide, since most relaxation processes are thermally activated. At the same time, these possibilities in the study of ferroelectric polymers have not yet been detected and have not been used. This relates to the study of the effective conductivity dynamics in the process of the thermoelectret poling (TEP), the comparison of TEP and TSD currents, the separation of homocharge and heterocharge contribution in the ferroelectric polymers to the relaxation current, the use of fractional TSD methods for separation of the pyroelectric and the relaxation components of the total current.

The aim of the present article was to obtain additional experimental results by using above mentioned methods for clarifying physical processes responsible for formation of the residual polarization in PVDF films, being a typical representative of the ferroelectric polymers. Another part of investigation is related to finding interrelation between TSD currents and the pyroelectricity in this class of materials. We also performed some experiments on P(VDF-TFE) copolymer and composites of PVDF with the inorganic crystals like barium titanate ($BaTiO_3$) and lead zircanate titanate (PZT) in order to make the research more generalizing.

## 2. The features of the Thermally Stimulated Depolarization Current method

In the case of thermoelectret poling (TEP), unlike isothermal poling, the sample was initially kept at room temperature at constant voltage for some time necessary to decrease the absorption current to values of the order of $10^{-11}$ A, and then the temperature was linearly increased to 150 °C at a constant rate (0.5−4 °C/min) with continuous measuring of the poling current. After completion of poling, the samples were quickly cooled down without switching off the applied voltage.

Since many relaxation processes have the thermally activating nature, the TSD method [12] was used to predict the space charge and the residual polarization stability and to study the mechanism of their formation. The sample electrodes were connected with each other through the electrometer having a sensitivity of $10^{-14}$—$10^{-16}$ A by current and the current recording device. The temperature was raised at a constant rate in the range of 15—150 °C and measured by a chromel-copel thermocouple. The principle of the current TSD method is shown in Fig. 1.

We measured the thermally stimulated decrease in the electret potential by the vibration electrode method (Kelvin's method). Measurements in the range from -100 °C to +180 °C was carried out by using the relaxation spectrometer Solomat-91000. Studies in the field of low temperatures from -170 to + 40 °C were performed by using the Kithley-6100 electrometer on the samples that were originally cooled in liquid nitrogen.

## 3. Relaxation of homocharge and heterocharge during the TSD currents measurement

A number of properties of the ferroelectric polymers can be explained within the framework of the modern theory of polar electrets [13] considering relaxation of the homocharge and the heterocharge in a self-consistent regime without taking into account the ferroelectricity in the crystalline phase. We consider the interrelation between the homocharge and the heterocharge taking into account experimental data on thermally stimulated and isothermal relaxation of PVDF films poled in the corona discharge.

Four types of depolarization varieties were used to study the relaxation processes, namely thermally stimulated (T) and isothermal (I) depolarization of short-circuited samples (S) and depolarization in the open circuit mode (O). The modes were denoted as TS, TO, ISO, and IO, where the first letter indicated the temperature mode (thermally stimulated or isothermal), and the second was related to the electrical state (short-circuit or open circuit). Additional experiments on the thermally stimulated electret potential (TP) kinetics were also performed after 24 hours of keeping in the open circuit configuration. A film of polytetrafluoroethylene (PTFE) with a thickness of 10 μm was used as a dielectric gap in TO and IO modes. All thermally stimulated experiments were performed at a constant heating rate of 3 °C/min. In isothermal experiments, the temperature was maintained constant after the required temperature value was achieved by rapid heating. The electret potential in the TP method was measured by the Kelvin method and was continuously recorded.

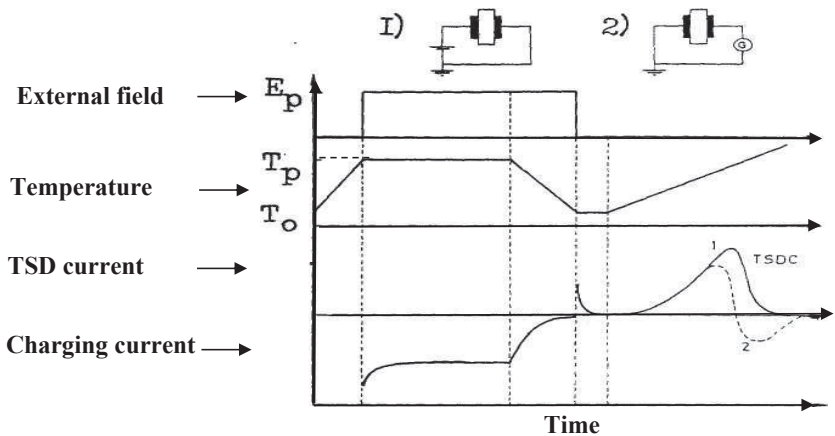

Fig. 1. The principle of the thermoelectret poling (1) and depolarization (2) methods

During poling in a corona discharge, the excess charge is localized on the surface of the sample forming a homogeneous charge $\sigma$, which creates a homogeneous field in the sample volume, in which the internal dipolar polarization (heterocharge) is formed characterized by the surface density of the bound charge $P$. In the case of the sample short-circuiting without a gap, only the heterocharge relaxes [12], and the equality $\sigma = P$ and the zero internal field ($E = 0$) is supported due to the current redistribution in the external circuit. In the presence of a dielectric gap and in the open circuit mode, the relaxation currents of the homocharge and the heterocharge flow in opposite directions.

In order to find separately components corresponding to the decay of the homocharge $\sigma$ and the heterocharge $P$ of the full depolarization current in the open circuit mode, we present the current density $i(t)$ and the surface potential $V(t)$ as

$$i(t) = s\left[\frac{dP(t)}{dt} - \frac{d\sigma(t)}{dt}\right], \tag{2}$$

$$V(t) = \frac{sx_1}{\varepsilon_0\varepsilon_1}\left[\sigma(t) - P(t)\right], \tag{3}$$

$$i(t) = -\frac{\varepsilon_0\varepsilon_1}{x_1}\cdot\frac{dV(t)}{dt}, \tag{4}$$

where, $t$ is time, $\varepsilon$ and $x_o$ are the dielectric constant and the thickness of the sample; $\varepsilon_1$ and $x_1$ are similar parameters of the dielectric gap. For the conductivity current density, it is possible to write down

$$i_C(t) = \frac{g}{x_0}V(t) = -\frac{d\sigma(t)}{dt} ,\tag{5}$$

where $g = g_0 \exp(-Q/kT)$ is the specific conductivity, $k$ is Boltzmann's constant, $T$ is temperature, $Q$ is the activation energy of the intrinsic conductivity, $g_o$ is a pre-exponential factor. Integrating (4) over time and replacing the variable $t$ by $T$ taking into account the linear heating $T = T_0 + \beta \cdot t$ where $\beta$ is the heating rate, $T_o$ is the initial temperature of the experiment, we obtain from (5) the temperature dependences of the depolarization currents and the electret potential

$$i_1(T) = \frac{d\sigma}{dt} = -\frac{x_1 g_0}{bT_0 x_0 \varepsilon_0 \varepsilon_1}\exp\left(-\frac{Q}{kT}\right)\int\limits_T^\infty i(T)dT' ,\tag{6}$$

$$i_2(T) = \frac{dP}{dt} = \frac{i(T)}{s} + \frac{d\sigma}{dt} ,\tag{7}$$

$$V(T) = \frac{x_1}{bT_0 \varepsilon_0 \varepsilon_1}\int\limits_T^\infty i(T)dT' .\tag{8}$$

All values in the right-hand sides of the equations (6–8) are known from the experiment. Thus, this technique allows to differentiate processes of homocharge and heterocharge relaxation in different modes of TSD by using experimental $i(T)$ curves.

### 4. Application of thermoelectret poling of PVDF films

The thermoelectret poling method (TEP) has several advantages over isothermal poling, because it allows to obtain additional data on the mechanism of the polarization formation. It is also possible to determine the optimal temperature of poling, as well as to find the temperature, at which the ohmic conductivity becomes significant.

From Fig. 2 and Table 1 one can distinguish the following features:

1) The TEP curves contain three characteristic areas: a) the growth of the current irrespective of the corona discharge polarity; b) decrease of the current (negative temperature coefficient of conduction); c) increase of the current at high temperatures.

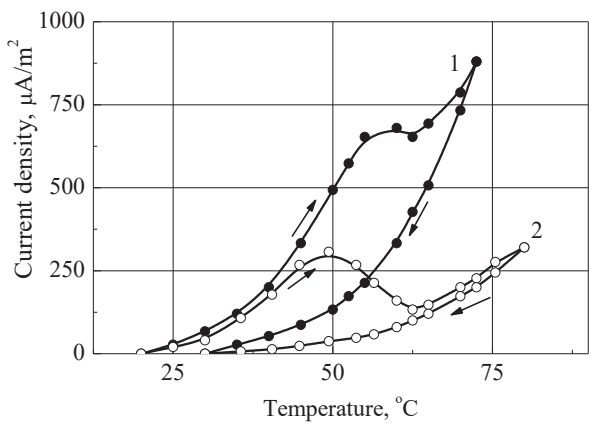

Fig. 2. Dependence of current on temperature during thermally stimulated poling in positive (1) and negative (2) corona discharge. Heating rate is 2 °C/min

**Effective activation energy during TEP in different temperature ranges
(from the inclination of linear parts in ln$i$-1/$T$ curves) [9]**

| Charge Polarity | Heating | | | Cooling | |
|---|---|---|---|---|---|
| | Activation energy, eV Temperature range, °C | | | Activation energy, eV Temperature range, °C | |
| + | 0.87 20−40 | 0.82 40−50 | 0.82 50−65 | 0.92 60−45 | 1.06 45−25 |
| − | 0.87 20−35 | 1.1 35−45 | 1.12 65−80 | 1.19 70−55 | 0.83 55−35 |

2) With negative corona discharge polarity, the second region is more pronounced than with positive polarity, and the decrease in conductivity begins at lower temperatures. The difference in the currents of the TB positively and negatively poled samples indicates different injection levels. In a positive corona, it is likely that both positive charges and negative ones (from the rear electrode) are injected. In the case of a negative corona, positive charge carriers from the metal electrode are not injected.

3) When thermoelectret poling of PVDF films occurs, the irreversible decrease in the effective conductivity is due to the fact that the value of the thermal current when cooled is much smaller than the current during heating.

In PVDF, the current peak during TEP (Fig. 2) cannot be considered to be due to polarization, as in the case of linear polar polymers, because its integration gives an unrealistically great value of the polarization. Probably, the conductivity current in TEP ferroelectric polymers is much larger than the polarization component of the current, therefore the graphs in Fig. 2 reflect the nature of changing in the films effective conductivity.

The mechanism and nature of conductivity in PVDF are unknown, but presence of the negative temperature coefficient of conductivity sections in curves Fig. 2 suggests that, along with the thermally activated increase in the number of moving carriers, there is also likely to be a trapping by deep traps, and with certain ratios of field strength and temperature the second process prevails over the first process. The trapped charges play an important role by neutralizing the depolarizing field and contributing to the preservation of the residual polarization.

High polarization in a poled film and a sharp decrease in conductivity (the current passes through the maximum) are probably interconnected. The ferroelectric polarization in crystallites creates conditions for trapping of charges at their boundaries. The field of the trapped charges screens the polarization and contribute to its stabilization. Thus, the processes of the polarization development and the charge trapping are interconnected and interdependent.

At the current curve of TEP from room temperature to 30–40 °C, the dependence $i(T)$ is exponential regardless of the corona polarity. This can be related to the thermal generation of charge carriers in the volume. The further course of the current graphs corresponds to the proposed model for the polarization formation and the charge trapping. There is again increase of the current in the third section that may be due to the internal thermoelectric detrapping of the previously trapped charges with partial destruction of the already formed polarization. If this assumption is valid, then it is an important for practice conclusion that it is impractical to heat PVDF during TEP above the minimum temperature on the current-temperature curve.

Decrease of the current and the effective conductivity in the second section of the TEP curve may be due to the following reasons:

1) Depletion of the stock of own carriers due to their migration in the external field and trapping near the electrodes (electrode polarization);

2) Irreversible changes at the electrodes or near to them leading to limiting of the charge injection;

3) Generation of regions in the volume that do not conduct current, for example, polarized crystallites with layers of the trapped charges;

4) Changing the equilibrium between free and trapped charges due to the formation of new traps at the boundaries of the polarized crystallites and macroscopic polarized regions.

The difference between TEP currents in positively and negatively charged samples is against the migration mechanism causing the reduction in conductivity. The second reason is also unlikely, because due to ionic impurities the field on the electrodes should increase, which together with the effect of the temperature would lead to increase in the injection level. The third reason is more probable, because the ordering in the crystalline phase of the polymers leads to decrease in localized states the density and decrease in conductivity (the band gap width is of the order of 6—9 eV). However, the ordering of the preferred orientation of dipoles in crystallites in the external field cannot substantially change their conductivity, because they already have the spontaneous polarization as the higher degree of the internal ordering.

Most likely, the conductivity of PVDF decreases as a result of the intensive trapping of carriers at the boundaries of polarized crystals and macroscopic polarized regions, which create favorable conditions for localization of the charges. We have found that similar processes at low temperatures and high fields lead to appearance of areas of negative dynamic resistance on volt-ampere characteristics.

If the proposed hypothesis is correct, then there the relationship between the temperature and the field strength must be observed, in which high polarization is formed and the irreversible decrease in conductivity appears. Dependence of the effective conductivity on temperature in TEP mode at different constant voltages is shown in Fig. 3, from which it is seen that the beginning of the negative temperature conductivity section moves to lower temperatures with increasing the polarizing voltage. It is known that the high polarization in ferroelectrics occurs in fields above the coercive one. From the data of Fig. 3 it follows that the value of the coercive field decreases with increasing temperature. From Fig. 3 and Table 1 it is evident that the activation energy that provides the effective conductivity in not polarized and polarized samples is practically the same. At the same time, the conductivity of the polarized films, and hence the concentration of free carriers in them is almost 100 times smaller than in not polarized films. Consequently, in the process of poling, there are redistribution of carriers and their additional trapping at the newly created traps.

Let us analyze the shape of the TEP current curve taking into account that the conductivity current IS is proportional to the concentration of free charge carriers

$$i_c = \mu \cdot n_c e \frac{U}{x_0} , \qquad (9)$$

where $\mu$ is the mobility; $U$ is the applied voltage; $x_o$ is thickness of the sample. In the presence of localized states in the forbidden zone, the concentration of trapped charges is determined by the Fermi-Dirac formula

$$n_t = N_t \{1 + (1 / g / \exp[-(F - E_t) / kT])\}^{-1}, \qquad (10)$$

where $N_t$ is the density of localized states; $g$ is their statistical weight; $E_t$ is the localized state energy; $F$ is a quasi-level Fermi based on its own and injected carriers. We assume that the total concentration of carriers and the Fermi level remain constant.

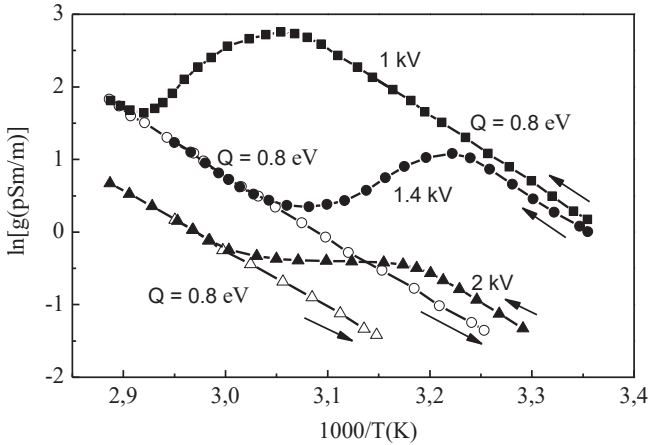

Fig. 3. Temperature dependence of the effective conductivity during thermally stimulated poling in a corona discharge under different voltages at a control grid (electret potential)

Increase in the density of the localized states with increasing polarization can be represented as a linear function where the polarization $P$ is a function of the field strength $E$. For ferroelectrics, the function $P(E)$ can be approximated by three rectilinear sections

$$P(E) = \begin{cases} 0, E < E_c, \\ (E - E_c) P_s / (E_s - E_c), E_c < E < E_s, \\ P_s, E_s < E, \end{cases} \qquad (11)$$

where $E_c$ is the coercive field; $E_s$ is the field strength at which polarization reaches saturation. Consider the decrease of the coercive field with increasing temperature

$$E_c = E_0 - \gamma \cdot T \ . \tag{12}$$

We will assume that the dynamic permittivity at $E_c < E < E_s$ does not depend on $T$, that is equivalent to the constancy of the difference $\Delta E = E_s - E_c$. From (10)−(15), taking into account the made assumptions, we obtain the dependence of the conductivity current on the temperature

$$i_c = (\mu \cdot e \cdot V \ / \ x_0) \left\langle n_0 - N_t \{1 + (1 \ / \ g) \exp[-(F - E_t) \ / \ kT]\}^{-1} \right\rangle , \tag{13}$$

where

$$N_t = N_0; T < T_1 = (E_{c0} - V \ / \ x_0) \ / \ \gamma,$$
$$N_t = N_0 + \alpha P_s; T > T_2 = (\Delta E + E_{c0} - V \ / \ x_0) \ / \ \gamma,$$
$$N_t = N_0 + (\alpha P_s \ / \ \Delta E)[(V \ / \ x_0) - (E_{c0} - \gamma T)], T_2 > T > T_1 \ . \tag{14}$$

It follows from expressions (13) and (14) that with increasing temperature in the range $T < T_1$ the current $i_c$ increases. $N_t$ begins to increase at $T > T_1$ provided

$$\alpha P_s \gamma \ / \ \Delta E > (Q \ / \ gkT^2) \exp(-Q \ / \ kT) , \tag{15}$$

where $Q = F - E_t$.

There is a decrease in current $i_c$ despite the increase in temperature. The current increases again, if the saturation of polarization is reached, or if the condition (15) is violated.

Reducing of the effective conductivity during TEP indicates the importance of volume-charge processes, since the charge trapped on the boundaries of the polarized regions compensates the depolarizing field and contributes to the long-term preservation of the residual polarization. A similar relation was established during isothermal poling in high fields [14].

Thus, the increase of temperature and the field strength equally influences on the generation and injection of moving charges, the large concentration of which is a prerequisite for the emergence and development of the high local polarization. As the polarization is formed, the conductivity of ferroelectric polymers is irreversibly reduced due to the trapping of the injected charge carriers at deep traps formed by the polarization of crystallites. These trapped charge carriers stabilize the residual polarization by compensating local depolarizing fields.

## 5. Thermally stimulated depolarization currents in PVDF

Measurement of TSD currents is a powerful tool for studying relaxation processes [12]. Although the theory of TSD currents was developed only for the thermally frozen dipole polarization, this method is widely used to study the ferroelectric polymers as well. In PVDF, two peaks are the most important. One of them, related to the glass transition in the amorphous phase, is always observed at a temperature of about -45 ˚C and it is well-studied. The nature of the second peak in the range of 50−80 ˚C (Fig. 4) is not fully understood, although it is clear that several processes, such as the reorientation of dipoles in the amorphous phase, the relaxation of the ferroelectric polarization, the displacement of the space charge, as well as interphase and piezoelectrode processes can be responsible for this peak. It is established that the temperature of about 60 ˚C is characteristic for PVDF, but its nature is not completely clear. Many researchers associate a peak at this temperature with so called $\alpha_c$ relaxation. Lacabane et. al. [15] explain the appearance of the peak by shrinkage, that is, by a partial restoration after stretching carried out for obtaining the ferroelectric β-phase in PVDF. We believe that this peak is associated with polarization in the amorphous phase.

Ferroelectric polymers have the properties of ordinary polar electrets in addition to the ferroelectricity. Therefore, one can expect the presence of two components of the residual polarization: one associated with the ferroelectricity in the crystalline phase, and another related to the amorphous phase, although there is currently no direct experimental confirmation of this phenomenon.

Analysis of the relationship between TSD currents in PVDF and pyroelectricity was carried out in the work of von Seggern and Fedosov [5]. They have found that the residual polarization decreases after heating to 60 ˚C, while the pyrocoefficient remains unchanged. They concluded that the ferroelectric polarization in the crystalline phase is partially offset by localized charges and partly by polarization in the amorphous phase. Therefore, in the formation of the TSD peak in the ferroelectric polymers, several currents are involved caused by relaxation of the electret and the ferroelectric components of the residual polarization, and associated with the space charge.

We investigated polarized specimens subjected to TSD either in short-circuit mode or in open-loop with PTFE film as a dielectric gap between the free surface of the sample and one of the electrodes [149−16]. The period of time after poling to the TSD measurement was either one day or 16 months. The samples were named "fresh" and "old" accordingly. Similarly to the data reported in other papers, we observed one broad peak in the mode of the short circuit on fresh samples (Fig. 4).

The direction of current at this peak corresponded to the residual polarization relaxation. Because the crystallinity of PVDF is about 50 % and most of the molecular dipoles in the crystalline regions are in the ferroelec-tric β-phase, the contributions of the electret and the ferroelectric components to the formation of this peak in fresh samples can be compared. As for the bulk charging component, it is known that it either does not contribute to the TSD current in the mode of the short circuit, or its direction coincides with the depolarization current component.

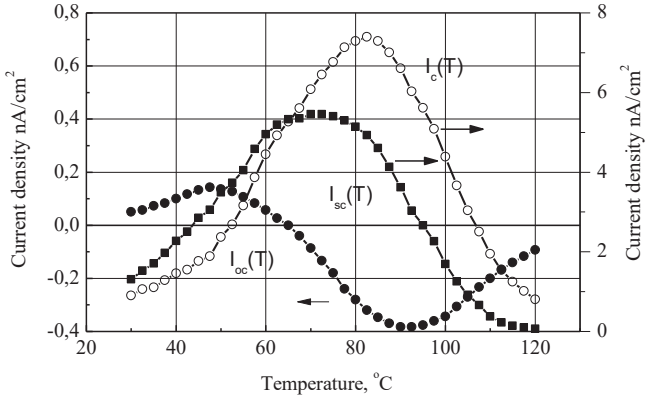

Fig. 4. TSD currents $I_{sc}(T)$ and $I_{oc}(T)$ measured on fresh polarized samples in short-circuit and open circuit modes. The curve $I_c(T)$ corresponds to the volume-charge current

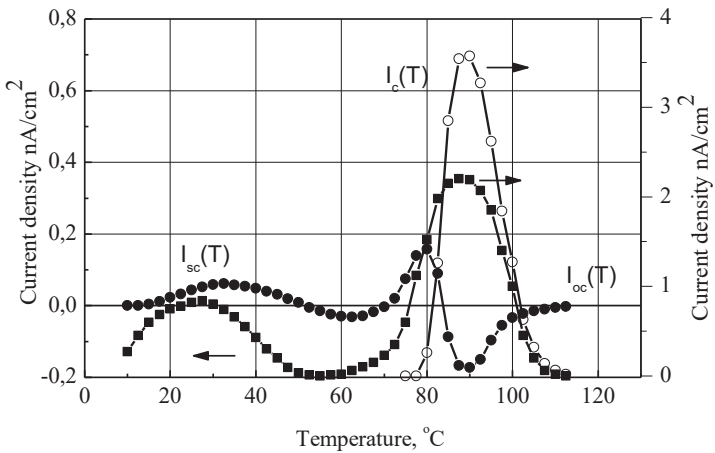

Fig. 5 TSD currents $I_{sc}(T)$ and $I_{oc}(T)$ measured in polarized samples in short-circuited and open circuited modes after exposure for 16 months. The curve $I_c(T)$ corresponds to the volume-charge current

Comparing the TSD currents of fresh and aged polarized samples, we observed a new phenomenon. One broad TSD current peak in the mode of the short circuit was divided during the aging in two narrow peaks completely separated from each other. At the same time, two pairs of the oppositely directed peaks appeared in old samples instead of one pair of peaks typical for fresh samples (Fig. 5). This feature is likely to be common to all ferroelectric polymers and does not depend on the polarization conditions, because similar results were also obtained by us on samples poled by a non-focused electron beam at the accelerating voltage of 20 kV and in P(VDF-TFE) and PVDF films poled through a lime glass at the voltage of 7 kV.

The depolarization current in the open circuit mode remains unchanged, while the TSD current due to the charge changes the direction to the opposite. Therefore, the two peaks shown in Fig. 4 can be explained as the result of two partially overlapping and oppositely directed currents arising as a result of the relaxation of polarization and space charge.

In order to separate the depolarization current $I_p(T)$ from the space charge current $I_c(T)$, it is reasonable to assume that the polarization is homogeneous in the direction of the thickness. Since the compensating charges trapped near the surface do not generate any current in the short circuit mode, then $I_{sc}(T) = I_p(T)$, where $I_{sc}(T)$ is the experimentally measured TSD current in the short circuit mode. The current $I_c(T)$ can be calculated from the experimental curves $I_{sc}(T)$ and $I_{oc}(T)$ shown in Fig. 4

$$I_c(T) = 1_{0c}(T)\left[1 + \frac{\varepsilon_1 x_2}{\varepsilon_2 x_1}\right] - I_{sc}(T) , \qquad (16)$$

where $\varepsilon_1$, $x_1$, $\varepsilon_2$ and $x_2$ are dielectric permittivity and thickness of the sample and the dielectric gap, respectively. In our calculations, we used $\varepsilon_1 = 12$, $\varepsilon_2 = 2.1$, $x_1 = 20$ μm, $x_2 = 25$ μm. It is noteworthy that the peak $I_c(T)$ is at the higher temperature than the peak of the depolarization, indicating that the trapped charges are more stable than the residual polarization.

The obtained results can be explained qualitatively taking into account the different nature of the three components of the TSD current. The electret polarization accounting for almost 50 % of the residual polarization in fresh samples decays in time faster than the ferroelectric component. That is why the two peaks are overlapped in fresh samples, become completely separated in the old films, as if the slow redistribution of residual polarization is going on for a long time after the completion of poling.

Observed and calculated peaks are difficult to process quantitatively, since there is no TSD currents theory in ferroelectric polymers. However, as evident from the shape of the peaks, all three relaxation processes differ significantly from the ideal Debye case, corresponding to the absence of the relationship between the relaxing dipoles. This feature can be taken into account considering that the polarization relaxes over time in accordance with the law of the expanded exponent

$$P(t) = P_0 \exp\left(-\frac{t}{\tau}\right)^\alpha \quad 1 \geq \alpha \geq 0 \ , \tag{17}$$

where $\tau$ is a time constant, $P_o$ is the initial polarization. If the sample is linearly heated at the rate $\beta = dT/dt$, then

$$P(T) = P_0 \exp\left\{-\left[\left(\frac{1}{\beta}\right)\int_{T_0}^{T}\left(\frac{1}{\tau(T')}\right)dT'\right]^\alpha\right\}, \tag{18}$$

where $T_o$ is the initial temperature. It is reasonable to assume that the temperature dependence of $\tau$ corresponds to the Arrhenius law

$$\tau(T) = \tau_0 \exp\left(\frac{A}{kT}\right), \tag{19}$$

where $A$ is the activation energy, $k$ is the Boltzmann constant, $\tau_o$ is the characteristic time. The expression for the TSD current density is derived from the equations (17)−(19)

where

$$i(T) = -\left(\frac{\alpha P_0}{\tau_0}\right)\exp\left(-\frac{A}{kT}\right)[s(T)]^{\alpha-1}\exp\left\{-[s(T)]^\alpha\right\}, \tag{20}$$

$$s(T) = \left(\frac{1}{\beta\tau_0}\right)\int_{T_0}^{T}\exp\left(-\frac{A}{kT}\right)dT' \ .$$

The results of computer fitting of the experimentally observed and calculated TSD peaks in equation (20) confirmed our assumptions about the nature and the thermal stability of the relaxation processes. They showed that the depolarization peak in fresh samples where the electret and the ferroelectric components are mixed, is wide ($\alpha = 0.24$), because the two relaxation processes responsible for its formation are very different. The ferroelectric polarization is quite stable ($A = 2.7$ eV), and the TSD peak due to its relaxation is relatively narrow ($\alpha = 0.52$). The parameters of the space charge peaks in the fresh and old samples are completely different, as if there are two types of the space charges, one probably associated with the ferroelectric polarization, and the other one with the electret component. It is also likely that the small peak that occurs near the electret depolarization peak in open mode (Fig. 5) is due only to the electret component of the volume charge. Since the glass transition temperature is -45 °C in PVDF, the ordering of the dipoles in the amorphous phase is not thermally frozen, as in the ordinary polar electrets. The dominant orientation of dipoles in these conditions may be supported by the field of the trapped charges.

Thus, it has been shown that in corona poled films of the ferroelectric polymers, there are two components of polarization, and both components are accompanied by corresponding space charges. The electret-type thermodynamically unstable component relaxes as long as the broad TSD peak observed in fresh polarized samples is not transformed into two completely separated narrow peaks.

The unstable electret component of the residual polarization can be removed by heating the poled sample to a specific temperature (about 60 °C in the case of PVDF). Apparently, the trapped charges always accompany the dipolar polarization regardless of its nature.

### 6.1. Pyroelectric effect in ferroelectric polymers and its nature

Pyroelectric effect in PVDF films was discovered more than 40 years ago. However, despite the large number of works, the nature of pyroelectricity in PVDF still remains unclear. A series of papers was devoted to the pyroelectric properties of the ferroelectric polymers. It is assumed in the first of the three most popular models that pyroelectricity results from the contribution of electrostriction, dipole fluctuations and changes in the size with temperature. In the second model, only the change in the dimensions of the sample is considered and the crystals, while in the third model, the pyroelectricity is attributed to electrostrictions and to the change in size when temperature changes.

Under the pyroelectric effect, one means the range of phenomena associated with reversible changes in the electric displacement vector (induction) when the temperature changes. In the case of a free sample, the pyrocoefficient is determined by the following expression [13]

$$p_i = \left(\frac{\partial D_i}{\partial T}\right)_{H,E} = \left(\frac{\partial D_i}{\partial T}\right)_{U,E} + \left(\frac{\partial D_i}{\partial U_{i,j}}\right)\left(\frac{\partial U_{i,j}}{\partial T}\right)_{H,E}, \qquad (21)$$

where $D_i$ is the component of the induction vector; $U_{i,j}$ is the deformation tensor; $H$ is the mechanical stress; $E$ is the field strength, $\left(\dfrac{\partial D_i}{\partial U_{i,j}}\right)$ is the piezo modulus; $\left(\dfrac{\partial U_{i,j}}{\partial T}\right)$ is the thermal expansion coefficient. The first term in (21) corresponds to the primary or true pyroelectric effect measured on the compressed sample, and the second term characterizes the secondary pyroelectric effect being the result of the piezoelectric induction changes due to the thermal expansion.

Since the pyroelectric effect depends both on the internal polarization and on the space charge, in principle, it can be caused by the temperature dependence of both quantities. If we neglect the influence of the space charge, then for the case of a flat short-circuited sample with homogeneous polarization $P$ we obtain

$$p_0 = \frac{\partial D}{\partial T} = \frac{\partial P}{\partial T} = \frac{\partial \sigma}{\partial T} = \frac{\partial (q/S)}{\partial T}, \qquad (22)$$

where $q$ and $\sigma$ are magnitude and density of the bound surface charge; $S$ is the surface area.

In the experimental conditions, the current $I(T) = \dfrac{dq}{dt}$ occurring when the temperature change ($dT/dt$) is measured, and the pyrocoefficient is considered to have the following value

$$p = \frac{1}{S}\frac{dq}{dT} = \frac{1}{S}\frac{I(T)}{dT/dt}. \qquad (23)$$

Because $p_o \neq p$ there are differences in the values of the theoretically calculated and experimentally measured pyrocoefficients. It has been proved that the pyroelectric effect can only be caused by a nonuniform distribution of the space charge (without taking into account polarization).

Investigating the pyroelectric effect in PVDF, Lines and Glass [17] came to the conclusion that this is a real pyroelectricity, but not a depolarization effect observed in many polar electrets, because the crystalline phase of PVDF completely corresponds to the definition of a ferroelectric, as a pyroelectric with reversible spontaneous polarization under application of the electric field. A fundamental question was posed that has not been solved for the time being: is the pyroelectricity an equilibrium property of PVDF or a result of non-equilibrium polarization, that is, in some way it is a fixed orientation of dipoles?

In early works on PVDF, the effect of volume charge on the pyroelectric effect was considered decisive, but after the proof of the ferroelectric na-ture of PVDF crystallites, the pyroelectric was more often associated with the spontaneous polarization. So, in the model of Broadhurst and Davies [18] the behavior of rigid dipoles in thin crystalline plates (lamellae) distributed in the amorphous phase is considered. In the model of Wada and Hayakawa [19], the presence of spherical ferroelectric particles scattered in the amorphous phase is assumed. Both models predict the influence of thermal expansion (dimensional effect) on the pyroelectric effect, as well as temperature dependence of the dielectric constant and the spontaneous polarization $P_{sc}(T)$.

At the same time, there is no satisfactory correspondence between calculated and experimental data. drew Attention was attracted to the fact that the models of Broadhurst [18] and Wada and Hayakawa [19] ignored the contribution of the volume charge to the pyroelectric effect. An attempt to take into account the volume charge led to contradiction with the obtained data [13]. Estimated calculation of Lines and Glass [17] showed that the theoretical pyrocoefficient even at 100 % orientation of dipoles in PVDF is several times lower than the value measured in the experiment.

We believe that not only the crystalline, but also the amorphous phase contributes the pyroelectricity. Elling et al [20] found that the pyrocoefficient value is affected not only by the residual polarization, but also by the supramolecular structure on which the mechanical properties of PVDF depend.

It is shown in the work of Fedosov and Sergeeva [21] that one of the components of the pyroelectric effect in the ferroelectric polymers is the electret component, that is, the pyroactivity of PVDF is due to the reversible temperature changes of the residual polarization closely related to those in equilibrium with trapped charges.

Fedosov and von Seggern [3, 4, 6] proved that compensating charges localized on the surface of crystallites are very important in two-component ferroelectric polymers of the PVDF type for obtaining high and stable polarization. With periodic increase and decrease of temperature [21], the pyrocoefficient irreversibly decreases at temperatures much lower than the Curie point indicating the possible effect of charges.

It is generally accepted that the pyrocoefficient in PVDF is directly proportional to the value of the residual polarization. However, this relationship is more complicated, because the pyrocoefficient usually increases nonlinearly with increasing temperature, while there is no increase in polarization occurs in this case.

Summarizing the above data, we can conclude that the pyroelectricity in PVDF is usually considered in isolation from other processes. However, to understand the nature of this phenomenon, it is of interest to experimentally study the dynamics of its formation and changes simultaneously with other isothermal and thermally simulated processes, such as measurement of volt-ampere characteristics, thermoelectret poling and depolarization.

### 6.2. Measuring the pyroelectric coefficient in poled PVDF films

The pyroelectric effect is usually investigated in quasi-static or dynamic mode. In the first case, the pyroelectric current is measured during the slow heating of the short-circuited sample, while in the second case, the variable component of the current is studied during a rapid change of temperature. The main difficulties of the quasi-static method are the separation of the pyroelectric (reversible) component of the thermal shock from the relaxation (irreversible) component in the TSD current.

We measured the pyroelectric dynamic coefficient by the thermal pulse method developed by Collins [22] and used in a number of other papers.

The light pulse of 50 µs duration was generated using the Metz 45 CT-3 flashlight and was used as a reproduced heat source that penetrates the surface of the poled films. The pyroelectric signal was recorded using a broadband oscilloscope. This method is the dynamic one.

With the help of a highly sensitive pyroelectric sensor it was established that light pulses are characterized by a rather high reproducibility. The average energy scatter in measuring of 200 consecutive pulses was 2.4 %. The magnitude of the pyroelectric coefficient was judged by the maximum value of the electric signal; thus the results were obtained in relative units.

Measurement of the pyrocoefficient by a quasi-static method was carried out by linear heating and cooling of polarized samples. Dependence of the pyrocoefficient on temperature was calculated by the following formula

$$p(T) = \frac{I_p(T)}{A\beta_c}, \tag{24}$$

where $I_p(T)$ is the pyroelectric current measured during cooling, $\beta_c$ is the cooling rate, which is a derivative of the temperature over time, $A$ is the sample surface area. The heating rate was maintained constant 3 K/min, while the cooling rate depended on time and temperature.

### 6.3 Switching of polarization and pyroelectric activity of PVDF films

Pyroelectric studies of PVDF films have an independent value, since PVDF is widely used in pyroelectric sensors. However, it is interesting to study pyroactivity in conjunction with the residual ferroelectric polar-ization, because it will allow to clarify the nature of the pyroelectricity in PVDF, and to ensure its stability.

Measurement of the pyroactivity by the Collins method was carried out immediately after the polarization switching. Fig. 6 shows how the pyroelectric signal changes when the polarization is fully switched from a fully polarized state. Although the value of the pyrocoefficient can only be judged in relative units, it is evident that the sensitivity of the method is rather high and the signal a completely symmetric after the full switching. It appeared that full switching occurs only if the voltage pulse duration exceeds 100 s. At a shorter duration of the voltage pulse, there is only a partial switching of polarization judging from the data of Fig. 7.

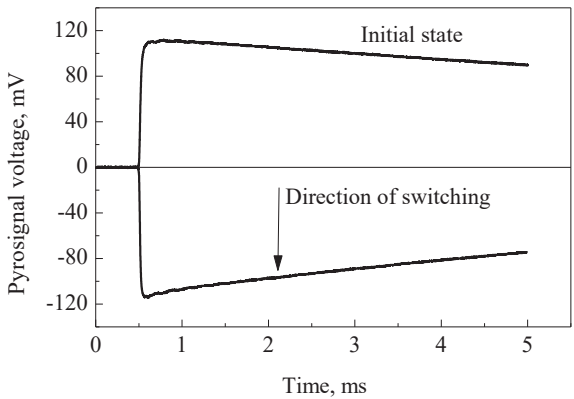

Fig. 6. The pyroelectric signal after the polarization switching of PVDF film by applying 2 kV voltage for 50 seconds. Pyroelectricity was measured after 1.5 min after the voltage switching off

Fig. 7 shows the results of four series of experiments, in which the polarization switching was performed at different durations of the voltage pulse, but with the same magnitude in each series. At a voltage of 0.5 kV (Fig. 7) that provides a field strength of about 40 MV/m, being in the same order as the coercive field, even with a pulse duration of 50 s, only 6.4 % of the polarization is switched, which in principle can be switched, and if the pulse duration is shorter than 50 ms, no switching is practically happening.

At a voltage of 1 kV applied for 50 s, 44.4 % of the residual polarization is switched, that is, the sample is almost converted to the state with zero mean polarization. At this voltage, the 2.2 % polarization is switched even within 50 μs of the switching voltage application. Increasing the voltage to 1.5 kV leads to the switching of 79.4 % of the residual polarization by a 50 s application of voltage. At a voltage of 2 kV for 50 s, the polarization is completely switched.

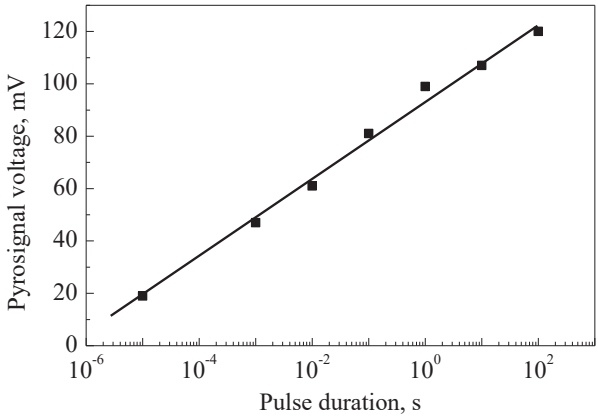

Fig. 7 Dependence of the pyroelectric signal on the duration of the polarizing pulse in the range from 10 μs to 100 s during initial poling of the PVDF film by 2.5 kV voltage

It is interesting to note the specific shape of the pyroelectric signal when switched polarization is more than 50 %, that is, when the direction of the average predominant orientation of the dipoles changes to the opposite direction.

In the electrode zone, which the thermal pulse passes during $t_o = 0.2$ ms, when the polarity direction changes to the opposite, a non-symmetric in shape pyroelectric signal is formed in relation to the initial one. In the vicinity of the electrode, the direction of the pyro-signal change is maintained during the switching polarization indicating the existence of a near-to-electrode layer of thickness about $x = \sqrt{\lambda t_0}$ where $\lambda$ is the thermal conductivity of PVDF.

According to the literature, the coefficient of thermal conductivity of PVDF is $\lambda = 6 \cdot 10^{-8}$ m²/s, thus the thickness of the electrode layer is of the order of 3 μm. We believe that the feature revealed by us is due to the fact that the originally formed polarization in this layer does not switch even in high fields.

It is natural to assume that polarization near the electrode does not increase sharply, but there is some transition layer in which the polarization grows from zero at the electrode to a maximum uniform value in the volume of the film. According to the Poisson equation, inhomogeneous polarization in any layer can be stable only with the presence of a compensating charge in this layer. Apparently, this charge was trapped by deep traps and not released during the polarization switching. The revealed phenomenon is similar to the established by us feature about impossibility of improving the polarization uniformity if its initial formation took place in weak or medium fields.

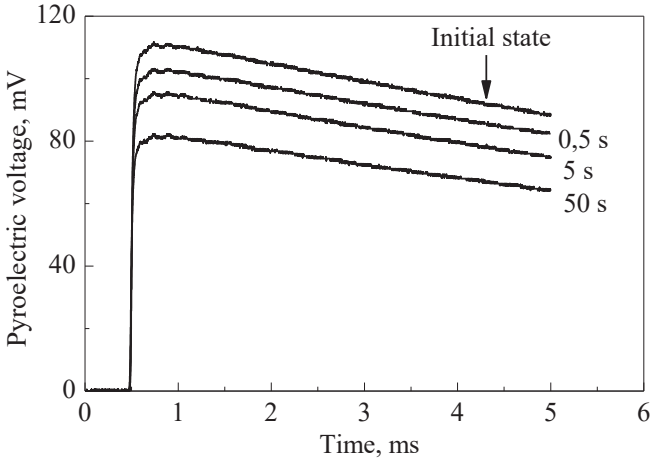

Fig. 8. Pyroelectric signal at sequential polarization switching in PVDF films by pulses of 0.5 kV voltage with duration from 5 ms to 50 s. The duration of the voltage pulse is indicated near the curves

It was found that polarization switched under the action of several successive short voltage pulses is much smaller than the polarization switched by one pulse of the duration equal to the total time of several short pulses. This indicates that there is some distribution of switching times, i.e. some dipoles are easily switched, while others require more time to be switched. Under the influence of short voltage pulses, only «fast» dipoles are switched, while during the continuous voltage application both «fast» and «slow» dipoles are switched, so the total switched polarization significantly increases.

In conclusion, polarization switching at different time and field strength is compared with the values of the pyrosignal under the same switching conditions (Fig. 9). The absolute similarity of the above experimental graphs indicates that there is a direct proportional relationship between the residual ferroelectric polarization and the value of the pyroelectric coefficient. This provision makes it possible to use the technically simple pyrocoefficient measurement to evaluate the polarized state of poled PVDF films, that is, to estimate the magnitude and the direction of the residual polarization.

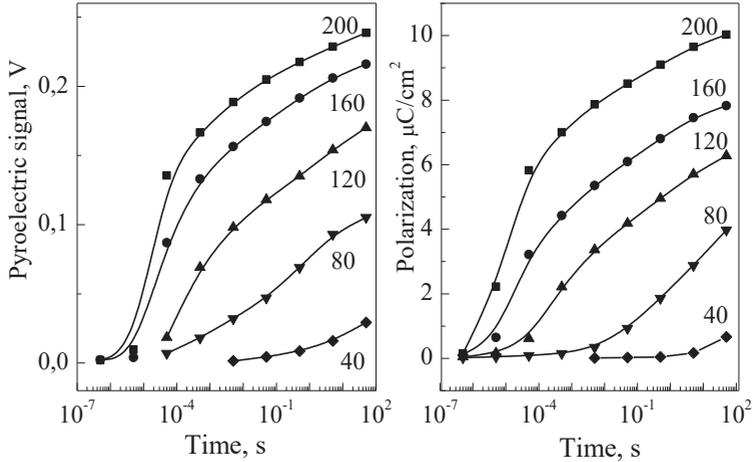

Fig. 9. Evolution of pyroelectric activity and stable ferroelectric part of polarization obtained by sequential application of switching voltage pulses with increasing duration from 0.5 µs to 50 s and at different field strength

## 7. Separation of TSD current components in PVDF

Despite the fact that PVDF is considered as a ferroelectric polymer, some of its electrical properties can be explained within the framework of the theory of polar electrets. The phenomenological model of Gross-Swan-Gubkin [23] suggests the presence of two types of charges in the electret, namely, the homocharge $\sigma(t)$, whose sign coincides with the polarity of the electrodes during poling, and the heterocharge $P(t)$ (internal polarization), which is the result of the micro — and macro- displacements of own charges in the dielectric under the field action. In the case of PVDF, the heterocharge is the dipole polarization, and the homocharge is formed by charges trapped on or near the surface [4].

Stability of the electret state in a polar dielectric depends on the mutual relaxation of the homocharge and the heterocharge.

Since the heterocharge (polarization) is usually the most important in PVDF, the role of the homocharge has not paid enough attention to the present, although the stabilizing effect of the space charge on the residual polarization has already been discussed [4, 6].

Thermally stimulated depolarization (TSD) is a method used to identify relaxation processes in polymer electrets. However, it is very difficult to divide the effect of the homocharge and the heterocharge on TSD currents especially if the corresponding peaks are superimposed on each other in a wide range of temperatures.

We have developed a method for separating homocharge and heterocharge currents [24] by solving the inverse problem. In addition, it has been shown that the application of various modifications of the TSD method, complemented by isothermal depolarization currents allows us to find such important parameters of relaxation processes as the activation energy, characteristic frequencies and the time constant.

The uniaxially oriented 25 μm thick PVDF films metallized on one side were poled in corona triode at the control grid voltage of 3 kV at room temperature and constant poling current density of 90 μA/m² for 30 min and then shortened and held at room temperature for 24 h (except for specimens for measuring the electret potential kinetics).

Four modifications of the TSD method were used, namely thermally stimulated (T) and isothermal (I) depolarization of short-circuited (S) and open circuit (O) samples. Thus, the modifications are named TS, TO, IS and IO where the first letter indicates the temperature mode (thermally stimulated or isothermal), and the second indicates the electric state of the sample (short-circuit or open circuit).

Additional experiments on the thermally stimulated electret potential (TP) kinetics were performed after 24 h of being in the open circuit state. As a dielectric layer in TO and IO modifications, PTFE film of 10 μm thickness was used.

Thermally stimulated experiments were performed at a heating rate of 3 K/min. In isothermal experiments, the temperature was maintained constant after its required value was achieved by rapid heating. The electret potential in TP modifications was measured by the Kelvin method and continuously recorded.

The main features of the experimental curves shown in Fig. 10 and 11 are as follows:

− The depolarization current in the TS modification forms a broad «non-classical» peak with a maximum of 65 °C;

− There is an inversion of the TSD current in the TO modification, while the current direction coincides with the direction of the current in the TS modification in the initial heating stage;

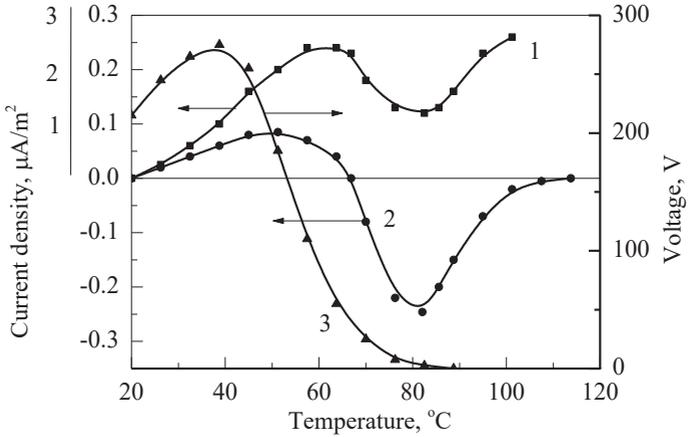

Fig. 10. Thermally stimulated currents in the TS modification (1) and in the TO modification (2), as well as the electret potential in the TP modification

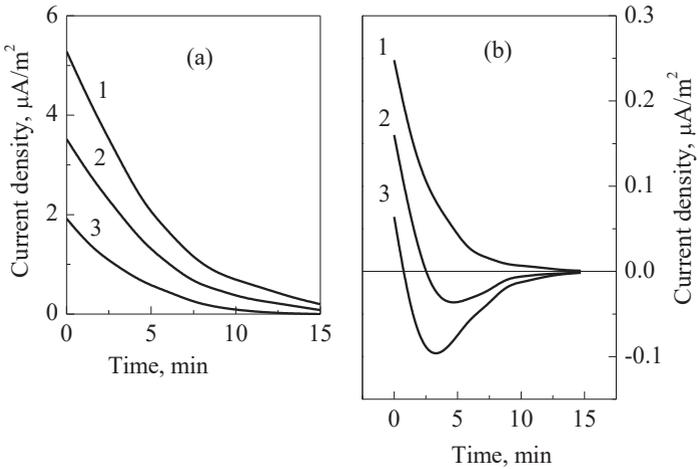

Fig. 11. Isothermal transient currents at different temperatures in the IS mode (a) and in the IO mode (b); 1–45 °C, 2–55 °C, and 3–70 °C.

– The electret potential in TP modification has a maximum at 40 °C.
– The current slowly decreases over time in the IS modification at all temperatures, while the isothermal current changes the direction in IO mode at elevated temperature.

These features can be explained within the framework of the model, which implies existence in the samples of the homocharge and the heterocharge. First, consider the processes of poling and relaxation qualitatively. It is reasonable to assume that negatively charged particles (ions and/or electrons) generated by corona discharge are adsorbed and thermalized on the surface of the sample due to their low (thermal) energy. Excessive charge in the near-to-surface layer or on the surface forms a homocharge that has a certain superficial density $\sigma$ and creates a homogeneous field $E$ in the volume of the sample. The high electron affinity of fluorine atoms facilitates the trapping of charges at traps and formation of the stable homocharge.

Homogeneous internal polarization $P$ (heterocharge) is formed as a result of dipoles -$CH_2$-$CF_2$- orientation in the field created by a homocharge. The formation of polarization is equivalent to the formation of a bound surface charge $P$, which has a sign opposite to the sign of the homocharge $\sigma$. Of all the polarization processes in PVDF, the orientation of the -$CH_2$-$CF_2$- dipoles is the most significant due to their large dipole moment of 2.1 Debye [13].

If the polarization $P$ is zero, then the field is created by a complete superficial charge $\sigma$. When $P$ begins to grow, the depolarizing field appears which is "neutralized" by a part of the surface charge. Thus, the field in volume is created by the difference ($\sigma - P$) between the surface charge and the polarization. Consequently, the surface charge $\sigma$ consists of two parts $\sigma = \sigma_1 + \sigma_2$, the first of which is a charge that provides compensation of the depolarizing field ($\sigma_1 = P$), and the second $\sigma_2 = \sigma - P$ creates the electric field in the volume of the sample.

After the short circuiting of the poled samples (in TS and IS modes), the "excess" charge $\sigma_2$ disappears.

The equilibrium between the homocharge and heterocharge ($\sigma = \sigma_1 = P$), as well as the zero internal field ($E = 0$) are supported by the current in the external circuit, so that the measured current corresponds to the relaxation of the heterocharge.

However, if after the short circuiting and the formation of equilibrium $\sigma = \sigma_1 = P$, a non-conductive dielectric insert (in TO and IO modes) is introduced between one of the electrodes and the surface of the sample, then one can observe the relaxation currents both the heterocharge and the homocharge flowing in the opposite directions. The field in the volume is no longer zero, so that the surface charge (homocharge) drifts in its own field through the entire thickness of the sample, or it is slowly neutralized by charge carriers responsible for its own conductivity.

In any case, the relaxation of the heterocharge occurs in a field other than zero and caused by thermal disordering of oriented dipoles [12, 13].

We will show that both components of the depolarization current can be found from the dependence of $i(T)$ in the TO mode (Fig. 10, curve 2). It is known [12, 13] that the TSD current $i(t)$ and the electret potential $V(t)$ in experiments with nonconductive insertion between the surface of the sample and the electrode, depend not only on the relationship between the homocharge and the heterocharge, but also on their derivatives, so

$$i(t) = s\left[\frac{dP(t)}{dt} - \frac{d\sigma(t)}{dt}\right], \qquad (25)$$

$$V(t) = \frac{sx_1}{\varepsilon_0\varepsilon_1}[\sigma(t) - P(t)], \qquad (26)$$

$$i(t) = -\frac{\varepsilon_0\varepsilon_1}{x_1} \cdot \frac{dV(t)}{dt}, \qquad (27)$$

where $s = x_0\varepsilon_1 / (x_1\varepsilon + x_0\varepsilon_1)$, $t$ is time, $\varepsilon$ and $x_o$ are the dielectric constant and the thickness of the sample, $\varepsilon_1$ and $x_1$ are corresponding values of the dielectric gap, $\varepsilon_o$ is the permittivity of a vacuum.

The full component $i_c(t)$ can be represented as

$$i_C(t) = \frac{g}{x_0}V(t) = -\frac{d\sigma(t)}{dt}, \qquad (28)$$

where $g = g_0\exp(-Q/kT)$ is the own conductivity, $k$ is the Boltzmann constant, $T$ is temperature, $Q$ is the activating energy of its own conductivity, $g_o$ is a pre-exponential factor. Integrating (27) and replacing the time $t$ with temperature $T$ in (25)−(28) according to $T = T_0(1 + bt)$, where $b$ is the heating rate, $T_o$ is the initial temperature, we obtain the expressions for the temperature dependences of the homocharge current $i_1(T)$ and the heterocharge current $i_2(T)$, as well as the voltage on the sample (potential) $V(T)$

$$i_1(T) = \frac{d\sigma}{dt} = -\frac{x_1 g_0}{bT_0 x_0\varepsilon_0\varepsilon_1}\exp\left(-\frac{Q}{kT}\right)\int_T^\infty i(T')dT', \qquad (29)$$

$$i_2(T) = \frac{dP}{dt} = \frac{i(T)}{s} + \frac{d\sigma}{dt}, \qquad (30)$$

$$V(T) = \frac{x_1}{bT_0\varepsilon_0\varepsilon_1}\int_T^\infty i(T')dT'. \qquad (31)$$

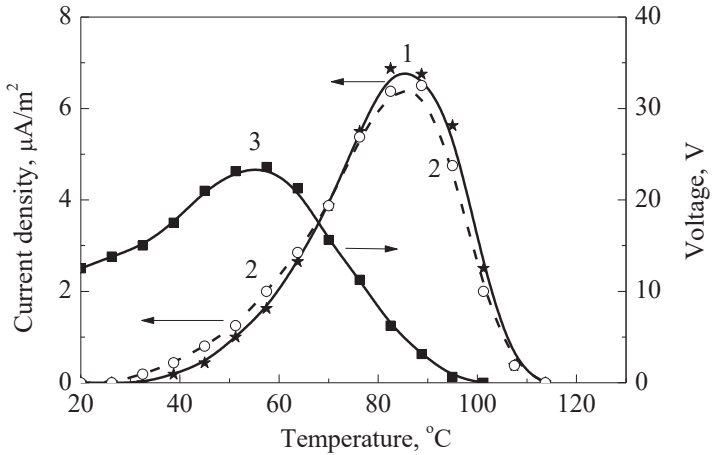

Fig. 12. Temperature dependences of the homocharge (1) and heterocharge (2) relaxation currents, as well as of the voltage on the sample (3) calculated according to the model

All quantities at the right side of the equations (29)−(31) are known, or can be obtained experimentally. The results of calculations according to the equations (29)−(31) based on the data of Fig. 10, are shown in Fig. 12. The values of the activation energy $Q = 0.76$ eV and the factor $g_o = 0.18$ Sm/m were obtained from constant values of the isothermal poling current and voltage.

As one can see in Fig. 12, the homocharge and the heterocharge form two broad peaks with almost identical maxima. The heterocharge relaxes faster in the low-temperature region where the homocharge is relatively stable. This is probably the reason for the initial increase of the thermally stimulated potential (see curve 3 in Fig. 10 and curve 3 in Fig. 12). The current inversion in TO and IO modes is caused by a change in the ratio between homocharge and heterocharge at high temperatures (curves 1 and 2 in Fig. 12).

It is known that the inversion of the TSD current can be caused by the re-polarization, that is, it arises as a result of the appearance of an additional heterocharge in the field of a homocharge, and the voltage in this case should decrease [12].

However, this was not observed in our case (Fig. 10). On the other hand, the initial growth of the electret potential during heating cannot be caused by increase of the surface charge density $\sigma$, since charges in this case would have to move against the electric field created by these charges, which is impossible. Therefore, the first peak of the TSD current and the increase of the electret potential (Fig. 10) are due to the faster disintegration of the heterocharge (polarization) compared with the homocharge. It is possible that in PVDF in the first stage of heating, not all polarization is destroyed, but only its least stable part.

Thus, the long-term conservation of the heterocharge in PVDF films is possible only in presence of the stabilizing field created by homocharge. We believe that many special properties of PVDF are associated with successful combination of a large dipole moment of -$CH_2$-$CF_2$- (2.1 D) that contributes to formation of heterocharge, and the high electron affinity of fluorine atoms (3.37 eV) that contributes to the creation of the stable homocharge. Although the electret state in the PVDF is unstable, the self-balanced relaxation of the homocharge and the heterocharge is slowed down due to the stabilizing effect of the homocharge.

In the theory of electrets [23], it is assumed that homocharge and heterocharge decay by the exponential law with the temperature dependent time constants. Therefore, such expressions should be valid for IO and IS modes

$$i_1(t) = -\frac{s\sigma_0}{\tau_1}\exp\left(-\frac{t}{\tau_1}\right), \tag{32}$$

$$i_2(t) = -\frac{P_0}{\tau_2}\exp\left(-\frac{t}{\tau_2}\right), \tag{33}$$

$$\tau_1(T) = \frac{\varepsilon_0\varepsilon}{g_0}\exp\left(\frac{Q}{kT}\right), \tag{34}$$

$$\tau_2(T) = \tau_0\exp\left(\frac{W}{kT}\right), \tag{35}$$

where $W$ is the activation energy of the heterocharge relaxation, $\tau_1$ and $\tau_2$ are the corresponding time constants.

Applying the equations (32)−(35) to the experimental curve in Fig. 11, we calculated the following relaxation parameters for homocharge and heterocharge: activation energies ($Q = 0.76$ eV and $W = 0.54$ eV), characteristic frequencies ($f_2 = 1/\tau_0 = 7.4$ MHz and $f_1 = (g_0/\varepsilon_0\varepsilon) = 1.7$ GHz, time constants at 20 °C ($\tau_1 = 31000$ s and $\tau_2 = 2800$ s). The results indicate that the homocharge is more stable than the heterocharge.

Thus, we have developed a method for separating the depolarization currents of the homocharge and the heterocharge from the measured TSD current, and revealed the relaxation behavior of the both components.

Application of various TSD modifications complemented with isothermal depolarization currents allowed to find the most important parameters of the relaxation processes.

The developed method allows us to analyze the relationship between the homocharge and the heterocharge not only in PVDF but also in other polar dielectrics. The introduction of polar groups with the simultaneous creation of deep traps could contribute to increasing of the residual polarization stability in polar polymer dielectrics. Therefore, if there are appropriate conditions for creating a homocharge, then a high level of the residual polarization can also be provided for a long time.

## 8. Thermally stimulated and isothermal processes in composites

Composite materials based on polymers with impurities of ferroelectric ceramics have a number of significant advantages over conventional ferroelectric ceramics, but the possibilities of using composite materials as active elements of piezoelectric and pyroelectric converters are not fully implemented.

It is known that most of the polarization in ferroelectric ceramics immediately switches back to its original state after switching off the applied voltage, and only $25-30$ % of the domains remain oriented if no special actions are taken. Therefore, the dominant orientation of domains should be somehow fixed. A similar problem exists in ferroelectric polymers, in which ferroelectric crystallites are distributed in the amorphous phase. This structural similarity between composites and ferroelectric polymers can also determine the similarity of the electrical relaxation processes in these two classes of materials.

PVDF data to verify the applicability of the concepts already proven for the case of the ferroelectric polymers. In addition, concrete data on the parameters of the electrical relaxation in the specified composites were obtained.

We considered the PVDF-BaTiO$_3$ composite as a model material. The obtained results were compared with ыamples of PVDF-BaTiO$_3$ composites with a thickness of 300 µm containing 0 %, 40 %, 50 % and 70 % of BaTiO$_3$ were produced by hot pressing of a mixture consisting of PVDF powder and BaTiO$_3$ particles with an average size of 10 µm. The composites were annealed at 140 °C and examined using a Solomat 91000 spectrometer for obtaining the general spectrum of TSD currents in the range from -80 °C to +180 °C (Fig. 13).

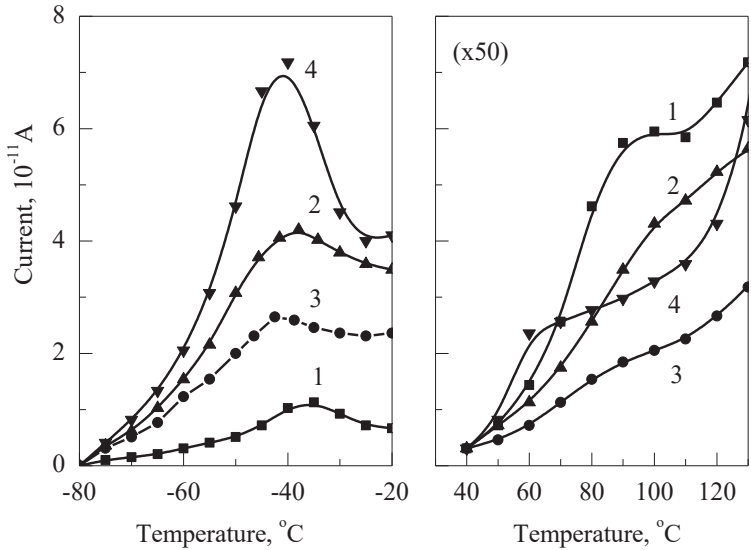

Fig. 13. TSD current curves of poled PVDF-BaTiO$_3$ composites with different content of BaTiO$_3$: 0 % (1), 40 % (2), 50 % (3) and 70 % (4)

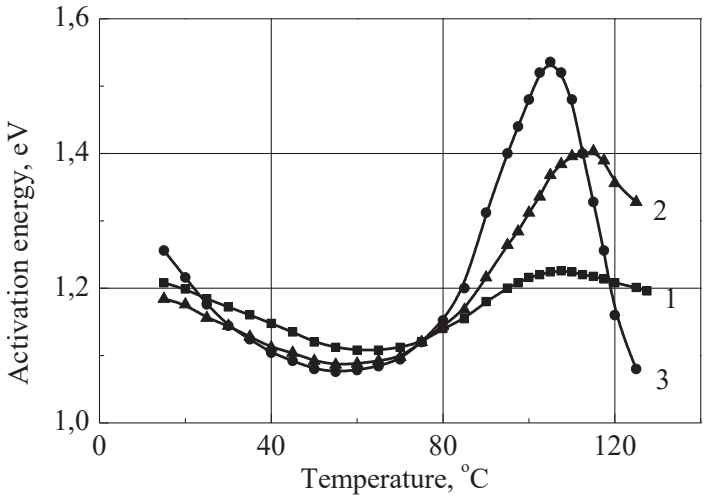

Fig. 14. The activation energy of relaxation processes in PVDF-BaTiO$_3$ composites containing 40 % (1), 50 % (2) and 70 % (3) of BaTiO$_3$

The samples were prepoled at 150 °C in the electric field of 1.25 MV/m for 15 min, and then cooled to −100 °C without disconnecting the electric field. The samples were then depolarized by heating in a short-circuit mode at the rate of 7 °C/min.

The fractional analysis of relaxation processes was carried out by the method of thermal windows. The polarization temperature increased every time for 5 °C from 20 °C to 150 °C. The equivalent frequency of experiments was about $2 \cdot 10^{-4}$ Hz. From these experiments, the activation energy of the relaxation processes was calculated (Fig. 14).

It was found that thermal activation of the polarization process is necessary, since polarization is not formed at room temperature even in high electric fields of about 20 MV/m. This fact is confirmed by the lack of the TSD current after poling of specimens at room temperature. In addition, the VAC at 20 °C was superficial and typical for the space charge limited currents, but not *N*-shaped, as in the case of PVDF.

In all samples, including PVDF without ceramic additives, well-expressed low-temperature peaks near -40 °C can be seen on TSD curves (Fig. 13). This peak is near the glass transition temperature of the amorphous phase in PVDF and is usually attributed to the β-relaxation associated with the micro Brownian motion of molecular chains in amorphous regions. Neither the peak nor its magnitude correlates with the amount of the filler in the composite, which indicates that this peak is associated with the properties of the polymer. The peak in the range 80−120 °C is structurally good only in the case of PVDF, but suppressed in composites by the exponentially increasing leakage current of unknown nature. To eliminate the parasitic currents, we periodically included a capacitor in series with the sample. But even in this case, the unambiguous interpretation of the peaks was difficult, because the theory of TSD currents in composites has not yet been developed.

It is assumed that in the thermal windows method each individual peak corresponds to a single Debye relaxation process. Then the peak analysis gives the temperature-dependent relaxation time $\tau(T)$, which can be approximated by the Arrhenius equation

$$\tau(T) = \tau_0 \cdot \exp(Q / kT) , \qquad (36)$$

where $\tau_o$ is the pre-exponential factor; $Q$ is the activation energy; $k$ is Boltzmann's constant.

As can be seen from Fig. 14, the activation energy slightly decreases in the range of 20−80 °C from 1.17 eV to 1.09 eV regardless of the samples composition. Then it sharply increases reaching the maximum values of 1.23—1.55 eV at 105—110 °C. The amount of the activation energy correlates with the concentration of the ceramic filler and equals 1.23 eV at 40 %, 1.4 eV at 50 % and 1.55 eV at 70 % of BaTiO$_3$ in the composite. In addition, the peak temperature in the Fig. 5.10 is very close to Curie point of BaTiO$_3$ confirming the fact that relaxation behavior of the composite near this temperature is determined by ceramics.

It was found that the maximum temperature of the thermal window peak was about 15 °C above the polarization temperature for all fractions, regardless of the composition of the sample.

The dielectric constant of the composites increased with temperature and was in a certain ratio with the percentage content of the filler equaling 20—250 at 40 %, 30—400 at 50 % and 40—1100 at 70 % of BaTiO$_3$ in the composite. It is known that the dielectric constant of pure PVDF was about 10—12, and in BaTiO$_3$ it was equal to 1500—7000. The polarization field applied to the composites in the experiments (1.25 MV/m) was higher than the coercive field of pure BaTiO$_3$ estimated as 0.3 MV/m, but it is unclear whether the ferroelectric polarization occurs, because the resistance the polymer matrix is much higher than that of ceramics.

Thus, it was established that the processes of the polarization formation and electrical relaxation in PVDF-BaTiO$_3$ composites are similar to similar processes in the ferroelectric polymers. This can serve as a prerequisite for the creation of a generalized model that not only explains, but also predicts the electrical behavior of polymer-ceramic composites.

We have established the influence of the polymer matrix conductivity and poling regime on the effective conductivity of the PVDF-PZT composites. The research was carried out on flat plates of PVDF-PZT composite, made by hot pressing of a mixture of PVDF powders and PZT ceramics taken in a volume ratio of 60:40. Two types of the PVDF powder differing in concentration of ionogenic end groups that contribute to dissociation of impurities, and therefore have a specific resistance at room temperature $10^{10}$ $\Omega \cdot$m and $10^{12}$ $\Omega \cdot$m, were used to study the influence of the properties of polymer matrix. Specific resistance of the PZT had an order of $10^{10}$ $\Omega \cdot$m.

Poling of the samples was carried out by the thermoelectret method. The samples were kept for 50 min at high temperature in the outer field, and then cooled without removal of the field. As changing parameters, we used the poling temperature (70—130 °C) and the conductivity of the polymer component. It was assumed that there are different conductivities and dielectric permittivities in the layers. In real ferroelectrics polymers the phenomenon of percolation and injection of carriers in volume should be taken into account. From the theory of percolation, it is known that for three-dimensional two-phase systems the leakage threshold depending on the structural features of the phases is in the range of 0.05−0.6. In the case of conventional ferroelectric polymers with the concentrations of the filler or crystalline ferroelectric phase of the order of 0.4−0.5 it is very likely to find the mixture either in the critical region or in the region where the infinite cluster is formed. Therefore, known formulas for generalized electrical characteristics of mixtures exopessed by formulae of Lichteneker, Landauer-Brugemann, Odelevsky and others are unsuitable for ferroelectric polymers and composites because they assume relative proximity of the components properties and the small volume fraction of one of them.

It is obvious that in the presence of contacts between particles of crystallites or ceramics, equivalent circuit diagrams should take into account not only sequential combinations of layers, but also parallel ones. Consideration of the injection based on the Poisson equation should lead to the field heterogeneity in the thickness of the sample. The mentioned effects in ferroelectric polymers and composites have not yet been studied and the theory of these phenomena is absent.

As can be seen from Fig. 15, change in the poling temperature affects the temperature dependence of the conductivity. The activation energy increases with increasing temperature both in low-conductive and high-conductive composites, and the value of the conductivity decreases. This corresponds to the proposed hypothesis that explains decrease of the conductivity by trapping a part of the carriers at the boundaries of the polarized crystallites. Indeed, residual polarization increases with increasing temperature and the specific conductivity decreases.

The degree of the residual polarization and its stability in a ferroelectric ceramic essentially depend on the magnitude of the injected space charge that apparently compensates the depolarizing field occurring when dipoles in crystallites are oriented. Similar processes occur in the ferroelectric polymers. However, in view of the morphological features, the conditions for maintaining the stable polarization in ferroelectric polymers are better than in ferroelectric composites where the incomplete polarization occurs due to boundaries scarcity, mechanical stress and restriction in free volume. Therefore, some of the residual polarization immediately relaxes after removing the external field. In the ferroelectric polymer, the ferroelectric particles are free that creates favorable conditions for trapping the charge at their borders. Although the particles are in contact with each other, they do not form a rigid grid and easily allow for volume changes during poling Large-scale

**173**

potential changes during poling contribute to deep trapping of charges, as well as to reduced molecular mobility in the interphase layer That is why the piezoactivity of polymer in ferroelectric polymers is higher than that of ceramics used as a filler.

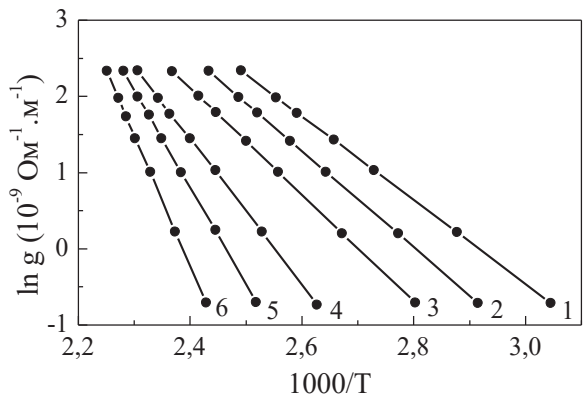

Fig. 15. Temperature dependence of the specific conductivity of PVDF-PZT samples poled at 70 °C (1.4), 100 °C (2.5) and 130 °C (3.6). Specific resistance of the polymer is $10^{10}$ Ω·m (1, 2, 3) and $10^{12}$ Ω·m (4, 5, 6)

In «polymer-ferroceramics» composites as in the ferroelectric polymers, one should not contradict the role of space charge and polarization in the appearance of high pyroactivity, but consider them in a relationship. In the composites like as in PVDF, irreversible relaxation processes and reversible (pyroelectric) are interconnected.

It is known that pyroelectric currents are reversible, that is when switching from heating to cooling they must change the direction to the opposite. However, as can be seen from Fig. 16, this is not always observed.

The imbalance of direct and reciprocal current is due to the influence of the relaxation component, which does not diminish instantaneously to zero with the termination of heating, but it relaxes with the time constant of order of tens and hundreds of seconds. As a result, there is a delay in the pyroelectric current, which is observed in Fig. 16.

With repeated heating and cooling of the samples, along with decrease of the current in the forward direction as a result of the relaxation processes annealing, the symmetry of the direct and the reverse current appears for the same reason (Fig. 17) indicating predominance of the pyroelectric component over the relaxation component.

**174**

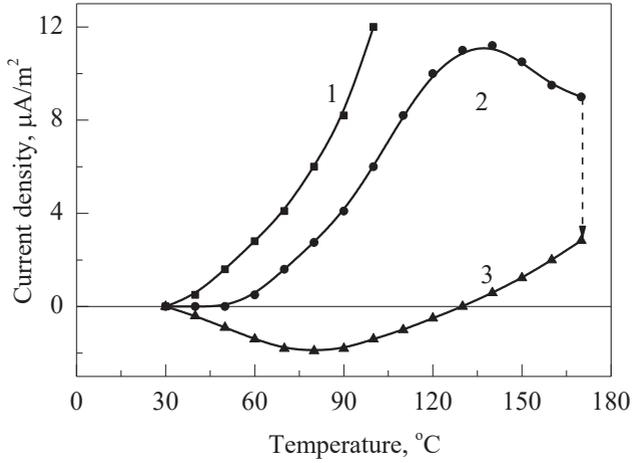

Fig. 16. Thermal currents during primary (1) and repeated (2) heating of poled PVDF-PZT composites samples, and also cooling after the reheating (3). The heating rate is 3.5 °C/min. The thickness of the samples is 280 μm. Piezo modulus is 8 pC/N

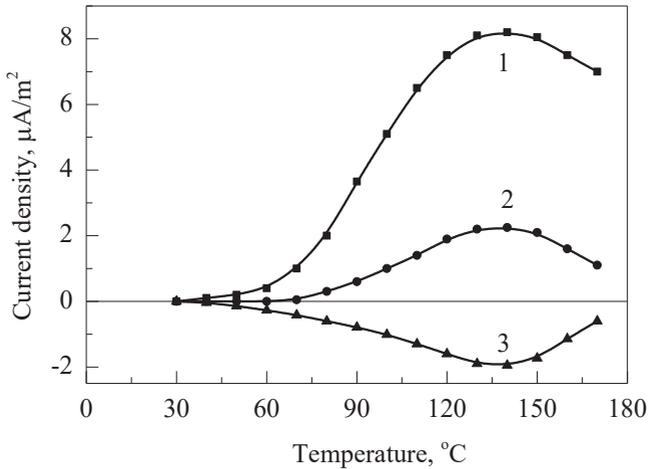

Fig. 17. Thermal currents when heated for the first (1) and the third (2) times, and also after cooling after the third heating (3) of PVDF-PZT samples poled by the thermoelectret method. The heating rate is 3.5 °C min. The thickness of the samples is 240 μm, the poling temperature is 100 °C

It is interesting to note that in the polymer-ceramic composite, as in PVDF, the maximum of pyroactivity coincides with the position of the TSD current peak indicating the interrelation of these processes, and possibly also their general nature.

### Conclusions

Application of the corona triode in most of our studies allowed to make the poling process fully controlled, to optimize the magnitude of the resulting polarization and to perform a virtual short circuiting after the completion of poling. Based on the multifactorial experiment, the best correlations of parameters such as temperature and time of poling, as well as the potentials of the corona electrode and the grid are established. A new technique for studying the relaxation of homocharge and heterocharge processes in the ferroelectric polymers was developed. The technique is developed for separation of the complete electrical displacement components during PVDF films poling by voltage pulses for allocation and analysis of the polarization components and kinetics of their formation.

The commonality and similarity of electrophysical and polarization processes in ferroelectric polymers and composites have been experimentally proved considering their two-phase structure and the need to neutralize the depolarizing field by trapped charges at the interphase boundaries.

Phenomenological models of the polarized state formation and relaxation processes under different conditions were proposed and calculated taking into account and explaining polarization heterogeneity, nonlinear dependence of polarization on the field and trapping of carriers at the boundaries of polarized regions.

A phenomenological model for the polarized state formation a ferroelectric polymer subjected to constant current poling was developed and analyzed, in which an important role is assigned to injection of charges, which create a heterogeneous distribution of the space charge, the field strength and the residual polarization. Three-stage nature of the poling process of the ferroelectric polymer films is explained. Comparison of experimental and calculated kinetics of the electret potential showed their high degree of conformity that allowed considering the reasonable assumption about deeply trapped injected charges, on the basis of which the model was constructed.

A model of the polarization switching in PVDF in the mode of the constant applied voltage has been developed that took into account the following features:

— Two-phase structure of the polymer,

— Presence of the intrinsic conductivity and injection of charges from the electrodes,

— Trapping of charges at the boundaries of polarized crystallites and their release depending on the stage of the process,

— Partial recombination of the released charges and their secondary trapping,

— Dependence of the polarization switching time on the field strength,

— Nonlinear dependence of quasi-stationary polarization in crystallites on the field strength.

A system of differential equations describing the process of the polarization switching was formulated and solved in which the following parameters were used as alternating variables:

— Field strength in amorphous and crystalline phases,

— Polarization in crystallites,

— The effective conductivity and the surface charge density at the interphase boundaries.

From comparison of the experimental polarization switching curve with the calculated curve, such parameters as the effective mobility, the characteristic polarization switching time and the activation field are found. Based on the model, the difference between the initial polarization formation in a two-phase polymer ferroelectric and the polarization switching was explained.

A model for explaining polarization profiles in PVDF films in the mode of the constant voltage creating either the middle field close to the coercive field, or the high field substantially exceeding the coercive field was developed and analyzed. The model took into account the monopolar injection of charges from a negative electrode, the nonlinear dependence of the quasi-stationary ferroelectric polarization on the field strength, the Poisson equation on interrelation between charges and the gradient of the field strength. The character of the injected charges front motion was calculated, as well as the time dependence of the field strength in the zone adjacent to the positive electrode. Formation of the inhomogeneous polarization in the case of the middle fields was explained, as well as formation of the deeply trapped charge layer at the boundary of the polarized region. This layer is stable even when the polarization is switched leading to distortion of the polarization uniformity profile and impossibility of its improvement by application of very high fields. It is shown why the uniform residual polarization is formed in the case of high applied fields during initial poling.

On the basis of the conducted research, practical recommendations for the modes of ferroelectric polymers and composites poling are developed, which provide high and stable residual polarization.

# DISTRIBUTION OF FERROELECTRIC POLARIZATION IN POLED PVDF AND P(VDF-TFE) FILMS

*Fedosov S. N.*

*У статті наведено результати експериментального дослідження рівномірності розподілу поляризації у сегнетоелектричних полімерних плівках за товщиною зразків. Об'єктами дослідження обрані типові полімерні сегнетоелектрики — полівініліденфторид (ПВДФ) та його сополімер з трифтороетиленом П(ВДФ-ТФЕ). Вимірювання виконані сучасним чутливим методом п'єзоелектрично генерованої сходинки тиску.*

*Встановлено, що розподіл поляризації істотно залежить від величини прикладеної напруги у процесі первинної електризації плівок. У разі слабких та середніх полів, близьких до коерцитивного, розподіл є неоднорідним з максимумом поблизу позитивного електрода. При цьому рівномірність поляризації не можна поліпшити шляхом подальшого застосування дуже сильних полів.*

*Якщо первинна електризація проводиться у сильних полях, то розподіл поляризації однорідний. Досліджено особливості сополімера, електризованого в коронному розряді. Розроблено феноменологічні моделі процесів, що відбуваються при формуванні поляризації в сегнетоелектричних плівках та сформульовані практичні рекомендації.*

*This article presents the results of experimental study of the polarization distribution uniformity in ferroelectric polymer films over the thickness of the samples. Typical polymeric ferroelectrics — polyvinylidene fluoride (PVDF) and its copolymer with trifluoroethylene P(VDF-TFE) were selected as objects of research. The measurements were carried out by a modern sensitive piezoelectric generated pressure step method.*

*It was found that the distribution of polarization substantially depends on the value of the applied voltage during the primary electrification of the films. In the case of weak and medium fields close to coercive, the distribution is inhomogeneous with a maximum near the positive electrode. However, the uniformity of polarization cannot be improved even by the subsequent application of very strong fields.*

*If the primary electrification is carried out in strong fields, then the polarization distribution is uniform. The features of the copolymer electrified in corona discharge have been investigated. Phenomenological models of the processes occurring during the formation of polarization in ferroelectric films have been developed and practical recommendations have been formulated.*

## 1. Introduction. State of the problem

Spatial distribution of polarization in PVDF films is extremely important both from scientific and practical points of view. Even in earlier works [1; 2] it was noted that the piezoelectricity and pyroactivity in PVDF films near the positive electrode are higher than near the negative one that was erroneously associated with injection of holes and the formation of a non-uniformly distributed positive space charge. Further studies [3—5] showed that in some cases not only the space charge, but also the polarization are distributed non-uniformly in the thickness direction.

For the first time, heterogeneity of polarization in PVDF was detected by Day et al. [1] by different values of the pyroelectric activity near two sides of polarized films. Sassner [2] found that from tightly pressed to each other three films only the film adjacent to the positive electrode was highly polarized It was found [6] that in the high field ($E > 150$ MV/m), the polarization is almost homogeneous, while in the middle fields the maximum of polarization is either in the center of oriented films, or near the anode in the unoriented ones.

Gerhard-Multhaupt et al. [7] investigating the distribution of piezoactivity by the method of a pressure pulse generated by a powerful laser confirmed that during thermoelectret poling in the middle fields, the maximum polarization is near the anode, as in the case of poling in a corona discharge.

Authors of [8] believe that because of the high conductivity of PVDF, areas of excess charge cannot exist, and charge-compensated polarization zones are formed. At the same time, we have calculated Debye's length of screening $L_D$ for the following PVDF characteristics: temperature $T = 300$ K, the dielectric permittivity $\varepsilon = 10$, the mobility of charge carriers $\mu = 10^{-12}$ m²/(V·s), the specific conductivity $g = 10^{-12}$ Sm/m.

We obtained $L_D = 170$ μm that is much larger than the typical thickness of the films (10—50 μm). Therefore, the effect of screening by the space charge should be weakly expressed in PVDF.

Mopsik and de Reggi [9] found an increased value of the coercive field strength near the surface of the PVDF, and de Reggi and Brodhard [10] found that, in spite of the displacement of polarization to the anode, it is zero near the electrode. Sessler and Berraissoul [11] found that the piezoactivity near the electrodes and, consequently the polarization is very small.

The weakening of the field, in our opinion, is an indication of the injection of charge carriers and, on the contrary, the field and polarization near the blocking electrode are increased.

In [4], the maximum polarization near the anode is reported after poling of PVDF in a positive corona discharge, although in the other work of the same authors it is indicated that the polarization in this case is concentrated in the central zone [6]. Bichler et al [3] found that the position of the maximum polarization depends on the mode of thermal and mechanical processing of PVDF. The authors of the paper [3] concluded that the polarization in the PVDF containing the $\alpha$-phase was shifted to the anode, while the polarization was uniform in the presence of the $\beta$-phase although the heat-treatment and stretching changed both the crystalline structure of the films and their other properties. In [12], the polarization attenuation near the anode is reported in corona poled PVDF films. Such contradictory data are explained by the fact that the selection of films for the study was random (various thicknesses, regimes of poling, annealing, and mechanical pre-treatment). In addition, the research methodology was imperfect in some cases, so that there was a subjective factor in interpreting the measurement results. Direct measurements of the polarization profile are possible only by the method of the pressure step, while all other methods should be considered as non-direct ones.

The dynamics of the polarization profile in PVDF was studied only in a few works. Thus, it was established [13] that polarization develops in the central zone in biaxially oriented PVDF films containing 70 % $\beta$-phase in the case of middle field (60 MV/m). However, when changing the polarity of the voltage, the complete switching near the anode does not occur and a bimorph structure is formed.

Fedosov and Sergeeva, investigating the distribution of polarization in films electrically charged in a corona discharge [5; 14—16], found that the maximum polarization is near a positive electrode with a negative polarity of the corona discharge.

This indicates that injection of negative carriers takes place, while the positive electrode is blocking. When charged in a positive corona there is a double injection: positive charges from the corona and negative charges from the electrode. It was found that the free surface of the film exposed to the corona discharge can be regarded as a virtual injection electrode.

In a number of studies, the thermal stability of polarization and its distribution in the thickness direction in PVDF and P(VDF-TrFE) [10], as well as in P(VDF-TFE) [16] were investigated. The charge and polarization distribution in PVDF films poled by the electron-beam method [17] was studied by the laser induced pressure pulse (LIPP) method.

Fedosov and Sergeeva found that the trapped electrons in the volume are not concentrated in a thin layer at a certain depth, but are distributed with uneven density in a zone of the finite thickness [18].

These electrons form a virtual negative electrode, from which injection of charges in a non-irradiated region takes place leading to increase in the imaginary penetration depth of the electrons and to inhomogeneity of polarization in the non-irradiated region.

The importance of the injection processes, but not the distribution of intrinsic charges in volume was evidenced by Mitsutani and Ieda [19]. They found that the poling current increases in 200 times, if a corona discharge is used instead of a metal electrode. The experimental results in papers [19; 20] have been explained by the injection of charge carriers. At the same time, some authors believe that the corona discharge forms an ideal blocking contact with any dielectric [21]. Thus, the question of the injection of charge carriers and their role in the formation of polarization remains open and controversial.

The reasons of the fragmentary and contradictory nature of the literature data on the polarization profiles in PVDF are the complexity of experimental methods and ambiguity in interpretation of the results.

The aim of this study is to clarify the situation with importance of the uniformity of the polarization distribution in polymer ferroelectrics by performing additions experiments and developing the corresponding phenomenological models.

## 2. Methods for studying polarization profiles in ferroelectric polymers

Profile of polarization and space charge in thin polymer films is studied by one of the following three methods is used:

1) The method of the piezoelectrically generated pressure step (PPS);

2) The method of the thermal wave induced by a modulated laser (LIMM);

3) The method of the laser induced pulse pressure (LIPP).

The analysis of the efficiency, sensitivity and resolution of these methods showed that the measured current in the PPS method is proportional to the polarization or gradient of the space charge, whereas in the LIPP method the measured signal is proportional to the charge (if any) and the gradient of polarization. Given the features of the LIPP method, only zones where polarization changes in thickness direction is detected. Therefore, a high homogeneous polarization and complete absence of the polarization give the same signal. Despite the resolution of the LIPP method is practically the same as the resolution of the PPS method, it should be recognized that the method of the pressure step (PPS) is more reliable and informative.

In the LIMM method, the depth of the thermal wave penetration decreases with the increase of the modulating frequency, so at very high modulation frequencies above 10 MHz it is possible to investigate very thin near-to-electrode layers of less than 1 μm in thickness.

However, the resolution of the LIMM method drastically decreases with increasing the distance from the surface to the depth of the sample, and this is a significant disadvantage of the LIMM.

Taking into account the results of the analysis, we have chosen the PPS method, which provides a fairly high resolution of about 2—4 μm and the possibility of observing the real-time polarization profile on oscilloscope's screen. It does not require complex calculations, assumptions, and solutions of incorrect inverse tasks as in the case of the LIMM method. The colossal advantage of the PPS method is the ability to study the dynamics of the polarization profile "*in situ*" directly in the process of poling, polarization switching, and short-circuiting of the samples.

Most methods for measuring polarization and space charge profiles in dielectrics are applied to already polarized samples providing the information about the final state, while the process of the polarization development and its profile remained inaccessible for a direct experimental study. In this sense, a unique possibility is provided by the piezoelectrically generated pressure step (PPS) method, in one of the modifications of which there is the ability to measure the profile "*in situ*" directly when the polarizing voltage is applied or the specimen is short-circuited. Such measurements of the dynamics of the polarization profile are extremely important for understanding the physical processes occurring in ferroelectric polymers during their poling and switching of polarization.

The PPS method developed by Eisenmenger et al. [6] was applied by us, and all measurements were performed in the laboratory of Prof. W. Eisenmenger at the Department of Physics of the Stuttgart University. The method is based on the generation of an electric signal (a current pulse) when a pressure step generated by a piezoelectric crystal passes through the sample. The piezoelectric pressure step results from a voltage step with a very steep front. The front of the pressure wave extends with the sound speed of 2250 m/s creating a current pulse in the short-circuited sample. It has been proved that the shape of the current pulse repeats the profile of the polarization distribution in the thickness of the film.

The block diagram of the installation using the PPS method is shown in Fig. 1. The voltage step is formed by means of a constant voltage source (500 V) loaded with connected in series a resistance and a capacitor. This chain with a frequency of about 100 Hz is locked to a resistor of 50 Ω, in parallel with which a quartz crystal is connected. A sample is pressed to the back side of the piezoelectric crystal, to which a grounded copper electrode is connected.

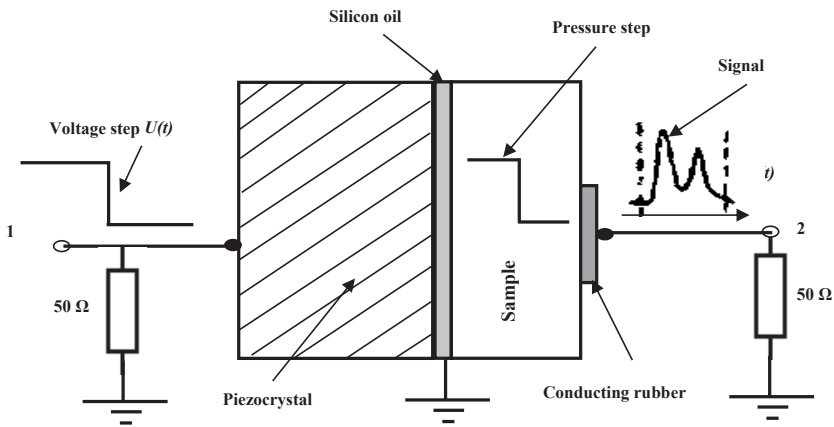

Fig. 1. Schematic diagram of the polarization profile measurement by the piezoelectrically induced pressure step (PPS) method

For better transferring of the pressure wave from quartz to the specimen, a thin layer of silicone oil is applied between them. The electric signal is taken from the rear side of the sample with the help of a clamping conductive rubber electrode. The load is a 50 Ω resistor, the signal from which is fed to the broadband amplifier and then either to the spectrum analyzer, or to the oscilloscope. In the case of thin specimens, a 23-μm thick polypropylene gasket was used to reduce the capacity.

Measurements have shown that the steepness of the pressure step front is of the order of 0.4 ns, while the sound speed in PVDF at room temperature is about 2250 m/s. Thus, the time of the pressure wave passage through the sample at its thickness of 20 μm is of the order of 10 ns. The duration of one voltage pulse was 100 ns, that is, a step-by-step mode was implemented.

All used devices (amplifier, spectrum analyzer, and oscilloscope) had a bandwidth of more than 1 GHz. The sensitivity of the PPS method was about 2 μm and was limited by the steepness of the pressure step. An electrical signal in the spectrum analyzer was converted into a digital code to allow computer processing of the data.

To investigate the polarization profile, several series of experiments were performed on PVDF films. Aluminum electrodes of 5 mm in diameter on both sides were pre-deposited at the samples by vacuum evaporation. Initial poling and the polarization switching were carried out at room temperature by applying a constant voltage of certain magnitude and polarity.

The magnitude of the voltage was chosen so that average field was 60 MV/m in a series of polarizing and switching, which is slightly higher than the coercive field of PVDF according to the literature data [12]. Such a mode was named as «middle fields».

In another series, the polarizing field strength was 160 MV/m being much higher than the coercive value. Such regimes were classified as «high fields». At the middle fields and at the high fields, full polarization cycles were investigated. After each phase of polarization or switching, the specimen was short-circuited for a time sufficient to establish a quasi-stationary state (from 200 to 2000 s).

The polarization profiles were measured about 100 times per second and recorded from the oscilloscope screen. Then the analog information of selected frames was converted into digital and entered into the computer for further processing. All results are presented in the form of graphs of dependence of polarization on the distance in the sample from its surface.

### 3. Distribution of polarization in thin PVDF films

Polarization profiles and the space charge give important information about their interrelation in the process of the polarized state formation and in ensuring of its stability. This allows us to move from hypotheses and assumptions to concrete experimental facts, the analysis of which contributes to the deeper understanding of the ferroelectric polymers characteristics.

That is why special attention was paid to the dynamics of polarization profiles in PVDF films not only during the process of poling but also during the polarization switching and short-circuiting both in the middle fields close to the coercive (50—60 MV/m) and in high fields with a strength of about 160 MV/m.

The obtained results allowed constructing models, which take into account the relation between injection and separation of charges, presence of deep charge trapping zones and its interrelation with the residual polarization. The impossibility of a complete switching of an inhomogeneously polarized ferroelectric polymer [22] discovered by us is of great practical importance for the choice of poling modes and shows how strong is relation between the ferroelectric polarization and the surface charge.

### 3.1. Poling field near the coercive value

From Fig. 2 it can be seen that, with the average field strength close to the coercive value $E_c = 50$ MV/m [23], the dynamics of the polarization profile is characterized by the following: At the initial stage of poling, af-ter 8 seconds after the voltage application, the polarization distribution is uniform, but its value is very low ($0.5\ \mu C/cm^2$). Over time, the distribution of polarization in in the thickness direction becomes non-uniform with the maximum near a positive electrode. After poling of PVDF in the average field for 2000 s, a sharply heterogeneous asymmetric distribution of the residual ferroelectric polarization appears with a layer of about 5 μm thickness near the negative electrode, in which the residual polarization is zero. When the voltage is disconnected and the sample is short-circuited, the character of the polarization distribution does not change and the polarization remains heterogeneous, but its magnitude decreases from 3.31 to $1.71\ \mu C/cm^2$ in the region of the maximum.

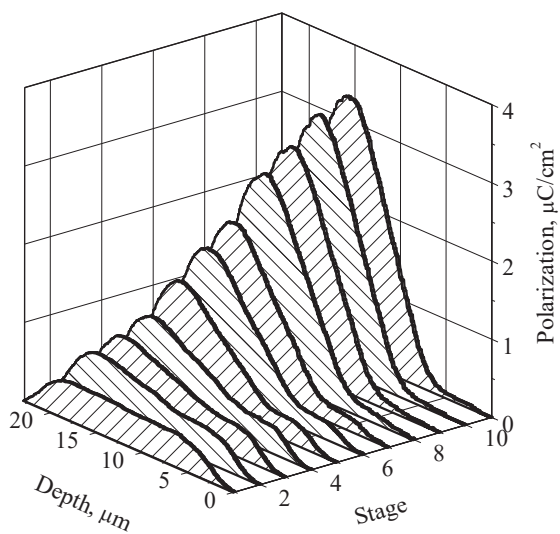

Fig. 2. Distribution of polarization in P(VDF-TFE) film during its poling in the field of 60 MV/m. The stage number corresponds to different times after the starting of poling: 1−8 s, 2−70 s, 3−100 s, 4−150 s, 5−250 s, 6−350 s, 7−450 s, 8−750 s, 9−1000 s, 10−1510 s, 11−2000 s

The resolution of the method for measuring the polarization profile is of the order of 2−3 μm, which leads to appearance of smooth polarization profiles in near-to-electrode regions and in other places of the virtually sharp polarization change. For example, it was shown [5] by measuring the polarization profile near the electrode by the LIMM method having the resolution near the electrodes of the order of 0.1 μm that the polarization changes sharply from the maximum value to zero within about 1 µm in the vicinity of the positive electrode. In the vicinity of the negative electrode where polarization is absent, distortion of the profile due to the finite resolution does not occur.

The effect of resolution, probably, also affects the boundary between the first and the second zones where the polarization changes more sharply than it follows from the curves in Fig. 2.

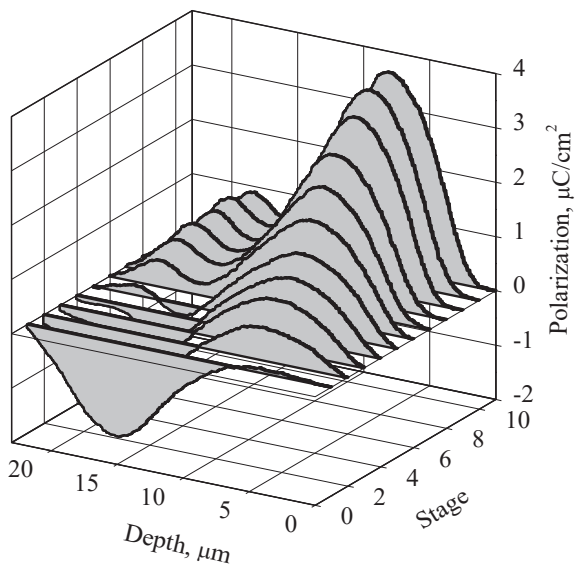

Fig. 3. Polarization profiles in the P(VDF-TFE) film during the polarization switching in the field 60 MV/m after initial poling and the short-circuiting. The stage corresponds to different times from the starting of the switching: 1−0 s, 2−0.2 s, 3−0.5 s, 4−1 s, 5−5 s, 6 −50 s, 7−200 s, 8−500 s, 9−1000 s, 10−1500 s, 11−2000 s

After changing the voltage polarity (Fig. 3) a minimum is formed in the place of the former maximum at a depth of about 16 µm, because the polarization in this place is not completely switched, but even does not reach zero. At the same time, the oppositely directed polarization is formed to the right and to the left of this intersection.

High polarization is formed again only near the positive electrode now connected to the opposite side of the film. It should be noted that the polarization switching is faster than initial poling, and the residual polarization in the peak area is almost 1.5 times greater than after initial poling.

Comparison of polarization profiles after several switchings showed that with even number of switchings, practically identical profiles are appeared. In the case of odd number of switchings, the profiles are also the same, except for the profile after the first charging when there is no reverse polarization near the negative electrode.

Thus, the profile is determined by whether the number of switchings is either even, or odd. In both cases, the distribution of polarization is sharply heterogeneous and asymmetrical with respect to the center of the sample. The main polarization maximum, regardless of the parity of the phases, is always near the electrode, which was positive in the last previous experiment, and the magnitude of this maximum is almost 1.5 times greater than in the case of an odd number of switchings than with the even number.

The time for quasi-stationary state formation decreases with increase in the number of voltage switchings from 2000 s during initial poling to 250—500 s during the subsequent transitions.

In some studies, for example [10, 24], the information that the coercive field near the surface is greater than that in the volume are incorrect, in our opinion, because the incomplete switching is, most likely due to heterogeneity of the field. Due to injection of charge carriers, the field near the surface is smaller than in the volume, and therefore the polarization is poorly switched there.

### 3.2. Attempts to improve properties of non-uniformly polarized films

It is seen from Fig. 2 and Fig. 3 that in the middle fields (60 MV/m) polarization is heterogeneous at any polarity of the voltage, and the complete switching does not occur in any section of the sample. The formed bimorph structure is stored regardless of the direction of the external switching field. At the same time, as will be shown below, homogeneous polarization is formed in high fields (160 MV/m), which then remains homogeneous with any changes in the magnitude and sign of the applied voltage up to complete depolarization. In this regard, it was interesting to investigate behavior of the polymer ferroelectric films in high fields, originally poled in middle fields.

If the initial polarization inhomogeneity is due to the fact that the field is not high enough, homogeneity should increase after applying a high field, due to expansion of the polarized region. However, we have found that the polarization does not become homogeneous, and the polarized region does not expand under the action of a high field.

Experiments were carried out as follows. PVDF films were placed in a field of 60 MV/m and polarization profiles were measured. As can be seen from Fig. 2 and 3, the spatial distribution of polarization was sharply heterogeneous. Further, without interrupting the measurements of the polarization profile we increased the voltage to 3.2 kV by steps of 200 V. The field strength at 3.2 kV is several times greater than the coercive value.

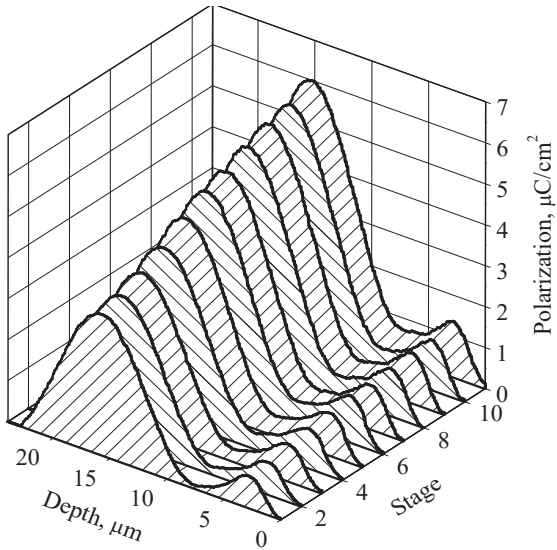

Fig. 4. Polarization profiles in the P(VDF-TFE) film during the stepped voltage increase from 1.2 kV (primary poling) to 3.2 kV. The value of the voltage step is 0.2 kV, the exposure time at each voltage is 50 s

However, as it follows from the graphs in Fig. 4, the increase in the field strength did not lead to the expected improvement of the polarization uniformity. Only the magnitude of the maximum increased, while the non-uniform character of the polarization distribution remained unchanged.

Thus, in order to obtain the high and homogeneous residual polarization, it is not enough to apply a high field. It is necessary to take into account the conditions in which the sample was poled for the first time. If initial poling was carried out in the high field, then the residual polarization will be homogeneous at any applied forward poling or switching voltage.

If initial poling was carried out in medium or weak fields, then the heterogeneity of the residual polarization cannot be corrected or eliminated by applying the high field. In this case, for obtaining the uniform profile of the residual polarization, we recommend a complete thermal depolarization of the sample and its annealing in the short-circuited condition at about 160 °C for several hours, so that the trapped charges in the volume will be completely dissipated. After cooling the sample, it is necessary to re-pole it, but necessarily in the high field.

### 3.3. Poling and switching of polarization in high fields

In the case of high fields (160 MV/m), (Fig. 5), polarization is much more uniform than in the case of middle fields, and the polarization uniformity appears even after initial poling. When polarity of the polarizing voltage changes, the symmetric switching of polarization occurs. By applying the voltage of the opposite polarity, and by increasing it in small steps, it is possible to almost completely depolarize the sample. The field strength at which this occurs corresponds to a value of 60 MV/m. Namely this value can be considered as a real coercive field.

It is interesting to note that the subsequent application of an external field of 60 MV/m of any polarity provides homogeneous residual polarization, which cannot be obtained after initial poling in such a field. The complete depolarization of a highly polarized sample irrespective of the polarity of the external field occurs at field strength of 60 MV/m, which indicates the symmetry of the hysteresis loop if initial poling was carried out in high fields.

Thus, the main features of poling and switching in high fields are as follows:

1. Polarization in the sample volume is homogeneous and symmetric with respect to the central section;

2. There is no difference in the shape of the profile and the magnitude of polarization at different polarizing voltages;

3. Polarization is easily switched over the entire volume, and full depolarization is possible;

Homogeneity of polarization is stored not only in high but also in middle fields.

Of great importance to practice, we have the effect of "formatting" or "conditioning" in a high field, after which homogeneous polarization is provided at any field strength, including the coercive field.

This enables, if necessary, to change the magnitude and sign of the residual polarization in a wide range from zero to saturation that cannot be achieved without this formatting.

## 4. Phenomenological model of polarization profile formation at constant poling field

It follows from Fig. 2 that in the initial stage of poling at 8 seconds after the application of a constant voltage creating average field strength of 60 MV/m the polarization is uniform and corresponds to about 0.5 μC/cm².

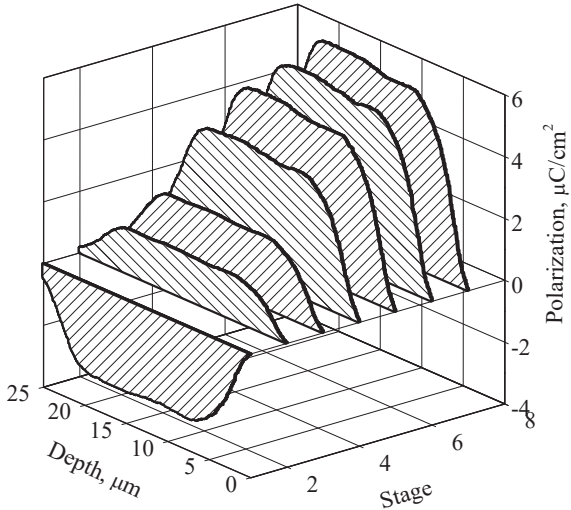

Fig. 5. Polarization profiles in the P(VDF-TFE) film initially poled at 3 kV after applying the opposite polarity voltage of different values. Stages: 1 — initial state, 2 — 1.2 kV, 3 — 1.6 kV, 4 — 2.,0 kV, 5 — 2.4 kV, 6 — 2.8 kV, 7 — 3.2 kV

This indicates a uniform distribution of the field strength and absence of injected charges [25]. At the same time, the stationary value of polarization, can be calculated by the formula corresponding to initial poling [26] taking into account its non-linear dependence on the field strength, the presence of the coercive value $E_c$ and at 50 % crystallinity

$$P_{st} = \frac{P_r}{2 \cdot (E_s - E_c)} (E - E_c) . \tag{1}$$

Substituting in (1) the value of $P_r = 13$ μC/cm² [27], $E_s = 200$ MV/m [23], $E_c = 50$ MV/m, $E = 60$ MV/m, we obtain $P_{st} = 0.43$ μC/cm². The reversible polarization component $P_{cap}$ is proportional to the field strength and the dielectric permittivity

$$P_{cap} = \varepsilon_o (\varepsilon - 1) E. \tag{2}$$

Assuming $\varepsilon = 10$ [21] and taking into account that $\varepsilon_o = 8.85 \cdot 10^{-12}$ F/m and $E = 60$ MV/m, we obtain $P_{cap} = 0.3$ μC/cm². From the graph of the polarization profile dynamics, it is seen that the value of polarization after 8 s of the voltage action is about $0.5$ μC/cm². Thus, all polarization is reversible, that is, the ferroelectric component during this time is not yet formed.

With further application of the field, the polarization becomes non-uniform (Fig. 2) indicating appearance of inhomogeneous distribution of the field strength with its weakening near the negative electrode and the increasing near the positive electrode. According to the Poisson equation, inhomogeneous polarization of this kind is possible only in the presence of excessive negative charge in the place of the field heterogeneity

$$\varepsilon_0 \varepsilon \frac{\partial E(x,t)}{\partial x} = \rho(x,t) \ . \tag{3}$$

This charge is likely to be injected from a negative electrode and is present near this electrode extending with time to the sample depth. Without taking into account the formation of the ferroelectric polarization, it can be assumed that the charge distribution is close to the rectangular [13], and the speed of the charge front motion is determined by mobility $\mu$ and the field strength $E_1$ at the boundary $x_1$ between the zone with the space charge and the zone free of excess volume charge

$$v(t) = \mu E_1 \big[ x_1(t), t \big] \ . \tag{4}$$

Since the applied voltage remains constant ($U_o = \text{const}$), the normalization condition is fulfilled

$$\int\limits_0^{x_0} E(x,t)dx = U_0 \ . \tag{5}$$

The expression (5) is simplified when the rectangular distribution of the injected charge is assumed, since in the region from 0 to $x_1$, the field according to the Poisson equation (3) depends linearly on the coordinate, while in the other part of the sample it is constant and equal to $E_1$. Thus, for finding the $x_1(t)$ function and the field strength $E_1(t)$ we have the following system of equations

$$\frac{dx_1(\text{t})}{dt} = \mu E_1(t) \ , \tag{6}$$

$$E_1(t) \left[ x_0 - \frac{1}{2} x_1(t) \right] = U_0 \ . \tag{7}$$

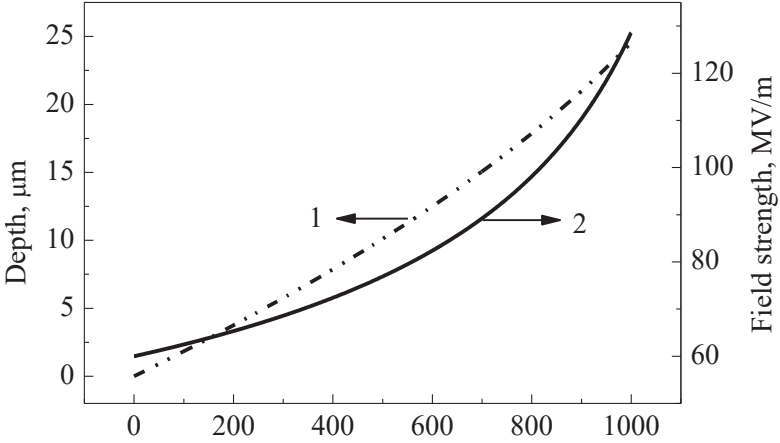

Fig. 6. Estimated graph of the front of injected charges motion after application of a constant voltage to a PVDF film under the charge mobility of $3\cdot10^{-16}$ m$^2$/V·s (1) and the time dependence of the field strength at the boundary, to which the front of the injected negative charges reached (2)

The equations (6) and (7) were solved by numerical methods, and the graphs $x_f(t)$ and $E_f(t)$ are shown in Fig. 6 at $x_o = 23$ µm, $\mu = 3\cdot10^{-16}$ m$^2$/V·s, $U_o = 1380$ V.

From the above graphs it follows that there is some acceleration of the motion of the injected charges front in time. At the same time, the field strength in the part of the sample which the injected charges have not yet reached increases with time exceeding the initial strength more than 2 times after 1000 s of poling.

The polarization switching time depends on the field. Let us explain this in more detail. If we consider that the switching time of the ferroelectric polarization in PVDF is about 5 µs at $E = 200$ MV/m, and the dependence of the switching rate on the field strength is of the following form

$$\tau = \tau_0 \exp\left(\frac{E_A}{E}\right),\tag{8}$$

where $\tau_o$ has an order of 20 ns, then for the activation field $E_A$ we will get the following value:

$$E_A = E \ln\frac{\tau}{\tau_0} = 1.1 \text{ GV/m}.\tag{9}$$

The polarization switching time in the field of 60 and 120 MV/m should be $\tau_{60} \approx 2$ s, $\tau_{120} \approx 2\cdot10^{-4}$ s in accordance with the formula (8).

Thus, the increase of the field strength by 2 times leads to decrease of the switching time by 4 orders of magnitude, but both values are small comparing to the time scale of the experiments. This allows assuming that the process of the ferroelectric polarization formation is quasi-stationary. In this case, we can disregard the dependence of the switching time on the field strength, but use the field dependence of the ferroelectric polarization (1) assuming that at any given time $P_{fe} = P_{st}$.

It was shown [16] that in the PVDF, in addition to the capacitive $P_{cap}$ and the ferroelectric $P_{fe}$ component of polarization, there is also a reversible component $P_{rev}$ of the definitely not established nature, the presence of which is associated with dipole polarization in the amorphous phase of the polymer. The correlation between the components of polarization can be established by analyzing the evolution of the polarization profile after the voltage is switched off and the sample is short-circuited. In the process of poling, when the voltage is applied, there are all three components of polarization

$$P_1 = P_{cap} + P_{fe} + P_{rev}, \tag{10}$$

where $P_1 = 3.31$ $\mu C/cm^2$ at the point of maximum polarization. In the case of the shortening, the components of $P_{cap}$ and $P_{rev}$ disappear and only the ferroelectric component remains, that is, $P_2 = P_{fe}$ with $P_2 = 1.71$ $\mu C/cm^2$. Since the polarization formation process is rather slow, the experiment time is much greater than the Maxwell relaxation time

$$\tau_M = \frac{\varepsilon_0 \varepsilon}{g} \approx 3s . \tag{11}$$

At own conductivity of PVDF $g = 3\cdot10^{-11}$ Sm/m [16], there is no reason to assume that there is a partial back switching of the ferroelectric polarization due to insufficient compensation of the depolarizing field. Therefore, $P_{fe}$ does not change with the short-circuiting. By the formula (1) we can find the corresponding polarization maximum of ($P_{st} = 1.71$ $\mu C/cm^2$) and the field strength $E = 89.5$ MV/m. The capacitive component of the polarization $P_{cap} = 0.79$ $\mu C/cm^2$ according to the formula (2). Then the reversible component of the polarization will be equal to $P_{rev} = P_1 - P_{cap} - P_{fe} = 0.81$ $\mu C/cm^2$.

Since the reversible polarization most likely, is due to the dipole structure of the amorphous phase, it can be taken into account by introducing the effective dielectric permittivity of the amorphous phase, which includes all the reversible processes.

The obtained value of the same order (19.6) was used in the works of von Seggern and Fedosov [15; 28] in calculations of the two-stage formation of the ferroelectric polarization in PVDF. The dynamics of polarization in four cross sections (Fig. 7) shows that the degree of heterogeneity increases, because the polarization of about 0.5 $\mu C/cm^2$ near the negative electrode at the depth of about 5—6 $\mu m$ does not increase in time, as it does in the second part of the sample, but it gradually decreases to zero (Fig. 7). At the same time, the maximum near the positive electrode located initially at a depth of 17.5 $\mu m$, and then shifted to a coordinate of 15.4 $\mu m$ increases with time.

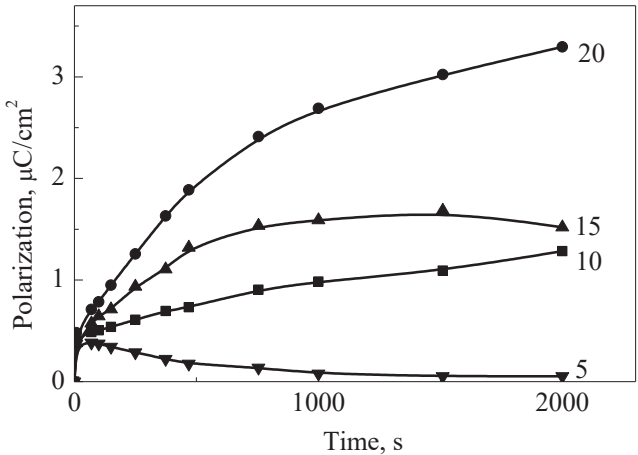

Fig. 7. Dynamics of polarization in the P(VDF-TFE) during initial poling in the field of 60 MV/м at different distances from the film surface (depth): 5, 10, 15 and 20 $\mu m$.

This suggests that the ferroelectric polarization is not formed near the negative electrode, but there is a decrease of the field and polarization to zero. It is known from the theory of injection currents [25] that the field strength at the injecting electrode is very small or equal to zero, but near the electrode where excessive charge is located, the field is non-zero increasing linearly in the case of the homogeneous charge distribution, as follows from the formula (6). If the charge density decreases in the direction of depth, the graph of the *E(x)* dependence will be convex. That is, only injection of charges cannot explain the presence of a zone in the thickness of about 5–6 μm, in which the field and polarization are almost zero. Obviously, there is another phenomenon that leads to decrease of the field strength and polarization near the negative electrode.

$$\varepsilon_a = (P_{cap} + P_{rev}) / \varepsilon_0 E = 20.2 .$$ (12)

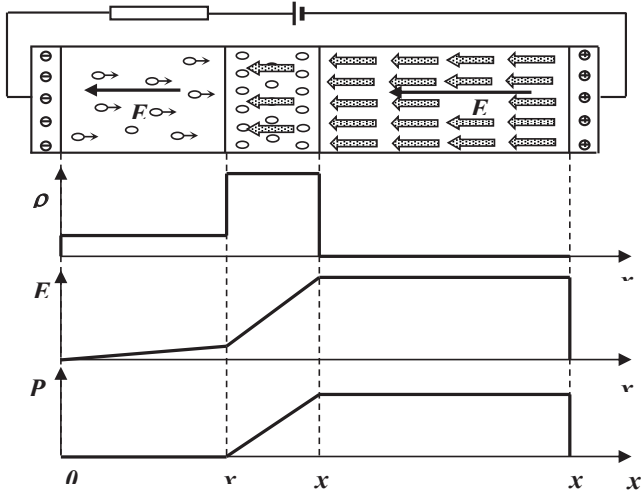

Fig. 8. Scheme of processes occuring during poling of PVDF films in middle fields and leading to formation of a heterogeneous three-layer structure. Also the distribution of the injected charge along the thickness, the field strength and polarization are shown

It is known that the effective conductivity decreases sharply during formation of the ferroelectric polarization in PVDF [17], that is, it can be assumed that the conductivity of the polarized part of the sample is much smaller than that of the not polarized part. In this case, the distribution of the total applied voltage $U_o$ between the not polarized and polarized parts occurs as between two connected in series resistors of different values. The voltage, and hence the field strength, is small in the not polarized part, it is higher in the polarized part. This distribution of voltages contributes to the formation of even greater heterogeneity of the residuals polarization.

Thus, equations (6) and (7), which do not take into account the dependence of the effective conductivity on the ferroelectric polarization $P_{fe}$, only valid until the beginning of the $P_{fe}$ formation. After this, the uniform motion of the injected charge is stopped, because its localization on the boundary of the polarized and not polarized regions occurs in accordance with the Poisson equation (3). Chargers only partially penetrate into the polarized region or do not penetrate at all, that is, the effective conductivity of the polarized region decreases. There is a redistribution of the applied voltage, so that it decreases at the not polarized part, and increases at the polarized part. As a result, the field strength and reversible polarization decrease in the adjacent to the injection electrode region, while they increase in the polarized region. Exactly this is observed on the experimental curves of the polarization profile evolution (Fig. 2).

This phenomenon leads to the formation of a three-layer structure, schematically shown in Fig. 9.

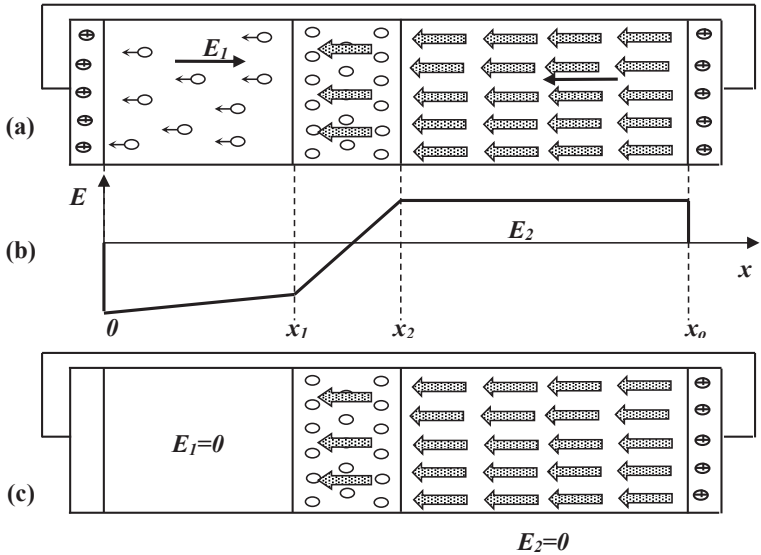

Fig. 9. Scheme of processes occurring in PVDF poled in middle fields at the moment of the short-circuiting (a), distribution of the field strength at the moment of cut-off (b), and state of the sample after aging in the short-circuited condition

In the area adjacent to the negative electrode, there is a high concentration of injected charges, high conductivity, low field strength and very low reversible polarization. At the boundary of the not polarized and polarized regions, a layer of localized negative charges is formed. Within this layer, the field is heterogeneous and polarization sharply increases from zero to the maximum value. In the third zone, there is the homogeneous field and the homogeneous polarization.

When the sample is short-circuited after the completion of poling, the average field strength becomes zero. At the same time, the direction of the field strength vector in the not polarized part of the sample $E_1$ (Fig. 9) becomes such that excess free injected charges from the first zone are "blown" through an electrode that was negative during poling. The reversible components of polarization in all zones are reduced to zero, the excess non-localized charges are dispersed due to their own conductivity, and the field strength at all points of the sample becomes zero with a time constant equal to the Maxwell's relaxation time (11), as shown in Fig. 9.

According to the experimental data of Fig. 2, the first zone occupies an area from $x = 0$ to $x_1 = 6$ μm, the second zone is from $x_1 = 6$ μm to $x_2 = 14$ μm, and the third zone is from $x_2 = 14$ μm to $x_o = 23$ μm. As it follows from the experimental polarization profiles, the boundaries of the zones do not change with time. In the second zone, the excess negative charge is distributed almost uniformly. This is confirmed by the presence of practically linear sections of the polarization profile in this zone.

It is essential that the polarization profile changes with time at a constant voltage. This indicates that slow processes of transfer and redistribution of the space charge are involved in formation of the polarization and in its switching. In the general case, as follows from our data, the polarization is a complex function of the field, coordinates in the volume of the dielectric and time.

The obtained results correspond to the model, which provides an important role of the volume charge in the formation of the polarized zones in PVDF and the injection of charge carriers from electrodes. Experimental data indicate that the level of injection of negative charges from the metal electrode is higher than from the positive electrode. It is also possible that the mobility of injected negative charges is much higher than positive charges. Intrinsic free carriers play a minor role in this case.

Thus, in the case of initial poling, the homogeneity of the field is disturbed by injection of negative charges. In a large part of the sample near the injectable electrode, the field is attenuated and smaller than the coercive field, so the ferroelectric polarization is not formed there and the residual polarization is zero. At the same time, the field exceeds the coercive value near the positive electrode, and the high residual polarization is formed there. The polarization heterogeneity is fixed by negative charges trapped in the region where the gradient of polarization exists. The depolarizing field on the opposite side is compensated by positive charges located either at the electrode or in the near-to-surface layer.

After poling and short circuiting, the field in the peak area supports polarization. There is also a redistribution of moving charges: in the vicinity of the negative electrode, they move in the opposite direction to the injection until the field at all points of the volume becomes zero (Fig. 9). Uncompensated trapped charges remain only at the slopes of the polarization peak.

The massive trapping of injected charges at the boundary of the polarized region begins immediately as soon as a zone of the high polarization appears. This charged layer divides the volume into two parts. In the first part, adjacent to the negative electrode, there is no high polarization, and the concentration of free injected carriers is rather high. This results in a high apparent conductivity and, accordingly, in a weakened field in this zone in the process of poling. At the same time, the polarized region appears separated from the injection electrode by a layer of the trapped charge carriers, and its apparent conductivity becomes considerably smaller than in the first zone. This phenomenon can be considered as the Maxwell-Wagner effect induced by the non-homogeneous polarization, which leads to increasing field in the polarized region. That is why, with the passage of time, the polarized region does not expand, and the value of polarization increases. Phenomenologically, trapping of charges and division into two zones is manifested in reducing the charging current at constant voltage, that is, in reducing the effective conductivity.

When polarity of the applied voltage is changed, the preferred injection of negative charge carriers is again takes place, but they are injected from the opposite electrode. As a result, the region of the high field appears where the residual polarization was zero. This leads to the high polarization formation in new direction in this zone. In the area where residual polarization was strong, the field is weakened due to the injection of negative charges and presence of the negative bulk charge. Therefore, switching of polarization does not occur here, but a part of the residual polarization of the former direction remains.

In the zone of the negative space charge localization (8−15 μm), the direction and value of the polarization gradient do not change. This indicates that the negative charges trapped during initial poling are still in place, despite the fact that the polarization direction in the zone where they are located changes to the opposite direction. This unusual phenomenon is completely consistent with the Poisson equation for the case of a zero field. It is also possible that there is a delocalization of previously trapped carriers and their re-trapping without a significant change in the spatial position of localization.

Thus, after the switching, an asymmetric bimorph structure is formed, which is stored at subsequent transitions. The negative charge layer, judging by the polarization gradient in the region of 8−15 μm, is preserved as if it is fixed in the sample volume during all its transitions. The presence of this layer explains the faster formation of the polarization profile during switching compared to initial poling when this layer is not yet present. This same layer prevents formation of the homogeneous polarization even in the case of high applied fields.

The effect of impossibility to improve the profile of polarization by increasing the applied voltage to the films initially poled in the middle fields can be explained by the influence of the injected and trapped charges. From Fig. 7 it is seen that the polarization gradient at the boundary of the polarized region in the sample volume does not change the sign when the polarity of the switching external voltage changes. This indicates that in the volume there is a layer of deeply trapped negative charges, which plays the role of a barrier preventing the expansion of the polarized region. This layer is stable because it compensates for the depolarizing field in the regions lying at one and the other side of it (at different polarities of the external field). Since this layer obstructs the free motion of injected negative charges, their concentration in the region between this layer and the cathode is much greater than between the charged layer and the anode. Accordingly, the effective conductivities of these regions are different, and the applied voltage is distributed unevenly, so that a significant part of the voltage is applied to the already polarized region.

Thus, increase in voltage cannot widen the polarized region because of the blocking layer, that is, it does not improve polarization homogeneity. So, the initial inhomogeneous polarization remains inhomogeneous in high fields of any magnitude and polarity.

The phenomenon discovered by us is of fundamental importance from scientific, methodological and practical points of view. First, it further clarifies the mechanism of interrelation of the polarization with the trapped space charge. It would seem that a high field 3 times higher than the coercive field, should provide uniform polarization regardless of the initial conditions. However, this is not the case.

The influence of the trapped charge is so significant that even high fields cannot suppress it. Secondly, in the study of switching and hysteresis phenomena in ferroelectric polymers, it seemed self-evident to start electrifying from a weak field gradually increasing the applied field. That is how the hysteresis measurements are performed "at the infra-low frequencies". Taking into account our data it turns out that such measurements are incorrect, because the magnitude and, most importantly, the profile of polarization, depends not only on the field strength, but also on the pre-history of the sample.

It was established by studying poling and switching in high fields that polarization is homogeneous in this case and it is easily switchable over the entire volume. In the case of high field initial poling, a complete depolarization is possible, and the polarization homogeneity persists not only in high but also in middle fields. These features can be explained by the fact that, given the presence of a high field, the polarized region quickly occupies almost the entire volume, which leads to blocking the movement of charges and a sharp weakening of the injection of charges role. The processes of compensation and neutralization of the depolarizing field occur in this case, either on the electrodes, or near the surface, so that the entire main volume remains free of injected and trapped charges, which could disrupt the field's uniformity and polarization. The change of the polarization gradient near the electrodes during the polarization switching indicates that the sign of the trapped compensating charges also changes. This is only possible if these charges are not trapped too deep, so they can be "shaken" from their traps under action of a high field with the subsequent localization in the same region.

## 5. Uniformity of polarization in corona poled P(VDF-TFE) copolymer

Polarization profiles in P(VDF-TFE) have not been studied before, and the data obtained on other corona-charged ferroelectric polymers are rather fragmentary. For example, it was found that polarization occupies the central zone of a positively charged PVDF [29]. In another sample of PVDF, which was in similar conditions, the peak of polarization was found near the positive surface, while the biaxially stretched PVDF showed more or less uniform polarization [30]. The polarization profiles in polarized PVDF films that were poled in a negative corona turned out to be bell-shaped [31], while a significant decrease in polarization was observed near the positive side of a biaxially stretched PVDF poled in a positive corona [32]. Distortion of polarization homogeneity is usually considered as a consequence of the injected charge presence [29–32], but the details of this mechanism are still only partially clear.

In [5], we report on measurements of polarization profiles obtained by applying a piezoelectrically generated pressure step (PPS) method to films

P(VDF-TFE) that were charged in a negative corona discharge under different conditions.

The samples were films from experimental batches of 20 μm thick P(VDF-TFE) consisting of 95 % VDF and 5 % TFE. The films were extruded from the melt and stretched unilaterally by the supplier (Plastpolymer, Russia) and contained approximately 90 % of the ferroelectric β-phase crystals according to the IR spectroscopy measurements. Aluminum electrodes with a diameter of 20 mm and a thickness of 150 nm were deposited at one surface of the samples by thermal evaporation in vacuum. Non-metallised films were also sometimes used.

Poling was carried out in a corona triode [33] with a bare surface of the sample subjected to a negative corona discharge initiated by a sharpened tungsten electrode. The ions and electrons passed through a control grid, which was held at a constant negative potential in relation to the grounded rear electrode. The polarization field was generated by charges adsorbed on the surface of the sample. The grid was made vibrating to allow simultaneous measurement of the surface potential by the Kelvin method and the DC poling current.

Six combinations with three poling parameters were investigated by maintaining the field strength at two levels (50 MV/m and 100 MV/m), temperatures (25 °C and 85 °C), and electrical mode (constant current and constant voltage). Moreover, we conducted experiments with a multi-layered sample formed from identical films, in which only the lowest film was metallized that was in contact with a positive electrode. Immediately after completion of poling all samples were short-circuited for 15 minutes. The short circuiting was carried out by the non-electrode grounding of the bare sample surface. To do this, polarity of the corona was changed from negative to positive with the simultaneous grounding of the control grid. Thus, the sample was short-circuited, because its upper surface, now bombarded with positive corona discharge ions, received a grid potential equal to the potential of the rear electrode. The duration of the short circuit was long enough to provide a zero field everywhere in the main part of the sample. After the short circuiting, the samples were stored in an open circuit conditions.

Polarization profiles were measured at room temperature using a piezoelectric-induced pressure step (PPS) method. The full description of the method is given elsewhere [29], and only its basic principle is described here. The pressure step is generated by an electrically controlled quartz crystal connected to the sample. Pressure waves propagate at a sound speed (~ 2000 m / s) through a sample in the direction of the thickness causing an electrical signal measured by an oscilloscope with a bandwidth of 1 GHz and then digitized for further processing. It was shown [30; 34] that the reaction of the short-circuit current to the pressure step provides a direct image of the spatial distribution of the piezoelectricity in the sample. It is also known [26] that piezoelectric coefficients in ferroelectric polymers are proportional to the level of the residual polarization. Thus, the magnitude of the current at any time was proportional to the residual polarization in the corresponding point of the sample. Therefore, the measured signal was calibrated directly in polarization units. A 23 μm polypropylene film was inserted between the sample and the measuring electrode to reduce the input capacitance. To obtain reliable data, we measured the polarization profiles twice on each sample.

We found that the residual polarization is distributed non-uniformly in P(VDF-TFE) films under constant current conditions, regardless of temperature, as can be seen from Fig. 10 and 11. The polarization peaks in the samples poled at room temperature are shifted to the positive side leaving almost half of the thickness not polarized (Fig. 10). In samples heated to 85 ˚C, the peak is higher and closer to the positive surface than at room temperatures (Fig. 11). Another small peak is observed near the surface bombarded with corona discharge ions in all samples poled by the constant current, as shown in Fig. 10 and 11.

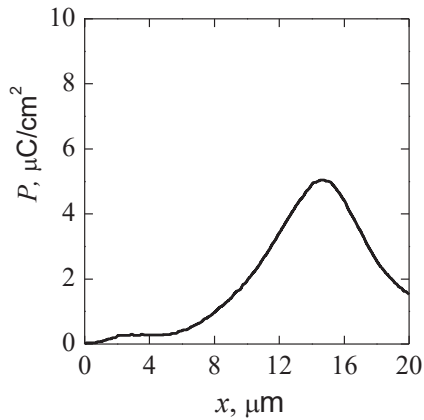

Fig. 10. Distribution of polarization in P(VDF-TFE) films after poling at 25 ˚C and the DC current density of 80 μA/m$^2$ for 15 min. The field at the end of poling was 100 MV/m. The coordinate $x = 0$ corresponds to the sample surface bombarded by negative corona ions

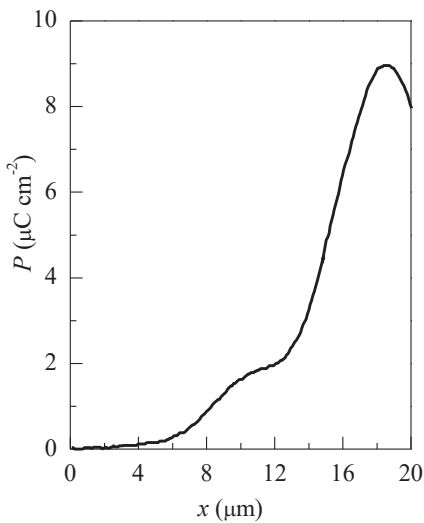

Fig. 11. Distribution of polarization in P(VDF-TFE) films after poling at 85 °C and the DC current density of 160 μA/m² for 15 minutes and cooled to 25 °C in the applied field of 100 MV/m. The $x = 0$ coordinate corresponds to the negative side of the sample

Multilayer samples were poled at constant voltage. The field strength was either moderate (50 MV/m) or high (100 MV/m). The first value was close to the coercive field of PVDF. The polarizing field in the multilayered samples was not the same in two-layer films from which the sample was composed. At the average field of 50 MV/m, only the film having direct contact with the positive rear electrode had residual polarization (Fig. 12). The upper film did not show any residual polarization indicating that the voltage was applied mainly to the lower film.

However, both films are polarized in the case of a high field, as can be seen from Fig. 13 The distribution of polarization in the lower («positive») film is rather uniform (Fig. 13 (b)), whereas in the upper film there are two asymmetric peaks with the higher one located near the surface that was bombarded by the corona ions (Fig. 13 (a)). Three-layer and four-layer specimens were poled in the average nominal field of 50 MV/m. The results shown in Fig. 14 and 15 differ significantly from the results obtained on the two-layered samples (Fig. 12 and 13). Of the three films in the sample, the upper film, which was subjected to the action of ions, was not polarized at all. The distribution of polarization in the film attached to the positive electrode is not uniform and similar to that in the case of constant current poling (Fig. 10 and 11), while in the middle film there are two symmetric peaks separated by a saddle (Fig. 14 (a)) In the case of four films, only two films at the positive side of the sample are polarized, but not uniformly (Fig. 15). Polarization peaks in both films are shifted to the positive side, and the magnitude of the polarization is much higher in the film, which contacts the electrode (Fig. 15 (a) and (b)).

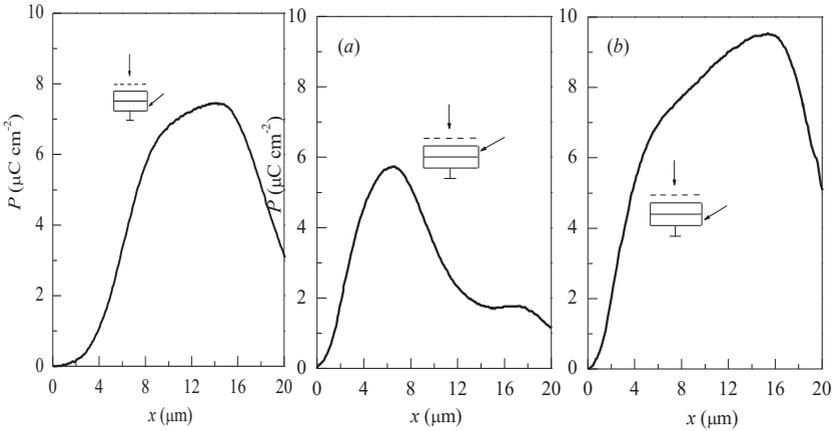

Fig. 12. Distribution of polarization in the lower film of a two-layered sample poled at 85 °C in the constant field with the average intensity of 50 MV/m for 15 minutes and cooled to 25 °C in the applied field. Coordinate $x = 0$ corresponds to the negative side of the film. The upper film bombarded by corona discharge ions did not have any residual polarization

Distribution of polarization in the middle and the lowest films of a three-layer sample poled at 85 °C in the average field of 50 MV/m and cooled to 25 °C in the applied field have shown the similar results. Coordinate $x = 0$ corresponded to the negative side of each film. The upper film that was bombarded with corona discharge ions did not have any residual polarization.

From our experiments on multilayer samples, it should also be anticipated that the polarization peak near the positive electrode with a large depleted polarization region near the negative electrode occurs in the case of one thick film. A similar phenomenon was observed in corona poled PVDF films with a thickness of 120 µm [30].

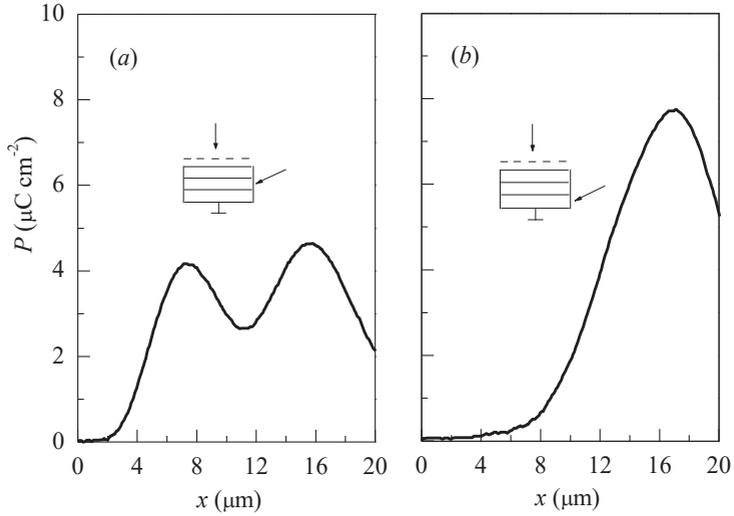

Fig. 13. Distribution of polarization in (a) upper and (b) lower films of a two-layer sample poled at 85 °C with the constant average field of 100 MV/m for 15 minutes and cooled to 25 °C in the applied field. Coordinate $x = 0$ corresponds to the negative side of each film

### Polarization and injection of charges

The heterogeneity of polarization in the direction of thickness in homogeneous specimens may obviously be due to the non-uniform distribution of the applied field. According to the Poisson equation, the inhomogeneity of the field strength $E(x,t)$ is due to the presence of either a real uncompensated charge $\rho(x,t)$ or the polarization charge $dP(x,t)/dx$:

$$\varepsilon_0\varepsilon\left[\partial E(x,t)/\partial x\right] = \rho(x,t) - \partial P(x,t)/\partial x , \qquad (13)$$

where $\varepsilon$ is the dielectric constant, $\varepsilon_o$ is the permittivity of a vacuum, $P$ is the ferroelectric polarization, $x$ is the coordinate in the direction of the film thickness, $t$ is time. Since the polarization $P$ itself depends on the field strength $E$, the initial heterogeneity of the poling field should be attributed only to the effect of real charges.

There are two main sources of the space charge in a dielectric. It can be caused by the spatial separation of already existing intrinsic positive and negative charge carriers, or by injection of charges to the volume from the outside. To show how to use equation (13) to distinguish the effects of injected and internal carriers, we first assume that the external voltage $V$ is applied to a sample of thickness $x_o$, when the density of the injected charges is much lower than that of the intrinsic carriers.

From equation (13) it follows that the field increases near both surfaces of the sample and accordingly decreases in the central region. If the applied field $E_p = V/x_o$ is equal to or close to the coercive value $E_c$, then two peaks of the ferroelectric polarization will appear in front of the electrode sections separated by a non-polarized zone, as shown schematically in Fig. 16 (b).

Now suppose that the same voltage $V$ is applied to another sample where monopolar injection of negative carriers takes place, and their density is much higher than that of the intrinsic charges. The injection charge does not affect the average $E_p$ field, but creates heterogeneity of the field, as shown schematically in Figure 16 (a). The field at the injecting electrode is almost zero, but it increases in the direction of $x$ in accordance with equation (13) until it reaches the $E_c$ value at a certain depth. It is clear that the peak of the residual polarization will be shifted to a positive electrode.

Thus, one can determine the dominant phenomenon from the position of polarization peaks. For example, profiles in Fig. 13 (a) and 14 (a) indicate that the level of injection was low in these samples. However, injection of negative charges in many cases is more important than the separation of the intrinsic carriers. The consequence is visible, for example, in Fig. 10, 11, 14 (b), 15 (a) and 16 (b), in which polarization peaks in all these cases are observed near the positive electrode.

Spatial neutrality inside the sample will be distorted due to the predominant movement of positive charges. The charges injected during poling do not remain there after a short circuiting. They form a spatial charge that corresponds to the slope of the residual polarization profile, since $dP(x)/dx = \rho(x)$. $E(x) = 0$ under short circuit conditions.

The contribution of the space charge to the measured signal cannot be experimentally separated from the residual polarization, but it can be considered as insignificant, since the piezoelectricity in ferroelectric polymers, to which the PPS method is sensitive [30; 34] is caused by the residual polarization, but not by the space charge.

Our results obtained for P(VDF-TFE) films are consistent with the data on the polarization profiles observed in the case of other ferroelectric polymers that have been poled in a corona discharge. The depletion of polarization near the negative side due to the charge injection was detected in PVDF and P(VDF-TrFE) [30−32]. Moreover, a similar phenomenon was observed in ferroelectric polymers, poled by the thermoelectret method [29; 30], by direct application of a high field and the electron-beam polarization.

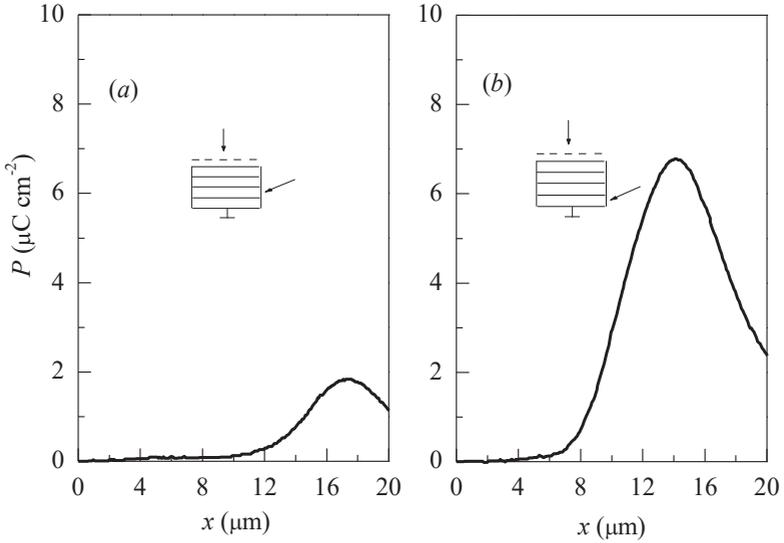

Fig. 15. Distribution of polarization in (a) penultimate lowest and (b) the last lowest film of a four-layer sample that was poled at 85 °C in the average field of 50 MV/m for 15 minutes

The charge injection is most likely appears from a virtual electrode formed on a surface bombarded by electrons and ions. Our results show that homogeneity of polarization is more severely distorted by injection, if low or moderate fields are used. For example, the field in the case of a constant voltage gradually increases from zero to about 100 MV/m, and the resulting polarization distribution is highly heterogeneous (Figs 10 and 11). Almost a quarter of the sample thickness is not polarized, since the field in this area is too low for the formation of the ferroelectric polarization. In the case of two-layer samples poled in the average field of 50 MV/m, charges are mainly injected into the film under action of the corona that causes increase of the film conductivity. We assume that the conductivity corresponds to the following equation

$$g = e\left[n(\mu_+ + \mu_-) + n_+ \mu'_+ + n_- \mu'_-\right], \tag{14}$$

where $e$ is the elementary charge, $n$ is the density of the carriers, $n+$ and $n-$ are injected charge densities, $\mu+$, $\mu-$, $\mu'+$ and $\mu'-$ are mobilities of intrinsic and injected carriers (they may be different). This expression implies that the conductivity increases if injection takes place. As a result, the applied voltage is redistributed, so that its main part is applied to a film attached to the positive electrode. The distribution of polarization in such a film is quite homogeneous (Fig. 12), although the effect of the negative charge injection is still considered as a thin non-polar layer near the negative side of the sample. The top film was completely not polarized because there was a very low field. Similarly, one film in a three layer and two films in four-layered experiments are also not polarized.

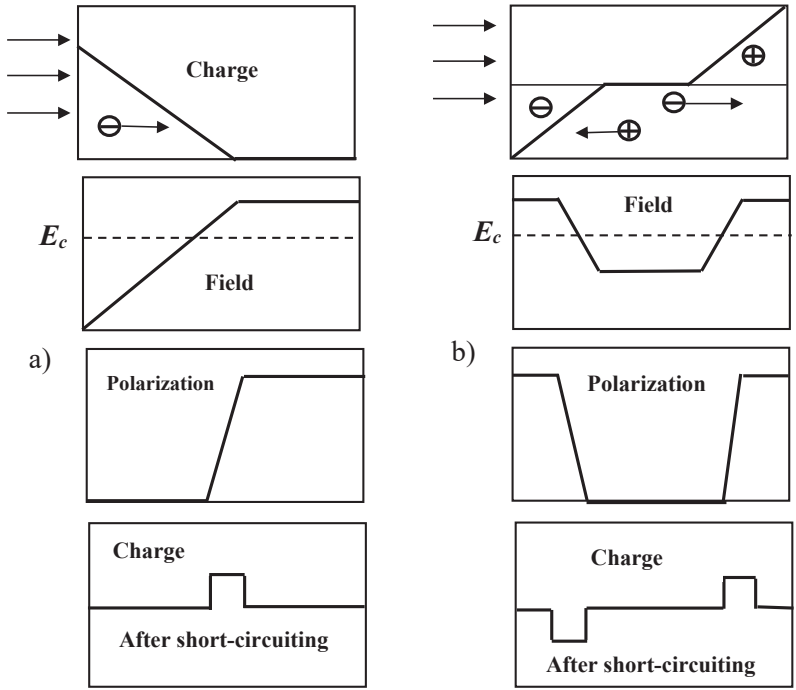

Fig. 16. Schematic diagram showing distribution of volume charge, field strength and ferroelectric polarization during corona poling in the case of (a) monopolar injection of negative charges and (b) separation of internal positive and negative charge carriers. The average field strength is equal to the coercive field $E_c$. Also shown is the distribution of localized charge after completion of poling and short circuiting of the sample

The results of our measurements on P(VDF-TFE) films coincide with the results obtained from multilayered experiments on PVDF films poled by the thermoelectret method [35], but the explanation of this phenomenon is different. The increase of the pyroelectric and piezoelectric activities near the positive electrode was attributed [35] to the effect of the positive charge injection. We believe that, according to the theory of injection currents [25], the heterogeneity of the field and hence polarization is due to injection of the negative charges, but not the positive ones, as previously thought [35].

This is considered normal if the charge is injected either from a real metal electrode, or from a virtual electrode formed on the surface of the sample bombarded by electrons and ions. However, our results indicate that the virtual injecting electrode can also be formed on a surface that was neither metallized nor bombarded by ions. Exhaustion of polarization at the negative side of the samples shown in Fig. 12, 14 (b), and 15 (b) proves that in all these cases, a negative charge is injected.

### Transition zones

It is worth analyzing the behavior of the space charge after the completion of poling. Immediately after a short circuiting, the average field in the sample becomes zero, but the local field still exists. Therefore, mobile charges are redistributed under the action of this field until the field becomes zero at any point of the sample. The characteristic Maxwell relaxation time for this process is given as

$$\tau = \varepsilon_0 \varepsilon / g \, , \tag{15}$$

where $g$ is the explicit conductivity. Considering the typical values of $g = (10^{-11}-10^{-12}$ Sm/m [12]) for PVDF and its copolymers, we obtain $\tau \approx 10-100$ s. The real value of $\tau$ is even lower, since additional carriers are introduced during poling, and the apparent conductivity increases accordingly, as can be seen from equation (14).

From equation (13) it follows that under conditions of equilibrium ($E(x) = 0$), the spatial charge $\rho(x)$ can be localized only at the boundaries of the polarized zones where the derivative $dP/dx \neq 0$.

$$\rho(x) = dP(x) / dx \quad (t > \tau) \, . \tag{16}$$

It is clear from equation (16) that thickness of the transition zone where the polarization decreases from its maximum value to zero, depends on the density of the charge, therefore, the higher the density of the charge, the narrower the transition zone.

The thickness of the transition zone cannot be measured with a high precision by the PPS method, since its resolution (2 µm) is comparable to the thickness of the zones. However, these values can be estimated by compar-ing the growth time of the measured electrical signal and the pressure step. The first in all cases was longer than the last, indicating that the transition zones are thicker than 2 microns. For example, the most delicate transition zones (4–5 μm) are shown in Fig. 12 and 13 (b). The corresponding times of the electrical signal grows and the pressure step are 2–3 ns and 1 ns, respectively.

It is known that any polarization heterogeneity creates a polarization charge with the density of $dP(x)/dx$. This charge creates a depolarizing field, which tends to switch the ferroelectric polarization back to its original state after the completion of poling. The residual polarization can be stable only if the depolarization field is compensated or neutralized. We believe that in the case of the ferroelectric polymers, the compensation is carried out by the spatial charge $\rho(x)$ trapped in the transition zones, by which the polarized part of the sample is separated from the not polarized part. Since the polarization charge $dP(x)/dx$ and the real charge $\rho(x)$ are equal to each other (according to equation (16)), the depolarizing field is completely compensated, so that $E(x) = 0$ everywhere in the sample. We consider the existence of the transition zones in conjunction with compensating spatial charges as a general feature of poled P(VDF-TFE) and, probably, of all other ferroelectric polymers. Presence of the spatial charge in the transition zones is a guarantee of a high stability of the residual polarization.

### Near-to surface regions

We observed two types of polarization profiles in near-to-surface zones of P(VDF-TFE). The residual polarization was zero to a certain depth, as can be seen in Fig. 10, 11, 14 (b), 15 (a) and 15 (b) near the negative surface. In other cases, $P_r = 0$ near the positive surface, as shown in Fig. 10, 11, 12, 13 (b), 14 (b), 15 (b) or near the negative surface in Fig. 12, 13 (a), 13 (b). It is clear that zones of the first type are created due to the massive injection of negative charges during poling, because the field near the injection surface is reduced and the ferroelectric polarization is not formed. The thickness of the not polarized zone depends on the depth of the injected carriers' penetration. The zones are particularly wide in the case of moderate poling fields, as can be seen from Fig. 10, 11, 15 (a) and 15 (b). On the other hand, if the polarizing field strength is high, the near-to-surface zones are very narrow if the negative charges are not injected deeply into the volume (Fig. 12, 13 (a), 13 (b), 14 (a)).

In some cases, the separation of the intrinsic charge carriers dominates over the external injection. Then there are two polarization peaks at the two sample surfaces. Not polarized near-to-surface zones are either very narrow or not observed at all (Fig. 13 (a) and 14 (a)). According to our results, it can be concluded that a certain time is required for the injected charge for deep penetration into the bulk. This can only be done if the pre-poled regions are not polarized, for example in the case of low or moderate electric fields. However, if the ferroelectric polarization is already formed near the surface, as in the case of a high field, the injected charge cannot easily pass through a polarized region. It seems that the effective conductivity of the polarized regions is much lower than that of not polarized ones. The second type of near-to-surface zones with rather high polarization can be seen near a metallized surface attached to a positive electrode during poling. The role of a positive electrode in the accumulation and distribution of polarization has been widely discussed since the discovery of the inhomogeneous distribution of piezoelectricity and pyroelectricity in PVDF [35], and many contradictory explanations of this phenomenon were proposed. Our measurements show that the maximum polarization appears in all metallized samples near the positive electrode. This means that the conditions are favorable both for the rapid development of ferroelectric polarization, and for its stabilization. Positive charges are either not injected or deeply trapped very closely to the surface creating good conditions for compensating the depolarizing field. At the same time, the trapped charges do not allow the attachment of a highly polarized zone directly to the electrode.

The PPS method with a resolution of about 2 μm cannot provide more information about the fine structure of near-to-surface zones, but this can be achieved by using the LIMM method [36]. Polarization profiles in uniformly electrified P(VDF-TFE), measured by this method, are presented in Fig. 17. We used the same specimens as those for which the results obtained by the PPS method are shown in Fig. 13 (b). It is known that the resolution of the LIMM method is about 0.1 μm near the illuminated electrode [37].

As can be seen from Fig. 17, the polarized zones are not directly attached to the positive and negative surfaces. The transition zones consist of thin layers where the polarization drops from maximum to zero and is supplemented by a completely not polarized layer of about 0.5 μm thickness.

The applied voltage is distributed non-uniformly between layers in the case of multilayered samples poling. Only the film near the positive electrode shows a high and fairly uniform polarization, while the upper films remain not polarized indicating that the injected charge permeates the entire thickness of the film. Therefore, to obtain a uniformly polarized P(VDF-TFE) copolymer film in a moderate field, it would be advisable to cover the main sample during poling with another auxiliary film. In the case of high poling fields, the residual polarization is homogeneous, since injections of charges are suppressed. But even in this case, the polarized part of the volume is separated from the sample surfaces by transition zones where compensating charges are trapped. A thin layer of about 0.5 µm thickness always remains completely not polarized near the sample surfaces.

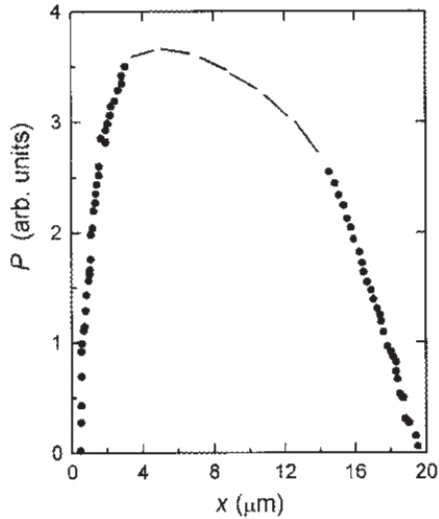

Fig. 17. Distribution of polarization measured by the method of the modulated intensity of laser radiation in near-to-surface regions of a nominally well-poled P(VDF-TFE) film. The conditions for poling were the same as for the sample shown in Fig. 5. Coordinate $x = 0$ corresponds to the negative side of the film

## 6. Effect of temperature on distribution of ferroelectric polarization

Recently, it has been shown that a high stability of the residual polarization in PVDF and P(VDF-TrFE) is due to interaction of the polarization with the injected charge trapped at the boundaries of crystallites or macroscopic polarized regions [26].

In both cases the polarization and the space charge form a stable and a self-consistent system in which the latter plays a decisive role. Assuming Debye's approximation for relaxation and the continuous distribution of activation energies for the charge trapping, Eisenmenger et al. received such a distribution for PVDF and P(VDF-TrFE) [38].

Our purpose was to find out how the bulk charge affects the thermal stability of the residual polarization in P(VDF-TFE) copolymer which also belongs to the class of the ferroelectric polymers, but is much less studied than PVDF and P(VDF-TrFE).

To do this, we measured the polarization profiles in P(VDF-TFE) samples as a function of temperature by performing the linear heating from 20 °C to the melting point of crystallites. The activation energy was calculated by applying our experimental data and the theoretical model proposed in [38]. The obtained results were compared with those that are known for PVDF and P(VDF-TrFE) copolymer.

The samples were cut from experimental batches of P(VDF-TFE) 20 μm thick copolymer films containing more than 90 % of the ferroelectric β-forms in the crystalline phase. Aluminum electrodes with a diameter of 5 mm and a thickness of 0.15 μm were deposited on both sides of the samples by thermal evaporation in vacuum. Poling was carried out either by direct application of high voltage (3.2 kV at 20 °C for 2 min) or by thermoelectret method (2.5 kV at 85 °C for 10 min and fast cooling to 20 °C under the applied voltage).

The polarization of the field in both cases was three to four times greater than the coercive field, which is 35–40 MV/m in P(VDF-TFE). The residual polarization profiles in the direction of the sample thickness were measured with a repetition rate of about 100 Hz using the PPS method, while the temperature was linearly increased at a rate of 3 K/min from 20 °C to the melting point of the crystallites, which turned out to be $134 \pm 2$ °C.

It is established that the distribution of polarization in the thickness direction is rather uniform, except for areas close to the electrode zones, where the polarization decreases from the maximum to the low value (Fig. 18). The profiles of polarization at all temperatures were slightly asymmetric, with a maximum located near the positive electrode, similarly to that observed in other ferroelectric polymers. We used these maximum values for evaluating the thermal stability of the residual polarization. From the data presented in Fig. 19, it is clear that the polarization in P(VDF-TFE) breaks down with the temperature throughout the studied range almost linearly decreasing from the maximum at room temperature to zero at a melting point (134 ° C). This behavior is significantly different from PVDF and P(VDF-TrFE) where the polarization decreases only when the temperature exceeds 90 °C.

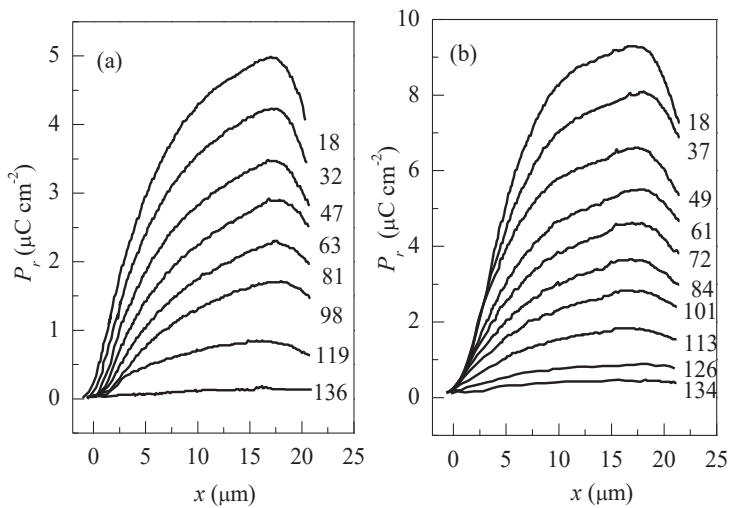

Fig. 18. Spatial distribution of polarization at different temperatures in P(VDF-TFE) films poled (a) by direct application of the high field and (b) by the thermoelectret method. Zero on the axis of the thickness corresponds to the negative surface of the sample during processing

It is advisable to use the Debye approximation for the relaxation of polarization with the temperature dependence of the decay constant corresponding to the Arrhenius law. Then, the current of depolarization $i_a(T)$ in the case of one activation energy is [38]:

$$i_a(T) = -dP / dT =$$
$$= \left[ h \cdot f_0 \exp(-a / T) \right] \cdot P_0 \exp\left[ -h \cdot f_0 \int_{T_0}^{T} \exp(-a / T) dT \right], \qquad (17)$$

where $h$ is the heating rate, $f_o$ is the proper frequency, $P_o$ is the initial value of polarization at $T_o$, $a = A k$, where $A$ is the activation energy, $k$ is Boltzmann's constant. Taking into account that energy is continuously distributed on the surface of the polarized crystallites surrounded by a disordered amorphous phase, one can conclude that the energy spectrum of traps is, most likely, continuous, rather than discrete.

Therefore, the total depolarization current $i(T)$ is a superposition of all relaxation components:

$$i(T) = \int_0^\infty g(a) i_a(T) da , \qquad (18)$$

where *g(a)* is the distribution function of activation energies. It was shown that the depolarization current *i(T)* calculated from the experimental curve *P(T)* is an image of the distribution function *g(a):*

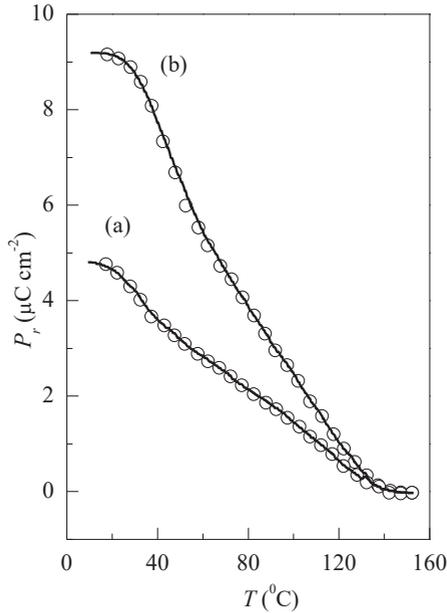

Fig. 19. The value of residual polarization in P(VDF-TFE) obtained experimentally (points) and theoretically (solid lines) depending on (a) poling in a high field and (b) b the thermoelectret method

The results of calculations based on experimental data of Fig. 19 are shown in Fig. 20. It is clear that the high-temperature behavior of the P(VDF-TFE) copolymer is regulated by two relaxation processes characterized by significantly expanded energy levels. The low-temperature peak in thermoelectret samples is slightly shifted to lower energy, whereas the high-temperature peak does not affect by the heat treatment. The relationship between the values of the two peaks in P(VDF-TFE) differs from PVDF [38] where the second peak is more advanced than the first one. Comparing the curves in Fig. 20 (a) and 20b, one can see that there is no significant difference in values and distribution of the activation energies in the samples polarized in a high field strength and by the thermoelectret method. This indicates that the residual polarization in P(VDF-TFE) is not thermally frozen, as in the case of ordinary polar thermoelectrets, but it is stabilized, most likely, by the field of the trapped charges. P(VDF-TFE) has two components of polarization, namely: a ferro-electric component and an electret component. The first one is concentrated in the crystalline phase, and the second is localized in the amorphous phase. Therefore, the two peaks shown in Fig. 20 can be related to the relaxation of these polarization components. A similar behavior was observed in PVDF and P(VDF-TrFE) copolymer [38] indicating that this phenomenon is likely to be common in the whole class of the ferroelectric polymers.

$$i(T) \propto g(mT) \,, \tag{19}$$

where $m$ is a constant value.

It is known that the ferroelectric polarization is stable only when the de-polarizing field is somehow neutralized or compensated. In the ferroelectric polymers, this compensation is performed by trapped charges [38].

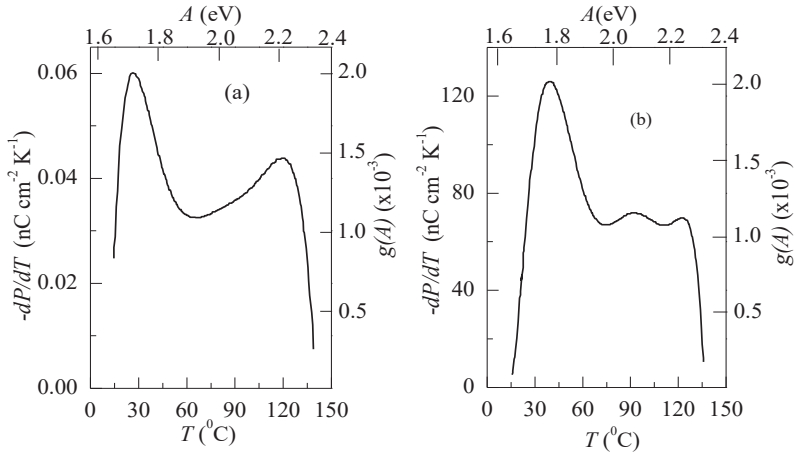

Fig. 20. The temperature dependence of the depolarization current i (T) = dP / dT and the distribution function of the activation energy in P(VDF-TFE) calculated depending on (a) poling in a high field and (b)

Since the charge is deeply trapped and the binding energy is in the range of 1.65–2.35 eV, as can be seen from Fig. 20, the spatial charge effectively compensates the depolarizing field in the crystals. This explains the high polarization stability at room temperature. According to our calculations, based on the data of Fig. 20, polarization is expected to decrease to 90 % of its initial value for about one year.

The model of the continuous distribution of activation energies was verified by calculating the dependence $P(T)$ by using the data given in Fig. 20. The results of the calculations shown in Fig. 19 by solid lines agree with the experimental data indicating that the thermal stability of the residual polarization in P(VDF-TFE) and, probably, in other ferroelectric polymers, is indeed controlled by the trapped volume charge.

**Conclusion**

Results of experimental study of the polarization distribution uniformity in ferroelectric polymer films over the thickness of the samples are presented in this article presents. We selected typical polymeric ferroelectrics, namely, polyvinylidene fluoride (PVDF) and its copolymer with trifluoroethylene P(VDF-TFE) as objects of the study. The measurements were carried out by a modern sensitive piezoelectric generated pressure step method.

It has been found that the polarization uniformity substantially depends on the value of the applied electric filed during initial poling of the films. the distribution of polarization was inhomogeneous with a maximum near the positive electrode in the case of weak and medium applied fields close to coercive value of the field.

A very important feature has been discovered. It appeared that the uniformity of polarization cannot be improved even by the subsequent application of very strong fields.

However, if the initial poling of a fresh sample has been carried out in strong fields, then the uniform polarization distribution has been formed. The features of the P(VDF-TFE) copolymer electrified in corona discharge have been investigated as well. Phenomenological models of the processes occurring during the formation of polarization in ferroelectric polymer films have been developed for clarifying physical processes responsible for the formation of the polarization distribution.

## Розділ II

## АВТОМАТИЗАЦІЯ ТА УПРАВЛІННЯ ТЕХНОЛОГІЧНИМИ ПРОЦЕСАМИ

## ТЕХНОЛОГІЧНИЙ РОЗВИТОК СУДНОПЛАВСТВА, СИСТЕМ ШВАРТУВАННЯ СУДНОПЛАВСТВА МАЙБУТНЬОГО

*Пунченко Н. О., Цира О. В.*

*Водний транспорт був і існує як провідний елемент системи світової економіки. Існує постійне збільшення морського та річкового перевезення, а також вимоги до якості вантажного перевезення шляхом водного транспорту (своєчасність, безпека, надійність), які змінюються у напрямку поліпшення. Розвиток відбувається з метою використання автономних систем, які є однією з найбільш вагомих змін, що спостерігаються в морській промисловості. У роботі наведено огляд систем управління безекіпажними суднами (кораблями), де людський фактор не впливає на рішення, що приймаються. Представлені переважаючі складові інтелектуальної системи. Наведено види автоматизованих систем швартування: лазерні, вакуумні, магнітні. Системи швартування знижують ризик перевищення швидкості судна за рахунок зниження впливу людського чинника при швартуванні, знижують вплив оцінки поточної ситуації зближення судна з причалом, обирають режими і наочно відображають робочий процес за визначеною ситуацією, підвищуючи тим самим ефективність системи в цілому.*

*В результаті огляду зрозуміло, що у світі немає галузі економіки, яка в останні роки не вплинула б на цифрову трансформацію, а телекомунікаційні компанії зробили достатньо, щоб розширити цей спектр послуг. Тим часом у морській промисловості у світі цифрова трансформація робить лише перші кроки. Основою для цього є канали зв'язку, без яких передача даних у принципі неможлива. На воді це завдання є найбільш складним, оскільки волоконно-оптичний кабель не може бути підведений до судна. Тому на річці та морі є необхідним супутниковий зв'язок, існує суттєва потреба його використання не тільки для комунікацій та цифрових розваг на борту, а й для моніторингу стану судна та вантажу, можливості дистанційного управління, контролю бункерних суден, питання безпеки. Звідси випливає, що безекіпажний флот буде використовувати інтегровані автономні системи управління, які можуть керуватися оператором на березі.*

*Water transport has been and exists as a leading element of the world economy. There is a constant increase in sea and river transport, as well as requirements for*

*the quality of freight transport by water transport (timeliness, safety, reliability), which are changing in the direction of improvement. The development is taking place with the aim of using autonomous systems, which is one of the most significant changes observed in the maritime industry. The paper provides an overview of control systems for unmanned vessels (ships), where the human factor does not affect the decisions made. The predominant components of the intellectual system are presented. The types of automated mooring systems are given: laser, vacuum, magnetic. Mooring systems reduce the risk of over-speeding of the vessel by reducing the influence of the human factor during mooring, reduce the impact of assessing the current situation of the ship's approach to the berth, select modes and visually reflect the working process according to the current situation, thereby increasing the efficiency of the system.*

*As a result of the review, there is no industry in the world that has not affected digital transformation in recent years, and telecommunications companies have done enough to expand this range of services. Meanwhile, in the marine industry, in the world, digital transformation is only taking its first steps. The basis for this is communication channels, without which data transmission is, in principle, impossible. On the water, this task is most difficult because the fiber-optic cable cannot be connected to the vessel. Therefore, on the river and sea, satellite communication is necessary, its use not only for communications and digital entertainment on board, but also for monitoring the condition of the vessel and cargo, remote control capabilities, control of bunker vessels, security issues. It follows that the unmanned fleet will use built-in autonomous control systems that can be operated by the operator ashore.*

Для програми розвитку перспективних шляхів підвищення загальної безпеки мореплавання, в умовах зростання інтенсивності морського судноплавства спостерігаються тенденції збільшення кількості смертельних випадків від морських аварій [2]. Однією з причин цього є людський чинник. Оскільки в інноваційному суспільстві така галузь як судноводіння при зародженні визначила себе як інноваційна. Такому визначенню є підтвердження, а саме група MariNet, яка створена в рамках Національної технологічної ініціативи. Група змогла об'єднати великі компанії і невеликі стартапи у галузі морських високих технологій, наукові центри, офіційні органи і внз [3], що представлені соціуму як інтелектуальні автономні системи, які є радикальними змінами в судноплавній індустрії. Це інтелектуальні системи, які приймають рішення без втручання ззовні. Інтелектуальні системи стали базисом для створення такого напрямку як безекіпажне судноводіння, де використовується комбінація дистанційного й автономного управління, яке зводить до мінімуму людський чинник у безпеці судноводіння.

Основи теоретичних та практичних наукових досліджень у галузі інформаційних технологій та систем судноплавства дуже детально представлені в роботах таких вчених: А. Е. Сазонова (математичне та програмне забезпечення автоматизованих систем управління суднами), С. В. Смоленцева (основи будівельних систем інтелектуального управління), С. В. Руда (системи моніторингу та управління суднами технічного та допоміжного флоту), І. Г. Малюгіна, В. І. Комашинського (питання будівництва транспортних систем), Д. А. Скорошодова (інтегровані системи управління судном), А. А. Сикарева (стійкі системи радіозв'язку), А. А. Диди (складні системи), а також іноземних вчених, таких як Ch. Liu (автоматизовані системи управління кораблем), М. Хойхтяя (автономні системи управління, супутникові зв'язки), Е. Топп (дистанційне управління системами), Р. Польвара (системи технічного бачення) та інші. У своїх роботах автори заклали теоретичну та практичну базу, яка сприяє підвищенню ефективності функціонування та розвитку водного транспорту.

Вважається, що судна без екіпажу будуть дешевшими, безпечнішими і будуть менше забруднювати навколишнє середовище. А ідея і обгрунтування ідей їх створення базується на кількох основних положеннях:

Положення 1 — за відсутності екіпажу вартість підтримки судна може бути зменшена на 30—40 %;

Положення 2 — за відсутності екіпажу макет і архітектура судна значно спрощуються, що тягне за собою зменшення вартості будівництва та обслуговування судна;

Положення 3 — вартість підготовки фахівців судна може бути значно зменшена;

Положення 4 — досягнення науки та технологій свідчать, що з технічної точки зору фундаментальних обмежень завдання не має;

Положення 5 — розробка та реалізація забезпечення кораблів належить до нового напрямку науки і техніки, який об'єднує та сприяє зростанню творчої діяльності наукових, дизайнерських, освітніх та промислових організацій галузі;

Положення 6 — будівництво та реалізація військових суден, здається, є довгим багатоступеневим процесом, що включає поступову зміну структури флоту, яка складається з традиційних суден, що обслуговуються екіпажами, і що є контрольованими віддалено або повністю автономно заданою програмою.

Перевага — це давня професія, яка має багато століть. Зрештою, в доісторичні часи люди подорожували до інших берегів на човнах, таким чином поступово зміцнюючись на землі. Основи доставки стародавніх артефактів були на практиці, тому що у них не було сучасної теоретичної бази, підручників та карт. Першими людьми-навігаторами були фінікійці, знання яких починають свій шлях з 15 століття, наприклад, лоцманом Васко да Гамы Ибн Маджид був накопичений досвід: кожен, хто хоче впоратися з елементами моря, повинен розбиратися в румбах та фазі місяця, відстанях та напрямках.

Величезна частина нашої планети покрита водою. Тому можна впевнено сказати, що морський транспорт ніколи не стане застарілим, незалежно від того, як розвивається наземне та повітряне обладнання. Щоб подолати моря та океани, потрібні грамотні судноводії, яким добре відомі пристрої їхнього судна, а також характер нестійкого елемента води.

В даний час безпілотні судна відповідно до міжнародних конвенцій є незаконними. Щоб підтвердити таке твердження, ми дамо невелику оцінку міжнародно-правовим актам, які встановлюють вимоги до стану транспорту та флоту, а також його експлуатації.

Найбільші зусилля в цьому напрямку були зроблені Міжнародною морською організацією та Міжнародною організацією праці. Починаючи з сорокових років, ці організації розробили цілий ряд міжнародних конвенцій безпеки. При розробці і застосуванні зазначених конвенцій функції управління залежать від двох відповідних вимог. Перш за все, це стан плавання судна в залежності від прапора. Другий координуючий орган повинен бути спеціалізованою організацією — класифікаційним товариством. Відповідальність за виконання вимог конвенції покладається на власника судна. Однак практика показала, що всі суворі вимоги до суден були розроблені міжнародними організаціями, ці вимоги не були виконані. Держава прапора з одного боку, зацікавлена особа і намагається забезпечити судновласників найбільш сприятливими умовами праці з економічної точки зору, з іншого боку, держава прапора не завжди має можливість здійснювати ефективний контроль стану її судна і його роботи. Відповідно до вищесказаного був введений нагляд з боку класифікаційних товариств різних країн. Для цієї мети був розроблений Міжнародний кодекс з управління безпекою (ISM-Code). Його головна мета полягає в тому, щоб забезпечити безпеку на воді.

***Міжнародні організації, що регулюють безпеку судноплавства***

Після Другої світової війни і створення ООН суспільство прийшло до висновку про необхідність впровадження авторитетної міжнародної організації в області безпеки судноплавства. І саме такою організацією в 1948 році стала Міжнародна морська консультативна організація. У 1973 році організація отримала назву Міжнародної морської організації (IMO). Вона функціонує в рамках ООН. Її штаб-квартира знаходиться в Лондоні.

Міжнародні організації безпеки судноплавства:

ILO — Міжнародна організація праці;

ICF — Міжнародна палата доставки;

ISF — Міжнародна федерація судновласників;

INSA — Міжнародна асоціація судновласників;

МАК — Міжнародна асоціація класифікаційних товариств;

IALA — Міжнародна асоціація маячних служб;

IPH — Міжнародна асоціація портів та гаваней;

MCE — Міжнародний телекомунікаційний союз;

CIRM — Міжнародний комітет морського радіозв'язку;

IMPA — Міжнародна асоціація морських лоцманів;

IFSMA — Міжнародна федерація асоціацій морських капітанів;

ICFTU — Міжнародна конфедерація профспілок вільної торгівлі;

WMO — Всесвітня метеорологічна організація;

ISO — Міжнародна організація стандартизації;

ICAO — Міжнародна організація цивільної авіації;

IMO — Міжнародна морська організація.

Верховний орган Міжнародної морської організації — Асамблея, яка регулярно відбувається кожні два роки і в якій беруть участь всі члени організації. Рада проходить між зборами. Виконує функції Асамблеї та має право надавати урядам рекомендації щодо безпеки доставки та запобігання забрудненню. Крім того, Рада організації має такі функції:

1) координація всіх органів організації;

2) надання робочих програм та бюджету для затвердження Асамблеєю;

3) приймає і удосконалює обов'язкові до виконання і рекомендаційні міжнародні конвенції, кодекси, резолюції, протоколи, циркуляри і рекомендації;

4) звернення Генерального секретаря до Асамблеї;

5) запрошення та організація зустрічі з представниками інших організацій для участі в Асамблеї.

Призначення до Ради Міжнародної морської організації відбувається відповідно до правил:

а) десять країн, що мають найбільший інтерес до міжнародної навігації;

б) десять країн, що демонструють найбільший інтерес до міжнародної морської торгівлі;

в) двадцять країн, які не включені до категорій а) та б), і ті, хто зацікавлені морським транспортом або навігацією та представляють всі географічні райони світу.

Комітети Міжнародної морської організації.

На постійній основі працюють такі комітети:

1-й Комітет з питань безпеки (MSC — Maritime Safety Committee) є найважливішим у Міжнародній морській організації, він включає всіх членів організації.

2-й Комітет з охорони навколишнього природного середовища (MEPC — The maritime Environment Protection Committee) включає всіх членів організації. Він був організований як збори, а в 1985 році отримав повний правовий статус незалежного.

Ці два комітети допомагають у роботі 9 підкомітетів:

— BLG — транспортування рідини та газів;

— DSC — перевезення небезпечних та загальних вантажів та контейнерів;

— FP — захист від пожежі;

— COMSAR — радіозв'язок, пошук та порятунок;

— Nav — безпечна навігація;

— модернізація та обладнання кораблів;

— SLF — стабільність, вантажний бренд, безпека риболовних суден;

— STW — стандарти навчання та навігаційні вимоги;

— FSI — впровадження держави.

3-й Юридичний комітет (Legal Committee) включає всі країни-члени Міжнародної морської організації.

4-й Комітет з технічного співробітництва (Technical Co-operation Committee) — включає всі країни-члени Міжнародної морської організації.

5-й Комітет формальностей (Facilitation Committee) — відкритий для всіх країн-членів Міжнародної морської організації.

Організовано в 1972 році як дочірнє відділення Ради з полегшення формальностей у міжнародній навігації. У 1991 році видав Додаток до

Конвенції Міжнародної морської організації як стандартного комітету. Однак додаток ще не набув чинності.

Секретаріат Міжнародної морської організації — складається з зовнішніх співробітників та 300 співробітників головного офісу в Лондоні.

2. IACS Міжнародна асоціація класифікаційних суспільств.

Міжнародну асоціацію класифікаційних суспільств організовано за рекомендацією Конвенції про вантажні морської конвенції 1930 року. У 1968 році вона отримала консультативний статус у Міжнародній морській організації як неурядової організації.

Члени Міжнародної асоціації класифікаційних товариств:

ABS — Американське бюро доставки;

BV — Бюро Верітас (Франція);

CCS — Китайська спілка класифікації;

DNV — Det Norske Veritas (Норвегія);

GL — Німецький Ллойд;

KR — Корейський реєстр суден;

LR — Регістр судноплавства Ллойда (Англія);

NK — Ніппон Кайджи Кіокай (Японія);

RINA — Італійський морський регістр;

RS — Російський морський реєстр судноплавства.

Асоційовані члени:

CRS — Хорватське судноплавства;

IRS — Індійський реєстр судноплавства.

Основні цілі:

— забезпечення безпеки людського життя в морі;

— забезпечення плавання в безпеці;

— забезпечення надійного перевезення вантажів морем та внутрішніми водними шляхами;

— запобігання забруднення навколишнього середовища.

Для досягнення цих цілей правила, засновані на наукових дослідженнях, розробляють та вдосконалюють вимоги міжнародних конвенцій та кодів.

3. ITU — Міжнародний телекомунікаційний союз. 24 травня 1844 р. Самуїл Морзе надіслав перше повідомлення телеграфною лінією між Вашингтоном та Балтімором. 17 травня 1865 р. Перша Міжнародна телеграфна конвенція була підписана в Парижі між 20 країнами. На підставі цієї Конвенції було створено Міжнародний телеграфний союз. Через заснування союзу ця конвенція зазнала значних змін — в 1876 році увійшов телефон, у 1896 р. — радіолокація. Зміни включені до Конвенції 1903 року. У 1906 р. Перша радіотелеграфна конвенція була підписана в Берліні на першій конференції з радіотехніки. У Додатку до цієї Конвенції прийняті перші правила для радіозв'язку, які пізніше були оновлені і тепер відомі як «Правила переміщення». У 1920 році почалися перші радіопередачі, а в 1927 році міжнародні радіослужби схвалили розподіл частот та правил між різними країнами та користувачами радіозв'язку. У 1932 році на Мадридській конференції були поєднані міжнародні конвенти. Об'єднана конвенція була названа міжнародною конвенцією телекомунікацій. Також з 1 січня 1934 року союз був перейменований на Міжнародний телекомунікаційний союз (МСЕ). У 1947 році на МСЕ (ITU) вступив до Егіди ООН, а головний офіс організації був переведений з Берна до Женеви.

У 1963 році був запущений перший супутник Synk-1, і почалася епоха супутникового зв'язку. На позачерговій конференції в Женеві було представлено комунікацію з космічного спілкування, де було передбачено не тільки частоти супутникового зв'язку, а й орбіти супутників зв'язку. У 1992 році ця комунікація була доповнена у зв'язку з новими особливостями цифрового зв'язку та використанням негеостаційних супутників. На пленарному засіданні додаткової конференції ITU була реорганізована. Тут трьома секторами стали: — телекомунікаційна стандартизація (ITU-T), — радіозв'язок (ITU-R), — розробка телекомунікацій (ITU-D). На Кіотській конференції 1994 року був затверджений перший стратегічний план розвитку союзу та комунікації у світі. Крім того, на цій конференції організовано глобальний телекомунікаційний політичний форум (WTPF) для вирішення політичних питань між країнами зв'язку. Перший WTPF пройшов у Женеві у 1996 році з питань глобальних мобільних супутникових комунікацій, а другий — у 1998 році для надання телекомунікаційного обслуговування.

4. ILO — Міжнародна організація праці.

Міжнародна організація праці була заснована в 1919 році на базі Версальського договору в інтересах соціальної справедливості. Ця організація має структуру тризіркової моделі, яку представляють уряди, роботодавці та працівники (профспілки). Цілі та завдання Міжнародної організації праці були підтверджені в Декларації у Філадельфії, прийнятій Конференцією Міжнародної організації праці в 1944 році. Декларація містить принципи:

— робота не є товар;

— свобода думки та право Асоціації є важливим елементом для підтримки процесу;

— бідність в будь-якому місці створює небезпеку для загального процвітання;

— всі люди незалежно від раси, переконань та статі мають право шукати матеріального добробуту та духовного розвитку в контексті поваги до свободи та гідності, економічної підтримки та рівних можливостей.

У 1946 р. Міжнародна організація праці стала першою спеціалізованою організацією, яка взаємодіє з ООН. У 1969 р. Міжнародна організація праці була нагороджена Нобелівською премією миру. Перша конференція Міжнародної організації праці відбулася у жовтні–листопаді 1919 року у Вашингтоні. Вона прийняла шість рекомендацій та 8 конвенцій, включаючи № 1 про 8-годинний робочий день.

Спочатку організація включала 42 держави, зараз — 175. Конференції проводяться щорічно. Кожні два роки конференція приймає дворічну програму та бюджет. Між конференціями адміністративна рада включає 28 представників урядів, 14 від працівників та роботодавців. Секретаріат та штаб-квартира розташовані в Женеві. Основні стратегічні цілі:

— розробка та впровадження норм та фундаментальних принципів та трудових прав;

— створення більш широких можливостей для жінок та чоловіків, щоб забезпечити гідну зайнятість;

— розширення охоплення та підвищення ефективності соціального захисту для всіх;

— зміцнення тристоронньої структури та підтримки соціального діалогу.

Способи досягнення цілей:

— розробка міжнародних заходів та програм для полегшення реалізації фундаментальних прав людини, вдосконалення роботи та життя, розширення можливостей зайнятості;

— розробка міжнародних стандартів праці (підтримка унікальної системи контролю за їх застосуванням), яка служить керівним принципом національних органів у здійсненні цих заходів;

— комплексна програма міжнародного технічного співробітництва, розроблена та впроваджена з активним партнерством із засновниками, щоб допомогти країнам у здійсненні цих заходів;

— підготовча, освітня та видавнича діяльність, сприяння реалізації всіх цих зусиль.

З 1919 р. Міжнародна організація праці прийняла 184 Конвенції та 194 рекомендації.

Основні міжнародні конвенції безпеки судноплавства.

Перший досвід у створенні міжнародних домовленостей виник на підставі правил запобігання зіткненням, що з'явилися на початку XIX століття. Пізніше, з розвитком флоту та глобальними перевезеннями, вони неодноразово переглядалися. Остання конвенція MPPSS-72, затверджена 20 жовтня 1972 року, набрала чинності лише 15 липня 1977 року.

Після трагічної смерті пасажирського лайнера «Титанік» була прийнята перша міжнародна конвенція про захист людського життя на морі 1914 року, потім 2-га конвенція про захист людського життя на морі — була прийнята в 1929 році, 3-тя у 1948 році, 4-ту прийнято 17 червня 1960 року — набрала чинності 26 травня 1965 року. Тепер чинна Конвенція Solas-74/78.

Вперше про забруднення навколишнього середовища світова спільнота підіймає це питання у другій половині XX століття. У 1954 р. з ініціативи Великобританії була проведена конференція з нафтового забруднення. Вона прийняла Конвенцію OILPOL-54, яка вступила в дію 26 липня 1958 року. Ця конвенція охоплювала два основних напрямки світового судноплавства, стосується тільки забруднення нафтою та її компонентами, корекція пройшла в 1962, 1969 і 1971 роках.

Після катастрофи танкера «Торрі Каньйон» в Ла-Манші в 1967 році світове співтовариство, оцінюючи суму збитку (попадання близько 120 000 тонн нафти в море), прийнли ряд різних міжнародних конвенцій:

Основні міжнародні конвенції з безпеки і запобігання забруднення навколишнього середовища:

СОЛАС-74/88;

SOLAS-74/88.

Міжнародна конвенція з охорони людського життя на морі 1974 з додатковим Протоколом 1988 року:

МАРПОЛ-73/78;

MARPOL-73/78.

Міжнародна конвенція з запобігання забруднення з суден 1973 року з додатковим Протоколом 1978 року:

ПДНВ-78/95;

STCW-95.

Міжнародна конвенція про підготовку і дипломування моряків та несення вахти 1978 року з Кодексом 1995 року:

МППСС-72;

COLREG-72.

Міжнародні правила запобігання зіткнення суден на морі — 1972 року:

САР-79;

SAR-79.

Міжнародна конвенція для збереження і порятунку — 1979 року:

КГМ-66/88;

LL-66/88.

Міжнародна конвенція про вантажну марку — 1966 року зі змінами 1988 року:

ФАЛ -65;

FAL-65.

Конвенція про формальності в Міжнародних морських перевезеннях вантажу — 1965 року:

КНА-88;

SUA-88.

Конвенція про боротьбу з незаконними актами, спрямованими проти безпеки морського судноплавства — 1988 року:

КСИ-89;

SALVAGE-89.

Міжнародна Конвенція про порятунок майна — 1989 року:

КГО -69;

CLC-69.

Міжнародна конвенція про цивільну відповідальність за шкоду, заподіяну забрудненням нафтою — 1974 року.

Конвенція СОЛАС-74/88 (SOLAS-74/88).

Конвенція була прийнята 1 листопада 1974 року на Міжнародній конференції з охорони людського життя на морі, а протокол до неї 10 листопада 1988 року на міжнародній конференції з гармонізованої системи експертизи та сертифікатів. Також 11 листопада 1988 року було прийнято низку виправлень до SOLAS -74.

Комітет з питань безпеки на морі постійно працює над покращенням та вдосконаленням SOLAS-74/78, внесення змін до неї.

Конвенція СОЛАС (SOLAS) та протокол 1988 року до неї були підписані англійською, іспанською, китайською, російською та французькою мовами, а всі тексти рівноцінні. Офіційною мовою за-лишається англійська, отже з розбіжностями англійська версія приймається як основа.

Конвенція СОЛАС (SOLAS) спочатку мала 8 розділів:

I — загальні положення

II-1 — конструкція — поділ на відсіки та стійкість, механічні та електричні установки

II-2 — конструкція — захист від вогню, виявлення та пожежогасіння. Нове видання було прийнято у грудні 2000 року, набрало чинності 1 липня 2002 року.

III — рятувальні інструменти та пристрої. Нове видання в 1996 році набрало чинності з 1 липня 1998 року.

IV — радіозв'язок. Нове видання затверджено в 1988 році. Введення чинності через запровадження ГМССБ з 1 лютого 1992 по 1 лютого 1999 року.

V — безпека мореплавання. Нове видання було затверджено в грудні 2000 року, набрало чинності 1 липня 2002 року (стосується AIC).

VI — перевезення вантажів.

VII — перевезення небезпечних вантажів. Затверджено Міжнародною морською організацією у 2002 році, набрала чинності з 1 січня 2004 року (була підсумована Кодексом перевезення небезпечних вантажів морем — UMDG-code).

VIII — ядерні судна — затверджені Асамблеєю Міжнародної морської організації в 1981 році.

Наступні глави були додані пізніше:

IX — управління безпечною експлуатацією суден. Затверджено в травні 1994 року, набуло чинності 1 липня 1998 року.

X — заходи безпеки для високошвидкісних суден. Затверджено в травні 1994 року, набуло чинності 1 січня 1996 року.

XI — спеціальні заходи щодо підвищення безпеки на морі. Затверджено в травні 1994 року, набуло чинності 1 січня 1996 року.

XII — додаткові заходи безпеки для навалочних суден. Затверджена в листопаді 1997 року, набуло чинності 1 липня 1999 року.

Основна мета конвенції SOLAS полягає в об'єднанні і зменшенні кількості стандартів для будівництва, обладнання та безпечного управління морських суден. Головне управління з виконання вимог Конвенції лежить на урядах країн, під прапором яких кораблі плавають. Конвенція СОЛАС також зобов'язує держави контролювати судна в портах навігації (Правило 4 глави XI).

MARPOL-73 / +78.

Міжнародна конференція з запобігання забруднення моря, скликана Міжнародною морською організацією в 1973 році, прийняла конвенцію з запобігання забруднення з суден, яку в 1978 році було змінено відповідно до Протоколу на Міжнародній конференції з питань безпеки та запобігання забрудненню танкерами. В результаті вона була названа: «Міжнародна конвенція з запобігання забруднення з суден 1973 року, змінений в 1978 році протокол» або скорочено МАРПОЛ-73/78. Конвенція набула чинності 2 жовтня 1983 (Додатки I і II).

Правила, що охоплюють різні джерела забруднення з суден, викладені в шести додатках до MARPOL -73/78:

I Правила для запобігання забруднення нафтою. Чинні з 2 жовтня 1983 року.

II Правила для запобігання забруднення шкідливими рідкими речовинами. Вступ в силу з додаванням 1985 на 6 квітня 1987.

III Правила запобігання забруднення шкідливими речовинами, що перевозяться морем в упаковці, вантажних контейнерах, знімних танках, автомобільних і залізничних цистернах. Чинні з 1 липня 1992.

IV Правила запобігання забруднення стічними водами із суден. Чинні з 27 вересня 2003 року.

V Правила для запобігання забруднення сміттям з суден. Чинні з 31 грудня 1988 року.

VI Правила запобігання забруднення атмосфери з суден. Затверджена в вересні 1997 року, але в силу ще не вступила.

Деякі важливі правила та положення Додатку I MARPOL (правила запобігання забруднення нафтою).

У правилі 1 надано визначення:

— «Нафта» означає мастило у будь-якій формі, включаючи сиру нафту, рідке паливо, мастило, що містить осади, масляні залишки та очищені нафтопродукти;

— «Нафто-вмісна суміш» означає суміш з будь-яким вмістом мастила;

— «Нафтове паливо» означає будь-яке мастило, що використовується як паливо для основних двигунів та допоміжних механізмів судна;

— Танкер нафти «означає судно, побудоване або адаптоване для транспортування нафти оптом у вантажних приміщеннях;

— «Комбінований вантажний корабель» означає судно, призначене для транспортування нафти оптом або твердого вантажу оптом;

— «Спеціальний округ» означає морську зону, де відповідно до визнаних технічних причин, що належать до його океанографічних та екологічних умов, специфіка доставки на ній вимагає прийняття спеціальних обов'язкових методів запобігання забруднення моря з мастилом. Спеціальні райони — це райони, перелічені у правилі 10 цього Додатку;

— «Інтенсивність миттєвого розливу» означає інтенсивність розливу нафти в літрах на годину в будь-який час, поділена на швидкість судна в вузлах;

— «Танк» означає закрите приміщення, призначене для транспортування рідин;

— «Бортовий танк» означає будь-який резервуар, що прилягає до бортової обрізки судна;

— «Центральний танк» означає будь-який резервуар, розташований всередині судна з поздовжнім пересуванням;

— «Стійкий грязьовий танк» означає будь-який резервуар, спеціально розроблений для збору залишків з танків, промивання води та інших масляних сумішей;

— «Чистий баласт» означає баласт у танку, який після останнього перевезення в ньому був очищений таким чином, що стік з цього танка, з нерухомого судна в чисту, спокійну воду в ясний день, не викликає видимих слідів нафти на поверхні води;

— «Ізольований баласт» означає водяний баласт, прийнятий у танку, який повністю відокремлений від вантажу або паливної системи;

— «Проникність приміщення» означає співвідношення об'єму приміщення, який може бути заповнений водою до загального обсягу приміщення.

Правило 9 показує обмеження для скидання нафти:

1. З урахуванням положень відповідно до правил 10 і 11 цього Додатка і пункту 2 цієї статті, заборонити скидання в море нафти або нафто-вмісних сумішей з суден, до яких цей додаток застосовується, за винятком випадків: а) з нафтового танкера, за винятком випадків, передбачених у підпункті (б) цього пункту: — танкер знаходиться за межами особливого району; — танкер знаходиться на відстані більше 50 морських миль від найближчого берега; — танкер на своєму шляху; — миттєва швидкість скидання нафти не перевищує 30 літрів на морську милю; б) з судна валовою місткістю 400 рег. т або більше тонн валової місткості, крім нафтових танкерів, а також від машинних приміщень нафтового танкера, відділень вантажного насоса, за винятком тих пір, поки вміст нафти в стоці не змішується з мастилом та залишками вантажу нафти:

— судно знаходиться за межами особливого району;

— судно знаходиться в дорозі;

— вміст нафти в стоці без розведення не перевищує 15 частин на мільйон;

— на борту експлуатується устаткування для фільтрування нафти, що задовольняє пункт 17 правила 16 цього Додатку.

2. Стосовно судна валової місткості менше 400 рег. т, крім нафтових танкерів, якщо судно знаходиться в особливому районі, адміністрація повинна забезпечити, щоб воно було обладнане, наскільки це доцільно і практично можливо, пристроєм для зберігання залишків нафтопродуктів на борту і їх скидання в приймальні споруди або в море відповідно до вимог пункту I (б) цього правила.

3. У всіх випадках, коли в безпосередній близькості від судна або його сліду на поверхні води виявлено видимі ознаки нафти, уряди Сторін Конвенції повинні без зволікання розслідувати цей факт.

6. Вуглеводневі радикали, які не можуть бути скинуті в море відповідно до пунктів 1 і 2 цього правила, зберігаються на борту і вивантажують в прийомні об'єкти.

У правилі 10 Додатку I наведені методи запобігання забруднення нафтою з суден при плаванні в особливих районах:

1. Це райони Середземного моря, Балтійського моря, Чорного моря, Червоного моря, район Затоки, Антарктична область і Аденська затока.

2. У спеціальній зоні забороняється викид в море нафти або суміші, що містить нафту, з будь-якого нафтового танкера або судна.

Додаток V конвенції MARPOL (правила запобігання забрудненню сміттям з суден).

У правилі 1 визначено:

— «сміття» означає всі види харчових, побутових та операційних відходів (усунення свіжої риби та її залишків), які утворюються в процесі нормальної роботи судна та підлягають постійному або періодичному видаленню, за винятком речовин, визначення або список яких наведено в інших додатках до цієї Конвенції.

— «найближчий пляж». Вираз «від найближчого берега» означає оригінальну лінію, з якої, за даними міжнародного права, відраховуються територіальні води відповідної території, за винятком пів-нічно-східного узбережжя Австралії, де початкова лінія наведена в Конвенції.

— «спеціальний округ» означає морську зону, де відповідно до визнаних технічних причин, що стосуються його океанографічних та екологічних умов, специфіка доставки на ній вимагає прийняття спеціальних обов'язкових методів запобігання забрудненню моря сміттям. Спеціальні райони — це райони, перелічені у правилі 5.

У правилі 3 додатку V MARPOL наводяться умови для видалення сміття за межами спеціальних територій:

а) заборонено викид у море всіх видів пластмас, включаючи синтетичні кабелі, синтетичні риболовецькі мережі та пластикові пакети для сміття, але не обмежуючись ними;

б) жодна шкідлива речовина, що перевозиться в упаковці, не може бути скинута за борт за жодних умов.

I. 25 морських миль від берегу для плаваючих і пакувальних матеріалів;

II. 12 морських миль для харчових відходів та іншого сміття, включаючи паперові вироби, ганчірки, скло, метал, пляшки, осколки та аналогічне сміття; в) кидати в море сміття, зазначене в підпункті (b) (ii) цього правила, може бути дозволено, якщо таке сміття проходить через подрібнювач, і це зроблено до меж від найближчого берега, але в будь-якому випадку заборонено, якщо відстань до найближчого берега становить менше 3 морських миль. Таке подрібнене сміття повинне проходити через поверхню з отворами не більше 25 мм.

Відповідно до пункту 1 правила 4, забороняється використовувати будь-які матеріали, які підлягають застосуванню зі стаціонарними або плавучими платформами, розробкою та пов'язаних з ними процесами обробки в морі морських притулків мінеральних ресурсів, а також від всіх інших суден, зв'язаних з такими платформами або знаходиться в межах 500 м від них.

Якщо сміття змішане з іншими відходами, видалення або скидання яких підпадає під інші вимоги, то жорсткіші вимоги. Скидання в море пластику й зол із пластику заборонене всюди.

У правилі 5, додаток V MARPOL наведені спеціальні зони для запобігання забруднення сміттям: 1. Для цілей цієї заявки спеціальні зони є областю Середземного моря, в районі Балтійського моря, акваторії Чорного, району Червоного моря і району Затоки, морські райони Північного моря і Антарктична територія, басейн Карибського моря, в тому числі в Мексиканській затоці і Карибському морі, визначення якого дано нижче:

a) район Середземноморського моря означає Середземне море із затоками і морями, розташованими в ньому, обмежено з Чорного моря з паралельною 41° північної широти, а на Заході — Meridian 5º36' Західної довготи перетину Гібралтарської протоки;

b) район Балтійського моря означає Балтійське море саме по собі з Botnik і фінських бухт і з проходом в Балтійське море, обмежене паралельно 57° 44,8' північної широти мису Скаген в Скагеррак протока;

c) район Чорного моря означає саме Чорне море, межує з Середземним морем з паралельно 41º північної широти;

d) район Червоного моря означає фактичне Червоне море з Суецьким, обмежене з півдня прямою лінією, що проходить між Рас-SI-ANS (12° 8,5' північної широти, 43° 19,6' східної довготи) і Husner Murad (12° 40,4 ' північної широти, 43° 30,2' східної довготи); (E) Район затоки означає морський район, розташований на північний захід від прямої лінії, що проходить між Рас-Ель-Хадда (22°30' північної широти, 59°48' східної Lension) і Рас-Ель-Fast (25°04' північної широти, 61° 25' східної довготи).

e) район зони Північного моря: в Північному морі обмежено: (I) від Північного моря на південь — паралелі 62° північної широти, а на сході — Meridian 4° західної довготи; (Ii) протоку Скагеррак, південна межа якого визначається паралельно 57° 44,8 північної широти на схід від мису Скаген; і (W) Манш і підходи на схід від меридіана 5° західної довготи і на північ від Parallel 48° 30' північної широти;

g) Антарктичний район означає морський район, розташований на південь від паралельної 60' південної широти;

h) район Карибського басейну, як визначено в параграфі I статті 2 Конвенції про захист та розвиток морського середовища Карибського басейну (Картахена та Індіас, 1998), означає Мексиканську затоку та Карибський басейн з бухтами та морями в них.

Перегляд, зміну та доповнення MARPOL-73/78 доручено Комітету захисту морського середовища.

STCW-78/95.

Конвенція STCW стала першим міжнародним документом з основних правил про підготовку та дипломування моряків та несення вахти. Вона була прийнята 7 липня 1978 року, набрала чинності 28 квітня 1984 року.

Стаття ІІ STCW-78 містить визначення, головні з яких:

— «Партія» означає державу, за яку набрала чинності Конвенція;

— «Адміністрація» означає уряд, під прапором якого корабель має право плавання;

— «Диплом» означає дійсний документ, незалежно від того, як це було названо, виданий адміністрацією або її повноваженними, або визнаний адміністрацією на право його власника займати позицію, зазначену в цьому документі або дозволену національними правилами;

— «Власник диплому» означає особу, яка володіє дипломом на правовій основі;

— «Організація» означає міжнародну морську організацію;

— «Морське судно» означає судно, відмінне від тих, що плавають виключно у внутрішніх водах у межах захищених вод або в безпосередній близькості від них, або в межах правил порту;

— «Риболовецьке судно» означає судно, що використовується для риболовлі, китів або інших живих ресурсів моря.

Правило І/1 Додаток до Конвенції 1978 року містить наступні визначення: «Капітан», «Старший помічник капітана», «Помічник капітана», «Механік», «Старший механік», «Другий механік», «Механік-статер», «Радіооператор», «Особа звичайного складу». Це правило повинно бути виключено для безекіпажних суден.

Меморандуми про взаєморозуміння.

Інститут контролю іноземних судів у портах, з метою встановлення дотримання цими судами до звичайних вимог, виник на початку 80-х років у формі регіональної угоди ряду країн (Паризький Меморандум взаєморозуміння щодо контролю суден державного порту).

Світ підписав та експлуатує наступні регіональні угоди про управління портом:

— Паризький меморандум — 01.07.82 у Парижі;

— Латиноамериканська угода — 5.11.92 у Vina Del Mar (Чилі) (Винья-дель-Мар);

— Меморандум Токіо — 01.12.93 в Токіо;

— Карибський меморандум — 09.02.96 в Крісчех (Барбадос);

— Середземноморський меморандум — 11.07.97 у Валетті (Мальта);

— Меморандум Індійського океану — 05.07.98 у Преторії (Південна Африка);

— Меморандум Центральної та Західної Африки — 22.10.99 в Абуджа (Нігерія);

— Чорноморський меморандум — 07.04.2000 в Стамбулі.

Регіональні угоди порту:

— перевірка іноземних суден у портах;

— використання ідентичних засобів керування;

— застосування узгоджених процедур контролю;

— застосування домовленостей;

— взаємний обмін інформацією.

Основні засоби контролю:

— SOLAS 74/88;

— MARPOL 73/78;

— STCW 78/95;

— COLREG-72;

— LL-66/88;

— CLC-69;

— Конвенція ILO.

Особлива увага приділяється наступним кораблям:

— пасажир, bulk;

— кораблі для перевезення небезпечних вантажів та забруднюючих речовин;

— відвідування порту вперше, або через 12 місяців та більшу перерву;

— прийшов з іншого порту з зауваженнями PSC;

— під прапором країн, що належать до «чорного списку».

«Чорний список» — це список країн, судна яких після перевірки PSC були затримані в портах, що дозволяє адміністрації порту не обтяжувати суда надмірними інспекціями. У той же час таки суди будуть перебувати під пильним контролем для запобігання можливим порушенням.

Постанова Міжнародної морської організації А.787 (19) «Процедури контролю судів державою порту» була прийнята 23 листопада 1995 року, і є основним документом, що регулює процедуру перевірки суден у портах.

Глава 1, Визначення, глава 2 регулює перевірки судових інспекцій у портах, глава 3 забезпечує процедури для більш детальної перевірки, глава 4 — арешти кораблів у порту навігації.

Міжнародна конвенція для пошуку та порятунку. SAR-79

Міжнародна система пошуку та рятування. Спочатку захоплене морем судно ставало видобутком прибережних мешканців і безжалісно грабувалося. Світова спільнота вперше прийняла міжнародну угоду про надання порятунку людям у морі в 1914 році в Конвенції про захист людського життя на морі, після катастрофи «Титаніка». Ця Конвенція була довірена судам, що знаходяться поруч. Тому частота 500 кГц та сигнал SOS були прийняті. Окрім того, 3 хвилини мовчання були встановлені кожні 30 хвилин у радіо, а світовий океан поділяється на 13 зон, через прослуховування ефіру в будь-який момент є безперервним. Наприкінці ХХ століття така система застаріла, крім того на кораблях та літаках з'явилося нове радіообладнання, набравши чинності ГМССБ. Тому постанова Міжнародної морської організації А.406 (Х) від 17 листопада 1977 року рекомедувала скликати конференцію для прийняття рятувальної конвенції. Конференція відбулася в Гамбурзі з 9 по 27 квітня 1979 року, де світова спільнота прийняла пошукові та рятувальні конвенції про море (SAR-79). За рішенням 69-го засідання КБМ у травні 1998 року було прийнято нову заяву до Конвенції САР-79, яка набрала чинності з 1 січня 2000 року. За даними Конвенції SAR-79 головна роль у пошуку та порятунку приділяється прибережним послугам — центрам рятувальних координацій (RCC), які повинні бути організовані в кожній країні.

Терміни, що використовуються в SAR-79: пошук, пошукові та рятувальні зони, центр порятунку, рятувальні підцентри, продукт та рятувальний інструмент, аварійний етап, стагінальна невизначеність, стадія тривоги, стадія катастрофи, координатор у пункті дії.

Глава 2 Додатків до Конвенції SAR-79 регулюються міжнародними стандартами для координаційних пошукових послуг та порятунку. При отриманні інформації про будь-яку особу, що страждає на катастрофу у морі, або, страждає на морі, влада повинна вживати термінових заходів для забезпечення необхідної допомоги. На підставі цієї допомоги вона зобов'язана самостійно організувати або разом з іншими державами пошукові та рятувальні послуги, які повинні мати наступні основні елементи:

1. Правова база;
2. Призначення відповідального органу;
3. Організація наявних коштів;
4. Засоби зв'язку;
5. Координаційні та виконавчі функції;
6. Процеси покращення послуг, включаючи планування, відносини на національному та міжнародному рівнях;
7. Підготовка персоналу.

Пошукові послуги та рятувальні роботи в рамках пошукового та рятувального майданчика, межі яких узгоджуються між зацікавленими сторонами. Сторони надають допомогу будь-якій людині, що потерпає від катастрофи у морі. Вони виконують це незалежно від національної приналежності або статусу такої особи або обставин, в яких ця особа знаходиться. Генеральний секретар надсилає інформацію про рятувальну службу, яка повинна бути своєчасно скоригована:

1. Інформація про національні органи, відповідальні за пошукові та рятувальні послуги на морі;

2. Розташування CRS або інших центрів для забезпечення координації пошукових та рятувальних операцій та засобів спілкування з ними;

3. Інформація про межі пошукової та рятувальної зони та прибережних комунікацій у катастрофі та безпеки;

4. Введення основних пошукових та рятувальних методів.

Для забезпечення ефективності сторони забезпечують найбільш повну координацію з повітряними послугами. Там, де можна створити RCC та JSC для цілей навігації та літака. Сторони забезпечують використання єдиних процедур для цілей як морського та повітряного пошуку, так і порятунку.

Координація в пункті дії, коли виникає інцидент: призначається координатор пошукових та рятувальних дій (SMC), який, як правило, діє з RCC Rescue Center або Центр RSC RSC, який забезпечує координацію. Обов'язки координатора (OSC):

— координує дії всіх пошукових та рятувальних інструментів у пункті дії;

— планує пошукові та рятувальні операції, якщо план не був отриманий від координатора дій (SMC);

— змінює план пошукових та рятувальних операцій відповідно до ситуації, інформує SMC;

— координує зв'язок у пункті дії;

— моніторинг виконання дій іншими засобами;

— забезпечує повне виконання операцій, приділяючи особливу увагу поділу всіх фондів як в ефірі, так і в морі;

— періодично передає повідомлення SMC, відповідно до стандартної форми SITREP:

— проводить детальний запис операції;

— інформує координатора дій (SMC) про можливість появи коштів, що більше не потрібні;

— повідомляє координатору дій (SMC) кількість збережених;

— надає координатору дій (SMC) імен та точок призначення коштів по збереженню людей;

— звіт, який зберігається по кожному крокові;

— запитує додаткову інформацію з (SMC). Відповідальність координатора на місці дії (OSC). Координатор на сайті Action (OSC) повинен отримати план дій якомога швидше від SMC. Тим не менше, OCC може розвивати свій власний план (залежно від обставин). Виконавці також повинні змінити план пошуку відповідно до екологічно змінної атмосфери, зокрема, коли вони виникають: прибуття додаткових засобів допомоги; отримання додаткової інформації; зміни погодних умов, видимості, умов освітлення тощо. Пошукові операції повинні починатися відразу після прибуття до місця порятунку.

У разі виникнення мовних труднощів слід використовувати міжнародні сигнали та стандартні фрази Міжнародної морської організації для спілкування на морі. Запитуючи обов'язки, OCB повинен інформувати відповідну прибережну радіостанцію (CRS) або службу управління повітряним рухом (ATS), а також координатора дій (SMC) з регулярними інтервалами або коли змінюється ситуація.

Національні пошукові послуги та рятувальні послуги. Кожна сторона розробляє відповідні процедури для загальної організації, координації та вдосконалення пошукових та рятувальних послуг. Для забезпечення ефективності пошукових та рятувальних операцій сторони повинні:

1. Забезпечити координацію використання наявних засобів;

2. Встановити тісну співпрацю між організаціями в таких сферах, як операції, планування та підготовка персоналу, навчання та дослідження.

Співпраця між державами. Сторони повинні координувати роботу своїх координаційних центрів, а також їх пошукові та рятувальні операції з сусідніми державами. При необхідності, координувати національні закони, сторони зобов'язані визнати їх територіальні води, територію та повітряний простір над ними для пошуку та порятунку людей. Сторонам, особливо якщо їх пошукові та рятувальні ділянки перекривають одна одну, необхідно укласти угоди про прийом рятувальних підрозділів на їх території або повітряному просторі. Якщо не існує ніякої угоди між сторонами, якщо це необхідно, надається запит, до якого сторони, відповідальні органи зобов'язані якомога швидше відповісти:

— Негайно підтвердити отримання запиту;

— Вказати умови, якщо такі є, згідно з якими рятувальні одиниці допускаються до території держави. Кожна сторона повинна авторизувати свої процедури пошуку;

— Просити допомогу з інших центрів координації, включаючи судна, авіацію, персонал, постачання тощо, які можуть знадобитися;

— Дати будь-який дозвіл на доступ до своєї території або повітряного простору таких суден, авіації, персоналу або пропозиції;

— Координувати з митними, імміграційними, санітарними та іншими органами влади необхідні заходи для прискорення рятування. Кожна сторона гарантує, що такі центри координації забезпечують негайну допомогу іншим центрам, включаючи допомогу авіаційних суден, персоналу, постачання тощо.

Системи суден для повідомлень. Сторони надають на цілодобовій основі швидке та надійне отримання сповіщень про лиха в межах пошуку та рятувальних районів. Будь-яка країна або декілька країн, що отримують повідомлення про страждання, зобов'язані:

— негайно транслювати повідомлення відповідному центру рятувальної координації або рятувального підключення, а потім, наскільки це можливо, допомагати забезпечити зв'язок у пошуках та порятунку;

— підтвердити сповіщення, якщо це необхідно. Для полегшення операцій пошуку та порятунку сторони можуть створювати систему суден, бажано на основі рекомендацій Міжнародної морської організації. Система повинна надавати користувачам інформацію про рух кораблів:

— план переходу;

— розташування судна;

— кінцеве повідомлення.

У випадку катастрофи:

— скоротити час між моментом втрати зв'язку з судном та початком пошуку та порятунку, без сигналу катастрофи;

— швидко визначити судна, які можуть бути залучені до допомоги;

— вміти встановити меншу область пошуку;

— сприяти наданню термінової медичної допомоги або консультації. Система суден повідомлень повинна задовольнити положення:

— надавати інформацію про місцезнаходження, плани переходу;

— дозволити відправити рух суден;

— отримувати повідомлення від учасників за певними інтервалами;

— бути простим у намірах і в операційних відносинах;

— дозволити стандартні формати, загальноприйняті на міжнародному рівні та стандартному порядку повідомлення.

Список національних контактних адрес. Контактні адреси національних центрів, відповідальних за безпеку моря та запобігання забрудненню від кораблів, публікуються Міжнародною морською організацією та оновлюються щорічно. Нова версія та адрес, як правило, затверджується на засіданнях Комітету з питань безпеки та Комітету з охорони морського середовища та поширюється циркулярно. Зазвичай список складається з двох частин:

1. Короткий список національних органів влади (раніше MSC/CHERC.630), місцеві підрозділи національних інспекційних послуг, офіційні послуги, що працюють від імені держави, а також органів, відповідальних за розслідування аварій (раніше MSC/Цик542), та секретаріату Меморандумів про взаєморозуміння щодо контролю суден державою порту.

2. Перелік контактних адрес існуючих національних центрів, відповідальних за прийом, передачу та обробку термінових повідомлень від кораблів зі шкідливими речовинами, включаючи мастило.

Список 1 повинен бути на кожному судні та в компанії в документації СУБ (план дій у надзвичайних ситуаціях). Список 2 має бути на кожному судні в аварійних засобах, щоб запобігти забруднення мастилом. Коректування контролюється реєстром з річним обстеженням.

Форма аварійного повідомлення про аварійні події в плані дій суден про надзвичайні ситуації:

План дій судна в надзвичайних ситуаціях.

Форма початкового повідомлення.

SS (координати, широта, довгота) / DD (відстань до прибережного знаку).

N C. град. хв. E W град. хв. / град. мор. міді.

EE курс град. / FF (швидкість, вузли) 1/10.

L L (передбачуваний шлях).

MM (слухаюча радіостанція).

NN (дата та час наступного повідомлення, UTC). День. Години. хв.

PP (вид та кількість вантажу/палива на борту).

QQ (коротка інформація про недоліки/пошкодження).

RR (коротка інформація про забруднення, включаючи оцінку втраченої кількості).

SS (коротка інформація про погоду та морський стан).

Напрямок, вітер швидкість (на масштабі Бофорта) / напрямок, висота хвиль (м).

TT (дані для зв'язку з судновласником / оператором / агентом).

UU (розмір і тип корабля).

Довжина: (м) / ширина: (м) / осад: (м) / тип.

X X (додаткова інформація).

Коротка інформація про інцидент.

Необхідність допомогти ззовні.

Дії вжиті.

Кількість екіпажу та інформація про будь-які тілесні ушкодження.

Інформація про страхову компанію.

Інша інформація.

Коли можуть виникнути труднощі та обмеження, що обумовлені нерозумінням мови, повинні включати англійську мову з використанням, коли це можливо, стандартних фраз міжнародної морської організації морських перевезень на морі. Для того, щоб передавати детальну інформацію, англійська мова може бути використана на свій розсуд, а також використані міжнародні сигнали. При їх використанні в тексті повідомлення відразу після літерного індексу необхідно внести відповідні вказівки про це.

Нижче наведені додаткові дані для заповнення спеціальної таблиці.

АА — назва судна, позивний або ідентифікаційні дані суднової радіостанції і прапор судна.

ВВ — група з 6 цифр, яка вказує день (перші дві цифри), години і хвилини (останні чотири цифри).

СС представляє собою групу з чотирьох цифр, що вказує на широту в градусах і хвилинах, а також знаки N (північ) або з S (південь), і групу з 5 цифр, яка вказує довготу в градусах і хвилинах, а також знаки E (схід) або W (захід).

DD є істинний пеленг (перші 3 цифри) і відстань в морських милях від чітко визначеної прибережної позначки (вказати берегову позначку).

EE — істинний курс.

FF — швидкість у вузлах і десятих вузла.

LL є оцінений шлях. При описі шляху необхідно давати широту і довготу кожної поворотної точки, як і в СС із зазначенням типу передбачуваного шляху між цими точками, наприклад, RL (по Loccodromia), ГК (на великій дузі окружності) або уздовж берегової лінії в разі прибережного плавання, очікуваної дати і часу характерних точок у вигляді групи з шести цифр, як у ВВ.

ММ — повністю вказати назви прослуховування станцій / частота.

NN — група із зазначенням дати і часу, як і в ВВ.

PP — найменування і кількість вантажу (бункер) на борту судна.

QQ — резюме несправностей / недоліків / пошкоджень. Короткі звіти про стан судна та можливості концентрації палива.

RR — коротка інформація про забруднення навколишнього середовища. Назва мастила або палива витоку в море; оцінка величини; оцінка руху скидання мастила / палива; Якщо це можливо, оцінити поверхню області розливу. Місце дається як в СС або DD.

SS є короткий опис переважаючих погодних і морських умов.

ТТ — ім'я, адреса, номер telemet і телефон судновласника і представника (фрахтувальник, власник або оператор судна або їх агент).

UU — інформація про довжину, ширину, осадку і тип судна.

ХХ — додаткова інформація:

короткий опис інциденту;

необхідність допомоги ззовні, допомога, яка була запрошена або була надана іншими суднами;

вжиті заходи щодо скидання і руху судна;

кількість членів екіпажу та відомості про будь-які тілесні ушкодження;

інформація про страхову компанію:

інша інформація.

Після передачі вихідного повідомлення в обсязі таблиці, додаткове повідомлення повинно бути передано так, що воно містить інформацію, важливу для безпеки судна і захисту морського середовища.

Наступна додаткова інформація повинна бути спрямована на судновласника або оператора в можливо короткий час після перших внесень інформації:

— додаткові деталі пошкодження судна і обладнання;

— вказується існуючий збиток;

— оцінка пожежної небезпеки і попереджувальних заходів, що вживаються;

— розміщення вантажу на борту і його номер;

— число нещасних випадків;

— пошкодження та збитки, завдані іншим суднам;

— час (GMT), коли була запрошена допомога, і час, протягом якого очікується допомога;

— ім'я рятувальника і тип аварійно-рятувального обладнання;

— чи було прохання про додаткову допомогу;

— вимоги до запасних частин та інших матеріалів;

— будь-яка інша важлива інформація.

Зв'язок в точці дії. Сигнал лиха:

— MAYDAY використовується для вказівки того, що судно знаходиться в стані загрози безпосередньої небезпеки і вимагає негайної допомоги;

— має перевагу перед усіма іншими повідомленнями.

Терміновий сигнал.

— PAN-PAN використовується, коли безпека мобільних засобів знаходиться під загрозою;

— термінoгенний сигнал PAN-PAN повинен бути використаний, коли існує небезпечна ситуація, яка, в кінцевому підсумку, може спричинити необхідність залучення допомоги;

— має перевагу над усіма повідомленнями, за винятком сигналу катастрофи.

Сигнал безпеки.

— SECURITE використовується для повідомлень, пов'язаних з безпекою судоводіння або передачею важливих метеорологічних попереджень.

Будь-які повідомлення, передані після цих сигналів, мають пріоритет перед звичайними повідомленнями. Як правило, сигнал повторюється тричі на початку повідомлення. Командир літака або капітан визначеного корабля повинен оголосити стан катастрофи за допомогою сигналу MAYDAY. Основні слова для радіопроцедур, пошуку, які рятувальні співробітники повинні використовувати та розуміти:

AFFIRMATIVE означає, що те, що передається, є правильним;

BREAK використовується для відокремлення частини повідомлення або одного повідомлення від іншого;

FIGURES вимовляються безпосередньо перед номерами в повідомленні;

I SPELL використовується для вимовляння слів по буквам;

NEGATIVE засобів немає;

OUT кінець передачі, коли відповідь не очікується або не потрібна;

OVER кінець передачі, коли очікується негайна відповідь;

ROGER означає, що прийняте повідомлення задовільне;

SILENCE вимовляється тричі і означає «зупинити негайно всі програми»;

SILENCE FINI означає скасування тиші, використовується для позначення кінця надзвичайної ситуації та відновлення нормального радіообміну;

THIS IS вимовляється до назви станції або позивного сигналу;

WAIT означає, що «я повинен призупинитися на кілька секунд, очікую подальшу передачу».

Більш детальний перелік процедурних слів наведено в «стандартних фразах міжнародної морської організації для спілкування на морі» та МСС. Передача повідомлення катастрофи від морського судна:

— 156,8 МГц (УКВ, канал 16);

— 156,525 МГц (УКВ ЦИВ 70 канал);

— 2182 кГц (радіотелефонія);

— ПВ/КВ ЦИВ (2187,55 кГц, 8414,5 кГц вахта несеться обов'язково) та ще на одній з частот 4207,5 кГц, 6312 кГц, 12577 кГц, 16804,5 кГц;

— Інмарсат 1644.3—1644,5 МГц (АРБ);

— Інмарсат 1626,5—1646,5 МГц — АРБ 406—406,1 МГц.

Якщо існують сумніви щодо прийому невідповідності повідомлення, його слід надсилати на будь-яку існуючу частоту, яку можна використовувати в місцевих районах, і на яких увага може бути отримана негайно. З метою створення сигналів катастрофи можна використовувати рятувальне радіо. Передача сигналів тривоги з літака здійснюється:

— зазвичай літак повідомляє блок керування рухом (АТС), який повинен повідомити RCC;

— 121,5 МГц;

— 4125 кГц (радіотелефонія);

— радар-респондент встановлюється при 7700 МГц;

— літак під час невизначеної катастрофи може використовувати будь-які засоби у своєму розпорядженні, щоб привернути увагу, повідомляти про своє місцезнаходження та допомогу.

Додаткове радіотехнічне обладнання, встановлене на морських та літальних апаратах відповідно до вимог Конвенції SOLAS-74/88 і з яким можна відправити невідповідність повідомлення:

— аварійний радіобуй (EPIRB), який, якщо вводиться або коли вмикається вручну, надсилає закодований сигнал, індивідуальний для кожного буя, на прибережні станції;

— радар-респондент (SART), після включення вручну, діє автоматично, приймаючи радіолокаційні імпульси РЛС.

Надсилає імпульси, які видно на екрані РЛС, як групу розширених точок, як сигнали респондентських маяків. Зазвичай на екрані РЛС респондент бачиться на 6—8 миль — на переносних УКВ-радіостанціях VHF. Повідомлення з судна про лихо повинно включати такі важливі компоненти:

— Ідентифікація та координати судна;

— Тип природної катастрофи та тип допомоги;

— Погода в безпосередній близькості, напрямок вітру, хвилі, видимість;

— Час залишення судна та кількість екіпажу, що залишився на борту;

— Кількість і тип рятувальних засобів;

— Надзвичайні інструменти для розміщення на рятувальному агенті або в морі;

— Кількість серйозно поранених.

У початковому повідомленні стільки інформації включено як практично доречну, але цілий ряд коротких повідомлень більш доцільніший, ніж одне довге. Сигнали візуальної катастрофи наведені в МППСС-72. Скасування повідомлення стихійного лиха повинно бути зроблено, як тільки буде надана допомога, або якщо допомога не треба. Будь-яке помилкове сповіщення слід скасувати, щоб не використовувати марно сили рятувальних послуг.

ISM-code. Світові стандарти. Незважаючи на покращений технічний стан флоту та сучасного навігаційного та радіотехнічного обладнання, відділ надзвичайних ситуацій світового флоту залишається на тому ж рівні. Вина покладається в основному на некомпетентність або непідготовленість екіпажів кораблів. За результатами розслідування надзвичайних справ на морі людство зазначило, що нещодавно «людський чинник» відіграє вирішальну роль. Тому було вирішено регулювати та стандартизувати людські відносини на борту суден.

Вимоги до безпечної експлуатації кораблів наведено в міжнародному кодексі з питань управління безпекою та забрудненням (ISM-code), який був прийнятий Міжнародною постановою Міжнародної морської організації А.741 (18) 4 листопада 1993 року. Вона включає в себе такі пункти:

1. Загальні положення;

2. Політика у сфері безпеки та охорони навколишнього середовища;

3. Відповідальність та повноваження компанії;

4. Призначена особа (особи);

5. Відповідальність та повноваження капітана;

6. Ресурси та персонал;

7. Розвиток планів проведення операцій на кораблях;

8. Готовність до надзвичайної ситуації;

9. Звіти про невідповідності, аварії, небезпечні ситуації та їх аналіз;

10. Технічне обслуговування та ремонт судна;

11. Документація;

12. Перевірка, огляд та оцінка, зроблена компанією;

13. Експертиза, перевірка та контроль.

Конвенція SOLAS-74/78.

Комітет з безпеки на морі в 1994 році прийняв постанову IX додавання до конвенції SOLAS-74/88.

Правило 1 «Визначення»:

1. ISM-code — означає Міжнародний код для управління безпечною експлуатацією суден та запобігання забрудненню, прийнятий Організацією А.741 (18) Постановою (18), з поправками, які можуть бути зроблені Організацією;

2. Компанія означає власника судна або будь-якої іншої організації, або людину, таку як менеджер, який взяв на себе відповідальність за роботу судна від власника судна, погодившись прийняти всі обов'язки та всю відповідальність, встановлену міжнародним кодом управління безпекою.

3. Обладнання — це судно, конструкція якого включає в себе один корпус, бортові шлери та бортові танки в вантажних приміщеннях і призначені переважно для транспортування насипних вантажів або рудозних або комбіновані судна.

4. Море-мобільна бурова установка — це судно, здатне виробляти бурові операції для розвідки або розвитку ресурсів, таких як рідкі або газоподібні вуглеводні, сірка або сіль.

5. Нафтовий танкер означає судно, побудоване або адаптоване, головним чином, для перевезення нафти оптом у своїх вантажних приміщеннях, і включає в себе комбіновані вантажні судна та будь-який танкер-хімічний транспорт, якщо він транспортує нафту оптом як вантаж або частину вантажів.

6. Швидкий корабель — це судно, здатне розвивати максимальну швидкість в метрах за секунду, рівну або більше: 3,7 V0,1667, де V 7 — розрахункове зміщення, м$^3$.

7. Танкер Himovo — означає вантажне судно, побудоване або пристосоване і використовуване для транспортування оптом будь-якого рідкого продукту, зазначеного в Міжнародному кодексі Хемноса.

8. Банзор — означає вантажний корабель, побудований або адаптований та використовуваний для транспортування оптом будь-якого скрапленого газу або іншого продукту, зазначеного в Міжнародному кодексі для газових транспортних засобів.

9. Риболовецьке судно — означає судно, що використовується для риболовлі, ловлі морських тварин та морепродуктів, рибного господарства.

10. DSC (доступний документ) — означає документ, виданий компанією, адміністрацією прапора, що підтверджує, що судно відповідає вимогам Кодексу.

11. SVUB (свідоцтво про управління безпекою) — це документ, виданий адміністрацією компанії, після генерації служби управління судном та підтверджуючий, що служба управління безпекою корабля відповідає вимогам правила 2 «Кодексу» — MCUB є Введено для всіх судновласників та кораблів, незалежно від дати будівництва, вчасно:

— 01.07.98 — швидкісні, масляні танкери, хімічні носії, газові носії, масові та вантажні високошвидкісні кораблі з валовою місткістю 500 або більше тонн;

— 01.07.02 — інші вантажні судна та морські бурові установки валовою потужністю 500 або більше тонн.

Ця глава не застосовується до державних суден, які працюють у некомерційних цілях.

Постанова Міжнародної морської організації А.787 (19) «Процедури контролю над кораблями держави порту». Судно повинно мати непрострочений сертифікат управління, виданий адміністрацією порту прапора судна. Якщо є підстави для більш детальної перевірки, зареєстровані наступні запитання ISM-code (посилання на елемент ISM-code).

ISM-code. Загальні положення (ISM-code). Основна ланка системи управління безпекою, відповідно до стандартів Міжнародної морської організації та Кодексу, є власником або оператором судна (компанії). Система управління безпекою (SUB) повинна бути реалізована у діяльності компанії для ефективних та професійних дій з інформацією судового управління та є невід'ємною частиною основної системи управління виробничою компанією. Код вимоги до системи управління безпекою компанії:

А) Стандарти якості безпеки та запобігання забрудненню:

— дотримання системи обов'язкових правил та стандартів;

— гарантії впевненості в тому, що система приймає кодекси, керівні принципи та стандарти, рекомендовані Міжнародною морською організацією та організаціями морської промисловості;

Б) Загальні цілі компанії:

— забезпечення якості наданих послуг;

— забезпечення безпечної експлуатації суден та безпечних умов роботи та навколишнього середовища;

— організація захисту від усіх виявлених ризиків;

— постійне вдосконалення навичок управління безпекою та судном, включаючи надзвичайну готовність, пов'язану з запобіжним засобами запобігання небезпеки і ризиків, пов'язаних із забрудненням.

В) Функціональні вимоги до системи управління безпекою:

— політика безпеки та екології;

— інструкції та процедури для забезпечення якості наданих послуг, безпечної експлуатації судна та охорони навколишнього середовища;

— кількість повноважень та зв'язків між узбережжям та персоналом судна, а також внутрішніми лініями спілкування на березі та з суднами;

— забезпечення взаємодії з радіостанціями та портів для організації надійних щоденних обліків кораблів компанії;

— процедури при аваріях та випадках невідповідності вимогам Кодексу;

— процедури підготовки до надзвичайних ситуацій та дій щодо їхнього усунення;

— процедури проведення внутрішніх аудитів та процедур розгляду керівництва. 17.ISM-code. Інститут у сфері безпеки та охорони навколишнього середовища (пункт 2 ISM-коду).

Кожна компанія повинна мати політику у сфері безпеки та охорони навколишнього середовища, яка полягає в:

— досягненні загальних цілей, передбачених Кодексом;

— забезпеченні безпечної експлуатації суден на рівні міжнародних та національних стандартів (правил та норм);

— підвищенні, на цій основі конкурентоспроможності своїх суден на світовому ринку.

У той самий час компанія проголошує свою прихильність і дає пріоритет, насамперед, забезпечуючи безпеку та запобігання забрудненню та повинна забезпечити основну мету політики.

Безпека на морі, запобігання смерті та травм людей, пошкодження навколишнього середовища, особливо морського середовища та майна, а також дотримання правил проведення комерційних операцій. Це досягається:

— дотриманням міжнародних та національних стандартів (правил та норм) щодо безпеки запобігання навігації та забруднення;

— стійким та надійним двостороннім спілкуванням кораблів з берегом;

— звітами капітанів за станами кораблів, проблем на борту та заходів для їх вирішення, необхідну підтримку узбережжя;

— наявністями взаємопов'язаних планів дій у надзвичайних ситуаціях та розробкою цих планів;

— здатністю компанії швидко і адекватно реагувати на небезпеку, яка може виникнути на кораблі;

— забороною приносити, зберігати та використовувати алкогольні напої та наркотики на борт суден;

— дослідженням аварій та надзвичайних ситуацій на кораблі та вживанням заходів щодо їх запобігання.

На підставі вищесказаного компанія здійснює:

— кадрову політику — збирання кваліфікованого персоналу;

— технічну політику — забезпечення проектно-технічної, технологічної та екологічної безпеки суден;

— соціальну політику — створення умов в інтересах персоналу для забезпечення безпечної експлуатації кораблів.

Політика затверджується підписом Ґенерального директора. Копія політики компанії, підписана Ґенеральним директором, викладена на визначеному місці в кожному підрозділі компанії, на судні — на місці, найбільш відвідуваному екіпажем, та в кабіні капітана. Всі співробітники компанії повинні бути знайомі з політикою компанії для її виконання.

Політика компанії складається з трьох рівнів документації.

Політика — цілі та завдання.

Загальна структура системи управління:

— опис стратегій системи управління та цілей;

— визначення діапазону системи управління;

— опис організаційної структури;

— визначення відповідальності потужних повноважень ключових працівників системи управління;

— перехресні посилання елементів посібника, використовуючи використані стандарти.

Процедури — що робити:

— метод управління системою;

— процедури, що описують перелік різних заходів щодо системи управління;

Інструкції — як це зробити:

— задокументовані завдання;

— опис робіт (інструкції з роботи);

— форми звітів та шаблонів, які використовуються в системі управління безпекою.

ISM-code. Відповідальність та повноваження компанії (пункт 3 ISM-коду). Стандартна структура СУБ будь-якої компанії повинна включати:

— Вище керівництво;

— Загальні збори акціонерів / засновників;

— Рада директорів;

— Генеральний директор;

— Рада, яку очолює Генеральний директор та включає всіх своїх депутатів та інших працівників, визначених зборами акціонерів. Прибережні одиниці:

— служба безпеки (СБМ);

— технічна експлуатаційна служба (корабель або МСС);

— відділ персоналу (ОК) або служба управління персоналом;

— видобуток;

— виробнича служба;

— комерційний відділ;

— юридичний відділ;

— служба зв'язку;

— департамент експлуатації;

— департамент логістики (ОМТС).

Мінімальні вимоги до структури компаній відповідно до рекомендацій галузевого стандарту є такими:

Маленька:

— Генеральний директор;

— призначена особа, за умови її заміни під час відсутності відповідного спеціаліста, прийнятого за угодою про зайнятість;

— технічний спеціаліст (механік, електромеханік), за умови її заміни на період відсутності фахівцем, прийнятим тимчасово. Примітка: договір для забезпечення системи управління безпекою невеликої компанії не звільняється від зобов'язання призначати осіб.

Середня:

— Генеральний директор;

— призначена особа;

— технічний фахівець:

— служба безпеки (один капітан-наставник для шести суден, включаючи призначену особу, та один фахівець з комунікацій та SPI на 12 суден);

— механік та судноплавна служба (1 менторний механік для шести суден, включаючи технічного спеціаліста).

Велика:

— Генеральний директор;

— призначена особа;

— головний інженер;

— служба безпеки (1 наставник для 6 кораблів);

— механік та судноплавна служба (1 наставник для 6 кораблів);

— радіотехнічне обслуговування (може бути частиною СБМ — 1 наставник для 12 кораблів);

— всі інші прибережні одиниці, наведені вище. Склад та кількість працівників кожного відділу визначаються керівництвом компанії.

ISM-code. Призначена особа. (Пункт 4 ISM-code). Безпека суден та якість послуг повинна бути під постійним контролем керівництва компанії. Для цих цілей керівник компанії встановлює призначену людину. Згідно з офіційною посадою, призначена особа будь-якої компанії може працювати в компанії на постійній основі або заступником керівника морської безпеки.

Під час відсутності призначеної особи обов'язки виконуються заступником. Призначена особа повинна бути затверджена:

— у великій компанії — у Державному комітеті;

— у середній та невеликій компанії — у ГА порту реєстру.

Призначена особа будь-якої компанії може бути фахівцем, який має морську базову освіту, диплом та всі сертифікати, включаючи

ISM-код, що і дозволяють капітану працювати на найбільшому кораблі компанії, щонайменше 3 роки.

Призначена особа виступає від імені керівника компанії її інструкцій щодо безпеки навігації та запобігання забрудненню, що є обов'язковими для всіх працівників компанії.

Щоб виконати це як частину системи управління безпекою, призначена особа:

— організовує та координує діяльність системи управління безпекою;

— підтримує цю систему, включаючи нормативні документи;

— підтримує постійне спілкування з суднами, контролює їхню безпеку та забезпечує їх прибережну підтримку, необхідну для використання безпечної експлуатації;

— має прямий доступ до ресурсів та управління компанією;

— забезпечує контроль за дотриманням стандартів (правил та норм) безпеки та ефективності системи управління безпекою;

— забезпечує судна та прибережні ресурси, що виділяються на безпеку;

— своєчасно і негайно реагує на повідомлення про невідповідності, небезпечні ситуації та аварії;

— організовує систематичні внутрішні та зовнішні перевірки системи управління безпекою, виправлення невідповідностей та виконання коригувальних дій;

— призводить до нормативно-правової документації (розповсюдження, налагодження, бюлетень тощо);

— проводить навчання для систематичних оглядів (аналізів) статусу безпеки в компанії та розробляє основні пропозиції щодо системи регулювання політики та системи управління безпекою.

ISM-code. Ресурси та персонал (пункт 6 ISM-code). До ресурсів безпечної роботи належать:

— Нормативні документи;

— Матеріальні ресурси;

— Довкілля;

— Фінанси;

— Підготовлений персонал.

Центральною ланкою системи управління безпекою є персонал, який має кваліфікований, компетентний та професійно підготовлений рівень. Весь прибережний персонал, що забезпечує систему управління безпекою, повинен мати морські назви та досвід роботи з командними позиціями не менше 3 років. Вимоги до персоналу передбачає посадовий опис, з якими вони знайомі до початку роботи ISM-code).

Готовність до надзвичайної ситуації (пункт 8 ISM-коду). Підготувати та забезпечити постійну готовність компанії та суден до надзвичайних ситуацій.

Компанія є операційним штабом з надзвичайних ситуацій, затверджених Генеральним директором, та на чолі з призначеною особою. Склад аварійної штаб-квартири компанії узгоджується з ГА портом реєстру.

ISM-код. Звіти про невідповідності, нещасні випадки, аварії, небезпечні ситуації та їх аналіз (параграф 9 МКУБ). СУБ компанії повинні забезпечити систему негайних звітів про всі інциденти, прямо або опосередковано впливати на безпеку навігації — звіти про невідповідності. Форму звіту про невідповідності наведено у документації суб'єкта. Звіти про невідповідності складають команду верфі з підписом капітана або головою прибережної одиниці у таких випадках:

— відбулися аварії, нещасні випадки;

— створені небезпечні, ризиковані та непередбачені ситуації;

— претензії, що виникли для рибного господарства, органів нагляду, портових органів;

— невідповідності (невідповідність вимогам) у системі управління безпекою;

— претензії та відгуки вимог до підсвічування;

— немає жодних пропозицій щодо модернізації та вдосконалення підтвердження: якщо невідповідність усунута самостійно, і допомога компанії не потрібна, звіт про невідповідність не складається. Звіт про невідповідність складається в 2 примірниках. 1-й направляється на ім'я призначеної особи відповідно до схеми суб'єкта, наведеної у документації, а другий залишається на кораблі / підрозділі компанії, яка написала звіт. Після отримання звіту про невідповідність служба безпеки морського моря назначає призначену особу, що повинна:

— зареєструвати доповідь, включаючи число на класифікацію документації суб'єкта;

— організувати дослідження та аналіз звіту;

— розробити рішення про це;

— моніторинг виконання коригувальних дій;

— проводити постійний рух доповіді та контролювати виконання коригувальних дій;

— встановити період для виконання коригувальних дій;

— призначити відповідальну особу за виконання коригувальних та профілактичних заходів.

Коригувальні дії зроблено у формі рішення про звіт про невідповідність, одна копія надсилається на адресу пристрою / судна, який написав звіт, а 2-га копія людині, відповідальній за виконання коригувальних дій. СБМ (призначена особа) контролює виконання цього рішення. Коригувальні та запобіжні заходи повинні бути спрямовані на забезпечення безпеки навігації та охорони навколишнього середовища, а ні в якому разі не зменшують рівень безпеки. Коригувальні дії:

— виправлення відповідних процедур та інструкцій;

— розробка нових процедур та інструкцій;

— розподілення досвіду серед суден та прибережного персоналу.

Звіт про невідповідність буде закрито після отримання призначеною особою інформації від голови підрозділу, який написав звіт, звітував про усунення невідповідностей у формі суб'єкта, наведеної у документації.

ISM-code. Документація (Пункт 11 МКУБ). Система управління безпекою будь-якої компанії регулюється безліччю документації. Її склад та порядок відліку кожна компанія встановлює самостійно, але це повинно охоплювати всі сфери компанії та кораблі. Кожне судно повинно мати повний пов'язаний з ним набір документації. За призначенням документація поділяється:

1) Розтягнутий — поставляється з суднобудівельного заводу. Мінімальна композиція безпеки Marigold включає:

— Технічний паспорт судна (vessel information book), що містить основні ТТД (основні зменшення, призначення, танкова ємність, кількість вантажів, ваги та інше).

— Інформація за стабільністю та схемами розрахунку (stability and trim book) — містить початкові дані для розрахунку стабільності та діаграм статичної та динамічної стабільності при різних завантажувальних діаграмах.

— Рисунки суднових конструкцій, механізмів та систем (drawings) є життєво важливими для забезпечення безпечної експлуатації судна.

2) Регуляторно-правова — це ключова документація. Це набір обов'язкових стандартів (правил та норм) для безпечної експлуатації судна.

19.0.МКУБ (ISM-code). Пункти 4—6 та 8 Кодексу. ISM-код. Призначена особа (особи) (Пункт 4 МКУБ). Безпека суден та якість послуг мають бути під постійним контролем керівництва компанії. Для цього керівник компанії своїм наказом засновує призначену особу. За службовим положенням, призначеною особою будь-якої компанії може бути людина, що працює в компанії на постійній основі, заступник керівника з безпеки мореплавання. Під час відсутності призначеної особи обов'язки виконує її заступник. Призначена особа має бути затверджена:

— у великій компанії — у Державному комітеті з рибальства;

— у середній та малій компанії — у ГА порту приписки.

Призначеною особою будь-якої компанії може бути фахівець, який має морську базову освіту, диплом та всі відповідні свідоцтва, включаючи МКУБ, які дозволяють працювати капітаном на найбільшому судні компанії, досвід роботи капітаном найбільшого судна компанії не менше трьох років.

Призначена особа діє від імені керівника компанії та її вказівки щодо безпеки мореплавства та запобігання забруднення обов'язкові для всіх працівників компанії.

Для виконання цього в рамках СУБ призначена особа:

— організує та координує діяльність системи управління безпекою;

— здійснює ведення цієї системи, зокрема нормативно-правових документів;

— підтримує постійний зв'язок із суднами, контролює їхню безпеку та надає їм берегову підтримку, необхідну для забезпечення безпечної експлуатації;

— має прямий доступ до ресурсів та керівництва компанії.

— забезпечує контроль за дотриманням стандартів (правил і норм) безпеки та ефективності системи управління безпекою;

— надає суднам та береговим підрозділам ресурси, виділені на забезпечення безпеки;

— своєчасно та оперативно реагує на доповіді про невідповідності, небезпечні ситуації та нещасні випадки;

— організовує проведення планомірних внутрішніх та зовнішніх аудиторських перевірок системи управління безпекою, виправлення невідповідностей та виконання коригуючих дій;

— веде нормативно-правову документацію (розподіл, коригування, розсилку тощо);

— проводить підготовку систематичних оглядів (аналізів) стану безпеки в компанії та розробку на їх підставі пропозицій щодо коригування політики та системи управління безпекою.

20.0. Відповідальність та повноваження капітана (Пункт 5 МКУБ). Відповідно до вимог МКУБ, КТМ та Статуту служби на суднах капітан є вищою посадовою та довіреною особою компанії на судні. Капітан керує судном на основі єдиноначальності, підпорядковується Генеральному директору та призначеній особі. Ніхто, ні суднова рада, ні судновий комітет, ні партійна організація, ні комітет з безпеки, ні будь-який працівник компанії, включаючи Генерального директора та призначену особу, не мають права скасувати рішення капітана з будь-якого питання виробничої та побутової діяльності судна. Усі члени екіпажу призначаються лише за згодою капітана. Капітан видає накази по судну та має право усунути будь-якого члена екіпажу від виконання його обов'язків або списати з судна, вказавши підстави у наказі. Капітан несе відповідальність за:

— підтримання та підвищення престижу та авторитету компанії;

— проведення на судні політики безпеки та розуміння її судновим персоналом;

— ефективне функціонування суднової СУБ;

— створення в судновому колективі моральних та матеріальних передумов для підвищення суднової СУБ;

— наявність та своєчасне підтвердження всіх суднових свідоцтв та документів суднового персоналу;

— організацію служби на судні, розподіл обов'язків, відповідальності та повноважень екіпажу, включаючи аварійні ситуації;

— складання та затвердження посадових інструкцій суднового персоналу, причому в праві відступити від статутних вимог підприємства;

— організацію зв'язку з компанією та внутрішньобортового зв'язку;

— передачу в компанію повідомлень про аварійні заходи та недотримання положень Кодексу;

— контроль за дотриманням персоналом судна міжнародних та національних стандартів, включаючи стандарти компанії, для забезпечення безпечної експлуатації судна;

— проведення занять, навчання та тренувань з відпрацювання судновим персоналом дій в аварійних ситуаціях, включаючи забруднення довкілля;

— ведення суднової документації та суднових журналів;

— надання до компанії оглядів (аналізів) щодо ефективності суднових СУБ та пропозицій щодо її вдосконалення. Капітан має виняткові повноваження в прийнятті рішень щодо забезпечення безпечної експлуатації судна та звернення до компанії за допомогою. Він не обмежений у праві прийняття рішень щодо забезпечення безпеки судна та суднового персоналу, запобігання забрудненню навколишнього середовища, збереження вантажу та майна і компанія підтримує його в цьому.

21.0.ISM-код. Ресурси та персонал (Пункт 6 МКУБ). До ресурсів для забезпечення безпечного ведення робіт належать:

— нормативні документи;

— матеріальні ресурси;

— довкілля;

— фінанси;

— підготовлений персонал.

Центральною ланкою СУБ є персонал, який має бути кваліфікованим, компетентним та професійно підготовленим. Весь береговий персонал, що забезпечує СУБ компанії, повинен мати морські звання та досвід роботи на командних посадах не менше трьох років. Вимоги до персоналу викладаються у посадових інструкціях, із якими вони знайомляться під розпис на початок роботи. Комплектування суднового персоналу здійснюється відповідно до чинного законодавства, з обов'язковим узгодженням із призначеною особою. Капітан зобов'язаний знати:

— національне та міжнародне законодавство та нормативно-правові документи;

— параметри непотоплюваності, міцності, стійкості, живучості судна та його особливості;

— міжнародні угоди щодо безпеки мореплавства;

— морське право, закони, правила та звичаї портів заходу;

— правила класифікаційного суспільства;

— правила, норми, рекомендації, інструкції компанії в частині експлуатації судна.

Судновий персонал повинен:

— мати морську базову освіту, дипломи, сертифікати та свідоцтва, що засвідчують його кваліфікацію;

— знати структуру судна та її особливості;

— мати достатній досвід роботи (при призначенні вперше на командні посади пройти відповідне стажування);

— вміти орієнтуватися в будь-яких умовах експлуатації, включаючи аварійні;

— знати умови експлуатації та галузь діяльності судна;

— знати режим роботи, робочі навантаження, розпорядок на судні;

— виконувати правила техніки безпеки;

— знати свої посадові обов'язки та суднову систему управління безпекою.

Береговий персонал повинен:

— мати спеціальну базову освіту, що відповідає призначенню;

— мати відповідні дипломи та свідоцтва, що підтверджують кваліфікацію;

— знати сферу діяльності підприємства та її СУБ;

— мати достатній досвід практичної діяльності;

— знати міжнародне морське право, відповідні міжнародні договори, національне морське законодавство;

— знати міжнародні та національні нормативно-правові стандарти з безпеки мореплавання та ПЗМ;

— знати міжнародні та національні правила ведення фінансових операцій;

— знати правила класифікаційних товариств.

Для підтримки кваліфікації персоналу на належному рівні компанія повинна здійснювати його планомірне навчання. Основними видами навчання суднового персоналу, передбаченими ПДМНВ-78/95, є:

1. Експлуатаційне навчання (in-service training) — проводиться на судні з виконання суднових операцій, але на березі перед призначенням на судно для підготовки та перевірки знань, майстерності, кваліфікації, компетентності та професійної підготовленості.

2. Сертифікаційне навчання (training for certification) — проводиться для підготовки та сертифікації командного складу за відповідними міжнародно визнаними стандартами, кваліфікованими інструкторами.

3. Виробничо-ознайомче навчання (shipboard familiarization) — проводиться з персоналом, який призначається на судно, з ознайомленням зі своїми обов'язками, влаштуванням судна та суднових приміщень, входів та виходів, включаючи аварійні, судновими при-строями, системами та обладнанням для нормальних та аварійних умов експлуатації.

4. Інше навчання (other training requirement applicable to all ships) — проводиться на всіх суднах з основним та тимчасовим судновим персоналом за способами та технікою виживання в аварійних ситуаціях, порядком залишення судна в кризових ситуаціях, а також методами індивідуального захисту (протипожежна безпека, техніка безпеки, перша невідкладна допомога тощо).

5. Спеціальне навчання (ship type specific training) — проводиться з судновим персоналом специфічних типів суден (добувні, обробні, приймальні, з небезпечними вантажами тощо).

6. Загальноосвітнє навчання — проводиться із судновим та береговим персоналом. Навчання проводиться на березі та на судні у вигляді лекцій, курсів підвищення кваліфікації, тренувань на тренажерах, стажувань як дублери тощо. Програми навчання на березі та на суднах узгоджуються та коригуються за результатами аварійних випадків, виявлених невідповідностей СУБ, зовнішніх та внутрішніх аудиторських перевірок. Відділ кадрів веде облік навчання кожного працівника. На судні навчання відображається в судновій документації та пред'являється наглядовим органам на їхню вимогу.

26.0. Готовність до аварійної ситуації (Пункт 8 МКУБ). Для підготовки та забезпечення постійної готовності компанії та суден до аварійних ситуацій створюються: у компанії — оперативний штаб з аварійних ситуацій, затверджений наказом Генерального директора та очолюваний призначеною особою. Склад аварійного штабу компанії узгоджується з ГА порту приписки. На судні — судновий комітет із безпеки (мінімальний склад капітан, старпом, стармех). Порядок дій аварійного штабу та суднового комітету з безпеки наводяться у взаємопов'язаних береговому (shore based emergency plan) та судновому планах дій в аварійних ситуаціях. Компанія має проводити підготовку до дій у потенційно можливих аварійних ситуаціях. Мета такої підготовки — постійна готовність компанії швидко та ефективно реагувати на аварійні ситуації, які можуть виникати на суднах. Відповідальним за готовність суден та їх екіпажів до дій у аварійних ситуаціях є призначена особа. Підготовка повинна передбачати:

— ідентифікацію та опис аварійних ситуацій, що можуть виникнути на суднах;

— розробку планів дій берегового та суднового персоналів у потенційно можливих аварійних ситуаціях;

— складання програм навчання та тренувань з відпрацювання береговим та судновим персоналом дій в аварійних ситуаціях, запобігання аваріям, локалізацією та зведенням до мінімуму наслідків (мають наводитися в положенні з тренувань та навчання).

— методи та підтримання контактів та зв'язку між судном та берегом, переданих в аварійних ситуаціях. Бажано використовувати рекомендації ІМВ (Резолюція А. 648(16) «Про основні засади системи суднових повідомлень та вимоги, що висуваються до них»). Плани дій у аварійних ситуаціях. Береговий повинен відображати:

— склад, посади, службові та домашні телефони основного персоналу штабу;

— порядок та місце збору штабу;

— обов'язки штабу та його взаємодію із зацікавленими партнерами, порядок запиту допомоги;

— методи та порядок повідомлень з судна на берег і назад;

— чек-листи, що ідентифікують аварійні ситуації, та буклети процедур для дій суднового персоналу у цих ситуаціях;

— взаємодія з оперативним штабом ГА порту;

— довідкова інформація про аварійно-рятувальні організації та центри в районах плавання суден компанії;

— порядок прийняття та виконання рішень та контроль їх виконання.

Суднові вимоги повинні додатково включати:

— склад, посади, телефони суднового комітету з безпеки;

— взаємодію та зв'язок із зацікавленими партнерами та суднами, що знаходяться в районі.

Відповідно до конвенції МАРПОЛ-73/78 на судні має бути «Судновий план надзвичайних заходів щодо боротьби із забрудненням нафтою (shipboard oil pollution emergency plan)». За структурою та побудовою він аналогічний до плану дій в аварійних ситуаціях. Компанія повинна проводити регулярні тренування та навчання суднового та берегового персоналу за вказаними вище планами. До програм підготовки повинні включатися:

— індивідуальні інструкції та навчання суднового персоналу щодо використання рятувальних та протипожежних засобів;

— заняття, тренування та навчання суднового персоналу щодо боротьби за живучість та дій у потенційно небезпечних аварійних ситуаціях;

— перевірки стану, надійності та готовності до дії суднового аварійного майна та обладнання, включаючи радіообладнання.

23.0. ISM-код. Доповіді про невідповідності, аварії, нещасні випадки, небезпечні ситуації та їх аналіз. (Пункт 9 МКУБ). СУБ компанії повинна передбачати систему негайних доповідей про всі події, що прямо чи опосередковано зачіпають безпеку мореплавства, — доповіді про невідповідності. Форма доповіді про невідповідність наводиться у документації СУБ компанії. Доповіді про невідповідності складаються судновим командним складом (обов'язково підписує капітан) або керівником берегового підрозділу у таких випадках:

— нещасних випадках, аваріях, аварійних пригодах;

— створених небезпечних, ризикованих та непередбачених ситуаціях;

— претензій рибоохорони, наглядових органів, влади портів;

— невідповідності (недотримання вимог) у системі управління безпекою;

— претензій клієнтури та зворотних претензій до субпостачальників;

— пропозицій щодо модернізації та вдосконалення СУБ, що з'явилися: якщо невідповідність усунена самотужки і допомоги компанії не вимагає, доповідь про невідповідність не складається. Доповідь про невідповідність складається у двох примірниках. 1-й прямує на ім'я призначеної особи за схемою, наведеною в документації СУБ компанії, а 2-й залишається на судні/підрозділі компанії, що написали доповідь. Після отримання доповіді про невідповідність служба безпеки мореплавства, а де її немає, призначена особа повинна:

— зареєструвати доповідь, надавши їй номер за класифікацією документації СУБ компанії;

— організувати вивчення та аналіз доповіді;

— виробити рішення щодо неї;

— проконтролювати здійснення коригувальних дій;

— вести постійний рух доповіді та контроль виконання коригувальних дій;

— встановити термін виконання коригувальної дії;

— призначити відповідальну особу за виконання коригувальних та запобіжних дій. Коригувальна дія оформляється у вигляді рішення за доповіддю про невідповідність і один примірник надсилається на адресу підрозділу/судна, що написав доповідь, а другий примірник особі відповідальній за виконання дії, що коригує. СБМ (призначена особа) контролює виконання цього рішення. Коригувальна та запобігаюча дія повинна бути спрямована на забезпечення безпеки мореплавства та захисту навколишнього середовища, та жодним чином не знижувати рівень безпеки. Коригувальні дії здійснюються шляхом:

— виправлення відповідних процедур та інструкцій;

— розробки нових процедур та інструкцій;

— поширення досвіду серед суднового та берегового персоналу. Доповідь про невідповідність закривається після отримання призначеною особою від керівника підрозділу, який написав доповідь. Донесення про усунення невідповідності за формою, наведеною в документації СУБ компанії.

24.0 МКУБ (ISM-code). Розробка планів проведення операцій на суднах. Пункт 7 Кодексу.

Основна відповідальність за розробку планів суднових операцій покладається на організацію. Керівництво компанії має визначити, які суднові операції найбільш важливі для функціонування її суден. Плани компанії:

— річний план, що передбачає огляд (аналіз) пропозицій промислової (судноплавної) діяльності;

— підготовчий план, який передбачає підготовку суден до рейсу, відповідно до завдання;

— експлуатаційний план, що передбачає здійснення промислу/вантажоперевезень, відповідно до рейсового завдання. Річний план виробничої діяльності, та аналіз його виконання здійснюється береговими службами підприємств. Підготовчий план здійснюється береговими службами підприємств, разом із судновим. Компанія розглядає та виконує різні договори/контракти (ремонт, сервісне обслуговування суднового обладнання, зв'язок, постачання, портові формальності та багато іншого), необхідні для успішної роботи судна в морі. На основі договірних (контрактних) умов основними видами підготовки судів є:

— Навігаційна — здійснюється службою безпеки щодо гарантування безпечного промислу (прогноз гідрометеобставин, забезпечення морськими картами, укомплектованість судна аварійним та рятувальним обладнанням, наявність планів дій в аварійних ситуаціях та підготовка екіпажів до дій в аварійних ситуаціях);

— Технічна — здійснюється механіко-судновою (технічною) службою (виконання планового технічного обслуговування, ремонту та

докування, забезпечення технічної та технологічної готовності до промислу, перевірка строків дії суднових документів, проведення чергових оглядів, організація бункерування суден, перевірка якості палива, організація матеріально-технічного постачання);

— Кадрова — здійснюється відділом кадрів (комплектація суднового персоналу, перевірка медичної придатності, перевірка дипломів та сертифікатів, організація заміни суднового персоналу);

— Експлуатаційна (промислова) — здійснюється службою мореплавства та відділом видобутку/комерційним/експлуатаційним (планування роботи в промислових районах, призначення агентів, забезпечення суден вантажною та експлуатаційною інформацією, забезпечення документами за правилами ведення промислу, організація зв'язку та диспетчерських зведень, перевірка знарядь постачання необхідного промислового постачання);

— Фінансова — здійснюється комерційним відділом та бухгалтерією, з виділенням повноважень капітана (забезпечення суден необхідними засобами, оплата експлуатаційних послуг, встановлення порядку використання виділених коштів, контроль фінансових операцій, дотримання комерційної таємниці підприємства, контроль руху готівки та майна);

— Страхування — здійснюється юридичним відділом (забезпечення всіх видів страхування, встановлення порядку надання доповідей про нещасні випадки та аварії, за результатами яких можливі ризики, статистичний облік збитків від виплат за претензіями та позовами).

— Суднові операції — під час експлуатації суден компанія здійснює розробку планів суднових операцій відповідно до ISM-code, враховуючи рекомендації IMO та ґрунтуючись на національній системі організації суднової служби багатьох країн. Суднові операції, за можливими наслідками, поділяються на:

— спеціальні — помилки у виконанні яких призводять до небезпечних ситуацій або виявляються після того, як аварія сталася;

— критичні — помилки у виконанні яких одразу породжують аварію або створюють загрозу для суднового персоналу, судна чи забруднення (наприклад: аварійні постановка та підйом знарядь лову, портові операції (лоцман, швартовка, якір та ін.), вантажні операції в морі та портах, бункерувальні операції, аварійні тощо). Критичні суднові операції мають виконуватися під суворим контролем. При цьому має бути повна переконаність у квалі-фікації, компетентності та практичній підготовленості суднового персоналу.

Суднові операції об'єднуються у послідовності процесу промислу та/або вантажоперевезень у такому порядку:

1. Загальні суднові операції: організація служби на судні — посадові обов'язки суднового персоналу — доповіді/рапорти суднового персоналу за підпорядкованістю — зв'язок судна з компанією — інспекції та контроль, що здійснюються капітаном та командним складом — суднова документація (склад, утримання, реєстрація) — медичне обслуговування — придатність до виконання посадових обов'язків та уникнення перевантажень суднового персоналу — алкоголь, медикаменти, наркотики (судова політика, контроль використання та обстеження) — організація технічного обслуговування та ремонту — інструкції з експлуатації та обслуговування суднового обладнання — охорона праці та техніка безпеки — запобігання забруднення навколишнього середовища — перевірочні листи — промисловий розклад.

2. Операції під час перебування судна у порту: судова вахтова служба (стоянкова) — взаємодія з владою порту — перевірка, випробування та підготовка до дії протипожежних засобів — навантаження та вивантаження — контроль розміщення вантажу, міцності та стійкості — вивантаження нафтомісних вод та шкідливих речовин на берег — організація здачі харчових відходів, сміття та стічних вод — ремонтні роботи в порту — випадкові розливи рідких вантажів із суднового бункера — відповідальність за випадки забруднення — дії, якщо судно тимчасово затримується в порту — отримання промислового та виробничого спорядження.

3. Операції з підготовки судна до рейсу: перевірка та реєстрація судна — перевірка міцності та стійкості — перевірка надійності закриття всіх люків та отворів у корпусі — перевірка надійності кріплення промислового устаткування — визначення/прогноз гідрометеообставин — підготовка навігаційних карт та планування переходу до підготовки документації — перевірка, коректура карт та посібників — бункерування судна — отримання продуктів, води та запасних частин — завершення ремонту та перевірка виконання — перевірка та підготовка ГД та механічного обладнання судна — перевірка та підготовка систем управління судном — перевірка та підготовка систем та механізмів забезпечення безпеки (засоби навігації, якір, навігаційні вогні та ін.) — перевірка та підготовка засобів зв'язку — перевірка та підготовка обладнання та пристроїв ПЗМ — перевірка та підготовка промислового та виробничого обладнання та пристроїв.

4. Операції при знаходженні судна в морі та на промислі: суднова ходова навігаційна вахта — спеціальні вимоги при плаванні в складних умовах — радіозв'язок — спостереження за навколишнім середовищем — спостереження за станом та режимами експлуатації судна та основного обладнання — готовність судна до маневрування — постановка знарядь лову — ведення промислових операцій — виробнича діяльність рибцехів та обробка риби — швартові операції (бункерування, розвантаження) — готовність до непередбачених/ екстремальних ситуацій.

5. Операції з підготовки судна до приходу в порт: перевірка ГД, рульового пристрою, засобів навігації та зв'язку, якірного пристрою — проводка судна (лоцманська) — зв'язок судна з портом та інформація — визначення/прогноз гідрометеообставин — обмеження з плавання в районі порту, сезонні таблиці та карти, настанови — баластування судна — контроль міцності, стійкості та водонепроникності — перевірка та підготовка швартовного пристрою судна. Документування суднових операцій здійснюється та оформляється у вигляді процедур та інструкцій. Процедура — це комплекс об'єднаний спільністю мети, дій (функцій), викладених у формі документа, що визначає призначення та завдання цього комплексу, склад, зміст та порядок виконання дій (функцій), що входять до нього, та їх кінцевий результат.

Основу процедур складають суднові операції. Процедура може відображати суднову операцію повністю або її складові. Інструкція — це розвиток та деталізація процедури. Вона є документом, що визначає технологію виконання передбачених процедурою дій (функцій). Основний склад процедур наводиться у положенні щодо процедур документації СУБ компанії. Вимоги міжнародних та національних нормативних документів виконуються в компаніях та на суднах без дублювання їх у документах нижчого рівня. У документації СУБ правомірно лише посилювати чи деталізувати вимоги документів найвищого рівня, прив'язуючи їх до конкретних особливостей роботи своїх суден. У процесі роботи документація СУБ має коригуватися та доповнюватися на основі розслідувань (аналізів) аварій, невідповідностей, нещасних випадків тощо. Найбільший відсоток важких аварій світового флоту падає на людський фактор, пов'язаний з навігаційною вахтою на містку та машині. Для полегшення дій судно-водія та механіка значного поширення набули суднові перевірочні листи (чек-листи). Ці чек-листи наводяться у документації СУБ кожної компанії та рекомендуються для використання при повсякденній виробничій діяльності судна. За рекомендацією ІМО (резолюція А.864(20) від 27 листопада 1997 року), при проведенні робіт, пов'язаних з підвищеним ризиком (на висоті та за бортом, у закритих та погано вентильованих приміщеннях, вогневі роботи тощо), судна повинні використовувати відповідні чек-листи. При використанні чек-листів під час вахти судноводій/механік повинен враховувати наступне:

— заповнений чек-листів судної операції не звільняє осіб, які несуть ходову навігаційну вахту, від відповідальності за невірні дії;

— чек-лист є юридичним документом і поряд із записами в судновому/машинному журналі може бути доказом правильних дій судноводія/механіка в екстремальних ситуаціях;

— про заповнення чек-листа необхідно зробити запис у судновому/машинному журналі;

— чек-листи заповнюються тільки ручкою синім або чорним чорнилом, забороняється використовувати олівець;

— на всі пункти чек-листа має бути відповідь ТАК;

— якщо якийсь із пунктів чек-листа не виконаний, суднова операція не повинна проводитися, а про ситуацію необхідно негайно доповісти капітану/старшому механіку за належністю;

— часто заповнювані чек-листи (зміна вахт, постановка трала тощо) кожен судноводій/механік заповнює один раз за рейс, а потім тільки робить відмітку в судновому/машинному журналі про його використання;

— чек-листи при плаванні у складних умовах необхідно заповнювати одразу;

— на кожному заповненому чек-листі повинні стояти дата, підпис та прізвище особи, яка заповнила його;

— всі заповнені чек-листи повинні зберігатися на судні щонайменше два роки. У разі невідповідності чек-листа судновим умовам необхідно направити призначеній особі компанії доповідь про невідповідність з обгрунтуванням коректури відповідного чек-листа. Список усіх чек-листів, використовуваних під час роботи судна, має бути вивішений на видному місці містка/машинного відділення (над штурманським столом, над пультом управління головним двигуном) [4—10].

На підставі огляду нормативних документів випливає, що, поки безпілотні судна не дотримуються правил Міжнародної морської організації, вони будуть розглядатися як не судоходні, як такі, що не підлягають страхуванню. Але час не стоїть на місці і вже є для безекіпажних суден ескізи регулювання правових відносин. Для повних автономних суден, якщо ступінь автоматизації судна дозволяє виконувати плавання без екіпажу на борту при постійному спостереженні за судном та управляти його рухом персоналом за межами судна, або без постійного моніторингу та керування персоналом поза судном.

Суднові документи. Оскільки багато суднових документів мають відношення до екіпажу судна та його функцій, необхідно замінити айсклоуз, що наявність або підтримка частини суднових документів для автономних суден у прийнятому тлумаченні повинні бути замінені. Для автономного судна необхідно консолідувати право не мати на борту суднових документів, а їх інспекційний контроль органи влади зможуть здійснити через судновласника в електронній та альтернативній формі.

Капітан судна, екіпаж судна. Існуючі стандарти передбачають, що основною функцією капітана судна є управління суднами у сенсі навігації. У випадку автономного судна функція управління навігацією судна автоматизована та забезпечується або повністю судновою технічною системою, або під керівництвом судновласного прибережного персоналу. Функції управління суднами, включаючи відправлення, щодо напівавтономного судна можуть бути виконані судовими автоматичними пристроями або фахівцями судновласників, розташованими за межами автономного судна.

Мінімальний судновий екіпаж. Для повністю автономних суден ця вимога не застосовується взагалі. Сертифікат мінімального складу напівавтономного судна повинен враховувати ступінь автоматизації (автономії) судна. Оскільки повністю автономне судно не має на борту екіпажу, документ, що встановлює кількість екіпажу є безглуздям.

Вимоги до кваліфікації персоналу. Управління автономним судном, а також у присутності екіпажу і при його відсутності слід підтримувати або здійснюватися за допомогою фахівців поза автономним судном. А до членів екіпажу автономного судна та фахівців з управління автономними судами повинні розробляти та встановлювати кваліфікаційні вимоги. Найбільш відповідним правовим інструментом для встановлення кваліфікаційних вимог до зазначених членів екіпажу та фахівців є положення про дипломи членів екіпажу морських суден.

Управління судном. Відповідальність за безпечне управління автономним судном може бути покладена на судновласника, який повинен мати спеціалістів, компетентних у сфері управління автономними суднами. Такі фахівці виходять за межі автономного судна, контрольованого ними (на березі або на іншому судні або кораблі), але повинні мати всі необхідні інструменти технічного та організаційного характеру для управління суднами. Судновласник також має обов'язок призначити особу, відповідальну за управління автономним судном щодо кожного автономного судна (прибережного капітана). Ця відповідальна людина може одночасно керувати кількома суднами. Оскільки управління автономними суднами є дуже конкретним завданням, яке вимагає концентрації спеціальних компетенцій, пропонується надати право судновласнику укласти угоду про управління автономним судном зі спеціалізованою організацією, компетентною в управлінні автономним суднам. У той самий час, відповідальність за безпечну роботу автономного судна, як і раніше, лежить на судновласнику.

Такі коригувальні коментарі для запровадження нових технологій вже введені в морське законодавство. У 2018 році DNV GL розробив документ «Autonomus and remotely operated ships», у 2020 р. РМРС видав «Положення щодо класифікації морських автономних та дистанційно керованих суден (МАНС)». Вимогами до систем автономних та дистанційно керованих суден є надійність, безпека та ступінь автоматизації, що є не гіршими, ніж у суднах з екіпажем. Автономні судна повинні мати мінімальну кількість споруд, а також корисний простір судна та рекреаційних відділів.

На автономних суднах ставка робиться на локальну мережу судна та системи зв'язку. Локальна мережа повинна бути реалізована з можливістю функціонування в будь-якому агрегаті, тоді як несправне обладнання повинно бути виключено з мережі під час усунення невдачі.

Однією з опцій інтегрованої системи є обчислювальна система, яка відповідає за управління суднами, повинна бути спроектована за мажоритарним принципом, з обчислювачем, який видає некоректні значення, що повинні бути відключені від загальної системи під час перезавантаження та перевірки, після чого він повинен бути реінтегрованим.

Неможливо відхилити такий аспект, як реалізація автономних суден з можливістю участі у рятувальних операціях, для яких необхідно обладнати судно штучним інтелектом у формі робототехнічних засобів.

У разі відмови працювати з автономним та дистанційно керованим судном воно повинно бути доставлене до найближчого порту або ремонтні роботи проводити на борту. У той самий час повинні бути вирішені такі завдання:

перевезення до найближчого порту;

організація доступу до судна;

оцінка доцільності ремонту.

Рішення цієї проблеми полягає у створенні товариства порятунку автономних та віддалено керованих суден, під егідою Міжнародної морської організації.

Наявність людини на борту віддалено керованого судна вимагає роботи професійного психолога з медичною освітою, вирішення психологічних проблем, таких як галюцинації, розвиток депресії, а також інші порушення людського організму.

Також не можна забувати про такий аспект морської діяльності, як піратство, який у різних формах існує з моменту зародження судноплавства.

У випадку фізичного захоплення центру дистанційного управління повинна бути реалізована система передачі повноважень управління на резервно-керовані структури, розташовані в інших центрах. Перехоплення суднового управління може здійснюватися на спеціально відібраних цілях — суднах. Крім того, такі атаки можуть бути замасковані як «піратська атака». Передача помилкових даних може бути безпосередньо спрямована на захоплення управління судном, але може побічно дозволити роботу судна на основі алгоритмів та неправильних рішень у центрі пульта дистанційного управління.

Особливості дизайну та експлуатації автономних та віддалено керованих суден вимагають:

— Впровадження суворих вимог до правових аспектів відносин транспортних операцій.

— Використання сертифікованого обладнання та програмного забезпечення.

— Впровадження модульного принципу обладнання.

— Можливості «гарячої заміни» обладнання.

— Уніфікація органів управління [11].

Але виробничі технології не стоять на місці, і багато іноземних компаній не очікують створення та затвердження нормативної бази для безпілотних суден та швидко розвиваються та створюють повну модель автономних суден та самого автономного судна. Такими компаніями є:

— Англійська Rolls-Royce Convice, яка разом з Finferries у грудні 2018 року, поблизу Турку, спеціально перетворені під автономний режим 53. 8-метрового типу порому «Falco». Проект називався SVAN (Safer Vessel with Autonomous Navigation). Тести були успішними, пором контролювався з командного пункту, розташованого за 50 км від експерименту.

— Норвезька компанія Yara та Kongsberg Gruppen у жовтні 2017 р. завершили розробку повністю електричної автономної вантажівки, що називається Yara Birkeland. Як пише газета The Wall Street Journal перший у світі автономний вантажний корабель Yara Birkeland було введено в експлуатацію наприкінці 2018 року. Після спуску на воду розробники тестували системи автопілотів приблизно півтора року. Цей процес відбудеться в три етапи. На першому етапі судном управляє команда на борту. На другому етапі оператор дистанційно контролює тести, які здійснюються на маршруті довжиною 32.5 милі. Наступний етап — електроход керується власним комп'ютером, використовуючи GPS та численні датчики, щоб визначити положення інших морських об'єктів, а також для безпечного причалювання. Все вищезгадане про безпілотні судна та їх випробування доводить, що прогрес у розвитку безпілотних технологій не стоїть на місці, але з розвитком будь-якої технології, особливо в галузі морського транспорту, потрібно ретельно розробити нормативну базу. Але, на жаль, про досконале опрацювання ще рано говорити.

Основною проблемою, яка вирішується для суден без екіпажів, є дотримання вимог безпеки навігації та запобігання забрудненню навколишнього середовища, а також відповідальність судновласників як частина використання безпілотних суден. В даний час безпілотні морські дослідження MAS проходять в «комфортних» умовах:

— тестування судноплавства дронів відбувається на достатній відстані від берега з впевненим, безперервним, безперебійним сегментом зв'язку з безпілотним судном;

— за відсутності близькості до інших суден;

— якщо є хороша зона супутникового покриття;

— з повною відсутністю тіньових секторів для суднових РЛС;

— забезпечено безперебійну роботу AIC, GPS, гірокомпаса, лага, ехолота, за винятком помилок у переданій інформації;

— за наявності достатніх орієнтирів, глибин, СНО можливо дублювати позиціонування суден у просторі;

— при низькій швидкості маневрування;

— з повною відсутністю складних гідрометеорологічних чинників (штормовий вітер, хвилювання моря з високою бальністю, сильна течія, наявність значного льоду, айсбергів, інородних вільно плаваючих предметів з низьким коефіцієнтом відбиваючого сигналу).

Варто звернути увагу на ряд проблем, які в майбутньому виникнуть з суднами без екіпажу:

1) як боротьба за життєздатність судна без екіпажу у разі аварії будь-якого виду (защемлення, зіткнення, пожежа, контроль води та інше);

2) у випадку повного блок-ауто відмова всього джерела живлення, включаючи резервні джерела живлення; хто може відремонтувати і за який час;

3) дії безпілотника в піратських водах;

4) екстрена віддача якоря, хто буде бігти на бак, щоб віддати якір;

5) плавання в льодових умовах (в льоду не йдуть прямо). Потрібні постійні маневри та часті реверси мають підтримувати безпечну відстань попереду рухомого судна, льодоколу, несподівано до відкритих перешкод.

6) виникнення ситуації невизначеності при дотриманні правил МППСС-72. Невизначеність є вираженою рисою морського транспорту. З усією ретельністю вивчення питань навігації фактор невизначеності присутній і буде присутній навіть при вищому ступені автоматизації. Таким чином, основні напрямки у сфері безпілотних технологій на морському транспорті повинні розглядатися як проблема вирішення невизначеності сприйняття інформації під час розбігу суден, котрі рухаються в складних умовах, у причалі, перегляду правил для запобігання зіткнення та законодавчої бази у разі аварійної ситуації.

На підставі вище викладеного розробка технології безекіпажного судноводіння та його застосування ведеться багатьма країнами, організаціями та компаніями, які здійснюють дослідження, технологічні та транспортні проекти. Інтенсивні розробки ведуть як установи, так і університети — Массачусетський інститут технології, Колумбійський університет, Плімутський університет, Учоанський університет технологій та транснаціональні корпорації та класифікаційні суспільства. Взаємне використання технологічних проектів може дозволити впровадженню у майбутньому повністю перейти до будівництва безекіпажних суден, а також значно розширює можливості для розвитку суднобудівної промисловості. Детальне вивчення проектів у сфері безекіпажного судноплавства дозволить створити техніко-економічне обгрунтування будівництва таких суден та визначити необхідний та остаточний набір технологій для їх реалізації. Телекомунікації технології, електронні датчики та системи, технологія електронної навігації вже вводяться у секторі морської промисловості. Економічні судна є найбільш складним технічним транспортом, в якому особлива роль призначена для автоматичної системи управління. Найближчим часом неможливо уявити цивільний флот без автономних безекіпажних суден.

В роботі розглянуто, насамперед, регуляторні та правові аспекти, але не можуть бути відкинуті технічні, пов'язані з описом та аналізом існуючих проектів безекіпажних суден. Безекіпажні судна відрізняються за ступенем обладнання засобами вимірювальної техніки та устаткування. На безекіпажних суднах не існує ходового містка, надбудови, житлових приміщень та системи життєдіяльності екіпажу.

У травні 2018 року в рамках секретаріату Міжнародної морської організації була створена міжгалузева цільова група на морських автономних поверхнях суден. На 99-й сесії Комітет морської безпеки розпочав обговорення постійного регулювання сегмента автономного перевезення, включаючи людський фактор, систему безпеки, взаємодію з портами, пілотною проводкою, ліквідацією наслідків аварій та захисту морського середовища для кораблів різних рівнів автономності.

Слід також зазначити, що в рамках єдиного морського європейського проекту до 2025 року планується створити єдину «екосистему» логістики автономної доставки. У той самий час проект реалізується на принципах державно-приватного партнерства за участю провідних гравців морської промисловості: Wartsila, Rolls Royce, Abb, Meyer Turku, Finnferries, Ericsson, Cargotec та ін. [12]

Дефіцит часу на прийняття рішення щодо забезпечення безпеки судна призводить до необхідності виділення тільки тієї інформації, яка потрібна для виконання основного завдання управління і при-йняття рішень. Виникає проблема попереднього відбору та аналізу інформації, необхідної для реалізації механізму логічного висновку і вироблення практичних рекомендацій [13]. З такою проблемою справляються інтелектуальні системи, розраховані на експлуатацію в контексті певних невизначеностей протягом тривалого періоду часу. Відзначається, що для того, щоб система відповідала інтелектуальній автономії, вона повинна володіти однією або декількома наступними можливостями: навчання; ситуаційна обізнаність; міркування; планування; людино-машинні інтерфейси; прийняття рішень; приведення в дію. Під «автономним судном» розуміється — «морське судно з датчиками, автоматизованою навігацією, руховими і допоміжними системами, з логікою прийняття рішень для проходження по планам місії, налаштуванням виконання місії і роботи без втручання людини», — представлено у звіті американського бюро судноплавства (ABS) про автономні судна (Autonomous Vessels: ABS 'Classification Perspective) за 2016 рік [14].

Відповідно до звіту ABS про автономні судна (Autonomous Vessels: ABS' Classification Perspective) за 2016 рік існують такі рівні автоматизації систем [14]: людський контроль (human control), деякі функції автоматизації (some functions automated), звичайні операції автоматизації, людина готова взяти на себе відповідальність (normal operations automated; human ready totake over), критичні функції безпеки автоматизації, людська присутність (safety-critical functions automated; human present), повна автономія критичних функцій безпеки і моніторингу навколишнього середовища на час рейсу (full autonomy of safety-critical functions and environmental monitoring for duration of trip), повна автономія без доступних для людини інтерфейсів управління (full autonomy with no human-available control interfaces).

В даний час безпілотні засоби — це концепт-проект, в якому увагу зосереджено на інтегрованій сенсорній технології і попередженні зіткнень. Основні елементи інтелектуальної системи, що використовується, забезпечують управління безекіпажним судном: автономна навігаційна система, система попередження зіткнень, система моніторингу та управління двигуном, автоматизовані системи швартування.

Швартовні операції суден (рисунок 1) входять до складу небезпечних процесів судноплавства. Прогрес в безпеці швартування суден досягається в результаті використання інноваційних техно-логій. На багатьох причалах існує досвід роботи таких технологій. Але статистика свідчить, що такі системи швартування ще не дійшли до такої досконалості, щоб повністю виключити небезпечні ситуації.

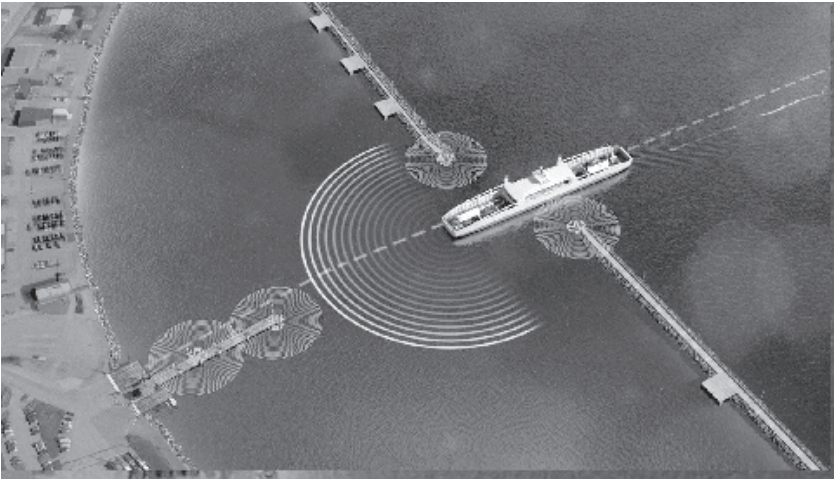

Рис. 1. Застосування інноваційних технологій швартування суден

Міжнародна морська організація схвалила вимоги до безпеки швартування суден та дизайну обладнання [10].

Нові вимоги включатимуть оцінку повної лінії причалів, включаючи новий режим технічного обслуговування та швартове обладнання.

На 102-му засіданні Комітету з питань безпеки моря Міжнародної морської організації було прийнято пакет обов'язкових вимог, у тому числі щодо безпеки еквівалентних операцій, повідомляє прес-реліз Міжнародної морської організації.

На засіданні Міжнародної морської організації одна із прийнятних вимог пов'язана з причалами, що повинні і забезпечити захист праці та безпечне пришвидшення суден, а також зменшити кількість наслідків аварій, які відбуваються під час роботи. Можливо, що результати цієї зустрічі матимуть значний вплив на проектування судна, зокрема на пристрої швартових лебідок та відповідного обладнання на палубі.

В даний час інструкції SOLAS (II-2 / 3−8) вважаються правилами для палубного обладнання, що використовується для причальних операцій, шляхом визначення максимально допустимого навантаження для кожної одиниці обладнання та оснащення.

Інструкції спрямовані на запобігання обмеженню доступу до робочої області та мінімізацію обмеження видимості пропорційної зони, щоб уникнути впливу динамічних навантажень швартування персоналу, що бере участь у причалюванні.

Це правило II-2 / 3—8 додає нові предмети до вимог дизайну. Спеціальна інформація повинна бути включена в так званий план буксирування та причалювання, описаний у нових інструкціях для проектування Міжнародної морської організації 1/1620 «Рекомендації щодо проектування причальних пристроїв та вибору відповідного швартового обладнання та оснащення для безпечного причалювання». У той самий час затвердження плану не вимагається адміністрацією порту прапора.

Що стосується інспекції та технічного обслуговування, Комітет прийняв нові положення для всіх суден, незалежно від розміру та дати будівництва судна, вимагаючи перевірки обладнання причалу, включаючи кабелі та мотузки. Перевірка Shridge зараз включає кількість, силу, розмір, довжину, характеристики та обмеження. Подальші стандарти містяться в новому посібнику MSC.1 / CHERC.1621 «Рекомендації щодо інспекції та обслуговування вологого обладнання, включаючи швартування».

Вимоги до швартовних пристроїв є особливо актуальними для конструкторів суден та суднобудівників, і їх слід обговорювати з клієнтом-судновласником. Вимоги до перевірки технічного обслуговування та заміни зіпсованого обладнання в основному важливі для судновласників та операторів.

Поправки набирають чинності з 1 січня 2024 року.

Процес швартування можливо розділити на кроки:

1) підхід судна на бажану відстань до причалу з поворотом у правильному напрямку;

2) підхід судна за допомогою буксирів до певної позиції по відношенню до причалу та утримування у цьому положенні під час подання швартування;

3) затягування судна до причалу за допомогою спеціального обладнання [15]. Кожен з цих етапів причального процесу індивідуальний.

Захід судна в порт призначення, з точки зору складності навігації, забезпечення радіоелектронного контролю та контролю інформацій-ного навантаження, може бути диференційованим на ряд етапів, які можна назвати фазами судна. Такий поділ дозволяє більш чітко висунути вимоги до точності визначення місця розташування судна у порту, а також дозволить оцінити безпеку суднобудівництва та заходи щодо зменшення впливу на можливі надзвичайні ситуації на кожному етапі судноплавства. Таке розділення забезпечується в 1983 році постановою Міжнародної морської організації А.529 (13), але це стосується вузького аспекту судноплавства: точність розташування [15]. У той же час безпека суднозаходу залежить від великої кількості факторів. Виходячи з вищезазначеної практики суднових вод, можливо окремо розібрати чотири етапи суднозаходу.

Перший етап суднозаходу.

Відповідно до Резолюції А.529 (13) «Компенсаційні стандарти» рейс судна можна розділити на вхід до гавані та підходи до неї, а також воду, яка має обмеження маневру та інші води. Для цього поділу майже всю відстань першого етапу суднозаходу відносить до етапу «інших вод». Для цього місця шлях руху точність навігації не повинна бути гіршою, ніж 4 % від відстані небезпеки, але не більше 4 морських миль [15]. На цьому етапі рейсу говоримо про відстані до небезпеки, що розраховуються десятками миль. Тоді порядок необхідної точності є долі та одиниці миль.

Саме цей етап суднозаходу, як показують статистичні дані, характеризується відносно низькою ймовірністю інцидентів та катастроф, що пояснює вказаний знижений рівень запиту до точності визначення координат судна.

Слід зазначити, що перераховані параметри систем спостерігаються в ідеальних умовах. Але на практиці існують особливості, які обмежують можливості перелічених систем, такі як обмеження здатності діапазону узбережних РЛС ($\pm$ 75 м), наявності зон та «радіотіні», а також екранування великим судном навколо розташованого невеликого судна.

Аналіз зазначеної інформації, а також параметрів суднових навігаційних інструментів дає можливість зробити висновок: у першій інформаційній фазі суднозаходу перераховані засоби навігації, залежно від їх ідеальної роботи, задовільно забезпечують необхідну точність визначення розташування судна, що заходить у порт.

Другий етап суднозаходу.

Другий інформаційний етап суднозаходу відноситься до стадії рейсу, визначеного в «стандартах оцінки» [15], як «вхід до гавані та підходів до неї, а також воду, в якому свобода маневру обмежена». Відповідно до [15], «вартість допустимої похибки місця залежить від місцевих умов, а його визначення є функцією відповідних адміністрацій».

Зрозуміло, що в умовах обмеженого маневру підходу до воріт фарватера потрібна більша точність місцезнаходження, ніж це було на першому етапі. Ця заява є актуальною, оскільки періодично спостерігаються навали суден на воротах фарватера.

Таким чином, з точки зору інформаційного модуля для визначення точності навігації на третьому етапі суднозаходу, як у другому, існуючі системи задовільно виконують свої функції, але в критичних ситуаціях їх ненормального функціонування на судні необхідно мати резерв для автономного визначення простору з точністю щонайменше 10—15 л.

Четверта фаза суднозаходу.

Ця фаза характеризує причальний процес швартування судна. Особливість цієї фази носія судна полягає в тому, що підхід судна до причалу здійснюється, як правило, коли двигун швартовного судна вимкнено, що призводить до повної практично неконтрольованої поведінки. Тому завдання безаварійного швартування багато в чому визначається оператором, який контролює технологію швартування [16; 17].

Більшість надзвичайних ситуацій у причалі пояснюються відсутністю технічних засобів об'єктивного контролю підходу судна до пристані. Аналізуючи навали суден на причали та їх об'єктах, можна стверджувати, що існує точне знання не тільки розташування судна щодо причалу, а й облік впливу найбільш складного компонента будь-якого технологічного процесу.

Зростаючі вимоги до безпеки навігації на кожному з етапів суднозаходу висуваються не тільки для покращення технічного обладнання суден, а передусім зміцнення технічного контролю за діями оператора [18; 19]. Тому Міжнародна морська організація та адміністрація морських портів світу в останні роки здійснюють активну роботу зі створення [20]:

— розділення шляхів суднопроходів в місцях з інтенсивним рухом;

— зон з обов'язковими або добровільними радіоповідомленнями між суднами, коли вони наближаються один до одного або проходять;

— удосконалення системи управління рухом суден (СУРС) у портах та підходів до них з поступовим збільшенням автоматизації контролю за якістю судноплавства, доставки в морських районах;

— забезпечення засобами високоточного розташування суден у прибережних водах, використовуючи контрольні та коригувальні диференціальні станції глобальних навігаційних супутникових систем (ДГНСС) ГЛОНАСС та GPS-типу;

— суцільного радіочастотного покриття (виключити тіньові зони) прибережні смуги морських територій — глобальної морської комунікаційної системи та забезпечення безпеки (ГМСЗБ) з цілодобовим надійним УКВ-зв'язком;

— супутникової морської системи зв'язку ІНМАРСАТ, що забезпечує глобальне та ефективне спілкування з суднами, розташованими в будь-якій частині світу.

Як частина роботи, проведеної в Міжнародній морській організації, щоб переглянути главу 5 «Навігаційна безпека» Конвенції про захист людського життя на морі (SOLAS), передбачається найближчим часом вставити принципово нову автоматичну інформацію (ідентифікацію) системи (AIS) на морському флоті. Перша версія AIS, введена у всьому світі, виконує три основні функції:

— автообмін навігаційними даними між суднами, коли необхідна розбіжність у морі;

— передача даних про судно та його вантаж до прибережних послуг, коли воно плаває в контрольованих областях, з обов'язковими повідомленнями;

— перенесення навігаційних даних з судна до прибережних СУДС, забезпечуючи більш точну та надійну проводку в зоні дії системи.

Таким чином, автоматична інформаційна система — це морська навігаційна система, в якій взаємний автоматизований інформаційний радіообмін використовується як між суднами, так і між суднами та прибережними службами, під час яких вони передають інформацію про позивний та назву кожного судна (для їх ідентифікації), їх координати, параметри (розміри, навантаження, осадка тощо), цілі рейсу, параметри руху (курс, швидкість тощо) для вирішення проблем попередження зіткнень суден, моніторингу дотримання режиму плавання та загального моніторингу статусу безпеки в контрольованому морському районі.

Серед найважливіших компонентів розвитку мережі АІС слід розглядати введення служби диференціальної підсистеми глобаль-них навігаційних супутникових систем (ДГНСС) типу американського GPS, додавання яких діфпідсістемами вирішує проблему високоточного визначення місця судна з підводною точністю (d5 M).

Таким чином, незважаючи на широке впровадження високоточних навігаційних систем (ГНСС ГЛОНАСС, GPS), а також засобів автоматичної ідентифікації суден (AIC), проблема забезпечення безпеки швартування залишається на останньому етапі суднозаходу [16]. Четвертий етап швартування є найскладнішим та відповідальним, що вимагає безперервного радіоелектронного контролю процесу наближення судна до причалу, а вимірювання до причалу потрібно визначити з точністю долі метра.

Аналіз літератури показує, що основний резерв вдосконалення потребує вдосконалення інтелектуально-інформаційних систем швартування різних типів. Потрібно враховувати, що кожна швартова робота має свої особливості відповідно до погодних, кліматичних, структурних особливостей суден [21], незалежно від того, яка ситуація, надзвичайна ситуація або вантажна робота:

— швартування до судна, що лежить у дрейфі;
— швартування до судна, що рухається;
— швартування до судна, що стоїть на якорі;
— швартування кормою до причалу в портах.

Автоматизовані системи швартування раціоналізують експлуатацію причалу і забезпечують максимальну віддачу роботи порту. Швартування в автоматичному режимі гарантує надійне кріплення судна і надає переваги з точки зору технічної та екологічної безпеки. Функції сигналізації таких систем використовуються в режимі реального часу.

Для проникнення в сутність автоматизованих процесів швартування і стикування виділяють основні автоматизовані пристрої швартування і стикування: магнітний; лазерний і вакуумний.

Якщо аналізувати швартувальні операції, побачимо, що прийняті заходи, хоча зменшують рівень аварійності, але проблема безпеки залишається актуальною завдяки багатьом факторам, включаючи слабке інформування судноводія про поточні параметри швартування. Для покращення управління та прийняття належного рішення судноводію потрібні спеціальні технічні засоби швартування, що оперують даними вимірювань реальних параметрів розташування, швартувального судна за допомогою системи, яка отримує та обробляє поточну інформацію про зміну цих параметрів, що видає оператор судноводію в зоровій формі небезпечної ситуації та що сигналізує звуковим сигналом, кольоровим сигналом. Технічним варіантом такого рішення може бути система аналізу швартування та контролю над підходом судна до причалу з попередженням поточного середовища — індикатор безпеки, наприклад, який є технічним інструментом, який виконує прийом, обробку та зберігання інформації, необхідної для швартування. Відповідно до використання розробленої прогностичної математичної моделі та програмного забезпечення цей пристрій повинен відображати результати обробки даних на екрані індикатора, як у формі ймовірнісної картини підходу судна до причалу, так і рекомендації судноводію у формі резервного часу та прогнозованої відстані. Ці вимоги на вищому технічному рівні застосовуються до безекіпажного судноплавства.

Розглянемо характеристики систем швартування, які прийняті для безекіпажного судноплавства:

Магнітна система автоматичного швартування (рисунок 2) контролює процес за комплексом динамічних впливів. Система складається зі здвоєних мертвих якорів, носових і кормових, оснащених магнітними подушками для надійного і міцного кріплення до будь-якого корпусу, плоского або вигнутого, пофарбованого або покритого корозією.

Подушки можуть пересуватися за корпусом з урахуванням змін по висоті. Магнітна система має жорсткі обмеження: дорожнеча експлуатації (енергоспоживання, професійне обслуговування), додаткове навантаження на корпус (вітри можуть деформувати борт) — подушки потрібно ставити в районі шпангоутів, що означає пристосовуватися під конкретний корпус. У надійності є сумніви, оскільки трос, що лопнув, замінюється, а подушка, що вийшла з ладу, — не має заміни. Дана система знайшла своє застосування на швидкому прийнятті рішення судном — пороми. Є заборона використання системи автоматичного навантаження на танкерах. Така заборона випливає з того, що система при поздовжньому «протягуванні» корпусу вздовж причалу, під дією хвилі або проходить поруч судна, змінює навантаження в шпринг і поздовжніх з системою контролю натягу кінців. Така дія призводить до деформації корпусу.

Застосування технології вакууму в швартових операціях і її розробки ведуть багато компаній, які здійснюють науково-дослідні, технологічні системи швартування.

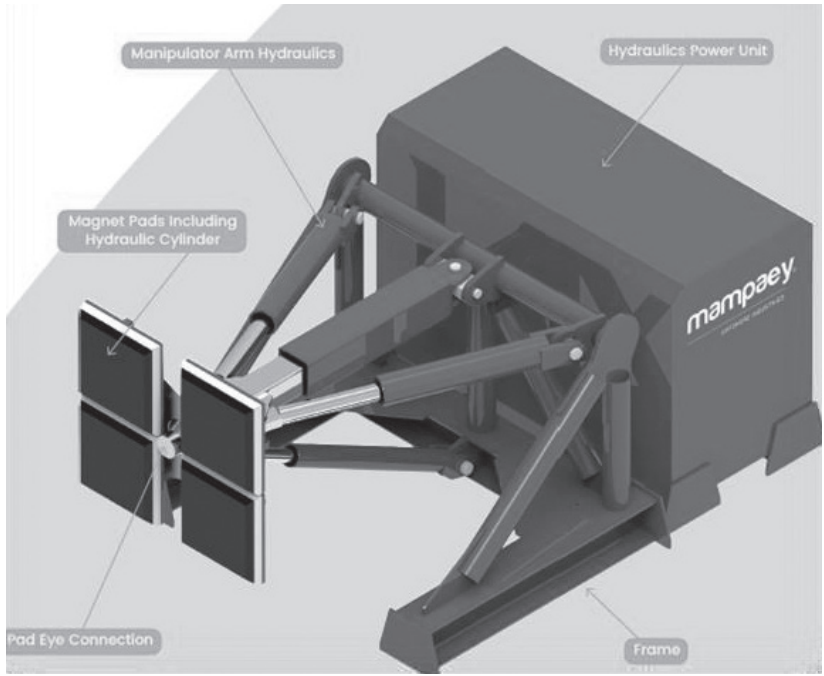

Рис. 2. Конструкція системи швартування «The intelligent Dock Locking»

Вакуумні системи автоматичного швартування (рисунок 3) для утримання судна біля причалу замість канатів використовують вакуумні подушки. У кожної подушки є своє контрольоване робоче навантаження, яке може забезпечити надійне фізичне з'єднання судна з причалом. Вакуумні подушки випробовуються і класифікуються під наглядом міжнародного класифікаційного товариства Det Norske Veritas (DNV), результати якого суміщені з сучасними тривимірними апаратними засобами, показують діапазон ходів і пружну еластичність автоматичних систем на рівні швартування за допомогою канатів. Інформація про навантажені утримання надходить від вимірювання рівнів вакууму і поперечних сил в носовій і кормовій тумбах. Оскільки вакуумна система може тримати судно ближче до причальної стіни, ніж перехрещений канат, ця система має причальну продуктивність. Маючи інформацію про всі умови швартування в режимі реального часу, оператор повністю контролює швартувальний стан судна.

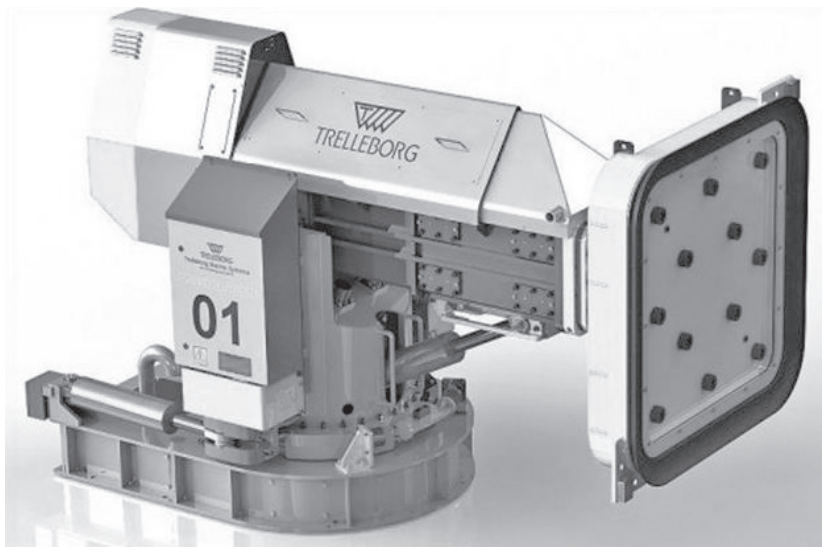

Рис. 3. Вакуумна система швартування «Auto Moor»

Такі системи високоефективні. Тисячоліття для швартування суден використовують канати, але протягом багатьох років процедура не стає менш небезпечною.

Автоматичні вакуумні швартові системи вирішують проблему натиском однієї кнопки. І менше ніж за 1/4 хвилини дозволяє спрацювати. Такі системи тримають судна всіх розмірів у одному місці в портах, де присутні брижі та довгі хвилі.

Лазерна система швартування (рисунок 4) яка відноситься до класу інструментальних систем, безперервно веде розрахунок дальності до судна кожним далекоміром. На основі отриманих даних система обрисовує візуальне положення судна з розрахунковим кутом щодо пірсу. Крім дальності і кута, система розраховує швидкість зближення або віддалення з пірсом як носа, так і корми. У разі наближення судна на близьку відстань з перевищенням зазначених в налаштуваннях швидкостей, відразу сигналізує про це через індикацію в інтерфейсі, а також через сирену на пірсі.

Оцінюючи характеристики систем швартування, розробники безекіпажних судів схильні до використання лазерних систем. Основною особливістю другого етапу швартування є зближення судна з причалом. При швартуванні часто неминучий контакт судна з при-чалом з ненульовою швидкістю, який називається навалом. Класифікують навали як навмисні, так і випадкові, що виникають при контакті судна з причалом, іншим стаціонарним об'єктом або з іншим судном, розташованим на відстані або паркуванні [15]. Оскільки розміри судна фіксуються, єдиним способом уникнути навалу судна на причалі є ретельне вимірювання швидкості підходу судна та відстані до причалу. Розміщення судна до причалу, як правило, здійснюється за допомогою буксирів [15; 16]. З початку етапу зближення вони розгортають судно правою стороною паралельно причалу. Завдання буксирів включає підводку судна близько до причалу і тримання його в цій позиції, доки не будуть заведені та обтягнуті швартові. Постановка судна до причалу є небезпечною технологічною операцією, яка вимагає кваліфікованих та своєчасних дій судноводія, який управляє технічними засобами судноводіння, доставки та буксирів при зближенні судна з причалом [15; 16].

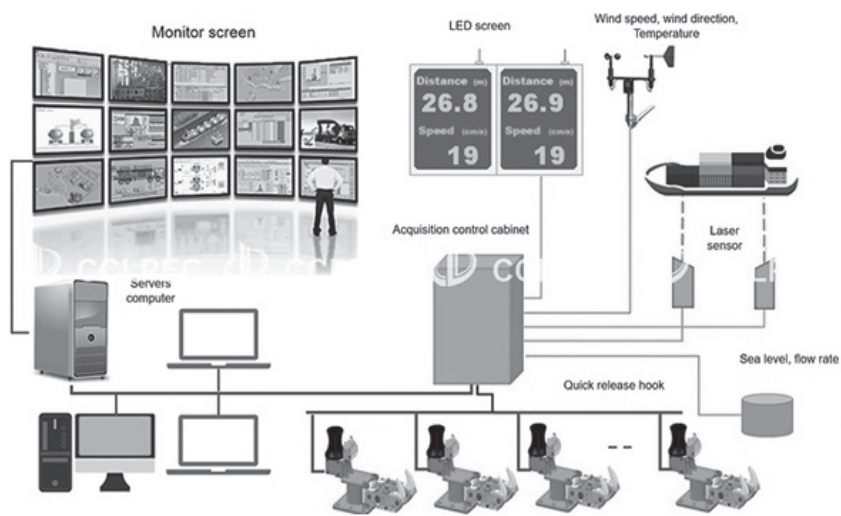

Рис. 4. Концепція лазерної системи швартування

Відмітна характеристика лазерної системи — сканування лазерного променя у вертикальній площині і отримання профілю відстаней з подальшою обробкою і визначенням відстані до причалу чи іншого об'єкта. Якщо аналізувати особливості лазерної дальнометрії, вона покаже, що експлуатація дальнометрії на причалах виявила свою на-дійність, легкість експлуатації, мінімум технічного обслуговування. Однак ця система не може вирішити всі проблеми порушення правил, які найчастіше встановлюються адміністрацією порту під час швартування. Тому необхідно більш глибоко розглянути можливість інформаційної підтримки причального процесу. Аналіз показує, що для забезпечення безпеки швартування використовується не весь інформаційний потенціал цих систем. Виявилося, що лазерна система швартування, що формується програмним забезпеченням для лазерної системи, дозволяє розширити інформаційну підтримку процесу швартування. Така система працює на підставі відстані з двох датчиків, розраховує швидкість і прискорення щодо цього об'єкта, а також можливо і визначення центру обертання судна. За рахунок застосування в пристрої датчика кута нахилу при розрахунках будуть компенсуватися качка корабля і різні розмірені параметри суден і причалів. У цих системах виробляються точні вимірювання в реальному часі, дані про відстані і швидкості судна. Інформаційні потоки даних за параметрами зближення судна з причалом, виміряні лазерною швартовою системою великотоннажних суден (ЛСШКС), передається до телеметричного пристрою за допомогою інтерфейсу модуля, призначеного для поєднання цих технічних засобів. Через передавальну антену поточна інформація надходить у радіопристрій, розташований на судні. Перетворена для подальшої обробки інформація через інтерфейс надходить у блок обчислення, де параметри розраховуються протягом всього швартування судна. Індикатор забезпечує такі режими візуалізації:

1. Відображає розраховану ймовірність перевищення швидкості судна під час контакту з причалом.

2. Хронологія зближення судна з причалом.

3. Комбінований режим відображення фактичної швидкості судна та ймовірність перевищення швидкості на момент контакту з причалом.

4. Автоматичний режим включення аналізу поточної ситуації з прогнозними оцінками підходу судна до причалу. Найбільш інформативним є автоматичний режим включення аналізу поточної ситуації з передбачуваними оцінками підходу судна до причалу, що дозволяє попередити судноводія про те, яка ситуація може виникнути під час першого дотику судна до причалу.

«Trelleborg», «Strainstall», «A. & Marine (Thai) Co., Ltd.» і «MARIMATECH» є ключовими розробниками автоматизованих

систем швартування в світі. Система «Smart Dock», розроблена «Trelleborg», складається з двох лазерних датчиків, контролера і центрального персонального комп'ютера. Дані про процес стиковки, а також аварійні сигнали при досягненні ризику критичних меж подаються декількома способами, в тому числі за допомогою великого екрану на причалі. Персональний комп'ютер в центрі управління реєструє дані і забезпечує графічне представлення всього процесу [22].

Система швартування «MARIMATECH» використовує два лазери, які встановлені на пристані і міряють відстань до сторони наближення суден, далі обчислює швидкість і кут нахилу судна. Концепція системи «MARIMATECH» заснована на дистанційній передачі даних. Дані відображаються на встановленому на причалі цифровому великому екрані, бездротових пристроях, таких як портативні пейджери або кишенькові персональні комп'ютери, на комп'ютерних моніторах диспетчерської.

Лазерна система швартування «Dock Aler» від «Strainstall» використовує блоки безпечного для очей лазера, встановлені по обидві сторони від головки причалу, для вимірювання відстані від носа до корми щодо причалу, а також забезпечує швидкість і кут нахилу судна до причалу. Дані від цих лазерів надходять в центральну систему управління, де вони можуть відображатися в диспетчерській, і передаватися переносним пейджерам, кишеньковим персональним комп'ютером і / або дисплеєм [23].

**Висновки:** Світові концерни, дослідницькі компанії роблять спроби для втілення концепції безекіпажного судноводіння в реальність. Для цього необхідно поєднати безаварійну експлуатацію судна та законів держави прапора, рішення портових органів загальної міжнародної юрисдикції. Застосовувані інновації, що використовуються в безекіпажному судноводінні, дали привід для дискусій в журналах, на конференціях і семінарах з розвитку судноплавства. У судноплавстві одна з фундаментальних змін — це реалізація концепт-проектів безекіпажного судноводіння. Цей напрямок включає одну з переваг — підвищення безпеки судноводіння за рахунок використання інновацій.

У роботі представлено огляд рівнів управління автономними системами в судноплавстві. Наголос зроблено на системи швартування. Наведено характеристики вакуумних, лазерних, магнітних систем. Безекіпажне судноводіння схиляється до використання систем ла-зерного швартування за рахунок розширення можливостей інформаційного забезпечення швартування із застосуванням системи, яка видає рекомендації у вигляді резервного часу і прогнозованої дистанції, що знижує ймовірність помилок. Як підсумок — безпека швартування підвищується.

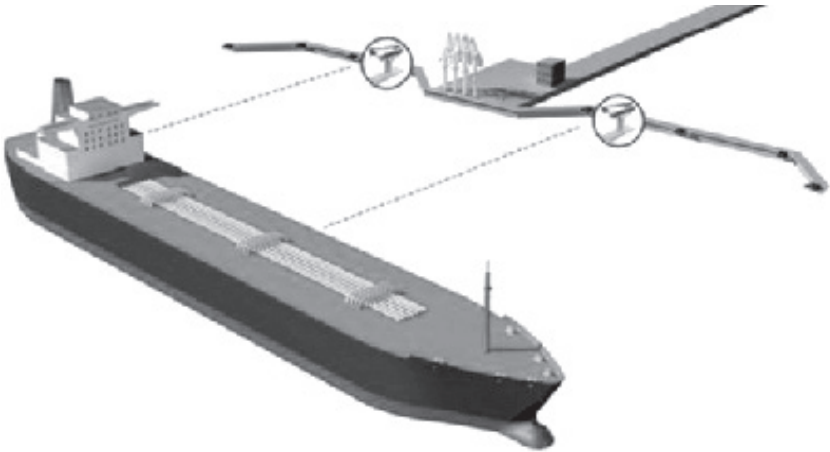

Рис. 5. Стикувальна система «Dock Aler» [23]

# АВТОМАТИЧНИЙ СИНТЕЗ МЕРЕЖ ПЕТРІ ПРИ РОЗРОБЦІ АЛГОРИТМІВ ЛОГІЧНОГО УПРАВЛІННЯ

*Гурський О. О.*

*У роботі розглядається актуальна задача, пов'язана з розробкою методів автоматичного синтезу мереж Петрі. Важливість розробки цих методів обумовлена розвитком інтелектуальних систем, що забезпечують автоматизацію трудомістких процесів.*

*Запропоновано принцип автоматичного синтезу мереж Петрі та певних алгоритмів логічного управління на основі функціонування штучної нейронної мережі. Представлений математичний опис методу зміни коефіцієнтів міжнейронних зв'язків мережі при синтезі мереж Петрі.*

*У програмному середовищі Matlab/Simulink 2012a були проведені експерименти, пов'язані зі спільним функціонуванням штучної нейронної мережі і мереж Петрі. Функціонування мереж Петрі в середовищі Matlab/Simulink було представлено за допомогою Statflow діаграм. У результаті експериментів були отримані часові характеристики функціонування штучної нейронної мережі, яка забезпечує композицію мереж Петрі. На основі часових*

*характеристик була встановлена принципова придатність застосування штучної нейронної мережі для забезпечення автоматичної композиції мереж Петрі.*

*В результаті було вирішено задачу, яка пов'язана з розробкою системи спільного функціонування нейронної мережі і мереж Петрі для формування алгоритмів та послідовних обчислень. Тим самим одержали подальший розвиток методика автоматичного синтезу мереж Петрі та методика розробки певних алгоритмів на основі функціонування нейронної мережі.*

*The important task was solved during the scientific research related to the development of the methods for automatic synthesis of Petri nets. The importance of development of these methods is due to the evolution of intelligent systems. These systems provide the automation of labor intensive processes.*

*The principle of automatic synthesis of Petri nets and the implementation of certain algorithms for tuning complex control systems based on the functioning of an artificial neural network are proposed. The mathematical description of the method for changing the coefficients in neural connections of network in the synthesis of Petri nets is presented.*

*The experiments were conducted in the Matlab\Simulink 2012a environment. These experiments were bound to the joint functioning of an artificial neural network and Petri nets. The functioning of Petri nets was presented in the Matlab \ Simulink environment using Statflow diagrams.*

*As a result of the experiments we have obtained the temporal characteristics of the functioning of artificial neural network providing the composition of Petri nets. The fundamental suitability of using artificial neural network to provide the automatic composition of Petri nets was determined on the basis of analysis of temporal characteristics.*

*The problem linked to the development of system for the joint functioning of neural network and Petri nets for the formation of algorithms and sequential calculations was solved in this work. Thus the method of automatic synthesis of Petri nets and the method of developing of the certain algorithms based on the functioning of a neural network were further developed.*

Мережі Петрі як прикладний математичний апарат досить відомі в області моделювання і аналізу дискретних динамічних або логіко-динамічних систем. Також мережі Петрі відомі як форми представлення паралельних алгоритмів і обчислень [1].

Актуальність розробки принципів автоматичного синтезу мереж Петрі лежить в області автоматизації процесу розробки алгоритмів логічного управління. Як приклад варто відзначити так звану задачу про «розумну мурашку», яка представлена в роботах [2; 3]. Мураха за допомогою проб та помилок, мутації будує автомат своєї поведінки. Але безсумнівно, що процес синтезу алгоритму носить інтелектуаль-ний характер, в даному випадку можливо задіяти відповідну інтелектуальну технологію, пов'язану зі штучними нейронними мережами і їх алгоритмами навчання [4].

Таким чином, нами вирішується задача, пов'язана з розробкою принципів синтезу алгоритмів і відповідних композицій мереж Петрі на основі певної інтелектуальної технології.

У роботі представлено етап розвитку певної інтелектуальної системи до застосування штучної нейронної мережі та алгоритми настроювання штучних нейронних мереж, пов'язаних з автоматичним синтезом мереж Петрі.

Відомо, що можливість автоматичного синтезу та існування методів автоматичної побудови мереж Петри відзначалися ще в роботі Джеймса Пітерсона [5]. З того часу з'явився ряд наукових публікацій, пов'язаних з автоматичною генерацією і композицією мереж Петрі, у яких відображаються особливості побудови мереж Петрі, засновані на певних методах [6—9]. Аналогічно з'явилася робота [10], у якій автоматичний синтез мереж Петрі здійснюється на основі методу перевірки досяжності дискретно-безперервних мереж [11; 12]. Якщо метод перевірки досяжності деякого стану в просторі змінних пов'язаний з перетворенням дискретно-безперервної мережі (ДБ-мережі) до одного переходу (елемента мережі Петрі), то при автоматичному синтезі мережі Петрі процес зворотний, з одного переходу перетворюється мережа Петрі, яка представляє досяжність деякого стану гібридної системи (з керованою структурою — СКС) [13; 14].

Досяжність стану СКС можна забезпечити при наявності відповідного алгоритму логічного управління. Такий алгоритм дозволить реалізувати $k$-процес функціонування СКС $\sum_{I+1}$. У даному випадку повинна існувати послідовність запусків переходів ДБ-мережі, що представляє модель системи. Таким чином, засоби ДБ-мереж дозволяють досліджувати досяжність системи на основі правил редукції мережі, а правила редукції мережі і методи перевірки досяжності відіграють важливу роль у розробці формуючого автомата, синтезуючого мережу Петрі.

Далі розглянемо формування алгоритмів на базі методу перевірки досяжності ДБ-мереж.

**Формування алгоритмів на базі методів перевірки досяжності дискретно-безперервних мереж.** Перевірка досяжності системи шляхом редукції безперервної і дискретної частин ДБ-мережі полягає в «згортці» мережі за певними правилами до макропереходу. Таким чином, при формуванні алгоритму дискретну мережу Петрі також можна розгорнути — сформувати аналогічно, як згорнути до макропереходу.

Для формування мережі Петрі, що представляє алгоритм логічного управління, були виділені такі правила формування матриці інцидентності $W$:

— рядок повинен починатися з $0$ або $+1$ і значення в рядках повинні чергуватися — $0, +1, 0, -1, 0$ і т. д.;

— поява хоча б однієї $+1$ у стовпці повинна супроводжуватися появою хоча б однієї $-1$ у тому ж стовпці;

— у рядку не може йти підряд дві та більше $+1$ або $-1$ навіть через нулі;

— для виключення формування занадто складних алгоритмів в одному рядку не може бути більше двох пар $+1, -1$.

Фрагмент алгоритму формування матриці інцидентності згідно з вищенаведеними правилами представлений у вигляді Stateflow діаграми на рисунку 1. З рисунка 1 видно, що автомат представлений паралельними станами StateC4, StateC5, StateC6 ... StateC$N$, де $N$ — кількість рядків формованої матриці інцидентності. Перехід з підстану State19 або State24 у підстани State20 або State25, тобто поява $-1$, супроводжується переходом з підстану State22 або State26 у підстани State23 або State27 (появою $+1$ у тому ж стовпці).

Слід зазначити, що перехід зі State19 або State24 в State20 або State25 може супроводжуватися залежно від умови data01$> -10$ з витримкою за часом або без умови. У цьому випадку формується мережа Петрі, у якій мають місце переходи з наступними умовами спрацьовування:

$$\forall p_i \in I(t_j) : \mu(p_i) = 1 \,\&\, J_{0j} < g \,\&\, t < t_k\,,$$

$$\forall p_i \in I(t_j) : \mu(p_i) = 1 \,\&\, t < t_k\,,$$

де $\mu(p_j)$ — маркування вхідних позицій переходу $t_j$; $g$ — граничне значення $J_{0j}$ деякого критерію якості роботи системи; $t_k$ — час витримки по спрацьовуванню переходу.

Переходи з різними умовами спрацьовування вибираються залежно від значення сигналу формування алгоритму data14 або data15 і т. д. Сигнали формування алгоритму $V1 .... Vn$ відіграють важливу роль у визначенні динаміки станів автомата формування матри-ці інцидентності мережі Петрі. Поява тієї або іншої одиниці *+1* реалізується залежно від значення сигналів *V1 .... Vn*. Наприклад, перехід зі стану State22 в State23 може здійснюватися за умовою [data24>0.5&data15>2.5], де data24 — локальна змінна Stateflow-діаграми, а data15 — значення сигналу *V1*.

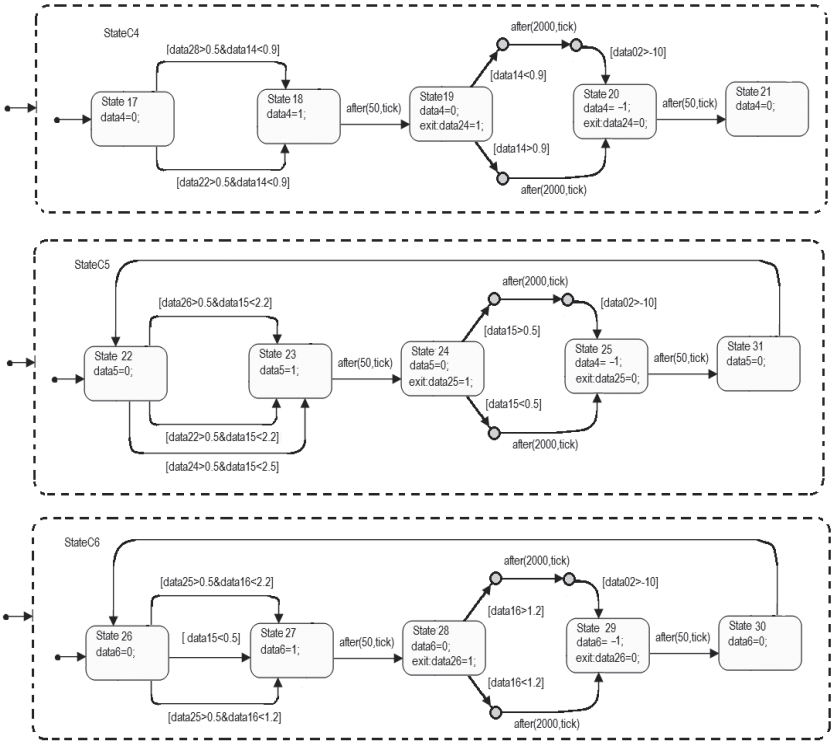

Рис. 1. Stateflow-діаграма, що представляє автомат формування матриці інцидентності (зв'язків між елементами мережі Петрі)

Таким чином, дана Stateflow-діаграма здатна представляти велику кількість усіляких алгоритмів, навіть можливо непередбачуваних експертом. Створення алгоритму визначається залежно від сигналів *V1 .... Vn*, а ці сигнали можливо коректувати, якщо алгоритм є незадовільним.

Як показано на рисунку 2, залежно від установлених сигналів *V1 .... Vn* і за значеннями показників — $J_{01}$, $J_{02}$, $J_{03}$ автомат формування матриці інцидентності виробляє послідовність значень, з яких складається матриця інцидентності мережі Петрі.

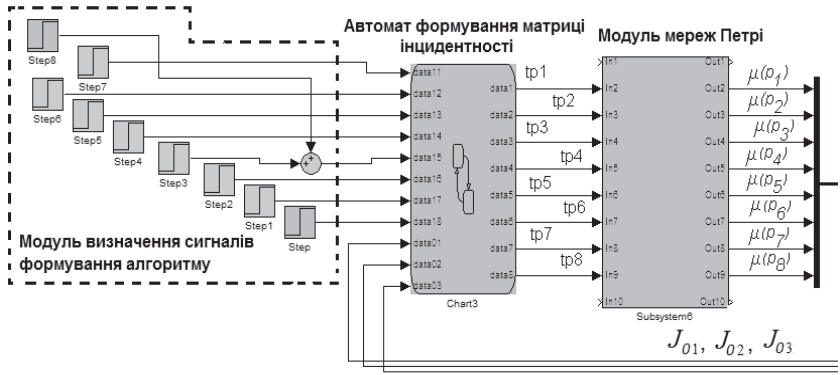

Рис. 2. Структурна схема, що відображає модель формування мережі Петрі

Однак виключення експерта в побудові мережі Петрі та у визначенні деякого алгоритму управління веде до того, що представляється відсутність прикладного характеру синтезованої мережі Петрі. У такому випадку в роботі [15] була запропонована автоматична композиція мережі Петрі на базі функціонування нейронної мережі, яка представляє інтелектуальну технологію у визначенні деякого алгоритму логічного управління.

Надалі модуль визначення сигналів *V1 .... Vn* буде представлятися нейронною мережею, а Stateflow-діаграми показані на рисунку 1, можуть бути представлені відповідними синхронно функціонуючими мережами Петрі.

**Формування алгоритмів на базі функціонування штучної нейронної мережі.** Таким чином, синтез мережі Петрі на основі функціонування штучної нейронної мережі становить область формування й автоматичного синтезу мереж Петрі [5; 8]. Як показано на рисунку 3, у цьому випадку нейронна мережа взаємодіє на принципах зворотного зв'язку із синхронно функціонуючими мережами Петрі. Із цих синхронно функціонуючих мереж можливо сформувати композицію, яка буде відображати певний алгоритм дій, реалізованих штучною нейронною мережею. Матриця коефіцієнтів міжнейронних з'єднань вихідного шару нейронної мережі має певну аналогію з матрицею інцидентності мережі Петрі. Таким чином, нейронна мережа генерує вихідні сигнали, аналогічні матриці інцидентності синтезованої мережі Петри.

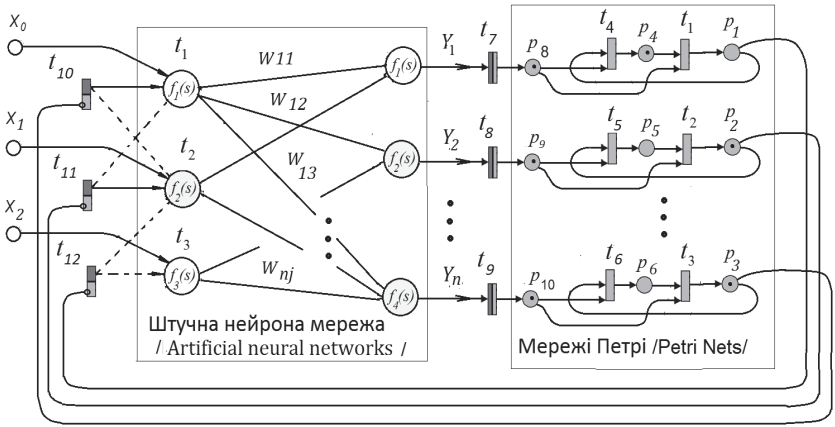

Рис. 3. Схема синтезу штучної нейронної мережі і мереж Петрі при формуванні відповідних алгоритмів логічного управління в системі, $w_{11}$... $w_{nj}$ — коефіцієнти міжнейронних з'єднань; $t_7$...$t_{12}$ — дискретно-безперервні переходи, що забезпечують зв'язок між штучною нейронною мережею і мережами Петрі

Функціонування мережі Петри можна описати рівнянням:

$$M_k = M_{k-1} + |A| \cdot U_{k-1},$$

де $M_k = |\mu(p_0), .... \mu(p_n)|^T$ — вектор маркування мережі Петрі на $k$-му кроці; $M_{k-1}$ — вектор маркування мережі Петрі на $k-1$ кроці; $|A|$ — матриця інцидентності, яка визначає взаємозв'язок позицій і переходів у мережі; $U_{k-1}$ — управляючий вектор.

При цьому для даного випадку, якщо $\mu(p_i) = 1$, де $i=1...n$, то змінюється значення відповідного параметра, наприклад, при настроюванні системи. Згідно зі схемою, представленою на рисунку 3, нейронна мережа представляє частину виразу $|A| \cdot U_{k-1}$ та генерує матрицю інцидентності $|A|$ синтезованої мережі Петрі.

Якщо певний алгоритм дій при функціонувані деякої системи незадовільний, то треба указати, на якому переході мережі Петрі була виявлена помилка в системі. Це необхідно для перенастроювання нейронної мережі.

Такий принцип автоматичного синтезу мереж Петрі неодноразово розглядався у наукових роботах [15—17], у яких показані різні експе-рименти для підтвердження принципової придатності такого методу формування певних алгоритмів.

Синтез мереж Петрі може бути заснований на композиції і декомпозиції мереж Петрі, тому що будь-яка мережа Петрі може складатися з однотипних функціональних підмереж [18]. Як показано на рисунку 4, при декомпозиції мережі Петрі $N_1$ можна виділити функціональні підмережі, які виконують різні логічні операції (АБО, І, операцію умовного переходу).

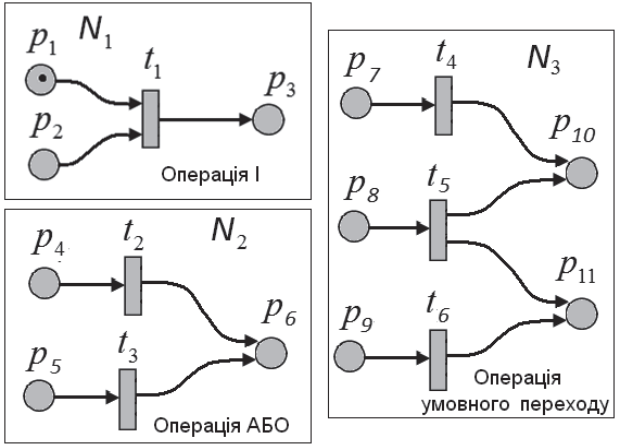

Рис. 4. Декомпозиція мережі Петрі $N$ на функціональні підмережі $N_1$, $N_2$, $N_3$

При реалізації зв'язків між функціональними підмережами формується мережа Петрі яка представляє певний алгоритм логічного управління. У відомих роботах синтез мереж Петрі реалізується на основі композицій і декомпозицій певних функціональних підмереж [18; 19].

Однак такі функціональні підмережі можуть функціонувати разом зі штучною нейронною мережею. У цілому штучна нейронна мережа, що взаємодії з функціональними підмережами, може представляти роботу різних мереж Петрі як композицію роботи окремих підмереж.

Таким чином, згідно зі схемою, представленою на рисунку 3, ДБ-мережа містить дискретні підмережі, пов'язані дискретно-безперервними переходами $t_i^3$, $t_i^4$, де $i=1...N$ [13]. Кожна така підмережа є мережею Петрі, яку можна розглядати незалежно від усієї ДБ-мережі.

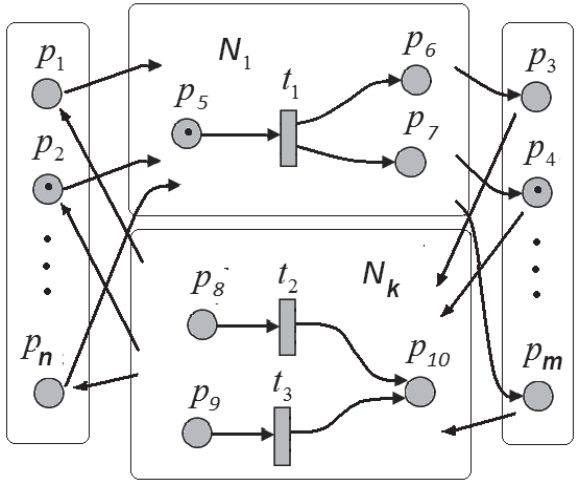

Рис. 5. Формування мережі Петрі при взаємозв'язку функціональних підмереж

Нейронна мережа формує сигнали $Vs=|V1 .... V5|^T$, згідно з якими здійснюється рух маркерів у мережах Петрі. При цьому рух маркерів носить погоджений характер. Наприклад, вихід маркера з позиції $P_2$ супроводжується появою маркера в позиції $P_3$.

Погоджений характер зміни маркування в мережах Петрі дає можливість виконати композицію цих мереж в одну загальну мережу Петрі. Як показано на рисунку 6 а, для композиції мережі Петрі необхідно об'єднати переходи, які одночасно спрацьовують у конкретний момент часу. На рисунку 6 а переходи $t_1$, $t_3$, $t_7$, $t_8$, $t_9$ поєднуються пунктирними кривими. Наприклад, переходи $t_3$, $t_7$ поєднуються в один перехід $t_{3,7}$, який спрацьовує в окремому випадку в момент часу $t_2$. За допомогою такого об'єднання можна перетворити різні мережі в одну загальну мережу Петрі.

Згідно з рухом маркерів у мережах Петрі забезпечується формування значень матриці інцидентності мережі Петрі, так само, як послідовна активізація підстанів Stateflow-діаграми, фрагмент якої представлено на рисунку 1. Як показано на рисунку 6 б, з кожним кроком формується мережа Петрі згідно із процесом формування матриці інцидентності. Слід зазначити, що в окремому випадку поява переходу $t_1$ була помилковою, тому що, наприклад, зміна маркування спричинила небажану зміну критерію якості роботи системи. У цьому випадку деякий блок автонастроювання повинен

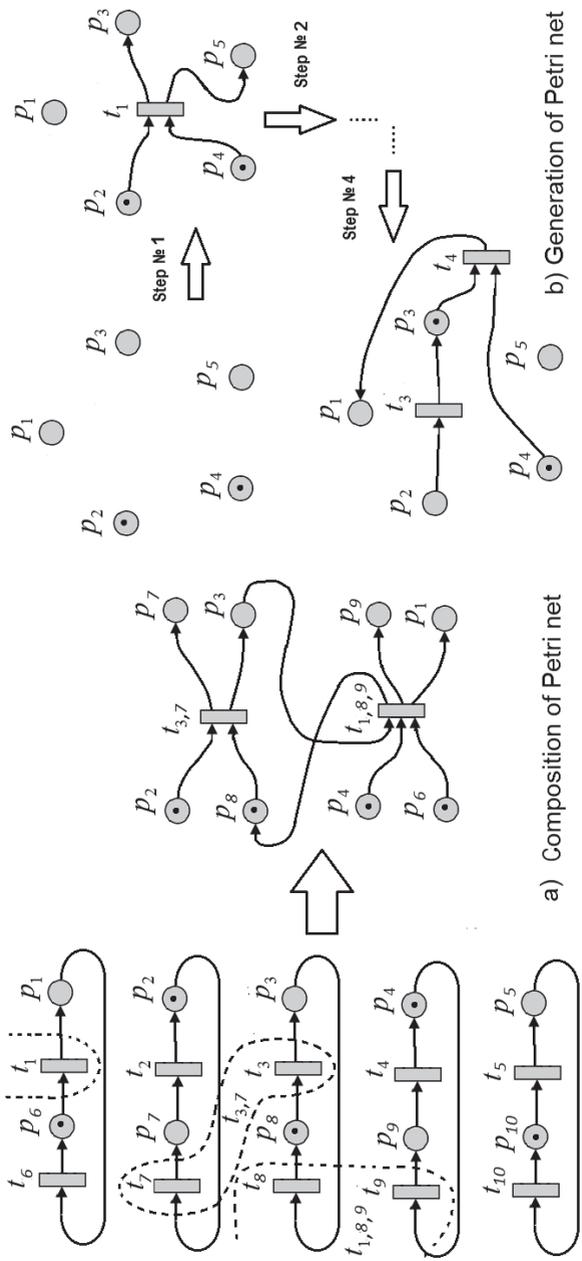

a) Composition of Petri net

b) Generation of Petri net

Рис. 6. Візуалізація автоматичного формування мережі Петрі випадковим образом змінити відповідні коефіцієнти міжнейронних з'єднань, зміна яких надалі дозволила би сформувати необхідну динаміку маркування, відповідну до мережі Петрі, організованої наприклад на 4-му кроці.

Формування алгоритму при роботі нейронної мережі із синхронно функціонуючими підмережами Петрі здійснюється при зміні коефіцієнтів міжнейронних з'єднань. Зміна коефіцієнтів міжнейронних з'єднань тягне за собою коректування відповідного алгоритму.

У цьому випадку деякий алгоритм має місце як при наявності певної локальної штучної нейронної мережі.

Слід врахувати, що може бути деяка кількість вихідних алгоритмів при синтезі системи логічного управління і відповідно можна виділити деяку кількість локальних штучних нейронних мереж, як показано на рисунку 7.

У процесі класифікації алгоритмів запускається відповідна локальна нейронна мережа, яка вже безпосередньо буде брати участь у формуванні певного алгоритму.

У такому випадку автоматичний синтез мереж Петрі можна представити у два етапи. На першому етапі вибір певного алгоритму і відповідної мережі Петрі з можливих варіантів. На другому етапі коректування обраного алгоритму і мережі Петрі. Послідовність формування такого алгоритму можна представити у вигляді рисунку 8, де відображені відповідні етапи синтезу мережі Петрі, що представляє формований алгоритм.

Ці два етапи, які представлено на рисунку 8, можна реалізувати за допомогою штучної нейронної мережі та її тренування. У цьому випадку необхідно представити інтелектуальну систему, що формує алгоритми логічного управління при автоматичному синтезі і композиції мереж Петрі. Надалі необхідно представити структурну схему такої інтелектуальної системи і при цьому визначити метод тренування штучної нейронної мережі при автоматичному синтезі мереж Петрі.

**Алгоритми настроювання штучних нейронних мереж при автоматичному синтезі мереж Петрі.**

Штучна нейронна мережа, як обчислювальна схема, що включає безліч обчислювальних одиниць — нейронів, може представити певну інтелектуальну систему при наявності певного методу зміни коефіцієнтів міжнейронних зв'язків. Саме подібна зміна коефіцієнтів міжнейронних зв'язків, яка має місце в процесі навчання нейронної мережі, може відігравати важливу роль у прояві інтелектуальних

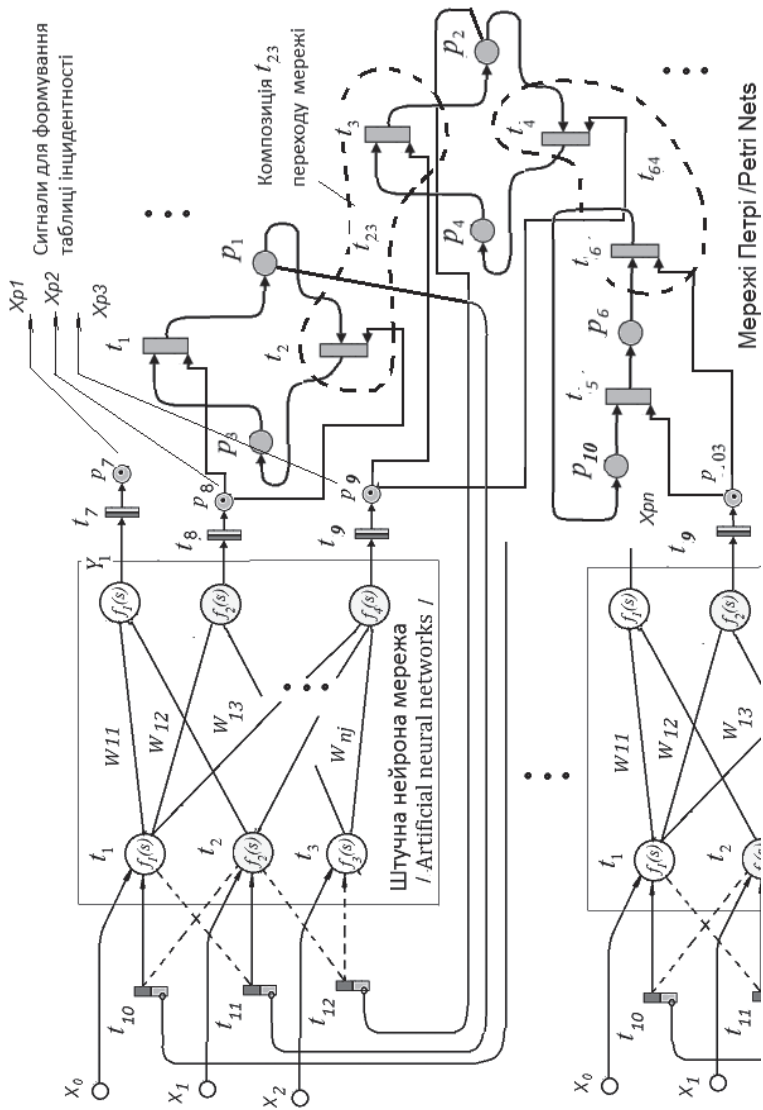

Рис. 7. Фрагмент схеми, що представляє візуалізацію автоматичного формування мережі Петрі на основі функціонування

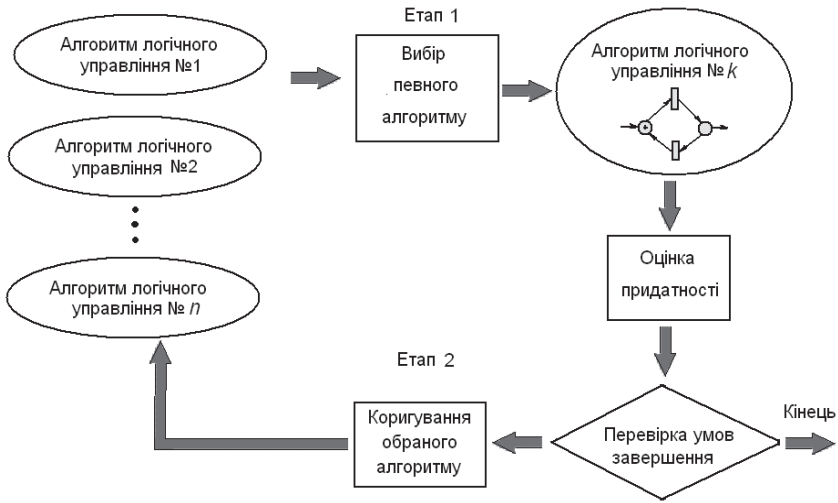

Рис. 8. Схема формування алгоритму логічного управління і відповідного синтезу мережі Петрі особливостей нейронної мережі при розв'язані певних задач. В окремому випадку такі задачі можуть бути пов'язані з автоматичним синтезом мереж Петрі, а також з синтезом алгоритмів логічного управління деяких об'єктів [3; 6].

Автоматичний синтез і композиція мереж Петрі припускає використання відповідних певних методів. Однак для того щоб сформована мережа Петрі була застосовна для розв'язку певного завдання, необхідне застосування інтелектуальних технологій, здатних замінити експерта в області формування певних алгоритмів. Наприклад, заміна експерта в області формування алгоритмів настроювання різних багаторівневих систем автоматичного управління [20].

**Особливості функціонування нейронної мережі при автоматичній генерації мережі Петрі.** Спочатку було встановлено, що автоматична генерація мережі Петрі повинна виконуватися при функціонуванні нейронної мережі, що визначає інтелектуальну технологію формування алгоритму управління деяким об'єктом.

Така нейтронна мережа є багатошаровою і прямонаправленою. Кожний шар нейронної мережі несе певне функціональне навантаження. Перший шар нейронної мережі забезпечує класифікацію

Рис. 9. Спрощена структурна схема системи, що формує алгоритми логічного управління при автоматичній композиції мережі Петрі на базі функціонуван-ня штучної нейронної мережі можливих алгоритмів. Внутрішній шар визначає алгоритми управління. Вихідний шар забезпечує формування певної матриці інцидентності мережі Петрі, що представляє алгоритм логічного управління об'єктом. Як показано на рисунку 9, нейронну мережу можна розділити на сектори, кожний сектор представляє певний алгоритм управління і відповідну матрицю інцидентності мережі Петрі. У зв'язку з тим, що формування алгоритму представляється як покроковий — поетапний процес, кожний крок при формуванні алгоритму супроводжується активізацією певного нейрона, названого початковим. Найперший початковий нейрон можна назвати командним, тому що після його активізації локальна нейронна мережа починає функціонувати. Вибір командного нейрона для його активізації здійснюється на етапі класифікації алгоритмів. Такий процес класифікації також можна реалізувати за допомогою штучної нейронної мережі. Слід зазначити, що відправні нейрони $W$ є в кожному секторі відповідної локальної нейронної мережі. Перенастроювання вагових коефіцієнтів міжнейронних з'єднань, пов'язаних з початковими нейроном, спричиняє коректування відповідного алгоритму, якщо він не задовольняє певним вимогам.

Нейронна мережа як обчислювальна схема визначає матрицю інцидентності мережі Петрі набором коефіцієнтів міжнейроних з'єднань $w_{ij(k)}$, пов'язаних з початковими нейронами. Матриця інцидентності розглянутої мережі, представленої на рисунку 9, має такий вигляд:

$$|W| =
\begin{array}{r|cccccc}
 & t_1 & t_2 & t_3 & t_4 & t_5 & t_6 \\
p_0 & w_{11(1)} & w_{12(2)} & w_{13(\#)} & w_{14(4)} & w_{15(5)} & w_{16(6)} \\
p_1 & w_{21(7)} & w_{22(8)} & w_{23(9)} & w_{24(10)} & w_{25(11)} & w_{26(12)} \\
p_2 & w_{31(13)} & w_{32(14)} & w_{33(15)} & w_{34(16)} & w_{35(17)} & w_{36(18)} \\
p_3 & w_{41(19)} & w_{42(20)} & w_{43(21)} & w_{44(22)} & w_{45(23)} & w_{46(24)} \\
p_4 & w_{51(25)} & w_{52(26)} & w_{53(27)} & w_{54(28)} & w_{55(29)} & w_{56(30)} \\
p_5 & w_{61(31)} & w_{62(32)} & w_{63(33)} & w_{64(34)} & w_{65(35)} & w_{66(36)} \\
p_6 & w_{71(37)} & w_{72(38)} & w_{73(39)} & w_{74(40)} & w_{75(41)} & w_{76(42)} \\
p_7 & w_{81(43)} & w_{82(44)} & w_{83(45)} & w_{84(46)} & w_{85(47)} & w_{86(48)}
\end{array} .$$

Кількість стовпців матриці інцидентності $|W|$ визначається кількістю початкових нейронів, а кількість рядків визначається кількістю вихідних нейронів $n$-го (вихідного) шару нейронної мережі.

Певні умови спрацьовування відповідних переходів мережі Петрі визначаються ваговими коефіцієнтами вхідних з'єднань початкових нейронів. Набір вагових коефіцієнтів вхідних з'єднань початкових нейронів також можна представити у вигляді матриці:

$$|N| = \begin{array}{c c c c c c c} & t_1 & t_2 & t_3 & t_4 & t_5 & t_6 \\ N_1 & w_{11(6)} & w_{12(12)} & w_{13(18)} & w_{14(24)} & w_{15(30)} & w_{16(36)} \\ N_2 & w_{21(2)} & w_{22(8)} & w_{23(14)} & w_{24(20)} & w_{25(26)} & w_{26(32)} \\ N_3 & w_{31(3)} & w_{32(9)} & w_{33(15)} & w_{34(21)} & w_{35(27)} & w_{36(33)} \\ N_4 & w_{41(4)} & w_{42(10)} & w_{43(16)} & w_{44(22)} & w_{45(28)} & w_{46(34)} \\ N_5 & w_{51(5)} & w_{52(11)} & w_{53(17)} & w_{54(23)} & w_{55(29)} & w_{56(35)} \\ N_6 & w_{61(1)} & w_{62(7)} & w_{63(13)} & w_{64(19)} & w_{65(25)} & w_{66(31)} \end{array}$$

У цій матриці $|N|$ кількість рядків $N_1$ .... $N_6$ відповідає різним умовам спрацьовування переходів, а кількість стовпців відповідає кількості кроків формування алгоритму.

Згідно з рисунком 3 процес формування елементів матриці інцидентності мережі Петрі здійснюється на базі вихідних сигналів $y_1...y_8$ нейронної мережі, що забезпечують паралельне — синхронне функціонування відповідних мереж Петрі. Це дає можливість представити деяку композицію з відповідних мереж Петрі, яка відображає алгоритм логічного управління певним об'єктом [15].

Якщо алгоритм логічного управління об'єктом не відповідає заданим вимогам, при яких здійснюється відповідна зміна значень критеріїв якості роботи системи, то блок автонастройки (Tuning unit) повинен змінити певні коефіцієнти міжнейронних з'єднань згідно з інцидентною матрицею синтезованої мережі Петрі.

**Особливості архітектури нейронної мережі, що здійснює синтез мереж Петрі і формування алгоритмів логічного управління.** Виходячи з вищерозглянутої схеми, представленої на рисунку 9, можна виділити деяку певну архітектуру нейронної мережі, яка може застосовуватися при формуванні алгоритму управління, відображеного відповідною мережею Петрі. Така архітектура нейронної мережі, зображена на рисунку 10, представляється двошаровою зі зворотними зв'язками і з елементами затримки сигналів.

Особливість функціонування такої нейронної мережі полягає в тому, що в будь-який момент часу може бути активним тільки один нейрон з $n$ можливих нейронів у вхідному шарі. При цьому активіза-ція лише одного нейрона у вхідному шарі може викликати активіза-
цію N нейронів у вихідному шарі.

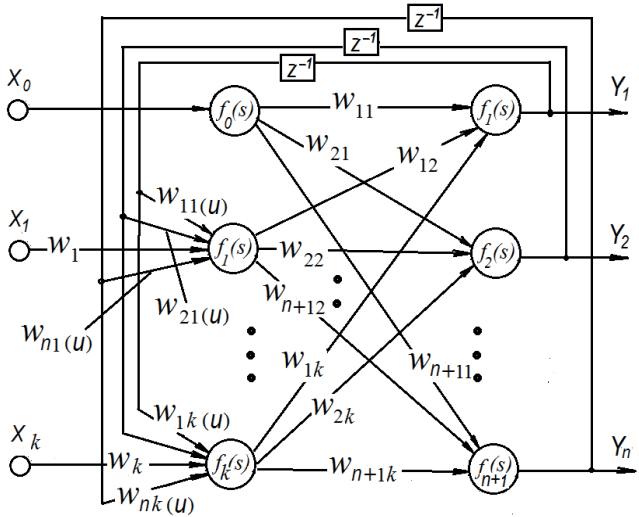

Рис. 10. Архітектура нейронної мережі, що використовується при формуван-
ні інцидентної матриці мережі Петрі

Розрахунки синаптичних коефіцієнтів вихідних нейронів обчис-
люються на основі синаптичних коефіцієнтів вхідних з'єднань почат-
кових нейронів мережі за формулою:

$$w_{nk(u)} = w_{nk} \cdot \left( \left| b_{k(u)} \right| - \sum_{n=1}^{m} w_{nk} \right) + w_{nk} - 1 \,,$$

де $w_{nk(u)}$ — синоптичний коефіцієнт $n$-го входу вхідного $k$-го нейрона;

$w_{nk}$ — синоптичний коефіцієнт $k$-го входу вихідного $n$-го нейрона;

$b_{k(u)}$ — величина зсуву вхідного $k$-го нейрона.

Ця формула розрахунків коефіцієнтів міжнейроних зв'язків ви-
значена з умови активізації лише одного вхідного нейрона з $n$ мож-
ливих нейронів, у будь-який момент часу. Кількість відправних ней-
ронів мережі повинна бути такою, щоб не повторювалися комбінації
значень сигналів виходів $Y_1 ... Yn$.

Для перевірки принципової придатності розглянутої архітекту-
ри нейронної мережі в програмному середовищі MATLAB/Simulink
була реалізована відповідна схема, що представлена на рисунку 11.

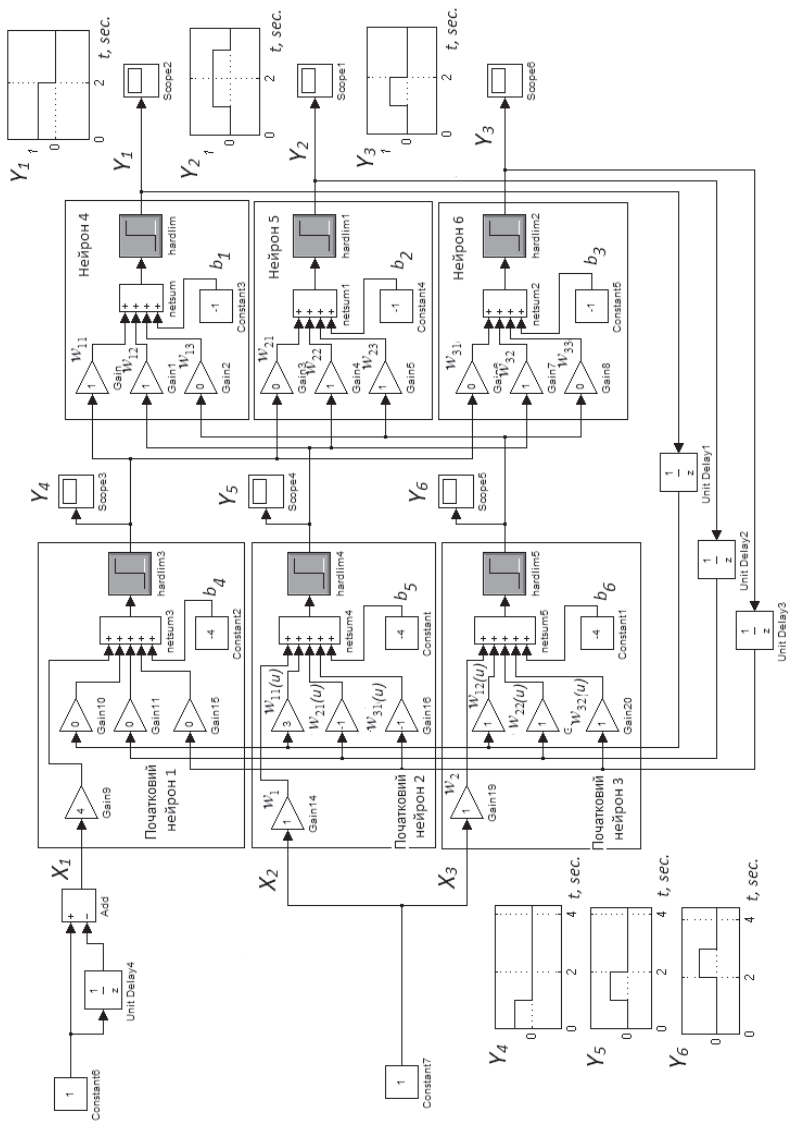

Рис. 11. Структурна схема нейронної мережі з початковими нейронами

Як видно з рисунку 11, представлена схема нейронної мережі, що складається з 3 відправних і 3 вихідних нейронів, має зворотні зв'язки з елементами затримки сигналу.

Згідно з вищенаведеною формулою представимо наступні розрахунки деяких коефіцієнтів міжнейронних з'єднань відповідної нейронної мережі:

$$w_{32(u)} = w_{32} \cdot \left( \left| b_{2(u)} \right| - \sum_{n=1}^{3} w_{n2} \right) + w_{32} - 1 = 1 \cdot (|-4| - (1+1+1)) + 1 - 1 = 1 \; ;$$

$$w_{31(u)} = w_{31} \cdot \left( \left| b_{1(u)} \right| - \sum_{n=1}^{3} w_{n1} \right) + w_{31} - 1 = 1 \cdot (|-4| - (1+0+0)) + 1 - 1 = 3 \; .$$

З відповідних часових діаграм активізацій нейронів, представлених на рисунку 11, видно, що в будь-який момент часу активний тільки один нейрон з 3 початкових, при цьому кількість активних вихідних нейронів різна. Тому що в кожний момент часу активний лише один нейрон у вхідному шарі, то розрахунки коефіцієнтів міжнейронних з'єднань були виконані вірно.

**Алгоритм настроювання нейронної мережі за дозволеними комбінаціями.** Розглянутий алгоритм настроювання нейронної мережі відповідає алгоритму навчання з послідовним підкріпленням знань, при якому мережі не надаються бажані значення вихідних сигналів, а замість цього мережі ставиться оцінка, гарний вихідний сигнал або поганий [21].

Передбачається, що в нейронній мережі в певний момент часу може бути активний лише один нейрон, названий початковим, з *n* можливих нейронів у шарі. Виходячи із цього, сутність настроювання нейронної мережі можна відобразити в такий спосіб: *якщо активність початкового нейрона привела до небажаної ситуації, то зв'язки із цим нейроном W повинні притерпіти зміни* (рисунок 12). При цьому зміна зв'язків з початковим нейроном повинна відбуватися таким чином, щоб не порушити правила формування мережі Петрі.

При обліку правил формування мережі Петрі коефіцієнти міжнейронних з'єднань, пов'язаних з вихідним нейроном, повинні змінитися певним чином залежно від сформованої матриці інцидентності. Отже, якщо при активізації певних нейронів була отримана небажана реакція деякої системи, то вагові коефіцієнти зв'язків цих нейронів $w_{ij}$ слід змінити у такий спосіб:

$$w_{ij(0)} = \eta \cdot \left( \sum_{j=1}^{n} w_{ij} \cdot S_{im} + \left( 1 - \sum_{j=1}^{n} w_{ij} \right) \cdot f_{ik} \right) \cdot F_i,$$

$$F_i = \begin{cases} 1 & npu \quad \left( \sum \overline{w}_{ij}^{T} \cdot \overline{S}_{im} \right) \cdot \left( 1 - \sum \overline{w}_{ij}^{T} \right) \cdot \overline{f}_{ik} > 0 \\ 0 & npu \quad \left( \sum \overline{w}_{ij}^{T} \cdot \overline{S}_{im} \right) \cdot \left( 1 - \sum \overline{w}_{ij}^{T} \right) \cdot \overline{f}_{ik} = 0 \end{cases},$$

$$w_{ij(i)} = \eta \cdot n_{ih},$$

де $|W| = \begin{array}{c} \\ p_0 \\ p_1 \\ \dots \\ p_{i-1} \end{array} \begin{array}{ccccc} t_1 & t_2 & \dots & t_j & \sum \overline{w} \\ w_{11} & w_{12} & \dots & w_{1j} & \sum w_{ij} \\ \dots & w_{22} & \dots & w_{2j} \Rightarrow & \sum w_{2j} \\ \dots & \dots & \dots & \dots & \dots \\ w_{i1} & w_{i2} & \dots & w_{ij} & \sum w_{ij} \end{array}$ — матриця інцидентнос-

ті формованої мережі Петрі і сформований вектор $\sum \overline{w} = \left| \sum w_{1j} \dots \right.$ $\dots \left. \sum w_{ij} \right|^{T}$; $\sum w_{ij}$ сума всіх значень $i$-го рядка матриці $|W|$;

$\overline{S}_{im}, \overline{f}_{ik}$ — вектори дозволених комбінацій ваг міжнейронних з'єднань, що не порушують правила формування мережі Петрі;

| | $\overline{S}_{i1}$ | $\overline{S}_{i2}$ | $\overline{S}_{i3}$ | | | $\overline{f}_{i1}$ | $\overline{f}_{i2}$ | $\overline{f}_{i3}$ | $\overline{f}_{i4}$ | $\overline{f}_{i5}$ | $\overline{f}_{i6}$ | $\overline{f}_{i7}$ | $\overline{f}_{i8}$ |
|---|---|---|---|---|---|---|---|---|---|---|---|---|---|
| $s_{1m}$ | 1 | 0 | 1 | | $f_{1k}$ | 1 | 0 | 0 | 0 | 0 | 0 | 0 | 0 |
| $s_{2m}$ | 1 | 1 | 0 | | $f_{2k}$ | 0 | 1 | 0 | 0 | 0 | 0 | 0 | 0 |
| $s_{3m}$ | 1 | 1 | 0 | | $f_{3k}$ | 0 | 0 | 1 | 0 | 0 | 0 | 0 | 0 |
| $|S_1| = s_{4m}$ | 1 | 0 | 1 | , $|f_1| = f_{4k}$ | | 0 | 0 | 0 | 1 | 0 | 0 | 0 | 0 ; |
| $s_{5m}$ | 1 | 0 | 1 | | $f_{5k}$ | 0 | 0 | 0 | 0 | 1 | 0 | 0 | 0 |
| $s_{6m}$ | 1 | 1 | 0 | | $f_{6k}$ | 0 | 0 | 0 | 0 | 0 | 1 | 0 | 0 |
| $s_{7m}$ | 1 | 1 | 0 | | $f_{7k}$ | 0 | 0 | 0 | 0 | 0 | 0 | 1 | 0 |
| $s_{8m}$ | 1 | 0 | 1 | | $f_{8k}$ | 0 | 0 | 0 | 0 | 0 | 0 | 0 | 1 |

$$
|S_2| = \begin{array}{c|ccc}
 & \overline{S}_{i1} & \overline{S}_{i2} & \overline{S}_{i3} \\
s_{1m} & 1 & 0 & 1 \\
s_{2m} & 1 & 1 & 0 \\
s_{3m} & 1 & 1 & 0 \\
s_{4m} & 1 & 0 & 1 \\
s_{5m} & 1 & 0 & 1 \\
s_{6m} & 1 & 1 & 0 \\
s_{7m} & 1 & 1 & 0 \\
s_{8m} & 1 & 0 & 1 \\
\end{array} \; , \quad
|f_2| = \begin{array}{c|cccccccc}
 & \overline{f}_{i9} & \overline{f}_{i10} & \overline{f}_{i11} & \overline{f}_{i12} & \overline{f}_{i13} & \overline{f}_{i14} & \overline{f}_{i15} & \overline{f}_{i16} \\
f_{1k} & 0 & 0 & 0 & 0 & 0 & 0 & 0 & 0 \\
f_{2k} & 1 & 0 & 0 & 1 & 0 & 0 & 0 & 0 \\
f_{3k} & 0 & 1 & 1 & 0 & 0 & 0 & 0 & 0 \\
f_{4k} & 0 & 1 & 0 & 1 & 0 & 1 & 0 & 1 \\
f_{5k} & 1 & 0 & 1 & 0 & 1 & 0 & 1 & 0 \\
f_{6k} & 0 & 0 & 0 & 0 & 1 & 1 & 0 & 0 \\
f_{7k} & 0 & 0 & 0 & 0 & 0 & 0 & 1 & 1 \\
f_{8k} & 0 & 0 & 0 & 0 & 0 & 0 & 0 & 0 \\
\end{array} \; ;
$$

$\overline{n}_{ih}$ — вектор комбінацій вагових коефіцієнтів вхідних зв'язків вихідного нейрона;

$$
|N_1| = \begin{array}{c|cccc}
 & \overline{n}_{i1} & \overline{n}_{i2} & \overline{n}_{i3} & \overline{n}_{i4} \\
n_{1h} & 1 & 0 & 0 & 0 \\
n_{2h} & 0 & 1 & 0 & 0 \\
n_{3h} & 0 & 0 & 1 & 0 \\
n_{4h} & 0 & 0 & 0 & 1 \\
\end{array} \; .
$$

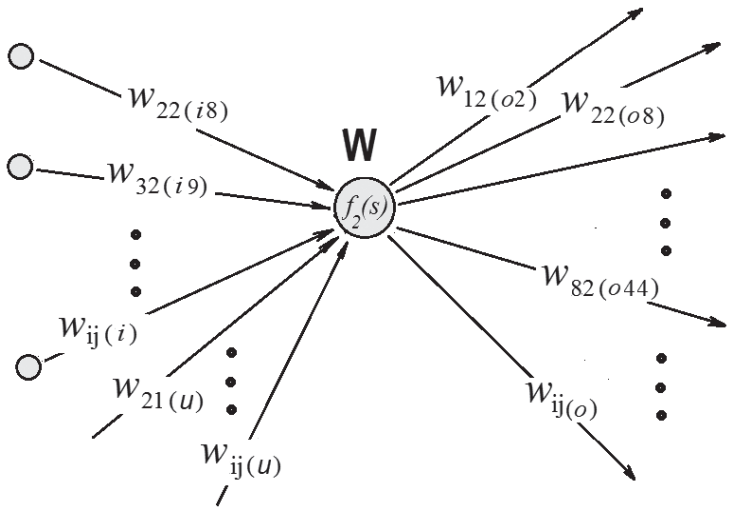

Рис. 12. Схема, що відображає зв'язки початкового нейрона з позначеними вагами міжнейроних з'єднань

Представлені вище вектори можливих комбінацій синоптичних коефіцієнтів зв'язків нейронів дозволяють нейронній мережі в процесі функціонування виконати композицію лише певних мереж Петрі, тому що не всі можливі комбінації коефіцієнтів міжнейронних зв'язків представлені. Очевидно, що ці мережі Петрі можуть відображати суто певні алгоритми управління об'єктами або, в окремому випадку, алгоритми настроювання певних систем управління. Однак слід зазначити, що розглянута архітектура нейронної мережі, що використовується при автоматичному синтезі мереж Петрі, має певну особливість, необхідну при формуванні різних алгоритмів управління.

Представлений нами алгоритм настроювання певної нейронної мережі з вихідними нейронами дає можливість виконати певний крок у розробці інтелектуальної системи, що формує алгоритми логічного управління об'єктами при автоматичному синтезі і композиції мереж Петрі. Автоматично синтезована мережа Петрі дозволяє представити результат формування алгоритму управління об'єктом і тим самим дозволяє фахівцеві виробити, при необхідності, потрібне коректування алгоритму.

Слід зазначити, що в даній інтелектуальній системі, яка пов'язана з автоматичним синтезом мереж Петрі, була представлена спроба використання принципу навчання з підкріпленням, яка відображає область штучного інтелекту, нейромережевого моделювання і управління, що активно розвивається. Тим самим інтелектуальна система, що розробляється, пов'язана із синтезом мереж Петрі і з формуванням, наприклад, алгоритмів настроювання особливого класу систем управління, достатньою мірою вписується в рамки сучасного розвитку інтелектуальних технологій, особливо актуальних в наш час.

Далі розглянемо ще один алгоритм настроювання нейронної мережі, що виключає наявність дозволених комбінацій коефіцієнтів міжнейронних зв'язків нейронної мережі.

**Алгоритм настроювання нейронної мережі по перебору можливих варіантів**. Принцип настроювання нейронної мережі з перебору можливих варіантів аналогічний вищенаведеному алгоритму за винятком його формалізації. У цьому випадку, якщо певний алгоритм дій незадовільний, то необхідно вказати, на якому з переходів мережі Петрі була виявлена помилка в системі. Це необхідно для перенастроювання штучної нейронної мережі. Так, якщо значення показників якості роботи деякої системи збільшується, то необхідно відповідно зв'язок між переходом $t_i$ і позицією $p_i$ ліквідувати. Але при цьому не-обхідно додати новий зв'язок між переходом $t_i$ і сусідньою позицією $p_{i+1}$. Таким чином, наприклад, мережа Петрі, що представлена на рисунку 13 а, змінюється в мережу Петрі, представлену на рисунку 13 b. Відповідно повинні змінитися коефіцієнти міжнейронних зв'язків штучної нейронної мережі.

Математично це можна формалізувати в такому виді:

$$w_{i,j}(t_{k+1}) = w_{i,j}(t_k) \cdot (1 - |w_{i,j}(t_k) \cdot w_{i,j+1}(t_k)| \cdot \delta_1) + w_{i+1,j}(t_k) \times$$
$$\times |w_{i+1,j}(t_k) \cdot w_{i+1,j+1}(t_k)| \cdot \delta_1 \qquad (1)$$
$$\text{при } i=2,4,6...$$

$$w_{i,j+1}(t_{k+1}) = w_{i,j+1}(t_k) \cdot (1 - |w_{i,j}(t_k) \cdot w_{i,j+1}(t_k)| \cdot \delta_1) + w_{i+1,j+1}(t_k) \times$$
$$\times |w_{i+1,j}(t_k) \cdot w_{i+1,j+1}(t_k)| \cdot \delta_1 \qquad (2)$$
$$\text{при } i=2,4,6...$$

$$w_{i,j}(t_{k+1}) = w_{i,j}(t_k) \cdot (1 - |w_{i,j}(t_k) \cdot w_{i,j+1}(t_k)| \cdot \delta_1) + w_{i-1,j}(t_k) \times$$
$$\times |w_{i-1,j}(t_k) \cdot w_{i-1,j+1}(t_k)| \cdot \delta_1 \qquad (3)$$
$$\text{при } i=1,3,5...$$

$$w_{i,j+1}(t_{k+1}) = w_{i,j+1}(t_k) \cdot (1 - |w_{i,j}(t_k) \cdot w_{i,j+1}(t_k)| \cdot \delta_1) + w_{i-1,j+1}(t_k) \times$$
$$\times |w_{i-1,j}(t_k) \cdot w_{i-1,j+1}(t_k)| \cdot \delta_1 \qquad (4)$$
$$\text{при } i=1,3,5...$$

де $w_{i,j}(t_k)$ — коефіцієнт міжнейронного з'єднання, який визначає відповідний зв'язок у мережі Петрі, що формується на кроці $t_k$; $w_{i,j}(t_{k+1})$, відповідний коефіцієнт міжнейронного з'єднання на кроці $t_{k+1}$.

Матриця коефіцієнтів міжнейронних зв'язків $N$ має певну аналогію з інцидентною матрицею мережі Петрі. Наприклад, з інцидентною матрицею $W_1$ мережі Петрі, представленої на рисунку 13 а. При помилці на переході $t_1$ інцидентна матриця $W_1$ змінюється відповідно в матрицю $W_2$ у такий спосіб:

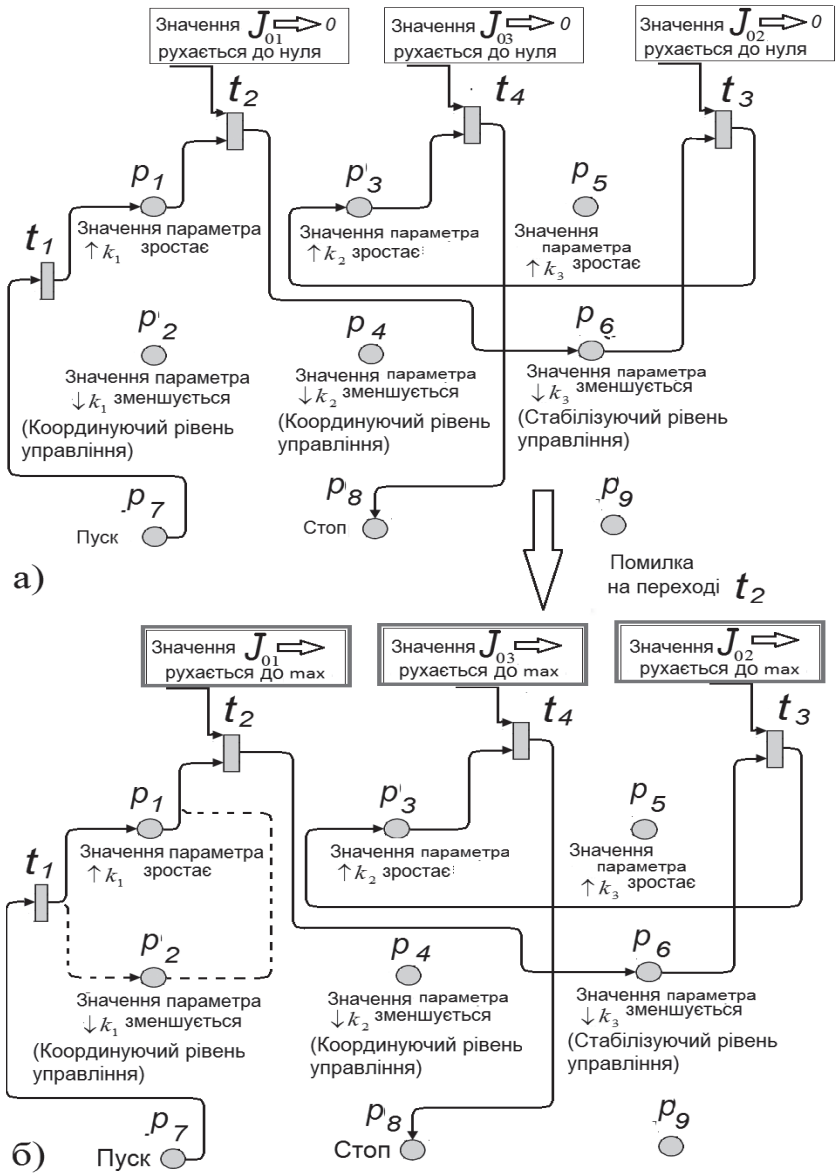

Рис. 13. Алгоритми логічного управління, що представлені мережами Петрі, які відображають процеси поетапного настроювання багаторівневої системи

$$
W_1 = \begin{array}{c} \\ p_1 \\ p_2 \\ p_3 \\ p_4 \\ p_5 \\ p_6 \\ p_7 \\ p_8 \\ p_9 \end{array}
\begin{array}{cccc}
t_1 & t_2 & t_3 & t_4 \\
+1 & -1 & 0 & 0 \\
0 & 0 & 0 & 0 \\
0 & 0 & +1 & -1 \\
0 & 0 & 0 & 0 \\
0 & 0 & 0 & 0 \\
0 & +1 & -1 & 0 \\
-1 & 0 & 0 & 0 \\
0 & 0 & 0 & +1 \\
0 & 0 & 0 & 0
\end{array}
\Rightarrow
W_2 = \begin{array}{c} \\ p_1 \\ p_2 \\ p_3 \\ p_4 \\ p_5 \\ p_6 \\ p_7 \\ p_8 \\ p_9 \end{array}
\begin{array}{cccc}
t_1 & t_2 & t_3 & t_4 \\
0 & 0 & 0 & 0 \\
+1 & -1 & 0 & 0 \\
0 & 0 & +1 & -1 \\
0 & 0 & 0 & 0 \\
0 & 0 & 0 & 0 \\
0 & +1 & -1 & 0 \\
-1 & 0 & 0 & 0 \\
0 & 0 & 0 & +1 \\
0 & 0 & 0 & 0
\end{array}.
$$

Рядки $p_1$ і $p_2$ матриці $W_1$ були змінені згідно з виразом (3). Відповідно, якщо на переході $t_i$ виявлена помилка, то $\delta_1 = 1$. У цьому випадку, якщо є зв'язок між переходом $t_i$ і позицією $p_i$ то $\left| w_{i,j}(t_k) \cdot w_{i,j+1}(t_k) \right| = 1$. Отже, згідно з виразом $\left( 1 - \left| w_{i,j}(t_k) \cdot w_{i,j+1}(t_k) \right| \times \delta_1 \right) = 0$ і коефіцієнт міжнейронного зв'язку, на кроці $t_{k+1}$ стане також рівним нулю $w_{i,j}(t_{k+1}) = 0$. Таким чином, відповідний зв'язок у мережі Петрі, що формується, зникне.

В іншому випадку, якщо відсутній зв'язок між переходом $t_i$ і позицією $p_i$, то $\left| w_{i,j}(t_k) \cdot w_{i,j+1}(t_k) \right| = 0$, $\left( 1 - \left| w_{i,j}(t_k) \cdot w_{i,j+1}(t_k) \right| \times \delta_1 \right) = 1$. При цьому, якщо був присутній сусідній зв'язок між переходом $t_i$ і позицією $p_{i+1}$, то $\left| w_{i+1,j}(t_k) \cdot w_{i+1,j+1}(t_k) \right| \cdot \delta_1 = 1$ і, отже, коефіцієнт міжнейронного зв'язку збільшиться на одиницю $w_{i,j}(t_{k+1}) = w_{i,j}(t_k) + 1$. Таким чином, з'явиться відповідний зв'язок у мережі Петрі, що формується.

Якщо на переході $t_i$ помилка не виявлена, то $\delta_1 = 0$, $\left( 1 - \left| w_{i,j}(t_k) \cdot w_{i,j+1}(t_k) \right| \times \delta_1 \right) = 1$ і, отже, $w_{i,j}(t_{k+1}) = w_{i,j}(t_k)$.

**Експерименти.** У програмному середовищі MATLAB/Simulink 2012 були проведені експерименти, пов'язані зі спільною роботою нейронної мережі з мережами Петрі. Функціонування мереж Петрі у програмному середовищі MATLAB/Simulink було реалізовано за допомогою Statflow-діаграм. Фрагмент Stateflow-діаграм, що представляють роботу мереж Петрі, представлено на рисунку 14. State1, State2, State3 і State4 є станами однієї мережі Петрі. Нейронна мережа пов'язана з роботою відразу трьох таких мереж Петрі. Це показано на рисунку 15.

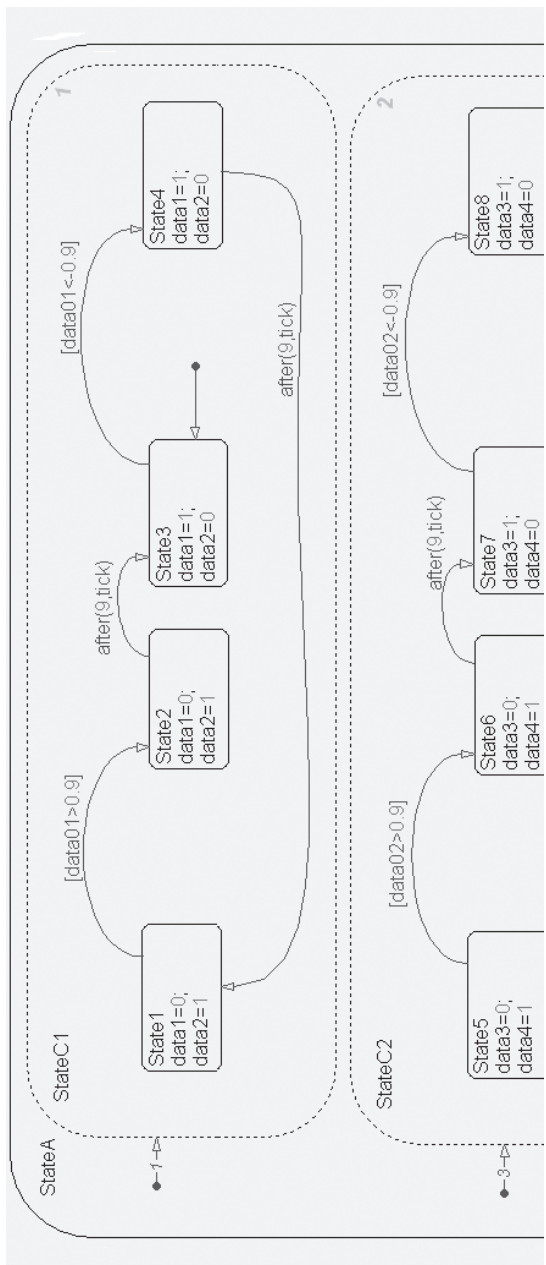

Рис. 14. Stateflow-діаграми, що представляють роботу мереж Петрі

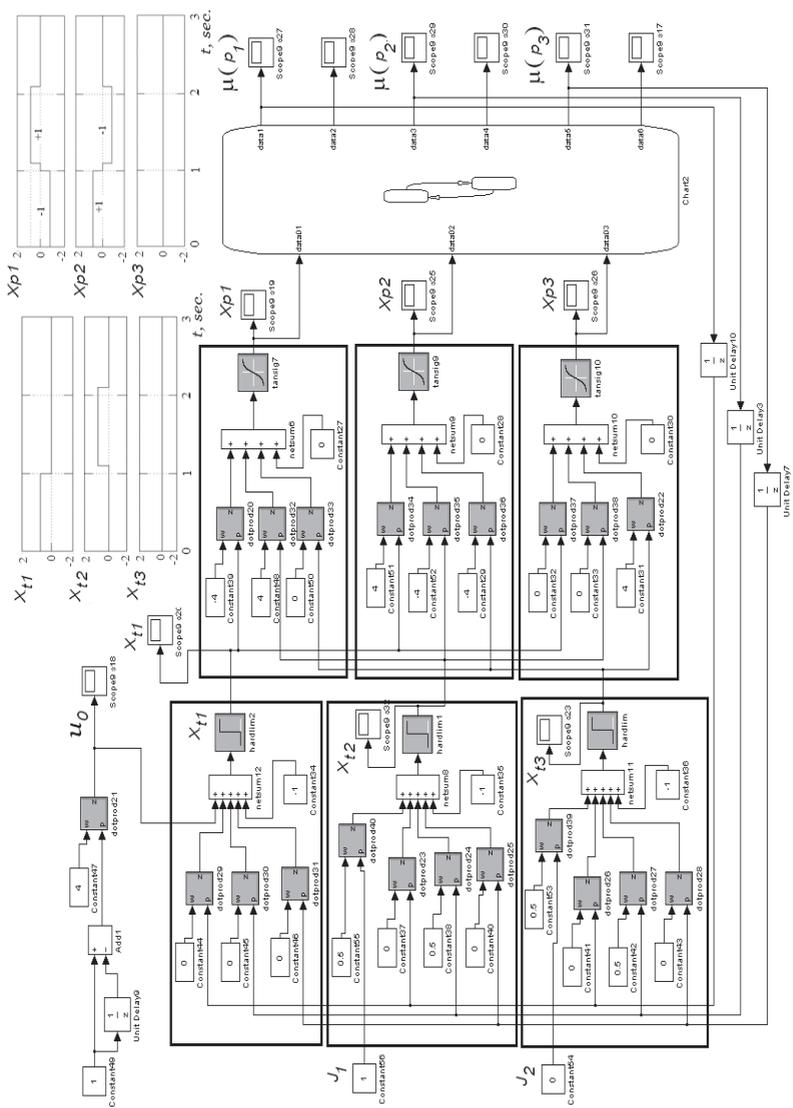

Рис. 15. Структурна схема нейронної мережі, що синтезує мережу Петрі

На рисунку 15 також представлена засобами середовища MATLAB/Simulink двошарова штучна нейронна мережа, що складається з шести нейронів із вихідними сигналами *Xp1, Xp2, Xp3,* пов'язаними з мережами Петрі.

Структурна схема цієї нейронної мережі і мереж Петрі аналогічна спрощеній схемі, яка представлена на рисунку 3. На рисунках 14 і 15 наведені всі необхідні параметри системи, яка представляє спільне функціонування нейронної мережі і мереж Петрі для автоматичного формування мереж Петрі і певних алгоритмів.

Система, структурна схема якої наведена на рисунку 4, здатна представити функціонування різних мереж Петрі.

Рівняння (2) описує таке спільне функціонування мереж Петрі і штучної нейронної мережі.

Якщо матриця інцидентності $|A|$ мережі Петрі має певну аналогію з матрицею коефіцієнтів міжнейронних з'єднань вихідного шару нейронної мережі, то штучна нейронна мережа генерує вихідні сигнали $\overline{V} = |A| \cdot U_{k-1}$, відповідні значенням матриці інцидентності мережі Петрі, що формується.

На рисунках 15 і 16 представлені часові діаграми функціонування мережі Петрі, що складається із трьох позицій і трьох переходів. Функціонування такої мережі Петрі представляє система спільної роботи штучної нейронної мережі і Stateflow-діаграм. Вихідні сигнали штучної нейронної мережі *Xp1, Xp2, Xp3* відповідають матриці інцидентності мережі Петрі, що формується. А вихідні сигнали $\mu(p_1)$ $\mu(p_2)$ $\mu(p_3)$ представляють зміну маркування мережі Петрі в часі. В окремому випадку представляється робота мережі Петрі, показаної на рисунку 16.

Як видно з часових діаграм, якщо присутній сигнал на вході $J_1$ ($J_{01} > 0$), то спрацьовує перехід $t_2$. Якщо з'являється сигнал на вході $J_2$ ($J_{01} > 0$), то спрацьовує перехід $t_3$. Одночасне спрацьовування переходів $t_2$ і $t_3$ відповідає конфліктній ситуації в роботі мережі Петрі.

Аналізуючи часові характеристики, наведені на рисунках 15 і 16, можна зробити висновок про принципову придатність розглянутої системи представляти роботу різних мереж Петрі. Важлива складова такої системи — це наявність штучної нейронної мережі, тренування якої пов'язане з автоматичним синтезом мережі Петрі. Таким чином, зміна коефіцієнтів міжнейронних зв'язків при тренуванні мережі пов'язана зі зміною синтезованої мережі Петрі.

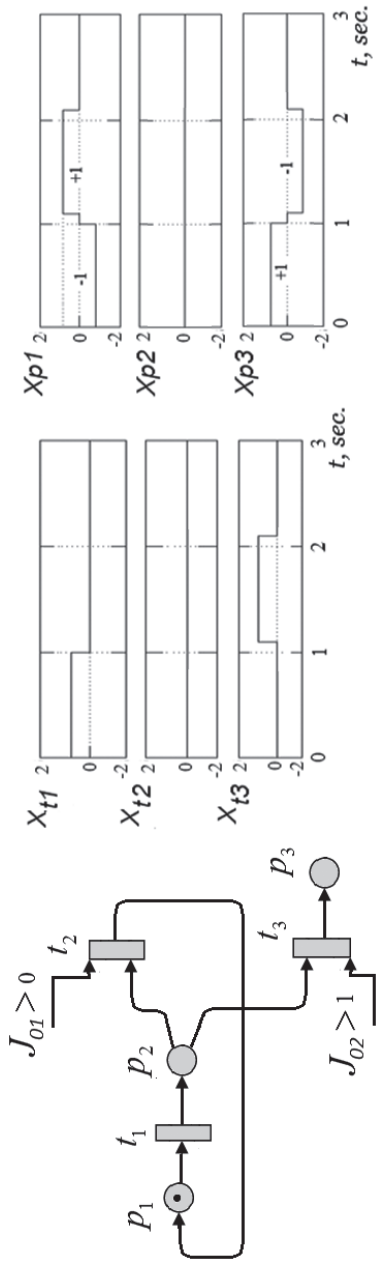

Рис. 16. Процес функціонування синтезованої мережі Петрі

Математичний опис зміни коефіцієнтів міжнейронних з'єднань при тренуванні мережі було представлено як одну зі спроб реалізації перебору можливих варіантів з'єднань у мережі Петрі. У цьому випадку сформована мережа Петрі є візуальним відображенням набору коефіцієнтів міжнейронних з'єднань у штучної нейронної мережі.

**Висновки**. Нами було вирішено задачу, пов'язану з розробкою системи спільного функціонування нейронної мережі і мереж Петрі для формування алгоритмів і послідовних обчислень. Тим самим одержали подальший розвиток методики автоматичного синтезу мереж Петрі і розробки певних алгоритмів на основі функціонування нейронної мережі. Був представлений математичний опис зміни коефіцієнтів міжнейронних зв'язків мережі при синтезі мережі Петрі.

Розроблені методики синтезу мереж Петрі дозволяють підійти до вирішення практичної задачі, пов'язаної з автоматизованим настроюванням складного класу багаторівневих автоматичних систем координуючого управління. У цьому випадку синтезована мережа Петрі дозволяє представити процес і алгоритм настроювання відповідної системи управління.

**Автоматизоване настроювання автоматичної системи координувального управління при синтезі мереж Петрі.** Практичне застосування розглянутих методів автоматичного синтезу мереж Петрі може бути в області автоматизації процесів настроювання певного класу багаторівневих автоматичних систем координувального управління [22; 23].

На основі відомих наукових праць можна зробити висновок, що спочатку при синтезі координувальну систему автоматичного управління треба розглядати як однорівневу, тобто необхідно виконувати синтез системи, починаючи з нижнього рівня, а потім переходити до синтезу верхніх рівнів [22–25]. Але в деяких випадках можливі й інші варіанти. Наприклад, синтез координувальної системи автоматичного управління (КСАУ) приводами робота-маніпулятора доцільно було починати з контуру регулювання верхнього рівня, потім необхідно було налаштувати нижній рівень, а потім параметри настроювання верхнього рівня необхідно було коректувати. Було встановлено, що можливі різні варіанти алгоритмів синтезу координувальних системи, а кожний з алгоритмів може приводити до різних результатів. Таким чином, виникає задача формування, отже, пошуку алгоритму, який дозволить досягти бажань значень показників якості роботи координувальної системи. У цьому випад-ку виникає задача, подібна до задачі досяжності системи в гібридному (дискретно-безперервному) просторі станів, яка розглядалася в роботах [14; 26].

В даному випадку розробляється система параметричного синтезу КСАУ на базі математичного апарата дискретно-безперервних мереж (ДБ-мереж), яка дозволяє досліджувати властивість досяжності системи шляхом редукції безперервної і дискретної частин мережі [12]. Перша спроба розробки такої системи була запропонована в роботі [27], у якій мережею Петрі представлявся алгоритм самонастроювання певних параметрів нейро-нечіткої системи управління. У цьому випадку запропоновано в системі, що розробляється, реалізувати формування алгоритму параметричного синтезу на основі методів перевірки досяжності ДБ-мережі. Виконана розробка автомата в середовищі MATLAB/Simulink формує матрицю інцидентності мережі Петрі дискретно-подійної частини ДБ-мережі, отже, формує мережу Петрі, що представляє алгоритм параметричного синтезу координувальної системи автоматичного управління.

Мета роботи є зниження необхідних обчислювальних і часових ресурсів на розробку складних багаторівневих систем автоматичного управління.

Для досягнення поставленої мети потрібно було розробити систему, яка здатна сформувати необхідний алгоритм параметричного синтезу і виконати необхідний порядок дій згідно зі сформованим алгоритмом.

Координувальна система автоматичного управління приводами робота-маніпулятора представляється як дворівнева система [28; 29]. Структурна схема моделі такої системи, що реалізована засобами середовища MATLAB/Simulink, представлена на рисунку 17. Верхній рівень управління системи пов'язаний з відпрацьовуванням помилок регулювання за положенням зхвату $L_m$ і за кутом повороту маніпулятора $\alpha_m$, а нижній рівень пов'язаний з відпрацьовуванням неув'язок співвідношень змінних, що представляють траєкторію руху зхвату в циліндричній системі координат. Дворівневий закон цієї системи управління можна представити так:

$$\overline{u} = \overline{u}_q + \overline{u}_p = \begin{bmatrix} u_{1Lm} & u_{2\alpha m} \end{bmatrix}^T ;$$

де $\overline{u}_q = \begin{bmatrix} u_{q1} \\ u_{q2} \end{bmatrix} = \begin{bmatrix} k_2 \cdot (1 + k_{21} \cdot p) \\ k_3 \cdot (1 + k_{31} \cdot p) \end{bmatrix} \cdot \psi(t)$ — закон управління нижнього рівня;

$$\overline{u}_p = \begin{bmatrix} u_{p1} \\ u_{p2} \end{bmatrix} = \begin{bmatrix} (L_{mz}(t) - L_m(t)) \cdot k_1 \cdot (1 + k_{11} \cdot p) \\ (\alpha_{mz}(t) - \alpha_m(t)) \cdot k_4 \cdot (1 + k_{41} \cdot p) \end{bmatrix} \quad - \quad \text{закон} \quad \text{управління}$$

верхнього рівня;

$\psi(t) = f(L_m) \cdot L_m(t) + k \cdot \alpha_m(t) - b$ — відхилення (неув'язка) від співвідношення параметрів у момент часу $t$;

$L_m(t)$, $\alpha_m(t)$ — регульовані зміни;

$f(L_m)$ — нелінійна залежність, що відображена в системі у вигляді ланки $NU$ (рисунок 17), що описує траєкторію руху зхвата в координатах $L_m - \alpha_m$;

$k_1$, $k_{11}$, $k_2$, $k_{21}$, $k_3$, $k_{31}$, $k_4$, $k_{41}$ — параметри налаштування системи які необхідно визначити з урахуванням прояву ефекту поділу руху;

$\alpha_{m.Z}(t)$, $L_{m.Z}(t)$ — задаючі впливи за кутом повороту і положенням маніпулятора в площині $F$;

$p$ — оператор диференціювання.

У даній роботі об'єкти системи координувального управління описуються такими передатними функціями:

$$W_1(p) = \frac{X_1(p)}{u_1(p)} = \frac{k_L}{p \cdot Q_L(p)},$$

$$W_2(p) = \frac{X_2(p)}{u_2(p)} = \frac{k_\alpha}{p \cdot Q_\alpha(p)},$$

у яких $k_L$, $k_\alpha$ — коефіцієнти передачі, $Q_L(p)$, $Q_\alpha(p)$ — деякі поліноми, такі, що $Q_L(0){=}1$, $Q_\alpha(0){=}1$.

Рівняння кінематики робота встановлює зв'язок між різними координатами. Рівняння $X_i{=}F(q_1, q_2)$, де $q_1$, $q_2$ — координати $L_m$, $\alpha_m$ відтворюючих систем, визначають математичну модель механічної частини робота.

Для параметричного настроювання даної координувальної системи автоматичного управління були реалізовані блоки формування значень наступних інтегральних показників якості роботи системи:

$$J_1 = \int_{t_0}^{t_1} (\beta_1 \cdot (L_{mz}(t) - L_m(t)^2) + u_{2L_m}(t))dt = \int_{t_0}^{t_1} (fL_m(t))dt ;$$

$$J_{01} = \left[ \int_{t_0}^{t_1} f_{L_m}(t)dt - \int_{t_2}^{t_3} f_{L_m}(t)dt \right] - \left[ \int_{t_1}^{t_4} f_{L_m}(t)dt - \int_{t_2}^{t_3} f_{L_m}(t)dt \right];$$

$$J_2 = \int_{t_0}^{t_1} (\beta_2 \cdot (\alpha_{m.z}(t) - \alpha_m(t)^2) + u_{2\alpha_m}(t))dt = \int_{t_0}^{t_1} (f\alpha_m(t))dt ;$$

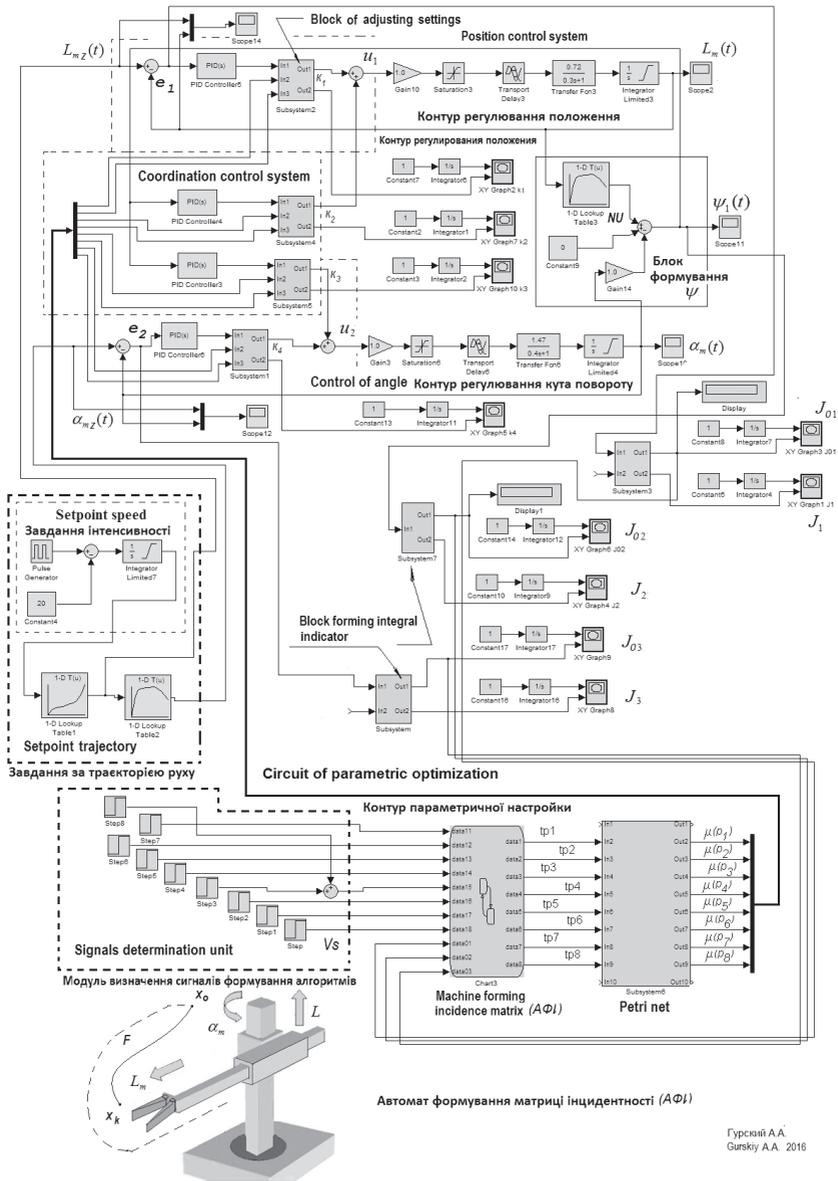

Рис. 17. Структурна схема моделі автоматичної системи координувального управління з контуром параметричного налаштування

$$J_{02} = \left[ \int_{t_0}^{t_1} f_{\alpha_m}(t)dt - \int_{t_2}^{t_3} f_{\alpha_m}(t)dt \right] - \left[ \int_{t_1}^{t_4} f_{\alpha_m}(t)dt - \int_{t_2}^{t_3} f_{\alpha_m}(t)dt \right];$$

$$J_3 = \int_{t_0}^{t_1} \psi^2(t)dt \; ; \; J_{03} = \left[ \int_{t_0}^{t_1} \psi^2(t)dt - \int_{t_2}^{t_3} \psi^2(t)dt \right] - \left[ \int_{t_1}^{t_4} \psi^2(t)dt - \int_{t_2}^{t_3} \psi^2(t)dt \right],$$

де $(t_1 - t_0) = (t_3 - t_2) = (t_4 - t_1) = (t_5 - t_3)$, $t_0 < t_2 < t_1 < t_3 < t_4 < t_5$.

За показниками $J_{01}$, $J_{02}$, $J_{03}$ оцінюється правильність напрямку зміни параметрів настроювання КСАУ [13]. При відсутності зміни параметрів $k_1$, $k_2$, $k_3$, $k_4$ значення показників $J_{01}$, $J_{02}$, $J_{03}$ будуть дорівнювати нулю.

Параметричний синтез КСАУ (визначення параметрів $k_1$, $k_2$, $k_3$, $k_4$) виконується згідно з певним алгоритмом. Задача формування алгоритму покладається в розробленій системі на формуючий автомат (ФА), який складається з модуля визначення сигналів формування алгоритмів (МВС) і автомата формування матриці інцидентності (АФІ). МВС визначає необхідний алгоритм синтезу, а АФІ формує матрицю інцидентності мережі Петрі, що представляє даний алгоритм у системі. Маркування мережі Петрі визначає зміну тих або інших параметрів настроювання КСАУ в блоках коректування параметрів.

Якщо розглядати координувальну систему автоматичного управління разом з контуром параметричної настройки, на базі алгоритму пошуку параметрів настроювань, то така система поводиться як логіко-динамічна. У цій системі як у логіко-динамічній та як у системі з керованою структурою [13] присутні як дискретні, так і безперервні сигнали, спостерігається багаторежимний характер функціонування (режим настроювання нижнього рівня, режим настроювання верхнього рівня при настроєному нижньому рівні) і т. д.

Відомо, що досить потужний математичний апарат для синтезу логіко-динамічних моделей та дослідження складних систем з дискретно-безперервним характером функціонування, заснований на базі засобів дискретно-безперервної мережі [13]. Ці засоби дозволяють досліджувати досяжність системи на основі правил редукції мережі, а правила редукції мережі і методи перевірки досяжності відіграють важливу роль у розробці формуючого автомата.

У нашій роботі модель КСАУ з системою настроювання розробляється на базі математичного апарата дискретно-безперервних мереж.

ДБ-мережу можна розділити на дві частини: неперервно-подійну частину (НПЧ) і дискретно-подійну частину (ДПЧ).

У нашому випадку НПЧ представляє координувальну систему управління, а ДПЧ алгоритм параметричного синтезу та настроювання КСАУ.

Дискретно-подійна частина ДБ-мережі описується рівнянням стану:

$$X_L^d(t_k) = X_L^d(t_{k-1}) + |W| \cdot v_L^d(t_k) + u_L^d(t_k) \,,$$

де $|W|$ — матриця інцидентності позицій і переходів ДПЧ, що визначає в даному випадку алгоритм настроювання системи; $v_L^d(t_k)$ — управляючий вектор, який формується залежно від умов спрацьовування переходів мережі Петрі, що представляє ДПЧ і відповідно залежно від значень показників $J_{01}$, $J_{02}$, $J_{03}$; $u_L^d(t_k)$ — вхідний вплив; $X_L^d(t_k), X_L^d(t_{k-1})$ — дискретні стани ДБЧ у моменти часу $t_k, t_{k-1}$, для даного випадку $X_L^d(t_k) = \left[ \mu(p_1^2), \mu(p_2^2), \ldots, \mu(p_8^2) \right]^T$ при цьому, якщо позиція $p_1$ маркірована — $\mu(p_1) = 1$, то проводиться збільшення значення параметра настроювання $\kappa_1$ координувальної системи управління. Отже наявність маркера в позиції $p_1$ визначає процес збільшення значення параметра настроювання $k_1$ до екстремального значення показника $J$ якості роботи КСАУ, тобто $\mu(p_1) = 1 \Rightarrow \uparrow k_1$, таким чином $\mu(p_2) = 1 \Rightarrow \downarrow k_1$, $\mu(p_3) = 1 \Rightarrow \uparrow k_2$, $\mu(p_4) = 1 \Rightarrow \downarrow k_2$, $\mu(p_5) = 1 \Rightarrow \uparrow k_3$, $\mu(p_6) = 1 \Rightarrow \downarrow k_3$, $\mu(p_7) = 1 \Rightarrow \uparrow k_4$, $\mu(p_8) = 1 \Rightarrow \downarrow k_4$.

**Аналіз процесів параметричного синтезу координувальної системи на базі сформованих алгоритмів.** Залежно від установлених сигналів формування алгоритму і за значеннями показників процесу синтезу $J_{01}$, $J_{02}$, $J_{03}$ автомат формування матриці інцидентності виробив послідовність значень, з яких складається матриця інцидентності мережі Петрі. Ця послідовність значень за період часу параметричної настройки КСАУ представлена на рисунку 18. На основі процесу, представленого на рисунку 18, можна скласти матрицю інцидентності і згідно з нею сформовану мережу Петрі, що відображає алгоритм параметричної настройки КСАУ.

Згенерована матриця інцидентності мережі Петрі, що представляє алгоритм параметричного синтезу координувальної системи, має 8 рядків згідно з кількістю виходів автомата АФІ і 10 стовпців згідно з кількістю кроків формування мережі Петрі. У даному випадку матриця інцидентності має такий вигляд:

$$|A_{L1}| = \begin{array}{c} \\ tp_1 \\ tp_2 \\ tp_3 \\ tp_4 \\ tp_5 \\ tp_6 \\ tp_7 \\ tp_8 \end{array} \begin{array}{cccccccccc} t_1 & t_2 & t_3 & t_4 & t_5 & t_6 & t_7 & t_8 & t_9 & t_{10} \\ +1 & -1 & 0 & 0 & 0 & 0 & 0 & 0 & 0 & 0 \\ 0 & +1 & -1 & 0 & 0 & 0 & 0 & 0 & 0 & 0 \\ 0 & 0 & 0 & +1 & -1 & 0 & 0 & 0 & 0 & 0 \\ 0 & 0 & +1 & -1 & 0 & 0 & 0 & 0 & 0 & 0 \\ 0 & 0 & +1 & -1 & 0 & +1 & -1 & 0 & 0 & 0 \\ 0 & 0 & 0 & +1 & 0 & -1 & +1 & -1 & 0 & 0 \\ 0 & 0 & 0 & 0 & 0 & 0 & 0 & +1 & -1 & 0 \\ 0 & 0 & 0 & 0 & 0 & 0 & 0 & 0 & +1 & -1 \end{array}.$$

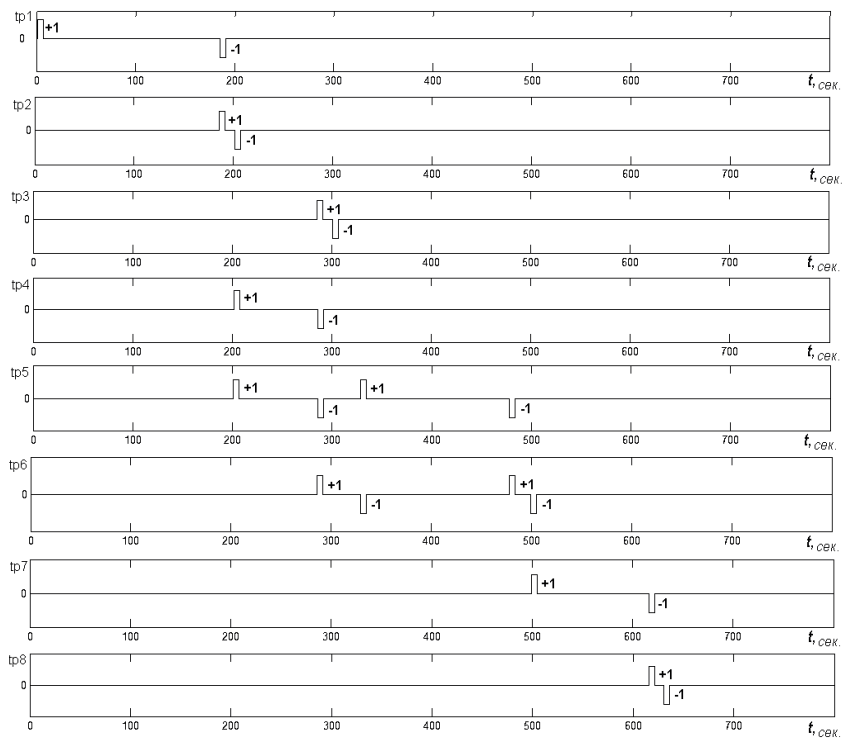

Рис. 18. Процес формування матриці інцидентності мережі Петрі, що представляє алгоритм параметричної настройки координувальної системи автоматичного управління

Таким чином, автомат формування матриці інцидентності з кожним кроком формує стовпець матриці інцидентності і тим самим розбудовує мережу Петрі, як показано на рисунку 19.

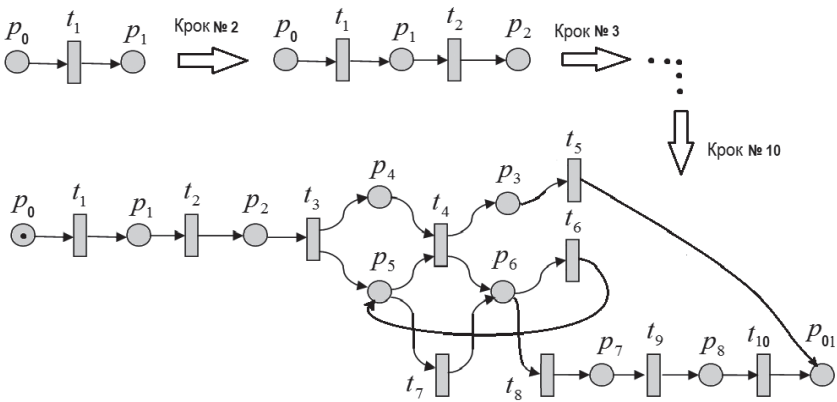

Рис. 19. Візуалізація формування мережі Петрі, що представляє алгоритм параметричної настройки координувальної системи автоматичного управління

З рисунку 19 видно, що сформована мережа Петрі «жива», однак у ній присутні дві конфліктні ситуації. Так, на 200-й секунді процесу параметричного синтезу виконуються умови спрацьовування одночасно для переходів $t_4$ і $t_7$, а на 300-й і 500-й секундах виконуються умови спрацьовування одночасно для переходів $t_6$ і $t_8$. Виходячи з алгоритму параметричної настройки, представленої на рисунку 20, і динаміки маркування, представленої на рисунку 21, отриманої в результаті функціонування мережі Петрі, що представляє алгоритм параметричного синтезу КСАУ, можна зробити висновок, що спочатку пріоритет у спрацьовуванні був за переходами $t_4$ і $t_6$.

Слід зазначити, що виділені правила побудови матриці інцидентності, на основі яких виконується синтез автомата формування матриці інцидентності, недостатні для генерації мереж Петрі з безконфліктними ситуаціями. Однак у даному випадку це не заважає виконати параметричний синтез КСАУ. Таким чином, згідно з динамікою маркування (рисунок 21) змінюються послідовно параметри настроювання КСАУ, як показано на рисунку 22. Параметр $k_1$ система налаштування припиняє збільшувати при досягненні значення нуля показника $J_{01}$ в точці А згідно з рисунком 23. Припинення зміни параметрів $k_2$ і $k_3$ відбувається в точці В, а в точці С припинення збільшення параметра $k_3$. І в остаточному підсумку припинення збільшення параметра $k_4$ в точці $D$ при досягненні області значення нуля показника $J_{03}$.

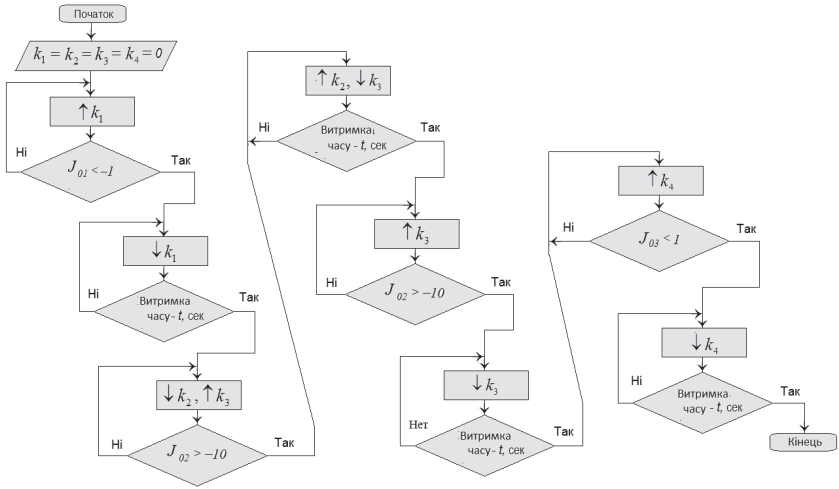

Рис. 20. Алгоритм параметричної настройки координувальної системи автоматичного управління: $\uparrow k_i$ — збільшення значення параметра $k_i$; $\downarrow k_i$ — зменшення параметра $k_i$, де $i = 1...8$

Для перевірки принципової придатності формуючого автомата сигнали формування алгоритму були змінені і тим самим задали алгоритм, суттєво відмінний від вищенаведеного на рисунку 20. Була сформована автоматом матриця інцидентності (рисунки 24 і 25) і відповідно їй мережа Петрі, представлена на рисунку 26, що відображає алгоритм параметричної настройки КСАУ.

Відповідно алгоритму параметричної настройки змінюється маркування мережі Петрі і відповідно їй коректуються послідовно параметри настроювання $k_4$, $k_2$ і $k_3$ координувальної системи автоматичного управління залежно від зміни значення показника $J_{02}$.

З рисунку 26 видно, що в цьому випадку при даних сигналах формування алгоритму сформована мережа Петрі «жива» і без конфліктних ситуацій.

Слід зазначити, якщо сигнали формування алгоритму не детерміновані, а випадкові, то формується випадковим образом і мережа Петрі. У результаті можна одержати безліч різних мереж Петрі й організувати пошук алгоритму не тільки в області параметричного синтезу.

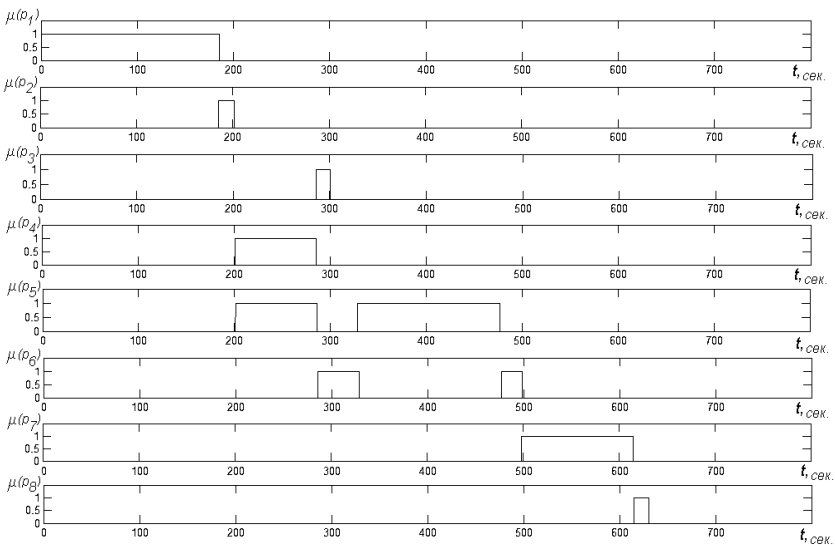

Рис. 21. Динаміка маркування, отримана в результаті функціонування мережі Петрі, що представляє алгоритм параметричної настройки координувальної системи автоматичного управління

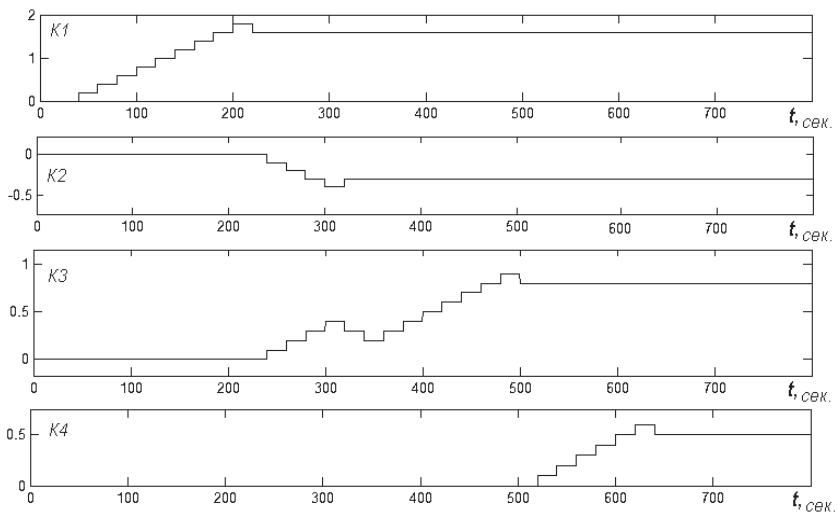

Рис. 22. Процес модифікації параметрів настроювання при параметричному синтезі КСУ згідно з алгоритмом, представленим синтезованою мережею Петрі

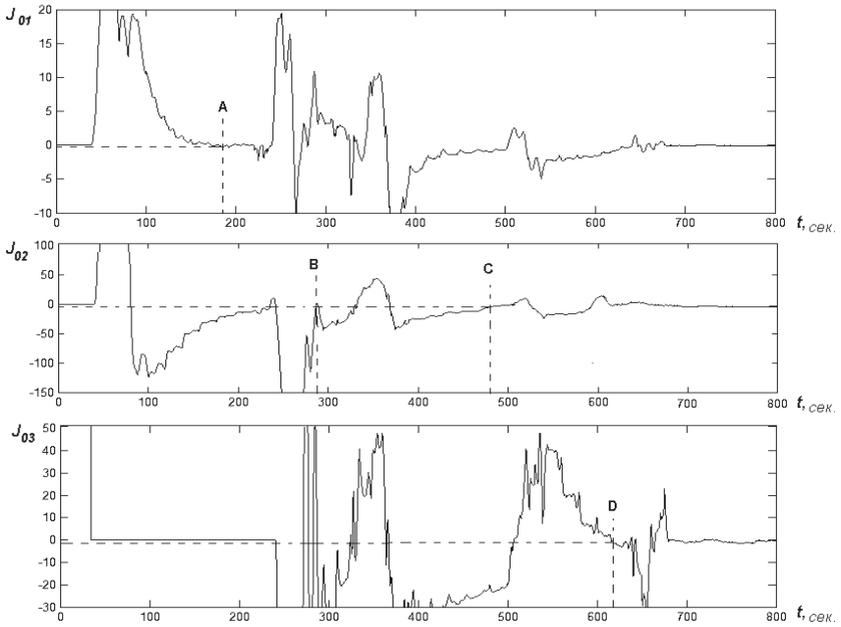

Рис. 23. Зміна показників якості роботи системи в часі

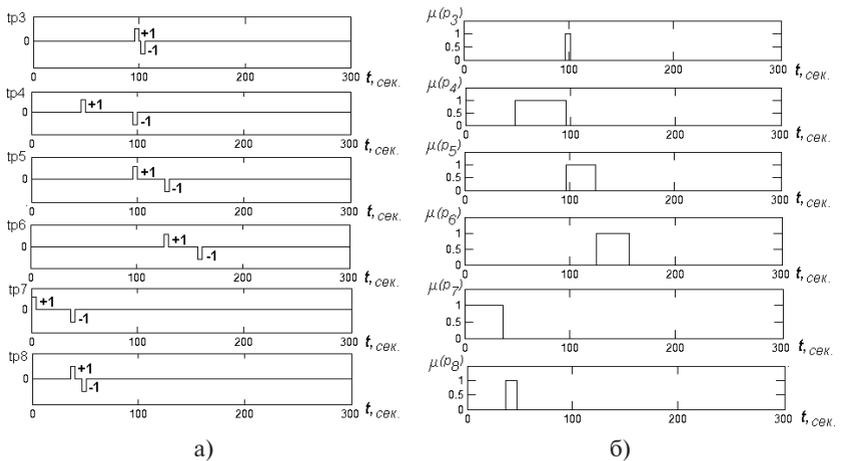

а)　　　　　　　　б)

Рис. 24. Процес формування матриці інцидентності (а) і динаміка маркування мережі Петрі (б), що представляє алгоритм параметричної настройки автоматичної системи координувального управління

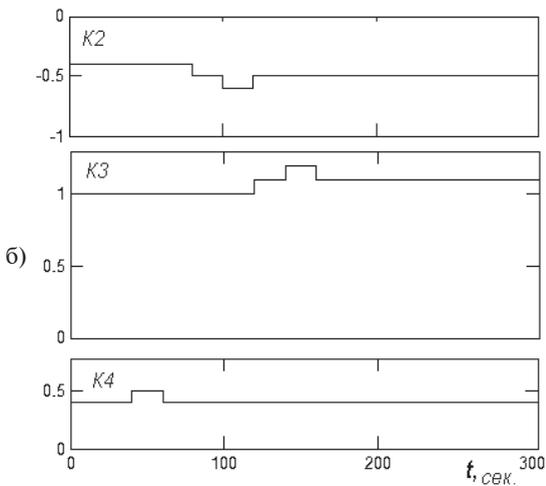

|        | $t_1$ | $t_2$ | $t_3$ | $t_4$ | $t_5$ | $t_6$ | $t_7$ |
|--------|-------|-------|-------|-------|-------|-------|-------|
| $tp_1$ | 0 | 0 | 0 | 0 | 0 | 0 | 0 |
| $tp_2$ | 0 | 0 | 0 | 0 | 0 | 0 | 0 |
| $tp_3$ | 0 | 0 | 0 | +1 | −1 | 0 | 0 |
| $tp_4$ | 0 | 0 | +1 | −1 | 0 | 0 | 0 |
| $tp_5$ | 0 | 0 | 0 | +1 | 0 | −1 | 0 |
| $tp_6$ | 0 | 0 | 0 | 0 | 0 | +1 | −1 |
| $tp_7$ | +1 | −1 | 0 | 0 | 0 | 0 | 0 |
| $tp_8$ | 0 | +1 | −1 | 0 | 0 | 0 | 0 |

а) $\left| A_{L2} \right| =$

Рис. 25. Згенерована матриця інцидентності мережі Петрі (а), що представляє алгоритм параметричної настройки КСАУ, і модифікація параметрів настроювання згідно з даним алгоритмом (б)

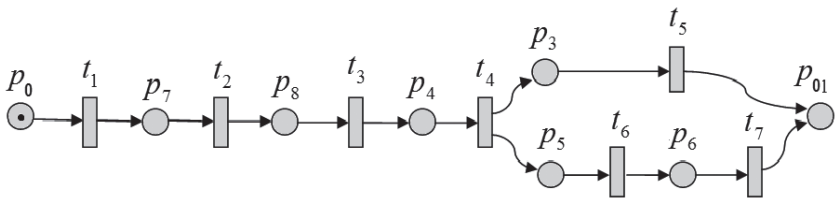

Рис. 26. Синтезована мережа Петрі, що представляє алгоритм параметричної настройки координувальної системи автоматичного управління

**Порівняльний аналіз результатів параметричного синтезу на базі сформованих алгоритмів з результатами параметричної оптимізації координувальної системи автоматичного управління за допомогою модуля Response Optimization середовища MATLAB/Simulink.** Для оцінки сформованого алгоритму параметричної настройки КСАУ на базі формуючого автомату доцільно порівняти результати синтезу з результатами параметричної оптимізації КСУ за допомогою модуля Response Optimization середовища MATLAB/Simulink. Модуль Simulink Response Optimization працює з лінійними та нелінійними, дискретними і безперервними моделями. Пакет виконує оптимізацію моделі, у циклічній послідовності порівнюючи отримані результати щодо коридору обмеження перехідного процесу. Ці особливості достатні для оптимізації КСАУ приводами робота-маніпулятора.

Очевидно, що при використанні методу gradient descent модуль буде виконувати цілий ряд недоцільних розрахунків, які відсутні в алгоритмі настройки, що був сформований на базі формуючого автомату. Однак, якщо встановити початкові значення параметрів ($k_1$=0,3; $k_2$=0; $k_3$=0; $k_4$=0,2), які дають стійкі перехідні процеси, але суттєво виходять за встановлені коридори обмежень, як показано на рисунку 27, то процес оптимізації дає бажані результати. Ці результати параметричної оптимізації, представлені на рисунку 28, можна порівняти з результатами параметричного синтезу КСАУ на базі формуючого автомату і зробити висновок, що модуль Response Optimization справляється з поставленим завданням досить якісно.

Однак, якщо при тих же настроюваннях модуля Response Optimization задати початкові значення параметрів настроювання КСАУ ($k_1$=$k_2$=$k_3$=$k_4$=0), які не дають стійкого процесу, то результати оптимізації, як показано на рисунку 30, будуть незадовільними. У цьому випадку система видасть повідомлення Optimization failed to converge — оптимізація не сходиться.

Виходячи з вищенаведених результатів, можна зробити висновок, що доцільно використовувати модуль оптимізації автоматичних систем після застосування алгоритму синтезу, який забезпечує досягнення відповідних бажаних значень показників якості роботи системи.

У результаті аналізу процесів параметричного синтезу на базі формуючого автомата було встановлено що:

— для одержання позитивних результатів, при різних початкових значеннях параметрів, алгоритм параметричного настроювання КСАУ повинен передбачати необхідний порядок настроювання рівнів;

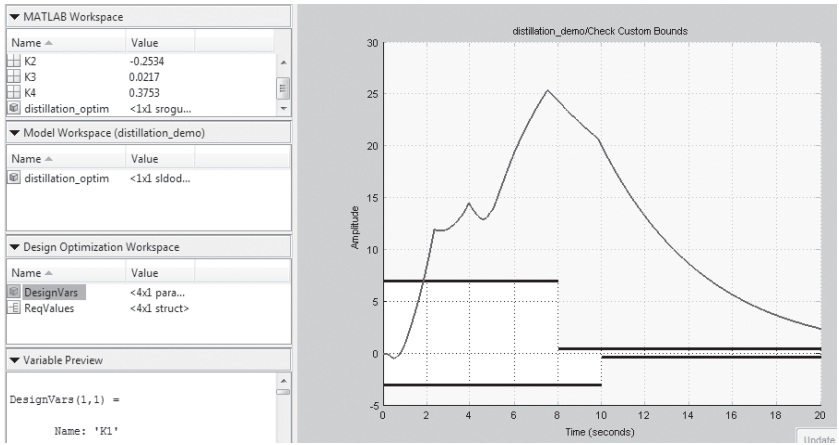

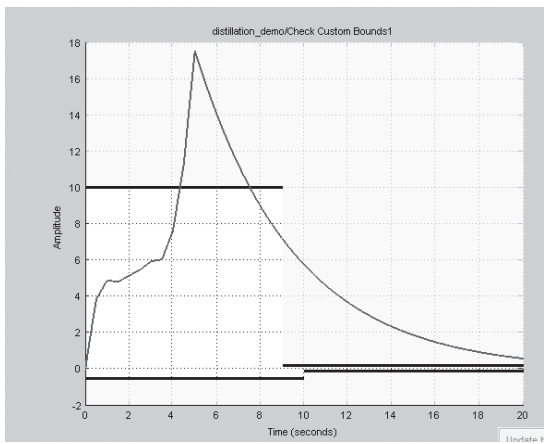

Рис. 27. Перехідні процеси щодо коридору обмежень при початкових значеннях параметрів $k_1$=0,3; $k_2$=0; $k_3$=0; $k_4$=0,2

– сформований алгоритм параметричного настроювання КСАУ формуючим автоматом виключає недоцільні розрахунки на відміну від традиційних методів випадкового пошуку тим самим зменшуючи обчислювальні витрати на процес синтезу системи;

– правила формування матриці інцидентності не виключають створення мереж Петрі з конфліктними ситуаціями, однак при встановлених пріоритетах на спрацьовуванні переходів алгоритми настроювання дають позитивні результати.

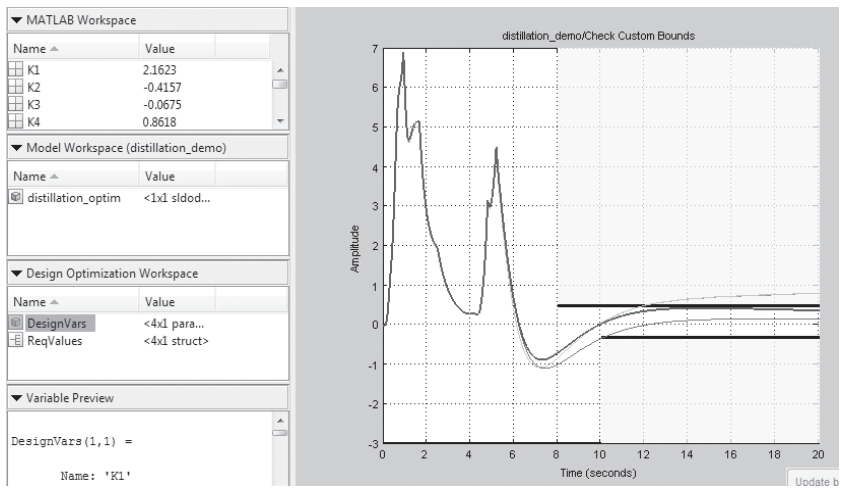

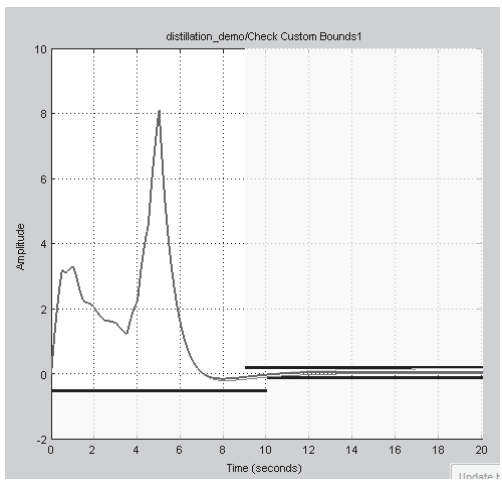

Рис. 28. Результати параметричної оптимізації за допомогою модуля Response Optimization середовища MATLAB/Simulink при початкових значеннях $k_1=0,3$; $k_2=0$; $k_3=0$; $k_4=0,2$

Подальший розвиток дослідження, пов'язаного з формуванням алгоритмів, дозволить створити на основі накопичених знань експертів інтелектуальну систему настроювання складних багаторівневих систем на базі завдань управління й особливостей технологічного об'єкта.

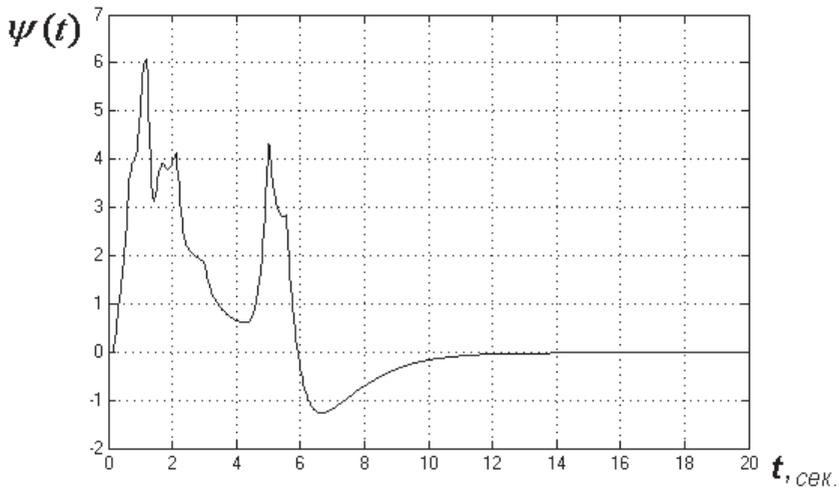

а)

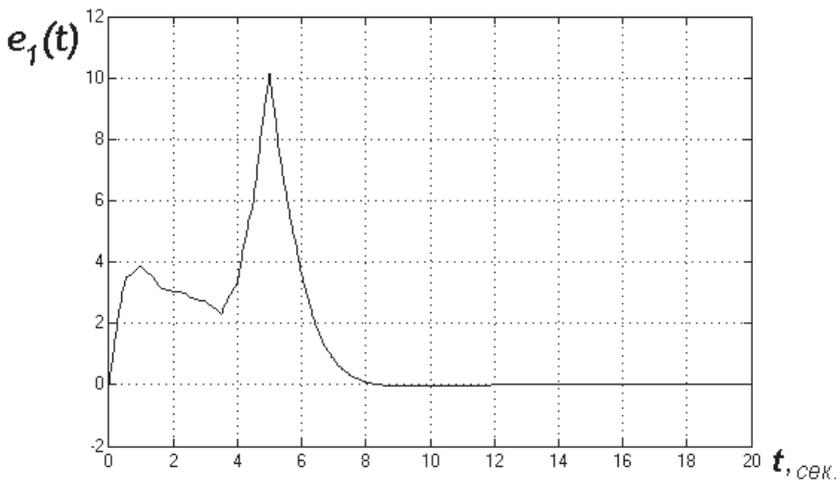

б)

Рис. 29. Результати параметричного синтезу за допомогою формуючого автомата, перехідні процеси за відхиленням від заданої траєкторії руху (а) і за сигналом неузгодженості в контурі регулювання положення захвату (б)

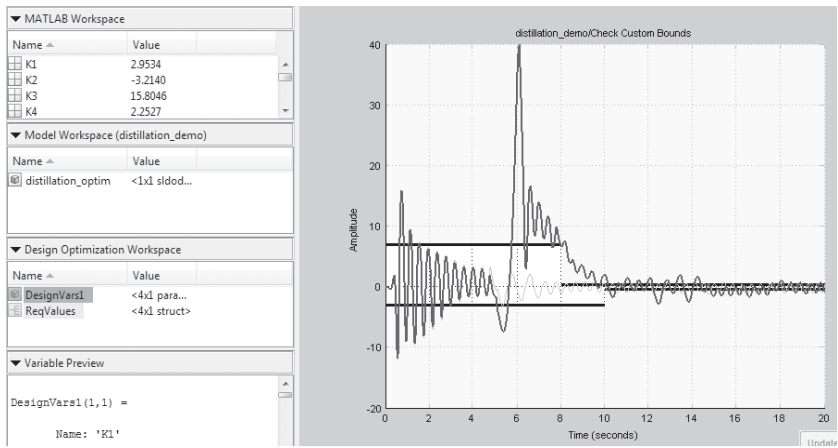

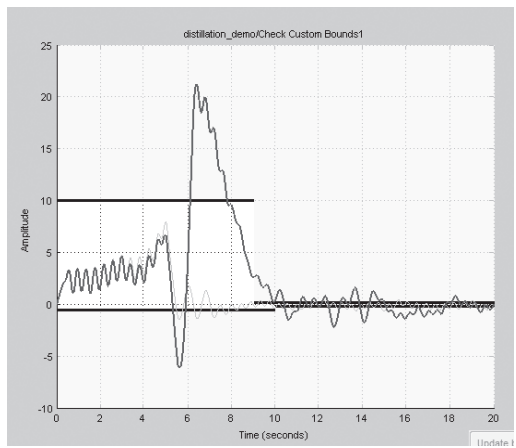

Рис. 30. Результати параметричної оптимізації за допомогою модуля Response Optimization середовища MATLAB/Simulink при початкових значеннях $k_1 = k_2 = k_3 = k_4 = 0$

**Особливості розробленої системи настроювання з урахуванням актуальності застосування засобів ДБ-мережі.** Подальша робота пов'язана з перетворенням системи, представленої на рисунку 31, з урахуванням додавання штучної нейронної мережі, що генерує сигнали визначення алгоритму.

Як і в попередньому випадку, розглядається розроблена модель координувальної системи з алгоритмом параметричного настроювання

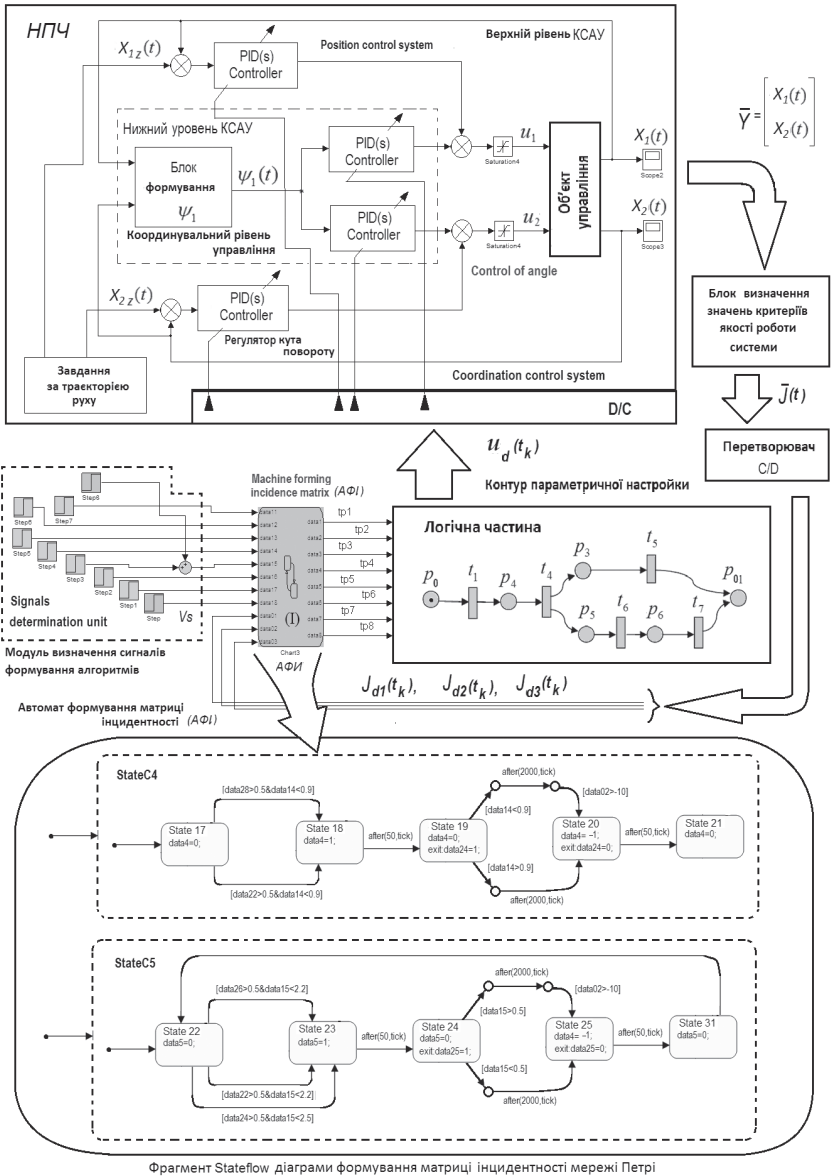

Рис. 31. Спрощена структурна схема координувальної системи управління приводами робота з системою автоматичного настроювання регуляторів, спрощена структурна схема якої представлена на рисунку 31 [14]. Як видно з рисунку, система складається із двох частин — неперервно-подійної частини і дискретно-подійної частини.

НПЧ представляє координувальну систему управління приводами робота, а ДПЧ представляє алгоритм упорядкування дій при синтезі КСУ і, у тому числі систему автоматичного настроювання.

Координувальна система автоматичного управління є дворівневою. Регулятори верхнього рівня відпрацьовують сигнали неузгодженості $e_1(t)=X_{1z}(t)-X_1(t)$ і $e_2(t)=X_{2z}(t)-X_2(t)$, де $X_{1z}(t)$, $X_{2z}(t)$, $X_1(t)$, $X_2(t)$ — задані і фактичні значення регульованих змінних. Регулятори нижнього рівня відпрацьовують відхилення від співвідношень змінних $\psi=f(X_1)\cdot X_1+k\cdot X_2-b$, де $X_1$, $X_2$ — регульовані змінні; $f(X_1)$ — нелінійна залежність, що відображена в системі у вигляді нелінійної ланки, що описує траєкторію руху робота як керованого об'єкта в координатах $X_1 - X_2$; $k$, $b$ — коефіцієнт і вільний член.

Закон управління можна представити у такому вигляді:

$$\overline{u} = \begin{bmatrix} k_2 \cdot (1+k_{21} \cdot p) \cdot \psi(p) + k_1 \cdot (1+k_{11} \cdot p) \cdot e_1(p) \\ k_4 \cdot (1+k_{41} \cdot p) \cdot \psi(p) + k_3 \cdot (1+k_{31} \cdot p) \cdot e_2(p) \end{bmatrix} = \overline{u}_q + \overline{u}_p = \begin{bmatrix} u_1 & u_2 \end{bmatrix}^T,$$

де $k_2$, $k_{21}$, $k_4$, $k_{41}$ — параметри настроювання нижнього рівня КСАУ; $k_1$, $k_{11}$, $k_3$, $k_{31}$ — параметри настроювання верхнього рівня КСАУ; $p=d/dt$ — оператор диференціювання; $\overline{u}_q$, $\overline{u}_p$ — вектори управління нижнього і верхнього рівнів; $u_1$ і $u_2$ — управляючі впливи.

Параметри настроювання $k_1$, $k_2$, $k_3$, $k_4$ КСАУ повинні бути визначені системою автоматичного настроювання з урахуванням часової співпідпорядкованості процесів регулювання.

Як видно з рисунку 31, процес формування елементів матриці інцидентності мережі Петрі на базі сигналів визначення алгоритму *Vs* описується за допомогою Stateflow-діаграми середовища MATLAB/Simulink [14]. Однак з урахуванням аналізу опису процесу формування матриці інцидентності Stateflow-діаграми надалі в минулому замінені на дискретно-безперервні мережі середовища DC-net.

Слід зазначити, що Stateflow-діаграма менш інформативна для розробки логічних модулів, наприклад, відсутня візуалізація зв'язків між паралельними станами State31 — State35, що викликає труднощі в розробці певних алгоритмів. ДБ-мережа в цьому випадку має деякі переваги у візуалізації алгоритму і процесу функціонування логічного модуля (І), що формує елементи матриці інцидентності, і також ДБ-мережа має переваги в аналізі алгоритму за допомогою використання методів аналізу мереж Петрі.

Отже, доцільно для моделювання і синтезу розглянутої системи управління використовувати програмне середовище DC-net, яке має у своєму розпорядженні засоби ДБ-мереж.

Програмне середовище DC-net спеціалізоване в напрямку аналізу і синтезу складних технологічних систем з логіко-динамічним характером функціонування. У нашому випадку, як і в попередньому, розглянута система управління і синтезу, спрощена структурна схема якої представлена на рисунку 31, за принципом функціонування аналогічна логіко-динамічній. У системі присутні як дискретні $J_d(t_k)$, $u_d(t_k)$, так і неперервні сигнали $X_1(t)$, $X_2(t)$, спостерігається багаторежимний характер функціонування. Таким чином, надалі необхідно визначити принципи формування алгоритмів на базі засобів DC-net.

**Принципи формування алгоритму за допомогою генерації мережі Петрі, на базі засобів ДБ-мереж.** Згідно з рисунком 31, послідовність операцій у системі при параметричному синтезі залежить від сигналів *Vs* завдання алгоритму. Ці сигнали *Vs* можуть бути вироблені нейронною мережею, адаптованою для виконання автоматичного настроювання певної системи, в окремому випадку — системи координувального управління.

В остаточному підсумку, спрощена структурна схема такої системи, побудованої на базі нейронної мережі з можливостями самонастроювання, буде мати вигляд, представлений на рисунку 32. У цій системі алгоритм параметричного настроювання рівнів управління визначається нейронною мережею, у якій коефіцієнти міжнейроних з'єднань змінюються в процесі функціонування блоком автонастройки.

Блок автонастройки вибірково змінює коефіцієнти міжнейроних з'єднань випадковим чином. При цьому вибір коефіцієнтів міжнейронних з'єднань для зміни здійснюється на базі значень критеріїв якості роботи системи, а також на базі аналізу згенерованої мережі Петрі, що представляє алгоритм параметричного настроювання системи.

У попередній системі, спрощена структурна схема якої представлено на рисунку 31, матриця інцидентності мережі Петрі формується за допомогою модуля, функціонування якого було представлено Stateflow-діаграмою. У розглянутій системі, реалізованій на базі нейронної мережі, замість Stateflow-діаграм використовуються мережі Петрі.

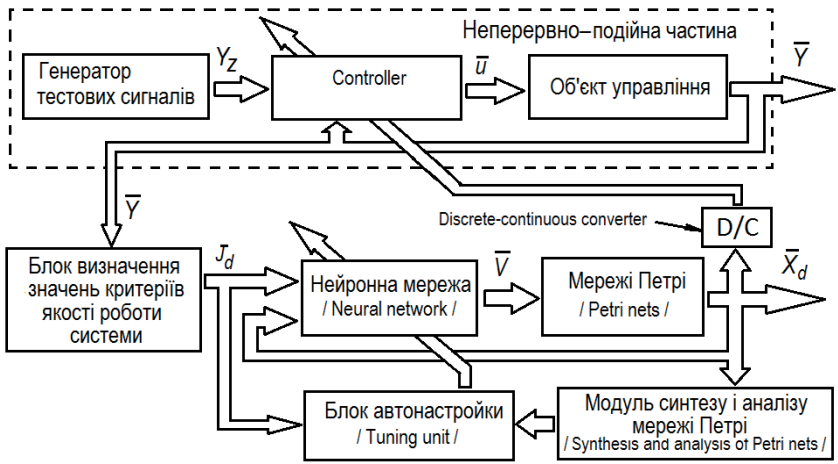

Рис. 32. Спрощена структурна схема системи, у якій здійснюється параметричне настроювання регуляторів координувального рівня управління

Нейронна мережа реалізує процес композиції мереж Петрі (рисунок 33). У результаті процесу представляється матриця інцидентності мережі Петрі, яка відображає алгоритм упорядкування дій при параметричній настройці координувальної системи.

У програмному середовищі DC-net була реалізована така система, у якій координувальний рівень управління синтезується на базі функціонування мереж Петрі і нейронної мережі. На рисунку 33 представлена схема даної системи засобами середовища DC-net.

Як видно з рисунка 33, нейронна мережа двошарова. Коефіцієнти міжнейронних з'єднань 2-го шару безпосередньо визначають інцидентну матрицю генерованої мережі Петрі. А вагові коефіцієнти 1-го шару визначають послідовність необхідних додаткових умов спрацьовування певних переходів мережі Петрі, пов'язаних з показниками якості функціонування системи $|\psi|$, $J$, де $J = \int\limits_{t_0}^{t_1} |\psi(t)| dt - \int\limits_{t_2}^{t_3} |\psi(t)| dt$, $(t_1 - t_0) = (t_3 - t_2)$, $t_0 < t_2 < t_1 < t_3$.

Відзначимо відмінні риси цієї системи, в окремому випадку за формуванням інцидентної матриці мережі Петрі, що представляє алгоритм настроювання системи. Якщо в попередньому випадку кожний рядок матриці інцидентності формується в процесі активізації різних станів Stateflow-діаграми, наприклад, поетапна активізація станів State17—State21 (рисунок 31), то в розглянутій системі ана-логічним чином формується рядок матриці інцидентності в процесі руху маркера в дискретних позиціях, наприклад, в позиціях $P_1$, $P_6$ (рисунок 33).

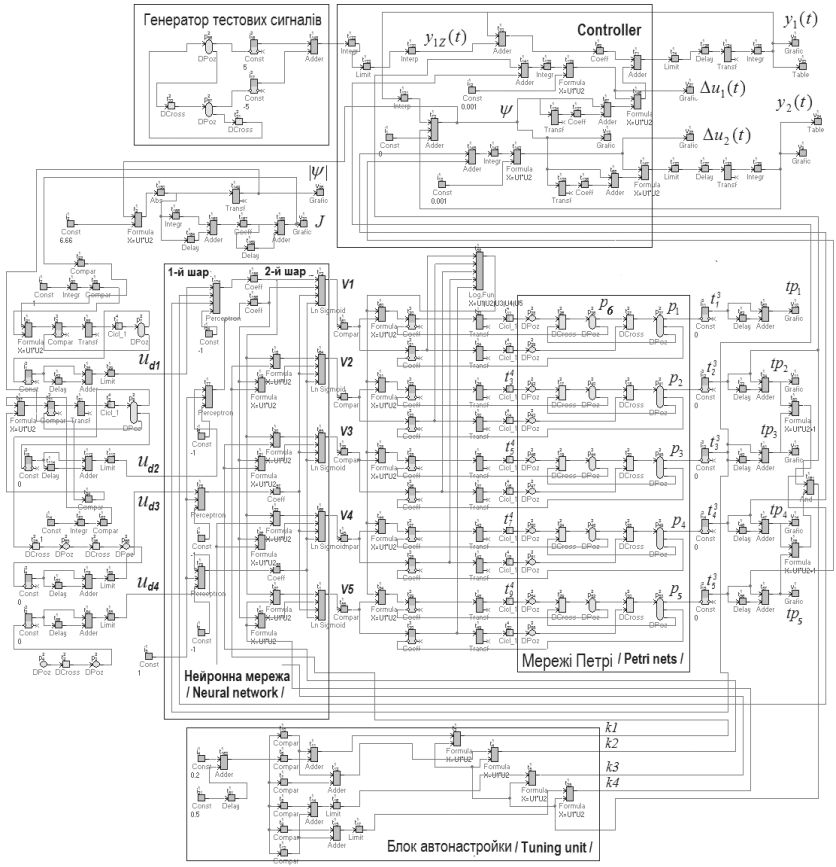

Рис. 33. Дискретно-безперервна мережа, що представляє систему управління і систему настроювання на базі нейронної мережі, що формує мережі Петрі

Залежно від сигналів завдання алгоритму *V1...V5,* які виробляє нейронна мережа, ДПЧ, що реалізована на базі засобів ДБ-мереж, генерує послідовність значень, з яких формується матриця інцидентності мережі Петрі. Така послідовність значень, як результат функціонування ДБ-мережі, представлена на рисунку 34.

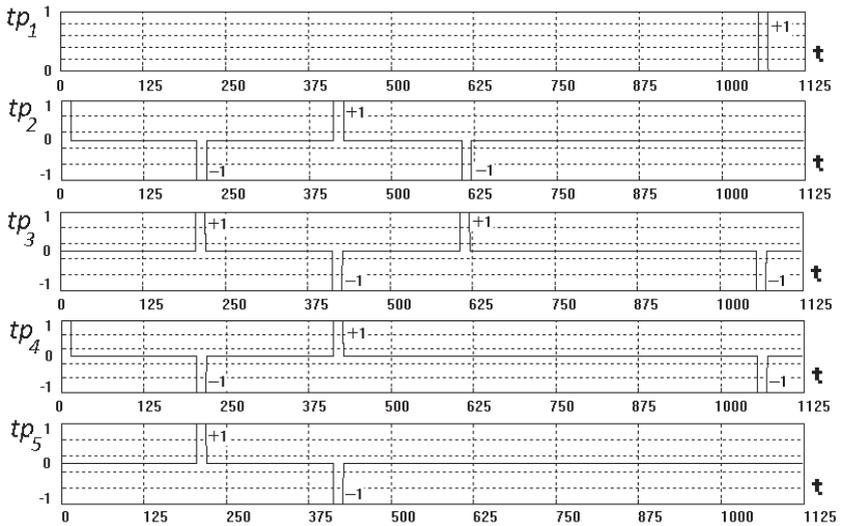

Рис. 34. Процес формування матриці інцидентності мережі Петрі

На основі процесів, представлених на рисунку 34, можна скласти матрицю інцидентності і згідно з нею сформувати мережу Петрі. Згенерована матриця інцидентності мережі Петрі має 5 рядків згідно з кількістю виходів *X1....X5* модуля і 4 стовпця згідно з кількістю кроків формування мережі Петрі. У цьому випадку матриця інцидентності має такий вигляд:

$$|A_L| = \begin{matrix} & t_1 & t_2 & t_3 & t_4 \\ tp_1 & 0 & 0 & 0 & +1 \\ tp_2 & -1 & +1 & -1 & 0 \\ tp_3 & +1 & -1 & +1 & -1 \\ tp_4 & -1 & +1 & 0 & -1 \\ tp_5 & +1 & -1 & 0 & 0 \end{matrix}. \tag{5}$$

Процес синтезу системи при даному експерименті буде здійснюватися поетапно. На рисунку 35 представлений процес зміни маркування синтезованої мережі Петрі, матриця інцидентності якої має вигляд (5). Згідно з початковим маркуванням збільшуються параметри настроювання $k_2, k_3$ координувального рівня до значення $|\psi| = 500$, після чого, у момент часу $t_1$, міняється маркування для повернення системи у вихідний стан (рисунок 36). У момент часу $t_2$, при досягненні значення критерію $J$=14, маркування також змінюється, і відповідно йому починають змінюватися параметри настроювання системи.

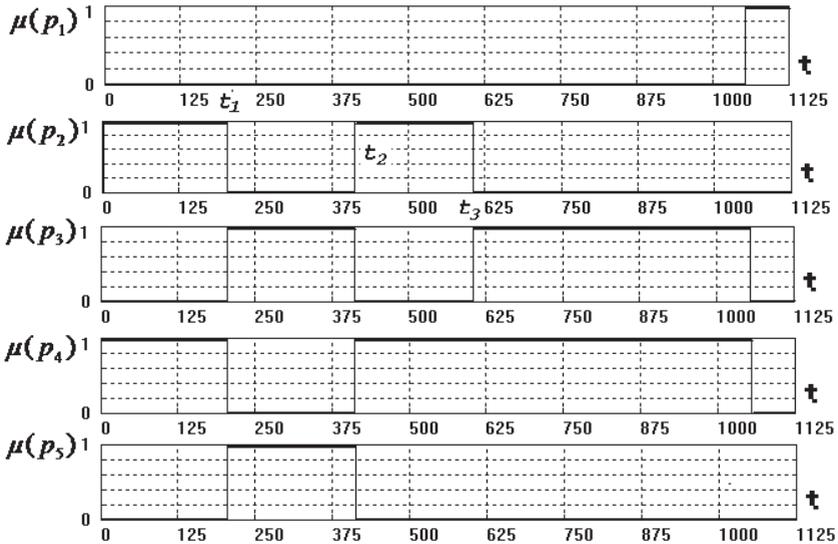

Рис. 35. Процес зміни маркування синтезованої мережі Петрі

При цьому в момент часу $t_2$ блок автонастройки змінює коефіцієнти міжнейронних з'єднань 2-го шару в рамках пошуку вірних кроків при синтезі системи. Починаючи з моменту часу $t_3$ значення критеріїв якості роботи системи зменшуються, що свідчить про вірний процес настроювання системи.

Процес формування алгоритму настроювання рівнів КСАУ шляхом визначення матриці інцидентності полягає в особливості функціонування реалізованої ДБ-мережі.

Згідно зі схемою, представленою на рисунку 33, ДБ-мережа містить дискретні підмережі, пов'язані дискретно-безперервними переходами $t_i^3$, $t_i^4$, де $i$=1...N. Кожна така підмережа є мережею Петрі, яку можна розглядати незалежно від усієї ДБ-мережі.

Нейронна мережа формує сигнали $Vs$=$|V1 .... V5|^T$, згідно з якими здійснюється рух маркерів у мережах Петрі. При цьому рух маркерів носить погоджений характер. Наприклад, вихід маркера з позиції $P_2$ супроводжується появою маркера в позиції $P_3$.

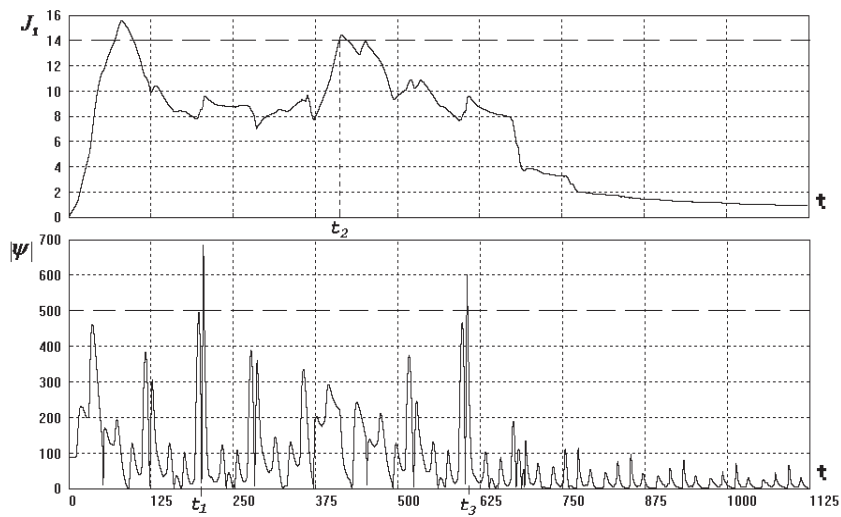

Рис. 36. Графіки зміни значень показників якості роботи системи в часі

Погоджений характер зміни маркування в мережах Петрі дає можливість виконати композицію цих мереж в одну загальну мережу Петрі. Це показано на рисунку 36.

У результаті проведених досліджень установлена принципова придатність розробленої системи, що формує композицію мережі Петрі на базі функціонування штучної нейронної мережі. Визначена схема перенастроювання окремих блоків нейронної мережі в процесі функціонування самонастроювальної системи. Однак слід зазначити, що цю систему необхідно розглядати як наближення до деякого оптимального варіанта, у якому повинен бути присутній алгоритм настроювання нейронної мережі, пов'язаний із завданням синтезу системи управління.

Розглянута система відображає основні особливості формованої інтелектуальної технології, спрямованої на заміну експерта при синтезі особливого класу систем управління. Така заміна дозволяє значно скоротити час на розробку системи управління, а також сформована технологія дасть можливість синтезувати ефективні алгоритми самонастроювання координувальних систем управління.

Наступний етап роботи безпосередньо повинен бути пов'язаний з розвитком інтелектуальної системи, пов'язаної з автоматизованим настроюванням складного класу багаторівневих систем управління.

Ця інтелектуальна система повинна забезпечувати настроювання систем на базі автоматичного синтезу мереж Петрі при функціонуванні штучної нейронної мережі.

**Принципові особливості розробленої системи, побудованої на базі штучної нейронної мережі, що забезпечує синтез мереж Петрі.** Система автоматизованого настроювання координувальних систем автоматичного управління була розроблена у програмному середовищі MATLAB/Simulink. Структурна схема системи автоматизованого настроювання КСАУ представлена на рисунках 37 і 38.

З рисунку 37 видно, що регулятори № 1 і № 2 відпрацьовують відхилення φ від заданого співвідношення значень регульованих змінних $X_1$ і $X_2$, де $\varphi = |A|^T \cdot \bar{X} + b$; $|A|^T = |a_1 \quad a_2|$ — матриця коефіцієнтів, що визначають задане співвідношення між регульованими змінними $X_1$ і $X_2$; $\bar{X} = |X_1 \quad X_2|^T$; $b$ — вільний член.

Регулятор № 3 відпрацьовує відхилення $e(t) = X_{Z3} - X_3(t)$ регульованої величини $X_3(t)$ від заданого значення $X_{Z3}$.

Об'єкт управління, у рамках лінійної моделі, можна описати сукупністю рівнянь:

$$X_1 = W_{11}(p) \cdot u_1 + W_{12}(p) \cdot u_2 + X_{01},$$

$$X_2 = W_{21}(p) \cdot u_1 + W_{22}(p) \cdot u_2 + X_{02}, \qquad (6)$$

$$X_3 = W_3(p) \cdot (W_{31}(p) \cdot X_1 + W_{32}(p) \cdot X_2 + X_{03}) + X_{04},$$

де $X_1$, $X_2$, $X_3$ — регульовані змінні; $u_1$, $u_2$ — управляючі впливи; $X_{01}$, $X_{02}$, $X_{03}$, $X_{04}$ — початкові значення регульованих змінних при $u_1 = u_2 = 0$; $W_{ij}(p)$ — передатна функція за відповідним каналом, де $i = 1...3$, $j = 0...2$.

При цьому можливі різні варіанти структурних схем моделей об'єктів управління. Наприклад, при $W_{12}(p) = W_{21}(p) = W_{31}(p) = 0$, $W_{32}(p) = W_3(p) = 1$, $X_2 = X_3$, $X_{03} = X_{04} = 0$.

Управляюча частина системи автоматизованого настроювання КСАУ пов'язана з нейропроцесором, відзначеним на рисунку 38 у вигляді блоку Subsystem 31 neuroprocessor. Цей процесор складається з сукупності локальних нейронних мереж, кожна з яких представляє той або інший алгоритм поетапного настроювання КСАУ. На рисунку 38 представлені локальні нейронні мережі у вигляді блоків Subsystem 1 і Subsystem 2 у рамках фрагмента структурної схеми процесора — Subsystem 31.

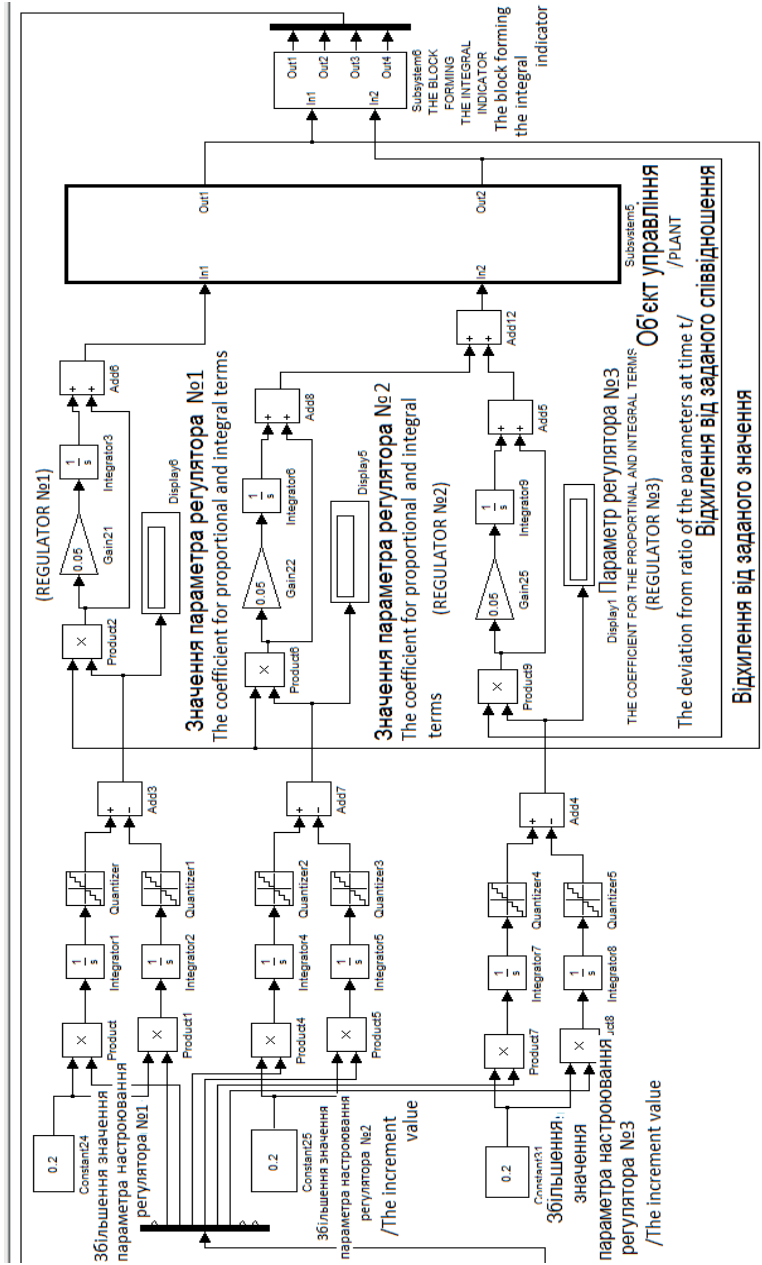

Рис. 37. Структурна схема системи автоматизованого настроювання регуляторів різних рівнів координувальної системи автоматичного управління

При настроюванні певної КСАУ необхідно встановити збільшення значень параметрів регуляторів № 1, № 2, № 3 у блоках Constant 24, Constant 25, Constant 31. Потім, виходячи з інерційності об'єкта управління, необхідно відзначити інтервали часу $t_1$ за оцінкою збільшень значень показників якості роботи КСАУ $\Delta J_{00}$, $\Delta J_{01}$, $\Delta J_{02}$, $\Delta J_{03}$. Таким чином, настроювання системи пов'язане з мінімізацією таких показників:

$$\Delta J_{00} = \left[ \int_{t_0}^{t_1} |e(t)| dt - \int_{t_1}^{t_2} |e(t)| dt \right] \to \min ,$$

$$\Delta J_{01} = \left[ \int_{t_0}^{t_1} |\varphi(t)| dt - \int_{t_1}^{t_2} |\varphi(t)| dt \right] \to \min ,$$

$$\Delta J_{02} = \left[ \int_{t_0}^{t_1} \left( |e(t)| + 3 \cdot |\varphi(t)| \right) dt - \int_{t_1}^{t_2} \left( |e(t)| + 3 \cdot |\varphi(t)| \right) dt \right] \to \min ,$$

$$\Delta J_{03} = \left[ \int_{t_0}^{t_1} \left( |\varphi(t)| + |e(t)| \right) dt - \int_{t_1}^{t_2} \left( |\varphi(t)| + |e(t)| \right) dt \right] \to \min ,$$

де $e(t)$ — відхилення регульованої змінної від заданого значення; $\varphi(t)$ — відхилення від заданого співвідношення значень регульованих змінних у часі.

Інтервали часу $t_1$ відзначаються в блоках Transport Delay 10, 15, 16, 12, показаних на рисунку 40. В остаточному підсумку необхідно встановити завдання процесору для запуску певної локальної нейронної мережі, яка буде формувати послідовність операцій поетапного настроювання КСАУ і відображати результати настроювання у вигляді синтезованої мережі Петрі.

Суть розглянутої процедури автоматичного синтезу мереж Петрі, як було визначено, полягає у виконанні двох послідовних етапів:

1) вибір з безлічі алгоритмів необхідного;

2) корекція обраного алгоритму і відповідної мережі Петрі.

Ці два етапи можна реалізувати за допомогою штучної нейронної мережі (ШНМ) і її тренування. Така ШНМ позначена нами у вигляді блоку Subsystem 31 (PROCESSOR).

Верхній рівень процесора або вхідний шар штучної нейронної мережі виконує функцію класифікації алгоритмів і вибору локальної нейронної мережі для запуску на функціонування.

Локальна нейронна мережа взаємодіє на принципах зворотного зв'язку з синхронно функціонуючими мережами Петрі, з яких можна реалізувати композицію формованої мережі Петрі. Це продемонстровано на рисунку 39.

Рис. 38. Структурна схема управляючої частини системи автоматизованого настроювання КСАУ

Як видно з рисунка, матриця інцидентності синтезованої мережі Петрі аналогічно відповідає матриці коефіцієнтів міжнейронних з'єднань 2-го шару нейронної мережі. Таким чином, якщо синтезована мережа відображає неадекватний алгоритм, то в штучній нейронній мережі повинні відповідно змінитися коефіцієнти міжнейронних з'єднань.

Нейронна мережа взаємодіє з мережами Петрі за допомогою структурно-керованих переходів $t_{17}$, $t_{18}$, $t_{19}$, $t_{20}$, вихідний сигнал яких залежить від маркування позицій $p_0$, $p_2$, $p_4$, $p_5$, у такий спосіб:

$$X_{out,t_{17}} = \begin{cases} 1 & npu \quad \mu(p_0) = 1 \\ 0 & npu \quad \mu(p_0) = 0 \end{cases}.$$

Переходи $t_{13}$, $t_{14}$, $t_{15}$, $t_{16}$ спрацьовують при наявності вихідних сигналів з ШНМ і при значенні яких предикатні функції $\Pi_i \rightarrow t_{ij}$ істинні. При спрацьовуванні переходів $t_{13}$, $t_{14}$, $t_{15}$, $t_{16}$ з'являються маркери у вихідних позиціях $p_{11}$, $p_{12}$, $p_{13}$, $p_{14}$.

Таким чином, у процесі функціонування локальна нейронна мережа буде генерувати послідовність вихідних сигналів відповідної інцидентної матриці синтезованої мережі Петрі. А в результаті синхронної роботи мереж Петрі можна виконати композицію мережі Петрі, яка буде відображати, в окремому випадку, алгоритм настроювання координувальної системи автоматичного управління.

З рисунку 41 видно, що композиція мереж Петрі реалізується при одночасному спрацьовуванні переходів. Наприклад, переходи $t_9$ і $t_6$ при одночасному спрацьовуванні поєднуються в перехід $t_1$, а переходи $t_7$ і $t_{12}$ аналогічно поєднуються в перехід $t_3$, і так далі можна виконати композицію відповідної мережі Петрі.

**Експерименти, пов'язані з настроюванням координувальної системи управління.** У програмному середовищі MATLAB/Simulink 2012 було реалізовано автоматизоване настроювання декількох різних координувальних систем автоматичного управління на основі розробленої системи автоматично формуючої мережі Петрі.

Результати експериментів, пов'язаних із процесом синтезу мереж Петрі, неодноразово представлялися в наукових працях [3; 8; 12]. Для підтвердження принципової придатності розробленої системи на рисунку 42 представлені процеси настроювання координувальної системи автоматичного управління охолодженням продуктів у холодильній тунельній камері.

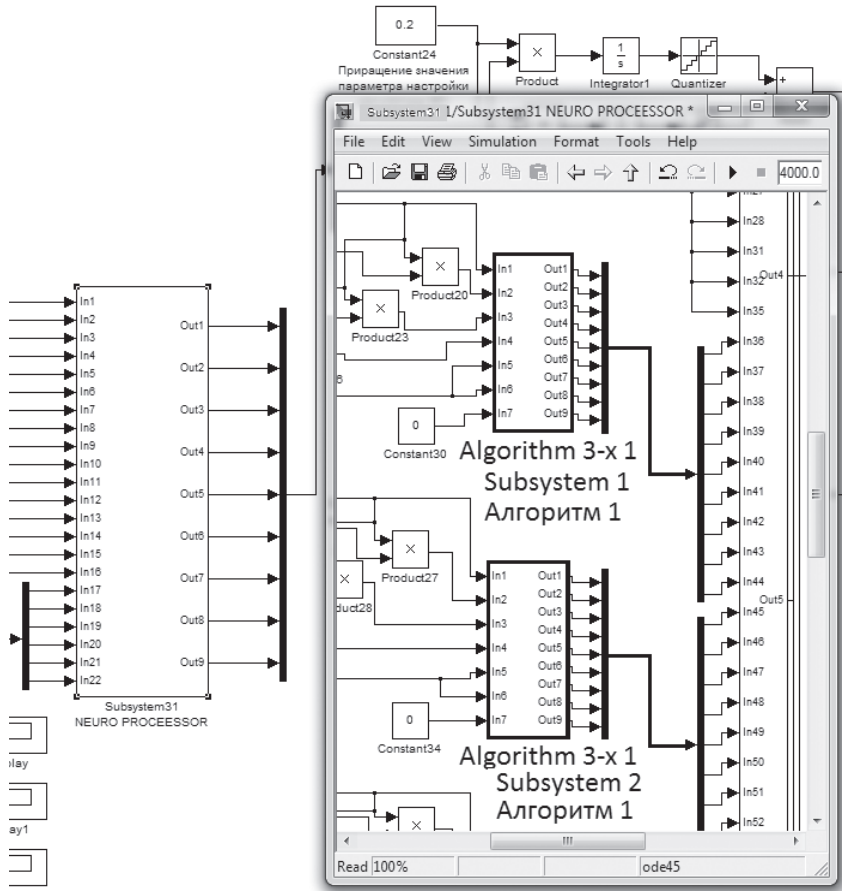

Рис. 39. Фрагмент структурної схеми блоку Subsystem 31 Neuro processor

Модель холодильної установки з тунельною камерою, як об'єкт управління, можна описати рівняннями (6) [30], де $W_{11}(p) = \dfrac{0,02 \cdot \exp(-1P)}{(15p+1) \cdot (15p+1)}$ — передатна функція за каналом управляючий вплив $u_1$ за швидкістю обертання двигунів нагнітувачів у тунельній камері — температура повітря на вході в тунельну камеру $X_1$, ℃; $W_{12}(p) = \dfrac{0,02 \cdot \exp(-1P)}{(90p+1) \cdot (90p+1)}$ — передатна функція за каналом управляючий вплив $u_2$ за швидкістю руху транспортера в тунельній

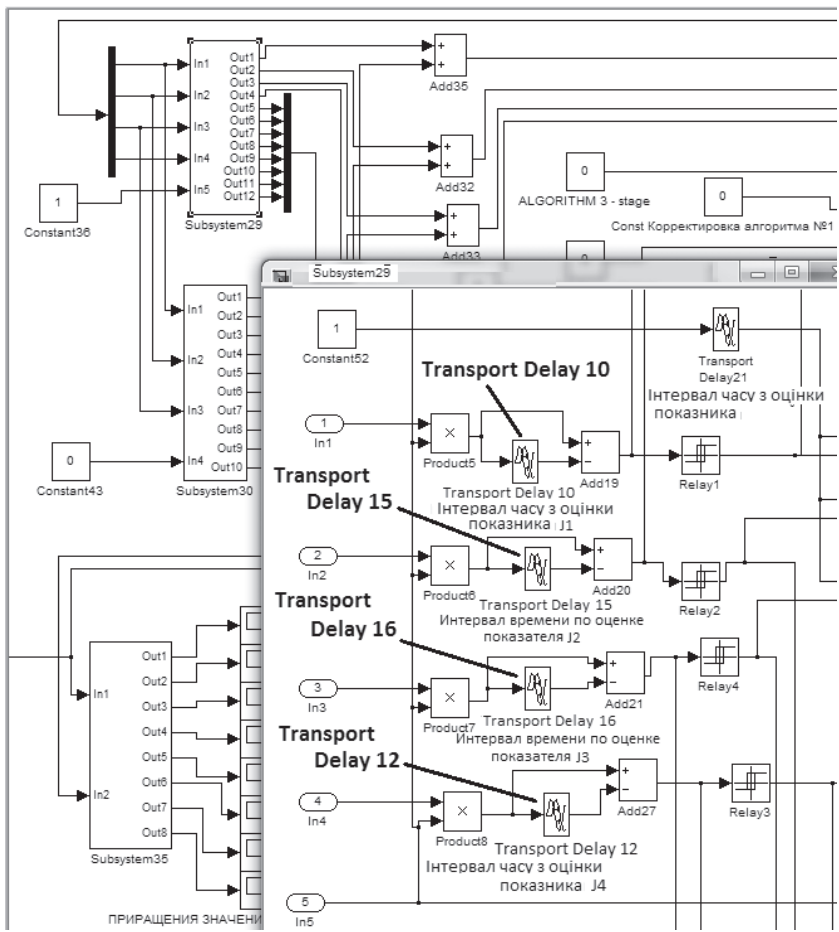

Рис. 40. Фрагмент схеми формування збільшень інтегральних показників якості роботи системи управління

камері — температура повітря на вході в тунельну камеру $X_1$, ℃;
$W_{21}(p) = \dfrac{0,04 \cdot \exp(-1P)}{(10p+1) \cdot (110p+1)}$ — передатна функція за каналом управ-

ляючий вплив $u_1$ за швидкістю обертання двигунів нагнітувачів у тунельній камері — температура повітря на виході з тунельної камери
$X_2$, ℃; $W_{22}(p) = \dfrac{0,04 \cdot \exp(-1P)}{(90p+1) \cdot (90p+1)}$ — передатна функція за каналом

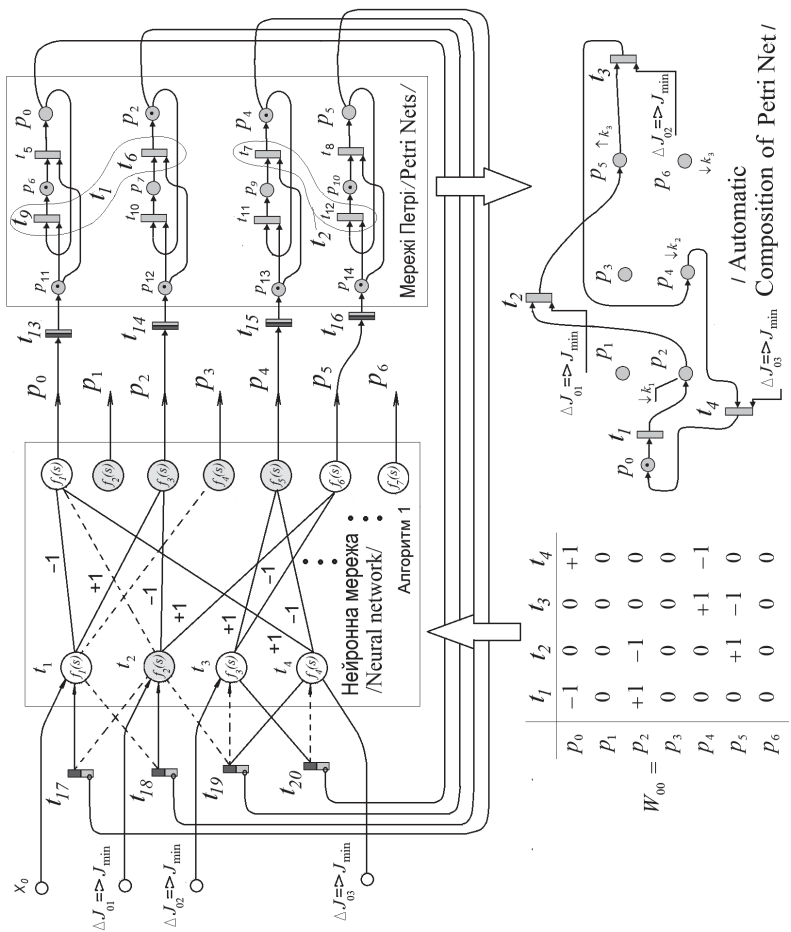

Рис. 41. Схема формування композиції мережі Петрі при функціонуванні штучної нейронної мережі

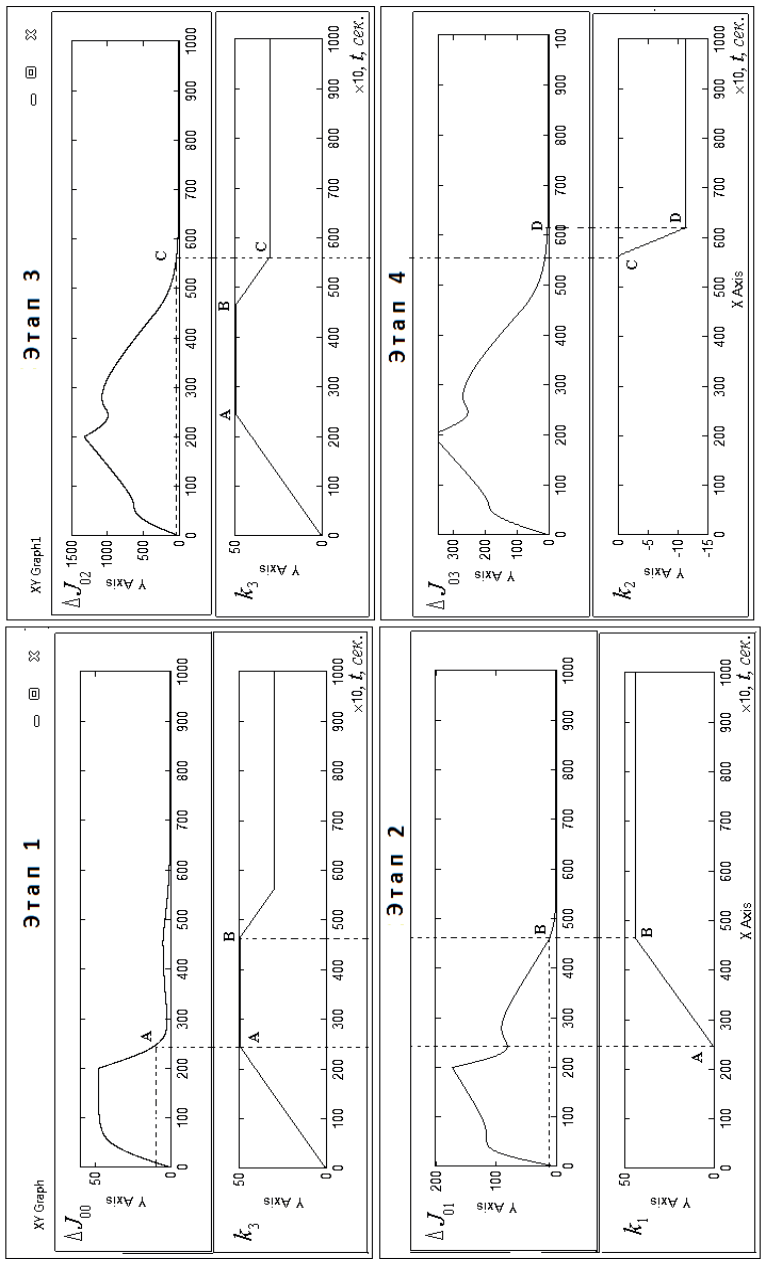

Рис. 42. Часові характеристики процесу настроювання координувальної системи автоматичного управління процесом охолодження продуктів у тунельній камері управляючий вплив $u_2$ за швидкістю руху транспортера в тунельній камері — температура повітря на виході з тунельної камери $X_2$, ℃; $X_{01} = -10$ ℃; $X_{02} = -4$ ℃; $W_{31}(p) = 0$, $W_{32}(p) = W_3(p) = 1$, $X_2 = X_3$, $X_{03} = X_{04} = 0$.

Закон управління розглянутої системи:

$$\overline{u} = \begin{bmatrix} u_1 & u_2 \end{bmatrix}^T ;$$

де $\quad u_1 = k_1 \cdot \left(1 + \dfrac{0{,}05}{p}\right) \cdot \varphi ; \qquad u_1 = k_2 \cdot \left(1 + \dfrac{0{,}05}{p}\right) \cdot \varphi + k_3 \cdot \left(1 + \dfrac{0{,}05}{p}\right) \cdot e ;$

$\varphi(t) = \begin{bmatrix} -1 & 3 \end{bmatrix} \cdot \overline{X} + b$ — відхилення від заданого співвідношення значень регульованих змінних; $b = 0$; $\overline{X} = \begin{bmatrix} X_1(t) & X_2(t) \end{bmatrix}^T$ — вектор регульованих змінних; $e(t) = X_{2Z} - X_2(t)$; $X_{2Z}$ — задане значення температури на виході з холодильної тунельної камери, ℃.

З рисунку 42 видно, що на кожному етапі настроювання приріст $\Delta J_{0i}$, $i=0...3$ відповідного критерію якості роботи системи (2) зводиться до мінімального значення при зміні певного параметра настроювання $k_j$, $j=1...3$. На часовому інтервалі від 0 до точки А настроюється регулятор № 3 верхнього рівня управління, на ділянці від А до В настроюється регулятор № 1 координувального рівня управління. Потім на ділянці В–С коректується параметр настроювання $k_3$ регулятора № 3 і в остаточному підсумку на 4 -му етапі настроюється регулятор № 2 координувального рівня.

Таким чином, система одержує кінцеві значення параметрів настроювання $k_1$, $k_2$, $k_3$ КСАУ, які забезпечують відповідну якість регулювання з урахуванням забезпечення режиму поділу руху [19].

Кожний етап процесу настроювання, представлений на рисунку 42, реалізується згідно зі згенерованою мережею Петрі, яка відображає процес настроювання КСАУ.

Штучна нейронна мережа реалізує формування і функціонування мережі Петрі покроково. На кожному кроці представляється мережа Петрі відповідно під номерами 1, 2 і т. д., як показано на рисунку 43.

Наприклад, згідно з рисунком 43, якщо перехід $t_1$ веде до небажаної ситуації — значення інтегральних показників якості системи за межами припустимого, то відзначається помилка на переході $t_1$ і змінюються коефіцієнти міжнейронних з'єднань таким чином, що синтезується мережа Петрі, представлена під номером 3. Отже, якщо були помилки на переходах $t_1$ і $t_2$, то сформується мережа Петрі під номером 4 і так далі, згідно з операціями, представленими на рисунку 43.

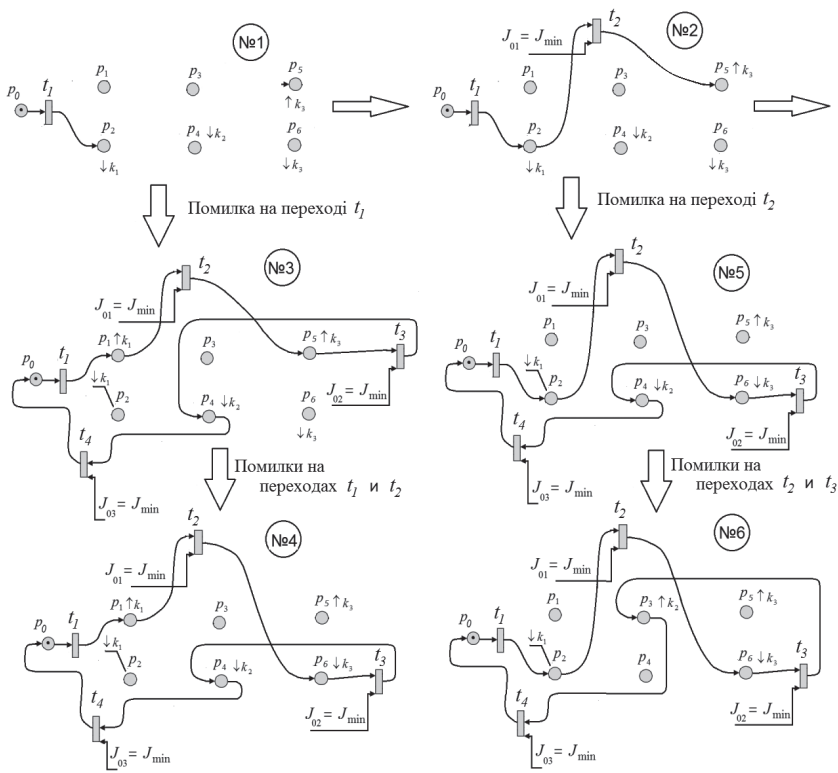

Рис. 43. Візуалізація процесів синтезу різних мереж Петрі, які відображають етапи настроювання координувальної системи автоматичного управління

Як видно з часових характеристик, представлених на рисунку 42, процес настроювання системи здійснюється за умови стійкої її роботи. Усі значення збільшень установлених критеріїв якості роботи системи рухаються до нуля, що вказує не тільки на стійку роботу синтезованої системи управління, але і на позитивний результат настроювання. Сформована мережа Петрі, що представлена на рисунку 41, відображає відповідний алгоритм настроювання системи, однак корегування цієї мережі Петрі користувачем або експертом спричинить впровадження в процес перенастроювання штучної нейронної мережі. Отже, процес узгодження таких дій, які тягнуть зміну мережі Петрі згідно з рисунком 43, важливо розглянути як окремий спосіб тренування штучної нейронної мережі з учителем.

**Висновки.** Нами було вирішено поставлену задачу, пов'язану з розробкою системи автоматизованого настроювання багаторівневих координувальних систем автоматичного управління. У процесі виконання поставленої задачі була встановлена принципова придатність систем, автоматично синтезуючих мережі Петрі на основі функціонування штучних нейронних мереж.

Робота була виконана в рамках наукових досліджень кафедри автоматизації технологічних процесів і робототехнічних систем Одеської національної академії харчових технологій в 2013—2020 рр.

### *СПИСОК ВИКОРИСТАНОЇ ЛІТЕРАТУРИ*

# СИСТЕМА АВТОМАТИЗОВАНОГО КОНТРОЛЮ ТЕХНІЧНОГО СТАНУ ТА ДІАГНОСТУВАННЯ ПОТУЖНИХ ОБЕРТОВИХ ЕЛЕКТРИЧНИХ МАШИН

*Граняк В. Ф.*

У роботі запропоновано перспективну апаратно-програмну реалізацію системи автоматизованого контролю та діагностування потужних обертових електричних машин. Реалізацію зазначеної автоматизованої системи пропонується здійснювати на основі нестандартної штучної нейроподібної мережі, що виступає в якості ключового елемента формування логічного висновку, як типового представника системи виключної складності. Показано, що запропоноване рішення може розглядатись як окремо взятий унікальний випадок, що має значну практичну цінність, оскільки може бути адаптованим для вирішення задач широкого класу.

Було показано, що одним з найбільш перспективних вхідних сигналів, що може використовуватися при побудові таких систем є вібросигнал. Зокрема, зазначений параметр володіє достатньо високою інформативністю та може бути виміряний безпосередньо в режимі реального часу роботи електричної машини без необхідності суттєвого втручання у її конструкцію.

*З метою підвищення ефективності роботи системи автоматизованого контролю та діагностування було визначено та теоретично обґрунтовано тривалість часових реалізацій вібросигналу, що доцільно використовувати при отриманні коефіцієнтів взаємокореляції вібросигналів у досліджуваних вузлах. А також запропоновано інтегральні високоінформативні числові критерії оцінки впливу нерівноваженості ротора, семетричного зростання напруженості основного електро-магнітного поля, асиметрії струмів у статорному колі та дефектів підшипників на коефіцієнти вейвлет-перетворення у вигляді середньоквадратичного значення вейвлет коефіцієнтів інформативних смуг частот при дослідженні часового інтервалу, що значно перевищує період обертання ротора електричної машини. Показано, що зазначені критерії мають понижену чутливість до впливу неінформативних одиничних збурень, які можуть виникати в процесі роботи електричної машини.*

*The paper proposes a promising hardware and software implementation of an automated monitoring system and diagnostics of powerful rotating electric machines. The implementation of this automated system is proposed to be carried out on the basis of a non-standard artificial neural network, which acts as a key element in forming a logical conclusion, as a typical representative of a system of exceptional complexity. It is shown that the proposed solution can be considered as a single unique case that has significant practical value, as it can be adapted to solve problems of a wide class.*

*It was shown that one of the most promising input signals that can be used in the construction of such systems is a vibrating signal. In particular, this parameter is quite informative and can be measured directly in real time of the electric machine without the need for significant intervention in its design.*

*In order to increase the efficiency of the automated control and diagnostic system, the duration of time realizations of the vibration signal was determined and theoretically substantiated, which should be used in obtaining the correlation coefficients of vibration signals in the studied nodes. Also, was proposed integrated highly informative numerical criteria for estimating the influence of rotor unbalance, symmetric increase of main electromagnetic field strength, asymmetry of currents in the stator circuit and bearing defects on wavelet transform coefficients in the form of root mean square value. exceeds the period of rotation of the rotor of the electric machine. It is shown that these criteria have a reduced sensitivity to the influence of uninformative single perturbations that may occur during the operation of the electric machine.*

Сьогодні склалася стійка тенденція до розвитку систем автоматизованого контролю та діагностування обертових електричних машин, що обумовлюється як збільшенням кількості відповідного обладнання, яке відпрацювало свій гарантійний строк, так і розвитком сучасної обчислювальної техніки та методів обробки вимірювальної інформації [1]. Проте на сьогодні не існує узагальненої теорії побудови таких систем, що суттєво ускладнює їх практичну реалізацію.

В процесі аналізу імовірності появи дефектів при роботі обертових електричних машин було показано, що для будь-якої електричної машини під час оцінювання надійності її роботи доцільне застосування методу слабких ланок, що дозволяє виокремити найбільш типові дефекти, які з прийнятною імовірністю варто очікувати при експлуатації такого обладнання [2]. Статистичний аналіз найбільш імовірних дефектів, що характерні для електричних машин, показав, що такими дефектами є: незбалансованість ротора, пошкодження підшипників, пошкодження ізоляції (пробій) обмоток ротора, пошкодження ізоляції (пробій) обмоток статора, пошкодження кріплень стержнів чи деформація статора та порушення механічної жорсткості кріплень [3].

Тож, на основі сказаного, можна дійти висновку, що розробка системи автоматизованого контролю та діагностування потужних обертових електричних машин, що базувалася б на розвиненій теоретичній концепції побудови такого обладнання, є актуальною науково-прикладною задачею.

**Розробка узагальненої апаратної структури системи автоматизованого контролю технічного стану та діагностування гідроагрегатів.** Реалізація системи автоматизованого контролю та діагностування потужних обертових електричних машин має характеризуватися гнучкістю та можливістю модернізації у широких межах, залежно від умов та особливостей експлуатації, а також необхідної ефективності [4; 5]. Тож, до загальних принципів побудови таких систем варто віднести модульний підхід до нарощування кількості вимірювальних каналів з можливістю відносно легкої модернізації шляхом підключення додаткових пристроїв та зміни програмного алгоритму роботи систем.

Крім цього є очевидною необхіднісь застосування дворівневої апаратної системи обробки вхідної інформації (результатів вимірювання). Зокрема перший рівень доцільно реалізувати у вигляді дискретних числових перетворювачів (мікроконтролерів), що здійснюватимуть формування пакетів вимірювальної інформації у придатному для подальшої обробки вигляді. Залежно від кількості вимірювальних каналів, складності вимірювальних алгоритмів та доступної апаратної продуктивності на першому рівні можуть застосовуватися на один агрегат один або декілька числових перетворювачів [4; 5].

Другий апаратний рівень, що доцільно реалізувати на основі штучної нейроподібної мережі, може бути представлений у вигляді високопродуктивного сервера, що здійснюватиме попередню обробку пакетів вхідних даних та розрахунок на їх основі високоінформативних критеріїв, що характеризують технічний стан електричної машини. З метою збільшення швидкості роботи алгоритму та враховуючи значну кількість інформації, що має передаватися від блоку попередньої обробки до штучної нейроподібної мережі (ШНМ), зазначені алгоритмічні операції доцільно виконувати у межах одного апаратного рівня [6].

У найпростішому випадку, при побудові системи автоматизованого контролю технічного стану та діагностування обертової електричної машини структура такої системи може бути подібною до [7; 8]. Структурну схему однієї з найпростіших систем автоматизованого контролю та діагностування електричної машини наведено на рис. 1.

Пристрій працює таким чином: n віброперетворювачів 11—1n здійснюють перетворення рівня віброприскорення, у n ключових вузлах електричної машини, в рівень постійної напруги, значення якої підсилюється до значення, придатного для роботи системи у n масштабуючих підсилювачах 61—6n; n смугових фільтрів 81—8n відфільтровують вищі гармоніки вхідного сигналу, що не досліджуються в процесі віброконтролю, пропускаючи на вихід лише ті гармонічні складові, за якими проводиться контроль вібраційного стану. Сигнал з виходів n смугових фільтрів 81—8n надходить на входи n елементів аналогової пам'яті 91—9n відповідно, де запам'ятовують у момент надходження з виходу формувача 7 одиничного сигналу, що відповідає повороту ротора електричної машини на визначений кут α. Цей же сигнал логічної одиниці з виходу формувача 7 поступає на перший вхід першого порту мікроконтролера 13 та служить сигналом початку операції вимірювального перетворення віброприскорення. Після цього на другому виході першого порту мікроконтролера 13 формується адресний сигнал, що відповідає першому інформаційному входу аналогового мультиплексора 11, що призводить до встановлення сигналу з його першого входу на його виході. Тоді на першому виході першого порту мікроконтролера 13 формується сигнал запуску аналого-цифрового перетворення, що поступає на другий вхід цифро-аналогового перетворювача 12, на перший вхід якого поступає сигнал з виходу аналогового мультиплексора 11, результат цифро-аналогового перетворення зчитується з виходу цифро-аналогового

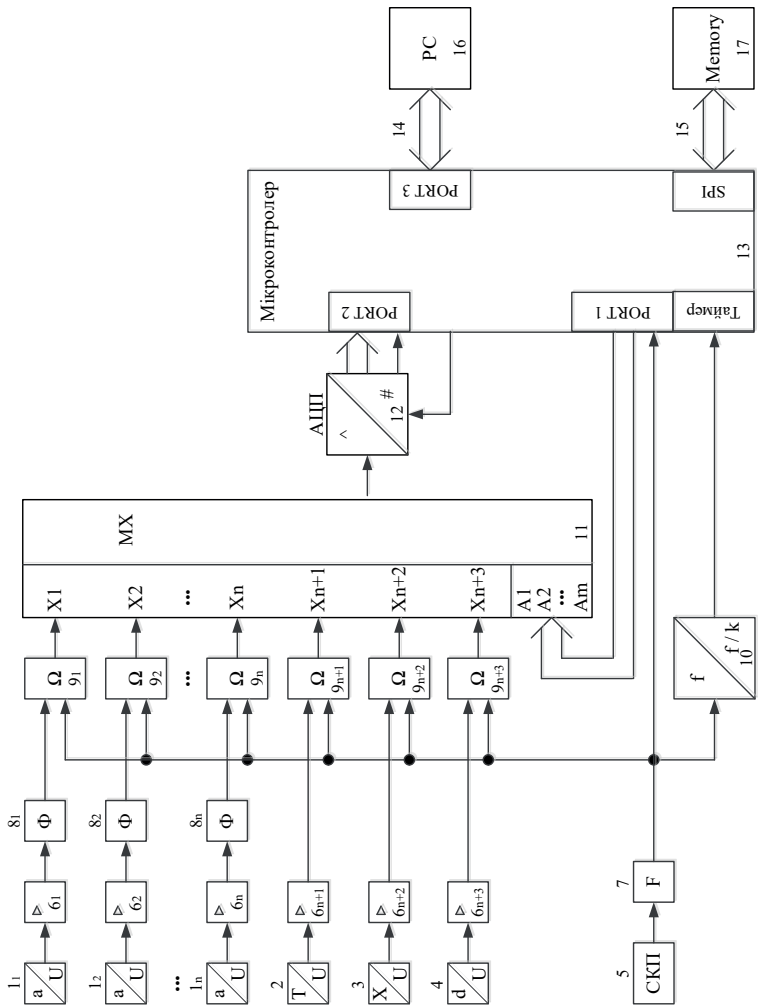

Рис. 1. Структурна схема однієї з найпростіших систем автоматизованого контролю та діагностування обертової електричної машини перетворювача 12 через перший вхід другого порту мікроконтролера 13 при приході на вхід другого порту мікроконтролера 13 сигналу закінчення вимірювального перетворення. Після цього на другому виході першого порту мікроконтролера 13 формується адреса наступного інформаційного входу аналогового мультиплексора 11. Решта операцій повторюється циклічно, доки не буде отримано цифрове значення сигналу на усіх входах аналогового мультиплексора 11, що відповідають рівням віброприскорення у всіх ключових точках агрегату, значенню температури поточної полюсної обмотки, поточному значенню осьового зміщення ротора та величини повітряного зазору між ротором та статором. Після завершення цих операцій вимірювальна система переходить у режим очікування наступного одиничного імпульсу з виходу формувача 7, а після його отримання операції повторюються циклічно.

На виході сенсора кутового положення 5 формується сигнал при повороті ротора електричної машини на заданий кут α, який поступає на вхід формувача 7. У формувачі 7 цей сигнал перетворюється у сигнал логічної одиниці та поступає, окрім других входів елементів аналогової пам'яті $81-8n+3$ та першого входу першого порту мікроконтролера 13, на вхід подільника частоти 10, на виході якого, при надходженні на його вхід k-го імпульсу, що відповідає коефіцієнту ділення частоти, формується сигнал логічної одиниці, який поступає на вхід таймера мікроконтролера 13, де служить сигналом запису поточного числа, відрахованого таймером мікроконтролера 13. При повороті ротора електричної машини на кут 360 градусів (повний оберт) на виході сенсора кутового положення 5 формується сигнал подовженої тривалості що у формувачі перетворюється на подовжений сигнал логічної одиниці, який слугує для мікроконтролера 13 маркером початку нового обороту ротора, що використовується для перевірки правильності роботи подільника частоти 10.

На виході безконтактного датчика температури 2 формується сигнал постійної напруги, що пропорційний температурі поточної полюсної обмотки ротора. Цей сигнал з виходу безконтактного датчика температури 2 надходить на вхід $n+1$-го масштабюючого підсилювача 6, де підсилюється до рівня, придатного для подальшої цифрової обробки. З виходу $n+1$-го масштабюючого підсилювача 6 підсилений сигнал надходить на перший вхід $n+1$-го елемента аналогової пам'яті 9, де запам'ятовується при надходженні на його другий вхід керуючого сигналу з виходу формувача 7.

На виході безконтактного датчика осьового зміщення ротора 3 формується сигнал постійної напруги, що пропорційний поточному осьовому зміщенню ротора. Цей сигнал з виходу безконтактний датчик осьового зміщення ротора 3 надходить на вхід n+2-го масштабуючого підсилювача 6, де підсилюється до рівня, придатного для подальшої цифрової обробки. З виходу n+2-го масштабуючого підсилювача 6 підсилений сигнал надходить на перший вхід n+2-го елемента аналогової пам'яті 9, де запам'ятовується при надходженні на його другий вхід керуючого сигналу з виходу формувача 7.

На виході безконтактного датчика повітряного зазору між ротором та статором 4 формується сигнал постійної напруги, що пропорційний поточному осьовому зміщенню ротора. Цей сигнал з виходу датчика повітряного зазору між ротором та статором 4 надходить на вхід n+3-го масштабуючого підсилювача 6, де підсилюється до рівня, придатного для подальшої цифрової обробки. З виходу n+3-го масштабуючого підсилювача 6 підсилений сигнал надходить на перший вхід n+3-го елемента аналогової пам'яті 9, де запам'ятовується при надходженні на його другий вхід керуючого сигналу з виходу формувача 7.

Виміряні значення віброприскорення у всіх ключових точках електричної машини, температури поточної полюсної обмотки ротора, поточного осьового зміщення ротора, повітряного зазору між ротором та статором, а також числовий код, відрахований таймером за час повороту ротора електричної машини на кут kα, передається через перший 14 та другий 16 пристрої перетворення інтерфейсу та лінію зв'язку на сервер 17. Додатково на сервер 17 поступає вимірювальна інформація від штатних сенсорів струму та напору. На сервері 17 здійснюється попередня обробка первинної вимірювальної інформації, прийняття рішення про наявність/відсутність дефектів, а також індикація результатів операцій контролю та діагностування.

Зовнішня пам'ять 15 застосовується для проміжного зберігання отриманих числових значень, пропорційних виміряним величинам, та, при потребі, програмного коду роботи мікроконтролера 13.

**Розробка та теоретичне обгрунтування концепції побудови нейроподібної мережі, як ключового елементу прийняття рішень системи автоматизованого контролю та діагностування.** Однією з головних тенденцій розвитку сучасної науки є збільшення питомої ваги систем, що можуть бути віднесені до класу систем з виключною складністю [9; 10]. Головною особливістю систем цього класу є наявність великої кількості зв'язків та (або) факторів впливу, класичний математичний опис яких є неможли-

вим або недоцільним внаслідок суттєвого зростання складності моделі, що робить її непридатною для практичного використання [10; 11].

Враховуючи масштаби суспільного запиту на наукові підходи, що можуть бути використані для розв'язання задач зазначеного класу, цілком логічний активний розвиток концепцій, які виходять за межі класичного математичного моделювання. І хоча на сьогоднішній день найбільш поширеним методом вирішення таких задач ще залишається метод експертного висновку [12], проте очевидно, що сучасний рівень розвитку науки та техніки потребує інших, більш швидких автоматизованих, а отже і менш трудомістких підходів, які можуть бути використані для побудови технічних систем оперативного реагування. До таких підходів можна віднести відносно нові напрямки нечіткої логіки та нейроподібного моделювання [13]. Проте суттєвим їх недоліком є відсутність чіткого алгоритму побудови термів чи вибору вхідних параметрів та типу структури нейроподібної мережі. Відтак, побудова кожної окремо взятої системи є унікальним технічним рішенням, що має значну науково-практичну цінність.

Однією із задач, що має значний практичний інтерес та пов'язана з необхідністю аналізу та формування логічного висновку у системі, що відноситься до систем виключної складності, є задача діагностування обертових електричних машин [14].

Одним з найперспективніших видів моніторингу технічного стану та діагностування електричних машин є вібродіагностування [14; 15], оскільки практично миттєва реакція вібросигналу на зміну технічного стану є незамінною якістю останнього в аварійних ситуаціях, коли визначальним чинником є швидкість постановки діагнозу і прийняття рішення. Крім того, віброакустичний сигнал має високу інформативність та при достатній кількості контрольованих точок дозволяє з високою вірогідністю не лише встановлювати факт наявності того чи іншого дефекту, а й потенційно виявляти місце його локалізації та прогнозувати час його розвитку [16].

Виключна складність при формуванні віброакустичних параметрів електричної машини пов'язана з динамічністю збурюючих впливів, обумовлених доволі складною конструкцією механічної частини електричної машини, що включаює у себе значну кількість просторово розподілених елементів з пружними та в'язкими зв'язками [17]. Крім цього, як було показано, аналіз одного лише вібросигналу не дозволяє забезпечити достатньо високу вірогідність контролю, а отже і досягти прийнятної ефективності системи діагностування. Тому су-часні концепції побудови таких систем передбачають аналіз не лише віброакустичного сигналу, а й інших додаткових параметрів, що неминуче призводить до підвищення складності об'єкта дослідження [6].

Оскільки побудова чіткої математичної моделі механічних зв'язків обертової електричної машини є практично неможливою, останню доцільно розглядати як «чорну скриньку». Тобто моделювати не її структуру, а зовнішнє функціонування [14]. Тому вирішення поставленої задачі доцільно здійснювати з застосуванням штучної нейроподібної мережі (ШНМ).

Оскільки діагностування неминуче передбачає необхідність прийняття логічних висновків, є очевидним, що зазначений об'єкт є класичним прикладом задачі формування логічного висновку в системах виключної складності за допомогою нейроподібної мережі. Тож, алгоритм вирішення цієї задачі та структура запропонованої нейроподібної мережі може розглядатися як окремо узятий унікальний випадок, що має значну практичну цінність, оскільки може бути адаптованим для вирішення задач подібного типу.

Для побудови ШНМ необхідно спочатку визначити, яка інформація може надходити на її входи і що необхідно отримати в результаті функціонування ШНМ.

Враховуючи аналіз, проведений раніше, побудову зазначеної системи діагностування пропонується здійснювати на базі вимірювальних каналів віброакустичного сигналу та вимірювального каналу температури, що здійснює послідовне вимірювання температури лобової частини кожної із полюсних обмоток ротора.

У системі передбачені також додаткові канали, що здійснюють вимірювання в режимі реального часу роботи машини таких параметрів, як потужність навантаження, частота обертання ротора та інших необхідних технічних характеристик.

Ці дані надходять до підсистеми поточного моніторингу, звідки, після попередньої обробки, передаються у підсистему діагностування. Попередня обробка сигналів включає в себе дискретне вейвлет-перетворення (ДВП) кожного із отриманих віброакустичних сигналів з використанням різних материнських вейвлетів, з подальшим розрахунком на основі отриманих коефіцієнтів ДВП високоінформативних критеріїв, максимально чутливих до інформативних чинників вібрації. На основі часової реалізації віброприскорення чи віброшвидкості, що типово вимірюються при роботі електричних машин [18], здійснюється також аналітичний розрахунок віброзміщення, методика якого детально описана у роботі [19]. Крім цього, у підсистемі поточного моніторингу проводиться розрахунок кутового прискорення ротора на основі миттєвих значень його швидкості обертання та інші додаткові параметри електричної машини, що можуть бути використані в якості вхідних величин ШНМ.

Отже, на вхід ШНМ мають надходити такі дані:

• всі значення віброзміщення, що перевищують допустиму норму по кожному із вібросенсорів за певний інтервал часу із часовою фіксацією цих значень на проміжок часу Δτ;

• значення високоінформативних критеріїв, розрахованих на основі віброакустичного сигналу;

• температура кожної з полюсних обмоток із фіксацією на один оберт для роторних обмоток;

• струми у фазах статора;

• кутове прискорення ротора;

• значення механічної напруженості у ключових вузлах опорних конструкцій;

• інші додаткові параметри технічного стану машини, що можуть використовуватися для підвищення ефективності роботи системи діагностування.

Враховуючи результати проведеного аналізу, запропоновано структуру ШНМ для реалізації задачі діагностування, яку можна представити у вигляді, наведеному на рис. 2.

Як видно з рис. 2, передбачається побудова чотиришарової неоднорідної нестандартної ШНМ.

Кількість вхідних нейронів нульового шару (на рис. 2 вони зображені колами) відповідає кількості додаткових параметрів, що подаються на входи ШНМ. Вхідні нейрони виконують функцію прийняття числових даних та їх сортування.

Перший шар ШНМ (нейрони якого позначені квадратиками із цифрою 1) містить $N \cdot M + K$ нейронів. Кожен із перших $N \cdot M$ нейронів отримує значення високоінформативного критерію, максимально чутливого до інформативних чинників вібрації. Особливу групу нейронів зазначеного шару формують K останніх нейронів, на входи яких надходить температура відповідної полюсної обмотки. Додатково на кожен із нейронів першого шару надходить інтегральна інформація (інегральний критерій), сформована нульовим шаром нейронів на базі внутрішнього аналізу вимірювальної інформації від додаткових вимірювальних каналів (кутового прискорення, потужності навантаження тощо).

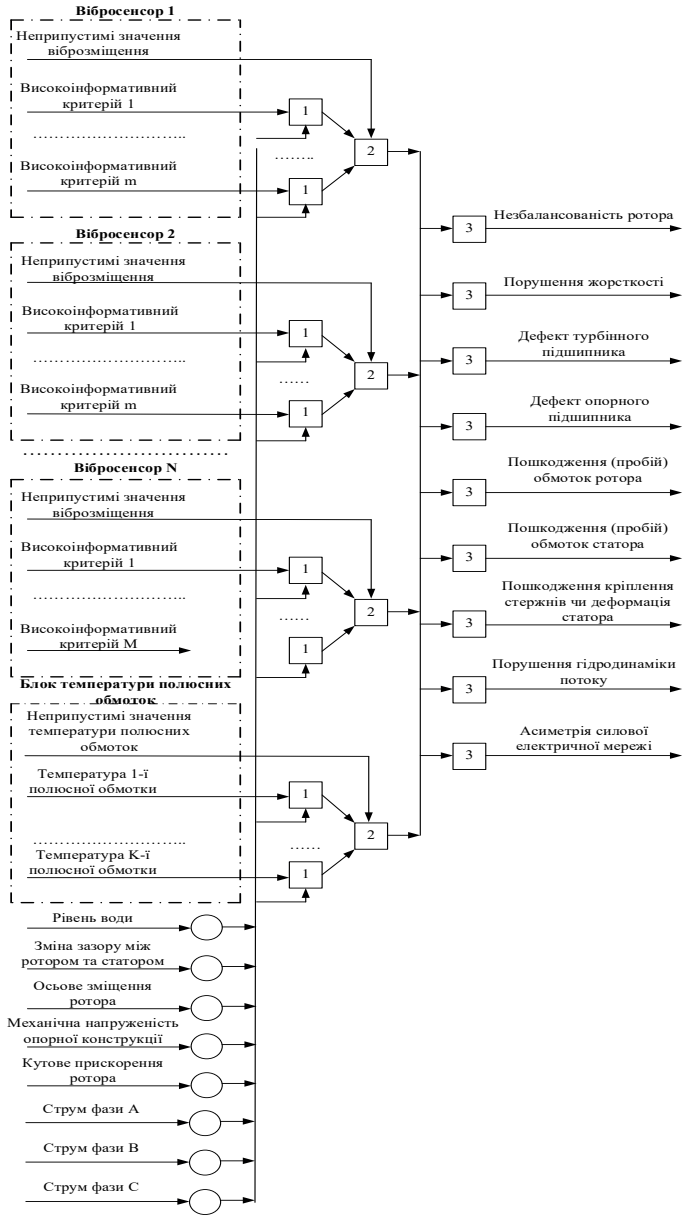

Рис. 2. Структура штучної нейроподібної мережі

Функції перетворення кожного із нейронів запропонованої штучної нейроподібної меежі формуються у результаті передексплуатаційного навчання на основі статистичної інформації про особливості роботи електричних машин дрослідджуваного класу. Нейрони першого шару призначені для коректування високоінформативних критеріїв та показників температури полюсних обмоток із урахуванням інформації від додаткових вимірювальних каналів. Такий підхід є виправданим, виходячи з того, що електрична машина є системою зі складними механічними зв'язками. Тож, при виникненні локальних збурень, обумовлених неінформативними факторами, результуюча дія яких має характеристики, подібні до впливів, обумовлених певними видами інформативних параметрів, в переважній більшості випадків суттєвого спотворення зазнаватиме вимірювальна інформація на виході лише окремо взятого сенсора чи групи сенсорів, локалізованих у певній області машини.

Другий шар нейронів ШНМ (позначений цифрою 2) містить $N+1$ нейрон, $N$ з яких отримує узагальнену критеріальну інформацію від кожного із сенсорів віброприскорення та здійснює їх інтегральну обробку. На цьому етапі реалізується попередня задача імовірнісного аналізу наявності інформативних параметрів та усувається надлишковість вимірювання, враховуючи неминучу залежність числових критеріїв, що подаються на вхід ШНМ, від понад одного інформативного параметру. Останній нейрон другого шару призначення для інтегральної обробки градієнта температур полюсних обмоток електричної машини. На нейрони цього шару, функції перетворення яких були сформовані на етапі передексплуатаційного навчання, додатково поступає інформація про перевищення рівня віброзміщення або недопустимого перегріву полюсних обмоток.

Спрацювання ШНМ відбувається лише у випадку, коли хоча б в одному із віброситналів міститься надмірне віброзміщення або має місце надмірний перегрів обмоток. В цьому випадку функція активації $N$ нейронів другого шару може бути представленою як:

$$\phi(a_i) = sign(a_i - a_0, \Delta\tau) \cdot \sum_{j=1}^{M} \psi(p_j), \qquad (1)$$

де $\Delta\tau$ — часова затримка на вимикання; $a_i$ — поточне значення віброзміщення, що поступає на відповідний нейрон; $a_0$ — порогове значення віброзміщення; $p_i$ — скоректований $j$-й інформативний критерій; $\psi(p_j)$ — функція впливу скоректованого $j$-го інформатив-ного критерію; $\text{sing}(a_i - a_0, \Delta\tau)$ — релейна функція з затримкою на вимикання.

Функція перетворення *N+1*-го нейрону формується аналогічно, з тією відмінністю, що спрацювання його релейної функції відбувається при перевищенні температури хоба б однієї полюсної обмотки деякого усталеного значення.

Третій шар ШНМ (позначений цифрою 3) містить 9 нейронів, кожен з яких відповідає одному з чинників, які є досліджуваними причинами виникнення вібрацій чи перегріву обмоток. У цьому шарі відбувається усереднення проміжних висновків, зроблених нейронами другого шару, на основі їх інтегрального аналізу.

Слід зазначити, що логічний висновок такої системи, сформований нейронами третього шару, носитиме ймовірнісний характер. Перевищення певного встановленого значення ймовірності для причин надмірного віброзміщення або надмірного перегріву обмоток, що відносяться до дефектів, формує логічний висновок про їх наявність.

**Розробка високоінформативних діагностичних ознак наявності найбільш поширених дефектів обертових електричних машин.** Серед існуючих достатньо описаних та вивчених підходів, придатних для аналізу часової реалізації вібросигналу, що може бути отриманий під час роботи реальної електричної машини, можна виділити перетворення Фур'є та дискретне вейвлет-перетворення (ДВП). Проте варто відзначити, що перетворення Фур'є математично є більш складним за ДВП, а отже при забезпеченні однакової швидкодії потребуватиме більших апаратних затрат, а також не передбачає можливості дослідження локалізованих збурень взагалі [20]. Зазначені особливості роблять його малоефективним для використання у сучасних системах аналізу вібраційних сигналів електричних машин. Тоді як ДВП, будучи у першу чергу адаптованим до виявлення саме локалізованих пікових збурень, не передбачає наявності готових інструментів, призначених для сепарації періодичної та аперіодичної складових. Виходячи зі сказаного, можна дійти висновку про необхідність розробки нових підходів до виявлення періодичних складових вібросигналу саме на основі ДВП, які можуть викликатися типовими дефектами обертових електричних машин.

За результатами статистичного дослідження причин виходу з ладу асинхронних електродвигунів (які є найбільш поширеним типом обертових електричних машин) встановлено, що у 79 % випадків при відмові останніх мав місце один із трьох типів аномального відхилення технологічних параметрів, а саме: механічний дебаланс ротора, пошкодження підшипників чи асиметрія струму у статорному колі [3]. При цьому результуючий вібросигнал на ранніх етапах розвитку дефектів, як правило, характеризується накладанням значної кількості рівноцінних збурюючих чинників, частина з яких має аперіодичний характер [14; 21].

Особливістю вібросигналів, обумовлених зазначеними типами дефектів, є їх квазіперіодичний характер [14; 21]. Це призводить до того, що при проведенні стандартного вейвлетаналізу наявність наведених дефектів на ранніх етапах їх розвитку не призводить до появи локального зростання амплітуди окремих коефіцієнтів ДВП, а отже буде малопомітною при аналізі результатів перетворення.

Одна з головних ідей вейвлетного представлення сигналів на різних рівнях декомпозиції (розкладання) сигналу полягає в розділенні функцій наближення до сигналу на дві групи: що апроксимує — грубу, з достатньо повільною часовою динамікою змін, і що деталізує — з локальною і швидкою динамікою змін на тлі плавної динаміки, з подальшим їх дробленням і деталізацією на інших рівнях декомпозиції сигналів. Це можливо як в часовій, так і в частотній областях представлення сигналів вейвлетами. В цьому випадку базисна вейвлет-функція дозволяє сконцентрувати увагу на тих або інших локальних особливостях аналізованих процесів. Причому за своєю суттю деталізація неперервного вейвлет-перетворення (НВП) є нічим іншим, як визначенням функції взаємокореляції між материнською вейвлет-функцією та досліджуваним сигналом, що випливає з математичної моделі такого перетворення [20; 22]:

$$W(a,\tau) = \int\limits_{-\infty}^{+\infty} f(t) \cdot \psi_{a,\tau}^{*}(t)dt \ , \qquad (2)$$

де $W(a,\tau)$ — функція деталізації (результат вейвлет-перетворення); $a$ — параметр масштабу; $\tau$ — параметр зсуву; $f(t)$ — функція, що аналізується; $\psi_{a,\tau}^{*}(t)$ — комплексно спряжена вейвлет-функція.

Враховуючи те, що обчислення при вейвлетперетворенні здійснюються шляхом зміни масштабу «вікна» аналізу, зсуву його в часі, множення на сигнал та інтегрування по всій осі часу [22; 23], то геометричний зміст такого перетворення можна представити як пошук ділянок аналізованої функції у часовій та частотній областях, що за своєю формою будуть корельованими з материнською вейвлет-функцією.

Аналогічний фізичний зміст зберігається й у ДВП, при здійсненні якого коефіцієнти деталізації можуть бути розраховані таким чином [20; 23]:

$$d_k^j = \sum_{n \in Z} g_{n-2k} \cdot c_n^{j+1},\tag{3}$$

де $d_k^j$ — $k$-й коефіцієнт деталізації $j$-ї частотної смуги; $g$ — коефіцієнт материнської вейвлет-функції; $c^{j+1}$ — апроксимуючий коефіцієнт попередньої частотної смуги, розраховуються так:

$$c_k^j = \sum_{n \in Z} h_{n-2k} \cdot c_n^{j+1},\tag{4}$$

де $h$ — коефіцієнт масштабуючої функції.

Для старшої частотної смуги в якості апроксимуючих коефіцієнтів використовується часова реалізації досліджуваного сигналу.

В такому випадку задача реєстрації зазначених дефектів може бути розбита на дві підзадачі: підбір материнського вейвлету, який був би максимально наближений до обумовленої дефектом складової вібросигналу, та розробка критерію, який би давав змогу кількісно оцінити вплив обумовленого наявністю дефекту коливання на коефіцієнти вейвлет-перетворення окремих частотних смуг та характеризувався б високою селективністю по відношенню до нього.

Аналіз наведених у літературі описів вібраційних сигналів, обумовлених неврівноваженістю ротора, показує, що зазначений дефект призводить до появи коливань, які містять гармонічну складову, локалізовану на роторній частоті обертання, а також її другій та третій гармоніці. Причому амплітуда коливань з переходом на другу та третю гармонічні складові різко зменшується [14; 21]. Зазначений факт обумовлює доцільність аналізу при його пошуку частотного діапазону, що включає у себе частоту обертання ротора та, меншою мірою, частотні діапазони, які відповідають подвоєній та потроєній роторній частоті. А підбір материнського вейвлету доцільно здійснювати виходячи з ознак, властивих одиничному гармонічному коливанню. Як показує дослідження літературних джерел, найбільш спорідненим з одиничним гармонічним коливанням серед типових материнських вейвлет-функцій можна вважати вейвлет Хаара та вейвлет Добеши 4-го порядку. Проте варто відзначити, що кожен із них має свої суттєві структурі відмінності [23].

Також, враховуючи періодичність вібраційного сигналу, обумовленого наявністю дебалансу ротора, і ту обставину, що при дослідженні кожне із гармонічних коливань представляється як окремий сплеск, варто очікувати періодичну зміну значень вейвлет-коефіцієнтів у часовій області в межах смуг частот, що включають у себе роторну частоту, а також її другу та третю гармоніки. Причому амплітуди таких періодичних змін будуть напряму пов'язані зі ступенем розвитку дефекту. Тож виконується нерівність [24]:

$$t_{cn} \gg T_p, \tag{5}$$

де $t_{cn}$ — тривалість часової реалізації досліджуваного сигналу; $T_p$ — період обертання ротора електричної машини.

Доцільним є застосування інтегрального підходу до аналізу коефіцієнтів вейвлет-перетворення. Відтак, у якості шуканого числового критерію наявності зазначеного дефекту може бути використано усереднене квадратичне значення вейвлет-коефіцієнтів досліджуваних частотних смуг у межах часового інтервалу, тривалість якого значно більша за період обертання ротора. Такий підхід дозволить врахувати наявність як додатних, так і від'ємних максимумів вейвлет-коефіцієнтів у межах досліджуваного часового інтервалу, а також характеризуватиметься пониженою чутливістю до неінформативних збурень, обумовлених аперіодичними збурюючими силами, що можуть виникати в процесі експлуатації електричної машини. Виходячи зі сказаного, математично числовий критерій оцінки впливу дебалансу ротора на коефіцієнти вейвлет-перетворення зазначених частотних смуг може бути представлений таким чином [24]:

$$k_{де6} = \frac{1}{n} \sum_{i=1}^{n} d_i^2 \ \textit{при умові } t_{cn} \gg T_p, \tag{6}$$

де $n$ — кількість коефіцієнтів вейвлет-перетворення досліджуваної частотної смуги; $d_i$ — $i$-й коефіцієнт вейвлет-перетворення досліджуваної частотної смуги.

З метою підтвердження неведених вище теоретичних міркувань було проведено експериментальне дослідження з використанням електричної машини в режимі холостого ходу з моментом інерції ротора — 0,002 кг·м², частотою обертання в режимі холостого ходу — 720 об/хв (12 Гц) та додатково внесеним дебалансом 0,002 кг·м. П'єзоакселерометр було закріплено на корпусі електричної машини таким чином, що вимірювальна вісь була напрямлена строго перпендикулярно ротору електричної машини. Частота дискретизації сигналу становила 232 Гц, довжина часової реалізації досліджуваного сигналу — $2^{14}$ значень.

При перетворенні отриманого сигналу віброприскорення за допомогою вейвлета Хаара та подальшого розрахунку запропонованого числового критерію для кожної із частотних смуг з та без використання дебалансу було отримано результати, наведені на рис. 3 та 4.

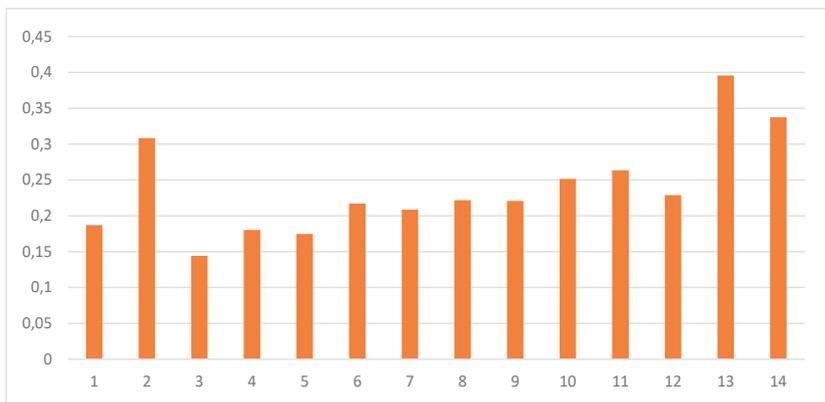

Рис. 3. Залежність усередненого квадратичного вейвлет-коефіцієнтів Хаара для кожної із частотних смуг вібросигналу без використання дебалансу

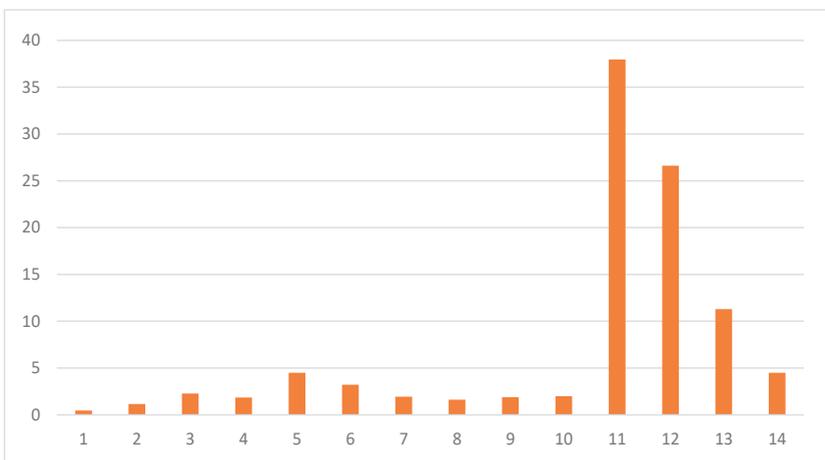

Рис. 4. Залежність усередненого квадратичного вейвлет-коефіцієнтів Хаара для кожної із частотних смуг вібросигналу при наявності дебалансу

Також було виконано аналогічне перетворення за допомогою вейвлета Добеши 4-го порядку. Результати розрахунку наведені на рис. 5 та 6.

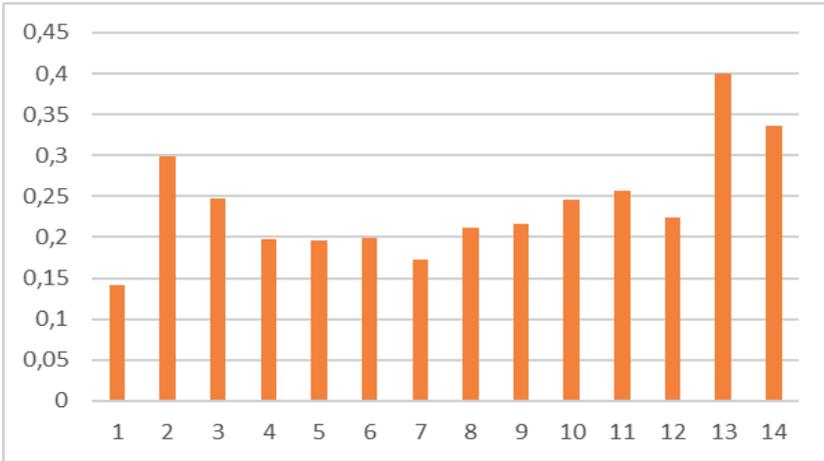

Рис. 5. Залежність усередненого квадратичного вейвлет-коефіцієнтів Добеши 4-го порядку для кожної із частотних смуг вібросигналу без використання дебалансу

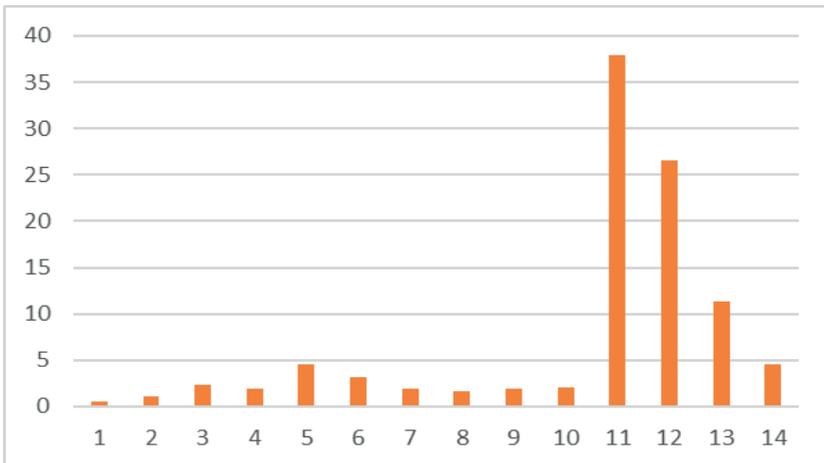

Рис. 6. Залежність усередненого квадратичного вейвлет-коефіцієнтів Добеши 4-го порядку для кожної із частотних смуг вібросигналу при наявності дебалансу

Як випливає з аналізу залежностей, наведених на рис. 3—6, найбільш інформативними для виявлення дебалансу ротора, як і очікувалося, є смуги частот, що відповідають частоті обертання ротора електричної машини та її другої та третьої гармоніки (відповідно 11, 12 та 13 частотні смуги). Порівняння ж результатів, отриманих при розкладанні сигналу на основі вейвлету Хаара та вейвлету Добеши 4-го порядку, показали, що обидва вейвлета характеризуються приблизно однаковою достатньо високою чутливістю до наявності досліджуваного дефекту. Тож, враховуючи той факт, що перетворення на основі материнської вейвлет-функції Хаара є математично більш простим (потребує меншої кількості математичних операцій) [20; 22; 23], можна зробити висновок, що використання саме останнього є більш ефективним для виявлення зазначеного дефекту.

Аналізуючи наведені у літературі описи вібраційних сигналів, обумовлених пошкодженнями підшипників обертової електричної машини, ми встановили, що зазначена група дефектів викликає доволі складний за формою квазіперіодичний вібраційний відгук, частота якого відповідає роторній частоті електричної машини [14; 21]. Враховуючи ту обставину, що вібраційний відгук, обумовлений пошкодженням підшипників, в межах одного періоду характеризуватиметься декількома піками, очевидно, що для виявлення такого пошкодження доцільним буде застосування материнських вейвлет-функцій старших порядків. Це пояснюється тим, що при зростанні порядку материнської вейвлетфункції типово зростає число її осциляцій. Тож, для вейвлет-функції $N$-го порядку буде справедливим вираз [20; 22; 23]:

$$\int_{-\infty}^{+\infty} t^k \psi(t)dt = 0, \quad k = 0,1,...,N-1. \tag{7}$$

Оскільки розрахунок коефіцієнтів переважної більшість дискретних вейвлет-функцій є доволі трудомістким [20; 22; 23], а форма вібраційного відгуку при пошкодженнях підшипників характеризується доволі складною структурою, яка однозначно не асоціюється з жодним із відомих вейвлетів, у якості базових вейвлет-функцій пропонується використання вейвлетів Добеши. Головною перевагою зазначеного сімейства вейвлет-функцій є можливість відносно простого аналітичного розрахунку їх коефіцієнтів для функції довільного порядку [25]. В такому випадку математичний числовий критерій оцінки впливу наявності дефектів підшипників на коефіцієнти вейвлетперетворення частотних смуг буде аналогічним критерію наяв-ності дебалансу ротора (6) за умови використання зазначених материнських вейвлет-функцій.

З метою підтвердження неведених вище теоретичних міркувань було проведено експериментальне дослідження з використанням асинхронної електричної машини АИМ90La6У2.5, номінальною потужністю 0,75 кВт та синхронною швидкістю обертання 1000 об/хв (16,67 Гц) при її експлуатації в режимі холостого ходу. Дефект підшипника було імітовано шляхом застосування підшипників при відсутності масляної плівки. Інші параметри експерименту були повністю аналогічними дослідженню впливу дебалансу ротора, описаному вище.

При перетворенні отриманого сигналу віброприскорення за допомогою вейвлета Добеши 4-го порядку та подальшого розрахунку середньоквадратичного вейвлет-коефіцієнтів для кожної із частотних смуг при роботі електричної машини у режимі холостого ходу було отримано результати, наведені на рис. 7 та 8.

Аналогічне перетворення отриманого вібросигналу було виконано за допомогою вейвлета Добеши 6-го порядку. Результати розрахунку наведені на рис. 9 та 10.

Як випливає з аналізу залежностей, наведених на рис. 7—10, найбільш інформативними для виявлення дефекту підшипників, як

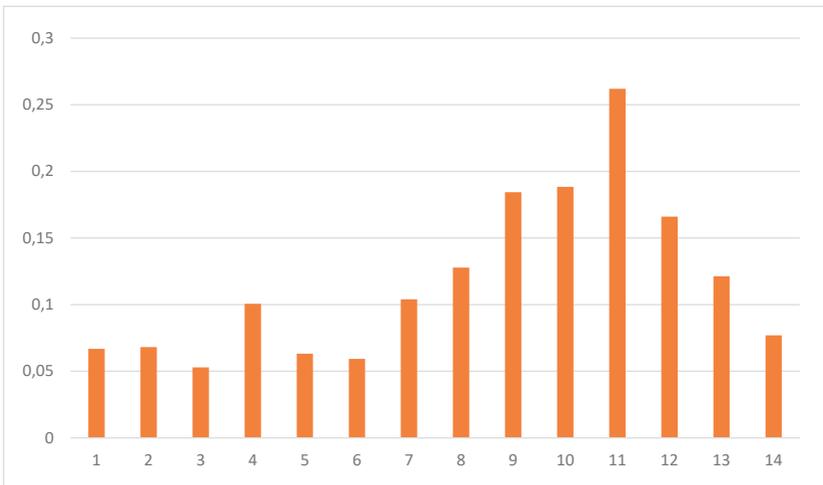

Рис. 7. Залежність усередненого квадратичного вейвлет-коефіцієнтів Добеши 4-го порядку для кожної із частотних смуг вібросигналу при наявності мастила у підшипнику

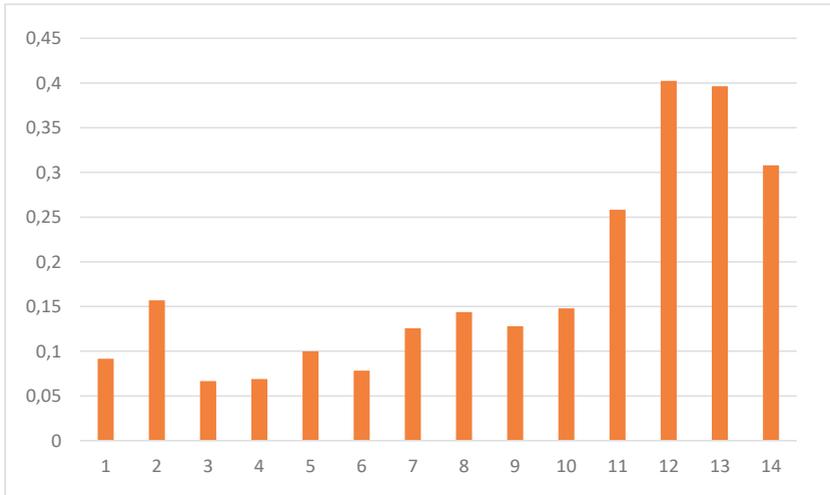

Рис. 8. Залежність усередненого квадратичного вейвлет-коефіцієнтів Добеши 4-го порядку для кожної із частотних смуг вібросигналу при відсутності мастила у підшипнику

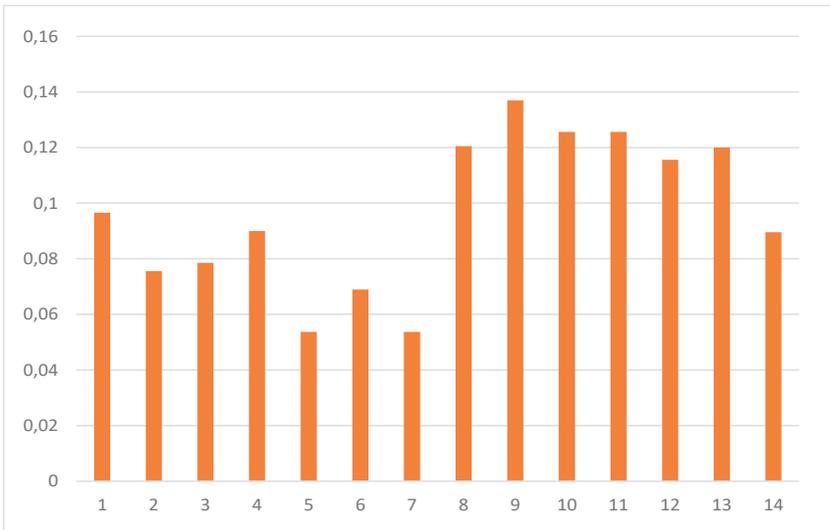

Рис. 9. Залежність усередненого квадратичного вейвлет-коефіцієнтів Добеши 6-го порядку для кожної із частотних смуг вібросигналу при наявності мастила у підшипнику

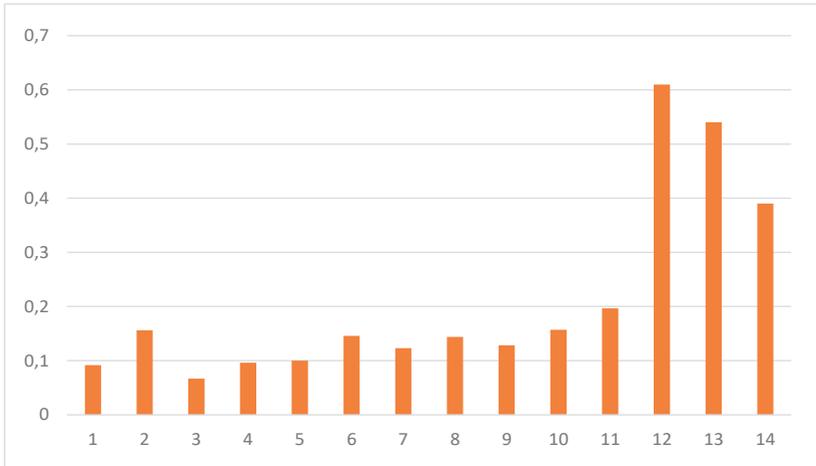

Рис. 10. Залежність усередненого квадратичного вейвлет-коефіцієнтів Добеши 6-го порядку для кожної із частотних смуг вібросигналу при відсутності мастила у підшипнику і очікувалося, є частотна смуга, що відповідає частоті обертання ротора (12 частотна смуга), та її друга (13 смуга частот) і третя (14 смуга частот) гармоніки. Порівняння ж результатів, отриманих при розкладанні сигналу на основі вейвлету Добеши 4-го, 6-го порядку, показали справедливість зроблених раніше припущень про доцільність використання для виявлення зазначеного дефекту вейвлетів старших порядків.

У свою чергу для вібросигналів, обумовлених асиметрією струму у статорному колі, є характерною наявність гармонічної складової вібросигналу, локалізованої на частоті напруги живлення електричної мережі [14; 21]. Зазначений факт обґрунтовує доцільність аналізу частотного діапазону, що включає у себе частоту напруги живлення, та використання вейвлетів Хаара та Добеши 4-го порядку виходячи з міркувань, наведених вище. В такому випадку числовий критерій оцінки впливу електромагнітної асиметрії статорного кола на коефіцієнти вейвлетперетворення зазначених частотних смуг може бути представлений так [26]:

$$k_{деб} = \frac{1}{n}\sum_{i=1}^{n} d_i^2 \ \text{при умові } t_{сп} >> T_{жс},$$ (8)

де $T_{жс}$ — період напруги живлення статорного кола.

З метою підтвердження неведених вище теоретичних міркувань було проведено експериментальне дослідження з використанням асинхронної електричної машини, описаної у попередньому досліді. Інші параметри експерименту були повністю аналогічними дослідженню впливу дебалансу ротора, описаному вище.

При перетворенні отриманого сигналу віброприскорення за допомогою вейвлета Хаара та подальшого розрахунку усередненого квадратичного вейвлет-коефіцієнтів для кожної із частотних смуг при роботі електричної машини у штатному режимі та обриві фази А було отримано результати, наведені на рис. 11 та 12.

Також було виконано аналогічне перетворення отриманого вібросигналу за допомогою вейвлета Добеши 4-го порядку. Результати розрахунку наведені на рис. 13 та 14.

Як випливає з аналізу залежностей, наведених на рис. 11—14, найбільш інформативною для виявлення асиметрії живлення, як і очікувалося, є частотна смуга, що відповідає частоті напруги живлення електромережі 50 Гц (13 частотна смуга). Порівняння ж результатів, отриманих при розкладанні сигналу на основі вейвлету Хаара та вейвлету Добеши 4-го порядку показало, що обидва вейвлета характеризуються приблизно однаковою достатньо високою чутливістю до наявності досліджуваного дефекту. Тож, виходячи з наведених вище міркувань, можна зробити висновок про доцільність використання саме вейвлета Хаара для вирішення поставленої задачі.

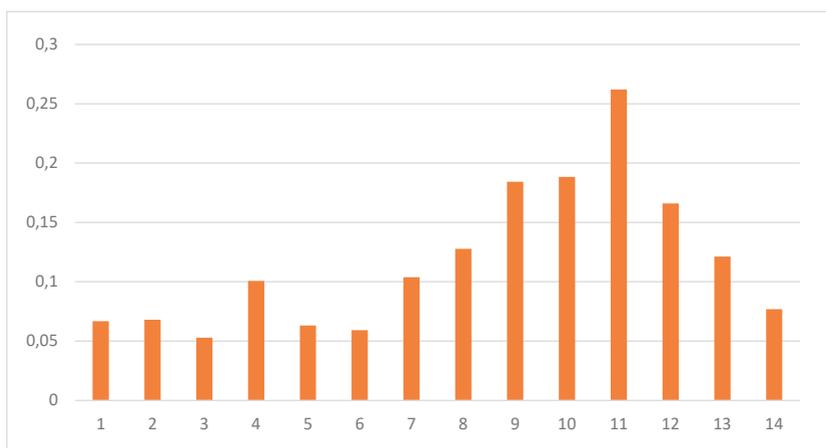

Рис. 11. Залежність усередненого квадратичного вейвлет-коефіцієнтів Хаара для кожної із частотних смуг вібросигналу при роботі електродвигуна у штатному режимі

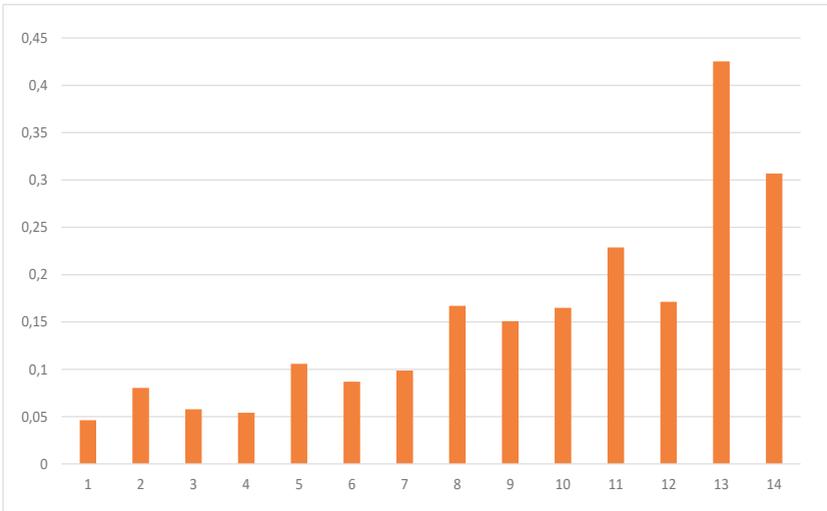

Рис. 12. Залежність усередненого квадратичного вейвлет-коефіцієнтів Хаара для кожної із частотних смуг вібросигналу при обриві фази А

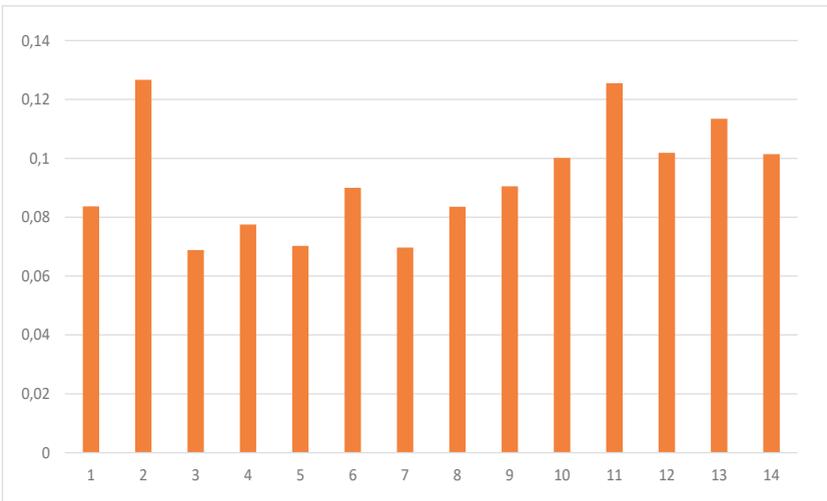

Рис. 13. Залежність усередненого квадратичного вейвлет-коефіцієнтів Добеши 4-го порядку для кожної із частотних смуг вібросигналу при роботі електродвигуна у штатному режимі

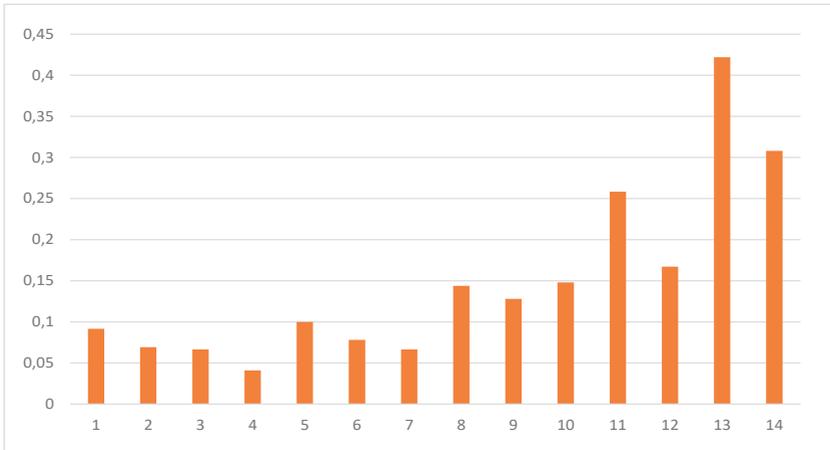

Рис. 14. Залежність усередненого квадратичного вейвлет-коефіцієнтів Добеши 4-го порядку для кожної із частотних смуг вібросигналу при обриві фази А

**Висновки.** 1. Запропоновано використання та розроблено структуру ШНМ в якості ключового елемента формування логічного висновку в системі діагностування гідроагрегатів, як типового представника системи виключної складності. Показано, що запропоноване рішення може розглядатись як окремо взятий унікальний випадок, що має значну практичну цінність, оскільки може бути адаптованим для вирішення задач широкого класу.

2. Набув подальшого розвитку алгоритм аналізу систем типу «чорна скринька» з розподіленими параметрами шляхом незалежної інтегральної обробки інформації на локалізованих ділянках, з подальшим її загальним аналізом у нейронному шарі верхнього рівня. Показано, що зазначений підхід дає змогу вилучити вплив неінформативних чинників, які за структурою своєї дії є подібними до інформативного впливу, проте носять локальний характер.

3. Визначено та теоретично обґрунтовано тривалість часових реалізацій вібросигналу, що доцільно використовувати при отриманні коефіцієнтів взаємокореляції вібросигналів у досліджуваних вузлах. Встановлено, що тривалість таких реалізацій повинна бути кратною частоті періоду обертання ротора електричної машини.

4. Запропоновано інтегральні високоінформативні числові критерій оцінки впливу неврівноваженості ротора, семетричного зростання напруженості основного електромагнітного поля, асиметрії струмів у статорному колі та дефектів підшипників на коефіцієнти вейвлет-перетворення у вигляді середньоквадратичного значення вейвлет-коефіцієнтів інформативних смуг частот при дослідженні часового інтервалу, що значно перевищує період обертання ротора електричної машини. Показано, що зазначені критерії мають понижену чутливість до впливу неінформативних одиничних збурень, які можуть виникати в процесі роботи електричної машини.

### *СПИСОК ВИКОРИСТАНОЇ ЛІТЕРАТУРИ*

# Розділ III

## НОВІ ІНФОРМАЦІЙНІ ТЕХНОЛОГІЇ В ОСВІТІ

## АВТОМАТИЗОВАНА ІНФОРМАЦІЙНА СИСТЕМА ОБЛІКУ ПІДВИЩЕННЯ КВАЛІФІКАЦІЇ ВИКЛАДАЧІВ

*Іванова Л. В., Котлик Д. О.*

*У статті розкриваються сучасний стан інформатизації закладів освіти, інформаційне забезпечення управлінської діяльності керівника закладу освіти. Наводиться порівняльний аналіз комерційних комплексних автоматизованих систем управління навчальним процесом із автоматизованими системами управління власних розробок закладів вищої освіти, а також їхні функціональні можливості в управлінні навчальним процесом та їхні особливості.*

*Представлені матеріали, що стосуються проектування автоматизованої інформаційної системи обліку підвищення кваліфікації викладачів закладу освіти. Встановлено, що реалізація цієї системи — це об'єктивна необхідність, яка обумовлена реформуванням діяльності освітніх закладів та доцільністю автоматизації оцінювання якості підвищення кваліфікації викладачів навчальних закладів для швидкого прийняття управлінських рішень щодо їх атестації та педагогічної діяльності.*

*У статті представлена автоматизована інформаційна система обліку підвищення кваліфікації викладачів закладу освіти, яка дозволяє зберігати, обробляти та контролювати результати підвищення кваліфікації та атестації викладачів. Визначено, що застосування комплексного підходу до зберігання результатів підвищення кваліфікації та атестації викладачів забезпечує: оперативний моніторинг інформації про підвищення кваліфікації викладачів для проходження чергової атестації; зручне заповнення та збереження інформації; можливість контролю підвищення кваліфікації викладачів з боку працівників методичного кабінету.*

*Визначено схему інформаційних потоків, спосіб зберігання і виведення даних, розглянуто сучасні технології, які дають можливість аналізувати і створювати візуальне представлення даних для здійснення процесів обліку, моніторингу та контролю. Спроектовані база даних та інтерфейс користувача інформаційної системи обліку підвищення кваліфікації та атестації викладачів закладу освіти.*

*Створена автоматизована інформаційна система обліку підвищення кваліфікації викладачів закладу освіти проходить апробацію в робочому про-*

*цесі методичного кабінету коледжу для обліку результатів підвищення кваліфікації та проходження атестації викладачів Відокремленого структурного підрозділу «Одеський технічний фаховий коледж Одеської національної академії харчових технологій»».*

*The article reveals the current state of informatization of educational institutions, information support of management activities of the head of the educational institution. The comparative analysis of commercial complex electronic systems of management of educational process with electronic systems of management of own developments of higher educational institutions, and also their functional possibilities in management of educational process and their characteristic is resulted.*

*Materials related to the design of an automated information system for accounting for professional development of teachers of educational institutions are presented. It is established that the implementation of this system is an objective necessity, which is due to the reform of educational institutions and the feasibility of automating the quality assessment of teachers by teachers to quickly make management decisions on their certification and teaching.*

*The article presents an automated information system for accounting for in-service training of teachers of educational institutions, which allows you to store, process and control the results of in-service training and certification of teachers. It is determined that the application of a comprehensive approach to the storage of results of in-service training and certification of teachers provides — operational monitoring of information about in-service training of teachers to pass the next certification; convenient filling and saving of information; possibility of control of advanced training of teachers by employees of a methodical office.*

*The scheme of information flows, the method of data storage and output are determined, modern technologies are analyzed, which give the opportunity to analyze and create a visual representation of data for the implementation of accounting, monitoring and control processes. The database and the user interface of the information system of the account of advanced training and certification of teachers of educational institution are designed.*

*The created automated information system of the account of advanced training of teachers of educational institution passes approbation in working process of a methodical office of college for the account of results of advanced training and passing of certification of teachers of Separated structural subdivision «Odessa Technical Applied College Odessa National Academy of Food Technologies».*

Освіта має винятково важливе значення для соціально-економічного розвитку та культурного збагачення суспільства, наділяючи людей відповідними знаннями та вміннями для покращення навичок, здатності до продуктивної праці в умовах подальшого глобального розвитку, домінантою якої є інтелектуальна економіка. Для України актуальним є завдання ефективного забезпечення та організації на-вчання учнів/здобувачів освіти у системі освіти, що трансформується, утворюючи нову інформаційну ментальність всіх зацікавлених сторін. Розв'язання цього завдання вимагає постійного моніторингу та оцінювання стану системи освіти, в основі яких лежать збір, опрацювання інформації, аналіз освітніх даних, необхідних для забезпечення ухвалення обґрунтованих управлінських рішень [1].

Одним із провідних напрямів розвитку сучасної освіти є її інформатизація. Реалізація цього напряму дає можливість зробити освіту більш ефективною, гнучкою, сучасною, такою, що відповідає міжнародним стандартам.

У концепції Національної програми інформатизації зазначено, що сутність інформатизації полягає в сукупності взаємопов'язаних організаційних, правових, політичних, соціально-економічних, науково-технічних, виробничих процесів, що спрямовані на створення умов для задоволення інформаційних потреб, реалізації прав громадян і суспільства на основі створення, розвитку, використання інформаційних систем, мереж, ресурсів та інформаційних технологій, створених на основі застосування сучасної обчислювальної та комунікаційної техніки [2; 3].

Однією з характерних особливостей модернізації українського суспільства є його глобальна інформатизація, яка зумовила масштабне запровадження інформаційно-комунікаційних технологій (ІКТ) в освітній процес, що дає змогу педагогу вирішувати методичні завдання на якісно вищому рівні. Вихідні концептуальні положення означеної проблеми викладені у Законі України «Про національну програму інформатизації», Концепції Національної програми інформатизації, Указі Президента України від 20.10.2005 р. № 1497 «Про першочергові завдання щодо впровадження новітніх інформаційних технологій», Державній програмі «Інформаційні та комунікаційні технологій в освіті і науці на 2006—2010 роки» та інших нормативних документах.

Інформатизація освіти залежить від об'єктивних умов та сучасних тенденцій розвитку інформаційного суспільства, до яких варто віднести такі:

— забезпечення мобільності інформаційно-комунікаційної діяльності користувачів в інформаційному просторі (Mobility), подальший розвиток мобільно орієнтованих засобів та ІКТ доступу до електронних даних;

— розвиток технології хмарних обчислень та віртуалізації, корпоративних, загальнодоступних і гібридних ІКТ-інфраструктур, а також

запровадження технології хмарних обчислень (Cloud Computing and Virtualization, Private, Public and Hybrid Clouds, ICT-infrastructures, Fog Computing);

— накопичення та опрацювання значних обсягів цифрових даних, формування та використання електронних інформаційних баз і систем (Big Data, Data Mining, Data Bases), зокрема електронних бібліотек (Electronic Libraries, Repositories) та наукометричних баз даних (Scientometric Data Bases);

— розвиток користувальних характеристик Інтернету людей (Internet of People — IoP), розгортання топології широкосмугових високошвидкісних каналів електронних комунікацій (Broadband Communication Channels), систем формування ІКТ-просторів бездротового доступу користувачів до електронних даних (Cordless Access to Digital Data, WiFi, Bluetooth, Cellular Networks);

— формування Інтернету речей (Internet of Things — IoT), розвиток його програмно-апаратних засобів, зокрема мікропроцесорних, та інтеграційних платформ, для забезпечення налаштування, управління та моніторингу електронних пристроїв за допомогою сучасних телекомунікаційних технологій;

— розвиток робототехніки (Robotics), робототехнічних систем, зокрема 3D-принтерів і 3D-сканерів;

— розвиток систем захисту даних в інформаційних системах та протидія кіберзлочинності (Data Security and Counteraction of Cybercriminality);

— розвиток індустрії виробництва програмних засобів (Software Development Industry), зокрема видання електронних освітніх ресурсів;

— забезпечення сумісності ІКТ-засобів та ІКТ-додатків, побудованих на різних програмно-апаратних платформах (Compatibility);

— розвиток мереж постачальників ІКТ-послуг (ринку ІКТ-аутсорсерів), передусім хмарних сервісів (Cloud Services), та мережі Центрів опрацювання даних (Computing Center Network) [24].

Невідкладного вирішення потребують проблеми розвитку та впровадження інформаційно-комунікаційних технологій у вітчизняній освіті, ключові з яких виокремлені в Національній доповіді 2016 р. «Про стан і перспективи розвитку освіти в Україні» [23].

Першою є проблема формування і широкого впровадження єдиного освітнього інформаційного простору України та забезпечення належного наукового супроводу цих процесів.

Другою є проблема розгортання та удосконалення необхідних елементів інфраструктури регіональних інформаційних і телекомуні-каційних мереж, взаємопов'язаних як між собою, так і з глобальною мережею Інтернет, що дозволить подолати «цифрову нерівність» у різних регіонах України, зокрема в сільській місцевості.

Третьою проблемою є низький рівень інформаційно-комунікаційно-технологічних компетентностей (ІКТ-компетентностей) та інформаційних компетентностей населення, застосування застарілих підходів у навчанні та низька мотивація суб'єктів навчального процесу щодо використання прогресивних ІКТ. Варто зазначити, що масштабний характер застосування засобів ІКТ в глобальній системі освіти зумовив появу нових методів і форм навчання (електронне навчання, мобільне навчання, застосування в освіті хмарних технологій, масових відкритих освітніх курсів тощо), що повільно запроваджуються в сучасній національній системі освіти України.

Четверта проблема — фактична несформованість цілісної національної політики застосування інформаційно-комунікаційних технологій в освіті, недосконала нормативно-правова база, що не забезпечує побудову інформаційного суспільства та, як наслідок, гальмує інформатизацію освіти в Україні. Завдання інформатизації освіти не знайшли належного системного відображення в чинних законах України з питань освіти та сучасних проектах. Суттєвим недоліком нинішньої освітньої політики є недооцінка важливості стимулювання ініціатив із запровадження інформаційно-комунікаційних технологій, що ініційовані закладами освіти, науковими установами, освітянами, громадськими організаціями та приватним бізнесом.

Останніми роками інформація стає одним з найважливіших виробничих факторів і одним з головних важелів управління будь-якої організації, в тому числі й освітнього закладу.

Інформатизація освіти — це процес зміни її змісту, методів і організаційних форм, спрямований на досягнення нової якості освіти на основі застосування інформаційних технологій. Вона повинна допомогти розв'язанню двох основних завдань школи: освіта — для всіх, і нова якість освіти — кожному.

Інформатизація управління освітньою установою пов'язана з прийняттям більш обгрунтованих управлінських рішень на основі автоматизованої обробки соціально-економічної, психолого-педагогічної та іншої інформації. Сьогодні інформатизація управління освітою розглядається на декількох рівнях, а саме:

— окремому — інформатизація охоплює управління окремими навчальними закладами;

— загальному — загалом охоплює кілька навчальних закладів одного району чи регіону, передбачається частковий інформаційний обмін між навчальними закладами та органами управління освітою;

— системному — поетапно охоплює всі освітні установи даної території з організацією повного інформаційного обміну на підставі єдиних інформаційних стандартів, що веде до формування єдиного інформаційного простору освіти.

Ефективність інформатизації закладу освіти значною мірою залежить від наукового обгрунтування цього процесу. Якщо узагальнити різні погляди науковців на інформатизацію закладу освіти, то можна зазначити, що більшість з них приділяють увагу такій функції, як підтримка управлінських рішень. Результати аналізу наукових праць та власного досвіду підтверджують, що інформаційні технології дають можливість підвищити ефективність усіх складових процесу розробки та реалізації управлінського рішення: отримання необхідної інформації, розробка управлінського рішення, доведення управлінського рішення до виконавців, контроль за виконанням управлінського рішення.

Однією з особливостей сучасної соціально-освітньої ситуації є самостійність освітніх установ. З одного боку, це активізує творчі сили педагогічних колективів, сприяє розвитку інноваційних процесів в освітніх установах. З іншого боку, процес управління освітніми установами значно ускладнився і вимагає його якісного перетворення. Якісне перетворення процесу управління освітнім закладом у свою чергу вимагає якісного зростання професійних фахівців, які здійснюють цей процес.

Численні можливості використання інформаційних технологій розглядали О. Андріянова, З. Пожидаєва, Н. Сомилкіна, Л. Забродська, Л. Жиліна, Е. Палат, М. Бухаркіна, М. Моїсеєва, А. Петров [4]. Л. Даниленко вважає, що застосування комп'ютерів в управлінській діяльності дає можливість забезпечити своєчасне надання оперативної інформації працівнику, який приймає рішення, з урахуванням її характеру; своєчасне надання аналітичної інформації; надання оптимального обсягу інформації; надання рекомендацій з вибору рішень та скорочення тривалості процесу вироблення рішення [5]. Погоджуючись з напрацюваннями науковців, зазначимо, що в швидкоплинних умовах актуальним стає прискорення використання інформаційних потоків у навчальному закладі. У наукових працях це питання висвітлюється фрагментарно, а питання комплексного застосування інформаційних потоків в навчальному закладі не висвітлено взагалі.

Результати аналізу наукових праць та власного досвіду підтверджують, що інформаційні технології дають можливість підвищити ефективність усіх складових процесу розробки та реалізації управлінського рішення: отримання необхідної інформації, розробка управлінського рішення, доведення управлінського рішення до виконавців, контроль за виконанням управлінського рішення.

За результатами презентації першого етапу проекту «Дебюрократизація управління освітою», реалізованого Міністерство освіти і науки України за підтримки ГО DOCCU і Швейцарсько-Українського проекту DECIDE у співпраці експертів Офісу ефективного регулювання BRDO, в якому проводилося дослідження системи управління загальною середньою освітою в Україні, розроблено проекти змін до 11 чинних НПА та створення двох нових НПА, які забезпечать цифровізацію шкільного діловодства, що потребує суттєвої модернізації.

Головною задачею проекту було надати експертну підтримку для розуміння того, що потрібно зробити, щоб спростити процедури звітування та діловодства, а також налагодити процеси обміну даними в галузі загально-середньої освіти.

В ході проекту було сформовано фокус-групи, учасниками яких стали освітяни, а також проаналізовано документи, які фігурують в діяльності шкіл. Це дозволило виявити основні проблеми та численні бюрократичні перепони, що стоять на заваді ефективного функціонування вітчизняної освітньої галузі.

Ось лише кілька цифр, які показово характеризують надмірну забюрократизованість процесів і процедур у цій галузі: 48 видів паперових документів школи зобов'язані вести та зберігати; 70 000 пачок паперу щомісяця витрачають українські школи на ведення цієї документації; 62 400 000 грн щороку коштує державі такий обсяг паперу.

Узагальнюючи такий поточний стан, ми можемо зробити такі висновки.

Звітність переважно дублюється у цифровому та паперовому варіантах для різних реципієнтів. Обмін інформацією між ними не налагоджений та потребує врегулювання.

Звітна інформація вноситься у ручному режимі. Це унеможливлює впровадження сучасних освітніх сервісів та не дозволяє повною мірою інтегрувати необхідні інформаційні освітні ресурси та забезпечити обмін даними з ключовими державними реєстрами.

Також було змодельовано нові процеси електронного обліку, вступу, відрахування, переводу учнів та дітей засобами модернізованої інформаційно-телекомунікаційної системи «Автоматизований інформаційний комплекс освітнього менеджменту» (АІКОМ). Нове технічне рішення дозволить впровадити нові цифрові підходи, що сприятимуть дебюрократизації та дерегуляції управління освітою.

Очевидно, що інформатизація системи освіти в Україні потребує суттєвої модернізації не тільки у сфері загально-освітньої підготовки, а і на інших рівнях освіти.

Актуальність проблеми інформатизації управління навчальним закладом полягає у створенні, впровадженні та розвитку комп'ютерно орієнтованого освітнього середовища на основі інформаційних систем, мереж, ресурсів і технологій. Головною метою є підготовка фахівця, в тому числі керівника закладу освіти до діяльності в умовах інформаційного суспільства, комплексна перебудова педагогічного процесу, підвищення його якості та ефективності. Вирішенню цього питання сприяє інформатизація навчального закладу.

Засоби інформаційно-комутаційних технологій, які застосовуються в управлінні освітнім закладом, повинні у сукупності представляти собою систему, засновану на використанні сучасних методів керівництва об'єктом системи освіти, застосуванні математичних моделей і методів у процесі прийняття рішень та створенні необхідної інформаційної бази на основі засобів комп'ютерної техніки і зв'язку, що забезпечує досягнення нової якості у підвищенні ефективності системи фахової передвищої та вищої освіти. Керівнику навчального закладу в умовах інформаційного суспільства важливо звернути особливу увагу на сучасні підходи у роботі з інформаційними матеріалами (збір, обробка, накопичення, зберігання, пошук і розповсюдження інформації), підготувати педагогічний колектив до реалізації засад «безпаперової інформатики» у побудові документообігу навчального закладу.

З вищесказаного можна зробити певні висновки: управління закладом освіти включає у себе велике коло питань: педагогічних, господарських, соціально-педагогічних, економічних, правових, фінансових. Інформатизація суспільства загалом і інформатизація освіти зокрема, привела ці системи у відповідність з потребами і можливостями сучасного інформаційного суспільства. Важливим фактором удосконалення управління є інформаційні технології, які надають масу нових можливостей, а саме: дозволяють накопичувати і поновлювати великі обсяги інформації, є інструментом оптимізації часу і коштів, що витрачаються на вирішення окремих задач управління, сприяють підвищенню якості прийнятих управлінських рішень за рахунок надання оперативної і достовірної інформації про стан керованого об'єкта.

Актуальність теми обумовлена низкою факторів, а саме:

— обсяг інформації про хід і результати освітнього процесу стає вищим, ніж рівень достатнього розуміння цієї інформації;

— механічна обробка без певного стандартного алгоритму не дає оперативних даних, що дозволяють приймати оптимальні управлінські рішення за результатами діяльності;

— робота закладу освіти в інноваційному режимі потребує багатогранного аналізу освітньої діяльності, оперативного простеження динаміки змін і своєчасного коригування;

— складні інформаційні моделі (автоматизовані системи управління закладом освіти), як правило, не виправдовують себе з фінансової точки зору, тому необхідно і доцільно комп'ютерні технології вводити там, де алгоритм управління досить простий і технічно здійсненний з відносно невеликими витратами.

Наша мета — представити автоматизовану комплексну систему обліку підвищення кваліфікації педагогічних та науково-педагогічних працівників закладу освіти, яка дозволяє зберігати, обробляти та контролювати результати підвищення кваліфікації та атестації педагогічних та науково-педагогічних працівників закладу освіти.

**Аналіз існуючих систем автоматизації моніторингу освітньої діяльності та управління закладом освіти.** Освітня управлінська інформаційна система — це не тільки набір формалізованих та інтегрованих операційних процесів, процедур та організаційно-правових заходів, але й повноцінна інституційна культура, яка забезпечує суспільство актуальними та достовірними даними про стан розвитку освіти.

Відомі системи управління освітнім процесом орієнтовані на підтримку основних функцій навчання (особливо в дистанційній формі) і не завжди охоплюють такі комунікаційні процеси планування, виконання, навчання, аналізу, звітування та контролю. Крім того, серед відомих систем управління навчанням як правило виокремлюють системи дистанційного навчання та моніторингу якості. На нашу думку, тільки комплексні електронні системи можуть стати ефективною інформаційною підтримкою для всіх категорій користувачів системи — від здобувача освіти до керівника навчального закладу. Саме це обумовило вибір та актуальність тематики досліджень.

Однак базовою для всіх інших функцій інформатизації навчального закладу є функція отримання, фіксації, зберігання та перетворення інформації. Ця функція є широкою у зв'язку з тим, що інформація потрібна не тільки для прийняття управлінських рішень. Саме ця функція створює умови для реалізації ще однієї важливої функції інформатизації навчального закладу — задоволення інформаційних потреб учнів, студентів, працівників, потенційних споживачів освітніх послуг, працівників інших освітніх установ та структур управління освітою.

Серед важливих компонентів інформатизації освіти є розроблення програмного забезпечення. Програми, які використовують у закладах освіти, поділяють на такі програмі засоби: навчальні (скеровують навчання з огляду на наявні знання та індивідуальні здібності студентів, а також сприяють засвоєнню нової інформації); діагностичні (тестові — призначені для діагностування, перевірки, оцінювання знань, умінь, здібностей студентів); тренувальні (розраховані на повторення та закріплення пройденого навчального матеріалу); бази даних (сховища інформації з різних галузей знань, у яких за допомогою запитів на пошук у різних галузях знань знаходять необхідні відомості); імітаційні (представляють певний аспект реальності за допомогою параметрів для вивчення її основних структурних чи функціональних характеристик); моделюючі (відображають основні елементи і типи функцій, моделюють певну реальність); інструментальні (забезпечують виконання конкретних операцій, тобто оброблення тексту, складання таблиць, редагування графічної інформації) [6].

Наразі Міністерство освіти і науки України активно приєднується до проєктів цифрової трансформації у ключових сферах. Система управління проектами від Мінцифри налічує вже 94 об'єкти цифровізації в Україні за різними напрямами. Цифровізація вже відіграла важливу роль у сфері освіти і науки України. Її мета — щоб всі освітні послуги стали більш доступними та контрольованими. Реалізація цих проектів передбачає впровадження цифрових сервісів в освіту і науку, автоматизацію освітніх та управлінських процесів

Серед презентованих програм, зокрема:

Цифровізація дошкільної, загальної середньої та позашкільної освіти (е-Школа). У межах проєкту е-Школа вже функціонує сучасний онлайн-ресурс «Всеукраїнська школа онлайн» для змішаного та дистанційного навчання учнів 5—11 класів із матеріалами, що відповідають державній програмі. А в найближчих планах — створення кабінету вчителя, де можна буде відстежувати навчальний прогрес, створювати та модифікувати уроки.

Цифровізація вищої, фахової передвищої та професійної освіти (е-Університет). У межах проєкту е-Університет заплановано:

— автоматизація вступної кампанії;

— автоматизація процесів набору та навчання (стажування) іноземців та осіб без громадянства;

— запровадження електронного ліцензування;

— модернізація Єдиної державної електронної бази з питань освіти;

— створення та модернізація єдиної електронної системи моніторингу працевлаштування випускників.

Модернізація систем подання документів та проведення державної атестації наукових установ і закладів вищої освіти в частині проведення ними наукової діяльності, розвиток репозитарію академічних текстів та підключення до нього локальних репозитаріїв і створення електронної системи доступу до нинішніх цифрових сервісів наукового призначення — серед цілей проєкту е-Наука.

Цифровізація фінансування та послуг у сфері науки (е-Наука).

Слід зазначити, що на сьогоднішній день існують діючі IC моніторингу діяльності навчальних закладів. Серед відомих систем управління навчальним процесом у закладі вищої освіти на ринку України можна відзначити такі:

— автоматизована система управління навчальним процесом для вищих навчальних закладів усіх рівнів акредитації АСК «Вищий навчальний заклад», розроблена у Науково-дослідницькому інституті (НДІ) прикладних інформаційних технологій, яка є частиною інформаційно-виробничої системи «Освіта» [7];

— система управління навчальним процесом для вищих навчальних закладів «Директива», розроблена у ТОВ «Комп'ютерні інформаційні технології» [8];

— пакет програм «Деканат», розроблений приватним підприємством «Політек-СОФТ», до складу якого входить модуль «ПС Студент» [9].

Поряд з цим у багатьох великих закладах вищої освіти функціонують і власні розробки подібних систем. До них можна віднести:

— електронну систему управління закладом вищої освіти «Сократ» Вінницького національного аграрного університету [10];

— інформаційно-аналітичну систему управління закладом вищої освіти «Університет» Херсонського державного університету;

— засоби автоматизації управління навчальним закладом, що діють в НУ «Львівська політехніка» та Львівському національному університеті імені Івана Франка;

— автоматизовану інформаційну систему «Електронний університет» [11], створену у Хмельницькому національному університеті.

Проведемо аналіз двох інформаційних систем автоматизації управління навчальним процесом у внз, а саме: автоматизована система управляння (АСУ) «ВНЗ», яку розробив Науково-дослідницький інститут прикладних інформаційних технологій в комерційних цілях та АСУ «Сократ», що самостійно розроблена та запроваджена у Вінницькому національному аграрному університеті. Насамперед потрібно сказати про загальні принципи роботи цих двох систем.

*Інформаційна система АСУ «ВНЗ»* побудована у вигляді WEB-додатка програми, тобто її робота вимагає підключення до всесвітньої мережі Internet. Всі дані зберігаються й обробляються на сервері, який фізично знаходиться в м. Києві, а не там, де ведеться експлуатація системи (у підрозділах університету в містах Львові, Харкові та Черкасах). Для роботи з системою не потрібно встановлювати спеціальне програмне забезпечення.

Структура системи АСУ «ВНЗ» реалізована на модульній основі, де кожен модуль може використовуватися самостійно (рис. 1). АСУ «ВНЗ» вирішує велику кількість автоматизованих функцій управління, у тому числі [7]:

— електронну реєстрацію, обробку даних та документообіг в єдиній інформаційній системі для кожного структурного підрозділу окремо і установи в цілому;

— планування, контроль та аналіз навчальної діяльності;

— оперативний доступ до інформації, що супроводжує навчальний процес;

— єдину систему звітів, як внутрішніх, так і за вимогами Міністерства освіти і науки (МОН) України;

— системи безпеки даних з урахуванням вимог законодавства;

— можливості безпосереднього обміну даними з інформаційно-виробничими системами «Освіта» та «Education».

Також система включає АС «Конструктор звітів», що дозволяє формувати інформацію і надавати її користувачам у найбільш зручному вигляді.

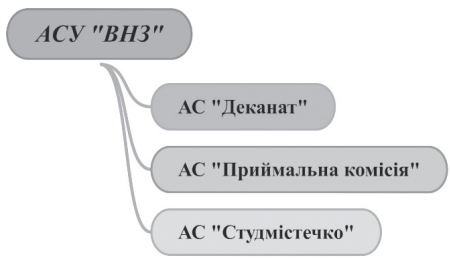

Рис. 1. Модулі АСУ «ВНЗ»

*Автоматизована система управління «Сократ»* — система управління якістю освітньої діяльності університету працює як відкрита інформаційна система, адаптована під потреби викладачів, студентів та адміністрації університету [10].

Натомість система АСУ «Сократ» власної розробки ВНАУ повністю автономна, сервер системи знаходиться на території навчального закладу та за необхідності може працювати у всесвітній мережі Internet. Для роботи з системою також не потрібно встановлювати спеціальне програмне забезпечення, вистачить будь-якого зручного користувачам браузера (Microsoft Internet Explorer, Mozilla Firefox, Google Chrome, Safari, Opera). Система надає доступ практично до всієї інформації з мережі Internet в будь-який час та можливість дистанційного навчання. Система потребує адміністрування у зв'язку з постійними нововведеннями та розширенням її структури та університету в цілому.

Розглянемо детальніше складові АСУ «Сократ».

Персональний кабінет викладача — це складова роботи викладача, за допомогою якої можливо проводити дистанційне навчання, ведення електронних журналів та змогу обміну інформацією з іншим користувачам системи.

Персональний кабінет студента — це навчальна складова студента, за допомогою якої можливо проводити дистанційне навчання.

Автоматизована система управління АСУ «Деканат» дає змогу: формувати та зберігати персональні дані студентів; формувати дані про успішність, модулі, заліки, іспити; ведення моніторингу якості знань; ведення контролю навчального процесу.

Бухгалтерія — це клієнт-серверна багатокористувацька програма, альтернатива до програми «1С:Бухгалтерія» яка зроблена за Web-технологіями.

Бібліотечна система «Софія» дозволяє зберігати і використовувати різного типу та змісту електронні документи та зручним способом для кінцевого користувача представляти їх.

Методичний кабінет проводить управління процесами створення та впровадження методичних розробок викладача.

Центр здоров'я дозволяє вести медичні дані про захворювання студентів в електронному вигляді, виписки медичних довідок та надання в автоматичному режимі інформації деканату.

Відділ кадрів викладачів займається веденням особових справ викладачів.

Відділ кадрів студентів займається веденням особових справ студентів.

Відділ діловодства займається всією документацією, листами та розсилкою їх структурам, до яких вони відносяться.

Дистанційне навчання дає змогу проходження тестів та перегляду інформації з Інтернету.

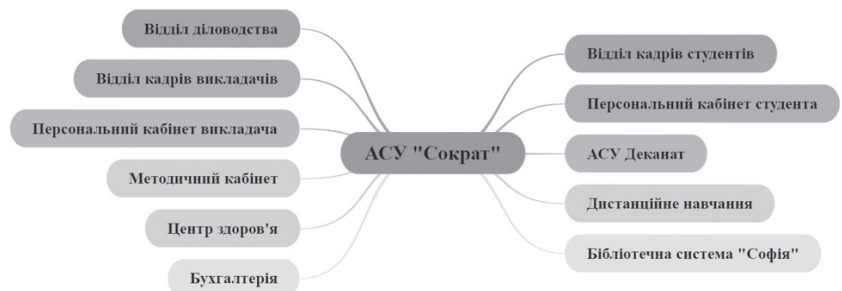

Рис. 2. Електронна автоматизована система управління (АСУ) «Сократ» та її складові

Головна відмінність АСУ «Сократ» — використання персональних кабінетів студента, викладача, співробітника, навколо яких формується інформаційне поле (рис. 3).

Персональний кабінет викладача передбачає:

— ведення електронного журналу (який можуть переглядати студенти);

— перегляд розкладу занять і навчальних планів з навчальної частини on-line;

— розклад занять на мобільному телефоні;

— створення тестів для студентів;

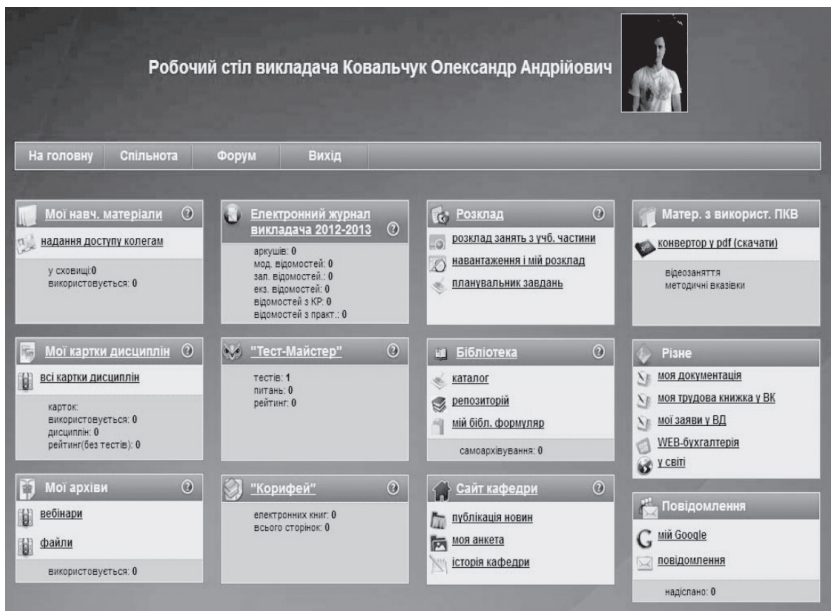

Рис. 3. Складова електронної системи «Сократ», персональний кабінет викладача

— доступ до автоматизованої бібліотечної системи «Софія»;

— можливість публікації власних методичних матеріалів для використання їх в навчальних картках дисциплін;

— можливість самопублікації власних наукових матеріалів у електронному репозиторії;

— доступ до WEB-чату, блогів, форумів всередині системи.

Персональний кабінет студента передбачає (рис. 4):

— перегляд своєї навчальної картки;

— пошук та перегляд інформаційних ресурсів в бібліотеці;

— проходження тестування;

— доступ до WEB-чатів, мікроблогів, студентських форумів.

Електронна система «Сократ» має зв'язки з мобільними технологіями та соціальними мережами. Так, викладачі мають можливість одержувати свій розклад на мобільний телефон. В системі «Сократ» можливо здійснити формування дистанційних курсів на основі матеріалів навчальної картки дисципліни, відеолекцій та вебінарів.

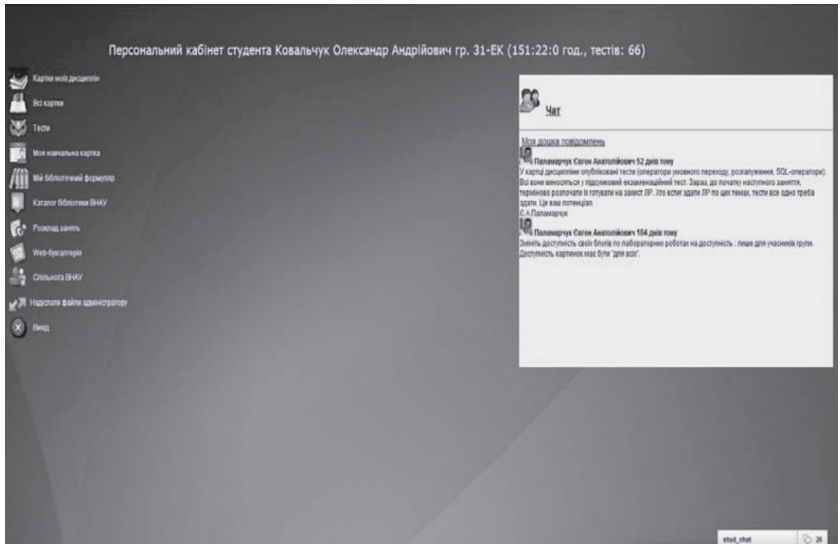

Рис. 4. Складова електронної системи «Сократ», персональний кабінет
здобувача освіти

Безперечна перевага системи АСУ «ВНЗ» — наявність блоку
«Приймальна комісія» та його зв'язок з міністерською програмою
«Вступ». Блок «Студмістечко» підтримує звітність гуртожитків. Взагалі АСУ «ВНЗ» орієнтована на формування звітності та виконання
функцій АСУ «Деканат», «Приймальна комісія», «Студмістечко» та
не передбачає інформаційної підтримки дистанційної форми навчання, формування банку методичних матеріалів, зв'язку з бібліотечними ресурсами.

Перевага електронної системи «Сократ» в орієнтації на студента,
викладача, методиста, адміністрації закладу, їх запитів з питань методичної інформаційної підтримки та контролю. Недоліком є необхідність створення нових модулів та інтеграції з загальнодержавними
програмними продуктами.

До популярних систем автоматизації та моніторингу в освітній діяльності можна віднести:

1. Інформаційно-виробнича система «Освіта», розроблена у Науково-дослідницькому інституті (НДІ) прикладних інформаційних
технологій [12];

2. Інформаційна система «Вступ.ОСВІТА.UA» [13];

3. Програмно-апаратний комплекс «Автоматизований інформаційний комплекс освітнього менеджменту», розроблений ДНУ «Інститут освітньої аналітики» [14];

3.1. E-Journal;

4.Програмний комплекс «КУРС: Освіта», розроблений ТОВ «Нові знання» [15];

4.1. «Курс:Школа»;

4.2. «Україна. ІСУО»;

4.3. Портал «NZ.UA»;

4.4. «Курс:Дошкілля»;

5. Автоматизована система підвищення кваліфікації педагогічних працівників закладів загальної середньої освіти, розроблена ТОВ «Активмедіа» [16].

Проведемо аналіз систем автоматизації та моніторингу в освітній діяльності.

*Загальнодержавна інформаційно-виробнича система «Освіта»* забезпечує єдине цілісне інформаційне середовище України в галузі освіти та являє собою сукупність адміністративних, правових, програмних та апаратних засобів.

Автоматизована система ідентифікації особистості в суспільстві та державі стає неодмінним атрибутом адміністративного функціонування цивілізованих країн. Базовою інформацією про людину в будь-якій системі ідентифікації є персональна інформація: прізвище, ім'я, по батькові, дата і місце народження, фотографія.

На етапі становлення єдиного інформаційного простору України стає необхідним впровадження єдиної загальної системи документального супроводу громадян, заснованої на невід'ємному поєднанні інформаційної системи обліку даних та системи виробництва документів. Відповідно до закону «Про освіту» отримання базової освіти є обов'язковим для всіх громадян, тому система, що передбачає введення в базу даних персональних даних громадянина з цифровою фотографією, повинна стати основою для присвоєння індивідуального ідентифікаційного податкового номера за погодженням з Державною податковою адміністрацією.

Така багатофункціональна інформаційно-виробнича система збору та обліку даних про фізичних осіб та їх документального супроводу розроблена Інститутом кібернетики і ЗАТ «НДІ прикладних інформаційних технологій» Кібернетичного центру НАН України спільно з Міністерством освіти і науки України.

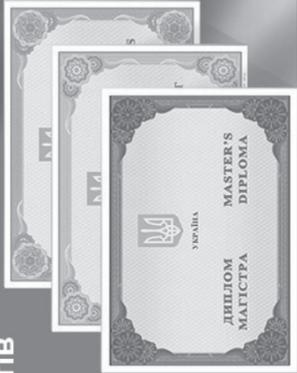
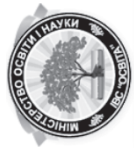

Рис. 5. Головна сторінка інформаційно-виробничої системи «Освіта»

На сьогоднішній день вже створена й успішно функціонує інфраструктура, яка забезпечує збір первинних даних і охоплює всю територію України до районного рівня.

Основні цілі створення ІВС «ОСВІТА»:

— аналіз кадрового потенціалу держави та прогнозування тенденцій відносно змін у структурі професійного складу;

— аналіз кількісних показників національних трудових ресурсів;

— створення типових інформаційних та програмних засобів для ефективного керування навчальним закладом;

— моніторинг діяльності навчальних закладів;

— аутентифікація документів про освіту державного зразка;

— впорядкування надання пільг учням та студентам.

Відповідно до призначення головними завданнями ІВС «ОСВІТА» є:

— створення ефективного автоматизованого комплексу для вироблення управлінських рішень у галузі освіти в Україні;

— створення ефективного автоматизованого комплексу для управління вищим навчальним закладом у складі типової АС «Вищий навчальний заклад»;

— створення та супровід єдиної бази даних навчальних закладів України;

— створення та супровід єдиної бази даних учнів та студентів в Україні;

— впровадження реєстру вищих, професійно-технічних та загальних середніх навчальних закладів;

— отримання та аналіз інформації щодо діяльності освітніх закладів України;

— підтримка ведення інформаційної бази Міністерства освіти і науки України щодо документів у галузі освіти, їх власників та методичної бази аналізу фахового складу населення;

— упорядкування процедур надання відповідно до чинного законодавства пільг учням та студентам;

— автоматизація збору та збереження інформації, що стосується документів у галузі освіти;

— автентифікація інформації на різних етапах збору і обробки даних в ІВС «ОСВІТА»;

— підвищення рівня захисту від підробок документів у галузі освіти;

— централізація та вдосконалення процесу виготовлення документів у галузі освіти, а також дублікатів їх пластикових карток у випадках втрати або пошкодження;

— забезпечення цілісності, достовірності та актуалізації інформації щодо виготовлених документів у галузі освіти;

— ідентифікація виготовлених ІВС «ОСВІТА» документів у галузі освіти;

— отримання статистичних даних, надання необхідної статистичної інформації державним органам та підприємствам;

— забезпечення захисту інформації від несанкціонованого доступу відповідно до вимог чинного законодавства.

Інформаційно-виробнича система ІВС «ОСВІТА» включає систему збору даних про фізичних осіб (включаючи цифрове фото) з використанням мережі Інтернет, центрального банку даних, виробничого комплексу з виготовлення та обліку документів. Інформаційна безпека системи реалізована з використанням сучасних програмних і криптографічних засобів, що дозволяє повністю виключити можливість несанкціонованого доступу до інформації.

ІВС «ОСВІТА» впроваджена в Міністерстві освіти і науки України для інформаційного та документального супроводу навчального процесу. Вона охоплює всі етапи навчання — від середньої школи до закладів вищої освіти, забезпечує збір даних про особу та централізоване виготовлення електронних шкільних квитків, електронних студентських квитків та документів про освіту. У 2000 році результатом функціонування системи стало виготовлення та постановка на облік понад 1,8 мільйона документів про освіту в Україні. На сьогоднішній день центральний банк містить інформацію про понад 39 мільйонів документів та їх власників.

Оскільки персональні дані людини, що містяться ІВС «ОСВІТА», використовуються в подальшій життєдіяльності людини в державі, система повинна бути тісно пов'язана з діяльністю не тільки в галузі освіти, а й у сферах відповідальності інших відомств і адміністративних органів України: Міністерства внутрішніх справ, Державної податкової адміністрації України, Міністерства оборони, Міністерства охорони здоров'я, Міністерства транспорту, обласних та міських державних адміністраціях. В даний час узгоджені принципи взаємодії і виробляються спільні роботи з Міністерством транспорту і Державною податковою адміністрацією України. Результатом цієї взаємодії є система обліку та надання пільг.

# Вступ.ОСВІТА.UA

Дані отримані з ЄДЕБО 18.11.2021 04:00
(наступне оновлення до 09:00 22.11.2021)

Закладки   FAQ   Про сайт   Форум

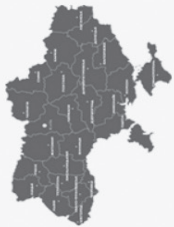

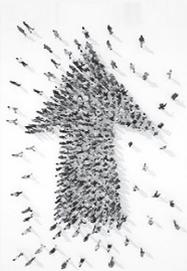

■ Дізнайся скільки балів потрібно щоби вступити на бюджет.
■ Отримай інформацію про спеціальності у кожному виші.
■ Зроби відповідальний вибір навчального закладу.
■ Дивись рейтингові списки онлайн у липні-серпні 2021 року.

## Пошук навчального закладу

Оберіть регіон   ▶

## Пошук спеціальності

Оберіть освітній ступінь   ▶

Рис. 6. Головна сторінка «Вступ.ОСВІТА.UA»

*Інформаційна система «Вступ.OCBITA.UA».* Інформація, що розміщена на сайті інформаційної системи «Вступ.OCBITA.UA», отримана з Єдиної державної бази з питань освіти (ЄДЕБО) на підставі укладених договорів із Державним підприємством «Інфоресурс», що є розпорядником ЄДЕБО.

Інформаційна система «Вступ.OCBITA.UA» здійснює інформування широкої громадськості про перебіг подання заяв щодо вступу, рекомендації до зарахування та зарахування до закладів вищої освіти під час проведення вступної кампанії з метою забезпечення відкритості та прозорості при проведенні прийому до закладів вищої освіти.

Також сайт створений з метою інформування абітурієнтів, їх батьків, освітніх експертів та інших зацікавлених осіб про умови вступу до закладів вищої освіти України та надання статистичної інформації про результати вступу до закладів вищої освіти у минулих роках.

На сайті розміщується інформація про всі діючі заклади вищої освіти в Україні, які здійснюють прийом абітурієнтів на навчання для здобуття ступеня молодшого спеціаліста, бакалавра або магістра.

По кожному закладу вищої освіти надається контактна інформація, інформація про спеціальності та освітні програми, за якими здійснюється підготовка молодших спеціалістів, бакалаврів та магістрів. Інформація включає дані про конкурсні предмети, ліцензійні обсяги, обсяги державного замовлення та інше.

*Програмно-апаратний комплекс «Автоматизований інформаційний комплекс освітнього менеджменту»* (ПАК «АІКОМ») — електронна система управління освітою, яка є модернізованою Державною інформаційною системою освіти (ДІСО), призначена для обробки державних електронних інформаційних ресурсів та персональних даних у сфері освіти в рамках єдиного інтегрованого середовища.

Основна мета — забезпечення переходу до електронного документообігу (звітність, комунікація, сповіщення, опитування, голосування, оперативне збирання даних) та оптимізація даних бізнес-процесів у сфері дошкільної, загальної середньої, позашкільної та професійної (професійно-технічної) освіти та управлінь освітою місцевого та обласного рівнів (створення відповідних модулів в ПАК «АІКОМ»), що дасть змогу суттєво підвищити достовірність освітньої статистичної та адміністративної інформації та покращити на цій основі якість управлінських рішень, зокрема щодо розподілу коштів освітньої субвенції та інших бюджетних коштів для фінансування освіти,

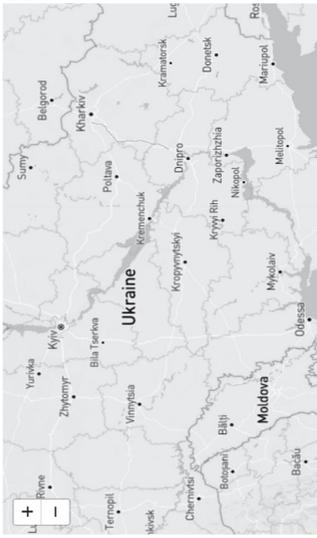

Рис. 7. Головна сторінка ПАК «АІКОМ»

забезпечить передумови для відмови від паперових документів в рамках загальної дебюрократизації.

Очікувана модернізація передбачає створення електронних кабінетів здобувача освіти та педагогічного працівника дошкільної, загальної середньої, позашкільної та професійної (професійно-технічної) освіти, що уможливить отримання ними доступу до усіх державних інформаційних освітніх сервісів та полегшить отримання зведених деперсоналізованих даних для формування в автоматичному режимі звітності, необхідної для Державної служби статистики України.

Серед інших цілей модернізації є також розвиток безкоштовного державного електронного щоденника та журналу, розгортання єдиної системи авторизації для державних освітніх сервісів, підключення альтернативних програмних рішень ринку освітніх інформаційних послуг до центральної бази даних з використанням API ПАК «АІКОМ» тощо.

Одним з популярних цифрових інструментів, який забезпечує базову цифровізацію системи загальної середньої освіти, є система електронних журналів. Такі цифрові інструменти створюють нові можливості для забезпечення безперервної взаємодії та ефективної співпраці між чотирма групами учасників процесу — адміністрація школи, вчителі, учні та батьки.

Починаючи з грудня 2020 року доступний для впровадження та використання в ЗЗСО безкоштовний державний сервіс електронних журналів на базі програмно-апаратного комплексу «Автоматизований інформаційний комплекс освітнього менеджменту» (EJournal), створений командою Міністерства освіти і науки України та Державної наукової установи «Інститут освітньої аналітики», який введено в дослідну експлуатацію наказом МОН від 22.12.2020 № 1545. Рішення про його використання приймає ЗЗСО на добровільній основі.

Переваги запровадження електронного журналу для різних груп учасників освітнього процесу:

*Для адміністрації школи:*

— можливість оперативно готувати освітню звітність, діаграми успішності по класах і предметах;

— аналіз результативності роботи педагогів;

— облік відвідування;

— можливість відслідковувати динаміку успішності учнів, класів, школи;

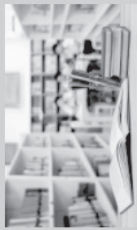

Рис. 8. Можливості системи E-Journal

— підвищення рейтингу закладу освіти.

*Для вчителів:*

— звільнення від надлишкової паперової роботи;

— простий доступ до актуального розкладу у смартфоні або комп'ютері;

— можливість завантажувати навчальні матеріали для ознайомлення та допомоги у підготовці домашніх завдань;

— економія часу для підготовки до уроків;

— зручний поділ класів на групи без паперових журналів;

— просте автоматичне формування складних звітів за підсумками семестру або навчального року;

— ефективна комунікація з учнями та батьками.

*Для учнів:*

— зручний доступ до навчальних матеріалів і домашніх завдань;

— участь в онлайн вебінарах і конференціях, організованих школою;

— можливість перегляду матеріалів уроків у зручний час (у т. ч. у період відсутності в школі);

— можливість віддаленої взаємодії з учителями;

— можливість самостійного контролю успішності навчання.

*Для батьків:*

— можливість оперативно отримувати інформацію про успішність та відвідування дітей;

— ефективний контроль засвоєння знань та виконання домашніх завдань;

— пряма комунікація з учителями;

— можливість брати участь в оцінці якості освітніх послуг.

*«КУРС:Освіта»* — це програмний комплекс, до складу якого входять наступні ключові складові:

Комп'ютерна програма «КУРС:ШКОЛА» для загальноосвітніх навчальних закладів та її складові — «КУРС:ШКОЛА +» для місцевих органів управління освітою (відділи, управління, департаменти освіти районних та міських органів влади) і «КУРС:САЙТ» для автоматизації передачі даних на WEB-портали верхнього рівня.

Дозволяє автоматизувати і якісно керувати навчальними процесами. Генерує обов'язкові форми звітності ЗНЗ-1 і 83-РВК, затверджені наказом № 766 МОНМС України від 02.07.2012 року, форму статистичної звітності 77-РВК, затверджену наказом Держкомстату № 317 від 06.08.2010 року і пересилає їх електронні версії згідно з підпорядкованістю. Допомагає встановити навантаження вчителям. Складає

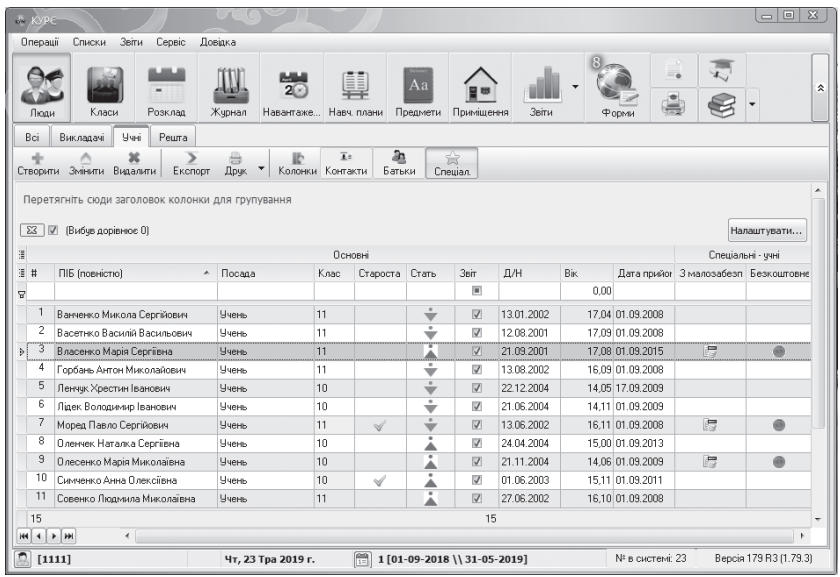

Рис. 9. Інтерфейс програми «КУРС:ШКОЛА»

розклад занять як в ручному, так і в автоматичному режимі. Має модуль електронних класних журналів. Програма «КУРС:ШКОЛА» може працювати в двох режимах: з доступом і без доступу до мережі Інтернет.

Портал *ІСУО* — інформаційна система управління освітою. Приймає і консолідує дані із закладів загальної середньої освіти, генерує обов'язкові форми звітності ЗНЗ-1, 76-РВК, 77-РВК, 83-РВК, Д-4, Д-5, Д-6, Д-7, Д-8, затверджені діючим законодавством, і пересилає їх електронні версії згідно з підпорядкованістю. Дозволяє здійснювати пошук інформації. Полегшує вибірку необхідних даних і складання користувацьких звітів. Має надійні алгоритми захисту інформації від несанкціонованого використання. Кожен регіон України має власне доменне ім'я і, відповідно власну Систему управління освітою регіону. Склад і функціонал може доповнюватися і нарощуватися в залежності від завдань і потреб.

Портал «NZ.UA» — це публічний сайт для всіх учасників освітнього процесу: учнів, батьків, вчителів та інших працівників освіти.

Дозволяє вчителям виставляти оцінки і вести облік відвідуваності в класних журналах та учнівських щоденниках. Зручно спілкуватися та обмінюватися інформацією всім учасникам освітнього процесу.

**411**

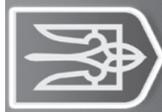
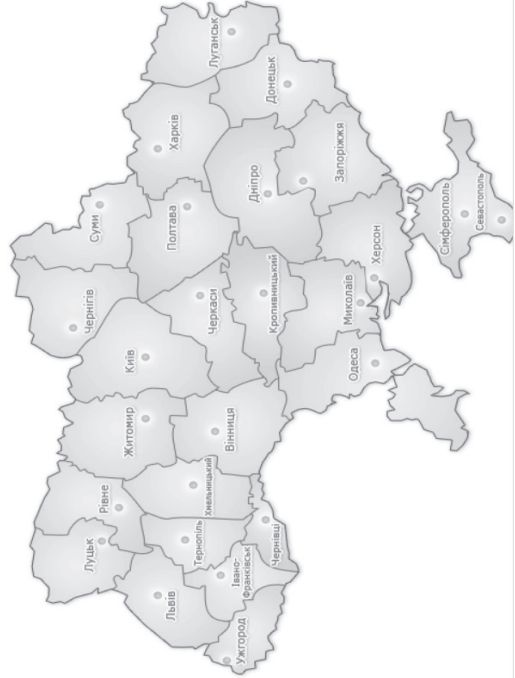

Рис. 10. Головна сторінка «Україна.ІСУО»

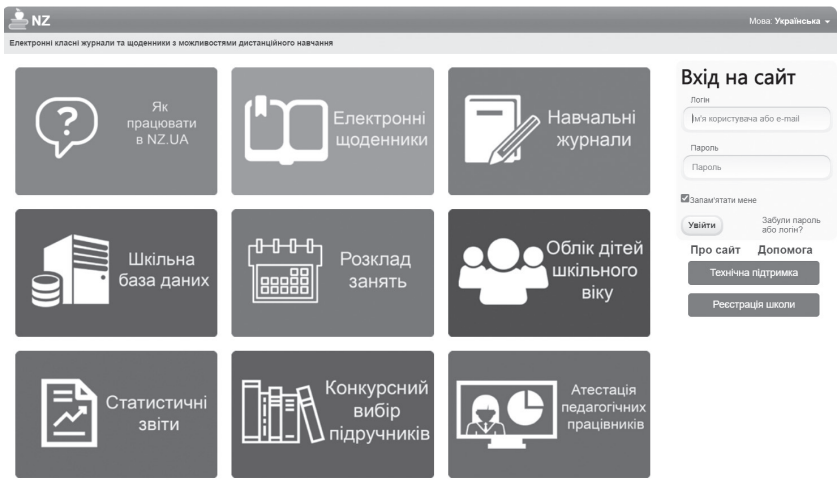

Рис. 11. Головна сторінка порталу «NZ.UA»

Публікувати домашні завдання як для всього класу, так і персонально для кожного учня. Батьки завжди будуть в курсі того, що відбувається в школі, навіть якщо вони дуже зайняті або живуть і працюють далеко від своїх чад. Існує можливість архівувати дані, аналізувати відвідуваність і успішність як вчителям, так і батькам. Дозволяє робити висновки та різні статистичні вибірки. Портал «NZ.UA» автоматично синхронізується з програмою «КУРС:ШКОЛА», що значно полегшує роботу користувачам.

**Комп'ютерна програма «КУРС:ДОШКІЛЛЯ».** Ця програма призначена для ведення єдиної бази даних дитячої дошкільної установи, управління процесами, обліку дітей дошкільного віку і автоматичного (натисненням однієї кнопки) складання обов'язкового статистичного звіту за формою 85-к. Програма «КУРС:ДОШКІЛЛЯ» враховує відомості про педагогічний склад, вихованців, їх батьків або опікунів, допомагає вести контроль і відображення відвідуваності протягом тижня, місяця, року для окремих груп так і для усього закладу в цілому, покращує ефективність роботи дитячої дошкільної установи, створює комфортніші умови для плідної роботи персоналу. Програма підтримує дві мови (українську і російську), розмежовує права і рівні доступу до даних. Надає можливість роботи в розрахованому на одного користувача і мережевому режимі з підтримкою персоналізації інтерфейсу. Використовує новітні інтелектуальні методи представлення даних.

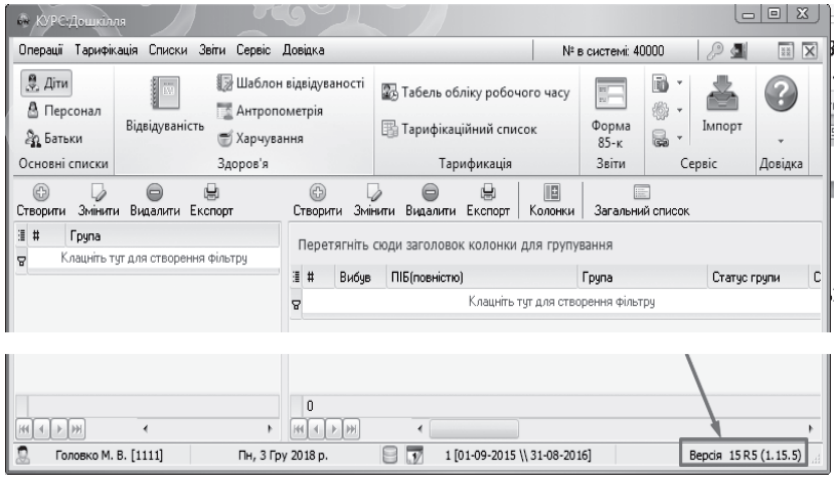

Рис. 12. Інтерфейс програми «КУРС:ДОШКІЛЛЯ»

*Автоматизована система підвищення кваліфікації педагогічних працівників закладів загальної середньої освіти (АПК)* призначена для підтримки процесу проходження навчання за програмами підвищення кваліфікації педагогічних працівників закладів загальної середньої освіти.

На новій платформі можна переглянути курси і програми підвищення кваліфікації, оформити замовлення на проходження навчання, створити власний кабінет для зберігання інформації про хід і результати навчання, здійснити оплату, отримати сертифікати встановленої форми про підвищення кваліфікації тощо.

Надання послуг здійснюється відповідно до вимог Постанов Кабінету Міністрів України № 800/1133 від 21 серпня і 27 грудня 2019 року щодо порядку підвищення кваліфікації педагогічних і науково-педагогічних працівників. Послуги надаються на підставі угоди, яка стає доступною зареєстрованому користувачу в процесі формування замовлення на обраний курс. Оплата за послуги здійснюється користувачем на підставі рахунку-фактури одним із доступних способів: на банківську картку, онлайн, банківським переказом. Після завантаження всіх передбачених для конкретного курсу документів і здійснення оплати користувач отримує в свій кабінет сертифікат встановленого зразка, а також акт про надання послуг.

ПІДВИЩЕННЯ КВАЛІФІКАЦІЇ:
АДМІНІСТРАТОР ПОСЛУГ

Рис. 13. Автоматизована система підвищення кваліфікації педагогічних працівників закладів загальної середньої освіти

**415**

Документами, на підставі яких видається сертифікат про підвищення кваліфікації, є:

для дистанційної форми навчання: електронний документ про проходження відповідного онлайн-курсу;

для дуальної форми навчання:

1) електронний документ про проходження відповідного онлайн-курсу;

2) довідка про проходження практики за місцем роботи за підписом керівництва закладу освіти.

Провівши аналіз інформаційних систем управління навчальним процесом та систем автоматизації та моніторингу в освітній діяльності, можна зробити висновок, що перевагами універсальних систем є простіше початкове налаштування, яке потребує лише наявності підключення до мережі Internet, відсутність потреби адміністрування бази даних з боку користувача. Технологічні процеси у таких системах вибудувані згідно з вимогами і нормативними документами Міністерства освіти і науки. Однак у жертву широкомасштабності та універсальності була віддана продуманість окремих специфічних функцій та зручність користувача. Натомість спеціалізовані електронні системи управління потребують спеціального адміністрування, але це дозволяє їм бути більш гнучкими до будь-яких потреб закладу освіти. Значною перевагою цієї системи є те, що всі модулі використовують єдину базу даних як викладачів, так і здобувачів освіти. Також система спонукає персонал до підвищення виконавчої дисципліни, тому що не дозволяє довільно змінювати вихідну інформацію, створює необхідність використання сучасних інструментів навчання та комунікацій (корпоративної пошти, чату, науково-освітньої спільноти тощо).

Результати дослідження дозволяють сформувати такі рекомендації:

1. Створення електронної системи управління навчальним процесом повинно враховувати цілі та завдання аудиторії користувачів (студентів, викладачів, співробітників, адміністрації закладу);

2. Впровадження комплексної електронної системи управління навчальним закладом буде дійсно ефективним при умові технічної, програмної та організаційної підтримки.

**Автоматизована інформаційна система обліку підвищення кваліфікації викладачів.** Нова реформа освіти в Україні та системи післи-дипломної педагогічної освіти (ППО) вимагає моніторингу процесу підвищення кваліфікації працівників навчальних закладів. Це забезпечить спостереження та аналіз впровадження відповідного процесу професійного розвитку в навчальних закладах з метою виявлення проміжних та кінцевих результатів. Загалом це сприятиме прийняттю відповідних управлінських рішень для регулювання та коригування навчального процесу для забезпечення його якості. Водночас ефективність моніторингу та своєчасність необхідних змін залежать від автоматизації обробки наявних даних про діяльність навчального закладу. Це визначає можливість використання відповідного програмного забезпечення, зокрема спеціалізованих інформаційних систем (ІС).

Судячи з розглянутих вище застосунків, можна зробити висновок, що існує багато варіацій систем автоматизації освітнього процесу. У кожного з них є свої переваги та недоліки, також варто відмітити, що більшість розглянутих рішень працюють в онлайн-форматі. Після їхнього аналізу можна виділити основну функціональність, яка повинна бути у розробленій системі, а саме:

зрозумілий інтерфейс;

швидкість роботи;

простота налаштування та запуску;

наявність різних типів доступу;

створення окремих груп користувачів;

інтеграція на різні середовища;

можливість реалізації додаткових модулів;

високий ступінь безпеки;

широка функціональність.

Усі інформаційні системи незалежно від архітектури і сфери використання складаються з двох частин: функціональної, до якої належать елементи системи, що визначають її функціональні можливості, призначення, функції управління, та забезпечувальної (рис. 14).

Функціональна частина складається з підсистем, комплексів задач, автоматизованих робочих місць. Функціональна підсистема — це самостійна частина системи, що виконує конкретні функції та завдання управління і характеризується відповідним цільовим призначенням, певною методикою проведення розрахунків економічних показників, підпорядкованістю та технологічними особливостями експлуатації. Перелік функцій конкретної інформаційної системи залежить від сфери її застосування, об'єкта управління. Відповідно до виділених функціональних підсистем і до фаз управління визначається комплекс задач, що здійснюється з урахуванням основних фаз управління: планування, обліку, контролю, аналізу, регулювання. Вибір і обґрунтування комплексу функціональних задач — один із найважливіших елементів створення інформаційних систем [22].

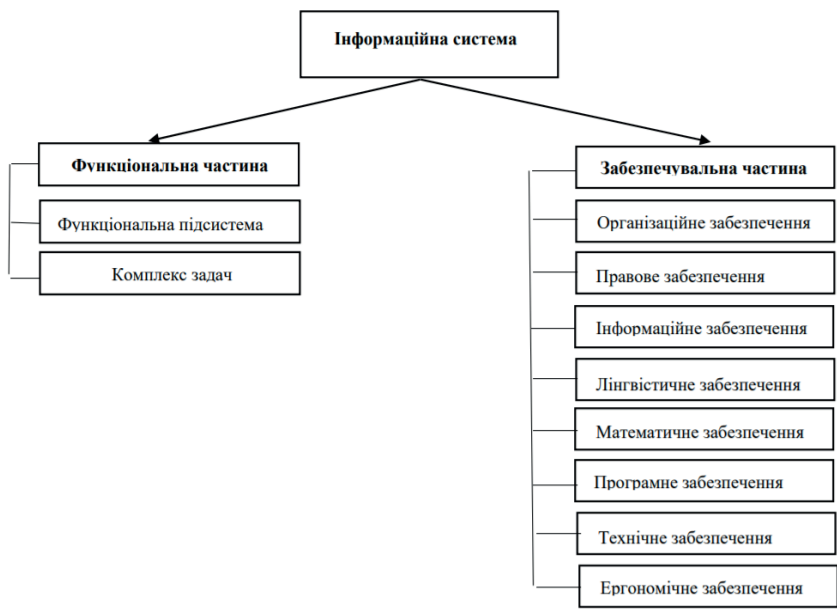

Рис. 14. Структурна схема інформаційної системи

Забезпечувальна частина інформаційної системи складається з підсистем, які є інструментами для реалізації функціональної частини. До їх складу входять організаційне, правове, інформаційне, лінгвістичне, математичне, програмне, технічне, ергономічне забезпечення.

Підсистема організаційно-методичного забезпечення є сукупністю правил, документів, інструкцій та положень, що забезпечують створення системи та взаємодію її складових частин у процесі функціонування. Організаційне забезпечення реалізує такі функції:

— аналіз існуючої системи керування організацією, де буде використовуватися інформаційна система, і виявлення завдань, що підлягають автоматизації;

— підготовка завдань до вирішення на комп'ютері, включаючи технічне завдання на проектування інформаційної системи і техніко-економічне обгрунтування її ефективності;

— розроблення управлінських рішень, методології вирішення завдань, спрямованих на підвищення ефективності системи керування.

Документи визначають технологію функціонування інформаційної системи, методи вибору і застосування користувачами технологічних прийомів для одержання конкретних результатів при функціонуванні інформаційної системи.

Підсистема правового забезпечення — сукупність законодавчих актів, правових норм і нормативів, що пов'язані з функціонуванням інформаційної системи. Нормативними документами визначається порядок одержання, перетворення й використання інформації. Правове забезпечення базується на юридичному підході до організаційних і функціональних аспектів розроблення системи. Конструктивно функції управління в правовій сфері реалізуються у вигляді нормативно-правових актів, планів, методик, обов'язкових для всіх державних стандартів. Нормативно-правові акти поділяються на законодавчі, нормативні акти міністерств і відомств, акти місцевих органів управління, локальні нормативні акти.

Інформаційне забезпечення — це сукупність методів і засобів розміщення й організації інформації, що включає єдину систему класифікації і кодування повідомлень, схем інформаційних потоків, а також методологію побудови баз даних. Інформаційне забезпечення включає сукупність показників, довідкових даних, класифікаторів та кодифікаторів інформації, масиви інформації на відповідних носіях і є однією з основних ланок функціонування інформаційної системи, що містить ресурси, заради використання яких і ведеться розробка системи. Засобами інформаційного забезпечення досягається інформаційна сумісність даних різних інформаційних систем. Для організації їх взаємодії дані повинні бути описані однотипово.

Для формування і ведення баз даних використовується спеціальний комплекс програм — система керування базами даних (СКБД). Цей комплекс забезпечує створення логічної структури бази даних, введення та редагування бази даних, пошук і збереження даних, доступ до окремих записів, полів тощо.

Формування інформаційного забезпечення є однією із найвідповідальніших ланок створення інформаційної системи. Викону-ються роботи з впорядкування інформаційної бази, розроблення єдиної системи класифікації і кодування інформації, визначення методів вводу, збереження, накопичення і відновлення інформації.

Математичне забезпечення — сукупність математичних методів, моделей, алгоритмів обробки інформації. Воно містить засоби забезпечення, методи вибору розв'язування, оптимізації, математичного програмування, математичної статистики, технічну документацію. Технічна документація з цього виду забезпечення передбачає опис задач, завдання з алгоритмізації, економіко-математичні моделі задач і програм для реалізації функціонування інформаційних систем, описи пакетів прикладних програм. Спеціалісти математичного забезпечення займаються постановкою задач та використання математичних аналітичних і числових методів.

Програмне забезпечення включає узгоджений комплекс програм, що уможливлює локальне і спільне функціонування всіх компонентів інтегрованого середовища, а також сукупність програм для стабільної роботи комплексів технічних засобів. До складу програмного забезпечення входять системні і прикладні програмні засоби, а також інструктивно-методичні матеріали. До системних належать програми, які забезпечують діяльність комп'ютерних операційних систем. Вони призначені для організації обчислювального процесу і вирішення стандартних завдань обробки інформації. Зокрема це операційні системи, антивірусне програмне забезпечення, командно-файлові процесори та ін. Прикладні програмні засоби — це сукупність програм, що розробляються при створенні системи конкретного функціонального призначення. Вони включають пакети прикладних програм для збору даних і їх обробки при вирішенні функціональних завдань. До таких засобів належать системи:

— підготовки документів (текстових, табличних, графічних);
— оброблення фінансово-економічної інформації;
— керування базами даних;
— підтримки прийняття рішень і експертні системи тощо.

Технічне забезпечення — це технічні засоби, які призначені для роботи інформаційної системи, а також відповідна документація на ці засоби і технологічні процеси. Створення інтегрованого інформаційного середовища пов'язано з розбудовою інформаційної інфраструктури, до складу якої входить:

— апаратне забезпечення (базис усього комплексу технічних засобів, призначений для оброблення і перетворення різних видів інформації);

— програмне забезпечення;

— засоби зберігання даних;

— комунікаційні засоби для забезпечення передачі інформації в межах системи й обмін даними із зовнішнім середовищем;

— комп'ютерна мережа.

Ергономічне забезпечення охоплює сукупність методів і засобів, що використовуються на різних етапах розроблення та функціонування системи, і призначені для створення оптимальних і безпечних умов високоефективної безпомилкової роботи персоналу і клієнтів.

Впровадження інформаційних систем застосовується з метою підвищення ефективності обробки інформації за рахунок не тільки опрацювання і збереження рутинної інформації, автоматизації звітів, але і за рахунок принципово нових методів управління, заснованих на моделюванні дій при прийнятті рішень.

Інформаційне сприяння створенню кадрових умов визначає необхідність отримання, фіксації та використання такої інформації: штатного розкладу та посадових обов'язків працівників; анкетних даних на кожного працівника; графіка підвищення кваліфікації працівників, графіка атестації працівників; бібліографічних та анотованих списків літератури; картотеки педагогічного досвіду; індивідуальних планів роботи викладачів, планів самоосвіти.

Варто відмітити, що наразі не існує спеціалізованих інформаційних систем для обліку, моніторингу процесу підвищення кваліфікації на рівні навчального закладу. З метою зменшення кількості документів в ході підвищення кваліфікації та атестації викладачів, які необхідно відстежувати та обробляти, пропонується створити автоматизовану інформаційну систему обліку підвищення кваліфікації та атестації науково-педагогічних та педагогічних працівників закладу освіти.

Мета створення інформаційної системи полягає у такому:

— забезпечення ефективності процесу підвищення кваліфікації та атестації педагогічних та науково-педагогічних працівників закладу освіти за допомогою інформаційної підтримки задач прогнозування і планування організації цього процесу;

— зручний автоматизований контроль за процесом підвищення кваліфікації педагогічних та науково-педагогічних працівників закладу освіти;

— контроль за виконанням нормативів по кількості годин підвищення кваліфікації педагогічних та науково-педагогічних працівників закладу освіти за розрахунковий період;

— представлення розрахункових показників (кількість годин підвищення по кожному викладачу, менше 150 годин підвищення, тощо);

— формування звітів щодо результатів підвищення кваліфікації та атестації педагогічних та науково-педагогічних працівників закладу освіти;

— аналітика результатів за напрямами щодо підвищення кваліфікації та атестації педагогічних та науково-педагогічних працівників закладу освіти.

Визначення діапазону дії і меж застосувань бази даних:

— навчально-методичні кабінети закладів освіти.

Визначення складу користувачів і областей застосування:

— працівники навчально-методичних відділів закладів освіти.

Визначення представлень користувачів, що підтримуються БД:

— зберігання і обробка інформації стосовно планування та результатів підвищення кваліфікації викладачів коледжу.

Вимоги користувачів:

— автоматизація обліку і контролю процесу підвищення кваліфікації та атестації педагогічних та науково-педагогічних працівників закладу освіти;

— автоматизація планування, фіксування і контролю підвищення кваліфікації педагогічних та науково-педагогічних працівників закладу освіти;

— автоматизація формування показників підвищення кваліфікації та атестації педагогічних та науково-педагогічних працівників закладу освіти.

Вимоги до інформаційної системи:

Система повинна забезпечувати можливість виконання таких функцій:

— ініціалізацію системи (введення інформації про педагогічних та науково-педагогічних працівників закладу освіти, перелік баз для підвищення кваліфікації і т. п.);

— введення і корекцію інформації про педагогічних та науково-педагогічних працівників закладу освіти та результати їх підвищення кваліфікації;

— одержання відомостей про підвищення кваліфікації та атестації педагогічних та науково-педагогічних працівників закладу освіти.

Система повинна забезпечувати можливість виконання таких запитів:

Запит, що виконує вибірку даних та повертає список педагогічних та науково-педагогічних працівників закладу освіти певної категорії (вибір здійснюється через форму «Пошук за категорією»).

Запит, що виконує вибірку даних та повертає список педагогічних та науково-педагогічних працівників закладу освіти певної циклової комісії/кафедри.

Запит, що повертає список педагогічних та науково-педагогічних працівників закладу освіти із педагогічним званням «викладач-методист».

Запит, що повертає список педагогічних та науково-педагогічних працівників закладу освіти в із науковим ступенем/вченим званням «доктор наук».

Запит, що виконує підрахунок загальної кількості годин з підвищення кваліфікації педагогічних та науково-педагогічних працівників закладу освіти.

Запит, що повертає список педагогічних та науково-педагогічних працівників закладу освіти із вченим званням «кандидат наук».

Запит, що виконує вибірку даних та повертає інформацію за датами підвищення.

Запит, що виконує вибірку даних та повертає інформацію результатів підвищення кваліфікації педагогічних та науково-педагогічних працівників закладу освіти по конкретному педагогічному та науково-педагогічному працівнику закладу освіти.

Запит, що виконує вибірку даних за роком атестації.

Запит, що виконує вибірку даних викладачів, які по підвищенню мають менше 150 годин за 5 років.

Вихідними даними є:

— інформаційні картки педагогічних та науково-педагогічних працівників закладу освіти з їх особистою інформацією;

— атестаційні картки педагогічних та науково-педагогічних працівників закладу освіти;

— плани підвищення кваліфікації педагогічних та науково-педагогічних працівників закладу освіти;

— поточні відомості про підвищення кваліфікації педагогічних та науково-педагогічних працівників закладу освіти;

— підсумки підвищення;

— різноманітні звіти та плани.

Вимоги до надійності, вимоги до забезпечення надійного функціонування програми:

— передбачити контроль введення інформації;

— передбачити блокування некоректних дій користувача при роботі із системою;

— забезпечити цілісність збереженої інформації.

Процес проектування ІС являє собою послідовність переходів від неформального мовного опису інформаційної структури предметної області до формалізованого опису об'єктів предметної області в термінах деякої моделі. Проектування ІС складається з таких етапів:

— системний аналіз предметної області;

— концептуальне проектування;

— логічне проектування;

— фізичне проектування.

Системний аналіз передбачає мовний опис реальних об'єктів предметної області, визначення зв'язків між об'єктами, дослідження характеристик об'єктів і зв'язків. Результати дослідження використовуються при концептуальному проектуванні БД.

Для визначення складу і структури предметної області застосовуються або функціональний, або предметний підходи. Функціональний підхід застосовує рух «від задач» і використовується у тих випадках, коли заздалегідь відомі функції майбутніх користувачів БД, а також відомі всі задачі, для інформаційних потреб яких створюються БД. В цьому випадку на основі виробничих документів, опитувань замовників можна чітко визначити мінімальний набір об'єктів предметної області та їх взаємозв'язок.

Предметний підхід застосовується у тому випадку, коли інформаційні потреби майбутніх користувачів чітко не визначені. В цьому випадку не можна чітко визначити мінімальний набір об'єктів предметної області. В опис предметної області включаються об'єкти та зв'язки, які є найбільш характерними та найбільш суттєвими для неї. БД називається предметною і може використовуватися для розв'язання задач, які заздалегідь не визначені.

У практичній діяльності використовується комплексний підхід, який, з одного боку, дозволяє розв'язувати конкретні інформаційні та функціональні задачі, а з іншого боку — враховує можливість додавання нових застосувань.

У загальному випадку існують два підходи до проектування БД: низхідне проектування і висхідне проектування.

Низхідне проектування починається з визначення наборів даних, потім визначаються елементи даних для кожного з таких наборів. Цей процес включає в себе ідентифікацію різних типів сутностей і визначення атрибутів кожної сутності. Низхідне проектування включає операції декомпозиції, що передбачає заміну вихідної множини відношень, що входять в схему БД, іншою множиною відношень, які є проекціями вихідних відношень.

Цей підхід рекомендується застосовувати у тих випадках, коли кількість, різноманітність та складність сутностей, зв'язків і транзакцій значна за розмірами. Найбільш поширеними моделями для цього проектування є моделі «сутність — зв'язок» (ER-моделі, Entity-Relationship model).

Висхідне проектування починається з виявлення елементів даних, які потім групуються в набори даних. Спочатку визначаються атрибути, які потім об'єднуються в сутності. Висхідне проектування включає операції синтезу, що передбачає виконання компоновки із заданої множини функціональних залежностей між об'єктами предметної області вихідних відношень схеми БД.

Цей підхід рекомендується застосовувати у тому випадку, якщо розробляється невелика БД з незначною кількістю об'єктів, атрибутів і транзакцій.

Концептуальне проектування полягає у створенні концептуальної моделі, яку відображає концептуальна схема БД. На цьому етапі визначаються об'єкти, зв'язки між об'єктами, атрибути, ключові атрибути.

Логічне проектування полягає у створенні логічної моделі на основі вибраної моделі даних. На цьому етапі необхідно вже знати, яка СУБД буде застосовуватися в системі (ієрархічна, мережна, реляційна, об'єктно орієнтована). Для перевірки вірності логічної моделі застосовується нормалізація. Крім того, логічна модель перевіряється на умову забезпечення всіх транзакцій користувачів.

Фізичне проектування полягає в описі засобів фізичної реалізації логічного проекту БД. Фізичні моделі визначають засоби розміщення даних в середовищі зберігання і засоби доступу до цих даних, які підтримуються на фізичному рівні.

Аналіз предметної області. Відповідно до законодавства педагогічні та науково-педагогічні працівники зобов'язані постійно під-вищувати свій професійний і загальнокультурний рівень та педагогічну майстерність. Але такий обов'язок урівноважується правом педагогічних працівників на вільний вибір освітніх програм, форм навчання, закладів освіти, установ і організацій, інших суб'єктів освітньої діяльності, що здійснюють підвищення кваліфікації. Пошук інформації про підвищення кваліфікації педагогічний працівник може здійснювати у будь-який спосіб — безпосередньо на сайтах суб'єктів підвищення кваліфікації, на різноманітних інформаційних чи спеціальних ресурсах, у тематичних групах, через запит необхідної інформації безпосередньо у суб'єкта підвищення кваліфікації тощо.

Заклади освіти, в яких працюють педагогічні та науково-педагогічні працівники, сприяють їхньому професійному розвитку та підвищенню кваліфікації. Виходячи з цього, Міністерство освіти і науки України покладає на керівництво усіх закладів освіти обов'язок здійснювати контроль та максимально активно сприяти професійному розвитку та підвищенню кваліфікації педагогічних та науково-педагогічних працівників на засадах, визначених законодавством, та за процедурами, визначеними Порядком підвищення кваліфікації педагогічних та науково-педагогічних працівників, затвердженим постановою Кабінету Міністрів України від 21 серпня 2019 р. № 800 «Деякі питання підвищення кваліфікації педагогічних та науково-педагогічних працівників». Активна підтримка педагогічних працівників адміністрацією закладу — роз'яснення нової процедури підвищення кваліфікації, допомога (у разі потреби) у визначенні компетентностей, удосконалення яких педагогічні працівники потребують найбільше, тощо — є запорукою формування педагогіки партнерства в закладі освіти, його сталого розвитку, збереження здорового мікроклімату в колективі, покращення діяльності закладу освіти та якості освіти загалом.

Для закладів вищої та фахової передвищої освіти підвищення кваліфікації є обов'язковою складовою системи забезпечення якості освіти. Як правило, підвищення кваліфікації здійснюється за програмою підвищення кваліфікації, у тому числі шляхом участі у семінарах, практикумах, тренінгах, вебінарах, майстер-класах тощо. Окремі види діяльності науково-педагогічних працівників (участь у програмах академічної мобільності, наукове стажування, самоосвіта, здобуття наукового ступеня, вищої освіти) можуть бути визнані підвищенням кваліфікації. Процедура зарахування окремих видів ді-яльності, їх результатів та обсяг підвищення кваліфікації науково-педагогічних працівників визначаються вченими (педагогічними) радами відповідних закладів освіти. Науково-педагогічні працівники самостійно обирають форми, види, напрями та суб'єктів підвищення кваліфікації.

Законодавство встановлює різні вимоги до періодичності та обсягів підвищення кваліфікації науково-педагогічних працівників, які здійснюють свою професійну діяльність на різних рівнях освіти. Так, науково-педагогічні працівники закладів фахової передвищої освіти зобов'язані підвищувати свою кваліфікацію щорічно, а загальна кількість академічних годин для підвищення кваліфікації упродовж п'яти років не може бути меншою за 120 годин, з яких певна кількість годин обов'язково має бути спрямована на вдосконалення знань, вмінь і практичних навичок у роботі зі студентами з особливими освітніми потребами та дорослими студентами. Крім цього, обсяг щорічного підвищення кваліфікації науково-педагогічних працівників закладів фахової передвищої освіти встановлюється засновником (або уповноваженим ним органом).

Обсяг (тривалість) підвищення кваліфікації науково-педагогічних працівників установлюється в кредитах Європейської кредитної трансферно-накопичувальної системи (один кредит ЄКТС становить 30 годин) за накопичувальною системою і для науково-педагогічних працівників закладів вищої та післядипломної освіти протягом п'яти років не може бути меншим ніж шість кредитів ЄКТС. Накопичувальна система передбачає можливість враховувати обсяги підвищення кваліфікації чи інших видів професійного удосконалення, які визнаються підвищенням кваліфікації і які здійснювалися науково-педагогічним працівником будь-коли впродовж міжатестаційного періоду. Науково-педагогічних працівникам закладів освіти вперше, призначеним на посади: керівника, заступника керівника закладу вищої, післядипломної освіти, керівника, заступника керівника факультету, інституту чи іншого структурного підрозділу, керівника кафедри, завідувача аспірантури, докторантури закладу вищої освіти — підвищення кваліфікації відповідно до займаної посади протягом двох перших років роботи є обов'язковим. Обсяги такого підвищення кваліфікації визначаються вченою (педагогічною) радою відповідного закладу освіти.

Результати підвищення кваліфікації враховуються під час: проведення атестації педагогічних працівників закладів вищої та післяди-пломної освіти, обрання на посаду за конкурсом чи укладення трудового договору з науково-педагогічними працівниками.

Атестація педагогічних працівників — це система заходів, спрямована на всебічне комплексне оцінювання їх педагогічної діяльності, за якою визначаються відповідність педагогічного працівника займаній посаді, рівень його кваліфікації, присвоюється кваліфікаційна категорія, педагогічне звання. Атестація може бути черговою або позачерговою. Чергова атестація здійснюється один раз на п'ять років. Умовою чергової атестації педагогічних працівників є обов'язкове проходження не рідше одного разу на п'ять років підвищення кваліфікації на засадах вільного вибору форм навчання, програм і навчальних закладів. Для організації та проведення атестації педагогічних працівників у навчальних та інших закладах, органах управління освітою щороку до 20 вересня створюються атестаційні комісії I, II і III рівнів [18].

У процесі вивчення професійної діяльності керівних кадрів навчальних та інших закладів атестаційна комісія з'ясовує:

— виконання програми розвитку навчального закладу та результати інноваційної діяльності;

— стан організації навчальної та виховної роботи, додержання вимог державних освітніх стандартів;

— результати державної атестації навчального закладу;

— результати перевірок, проведених Державною інспекцією навчальних закладів, місцевими органами управління освітою та іншими органами державного нагляду (контролю);

— додержання вимог до забезпечення безпечних та нешкідливих умов навчання учнів;

— підсумки моніторингу роботи з педагогічним колективом та іншими працівниками навчального закладу;

— ефективність взаємодії з громадськими організаціями та органами шкільного самоврядування;

— додержання педагогічної етики, моралі;

— звіти керівника про свою роботу на загальних зборах (конференціях) колективу навчального закладу;

— аналіз розгляду звернень громадян.

Визначення схеми інформаційних потоків автоматизованої інформаційної системи обліку підвищення кваліфікації викладачів.

Потік інформації базується на потоці паперових або електронних документів. Залежно від цього його можна виміряти або кількістю оброблених та переданих одиниць паперових документів, або загальною кількістю рядків у цих документах.

Автоматизована інформаційна система обліку підвищення кваліфікації викладачів отримує дані від джерела інформації. Ці дані надсилаються для зберігання або обробки в системі, а потім надсилаються споживачу (рис. 14).

Споживачем може бути людина, пристрій чи інша інформаційна система. Зворотний зв'язок може бути встановлений між споживачем та самою інформаційною системою (від споживача до одержувача інформації).

В нашому випадку джерелом інформації є педагогічні та науково-педагогічні працівники закладу освіти, а споживач інформації — керівні структури.

Автоматизована інформаційна система обліку підвищення кваліфікації викладачів включає вхідну інформацію (дані викладача, сертифікати, дипломи) та вихідну інформацію (звіти, плани, розрахунки) і функціонує в інформаційному середовищі. За допомогою засобів обробки інформації вхідна інформація перетворюється на вихідну, з якою потім може працювати завідувач методичного кабінету (рис. 15).

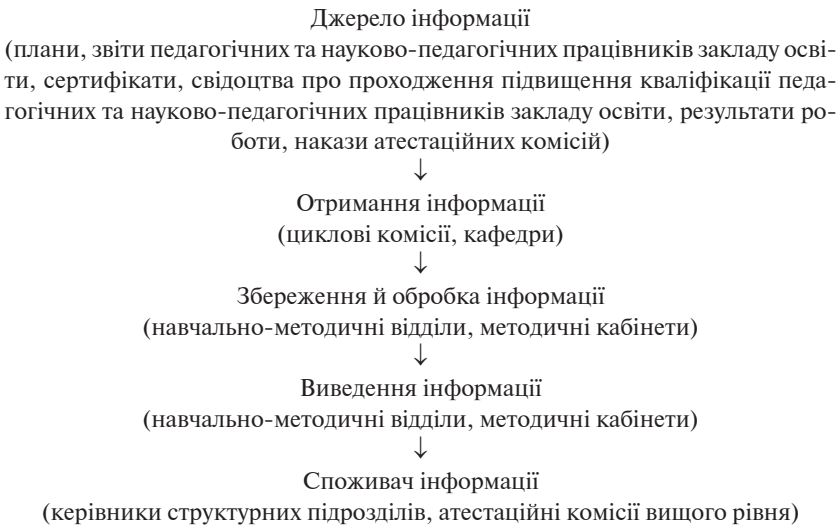

Джерело інформації
(плани, звіти педагогічних та науково-педагогічних працівників закладу освіти, сертифікати, свідоцтва про проходження підвищення кваліфікації педагогічних та науково-педагогічних працівників закладу освіти, результати роботи, накази атестаційних комісій)
↓
Отримання інформації
(циклові комісії, кафедри)
↓
Збереження й обробка інформації
(навчально-методичні відділи, методичні кабінети)
↓
Виведення інформації
(навчально-методичні відділи, методичні кабінети)
↓
Споживач інформації
(керівники структурних підрозділів, атестаційні комісії вищого рівня)

Рис. 15. Ланцюг надходження інформації

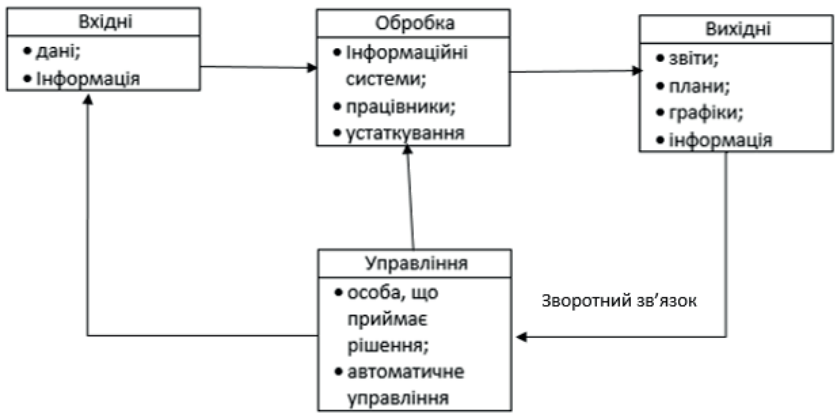

Рис. 16. Структурна схема інформаційної системи

Працівник методичного кабінету, взаємодіючи з інтерфейсом ІС, взаємодіє з базою даних і може редагувати будь-які дані, створювати звіти чи надавати інформацію для подальшого розгляду, опрацювання та корегування керівнику навчально-методичного кабінету.

Концептуальне моделювання. Для створення концептуальної та логічної моделі бази даних ІС була обрана реляційна модель, введена Е. Ф. Коддом в 1970 році як спосіб зробити системи управління базами даних більш незалежними від якогось конкретного додатка. Це математична модель визначена в термінах логіки предикатів і теорії множин, а використовувані нею системи застосовуються в системах мейнфреймів, персональних комп'ютерів і мікрокомп'ютерів. Продукти, які зазвичай називаються реляційними базами даних, фактично реалізують модель, яка є лише наближенням до математичної моделі, визначеної Коддом. Три ключових терміна широко використовуються в моделях реляційних баз даних: стосунки, атрибути і домени. Стосунок — це таблиця зі стовпцями і рядками. Іменовані стовпці стосунка називаються атрибутами, а домен — це набір значень, які можуть приймати атрибути.

Основною структурою даних реляційної моделі є таблиця, в якій інформація про конкретний об'єкт (скажімо, працівника) представлена в рядках (також званих кортежами) і шпальтах. Таким чином, «ставлення» в «реляційній базі даних» відноситься до різних таблиць в базі даних. Стосунок являє собою набір кортежів. У стовпчиках перераховані різні атрибути об'єкта (наприклад, ім'я співробітни-ка, адреса або номер телефону), а рядок є фактичним екземпляром об'єкта (конкретного співробітника), який представлений відношенням. В результаті кожен кортеж таблиці Employee представляє різні атрибути одного співробітника. Всі стосунки (і, отже, таблиці) в реляційній базі даних повинні дотримуватися деяких основних правил, які можна кваліфікувати як стосунки. По-перше, впорядкування стовпців в таблиці несуттєво. По-друге, в таблиці не може бути однакових кортежів або рядків. І, по-третє, кожен кортеж буде містити одне значення для кожного з його атрибутів.

Реляційна база даних містить кілька таблиць, кожна з яких схожа на таку в «плоскій» моделі бази даних. Одна із сильних сторін реляційної моделі полягає в тому, що в принципі будь-яке значення, що має місце в двох різних записах (що належать до однієї таблиці або до різних таблицях), має на увазі взаємозв'язок між цими двома записами. Проте для забезпечення явних обмежень цілісності відносин між записами в таблицях також можуть бути визначені явно, шляхом ідентифікації відносин батько — дитина, що характеризуються привласненням потужності (1: 1, (0) 1: М, М: М). Таблиці також можуть мати призначений єдиний атрибут або набір атрибутів, які можуть діяти як «ключ», і можуть використовуватися для однозначної ідентифікації кожного кортежу в таблиці. Ключ, який може використовуватися для однозначної ідентифікації рядка в таблиці, називається первинним ключем. Ключі зазвичай використовуються для об'єднання даних з двох або більше таблиць. Наприклад, таблиця Employee може містити стовпець з ім'ям Location, який містить значення, відповідне ключу таблиці Location. Ключі також важливі при створенні індексів, які полегшують швидкий пошук даних з великих таблиць. Будь-який стовпець може бути ключем, або декілька стовпців можуть бути згруповані разом в складений ключ. Немає необхідності заздалегідь визначати всі ключі. Стовпець можна використовувати в якості ключа, навіть якщо він спочатку не був призначений для цієї мети. Ключ, який має зовнішній, реальний сенс (наприклад, ім'я людини, ISBN книги або серійний номер автомобіля), іноді називають «природним» ключем. Якщо жодний природний ключ не підходить (подумайте про багатьох людей на ім'я Іван), може бути призначений довільний або сурогатний ключ (наприклад, вказуючи ідентифікаційні номери співробітників). На практиці більшість баз даних мають як згенеровані, так і природні ключі, оскільки згенеровані ключі можуть використовуватися всередині, щоб створювати зв'язки між рядками, які не можуть зламатися, в той час як природні ключі можна використовувати менш надійно, для пошуку і для інтеграції з іншими базами даних (наприклад, записи в двох незалежно розроблених базах даних можуть бути співставлені номером соціального страхування, за винятком випадків, коли номери соціального страхування невірні, відсутні або змінені).

З концептуального проектування починається створення концептуальної схеми бази даних, в основі якої лежить концептуальна модель даних. Концептуальна модель представляє загальний погляд на дані. Розрізняють два головних підходи до моделювання даних при концептуальному проектуванні: семантичні моделі; об'єктні моделі.

Семантичні моделі головну увагу приділяють структурі даних. Найбільш поширеною семантичною моделлю є модель «сутність — зв'язок» (Entity Relationship model, ER-модель). ER-модель складається із сутностей, зв'язків, атрибутів, доменів атрибутів, ключів. Моделювання даних відображає логічну структуру даних, так само, як блок-схеми алгоритмів відображають логічну структуру програми.

Об'єктні моделі головну увагу приділяють поведінці об'єктів даних і засобам маніпуляції даними. Головне поняття таких моделей — об'єкт, тобто сутність, яка має стан і поведінку. Стан об'єкта визначається сукупністю його атрибутів, а поведінка об'єкта визначається сукупністю операцій, специфікованих для нього.

Зближення цих моделей реалізується в розширеному ER-моделюванні (Extended Entity Relationship model, EER-модель).

ER-моделювання являє собою низхідний підхід до проектування бази даних, який починається з визначення найбільш важливих даних, які називаються сутностями (entities), і зв'язків (relationships) між даними, які повинні бути представлені в моделі. Потім в модель заноситься інформація про властивості сутностей і зв'язків, яка називається атрибутами (attributes), а також всі обмеження, які відносяться до сутностей, зв'язків і атрибутів. ER-модель дає графічне представлення логічних об'єктів і їх відношень в структурі БД. Сутність дозволяє моделювати клас однотипних об'єктів. Сутність має унікальне ім'я у межах системи, що моделюється. Оскільки сутність відповідає деякому класу однотипних об'єктів, то передбачається, що в системі існує багато екземплярів даної сутності. Об'єкт, якому відповідає сутність, має набір атрибутів, які характеризують його властивості. При цьому набір атрибутів повинен бути таким, щоб можна було розрізняти конкретні екземпляри сутності.

Для автоматизованої інформаційної системи обліку підвищення кваліфікації викладачів відповідно до предметної області визначено сутності: «Викладач», «Підвищення кваліфікації», «Атестація», «Напрям підвищення», «Бази підвищення кваліфікації», «Результати підвищення».

Сутність «Викладач» (атрибути: табельній номер, фото, ПІБ, циклова комісія/кафедра, номер телефону, E-mail, кваліфікаційна категорія, педагогічне звання, науковий ступінь/вчене звання, спеціальність за дипломом).

Сутність «Атестація» (атрибути: № п/п, ПІБ, кваліфікаційна категорія, рік попередньої атестації, запланована атестація, категорія, статус категорії, педагогічне звання, статус педагогічного звання).

Сутність «Заплановане підвищення» (атрибути: номер підвищення, табельний номер, ПІБ, код курсу, місце проходження, форма підвищення, вид підвищення, дата останнього підвищення, дата початку, дата закінчення, відмітка про виконання).

Сутність «Інформаційна база підвищення кваліфікації» (атрибути: код курсу, емблема, назва, адреса, телефон, E-mail, сайт).

Сутність «Напрями та форми підвищення кваліфікації» (атрибути: код напряму, назва напряму підвищення кваліфікації).

Сутність «Результати підвищення» (атрибути: код документу, тип документу, виданий кому, виданий ким, напрям, тема підвищення, дата видачі, кількість годин, кількість кредитів, додаткове (незаплановане).

Розроблений концептуальний проект було перевірено на надлишковість та на відповідність транзакціям користувачів.

Перевірка на надлишковість передбачає перевірку ER-моделі з метою виявлення надлишкових даних і вилучення їх, в тому випадку, якщо вони визначені. Надлишкові зв'язки виявляються в тому, що між двома сутностями є декілька шляхів і вони дублюють один іншого (це не відноситься до зв'язків, які представляють різні асоціації).

Перевірка моделі на відповідність транзакціям користувачів виконується на основі таких підходів:

— перевірка того, чи представляє модель всю інформацію (сутності, атрибути, зв'язки), яка необхідна для кожної транзакції;

— перевірка по ER-діаграмі маршруту кожної транзакції.

Перевірка моделі на надлишковість та на відповідність транзакціям користувачів дозволяє зробити висновок, що концептуальний проект відповідає всім необхідним вимогам.

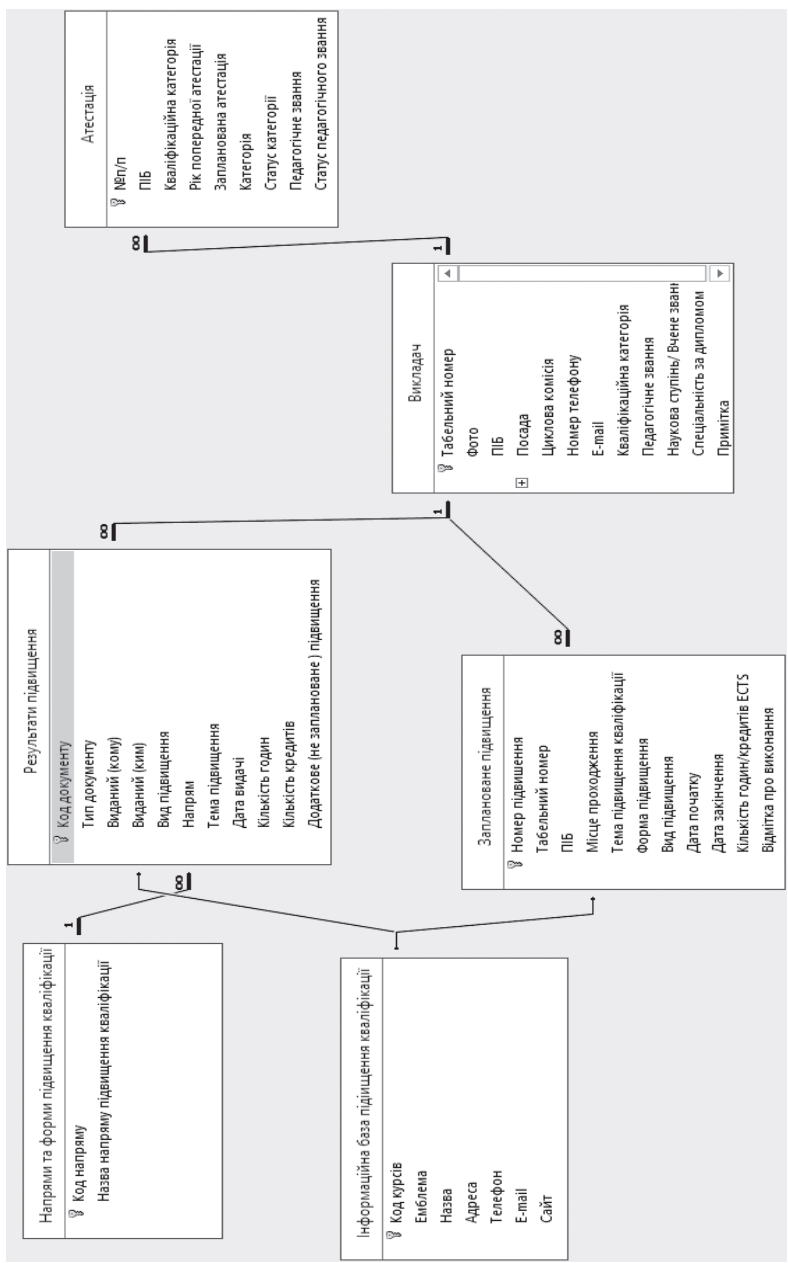

Рис. 17. Концептуальна модель (ER-модель)

**434**

Слід звернути увагу на те, що розроблений концептуальний проект не є єдиним проектом, який відповідає поставленій задачі. Можливі варіанти розробки системи із застосуванням інших зв'язків між сутностями або із застосуванням розширеної ER-моделі.

Застосування ER-діаграм дозволяє забезпечити просте і наочне уявлення про головні логічні об'єкти БД і про зв'язки, які між цими об'єктами існують. Також до переваг ER-діаграми слід віднести те, що вони добре інтегрують з реляційною моделлю.

Між сутностями встановлюються зв'язки, які вказують, яким чином сутності співвідносяться або взаємодіють між собою. Розрізняють такі зв'язки:

— між двома сутностями (бінарний зв'язок);
— між трьома сутностями (тернарний зв'язок);
— між N сутностями (N-арний зв'язок);
— між однією сутністю (рекурсивний зв'язок).

Найбільш поширеними є бінарні зв'язки. Зв'язок показує, яким чином екземпляри сутностей пов'язані між собою. Бінарні зв'язки бувають:

— 1:1 (один до одного);
— 1:M (один до багатьох);
— N:M (багато до багатьох).

Сутності «Інформаційна база підвищення кваліфікації» та «Заплановане підвищення» пов'язані зовнішнім ключем по полю «Код курсу» зв'язком «один до багатьох», бо одна база підвищення може бути декілька раз запланована у різних викладачів.

Сутності «Викладач» та «Заплановане підвищення» пов'зані зовнішнім ключем по полю «Табельний номер» та «ПІБ» (тобто таблиця «Викладач» є головною, а таблиця «Заплановане підвищення» — підпорядкованою) зв'язком «один до багатьох», бо для одного викладача може бути заплановано декілька підвищень. В таблиці «Заплановане підвищення» є штучний ключ «Номер підвищення».

Сутності «Викладач» та «Результати підвищення» зв'зані зовнішнім ключем по полю «Табельний номер» та «Виданий (кому)» (тобто таблиця «Викладач» є головною, а таблиця «Результати підвищення» — підпорядкованою) зв'язком «один до багатьох», бо один викладач може мати багато документів (результатів) підвищень. В таблиці «Результати підвищення» є штучний ключ «Код документа».

Сутності «Напрям та форми підвищення кваліфікації» та «Результати підвищення» пов'зані зовнішнім ключем по полю «Код напря-

му» та «Напрям» (таблиця «Напрям та форми підвищення» — головна, а таблиця «Результати підвищення» — підпорядкована) зв'язком «один до багатьох», бо один напрям буде багато раз обиратись викладачами.

Сутності «Напрям та форми підвищення кваліфікації» та «Результати підвищення» пов'зані зовнішнім ключем по полю «Код напряму» та «Напрям» (таблиця «Напрям та форми підвищення» — головна, а таблиця «Результати підвищення» — підпорядкована) зв'язком «один до багатьох», бо один напрям буде багато раз обиратись викладачами.

Сутності «Викладач» та «Атестація» пов'зані зовнішнім ключем по полю «Табельний номер» та «ПІБ» (таблиця «Викладач» — головна, а таблиця «Атестація» — підпорядкована) зв'язком «один до багатьох», бо один викладач буде неодноразово атестуватися (через кожні 5 років). У таблиці «Атестація» доцільно додати поле-лічильник «№ п/п», яке буде штучним ключем.

**Логічне проектування.** Логічне проектування виконується для певної моделі даних. Для реляційної моделі даних логічне проектування полягає у створенні реляційної схеми, визначенні числа і структури таблиць (табл. 1–6), формуванні запитів до бази даних (рис. 17), визначенні типів звітних документів (рис. 18), розробці алгоритмів обробки інформації (рис. 20), створенні форм для вводу і редагування даних в базі даних (рис. 19) і рішенні ряду інших задач. Концептуальні моделі за певними правилами перетворюються в логічні моделі даних. Коректність логічних моделей перевіряється за допомогою правил нормалізації, які дозволяють переконатися в структурній узгодженості, логічній цілісності і мінімальній надлишковості прийнятої моделі даних. Модель також перевіряється з метою виявлення можливостей виконання транзакцій, які будуть задаватися користувачами.

Створений на попередніх етапах набір відношень логічної моделі БД повинен бути перевірений на коректність об'єднання атрибутів у кожному відношенні. Перевірка виконується шляхом застосування до кожного відношення процедури послідовної нормалізації. Нормалізація гарантує, що отримана модель не буде мати протиріч і буде мати мінімальну надлишковість. Атрибути в результаті нормалізації будуть згруповані відповідно до існуючих між ними логічних зв'язків. Для забезпечення коретності логічної моделі, у разі виявлення відношень, які не відповідають вимогам нормалізації, необхідно повернутися на попередні етапи проектування і перебудувати помилково створені елементи моделі.

**Таблиця містить особисту інформацію про викладача і має наступні поля**

| Назва поля бази даних | Тип даних | Розмірність | Призначення |
|---|---|---|---|
| Табельний номер | Числовий | Довге ціле | Ключове поле таблиці |
| Фото | Поле об'єкта OLE | — | Фото викладача |
| ПІБ | Короткий текст | 60 | Призвіще, ім'я, по батькові викладача |
| Посада | Короткий текст | 20 | Для відображення посади викладача |
| Циклова комісія | Короткий текст | 255 | Циклова комісія, на якій працює викладача |
| Номер телефону | Короткий текст | 20 | Номер телефону |
| E-mail | Короткий текст | 60 | Електронна адреса викладача |
| Кваліфікаційна категорія | Короткий текст | 20 | Кваліфікаційна категорія, яку має викладач |
| Педагогічне звання | Короткий текст | 20 | Педагогічне звання викладача |
| Науковий ступінь / вчене звання | Короткий текст | 20 | Науковий ступінь / вчене звання викладача |
| Спеціальність за дипломом | Короткий текст | 255 | Спеціальність викладача за дипломом |
| Примітка | Короткий текст | 255 | Для додаткової інформації |

**Атестація містить дані про атестацію викладача і має наступні поля**

| Назва поля бази даних | Тип даних | Розмірність | Призначення |
|---|---|---|---|
| № п/п | Лічильник | Довге ціле | Ключове поле таблиці |
| ПІБ | Текст (числовий) | 60 | ПІБ викладача (дані беруться з таблиці «Викладач») |
| Кваліфікаційна категорія | Короткий текст | 30 | Дані про кваліфікаційну категорію, яку на даний момент має викладач |
| Рік попередньої атестації | Дата та час | Короткий формат дати | Дата проходження попередньої атестації |
| Запланована атестація | Дата та час | Короткий формат дати | Дата проходження запланованої атестації |
| Категорія | Короткий текст | 60 | Обрана категорія |
| Статус категорії | Короткий текст | 60 | Отримання, підтвердження або підвищення категорії |
| Педагогічне звання | Короткий текст | 60 | Обране педагогічне звання |
| Статус педагогічного звання | Короткий текст | 60 | Отримання або підтвердження |

**Заплановане підвищення містить інформацію про плани підвищення кваліфікації викладачів і має наступні поля**

| Назва поля бази даних | Тип даних | Розмірність | Призначення |
|---|---|---|---|
| Номер підвищення | Лічильник | Довге ціле | Ключове поле таблиці |
| Табельний номер | Числовий (дані беруться з другої таблиці) | Довге ціле | Ід викладача, з яким пов'заний запис |
| ПІБ | Числовий (дані беруться з другої таблиці) | Довге ціле | ПІБ викладача |
| Місце проходження | Числовий (дані беруться з другої таблиці) | Довге ціле | Інформація про місце проходження |
| Форма підвищення | Короткий текст | 20 | Форми підвищення |
| Вид підвищення | Короткий текст | 255 | Види підвищення |
| Дата початку | Дата та час | Короткий формат дати | Запланована дата початку підвищення |
| Дата закінчення | Дата та час | Короткий формат дати | Запланована дата закінчення підвищення |
| Кількість годин/кредитів ЄКТС | Короткий текст | 15 | Запланована кількість годин |
| Відмітка про виконання | Логічний | Так / ні | Пройдене підвищення чи ні |

**Інформаційна база підвищення кваліфікації містить інформацію про бази
для підвищення кваліфікації і має наступні поля**

| Назва поля бази даних | Тип даних | Розмірність | Призначення |
|---|---|---|---|
| Код курсу | Числовий | Довге ціле | Ключове поле таблиці |
| Емблема | Поле об'єкта OLE | - | Емблема бази |
| Назва | Короткий текст | 255 | Назва бази підвищення |
| Адреса | Короткий текст | 80 | Місце розташування бази |
| Телефон | Короткий текст | 50 | Телефонні дані |
| E-mail | Короткий текст | 150 | Електронна адреса |
| Сайт | Гіперпосилання | - | Посилання на сайт |

Таблиця 5

**Напрями та форми підвищення мають наступні поля**

| Назва поля бази даних | Тип даних | Розмірність | Призначення |
|---|---|---|---|
| Код напряму | Числовий | Довге ціле | Ключове поле таблиці |
| Назва напряму підвищення кваліфікації | Короткий текст | Довге ціле | Назва напряму |

Обмеження цілісності запобігають появі в БД суперечливих даних. Вирішення цієї проблеми на стадії проектування полягає у такому:

– наявність обов'язкових і необов'язкових значень даних для атрибутів (NULL, NOT NULL);

– наявність обмежень для доменів атрибутів (визначення області значень або діапазону значень);

– цілісність сутностей (обов'язкова наявність Primary Key в кожному відношенні);

– цілісність посилання (зв'язування таблиць за допомогою Foreign Key);

– обмеження предметної області (бізнес-правила), які реалізуються як засобами БД, так і на рівні застосувань.

Після створення логічної моделі даних реляційна схема аналізується на коректність об'єднання атрибутів в одному відношенні. Перевірка коректності виконується шляхом застосування послідовної нормалізації до кожного з відношень. Метою цієї перевірки є отримання гарантій того, що схема бази даних знаходиться щонайменше в 3-й нормальній формі або в нормальній формі Бойса — Кодда. Якщо ця умова не виконується, то необхідно повернутися на попередні етапи

**Результати підвищення мають наступні поля**

| Назва поля бази даних | Тип даних | Розмірність | Призначення |
|---|---|---|---|
| Код документу | Короткий текст | Довге ціле | Ключове поле таблиці |
| Тип документу | Короткий текст | Довге ціле | Назва типу документу |
| Виданий (кому) | Числовий | Довге ціле | Назва бази підвищення |
| Виданий (ким) | Числовий | Довге ціле | ПІБ викладача |
| Напрям | Числовий | Довге ціле | Назва напряму підвищення |
| Тема підвищення | Короткий текст | 255 | Тема підвищення |
| Дата видачі | Дата та час | Короткий формат дати | Дата видачі документу |
| Кількість годин | Числовий | 10 | Отримана кількість годин за підвищення |
| Кількість кредитів | Числовий | 10 | Отримана кількість кредитів за підвищення |
| Додаткове (не заплановане) | Короткий текст | 255 | Інформація про додаткове підвищення |

Рис. 18. Запити до бази даних

Рис. 19. Звітна документація

Рис. 20. Форми для введення та редагування даних

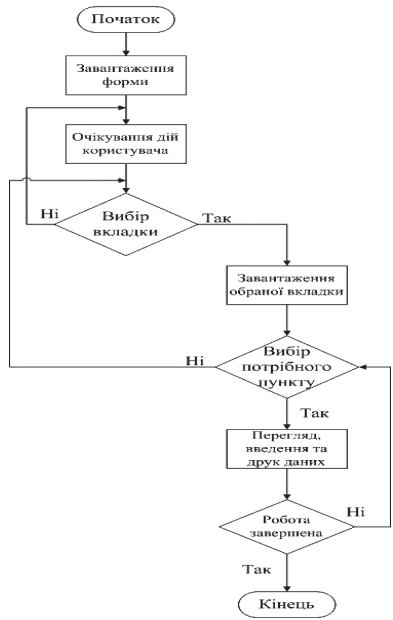

Рис. 21. Алгоритм роботи інформаційної системи

**443**

проектування і перебудувати помилково створені фрагменти моделі. Перевірка логічної моделі бази даних показує, що реляційна схема знаходиться в 4-й нормальній формі й корегування моделі не потрібно.

Після перевірки логічної моделі за допомогою правил нормалізації система аналізується на предмет виконання транзакцій користувачів, які задаються на початкових етапах проектування. У разі неможливості виконання певних транзакцій необхідне корегування моделі бази даних.

**Розробка застосувань.** Застосування — програма або програмна система, яка призначена для рішення деякої сукупності задач в даній предметній області, або яка являє собою типовий інструментарій, що застосовується в різних областях.

На цьому етапі вирішуються такі задачі:

— проектування транзакцій;

— проектування інтерфейсів користувачів.

Транзакція може складатися з декількох операцій по роботі з БД, які переводять БД з одного цілісного стану в інший. Розрізняють транзакції по отриманню певної інформації з БД і транзакції по зміні даних в БД (оновлення, вилучення, додавання). Транзакції також можуть бути змішані.

Інтерфейс користувача — сукупність функціональних компонентів, які забезпечують взаємодію користувача з системою.

Для зручного використання автоматизованої інформаційної системи обліку підвищення кваліфікації викладачів закладу освіти було розроблено максимально зручний та інтуїтивно зрозумілий інтерфейс.

У вкладці «Викладачі» користувач може переглянути, додати або ввести зміни в інформаційну картку викладача, переглянути, роздрукувати та зберегти звіти щодо наукового ступеня/вченого звання викладачів та різної звітної інформації.

На цій вкладці знаходяться три розділи:

— головна інформація;

— інформація щодо наукового ступеня та вченого звання викладачів;

— звітна інформація.

В розділі «Головна інформація» знаходиться інформаційна картка викладача та контактні дані викладачів. Інформаційна картка призначена для створення, редагування зберігання, пошуку головної інформації про викладача.

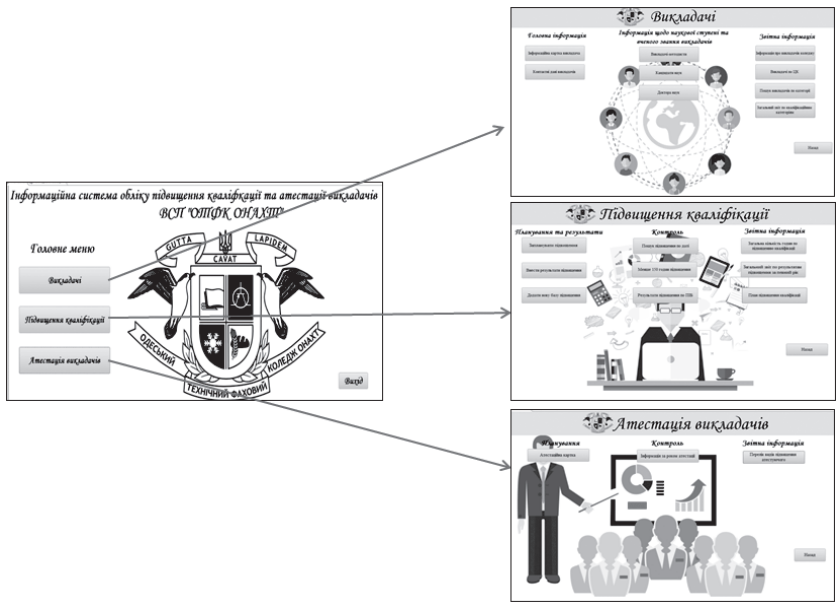

Рис. 22. Вкладки інформаційної системи

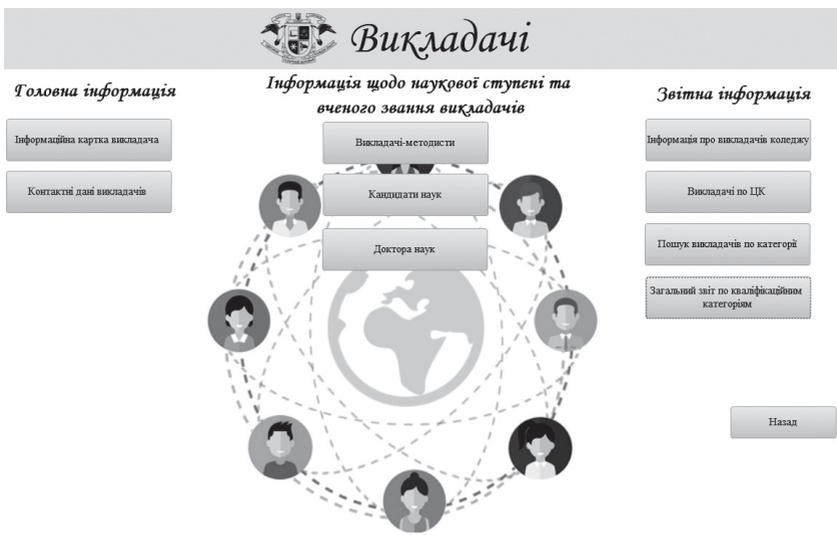

Рис. 23. Вкладка «Викладачі»

В розділі «Інформація щодо наукового ступеня та вченого звання викладачів» знаходяться три кнопки: «Викладачі-методисти», «Кандидати наук» та «Доктори наук», які призначені для вибору певної інформації.

В розділі «Звітна інформація» знаходяться звіти стосовно викладачів коледжу «Викладачі по ЦК», «Пошук викладачів по категорії», «Інформація про викладачів коледжу» або «Загальний звіт за кваліфікаційними категоріями».

У вкладці «Підвищення кваліфікації», користувач може запланувати підвищення, ввести результати підвищення, проконтролювати підвищення певних викладачів, а також переглянути, роздрукувати та зберегти звіти різної звітної інформації стосовно підвищення кваліфікації.

На цій вкладці знаходяться такі розділи:

— планування та результати;

— контроль;

— звітна інформація.

Розділ «Планування та результати» містить в собі три форми, які призначені для введення інформації.

Перша кнопка відкриває форму «Запланувати підвищення» — для планування підвищення кваліфікації викладачів.

Друга кнопка відкриває форму «Внести результати підвищення», яка призначена для внесення та зберігання даних про результати підвищення кваліфікації викладача.

Третя кнопка відкриває форму «Додати нову базу підвищення» — призначена для створення запису про можливі бази підвищення кваліфікації, де зберігаються контактні дані.

Розділ «Контроль» містить три запити для пошуку інформації.

Натиснувши на кнопку «Пошук запланованого підвищення по даті», відкриваємо нову форму для введення діапазону дат, після чого формується звіт.

Друга кнопка відкриває форму «Менше 150 годин підвищення» — ця форма служить для відбору інформації про викладачів, у яких за останні 5 років менше 150 годин підвищення кваліфікації.

Натиснувши на третю кнопку «Результати підвищення по ПІБ», відкриваємо вікно запиту, в яке потрібно ввести ПІБ викладача, інформація про якого нас цікавить, та натискаємо кнопку «ОК» для отримання результату.

Розділ «Підвищення кваліфікації» — це «Звітна інформація». В цьому розділі знаходяться три звіти.

# Підвищення кваліфікації

**Планування та результати**

Запланувати підвищення

Внести результати підвищення

Додати нову базу підвищення

**Контроль**

Пошук підвищення по даті

Менше 150 годин підвищення

Результати підвищення по ПІБ

**Звітна інформація**

Загальна кількість годин по підвищенню кваліфікації

Загальний звіт по результатам підвищення за певний рік

План підвищення кваліфікації

Назад

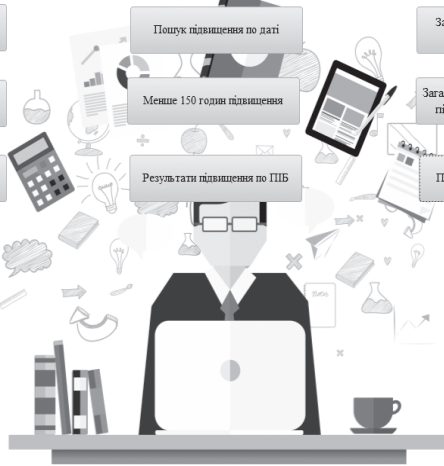

Рис. 24. Вкладка «Підвищення кваліфікації»

# Атестація викладачів

**Планування**

Атестаційна картка

**Контроль**

Інформація за роком атестації

**Звітна інформація**

Перелік видів підвищення атестуючого

Назад

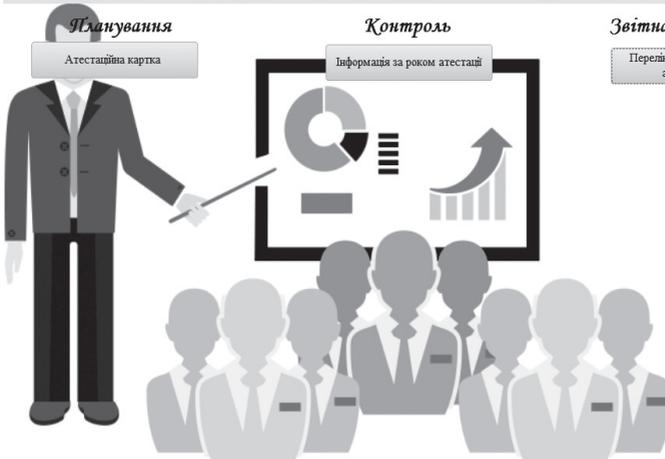

Рис. 25. Вкладка «Атестація викладачів»

В першому звіті «Загальна кількість годин з підвищення кваліфікації» ми бачимо ПІБ викладача, дату підвищення, кількість отриманих годин, кредитів та їх суму.

Другий звіт «Загальний звіт за результатами підвищення за певний рік» побудований за допомогою запиту на вибірку.

Третя кнопка «План підвищення кваліфікації» відкриває сформований план підвищення кваліфікації, дані для якого беруться з форми «Заплановане підвищення» та відбираються за вказаним роком — організовано за допомогою запиту на вибірку.

У вкладці «Атестація викладачів» користувач може створити або змінити атестаційну картку викладача, контролювати інформацію про атестацію за роком, переглянути, роздрукувати та зберегти звіти різної звітної інформації стосовно підвищення кваліфікації.

На цій вкладці знаходить три розділи:

— планування;

— контроль;

— звітна інформація.

У розділі «Планування» знаходиться одна кнопка «Атестаційна картка», призначена для планування атестації викладача.

У розділі «Контроль» знаходиться кнопка «Інформація за роком атестації», натиснувши на неї, бачимо віконце запиту для введення року атестації.

У розділі «Звітна інформація» знаходиться одна кнопка «Перелік видів підвищення атестованого», яка відкриває сформований документ про перелік окремих видів підвищення.

Висновки та перспективи запровадження автоматизованої інформаційної системи обліку підвищення кваліфікації викладачів закладу освіти.

Наразі автоматизована інформаційна система обліку підвищення кваліфікації викладачів пройшла тестування та апробацію в навчально-методичному кабінеті ВСП «ОТФК ОНАХТ» щодо відповідності реалізованих функцій системи поставленим задачам, а саме: забезпеченню ефективності процесу підвищення кваліфікації та атестації педагогічних та науково-педагогічних працівників закладу освіти за допомогою інформаційної підтримки задач прогнозування і планування організації цього процесу; зручному автоматизованому контролю за процесом підвищення кваліфікації педагогічних та науково-педагогічних працівників закладу освіти; контролю за виконанням нормативів по кількості годин підвищення кваліфікації педагогічних та науково-педагогічних працівників закладу освіти за розрахунковий період; представленню розрахункових показників (кількість годин підвищення по кожному викладачу, менше 150 годин підвищення, тощо); формуванню звітів про результати підвищення кваліфікації та атестації педагогічних та науково-педагогічних працівників закладу освіти; аналітиці результатів за напрямами щодо підвищення кваліфікації та атестації педагогічних та науково-педагогічних працівників закладу освіти. Зручний інтерфейс автоматизованої інформаційної системи обліку підвищення кваліфікації викладачів закладу освіти забезпечує швидкий пошук необхідної інформації, дозволяє одразу включитися в підготовку професійних документів, підвищує ефективність, якість роботи і скорочує терміни її виконання.

На основі запропонованої моделі автоматизованої інформаційної системи обліку підвищення кваліфікації викладачів можливе здійснення моніторингу процесу підвищення кваліфікації за допомогою реалізації таких запитів і звітів, як:

— визначення загальної кількості слухачів, які здійснили підвищення кваліфікації у закладі та за визначений навчальний рік зокрема;

— розподіл слухачів за джерелами оплати освітньої послуги з підвищення кваліфікації;

— визначення кількості та типу освітніх закладів, працівники яких підвищують кваліфікацію;

— розподіл слухачів за спеціалізаціями, рівнем освіти, стажем роботи, кваліфікаційною категорією в цілому та за певний навчальний рік зокрема;

— визначення обсягу годин підвищення кваліфікації певного слухача за відповідний проміжок часу: навчальний рік, декілька років, п'ятирічний термін тощо;

— порівняння результатів навчальної діяльності слухача у розрізі кваліфікаційної категорії: спеціаліст, друга, перша, вища категорії;

— порівняння результатів навчальної діяльності залежно від форми навчання: очна, заочна, очно-дистанційна;

— порівняння результатів навчальної діяльності у розрізі навчальних програм підвищення кваліфікації;

— визначення можливого комбінування різних навчальних програм для забезпечення ефективного здійснення підвищення кваліфікації викладачем з гармонійним розвитком його професійних компетентностей і дотримання вимог постанови № 800 Кабінету Міністрів України [17];

— визначення актуальної форми навчання (очної, очно-дистанційної, заочної), календарних періодів внаслідок здійснення аналізу розподілу слухачів відповідно до зазначених критеріїв за декілька років;

— відслідковування спеціалізації, спектра вчених ступенів та вчених звань, посад викладацького складу закладу, що опосередковано визначатиме якість результатів підвищення кваліфікації;

— моніторинг планування для викладача та виконання ним запланованого підвищення кваліфікації;

— відслідковування змін у тематиці та напрямів підвищення кваліфікації певного викладача за необхідний проміжок часу;

— виключення можливості запису викладача на один і той же напрям підвищення кваліфікації за певний атестаційний період;

— ефективне планування процесу підвищення кваліфікації педагогів у майбутньому.

В цілому зазначена модель автоматизованої інформаційної системи обліку підвищення кваліфікації викладачів має забезпечити:

— покращення якості цього процесу завдяки автоматизації планування курсів підвищення кваліфікації педагогів відповідно до існуючих потреб та аналізу результатів навчальної діяльності слухачів;

— зменшення часових затрат працівників закладу освіти на підготовку документів внутрішньої звітності (навчальної програми, журналу обліку курсів підвищення кваліфікації; календарних графіків тощо) за рахунок одноразового введення відповідних даних та їх багаторазового використання;

— зменшення часових затрат адміністрації закладу на моніторинг документації внутрішньої звітності;

— підвищення якості планування спільної діяльності закладу із закладами ППО, районними методичними кабінетами та відповідними представниками об'єднаних територіальних громад регіону щодо професійного розвитку педагогів у період завдяки проведенню різностороннього аналізу результатів навчальної діяльності науково-педагогічних та педагогічних працівників за допомогою методів математичної статистики тощо.

У перспективі інформаційна система буде зберігатися на віддаленому сервері для підвищення захищеності бази та можливості дистанційної роботи. Автоматизована інформаційна система обліку підвищення кваліфікації викладачів може бути рекомендована для роботи структурних підрозділів навчально-методичного напряму у закладах вищої та фахової передвищої освіти для моніторингу та обліку підвищення кваліфікації та атестації педагогічних та науково-педагогічних працівників.

### *СПИСОК ВИКОРИСТАНОЇ ЛІТЕРАТУРИ*

# СИСТЕМНИЙ ПІДХІД ПРИ ОРГАНІЗАЦІЇ НАВЧАЛЬНОГО ПРОЦЕСУ У ЗАКЛАДАХ ВИЩОЇ ОСВІТИ З ЗАСТОСУВАННЯМ НОВИХ ІНФОРМАЦІЙНИХ ТЕХНОЛОГІЙ

*Воінова С. О.*

*Розглянуто виклики часу, що постали перед вищою освітою України, та можливі шляхи вирішення поставлених ними завдань. Визначено актуальність використання нових методів навчання, що базуються на сучасних інформаційних технологіях. Розглянуто філософсько-методологічні аспекти використання інформаційних технологій в освітньому процесі.*

*Розкрито сутність системного підходу до навчального процесу у закладах вищої освіти із застосуванням сучасних інформаційних технологій. Розглянуто етапи розвитку інформаційних технологій. Проаналізовано риси сучасного етапу розвитку системи вищої освіти. Виділено зовнішні та внутрішні фактори використання дистанційного навчання у системі освіти.*

*Розглянуто дистанційні освітні технології та електронне навчання як основу дистанційного навчання. Виділено поняття та структуру дистанційного курсу як важливого елемента системи дистанційного навчання. Відзначено переваги та недоліки дистанційного очного та заочного навчання перед традиційним. Особливу увагу приділено модульній технології навчання. Розглянуто цілі, напрями, принципи інформатизації вищої освіти як пріоритетного напряму реформування вищої школи.*

*Висвітлено ресурс системного підходу до організації навчального процесу, інтегрованого застосуванням нових інформаційних технологій у закладах вищої освіти, що дозволяє усвідомлювати взаємозв'язок компонентів освітньої системи та ефективно реалізовувати основні її функції.*

*The challenges of the time, faced by the higher education of Ukraine, and possible ways of solving the tasks set by them are considered. The relevance of using new teaching methods based on modern information technologies is noted. Philosophical and methodological aspects of the information technologies using in the educational process are considered.*

*The essence of the systematic approach as applied to the educational process in institutions of higher education using modern information technologies is disclosed. The stages of information technologies development are considered. The features of the present-day stage of development of the education system are analyzed. External and internal factors of using distance learning in the education system are highlighted.*

*The distance learning educational technologies and e-learning as the basis of distance learning are considered. The concept and structure of a distance course as an important element of a distance learning system are highlighted. The advantages and disadvantages of the distance face-to-face teaching and the part-time education over the traditional learning are noted. Particular attention is paid to modular learning technology. The goals, directions, principles of informatization of higher education as a priority direction of reforming higher education are considered.*

*The resource of a systematic approach to the organization of the educational process, integrated by the use of new information technologies, in institutions of higher education is highlighted. This resource as emphasized allows to understand the relationship between the components of the educational system and effectively implement its main functions.*

Вища освіта в Україні має масовий характер. Згідно з аналітичною інформацією Міністерства освіти і науки України, рівень охоплення вищою освітою населення традиційного офіційного віку навчання є високим — 82,7 %. За цим показником Україна у Глобальному інноваційному індексі 2020 р. посіла 14-те місце зі 131 країни. Меншим є охоплення у Німеччині (70,2 %, 28-ме місце), Польщі (67,8 %, 34-те) та Великій Британії (60 %, 46-те місце) [1].

В Україні високий рівень якісного складу викладачів закладів вищої освіти. На початку 2019/1920 навчавльного року у загальній кількості викладачів університетів, інститутів, академій 47,3 % складали доктори філософії / кандидати наук, 11,7 % — доктори наук, 31,1 % — доценти, 9,3 % — професори.

В Україні мережа університетів — одна з найбільш щільних: на 1 млн населення припадає 6,7 університету. За останні п'ять років кількість університетів, інститутів, академій зросла на 1,4 %.

Розвиток інфраструктури вищої освіти в Україні демонструє ключову тенденцію: зростання чисельності університетів, академій й інститутів, навчання в яких орієнтовано на більш ґрунтовну підготовку. Така тенденція стала наслідком, перш за все, змін у структурі економіки країни, переходу до масової вищої освіти. Розвиненість і регіональна розгалуженість мережі закладів вищої освіти України надає можливість охоплення значної частини населення країни вищою освітою з подальшим просуванням ціложиттєвої освіти та перекваліфікації.

Освіта наразі дещо відстає від загальної діджиталізації, і необхідно докласти більше зусиль, щоб скористатися інструментами та сильними сторонами нових інформаційних технологій, одночасно вирішуючи проблеми щодо можливих зловживань, таких як кібервторгнення та проблеми конфіденційності.

Завдання розвитку законодавства у сфері освіти окреслюються Національною доктриною розвитку освіти, яка визначає систему концептуальних ідей та поглядів на стратегію і основні напрямки розвитку освіти до 2025 р. [2; 3].

У рамках підготовки звіту «Горизонт 2020» асоціацією EDUCAUSE залучені експерти з різних країн світу окреслили ландшафт і визначили найбільш впливові тенденції, що формують вищу освіту, викладання та навчання [4]. Ключові тренди були ідентифіковані у рамках п'яти категорій: соціальні, технологічні, економічні, вищої освіти та політичні. Зокрема категорія «Технологічні» включає розвиток штучного інтелекту, формування цифрового навчального середовища наступного покоління, проблеми аналітики даних та питання конфіденційності. Категорія «Вища освіта» включає альтернативні шляхи до освіти, онлайн-освіту.

Згідно зі Стратегією розвитку вищої освіти в Україні на 2021—2031 роки, в основу концептуальної моделі вищої освіти України має бути покладений кібернетичний принцип необхідного розмаїття [1].

Заклади вищої освіти мають створити власні або використовувати вже існуючі онлайн-платформи для поширення знань у професійній громаді, для професійної орієнтації тощо.

Актуальним є питання впровадження віртуальних університетів, які можуть здійснювати підготовку за багатьма або окремими спеціальностями, існувати як окрема онлайн-платформа або бути віртуальним дублером традиційного закладу вищої освіти. Віртуальні університети орієнтовані на розширення доступу до вищої освіти для різних категорій населення, включення до освітнього процесу нетрадиційного контингенту (дорослі здобувачі вищої освіти, люди протягом усього життя), підвищення кваліфікації, опанування додаткових навичок, актуалізацію знань та навичок, поширення кращих навчальних практик. Освітні програми мають відрізнятися максимальною гнучкістю, передбачати мікро- і нанокредитивні навчальні дисципліни. Віртуальні університети стануть головним провайдером екстернатної, неформальної та інформальної освіти. Форма навчання тут передбачається дистанційна [5].

Як найважливішу операційну мету запропоновано впровадження інноваційних інформаційних технологій і дистанційного навчання у вищій освіті. Як конкретні завдання для реалізації цієї мети запропоновано створення індустрії інноваційних інформаційних технологій та засобів навчання, що відповідають світовому науково-технічному рівню; діджиталізація усіх процесів у системі вищої освіти; унормування дистанційного навчання як форми здобуття вищої освіти [1].

Реформування освітньої галузі — це відповідь на суспільний запит, адже саме освіта забезпечує якість людського капіталу, який є основою економічного розвитку країни. Упродовж 2020/2021 н. р. Міністерством освіти і науки України було продовжено стратегічний курс на реформування усіх сфер освіти. Водночас у 2019/2020 н. р. з огляду на пандемію Covid-19 та запроваджені карантинні обмеження перед системою освіти постали нові виклики, пов'язані з забезпеченням неперервності освітнього процесу, спроможністю закладів усіх рівнів освіти забезпечити якість і сталість здобуття освіти в умовах карантинних обмежень, необхідністю розвитку дистанційної форми здобуття освіти.

Усе це спонукає посилювати складові реформи освітньої галузі, пов'язані з діджиталізацією освітнього середовища, передусім із забезпеченням закладів вищої освіти швидкісним доступом до мережі Інтернет, а здобувачів вищої освіти і педагогічних працівників — цифровими пристроями та електронними освітніми ресурсами [6].

Пандемія змусила все глобальне академічне співтовариство звернутися до нових методів навчання, включаючи дистанційне та онлайн-навчання. Аналіз вузівських практик показує, що в період віддаленої роботи склалося кілька режимів організації освітньої діяльності: асинхронний або заочний (здобувачі вищої освіти вивчають матеріал у зручний для них час, відповідно до встановлених викладачем термінів); синхронний (одночасна участь у занятті, наприклад, у форматі вебінара); змішаний (поєднання синхронної та асинхронної взаємодії залежно від характеру педагогічних завдань).

Досвід останніх двох років показав широкі можливості використання форматів та технологій дистанційної роботи для вирішення як традиційних, так і нових завдань університетів.

Незалежно від цього в останні десятиріччя виникла нова проблема розвитку системи освіти. Знання старіють кожні 3—5 років, а технологічні знання — кожні 2—3 роки. Мине ще трохи часу і це буде 1,5—2 роки, а необхідний обсяг знань для випускників освітніх закладів подвоюється кожні 3—4 роки. Якщо не змінювати освітніх технологій, то якість підготовки фахівців об'єктивно відставатиме від вимог на ринку праці. Засвоєння знань здобувачами вищої освіти за допомогою інформаційних та комунікаційних технологій за найнижчими оцінками є на 40—60 % швидшою або більшою в одиницю часу, ніж за звичайними технологіями (за один і той же період надається більше знань).

Одним із видів інновацій в організації вищої професійної освіти є запровадження дистанційного навчання.

У закладах вищої освіти сьогодні навчається нове, єдине поки що покоління, яке з перших днів життя стикається з інформаційними технологіями та мікропроцесорною технікою, зокрема комп'ютерами. Цифрові пристрої для них є такми ж звичними, як телевізор або холодильник для старшого покоління. Для колективів закладів вищої освіти немає більш нагального завдання, ніж пізнання культури, психології, цінностей та змін, які очікує на це нове покоління. Мережеве покоління обов'язково змінить сам спосіб виробництва, створить нову культуру праці. Представники нового покоління бажають активно використовувати у навчанні мобільні пристрої (смартфони, планшети, ноутбуки) та сервіси Інтернету, отже заклад вищої освіти має бути технологічно готовий надати такі можливості новому поколінню здобувачів вищої освіти [7].

Можливість та необхідність використання інформаційних технологій у закладах вищої освіти не викликає сумніву. Застосування мультимедійних програм, можливість візуалізації розрахунків, що проводяться, дозволяють зробити навчання більш наочним, значною мірою допомагають подолати традиційно надто формалізоване та абстрактне викладання багатьох університетських навчальних курсів. При цьому для багатьох дисциплін використання комп'ютерних засобів необхідно пов'язувати з комп'ютерним моделюванням. За багатьма розділами фундаментальних наук в Інтернеті накопичено величезну кількість корисної інформації, яку необхідно шукати і систематизувати із застосуванням пошукових систем і використовувати у процесі викладання.

Ефективне використання цих технологій потребує особливих компетенцій викладачів, управлінців, здобувачів вищої освіти, а також ефективних та зручних технологічних рішень, особливої організації навчального процесу. Без цього не можна говорити про повноцінну освіту у дистанційному форматі. Саме заклади вищої освіти покликані та мають найближчим часом стати колискою формування нової інтернет-орієнтованої свідомості молодих людей. Саме заклади вищої освіти мають забезпечити цей процес матеріально. Викладач у своїй діяльності має орієнтуватися як на традиційні, так і на нетрадиційні методи навчання. Головне — сформувати у здобувача вищої освіти інтернет-орієнтований спосіб мислення, навчити використовувати інформацію для самоосвіти, підвищення кваліфікаційного рівня, вирішення можливих виробничих проблем та завдань.

Практика використання інформаційних технологій в освітньому процесі вищої школи свідчить про наявність протиріч між такими обставинами:

— традиційними видами навчально-методичного забезпечення та потребою практики в інноваційних формах застосування інформаційних технологій у навчальному процесі закладів вищої освіти;

— процесом інформатизації освіти та відсутністю загального підходу до використання інформаційних технологій при навчанні здобувачів вищої освіти в сучасних освітніх установах;

— між пошуком та виявленням інформації, необхідної для організації навчального процесу та діджиталізованою інформацією, яка дозволяє застосовувати більш активні сучасні способи пошуку, сприйняття, обробки, використання та зберігання інформації;

— абсолютизацією структур та форм побудови навчально-методичних матеріалів для здобувачів вищої освіти та потребою практики в їх інноваційних структурах з розширеними функціональними та інформаційними можливостями.

Зазначені протиріччя сприяли пошуку шляхів застосування інформаційних технологій у навчанні здобувачів вищої освіти на основі системного підходу.

У сфері освіти інформаційні технології забезпечують збирання, обробку, надання та публікацію даних, що належать до навчання, і допомагають викладачам краще забезпечити навчальний процес необхідними матеріалами, виявити прогалини, адаптувати зміст та педагогічні підходи до конкретної групи.

Цінність інформаційних технологій для розвитку навчальної діяльності закладу вищої освіти полягає в такому:

— покращення якості навчання за допомогою більш повного використання доступної інформації, підвищення мотивації здобувачів вищої освіти та творчої активності викладачів;

— впровадження нових освітніх технологій — розвиваюче та проектне навчання, ділові ігри, візуалізація, імітаційне моделювання, дистанційне навчання;

— інтеграція різних видів діяльності (навчальної, навчально-дослідницької, наукової);

— зменшення залежності здобувача вищої освіти від викладача;

— покращення оцінки навчальних досягнень на основі комп'ютерного тестування.

Інформаційні технології змінюють роль викладача, який з єдиного носія знань перетворюється на навчального менеджера та наставника, спрямовуючи та контролюючи зусилля здобувачів вищої освіти з освоєння певної програми — через індивідуальні завдання, визначення відповідних навчальних ресурсів, створення спільних можливостей для навчання, а також надання свого розуміння матеріалу та консультаційної підтримки як під час очного процесу, так і в навчальних середовищах та віртуальній взаємодії. Викладач залишається, безумовно, ключовим, але все ж таки одним із учасників освітнього процесу, з пультом біля проектора або за комп'ютером в інформаційному середовищі. Успіх нового підходу залежить від людського фактора та готовності викладача увійти в віртуальні аудиторії та середовища. Викладач стане більшою мірою наставником, спрямовуватиме і навчатиме думати, досліджувати, вирішувати проблеми, а заклад вищої освіти — готувати здобувача вищої освіти до реальної професійної кар'єри.

Для забезпечення належного реагування на виникаючі проблеми заклади вищої освіти повинні зосередитися на якості, актуальності та оперативності.

Освітня діяльність та підготовка здобувача вищої освіти є комплексним поняттям, отже їхня ефективна реалізація повинна спиратися на системний похід [8].

Зміст поняття «якість освіти» у системі професійної підготовки змінюється разом зі змістом цієї підготовки. Ці зміни викликані такими причинами: по-перше, науково-технічними та інформаційними змінами виробничих технологій; по-друге, падінням попиту на некваліфіковану працю; по-третє, поширенням автоматизованих систем управління виробничими процесами, тобто зміною у змісті професій, що вимагає змін у змісті професійної підготовки, підвищення її якості, створення механізмів, що забезпечують її постійне настроювання на вимоги ринку праці, що динамічно змінюються, використання системного підходу при організації професійної підготовки.

У науковій літературі системний підхід — це методологічний напрям, який ставить завданням розробку принципів, методів і засобів вивчення об'єктів, що являють собою системи.

Системний підхід, будучи елементом діалектичного методу в цілому, є не тільки конкретизацією діалектико-матеріалістичного вчення про загальний зв'язок явищ, але і однією зі сторін діалектико-матеріалістичного вчення про розвиток. Принцип системності вимагає розглядати всі явища у взаємозв'язку, у взаємодії. Таким чином, в основі системних досліджень лежить положення про діалектичну єдність принципу системності та розвитку.

При системному підході виявляють та вивчають зв'язки та відносини між елементами (підсистемами) будь-якого об'єкта управління. Важливим моментом стає підпорядкування окремих, локальних завдань окремих підсистем загальній кінцевій меті. При цьому обов'язковою умовою є чітке формулювання єдиних цілей, завдань, а потім визначення шляхів найефективнішого розв'язання як для системи в цілому, так і для окремих її елементів.

Система — це сукупність компонентів, що знаходяться у певних відносинах і пов'язані один з одним, взаємодія яких породжує нову якість, не властиву цим компонентам окремо. У системі існують елементи (будь-які об'єкти) і певна структура як відносно стійкий спосіб зв'язку елементів того чи іншого складного цілого.

Педагогічна система — це певна сукупність взаємопов'язаних засобів, методів та процесів, необхідних для створення організованого, цілеспрямованого та спеціалізованого педагогічного впливу на формування особистості із заданими якостями [9].

З позицій системного підходу у наведеному визначенні виділяються такі внутрішні аспекти педагогічної системи: системно-компонентний, системно-структурний, системно-функціональний та системно-інтегративний.

Наявність структури — умова накопичення кількісних змін усередині системи, що є необхідною передумовою для її подальшого розвитку та перетворення.

Будь-яка система соціального порядку є активною і діяльною, що проявляється у її функціях. У свою чергу, функції системи є інтегрованим результатом певних дій компонентів, що її утворюють. По відношенню до системи вони мають доцільний характер, інакше компонент випадає із системи та стає стороннім тілом для неї. Зміни у природі компонентів, у характері їхньої взаємодії викликають відповідні зміни у функціях як самих компонентів, так і системи в цілому. Отже, системний підхід — це не просто виділення системного об'єкта суб'єктом пізнання, але й процедура його системного освоєння. Найбільш продуктивним шляхом до цього є конструювання педагогічної структури відповідно до потреб та цілей вищої освіти.

Основним завданням підходу, що розглядається, у вищій освіті є побудова оптимальної моделі, яка поєднує освоєння теоретичних знань і застосування їх у вирішенні практичних питань, що сприяє формуванню професійних компетенцій у майбутнього спеціаліста.

Розглядаючи проблему професійної підготовки здобувачів вищої освіти, її необхідно представляти як систему, до якої входять підсистеми, які підпорядковуються тим самим принципам, що і система: цілісності, структуризації, множинності. Однією з підсистем професійної підготовки, яка є також системою, що має всі її властивості, є навчальний процес. Він має певну цілісність, що дозволяє розглядати одночасно систему як єдине ціле й як підсистему для вищих рівнів. Структуризація дозволяє аналізувати елементи системи та їхні взаємозв'язки у межах конкретної організаційної структури. Множинність дозволяє використовувати безліч форм, засобів, методів для реалізації окремих елементів та системи в цілому. При цьому навчальний процес має синергію (від грецької — «разом діючий»), це пояснює більший сумарний ефект функціонування всіх складових елементів системи порівняно із сумою дії кожного з елементів.

Застосування у педагогічній практиці навчання системного підходу передбачає наявність взаємозв'язків між компонентами навчального процесу, кожен з яких може функціонувати з максимальною ефективністю, спираючись на внутрішні зв'язки в цій системі. Зміст матеріалу, що вивчається, є одним із структурних компонентів навчального процесу, засвоєння якого пов'язане з обраними методами, формами та засобами навчання. При цьому керує системою ви-кладач, будучи по суті її компонентом. Від того, які технології будуть використані викладачем, залежить ступінь ефективності функціонування даної системи.

Сьогодні широко використовуються інформаційні текхнології, які вбудовуються в усі компоненти навчального процесу, розробляються та впроваджуються нові.

Системний підхід вимагає розгляду філософсько-методологічних аспектів використання інформаційних технологій в освітньому процесі. Такий аналіз спрямовано на обґрунтування загальних світоглядних установок у дослідженні різноманітних можливостей застосування інформаційних технологій у сфері освіти [10].

Інформаційні технології, втілені в обчислювальній техніці, в наші дні глибоко проникають у структури людської діяльності, перетворюють зміст і характер праці та навчання, по-новому ставлять проблеми розвитку людського інтелекту та особистості, справляють серйозний вплив на світогляд людей та ідеологічні концепції, породжують нові способи та форми організації наукових досліджень. Ось чому соціально-філософські та філософсько-методологічні аспекти розвитку обчислювальної техніки, як результату розвитку інформаційних технологій, заслуговують на пильну увагу [11].

Ця вимога відноситься до різних галузей наукових знань, у тому числі і педагогічної науки, що відображає сутнісні характеристики предметної області свого дослідження — педагогічної діяльності.

Інтеграція педагогіки з іншими суспільними, природничими, технічними науками — це провідна тенденція вдосконалення системи науково-педагогічних знань. Зазначена тенденція проглядається і при вивченні проблеми взаємодії людини з комп'ютером у сфері освіти, яка за всієї своєї самостійності та важливості є, по суті, лише частиною більш загальної соціально-філософської проблеми «людина — машина» — центральної проблеми сучасної науково-технічної революції. При цьому категорія «машина» носить збірний характер, позначаючи будь-які технічні пристрої, що використовуються як засіб підвищення ефективності, доцільності людської діяльності.

В умовах прискорення соціально-економічного та науково-технічного прогресу, посилення уваги до комплексу питань, пов'язаних із трактуванням ролі та місця людського фактора в інтенсифікації суспільного виробництва, все більш актуальною є проблема прямого та безпосереднього взаємозв'язку педагогічних та власне технічних факторів. Цей взаємозв'язок знаходить своє відображення у процесі взаємодії людини з мікропроцесорною, комп'ютерною технікою — найвищим проявом технізації суспільно корисної людської діяльності. Посилення взаємодії суспільних, природничих та технічних наук підкреслює значущість системи філософських знань, які становлять ядро інтегративних, міждисциплінарних форм пізнання і мають вирішальне значення у розумінні процесу зближення наук про природу та суспільство на розвиток техніки.

Безпосереднім методологічним орієнтиром взаємодії наук про людину і техніку служить філософський принцип загального зв'язку та взаємозалежності предметів та явищ об'єктивної дійсності, на якому базується системний підхід. З діалектико-матеріалістичної точки зору, здатність речей, об'єктів, явищ, процесів, часто якісно глибоко різних, до різноманітних взаємодій є проявом принципу матеріальної єдності світу, що стверджує наявність законів, які поширюються на різні за своєю природою об'єкти. Водночас діалектична логіка вимагає, щоб поряд з констатацією принципових відмінностей між трудовою діяльністю людини та функціонуванням різноманітних машин було виявлено й об'єктивно існуючі аналоги між інтелектуальними та фізичними процесами, що супроводжують трудову діяльність людини, та процесами, що протікають у машинах. Більше того, життя вимагає створення таких концепцій, які дозволили б розглядати людину і машину з єдиної позиції, яка аж ніяк не означає, що при цьому стираються всякі якісні відмінності між людиною і роботом. «У строгому сенсі слова, жодна машина, навіть найдосконаліша, не працює і не може працювати. Вона лише є знаряддям праці, за допомогою якого людина впливає на природу, змінюючи останню відповідно до заздалегідь поставленої мети. Яких чудових успіхів не досягла б техніка, які б дивовижні автомати не створювалися, праця завжди була і залишається свідомою діяльністю людини, а людина — суб'єктом праці» [12].

Інформаційні технології можна подати як сукупність трьох основних способів перетворення інформації: зберігання, обробки та передачі [13].

Методами інформаційних технологій є методи обробки та передачі інформації.

Засобами інформаційних технологій є технічні, програмні, інформаційні та інші засоби, за допомогою яких реалізується ІТ.

Інформаційна технологія спрямована на доцільне використання інформаційних ресурсів та забезпечення ними всіх елементів організаційної структури.

Реалізація системного підходу в описі інформаційних технологій передбачає використання принципу цілісності, відповідно до якого при системному підході виділяються такі аспекти або підходи.

Сутнісний підхід полягає у розкритті сутності системи, якісної специфіки, властивих їй системних якостей. Виявлення сутності системи — найскладніший етап пізнання суттєвих ознак інформаційної технології, які відрізняють її від інших об'єктів та систем.

Аналіз наведених визначень понять «технологія» та «інформаційна технологія» дозволяє виділити ряд суттєвих ознак:

— процесний характер інформаційної технології — виявляється в тому, що сутність технології пов'язана з перетворенням властивостей, форми, змісту та іншої інформації, по-перше, і по-друге, з процесом організації інформаційної технології;

— формалізований характер інформаційної технології — представляється у різних формалізованих формах: у вигляді проекту, алгоритмів та програм, різноманітних математичних моделей та ін.;

— орієнтація на практику — завдяки запитам практики інформаційні процеси стали реалізовуватися у формі технологій;

— концентрація у собі наукових знань та досвіду реалізації інформаційних процесів;

— отримання ефективності, досягнення кінцевого результату — невід'ємні характеристики інформаційної технології. Головним критерієм соціальної ефективності інформаційної технології є вільний час людини. Інформаційна технологія забезпечує економію витрат праці, енергії, ресурсів;

— забезпечення заданого користувачем рівня якості реалізації інформаційних процесів.

Елементний аспект передбачає опис складу системи, кількісну та якісну характеристику частин, компонентів, їх координацію та субординацію, пріоритетну (лідируючу) частину системи.

Засоби забезпечення інформаційних технологій представлені методами, технічними засобами (апаратні засоби РС, оргтехніка та ін.), алгоритмічними та програмними засобами, інформаційним та методичним забезпеченням, комп'ютерними мережами та телекомунікаціями, персоналом та ін.

Функціональний підхід вимагає відповіді на такі питання. Які внутрішні та зовнішні функції інформаційної технології як системи? Як ці функції дозволяють досягати цілі системи? Яка активність, життєдіяльність системи?

Визначення функцій системи передбачає встановлення її мети. Інформаційна технологія є цілеспрямованою системою. Основна мета інформаційної технології полягає у формуванні якісного інформаційного ресурсу (нової інформації, знань), необхідного для підвищення ефективності системи, в якій вона функціонує. Декомпозиція загальної мети дозволяє побудувати дерево цілей, яке відбиває всі напрямки реалізації інформаційної технології.

Для досягнення цілей інформаційній технології необхідно виконати певні функції. Головними з них є формування концептуальної моделі інформаційної технології, перетворення даних, інформації та знань (збір, обробка, зберігання, передача, розповсюдження та ін.) та функції із забезпечення інформаційною технологією. Зовнішні функції реалізуються для задоволення потреб у якісних і ефективних інформаційних системах різноманітних елементів: держави, політичного середовища, соціальної та виробничої сфери, науки, економіки, технології та інших.

Структурний підхід дозволяє встановити внутрішню організацію системи, способи взаємозв'язку елементів, компоненти у системі, її структури. Виділені функції закріплюються у структурах, які є способами взаємодії елементів у системі.

Функціональна структура технологічного інформаційного процесу задається логікою реалізації процедур перетворення інформації та технологічними принципами. Конкретна інформаційна технологія має вписуватися у відповідну організаційну структуру управління інформаційною системою, технологічну систему.

Комунікативний підхід розкриває питання взаємодії системи із зовнішнім середовищем шляхом визначення матеріальних, енергетичних та інформаційних зв'язків. Цю властивість системи називають відкритістю [14].

Властивість відкритості інформаційної системи проявляється у взаємодії із зовнішнім середовищем шляхом постійного обміну з ним енергією, речовиною та інформацією. Тут основне значення має інформаційний аспект, хоча матеріальний та енергетичний обмін також відіграють не останню роль у функціонуванні та розвитку технології.

Інформаційна технологія як відкрита система має такі характерні властивості.

Перша — має властивість активності. Вона проявляється у цілеспрямованій взаємодії із зовнішнім середовищем для задоволення своїх потреб. Активність інформаційної технології пов'язана з наяв-ністю в ній ціленаправлених компонентів, головним елементом яких є персонал.

Зростання активної ролі інформаційних технологій пов'язане зі зміною характеру розвитку зовнішнього середовища та зі зростаючою складністю взаємодії зі споживачем. Інформаційна технологія (суб'єкт) за впливом на об'єкт має приводити його у той стан, який найбільшою мірою допомагає досягати мети об'єкта, тобто носити активний характер. Активність інформаційної технології передбачає передусім розширення її дій та функцій у процесі свого функціонування та розвитку. Тому активність інформаційної технології повинна мати певну стійкість.

У ході практичної реалізації поставленої мети інформаційній технології необхідно керувати своєю діяльністю відповідно до змін зовнішнього середовища. Отже активність інформаційної технології має передусім бути спрямованою на пізнання закономірностей розвитку зовнішнього середовища задля її подальшого активного впливу на нього.

Важливим у використанні інформаційних технологій є облік властивості гомеостатичності, яка забезпечує цілісність системи за умов постійно змінного стану зовнішнього середовища. Тут слід зазначити, що в різних станах довкілля істотні змінні системи залишаються стабільними чи змінюються у заданих межах, чим забезпечують рівновагу із зовнішнім середовищем. Такий стан характеризує систему як цілісність і не може бути приписаний жодній її частині (підсистемі).

У практичному плані властивість відкритості інформаційної технології реалізовано у розробці концепції відкритих інформаційних систем. Суть її коротко зводиться до такого: кожна відкрита інформаційна система призначена для розв'язання двох завдань (обробки та передачі даних) і складається з двох частин — прикладні процеси, призначені для обробки даних і, насамперед, для задоволення потреб користувачів; та область взаємодії, що забезпечує передачу даних між прикладними процесами, розташованими у різних системах. Головну роль у розробці відкритих систем грає Міжнародна організація зі стандартизації (International Organization for Standardization, ISO). Вона розробляє стандарти взаємодії відкритих систем (Open Systems Interconnection, OSI).

Інтегративний підхід виявляє системоутворюючі чинники, механізми забезпечення єдності системи, її цілісності.

Досліджуючи проблему цілісності, багато вчених дотримуються різних поглядів на це поняття. Перший підхід до проблеми цілісності пов'язують із наявністю у системи нових властивостей (неадитивності, емерджентності, інтегральності тощо), які не притаманні її елементам.

Другий підхід акцентує увагу на автономності, цілісності системи та протиставленості зовнішньому середовищу.

У третьому підході як критерій цілісності системи виділяють наявність певного ступеня впорядкованості, організованості елементів системи, взаємозв'язків та взаємодій, певної тісноти зв'язків; наявність такого поєднання елементів (підсистем), властивостей та зв'язків системи, яке найбільшою мірою відповідає її цілям функціонування та розвитку. Тут простежується тісний зв'язок властивості цілісності з організованістю системи. Четвертий підхід поєднує перший і третій [15].

Останнім часом у зв'язку з розвитком функціонального підходу у науковому пізнанні розвивається погляд на проблему цілісності з цих позицій. Виділення функціональної цілісності у пізнанні систем — ще один крок до вивчення цієї складної проблеми, що дозволяє враховувати властивість відкритості. Розглядаючи джерело цілісності систем, необхідно враховувати зв'язки із зовнішнім середовищем.

Інформаційна технологія дійсно володіє новими властивостями, які виявляються внаслідок її функціонування та розвитку. Особливий інтерес представляє поява серед нових властивостей таких, які не характерні жодному з елементів системи, тобто інтегральних (емерджентних тощо) властивостей.

На думку багатьох фахівців, причиною виникнення системи інтегральних властивостей є наявність різноманітних, стійких зв'язків як усередині системи, так і з зовнішнім середовищем. Саме зв'язки становлять той новий прихований доданок, який відрізняє ціле від суми частин.

Нові властивості інформаційної технології можуть бути найрізноманітнішими. Для цілеспрямованих систем (якою і є інформаційна технологія) важливе значення має виникнення нових властивостей, пов'язаних з їх цільовим призначенням, тобто тих властивостей, які визначають її якісну специфіку.

Інформаційна технологія як цілісна система у процесі свого функціонування та розвитку здатна на більше, ніж кожен із її ізольованих елементів або підсистем. Така нова властивість систем у теорії організації одержала назву синергетичного ефекту.

Отже властивість цілісності інформаційної системи можна конкретизувати і виразити через систему зв'язків між елементами та із зовнішнім середовищем, з одного боку, та інтегративністю — з іншого. З теорії систем відомо, що наявність зв'язків не є характерною ознакою лише систем. Для забезпечення цілісності інформаційній технології необхідно, щоб зв'язки між елементами мали стійкий характер і серед них знаходилися системоутворюючі зв'язки. Роль системоутворюючих зв'язків грають зв'язку управління, організаційні, функціональні, зворотні, технологічні та ін.

Історичний підхід визначає процеси виникнення системи, її становлення, функціонування, тенденції та перспективи розвитку.

Найбільший інтерес для наших завдань мають етапи розвитку інформаційнні технологій, пов'язані з розвитком електронно-обчислювальних машин.

1-й етап (до другої половини XIX ст.) — «ручні» технології: перо, чорнильниця, книга, елементарні ручні засоби рахування. Комунікації здійснювалися шляхом доставки кінною поштою листів, пакетів, депеш, у європейських країнах застосовувався механічний телеграф. Основню метою технологій було подання та передача інформації у потрібній формі.

2-й етап (кінець XIX ст. — 40-ві рр. XX ст.) — «механічні» технології: друкарська машинка, арифмометр, телеграф, телефон, диктофон, оснащена більш досконалими засобами доставки пошта. Основною метою технологій було подання інформації у потрібній формі зручнішими засобами, скорочення витрат на компенсацію втрат та виправлення спотворень.

3-й етап (40-ві — 60-ті рр. XX ст.) — «електричні» технології: перші лампові електронно-обчислювальні машини і відповідне програмне забезпечення, електричні друкарські машинки, телетайпи (телекси), ксерокси, портативні диктофони. Організація доставки інформації у заданий час. Акцент в інформаційних технологіях починає переміщатися з форми подання інформації на формування її змісту.

4-й етап (70-ті рр. — середина 80-х рр.) — «електронні» технології, основний інструментарій — великі електронно-обчислювальні машини та створювані на їх базі автоматизовані системи управління (АСУ) та інформаційно-пошукові системи, оснащені широким спектром базових та спеціалізованих програмних комплексів. Центр ваги технологій зміщується на формування змістовної складової інформації для управлінського середовища різних сфер громадського життя, особливо на організацію аналітичної роботи.

5-й етап (з середини 80-х рр.) — «комп'ютерні» («нові») технології, персональний комп'ютер із широким спектром стандартних та замовлених програмних продуктів широкого призначення. Створення систем підтримки прийняття рішень на різних рівнях управління. Системи мають вбудовані елементи аналізу та штучного інтелекту, реалізуються на персональному комп'ютері та використовують мережеві технології та телекомунікації для роботи в мережі.

6-й етап (з середини 90-х рр.) — «Internet/Intranet» («найновітніші») технології. Широко використовуються в різних областях науки, техніки та бізнесу розподілені системи, глобальні, регіональні та локальні комп'ютерні мережі. Розвивається електронна комерція. Збільшення обсягів інформації призвело до створення технології Data Mining (з англ.: видобуток даних) — автоматизований пошук даних, заснований на аналізі великих масивів інформації. За мету береться ідентифікація тенденцій та патернів, яка за звичайного аналізу неможлива. Для сегментації даних та оцінки ймовірності подальших подій використовуються складні математичні алгоритми [16].

Безперечним є твердження про початок переходу людської цивілізації в новий якісний стан — постіндустріальна чи інформаційна культура приходить на зміну індустріальній, яка, у свою чергу, замінила в середні віки аграрну. Кожній стадії розвитку суспільства відповідають свої форми та зміст процесу навчання нових поколінь, передачі їм накопичених знань, навичок, традицій.

Аграрна цивілізація, яка існувала від первісного та рабовласницького суспільства до середини XV століття, породила менторську школу, засновану на усному спілкуванні, а потім на створенні та використанні рукописних конспектів лекцій та протоколів наукових дослідів. Вона характеризувалася найвищою духовною близькістю вчителя та учня, низьким рівнем безпеки даних. Інформаційні втрати були катастрофічними. Але за час існування менторської школи людство нагромадило потенціал, що дозволив зробити прорив у індустріальне суспільство, яке породило нові форми освітнього процесу — ремісничу школу. Основою цього реформування стало книгодрукування, яке розширило аудиторію користувачів і підвищило безпеку інформації, вперше забезпечило масовий характер дистанційного навчання. Реміснича школа, система міст-університетів насамперед задоволь-няли потреби зростаючої промисловості. Аудиторне навантаження професорів у розрахунку на одного студента знизилося. Зменшився і світоглядний вплив викладача. І лише наприкінці XX століття нові інформаційні, комунікаційні та насамперед комп'ютерні технології перевернули уявлення про можливості інформаційного обміну. Це сталося завдяки п'яти сторіччям передачі знань у рамках ремісничої школи.

Новий етап розвитку системи освіти — етап виходу навчального процесу за межі конкретного навчального закладу. Стають загальнодоступними найкращі світові зразки викладання дисциплін, готові курси, програми. Надійність систем дублювання та збереження інформації стає абсолютною, як можливість віддаленого доступу до невичерпних ресурсів світових бібліотечних фондів, інформаційних баз даних, наукових результатів лідируючих лабораторій та інститутів у всьому світі. Корінним чином змінюється і форма подачі навчального матеріалу: стає ясно, що виникнення радіо, кіно і телебачення стало не народженням самодостатніх засобів комунікації, а лише попередником синтетичних способів віддаленого впливу на людський мозок за допомогою аудіо- і відеоінформації. Відбуваються зміни і у сфері міжособистісних відносин: знижуються корпоративність та колективізм навчальних груп, у масовому навчанні слабшають елементи духовного спілкування та виховання.

Основними рисами розвитку такого типу освіти в інформаційному суспільстві можна назвати такі.

1. Всесвітня інформаційна мережа — заміна письмового спілкування електронною поштою, колективних усних дискусій — чатами, відмова від поліграфічної форми підручників на користь електронних версій. Роль, місце та функціональні обов'язки викладача змінюються. Він повинен не тільки володіти всіма цими технологіями, вміти відбирати, оцінювати, застосовувати найцінніші освітні ресурси, але й допомогти здобувачеві вищої освіти не потонути в інформаційному океані.

2. Нові форми пред'явлення знань, умінь, навичок: інтерактивні живі тексти, гіпертексти, аудіовізуальні засоби, комп'ютеризовані практикуми — симулятори, віртуальні лабораторії. Тут викладач повинен як мінімум бути в змозі поставити завдання дизайнеру, програмісту, аніматору при створенні таких методичних матеріалів та застосувати вже створені професіоналами інтерактивні, мультимедійні чи віртуальні посібники у своїй педагогічній області.

3. Пролонгований у часі характер. Технічні можливості надання якісних освітніх послуг входять у суперечність із бажанням тривалого безпосереднього спілкування вчитель — учень. Інформаційний пресинг засобів масової інформації та Інтернет повністю змінили характер, глибину та швидкість сприйняття зовнішніх подразників новими поколіннями. Аудиторне навантаження знижується за прямими медичними показаннями. Спрощений варіант загальної вищої освіти поширюється одночасно з інформаційною інфраструктурою.

У цих умовах потяг молоді до Інтернету треба використовувати для грамотного, ввічливого та змістовного мережевого спілкування. Необхідно розпочати підготовку педагогів, здатних застосовувати найсучасніші інформаційні технології навчання, зберігаючи при цьому той безцінний досвід, знання та методики викладання, які їх носії мають передати їх новим поколінням. Також необхідно розвивати телекомунікаційну освітню інфраструктуру, випереджаючими темпами розробляти навчально-методичне та апаратно-програмне забезпечення для всіх форм очних занять, дублюючи його версії у локальному та дистанційному варіантах.

Особливу увагу необхідно приділити розробці освітніх серверів, сайтів, інших інтернет-ресурсів, які здатні охопити максимально широку аудиторію, створити та підтримувати єдиний освітній простір, забезпечити державні стандарти. Поступово необхідно трансформувати навчальні плани закладів вищої освіти у бік зменшення аудиторного навантаження викладачів, зміщення центру ваги на дистанційні форми поза жорстким розкладом, самостійну роботу студентів за консультаційної та методичної підтримки викладачів.

У перспективі слід прагнути до максимальної інтеграції провідних закладів вищої освіти на основі уніфікованих освітніх стандартів, надання загального мережевого сервісу за всіма формами навчального процесу та єдиної рейтингової системи оцінки якості навчання. Очевидно, що це вимагає від усього суспільства створення розгалуженої інфраструктури швидкого інформаційного обміну, заснованої на взаємодоповнюючих нових інформаційних та телекомунікаційних технологіях [17].

Серед нових інформаційних технологій можна назвати такі: штучний інтелект, нейрокомп'ютерна технологія, нейронні мережі, відеотехнологія, мультимедіатехнологія, об'єктно орієнтована технологія, інтернет-технологія, віртуальна реальність та інші. Нові інформаційні технології являють собою певні методики, що дозволяють на іншому, новішому рівні вирішувати освітні та виховні завдання; здійснювати прогноз та аналіз інформації; допомагати при прийнятті правильного та ефективного рішення.

З їх допомогою здійснюються пошук, збирання інформації, її переробка, зберігання, подання у доступному для здобувачів вищої освіти вигляді, а також актуалізація інформації, що пов'язано з розвитком науково-технічного прогресу, оновленням техніки та технологій, розвитком наукомістких виробництв. Масштабно використовувані цифрові навчально-методичні матеріали, у тому числі в мультимедійному поданні, бездротові технології, презентаційне обладнання, мережеві технології для доступу до ресурсів, впливають на інфраструктуру інформаційних технологій, на сервіси інформаційної системи закладу вищої освіти [18].

Використання інформаційних технологій у вищій освіті буде найбільш успішно здійснюватися якщо: в освітньому просторі закладу вищої освіти буде створено насичене мультисередовище, до якого залучаються різні канали сприйняття інформації — текстової, візуальної, аудіоінформації, що дозволить інтенсифікувати навчальний процес та готувати здобувача вищої освіти до майбутньої роботи у професійній сфері, з урахуванням відомого останнім часом посилення тенденції конвергенції сенсорних каналів сприйняття інформації.

У сфері освіти, як і в усіх інших, сучасне використання інформаційних технологій пов'язане з переходом до обробки, зберігання та обміну інформацією у мережі. Інтернет став основою XXI ст. як століття інформаційних технологій, будучи глобальною інформаційно-телекомунікаційною мережею, що пов'язує інформаційні системи та мережі електрозв'язку різних країн за допомогою глобального адресного простору та надає можливість реалізації різних форм комунікації.

Інформаційні технології полегшують та спрощують процес навчання і роблять його зручнішим і доступнішим, виконуючи при цьому три взаємопов'язані між собою функції: діагностичну, навчальну та виховну.

Реалізувати процес підвищення якості сучасної вищої освіти засобами інформаційних технологій можна за допомогою управлінського та вже згаданого системного підходів.

Суть управлінського підходу полягає у прийнятті управлінських рішень із використанням цифрових технологій.

Системний підхід ґрунтується на організації взаємодії викладача та студента як елементів єдиної вельми складної системи за допомогою цифрових засобів навчання.

Можливості цифрових технологій дозволяють залучати до освітнього процесу певного завкладу вищої освіти викладачів-практиків, лекторів, учених, які перебувають за межами міста чи навіть країни.

Останніми десятиліттями дистанційні освітні технології у світі набули інтенсивного розвитку. Настала епоха інформатизації освітнього процесу. Сучасну фазу розвитку можна характеризувати як телекомунікаційну. Це — фаза спілкування, фаза трансферу інформації та знань. Навчання та робота сьогодні — синоніми: професійні знання старіють дуже швидко, тому потрібне їх постійне вдосконалення — це і є відкрита освіта [19].

Існують два види факторів використання дистанційного навчання у системі вищої освіти: зовнішні та внутрішні фактори.

Зовнішні фактори. В інформаційному суспільстві пріоритетним стає високий рівень освіченості його членів. Тільки високоосвічені люди здатні ефективно використовувати інформацію як продуктивний ресурс. Ефект «інформаційного вибуху» вимагає від кожного члена суспільства постійного оновлення своїх знань. Людині недостатньо «освіти на все життя», їй необхідна «освіта протягом усього життя». Навчання об'єктивно стає безперервним. Істотно змінюється і характер процесу навчання. Ці питання є вельми актуальними у масштабах всієї Земної кулі. ЮНЕСКО як провідна установа Організації Об'єднаних Націй, яка займається питаннями освіти, спільно зі спеціалізованим Міжнародним інститутом планування освіти надає країнам технічну підтримку у плануванні та аналізі політики в галузі освіти. Інститут ЮНЕСКО з навчання протягом усього життя (lifelong learning education) відіграє ключову роль у підтримці держав-членів щодо розробки політики у цій галузі.

Термін «безперервна освіта» багатозначний. По-перше, безперервна освіта означає постійне, безперервне вдосконалення знань, умінь, навичок людини, пов'язане з необхідністю бути актуальним у сучасному середовищі (професійному, соціальному). По-друге, під цим терміном розуміють систему поглядів на освітній процес загалом. Ця система розглядає навчальну діяльність як невід'ємну та основну складову способу життя в будь-якому віці; передбачає необхідність добудови освітніх сходів новими сходинками, розрахованими на всі періоди життя людини. По-третє, безперервна освіта передбачає постійне збагачення творчого потенціалу особистості, розвиток людини як творчої особистості. Безперервна освіта — процес цілісний, який складається зі ступенів, що послідовно прямують один за одним, спеціально організованої навчальної діяльності, яка створює людині сприятливі умови для життя [20].

Також необхідно відзначити важливість впровадження інформаційних технологій у глобальному аспекті:

по-перше, інформаційні технології створюють нові можливості для освіти, дають можливість охоплення широкого кола населення та задовольняють потреби особистості у прагненні до знань, підвищенні кваліфікації в обраній галузі та професійній діяльності;

по-друге, усувають бар'єри, пов'язані з доступом до необхідної інформації, та скорочують витрати під час обміну інформацією;

по-третє, сприяють залученню в країну інвестиції та просуванню прогресивних технологій у виробництві, управлінні, освіті;

по-четверте, інформаційні технології підвищують ефективність економіки та суттєво прискорюють темпи глобалізації тощо.

Усе це робить подальше використання інформаційних технологій вельми актуальним питанням та необхідною умовою успішного застосування дистанційних технологій навчання.

Внутрішні фактори. Сучасні вимоги, продиктовані реформуванням економіки та суспільства, карантинними обмеженнями, частішими природними катаклізмами призвели до значного збільшення ресурсомісткості навчального процесу. Надання освітніх послуг та організації системи підвищення кваліфікації і перепідготовки кадрів без урахування сучасних вимог та умов, що змінилися, призводить до суттєвого обмеження зростання обсягу навчального контингенту, знижує доступність та ефективність освітніх послуг і, як наслідок, звужує коло їх потенційних можливостей.

Для зниження ресурсомісткості навчального процесу, забезпечення більшої доступності навчання в закладах вищої освіти, навчальні технології повинні стати максимально ефективними, тобто такими, що забезпечують високий рівень економічності навчального процесу при вищій якості навчання. Необхідне широке застосування інноваційних методів навчання, що інтенсифікують навчальний процес. Все це можна досягти широким впровадженням у навчально-освітній процес сучасних педагогічних та новітніх інформаційних технологій.

Керуючись наведеним вище, можна виділити декілька основних причин створення та впровадження технології дистанційного навчання в системі освіти:

— розширення системи підвищення кваліфікації та перепідготовки кадрів;

— підвищення вимог до якості освіти;

— необхідність широкого надання освітніх послуг особам з обмеженими можливостями (інвалідам), соціальної адаптації;

— необхідність реалізації вимог до комфортності навчання та викладання за рахунок можливості як здобувача вищої освіти, так і викладача проводити процес навчання у зручний для себе час, у зручному місці та темпі. Нерегламентований відрізок часу на освоєння курсу надає здобувачам вищої освіти можливість освоєння курсу за менший або більший час у порівнянні з жорстко регламентованим за часом традиційним курсом. Для викладача зменшується частка аудиторного навчального завантаження;

— створення конкурентного середовища в освіті між традиційною і дистанційною освітою, що неодмінно стимулює підвищення якості освіти.

Таким чином, новий, інформаційний етап розвитку світової системи освіти є об'єктивним і незворотнім. Використання інформаційних технологій у навчанні, що відповідають світовому рівню, — основний та ефективний шлях розвитку вітчизняної системи освіти [21].

Дистанційне навчання має тривалу історію становлення, зазнаючи значних змін з часом: від навчання, яке проводилося за допомогою кореспонденції, до поступового використання аудіо- та відеоматеріалів та, нарешті, перехід до активного впровадження у процес навчання комп'ютерних технологій та Інтернету. Традиційно зародження дистанційного навчання пов'язують зі становленням The Open University у Великій Британії 1960-ті роки.

Дистанційне навчання є універсальною гуманістичною формою навчання, що базується на використанні можливостей електронного навчання та дистанційних освітніх технологій, які створюють для здобувачів вищої освіти умови вибору освітніх дисциплін основної та додаткової освіти, діалогового обміну з викладачем, при цьому процес навчання не залежить від розташування слухача у просторі та часі.

Дистанційні освітні технології — освітні технології, що реалізуються із застосуванням інформаційних та телекомунікаційних техно-логій при опосередкованій (на відстані) взаємодії студентів та педагогічних працівників. Електронне навчання — організація освітньої діяльності із застосуванням інформації, що міститься у базах даних, і використовуваної під час реалізації освітніх програм та інформаційних технологій, які забезпечують її обробку, технічних засобів, а також інформаційно-телекомунікаційних мереж, які забезпечують передачу лініями зв'язку зазначеної інформації, взаємодію здобувачів вищої освіти і педагогічних працівників.

Електронне навчання як самостійну форму навчання або складову дистанційного навчання можна розглядати як процес навчання та викладання з використанням електронних технологій, який забезпечує гнучкий доступ до навчальних ресурсів експертам, колегам, освітнім сервісам та послугам, і розкриває потенціал комп'ютерних технологій у можливості зробити навчання доступним у будь-який час і в будь-якому місці. У світовому науковому співтоваристві ще зовсім нещодавно склалася єдина точка зору на розуміння сутності та особливостей навчання в електронному середовищі. Але ж фахівці продовжують дискутувати на цю тему.

Одні дослідники вважають, що немає значної різниці між навчанням в електронному середовищі та традиційними формами навчання як на самому етапі навчання, так і на етапі отриманих у результаті навчання знань. Але інші, яких на даний момент більшість, вважають, що навчання в електронному середовищі є абсолютно новою парадигмою освіти, яка формується на основі особливої культури навчання. Тому важливим елементом навчання в електронному середовищі стає його організація та методологія, які значно відрізняються від традиційних форматів та методів навчання.

Сучасне електронне навчання включає в себе on-line навчання (навчання за допомогою Інтернету), off-line навчання (навчання за допомогою електронних носіїв, наприклад, мультимедійних компакт-дисків), m-learning (мобільне навчання за допомогою мобільних телефонів, смартфонів, планшетів, ноутбуків та іншого, що дозволяє також використовувати Інтернет).

Можливість створення повноцінного віртуального освітнього середовища обумовлюється врахуванням таких положень:

1) акцент на залежності пізнання (та набуття знання) від соціального контексту;

2) наявність тісного зв'язку між навчанням («дозріванням») та розвитком, що передбачає необхідність пов'язувати (або вибирати)

ту чи іншу форму навчання залежно від рівня розвитку, потреб здобувачів вищої освіти тощо;

3) необхідність розглядати здобувача вищої освіти як центральну фігуру та елемент системи навчання;

4) ефективність спільнот, що навчаються, у яких соціальна, педагогічна і когнітивна присутність забезпечують плідне середовище для розвитку особистості та трансформації життєвих перспектив як тих, хто навчає, так і тих, хто навчається.

Таким чином, електронне навчання є порівняно новою формою навчання та сферою наукових досліджень. Електронне навчання являє собою процес навчання та викладання з використанням електронних технологій, що забезпечує гнучкий доступ до навчальних ресурсів, експертів, колег, навчальних сервісів і послуг і розкриває потенціал комп'ютерних технологій у можливості зробити навчання доступним у будь-який час і в будь-якому місці. В Україні сьогодні активно розвивається тема електронного навчання, що підтверджується відкриттям нових порталів, присвячених цій темі, державних програм.

Однак як результат впровадження електронного навчання в Україні необхідно не лише розглядати кількісні показники ефективності його впровадження, а й аналізувати психологічні складові електронного навчання. Жодна з програм електронного навчання не буде ефективною, якщо не враховано психологічний аспект, не створено психологічних умов для електронного навчання, у тому числі й психологічних умов автентичності порозуміння учасників освітнього процесу.

В Україні вже створено істотну основу майбутньої системи дистанційного навчання, як форми навчання з використанням комп'ютерних і телекомунікаційних технологій, які забезпечують інтерактивну взаємодію викладачів та здобувачів вищої освіти на різних етапах навчання і самостійну роботу з матеріалами інформаційної мережі [22].

Сьогодні в нашій країні, як і в усьому світі, багато аспектів нашого життя вже перенесено в мережу, що прискорює темпи розвитку інформаційного суспільства. У сфері освіти достатньо довгий час, ще з радянських часів існує заочна форма навчання студентів, але її можливості були дуже обмежені. Нові інформаційні технології, Інтернет дають змогу зробити заочне навчання повноцінним та всеохоплюючим і разом з очним навчанням забезпечити його цифровими за-собами надання навчального матеріалу здобувачу вищої освіти; цифровими засобами контролю успішності здобувача вищої освіти; цифровими засобами консультації здобувача вищої освіти програмою-викладачем; цифровими засобами інтерактивної співпраці викладача і здобувача вищої освіти; можливістю швидкого доповнення курсу новою інформацією, коригування помилок та ін.

У наш час в умовах економічних відносин і жорсткої конкуренції на ринку праці особливе значення мають знання, навички та досвід. Фахівець XXI століття — це людина, яка вільно володіє сучасними інформаційними технологіями, постійно підвищує і вдосконалює свій професійний рівень. Набуття нових знань і навичок, практично корисних і застосовуваних у роботі в епоху інформаційного суспільства, значно розширює можливості самореалізації і сприяє кар'єрному росту. Проте однією з головних перешкод, що виникає на шляху тих, хто бажає продовжити навчання (враховуючи, що більшість з них вже працює), є брак часу. Більшість не має можливості приїжджати кожного дня на заняття до навчального закладу. Іншою значною перешкодою є відстань. Якщо навчальний заклад розташований в іншому місті, часто відвідувати заняття також незручно і дорого.

«Класична» заочна форма навчання часто не виправдовує свого призначення. Знання, що отримує студент, часто є поверховими, а самі заняття непродуктивними. Крім того, навчальний процес продовжується досить довго.

Дистанційна заочна освіта має такі переваги перед класичною:

— гнучкість — можливість викладення матеріалу курсу з урахуванням підготовки, здібностей студентів. Це досягається створенням альтернативних сайтів для одержання більш детальної або додаткової інформації з незрозумілих тем, а також низки питань-підказок тощо;

— актуальність — можливість упровадження новітніх педагогічних, психологічних, методичних розробок;

— зручність — можливість навчання у зручний час, у певному місці, здобуття освіти без відриву від основної роботи, відсутність обмежень у часі для засвоєння матеріалу;

— модульність — розбиття матеріалу на окремі функціонально завершені теми, які вивчаються у міру засвоєння і відповідають здібностям окремого студента або групи загалом;

— економічна ефективність — метод навчання дешевший, ніж традиційні, завдяки ефективному використанню навчальних примі-

щень, полегшеному коригуванню електронних навчальних матеріалів та мультидоступу до них;

— можливість одночасного використання великого обсягу навчальної інформації будь-якою кількістю студентів;

— інтерактивність — активне спілкування між здобувачами вищої освіти і викладачем, що значно посилює мотивацію до навчання, поліпшує засвоєння матеріалу;

— більші можливості контролю якості навчання, які передбачають проведення дискусій, чатів, використання самоконтролю, відсутність психологічних бар'єрів;

— відсутність географічних кордонів для здобуття освіти. Різні курси можна вивчати в різних навчальних закладах світу.

На Заході ця форма з'явилася вже досить давно і має велику популярність серед студентів через її економічні показники і навчальну ефективність. Дистанційну форму навчання там називають, як вже зазначалося, «освітою протягом усього життя» через те, що більшість тих, хто навчається, — дорослі люди. Багато хто з них вже має вищу освіту, проте через необхідність підвищення кваліфікації або розширення сфери діяльності у багатьох виникає потреба швидко і якісно засвоїти нові знання і набути навички роботи. Саме тоді оптимальною формою може стати дистанційне навчання.

У системі дистанційного навчання виділено чотири типи суб'єкта:

1. Здобувач вищої освіти — той, хто навчається.

2. Тьютор (викладач) — той, хто навчає.

3. Організатор — той, хто планує навчальну діяльність, розробляє програми навчання, займається розподіленням студентів за групами і навчальним навантаженням на тьюторів, вирішує різні організаційні питання.

4. Адміністратор — той, хто забезпечує стабільне функціонування системи, вирішує технічні питання, слідкує за статистикою роботи системи.

Важливим елементом дистанційного навчання є дистанційний курс. Ще до початку навчання тьютори розробляють дистанційний курс за своїми предметами. В процесі навчання курси можуть змінюватися і доповнюватися. Кожний викладач має змогу сам вирішувати, як буде виглядати дистанційний курс і які мультимедійні елементи в ньому будуть застосовуватися. Міра і спосіб використання комп'ютерних технологій при підготовці дистанційного курсу значно впливають на ефективність його засвоєння. Світовий досвід показує, що використання динамічних об'єктів для створення наочних моделей процесів, адаптивне моделювання здобувача вищої освіти в багатьох випадках значно підвищує навчальний ефект.

Курс розбивається на розділи, які потрібно проходити у визначений час. За матеріалом розділів тьютори створюють і призначають тести і завдання, які також потрібно вчасно проходити. Тьютор має можливість призначати спеціальні перевірочні (граничні) тести за відповідними розділами курсу. Тьютор може призначати завдання для підгруп здобувачів вищої освіти, тоді завдання розв'язується колективно.

Взаємодія між суб'єктами системи дистанційного навчання здійснюється за допомогою системи індивідуальних гостьових книг, форумів, чатів та електронної пошти.

Для організації дійсно ефективного навчального процесу дистанційного навчання необхідна систематична робота з сторінкою як здобувача вищої освіти, так і тьютора майже кожного дня протягом всього терміну навчання.

В інформаційному суспільстві в умовах колосального інформаційного навантаження у сфері освіти системний підхід до організації навчального процесу може бути реалізований через модульну технологію навчання, яка повністю сумісна з сучасними інформаційно-комунікаційними технологіями [23].

Модульне навчання — це така інструментальна форма організації навчального процесу, коли здобувачі вищої освіти працюють із навчальним середовищем, складеним з навчальних модулів, у режимі активної самоосвіти за варіативними або індивідуальними освітніми маршрутами. Технологія модульного навчання є одним з напрямків індивідуалізованого навчання та дозволяє організувати процес саморозвитку та самонавчання, регулювати темп навчання та зміст навчального матеріалу.

Інформаційний навчальний матеріал у модульному навчанні має бути організований у вигляді чіткої ієрархічної структури та наданий здобувачам вищої освіти у всіх можливих кодах: графічному, числовому, символічному та словесному. На цьому фундаменті формується цілісне системне сприйняття навчальної інформації. У модульному курсі має бути організовано інструментальне навчальне середовище, що включає набір інформаційно-методичних матеріалів та інтерактивних навчальних моделей для організації самостійних навчальних дій.

При цьому інформаційний навчальний гіперпростір є посередником між викладачем та здобувачем вищої освіти. Таке навчальне середовище моделює діяльність педагога і забезпечує можливість організації навчального процесу в режимі самонавчання, саморозвитку, самоорганізації. Навички самоосвіти та самостійної роботи з різними джерелами інформації в інтерактивному навчальному середовищі є основою для розвитку здібностей до навчання. Таке навчання побудоване на самомотивації, є розвиваючим та формує у здобувача вищої освіти навички самоврядування навчальної діяльності. Тому універсальні навчальні дії максимально ефективно та швидко формуються у здобувачів вищої освіти саме у модульному освітньому середовищі.

У циклі модульного навчання здобувачі вищої освіти організують навчальну діяльність відповідно до поставлених особистих цілей, використовуючи для цього необхідне інформаційно-методичне забезпечення та рекомендовані алгоритми навчальних дій. Вихідні результати після вивчення чергового модуля стають вхідними під час переходу до наступного циклу навчальної діяльності. При цьому проводиться коригування особистих навчальних цілей.

Поняття зворотного зв'язку є фундаментальним у кібернетиці. Від якості зворотного зв'язку залежить ефективність управління навчальним процесом. У кожному навчальному модулі канали зворотного зв'язку можуть бути побудовані в кібернетичних системах «здобувач вищої освіти — знання» та «викладач — здобувачі вищої освіти». Функція зворотного зв'язку може бути вбудована в рейтингову електронну таблицю. При цьому здобувачі вищої освіти у навчальному процесі за програмою модульного курсу мають можливість самостійно вимірювати обсяг навчальних дій, враховувати результати самоконтролю на етапах сприйняття, обробки інформації та у процесі виконання вправ у режимі тренування. За підсумками цих записів з'являється можливість системного аналізу структури навчальних процесів. При цьому здобувачі вищої освіти вчаться використовувати канал зворотного зв'язку, формують у себе навички самоврядування навчальним процесом і ці навички управління автоматично переносяться на інші сфери життя (управління своїм здоров'ям, емоційним станом, часом, грошовими потоками, ресурсами, іншими людьми та ін.).

У рейтинговій таблиці також має бути передбачена можливість введення даних про результати виконання контрольних завдань для визначення рівня засвоєння знань та вмінь здобувачів вищої освіти. Це вже канал зворотного зв'язку для викладача.

При недостатності зворотних зв'язків між викладачем та здобувачами вищої освіти протягом семестру виникає взаємне нерозуміння та погіршення якості навчання. Викладач повинен почути здобувачів вищої освіти у разі виникнення складнощів сприйняття навчальної інформації чи неприйняття методики викладання. Поняття прямих та зворотних зв'язків можна співвіднести з вертикальною та горизонтальною комунікацією між учасниками навчального процесу у закладах вищої освіти [24].

У рамках навчального процесу закладу вищої освіти системи «викладач — здобувач вищої освіти» передача інформації йде за вертикальним принципом. Можливість горизонтальної передачі з'являється у зв'язку з розвитком інтернет-технологій та соціальних мереж.

Зворотній зв'язок від студентів можливий завдяки використанню спеціалізованих ресурсів, де здобувачі вищої освіти виставляють своїм викладачам оцінки за певними критеріями та пишуть відгуки. Все це сприяє прозорості взаємин між викладачем та студентами та перешкоджає зловживанням з боку викладачів.

При нерівномірності прямих та зворотних інформаційних потоків викладач навчає здобувачів вищої освіти протягом семестру, а потім організує проміжний контроль. Найчастіше при екзаменаційному випробуванні здійснюється визначення рівня засвоєння інформації як відповідність між змістом навчального матеріалу та його відтворенням здобувачем вищої освіти. В результаті частіше визначається ступінь запам'ятовування, але не розуміння здобувачами вищої освіти змісту навчальної дисципліни. Вирішенням цієї проблеми може бути зміна підходу до проміжного та підсумкового контролю, коли питання до іспиту формулюються таким чином, щоб здобувач вищої освіти продукував нове знання.

Системний аналіз навчальної діяльності проводиться на основі даних самомоніторингу самостійних навчальних дій та вимірювання навчальних досягнень за результатами виконання контрольних завдань. У результаті системного аналізу учні можуть отримати таку інформацію:

— фактичні дані про структуру та обсяг самостійних навчальних дій;

— дані про структуру витрат навчального часу;

— причинно-наслідкові зв'язки (навчальна діяльність — навчальні результати);

— графічне відображення даних системного аналізу в вигляді графіків, діаграм.

Такий підхід до організації навчального процесу, побудований на основі організації самостійних навчальних дій здобувачів вищої освіти, забезпечує можливість формування та розвитку у них системи універсальних навчальних дій.

Напрямки розвитку та вдосконалення електронних освітніх ресурсів, сумісних з модульним та дистанційним навчанням, можуть бути такими.

У модульному курсі має бути організовано системне інструментальне навчальне середовище, що включає набір інформаційно-методичних матеріалів та інтерактивних навчальних моделей для організації самостійних навчальних дій. Такий інформаційний навчальний простір є посередником між викладачем та здобувачем вищої освіти, який моделює діяльність педагога, що забезпечує можливість організації навчального процесу в режимі самонавчання, саморозвитку, самоорганізації.

Реалізація принципу системності означає, що навчальні модулі у складі навчального курсу повинні включати цифрові освітні ресурси в різних форматах та інтерактивні керуючі інструментальні засоби, що забезпечують можливість організації повного навчального циклу, реалізації всіх самостійних навчальних дій, необхідних для досягнення навчального результату.

У цьому навчальному циклі обов'язково має бути передбачено таке:

— сприйняття та обробка інформації;

— тренування з метою формування навчання;

— контроль навчальних результатів;

— можливість самоаналізу та самоврядування навчальною діяльністю.

В інструментальне навчальне середовище мають бути вбудовані інтерактивні або друковані форми для моніторингу та системного аналізу навчального процесу.

Необхідно, щоб інтерактивне навчальне середовище мало відкриту модульну структуру та забезпечувало можливість зміни інтерфейсів, модифікації всіх компонентів системи, розширення функцій, проектування різних варіантів освітніх маршрутів та ін.

Відкритість структури та інтерфейсу інструментального середовища дозволяє організувати колективну творчість викладачів та здобувачів вищої освіти. Це інноваційний підхід до впровадження методу проектів у навчальний процес у закладах вищої освіти.

Реалізація повного навчального циклу на принципах системності та модульності забезпечує створення повноцінного інтерактивного навчально-методичного продукту як інструментального навчального засобу, на основі застосування якого можливе впровадження елементів модульного навчання в освітній процес.

Комп'ютерні навчальні засоби, організовані на описаних принципах, забезпечують системне сприйняття здобувачами вищої освіти змісту навчального курсу та сприяють розвитку у них системного мислення, що є важливою метою в циклі навчання. На основі застосування подібних інтерактивних програм можна побудувати навчальний процес у режимі активної самоосвіти, саморозвитку, самоврядування.

Можна побудувати ефективніший навчальний процес, якщо комп'ютерна навчальна програма безпосередньо взаємодіятиме зі здобувачами вищої освіти і виконуватиме функції інтерактивного тренажера.

Модульний підхід до побудови навчального середовища дає можливість організувати колективну творчу діяльність здобувачів вищої освіти та за їх допомогою поступово наповнювати освітніми ресурсами структури навчальних модулів. При цьому в модульному навчальному процесі автоматично реалізується можливість застосування методу проектів у навчальний процес. Після розробки ієрархічного графа — структури навчального модуля — викладач має можливість на цій основі підготувати велику кількість технічних завдань на розробку міні-проектів та подати їх у формі навчальних завдань різного рівня складності.

Метод проекту освітньої технології, що розвивається, будується на активній освітній діяльності здобувачів вищої освіти, а не на пасивне сприйняття інформації. Проектування освітніх ресурсів організовано на сприйняття і самостійній обробці навчальної інформації, саме тому метод проектів сприяє формуванню у здобувачів вищої освіти системи універсальних навчальних дій, оскільки тільки в цій освітній технології забезпечений творчий системний підход до організації самостійних навчальних дій.

Використання системного підходу до освіти передбачає вдосконалення процесів на всіх рівнях навчання.

За нестачі системного підходу у закладі вищої освіти спеціаліст формується стихійно, часто відсутня самоідентифікація майбутнього спеціаліста. Тому профорієнтацію необхідно продовжувати під час навчання (при проведенні лекційних та практичних занять), інакше виникає безцільність навчання.

Під час навчання здобувачів вищої освіти на молодших курсах важливо дати їм правильний напрямок — «навчити вчитися». Необхідно розділити два поняття: «навчання» та «вчення».

Навчання — це конкретніша і краще опрацьована сфера діяльності, і спрямоване воно на здобувачів вищої освіти з боку викладачів. Це методики навчання та підходи до організації навчального процесу.

Методика вчення — це мало формалізована галузь і її можна віднести до мистецтва.

Виходом тут може бути спільна з викладачем діяльність із вироблення методики вчення. При розробці методики навчання для здобувачів вищої освіти, які не володіють методикою вчення, потрібно врахувати таке: завдання мають бути гранично конкретними, дуже докладними та адресними, тобто для кожного здобувача вищої освіти необхідно виділяти завдання, яке необхідно виконати в певний термін.

В результаті ефективної роботи закладу вищої освіти, узгодженої роботи викладача та здобувача вищої освіти, взаємодії здобувачів вищої освіти між собою у процесі навчання формуються чотири види інтелекту, якими володіє людина: фізичний інтелект (або інтелект тіла), ментальний інтелект, емоційний інтелект та духовний інтелект, і освітні технології повинні бути спрямовані на розвиток всіх видів інтелекту.

Потрібно визнати, що нинішні випускники наповнені інформацією, але не знають, як її застосувати на практиці.

Для досягнення високих результатів в освітній системі необхідний повний системний взаємозв'язок усіх етапів навчання у закладі вищої освіти.

В основі підготовки хороших фахівців має бути викладач, який інтегрує весь процес навчання (випускна кафедра, завідувач кафедри або гарант освітньої програми) — від прийому до закладу вищої освіти до працевлаштування випускника та подальшого його супроводу (наприклад, підвищення кваліфікації у процесі кар'єрного зростання).

Важливо відзначити, що універсального інструменту оцінки рівня діджиталізації освіти бути не може, якість освітнього процесу зали-шається незмінною в умовах формування компетентнісного підходу до підготовки професіонала.

Завдяки новим інформаційним технологіям застосовуються інтерактивні методи навчання, можливе індивідуальне, диференційоване, різнорівневе навчання. Різноманітні і засоби навчання, що застосовуються у навчальному процесі: це інтерактивні дошки, комп'ютери, проектори, програмні продукти, додатки. У професійній підготовці інформаційні технології проникли не тільки в область теоретичного навчання, а й у сферу практики, а також в область оцінки знань і умінь здобувачів вищої освіти.

Інформаційні технології у вищому навчальному закладі дозволяють найбільш ефективно організувати діяльність людей (навчальний процес) та доступ до цифрових даних. Виходячи з цього в будь-якому закладі вищої освіти можна виділити три основні компоненти ІТ-рішень, між якими існує тісний взаємозв'язок, — це люди, процеси та дані. З погляду управління, від того, наскільки добре вирішені та організовані процеси, забезпечений зв'язок людей та даних, багато в чому залежить успішна діяльність закладу вищої освіти [7].

Істотний вплив на впровадження нових інформаційних технологій здійснюють сучасні інформаційні технологічні тенденції, основні з яких такі:

— віртуалізація та «хмарні» технології;

— розширення використання сервісорієнтованих архітектур;

— впровадження мобільних пристроїв та рішень на корпоративному рівні для доступу до ресурсів та виконання корпоративних додатків;

— посилення диференціації користувацьких переваг;

— візуалізація, вебінари та відеоконференцзв'язок [25].

Одним із важливих шляхів забезпечення ефективного функціонування освітньої системи при впровадженні нових інформаційних технологій у навчальний процес у закладах вищої освіти є створення та використання електронних навчально-методичних комплексів. Усі документи у складі цього інформаційного комп'ютерного продукту — мультимедійні, у них завжди присутні елементи інтерактивності, вони можуть бути оформлені в вигляді набору веб-сторінок. Електронні навчальні комплекси можуть бути використані на лекційних заняттях (показ відеозаписів, інтерактивних моделей та анімацій), під час проведення віртуальних лабораторних робіт, атестації та самостійної роботи здобувачів вищої освіти. Таким чином, електронний навчально-методичний комплекс — це програмний мультимедіапродукт навчального призначення, що забезпечує безперервність та повноту дидактичного циклу процесу навчання та містить організаційні та систематизовані теоретичні, практичні, контролюючі матеріали, побудовані на принципах системного підходу, інтерактивності, інформаційної відкритості, дистанційності процедур оцінки знань.

Системний підхід при організації навчального процесу у закладі вищої освіти з використанням нових інформаційних технологій забезпечує проведення інформатизації закладу вищої освіти як здійснення комплексу заходів, спрямованих на покращення його діяльності як системи засобами інформаційних технологій [26].

Щоб підвищити ефективність роботи закладу вищої освіти, потрібно комплексно впливати на систему в цілому — стратегію, мережеву інфраструктуру, організаційну структуру, систему управління, систему мотивації до праці, корпоративну культуру. Для вирішення завдання інформатизації закладу вищої освіти необхідно створити його єдину електронну систему, яка б дозволила управляти знаннями, що забезпечило б розвиток інновацій, збільшення продуктивності праці шляхом скорочення часу пошуку потрібного рішення в управлінні та обсягу виконаних робіт, підвищення компетентності персоналу. В результаті користувачі отримають доступ до високоякісної інформації, а самі рішення в галузі інформаційних технологій будуть так вплетені в основні ділові процеси закладу вищої освіти, що персонал і здобувачі вищої освіти вже не зможуть обходитися без сервісів, що надаються інформаційним середовищем. При цьому підвищується ефективність виконання персоналом його посадових обов'язків, підвищується якість навчання здобувачів вищої освіти, що робить інвестиції в інформаційні технології економічно виправданими [7].

ІТ-місія закладу вищої освіти має бути базою для формування ІТ-стратегії та створення системи всередині ІТ-стандартів закладів вищої освіти. Під ІТ-стратегією розуміють формалізовану систему принципів, на основі яких формуються концепція інформатизації, основні вимоги та план розвитку інформаційних технологій у закладі вищої осіти. Стратегія забезпечує системний підхід до інформатизації та узгодження з пріоритетами розвитку закладу вищої освіти в цілому.

Основні стратегічні цілі інформатизації:

— забезпечити гідне положення закладу вищої освіти серед інших у галузі інформаційних технологій;

— розвинути нові форми та покращити якість освітніх послуг;

— підвищити віддачу від застосування інформаційних технологій в управлінні закладом вищої освіти та у навчальному процесі на основі узгодження бізнес-стратегії зі стратегією інформатизації, а також шляхом оптимізації інвестиційних, організаційних та технологічних рішень;

— знизити сукупну вартість володіння IT-ресурсами за рахунок покращення керованості ресурсами;

— підвищити ефективність управління закладом вищої освіти та покращити якість інформаційних сервісів, а також їх доступність для користувачів на основі моделі єдиної інформаційної системи закладу вищої освіти;

— знизити можливості для зловживання навчального персоналу щодо здобувачів вищої освіти та адміністративно-управлінського персоналу щодо викладачів та співробітників на основі впровадження систем комп'ютерного тестування, електронного документообігу, контролю за виконанням управлінських рішень, регламентованого доступу до управлінської та навчальної інформації;

— підвищити економічну ефективність застосування інформаційних технологій у вищому навчальному закладі [27].

До основних напрямів інформатизації можна віднести такі:

— IT-інфраструктура: обладнання, лінії та канали передачі даних, обчислювальна мережа, системне програмне забезпечення, бездротовий доступ до ресурсів;

— IT-рішення: комплексні проекти на основі інформаційнихз технологій, інформаційні системи та сервіси, інформаційні середовища, геоінформаційні технології;

— методологія застосування інформаційних технологій: інформаційні моделі бізнес-процесів у закладі вищої освіти, модель єдиної інформаційної системи, методика оцінки ефективності застосування інформаційних технологій, основні показники застосування інформаційних технологій, узгоджені з ключовими показниками результативності діяльності закладу вищої освіти, корпоративний стандарт на порядок розробки, впровадження та застосування інформаційних технологій у закладі вищої освіти, положення та регламенти;

— IT-служба: оргструктура, управління, взаємовідносини з іншими підрозділами [28].

До основних принципів інформатизації слід віднести:

— розвиток інфраструктури інформаційних технологій закладу вищої освіти як окремої хмари з віртуалізацією не лише серверів, а й клієнтів;

— розвиток інформаційного середовища на основі концепції інтеграції ресурсів та автоматизації бізнес-процесів;

— постійне вдосконалення процесів, що реалізуються ІТ-службою;

— постійне вдосконалення використання ІТ на основі оцінки ефективності їх застосування у закладі вищої освіти;

— фінансування інформаційних технологій визначається прийнятими стратегічними завданнями розвитку закладу вищої освіти;

— політика безпеки інформаційного середовища будується на основі принципу розумної необхідності;

— розвиток процедур забезпечення якості корпоративних даних [7].

Усі заходи, створені задля реалізації ІТ-стратегії, можна інтегрувати у єдиний проект створення єдиної електронної системи, ядром якої є формування єдиної інформаційної системи закладу вищої освіти. Процес формування єдиної інформаційної системи закладу вищої освіти включає комплекс заходів із впровадження в усі сфери діяльності закладу вищої освіти інформаційних технологій як сукупності програмно-технічних засобів обчислювальної техніки, а також прийомів, способів та методів їх застосування під час виконання функцій збирання, зберігання, обробки, передачі та використання інформації.

Можна виділити такі основні завдання, виконання яких спрямоване на формування єдиної інформаційної системи закладу вищої освіти:

— формування організаційної структури інформатизації;

— створення інформаційної інфраструктури закладу вищої освіти та автоматизація її управління;

— інформатизація процесів управління закладом вищої освіти, зокрема фінансами;

— інформатизація навчального процесу;

— інформатизація наукових досліджень та проектів;

— підвищення рівня компетентності персоналу у сфері інформаційних технологій [29].

При створенні єдиної інформаційної системи закладу вищої освіти слід забезпечувати розумний обсяг інновацій як у навчальній, так і в управлінській діяльності. Створення та організація єдиної інформаційної системи закладу вищої освіти — складне організаційне та технологічне завдання, що обумовлює доцільність поетапної розробки системи: розв'язання задачі отримання на кожному етапі закінченого продукту, який послідовно модифікуватиметься та нарощуватиметься від етапу до етапу. Взаємне ув'язування зазначених підсистем та інтеграція даних досягається на основі організаційної, функціональної, технічної програмної та інформаційно-лінгвістичної сумісності. Тільки на такій основі може бути забезпечене стійке функціонування єдиної інформаційної системи вищої освіти.

Інформатизація вищої освіти в Україні є одним із пріоритетних напрямків реформування вищої школи. На шляху інформатизації навчального процесу важливим є створення, впровадження та розвиток комп'ютерно орієнтованого освітнього середовища на основі нових інформаційних технологій, систем, мереж та ресурсів.

Це — комплекс перетворень, пов'язаних із насиченням освітньої системи інформаційною продукцією, інформаційними засобами, що ґрунтуються на мікропроцесорній техніці, та новими інформаційними технологіями при всебічному використанні можливостей системного підходу як методологічної бази.

Ресурс системного підходу, інтегрованого застосуванням нових інформаційних технологій у процесі професійної підготовки, дозволяє організаторам та учасникам навчального процесу чітко усвідомлювати взаємозв'язок усіх компонентів освітньої системи та більш ефективно реалізовувати основні її функції: організацію, керівництво, контроль.

# ПРОЕКТУВАННЯ ІНФОРМАЦІЙНИХ СИСТЕМ
# І ПРОГРАМНИХ КОМПЛЕКСІВ

## UKRVECTŌRĒS ТА VHEALTH:
## ІНТЕЛЕКТУАЛЬНІ СЕРВІСИ ПІДТРИМКИ ДИСТАНЦІЙНОЇ
## МЕДИЧНОЇ РЕАБІЛІТАЦІЙНОЇ ДОПОМОГИ

*Величко В. Ю., Малахов К. С.*

*Методологія реабілітаційних заходів в умовах пандемії має ряд суттєвих особливостей, пов'язаних з непередбачуваністю і високою швидкістю виникнення проблем високої складності, обмеженістю спілкування між реабілітологом і пацієнтом, необхідністю високої реактивності прийняття рішень і їх відповідністю, масштабністю процесу і пов'язаною з нею необхідністю використання масштабованих операційних засобів тощо. Одним з ефективних рішень в наданні медичної реабілітаційної допомоги є дистанційна пацієнтцентрична реабілітація, яка потребує online-засобів теледіагностики, телеметрії і втручання з орієнтацією на можливості пацієнта, розвинутої Internet-взаємодії, інтелектуальних інформаційних технологій і сервісів, ефективних методів когнітивної підтримки в системі «реабілітолог — пацієнт — мультидисциплінарна команда», статистичної обробки великих об'ємів інформації тощо. Звідси поряд з традиційними засобами реабілітації у складі системи Трансдисциплінарної інтелектуальної інформаційно-аналітичної системи супроводження процесів реабілітації при пандемії TISP з'явилася Smart-система телемедичного супроводження реабілітаційних заходів. В поєднанні з інтелектуальними дистанційними засобами біологічного зворотного зв'язку і ефективними мініатюрними приладами теледіагностики, телеметрії і відновлення такі системи мають великі перспективи, про що свідчить також і світовий досвід. Мета дослідження полягала в розробці формальної моделі, програмної реалізації та методологічних засад застосування сервісів дистанційної пацієнтцентричної Smart-системи надання медичної реабілітаційної допомоги пацієнтам при пандемії, зокрема нової коронавірусної хвороби COVID-19. Світове сучасне та загальноприйняте визначення поняття Телереабілітація або Е-реабілітація — це комплекс реабілітаційних вправ і навчальних програм, які надаються пацієнту дистанційно за допомогою телекомунікаційних комп'ютерних технологій переважно на амбулаторному етапі лікування. Бурхливий розвиток телереабілітації у світі та набуття цим напрямком медицини трансдисциплінарних зв'язків*

*з різноманітними предметними галузями, що виходять за рамки сучасної парадигми Е-здоров'я, призвів до появи найсучаснішого різновиду реабілітації — Гібридна Е-реабілітація. Цей різновид Е-реабілітації складається з ряду фундаментальних методів, підходів та технологій: телекомунікаційні технології, телеметрія, вбудовані системи та мініатюрні «розумні» прилади для носіння, біологічний зворотний зв'язок, віртуальні особисті помічники, методи, технології та програмні застосунки на основі штучного інтелекту для обробки великих баз даних. Розроблено формальну модель, програмну реалізацію та методологічні засади застосування сервісів (UkrVectōrēs та vHealth) дистанційної пацієнтцентричної Smart-системи надання медичної реабілітаційної допомоги пацієнтам при пандемії, зокрема, нової коронавірусної хвороби COVID-19.*

*The methodology of rehabilitation measures in a pandemic has several significant features associated with the unpredictability and high rate of emergence of problems of high complexity, limited communication between the therapist and the patient, the need for high responsiveness of decision-making and their compliance, the scale of the process and the associated need to use scalable operating tools, etc. One of the most effective solutions in medical rehabilitation assistance is remote patient/personal-centered rehabilitation. It requires online telediagnostic tools, telemetry and interventions focused on the patient's capabilities, developed Internet interaction, intelligent information technologies, and services. Patient/personal-centered rehabilitation also needs effective methods in the "Physical therapist — Patient — Multidisciplinary team" system, statistical processing of large volumes of data, etc. Therefore, along with the traditional means of rehabilitation, as part of the "Transdisciplinary intelligent information and analytical system for the rehabilitation processes support in a pandemic (TISP)" the Smart-system for remote support of rehabilitation activities and services appeared. Combined with intelligent remote biofeedback devices and effective miniature telediagnostics, telemetry and recovery devices, such systems hold great promise, as evidenced by world experience as well. Objective of the research was to develop a formal model, software implementation, and methodological foundations for the use of services of a remote patient/personal-centered Smart-system for providing medical rehabilitation assistance to patients in a pandemic, in particular, the new coronavirus disease COVID-19. The world modern and generally accepted definition of the concept of Telerehabilitation or E-rehabilitation is a complex of rehabilitation exercises and training programs that are provided to the patient remotely using telecommunication computer technologies, mainly at the outpatient stage of treatment. The rapid development of telerehabilitation in the world and the acquisition by this direction of medicine of transdisciplinary connections with various subject areas that go beyond the modern paradigm of E-health, led to the emergence of the most modern type of rehabilitation — Hybrid E-rehabilitation. This type of E-rehabilitation consists of a number of the following fundamental methods, approaches and technologies: telecommunication technologies, telemetry, embedded systems*

*and miniature smart wearable devices, biofeedback, virtual personal assistants, methods, technologies and software artificial intelligence applications for big data processing.*

*The formal model, software implementation, and methodological foundations for the use of services (UkrVectōrēs, vHealth) of a remote patient/personal-centered Smart-system for providing medical rehabilitation assistance to patients in a pandemic, in particular, the new coronavirus disease COVID-19, have been developed.*

## Перелік скорочень

API — Application Programming Interface

CBOW — Continous bag of words

DWIM — Do What I Mean

PMI — Pointwise mutual information

SPA — Single-page application

SVD — Singular value decomposition

TISP — Трансдисциплінарна інтелектуальна інформаційно-аналітична система супроводження процесів реабілітації при пандемії

БЗЗ — Біологічний зворотний зв'язок

БК — Біла книга з фізичної та реабілітаційної медицини в Європі

ЗПМ — Зростаючі пірамідальні мережі

МІС — медична інформаційна система

МКФ — Міжнародна класифікація функціонування, обмеження життєдіяльності та здоров'я

НФДУ — Національний фонд досліджень України

ПЗ — Програмне забезпечення

ФРМ — Фізична і реабілітаційна медицина

ШІ — Штучний інтелект

**1. Теоретичні засади методик мовного (дистрибутивно-семантичного) моделювання в математичній лінгвістиці.** Сучасний етап розвитку штучного інтелекту (ШІ) характеризують як вибух у сфері можливостей та перспектив упровадження технологій та інструментів ШІ. Очікується трансформація цілих галузей промисловості на основі цього впровадження [1].

Такі складні процеси не можуть базуватися тільки на розумінні та застосуванні інструментів ШІ як засобів стимулювання інновацій чи забезпечення збільшення прибутків. Потрібно розуміти також і обмеження не лише технологічні, а й організаційні. Технології ШІ та розроблені інструменти ґрунтуються на певних моделях, алгоритмах та їхніх програмних реалізаціях. Якщо результат роботи таких

інструментів — це певне передбачення, рекомендація чи рішення, що впливає на суспільство, то потрібна додаткова інформація для безпечного використання цих інструментів. Користувач повинен розуміти, на основі чого чи радше чому алгоритм дав такий результат, які чинники і як вплинули на результат. На основі цього формується довіра до результатів, але потрібно подивитися всередину «чорної скриньки» [2].

Природна мова завжди була предметом досліджень в ШІ, а протягом останніх років доволі успішно створювали технології, які забезпечують опрацювання, автоматичне розуміння та генерацію тексту [1]. Аналіз реального впровадження та використання цих технологій у промисловості засвідчив, що їх застосовують насамперед для текстової аналітики (81 %), аналітики соціальних мереж (46 %), а також у створенні чат-ботів (англ. *Chatbot*) для взаємодії з клієнтами (40 %), розумних помічників (23 %) та класифікації документів (30 %). Під час використання технологій опрацювання природної мови потрібно розв'язувати проблеми побудови таксономій та виконання тренувань у машинному навчанні, які необхідні для реалізації більшості сучасних алгоритмів. Проблема у створенні систем на основі машинного навчання полягає в необхідності забезпечити потрібний обсяг даних для тренування та їхню якість, особливо коли йдеться про навчання з учителем. Побудова таксономій вимагає встановлення ієрархічних взаємозв'язків між одиницями інформації. Залежно від галузі, де така таксономія застосовується, взаємозв'язки можуть змінюватися.

Спосіб представлення слова у вхідних даних та в моделях мови все ще залишається важливим у більшості завдань опрацювання природної мови. Донедавна в системах опрацювання природної мови слова кодували рядками умовно довільних символів, а корисну інформацію щодо подібності та відмінності між словами не завжди використовували.

Векторні моделі відомі та їх використовують в опрацюванні природної мови з 50-х років минулого століття. Розроблені протягом останніх років алгоритми, методи та засоби побудови векторних представлень — це, з наукової позиції, подальший розвиток дистрибутивної семантики, а з позиції практичного застосування інструмент, який використовують для розв'язання завдань видобування іменованих сутностей, маркування семантичних ролей, автоматичного реферування, встановлення взаємозв'язків між сло-вами, у системах питання — відповідь тощо. Векторні моделі — це вже загальноприйнятий метод як представлення одиниць мови, так і обчислення семантичної подібності між ними [3]. Широке використання векторних представлень для розв'язання завдань з опрацювання природної мови вимагає розуміння цих інструментів, їхніх можливостей та обмежень. Вивченням теоретичних та практичних засад векторного моделювання природної мови займається дослідницька область — *дистрибутивна семантика* — це галузь математичної (або обчислювальної) лінгвістики, метою якої є кількісне оцінювання семантичної подібності та категоризація мовознавчих елементів на основі властивостей їх розподілу у великих вибірках мовних даних.

**1.1. Поняття векторного представлення — Word Embbedings.** Векторне представлення, або подання/вкладання (англ. *Word embedding / Distributed word representation*) — це техніка, яка розглядає слова як вектори, відносна схожість між якими корелює з семантичною подібністю. Воно є одним із найуспішніших прикладів застосування навчання без учителя (англ. *Unsupervised learning*). Векторні представлення — техніка для опрацювання природної мови, альтернативна до традиційної, яка дозволяє відображати сутності (слова, словосполучення, терміни) зі словника на вектори дійсних чисел в малому щодо розміру словника просторі, а подібність між векторами корелює з семантичною подібністю між сутностями [1].

Значення сутностей, які трапляються (вживаються) в подібних контекстах, мають тенденцію до подібності. Такий зміст має формулювання (так звана дистрибутивна гіпотеза), яке запропонували у 50-х роках минулого століття Зелліг Саббеттай Гарріс (1954) та Джон Руперт Фірт (1957), коли розвинулися дистрибутивні методи, в яких значення сутності (в даному випадку сутністю є слово) обчислюється з розподілу сутностей навколо нього. Сутність в такому разі представляється як вектор (масив чисел), котрий обчислюється в певний спосіб [3].

Словник слів за такого підходу — це не множина слів, які представлені рядком символів із відповідним індексом, а множина векторів у просторі. Додавання нового слова в такий словник — це не просто додавання нового рядка, а складніший процес; звідси походить термін *Word embedding* — вбудовування (вкладання) сутності (слова) у векторний простір. Окреме слово проходить процес відображення з власного багатовимірного простору його контекстів у векторний простір малого розміру.

У найпростішому випадку дистрибутивну модель значення слова, або просто вектор слова, можна побудувати на основі того, як часто воно трапляється разом з іншими словами. Зручним способом представлення такої інформації є матриця (англ. *Co-occurrence matrix*). Така матриця матиме однакову кількість рядків і стовпців. У комірках матриці будуть числа, які визначають, скільки разів слово, якому відповідає рядок матриці, зустрічається разом зі словом, якому відповідає стовпець матриці, в корпусі текстів. Числові значення обчислюють на основі оброблення корпусу текстів. Можна порахувати, скільки разів слова зустрічаються разом у документі чи тексті або його частині (параграф, абзац), але переважно використовують контекстне вікно певного розміру. Наприклад, в таблиці 1 зображено контекстне вікно для п'яти слів із фрагмента корпусу проблеми поетики творчого доробку Олеся Гончара. Розмір цього контекстного вікна становить 11 слів (центральне слово та по п'ять слів перед та після нього).

Таблиця 1

**Контекстне вікно сутностей (слів)**

| Не можу ж я забути про | **товариша** | по роботі, в значній мірі вчителя |
|---|---|---|
| не подумай, що гумор і | **собрата** | по стремлінню |
| Посідаю посаду | **редактора** | групової стінгазети молодий журналіст |
| В одній групі зі мною | **Столяренко** | пам'ятаєш, торік з комуніста баба приїжджала |
| Багато пройшло | **часу** | відтоді як я тобі послав листа |

В таблиці 2 подано відповідний фрагмент матриці, яка представляє спільне вживання слів у корпусі проблеми поетики творчого доробку Олеся Гончара.

Таблиця 2

**Фрагмент матриці, яка представляє спільне вживання сутностей (слів) у корпусі текстів**

| | ... | собрата | ... | посаду | журналіст | Столяренко | ... |
|---|---|---|---|---|---|---|---|
| товариша | | 3 | | 1 | 2 | 3 | |
| собрата | | 0 | | 1 | 1 | 2 | |
| редактора | | 1 | | 3 | 3 | 2 | |
| Столяренко | | 2 | | 2 | 0 | 0 | |
| часу | | 1 | | 0 | 0 | 1 | |

Фрагмент матриці демонструє певну подібність між словами «собрата» та «товариша», оскільки слово «Столяренко» трапляється в контекстних вікнах цих слів. Особливу увагу потрібно звернути на те, що більшість значень у цьому фрагменті — нулі й ця тенденція зберігається для всієї матриці. Отже, довжина вектора для кожного слова буде дорівнювати розміру словника корпусу текстів і більшість елементів цього вектора будуть нулями. У фрагменті корпусу проблеми поетики творчого доробку Олеся Гончара розмір словника становить 15 000 сутностей — слів, а якщо брати національні корпуси текстів, то це значення збільшиться до десятків мільйонів. На практиці такі вектори використовувати складно не тільки через їхню розрідженість, а й через те, що абсолютні значення частоти є не надто інформативною мірою спільного вживання слів [3]. На практиці використовують міру на основі поточкової взаємної інформації (англ. *Pointwise mutual information, PMI*) або позитивної PMI (англ. *Positive PPMI*) та їхніх варіантів, що дозволяє записати в комірки матриці значення, які вказують, як часто два слова зустрічаються разом порівняно з тим, коли їх можна побачити незалежно одне від одного. Побудовані векторні представлення слів дозволяють оцінити їхню подібність на основі зіставлення їхніх векторів. Мірі подібності векторів відповідає косинус кута між векторами, і ця міра відома як косинусна подібність (англ. Cosine similarity). На рисунку 1 наведено вектори для слів «журналіст» та «собрат», та позначено кут між ними. Що менший кут між векторами, то більше значення має косинус цього кута, і слова, яким відповідають ці вектори, вважають більш подібними. Косинусна подібність може набувати значення в діапазоні від −1 до 1: якщо значення дорівнює −1, то вектори протилежні; 1 — вектори збігаються (повна ідентичність контекстів); 0 — вектори ортогональні (відсутні схожі контексти). Відомі та використовуються й інші міри оцінки подібності, але міра на основі косинуса кута між векторами набула найбільшого поширення.

Значний розмір векторів та їхня розрідженість обмежують їхнє практичне використання. Для зменшення розмірності векторів і кількості нульових елементів у векторі, тобто для ущільнення вектора, розроблені окремі групи методів. Класичний метод, який використовують для зменшення розмірності векторів — це сингулярний розклад матриці (англ. *Singular value decomposition, SVD*). Застосування цього методу дозволяє зменшити розмір векторів до значень від 500 до 5000, але цей метод потребує виконання значної кількості до-даткових обчислень, і для деяких завдань обсяг обчислень стає співмірний із використанням повної PPMI матриці [3].

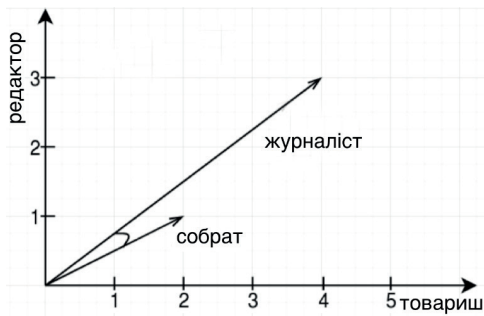

Рис. 1. Представлення векторів слів у двовимірному просторі

Вектори слів, які одержані в такий спосіб, представляють смислову та синтаксичну інформацію, але при цьому лишається багато проблем. Зокрема такі: значний розмір матриці ($> 10^6 \times 10^6$) та її розрідженість; складність внесення змін (додавання нових слів призводить до збільшення розміру матриці та повторного обчислення її елементів); висока обчислювальна вартість виконання SVD.

Одним з головних обмежень використання слів (векторних моделей слів у цілому) є те, що слова з кількома значеннями об'єднуються в єдине представлення (єдиний вектор в семантичному просторі). Іншими словами, багатозначність та омонімія не обробляються належним чином. Наприклад, в реченні «The club I tried yesterday was great!» не ясно, який сенс має термін «club»: «багатошаровий бутерброд», «бейсбольний клуб», «молитовня», «ключка для гри в гольф», чи будь-який інший сенс, який може мати слово «club». Необхідність розміщення декількох сенсів на слово в різних векторах (багатосенсові вкладення, англ. *Multi-sense embeddings*) стале мотивацією для розділення односенсових вкладень на багатосенсові.

**1.2. Методика та засіб мовного моделювання Word2vec.** Альтернативний підхід, який останніми роками бурхливо розвивається, передбачає використання нейронних мереж для моделювання природної мови. Модель мови на основі нейронної мережі дозволяє замість обчислення та зберігання величезних обсягів даних передбачати сутності — слова контексту для заданого слова і в процесі прогнозування одержувати щільні вектори слів. Word2vec — це найбільш відома й популярна технологія (набір методик та алгоритмів), що її побудував на основі такого підходу Томаш Міколов 2013 року та описав теоретичну і практичну частини у [4; 5]. Рисунок 2 ілюструє основну ідею Word2vec. Дано корпус текстів значного обсягу, і кожне слово зі словника цього корпусу представлене як вектор. Під час перегляду всіх текстів корпусу для кожної з позицій слова в реченні розглядають центральне слово (поточна позиція) та слово контексту. На основі подібності між векторами центрального слова та слова контексту обчислюють імовірність слова контексту для заданого центрального слова. Так само за заданими словами контексту обчислюють імовірність центрального слова. Основним завданням є підбір для слів таких векторних представлень, які максимізують ці ймовірності.

Корпус текстів (контекстне вікно = 5)                Пари слів

| листа | твого | одержав | з | трагічним | звітом | листа, твого; листа, одержав |
| листа | твого | одержав | з | трагічним | звітом | твого, ліста; твого, одержав; твого, з |
| листа | твого | одержав | з | трагічним | звітом | одержав, листа; одержав, твого; одержав, з; одержав, трагічним |
| листа | твого | одержав | з | трагічним | звітом | з, листа; з, листа; з, трагічним; з, звітом |

Рис. 2. Схема опрацювання корпусу текстів у Word2vec

Потрібно зауважити, що приймаються такі припущення: тексти в корпусі незалежні між собою; кожне слово залежить тільки від слів свого контексту; слова контексту незалежні одне від одного. Останнє припущення вважають недоліком технології Word2vec, оскільки не розглядаються відмінності в імовірності слова, якщо воно трапляються перед центральним словом і якщо це слово після центрального слова.

Прогнозування відбувається з використанням нейронної мережі. Здійснюється тренування простої нейронної мережі прямого поширення (англ. *Feedforward Neural Networks*) з одним прихованим шаром, але насправді мережу використовують з іншою метою. Метою тренування є отримання вагових коефіцієнтів прихованого шару, і ці коефіцієнти — це і є вектори слів.

У Word2vec реалізовано описаний вище підхід за допомогою моделей CBOW (англ. *Continous bag of words, CBOW*) та skip-gram [4; 6].

Skip-gram модель дозволяє отримати два окремі вектори для кожного слова: вектор для слова, як центрального слова контекстного вікна та вектор для цього самого слова, як слова контексту. Ці вектори формують дві матриці: матрицю слів та матрицю контекстів, які ви-користовують для розв'язання завдання прогнозування. Кожен рядок матриці слів — це вектор для слова зі словника слів корпусу текстів, а в матриці слів контексту вектором для цього ж слова буде відповідний стовпчик. За послідовного перегляду слів корпусу для кожного зі слів модель skip-gram дозволяє передбачити всі слова контекстного вікна, в якому поточне слово є центральним. Кожен такий прогноз можна розглядати як визначення ймовірності спільного вживання цих двох слів. Обчислення цієї ймовірності полягає в пошуку скалярного добутку двох векторів: вектора центрального слова і вектора слова контексту. Що більше значення скалярного добутку між векторами, то більш подібні вони між собою. Оскільки нормований скалярний добуток між векторами — це косинус кута між векторами, то його й використовують як міру подібності. Щоб зі скалярного добутку векторів одержати ймовірність, використовують нормовану експоненційну функцію softmax. Отже, модель skip-gram дозволяє обчислити ймовірність появи разом двох слів за допомогою знаходження скалярного добутку між векторами цих слів та перетворення його на ймовірність за допомогою нормованої експоненційної функції [7]. Описаний підхід має великий недолік: функція softmax потребує обчислення скалярного добутку вектора кожного слова зі словника зі всіма векторами інших слів словника. За використання корпусів належного обсягу зробити це безпосередньо практично неможливо. Модель CBOW, на відміну від skip-gram, дозволяє передбачити поточне центральне слово контекстного вікна на основі слів, які його оточують.

Вектори слів і контекстів формують за допомогою навчання без вчителя через максимізацію подібності між вектором поточного слова і векторами його сусідів та мінімізацію подібності з векторами інших слів. Для розв'язання завдання прогнозування, яке було розглянуто вище, ймовірність слова обчислюється як відношення скалярного добутку між вектором слова і вектором слова контексту до суми скалярних добутків векторів усіх слів. Замість знаходження величезної кількості скалярних добутків для обчислення знаменника в skip-gram використовують варіант skip-gram з негативною вибіркою (англ. *Negative sampling*), в якому знаменник обчислюється наближено [8].

На етапі тренування під час перегляду слів з корпусу для кожного слова вибирають слова з контексту як позитивні приклади, а для кожного позитивного прикладу вибирають також певну кількість прикладів шуму або негативних прикладів — слів, які не є сусідами поточного слова. Зокрема, якщо прийняти, що кількість негативних прикладів дорівнює двом, то для кожної з пар слово — слово контексту буде добрано по два слова шуму для кожного зі слів контексту. Наприклад, під час перегляду слів з наступного прикладу в таблиці 3 для поточного слова «вклали» буде дібрано шість негативних прикладів за умови, що контекстне вікно буде містити ще два слова зліва і два слова справа від цього слова.

Процес навчання починається з матрицями, значення в яких випадково згенеровані. Під час проходження по корпусу зміни значень у цих матрицях повинні забезпечити отримання такого вектора центрального слова, щоб його скалярний добуток з вектором кожного зі слів контексту був якнайбільшим. Додатково до цього потрібно, щоб вектори слів шуму мали малі значення скалярного добутку з вектором поточного слова. У такий спосіб відбувається генерація векторів. Результатом після тренування є вектори, які представляють семантичну та синтаксичну інформацію про слова.

Таблиця 3

**Формування негативної вибірки слів**

| Корпус | одержав я листа твого, прочитав і думав над ним цілий вечір; мені пригадався такий факт |
|---|---|
| Словник | одержав, я, листа, твого, прочитав, думав, цілий, вечір, мені, пригадався, факт |
| Пари слово — контекст | одержав (листа, твого, я) — одержав, листа; одержав, твого; одержав, я листа (твого, одержав, я, прочитав) — твого, одержав; прочитав, листа; одержав, я; я, прочитав |
| Негативна вибірка | одержав, вечір; одержав, пригадався; твого, мені ... |

Перевага техніки word2vec полягає в тому, що вона забезпечує високу ефективність обчислень. Програмний код є у вільному доступі, моделі швидко та ефективно тренуються, доступні вже готові векторні представлення слів для багатьох мов.

Відомі такі реалізації методів та алгоритмів для побудови векторних представлень:

— Оригінальна реалізація Word2vec; мова реалізації C; доступна для завантаження за посиланням [5];

— Medallia/Word2VecJava, мова реалізації Java; доступна для завантаження за посиланням [9];

— Spark MLLib Word2Vec; мова реалізації Java, доступна для завантаження за посиланням [10];

— Бібліотека Gensim Word2vec, FastText, мова реалізації Python, доступна для завантаження за посиланням [11];

— Google's TensorFlow Word2vec; мова реалізації Python; доступна для завантаження за посиланням [12];

— Бібліотека FastText; мова реалізації С++, доступна для завантаження за посиланням [13].

**1.3. Методика та засіб мовного моделювання fastText.** Створена лабораторією досліджень ШІ Фейсбук (англ. *Facebook's AI Research lab, FAIR*) бібліотека fastText — ще один серйозний крок у розвитку моделей дистрибутивної семантики природної мови (для навчання вкладень слів та класифікації тексту). В її розробці взяв участь Томаш Міколов, вже знайомий нам по Word2vec. Алгоритм fastText ґрунтується на працях [14, 15]. Для векторизації слів використовуються одночасно алгоритм skip-gram, алгоритм негативної вибірки Negative sampling та алгоритм безперервного мішка CBOW.

До основної моделі Word2Vec додана модель так званих символьних n-грам. Кожне слово представляється композицією декількох послідовностей символів певної довжини. Наприклад, слово they в залежності від гіперпараметрів може складатися з th, he, ey, the, hey. По суті, вектор слова — це сума всіх його n-грам.

Для отримання вектора слова для слова $w_t$ в даному наборі документів вводиться функція оцінки $s(w_t, w_t)$. Вона обчислює суму символів n-грам, помножених на навколишнє слово $w_c \in \{..., w_t - 2, w_t - 1, w_t + 1, w_t + 2, ...\}$ слова $w_t$. Кількість контекстних слів $w_t$ задається як довжина вікна виведення skip-gram. Skip-gram версія fastText розраховує умовну ймовірність [14] $p(w_c \mid w_t) = \dfrac{e^{s(w_t, w_c)}}{\sum\limits_{i=1}^{N} e^{s(w_t, w_c)}}$, замінивши результат скалярного множення між $u_c v_t^T$ на $s(w_t, w_t)$. Нехай $G$ словник n-грам і $G_t \subset \{1, ..., G\}$ це масив символьних n-грам для слова $w_t$. Тоді результат [14] $s(w_t, w_c) = \sum\limits_{g \in G_t} z_g^T c_c$ визначається для кожного контекстного слова $w_c$. Замість унітарно-кодованого слова вектора набір символів n-грам прогнозує контекстне слово у fastText. Усі вектори $z_g$ де $g \in G_t$ є символьними вкладеннями n-грам слова $w_t$. Їх набір будує представлення/вкладення слова $v_t$. Через те, що загальний набір n-грам для великого корпусу дуже великий, хешована версія символів n-грам використовується як вхідна інформація.

Точність fastText у завданнях обробки природної мови, таких як семантичний аналіз або класифікація за тегами, одна з найвищих серед найсучасніших методів.

**1.4. Відмінності моделей fastText та Word2vec.** Модель fastText є подальшим розвитком технології Word2vec, яка також дозволяє будувати векторні представлення. FastText ґрунтується на моделі skip-gram, яка реалізована в Word2vec. Основна відмінність моделі fastText від Word2vec полягає в тому, що в Word2vec кожне слово в корпусі розглядають окремо, як атомарний об'єкт, для якого будується вектор. У моделі fastText кожне слово розглядають як сукупність n-грам символів цього слова. Отже, вектор слова будується через суму векторів n-грам, з яких складається слово. Наприклад, при заданому мінімальному розмірі n-грами 3 і найбільшому розмірі n-грами 5, вектор слова «товариш» буде складатися з суми векторів таких n-грам: «^то», «тов», «ова», «вар», «ари», «иш^», «^тов», «това», «овар», «вари», «риш^», «^това», «товар», «овари», «ариш^».

Модель fastText дозволяє, на відміну від моделі Word2vec:

• Генерувати кращі вектори слів для слів, які рідко вживані. Навіть якщо слово нечасто трапляється в корпусі, то n-грами, з яких воно складається, можна побачити частіше як частини інших слів, що дозволяє згенерувати кращий вектор. Якщо вектор будується за допомогою Word2vec, то рідковживане слово (наприклад, 5 випадків уживання в корпусі) має меншу кількість сусідів порівняно зі словом, що трапляється частіше. В останньому є більше слів у контекстному вікні, а це забезпечує побудову кращого вектора для цього слова.

• Будувати вектори для слів, які не трапляються в корпусі. Вектор для такого слова буде складатися з n-грам символів, які є частинами інших слів, що наявні в корпусі.

• Будувати вектори слів для мов із багатою морфологією. Використання n-грам дозволить отримати точніші вектори для всієї морфологічної парадигми слова.

За умови використання моделі fastText велике значення має добір параметрів, зокрема мінімального й максимального розміру n-грам, бо це впливає на розмір корпусу. Оскільки побудова векторів слів відбувається за допомогою тренування на рівні n-грам, то збільшуються витрати часу порівняно з Word2vec.

**1.5. Застосування прогностичних моделей дистрибутивної семантики.** Сучасні векторні моделі дозволяють обчислити семантичну подібність між словами, реченнями чи документами, і саме на цих можливостях ґрунтується їхнє використання для розв'язання завдань опрацювання природної мови. Прогностичні моделі дистрибутивної семантики використовуються безпосередньо для вирішення широкого кола завдань, пов'язаних з семантичним моделюванням текстів природною мовою, а саме:

— виявлення семантичної близькості слів, словосполучень, текстів;

— розпізнавання іменованих сутностей;

— морфологічний аналіз слів;

— автоматична класифікація/кластеризація слів, словосполучень, текстів за ступенем їх семантичної близькості;

— автоматична генерація тезаурусів і двомовних словників;

— вирішення лексичної неоднозначності;

— розширення запитів за допомогою асоціативних зв'язків;

— визначення тематики тексту;

— класифікація/кластеризація текстових документів для інформаційного пошуку;

— отримання знань з неструктурованих джерел (текстових документів);

— автоматична побудова семантичних карт довільних предметних областей;

— визначення тональності текстів та висловлювань;

— моделювання селекційних обмежень слів.

Також із використанням векторних представлень вирішують завдання генерації текстів, машинного перекладу, виявлення парафраз, моделювання текстів.

Серед останніх відомих застосувань векторних представлень потрібно відзначити роботи [16; 17], в яких векторні моделі використовують для розв'язання завдань машинного перекладу. У роботах показано, як можна побудувати перекладний (англ. *Bilingual dictionary*) двомовний словник без використання паралельних корпусів текстів. Такий словник будують через вирівнювання векторних просторів за допомогою навчання без учителя. Для дванадцяти мовних пар розроблено словники досить високої якості, точність яких для окремих пар становить понад 60 %. Також указано, що штучно отримані словники успішно враховують багатозначність слів мовних пар. Векторні простори вирівнюють за допомогою пошуку відображення між незалежними векторними моделями для двох мов. Схематично цей процес автори роботи ілюструють так, як наведено на рисунку 3.

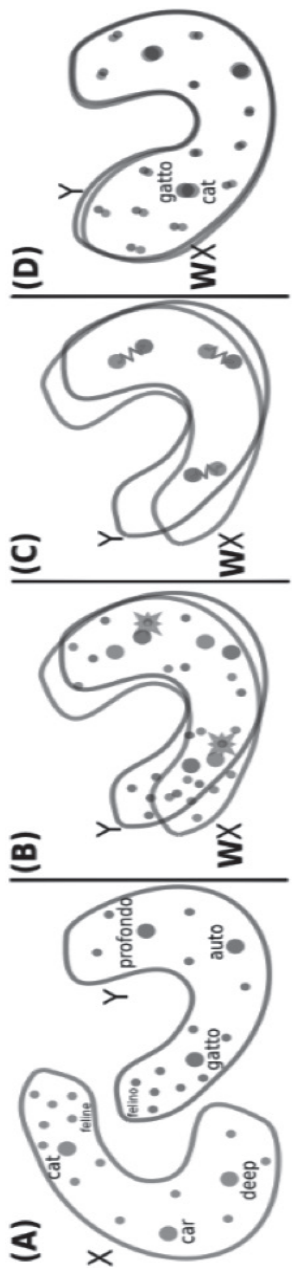

Рис. 3. Схема вирівнювання векторних просторів

Вирівнювання здійснюється між двома векторними просторами, які побудовані на основі моделі fastText із використанням Вікіпедії, як корпусу для тренування (A). Для побудови перекладних словників використовують тільки 200000 векторів найчастотніших слів. Кожне слово представлено на рисунку 3 точкою, а її розмір вказує на частоту слова в корпусі. Далі здійснюють пошук матриці повороту W, яка попередньо вирівнює два простори (B). Здійснюють пошук залежності, яка «притягує» слова з високою частотою вживання в корпусі, що дозволяє покращити вирівнювання (C). Знайдене відображення та додаткова метрика дає можливість здійснювати переклад слів (D).

Серед перекладних словників, які побудовані за допомогою векторних представлень, доступні також і українсько-англійський та англо-український словники. Обсяг цих словників становить 40722 та 47912 пар слів відповідно. Оскільки в такий спосіб перекладні словники ще не укладали, то для оцінки їхньої якості потрібно провести додаткові дослідження. Попередній аналіз було здійснено за допомогою перевірки наявності слів з цих словників у словнику проєкту ВЕСУМ — Великий електронний словник української мови [18]. Встановлено, що 30,7 % (12512) слів українсько-англійського словника відсутні у словнику ВЕСУМ, а для англо-українського словника ця частка збільшується до 48 % (19633). Такі результати частково можна пояснити наявністю в українсько-англійському та англо-українському словниках значної кількості власних назв, які записані з малої літери.

**2. Розробка мережевого засобу (Веб-сервісу) використання дистрибутивно-семантичних моделей векторного представлення сутностей природної мови — UkrVectōrēs**. Мережевий засіб UkrVectōrēs обчислює семантичні відношення між сутностями української мови в рамках обраної дистрибутивно-семантичної моделі векторного представлення сутностей. UkrVectōrēs — це інструмент дистрибутивного аналізу природної мови — це метод дослідження природної мови, заснований на вивченні середовища (дистрибуції, розподілу) окремих сутностей у тексті, та не використовує відомостей про повне лексичне або граматичне значення цих сутностей. В загальному випадку дистрибутивний аналіз використовує, базується та досліджує сутності природної мови, такі як слова або словосполучення.

В рамках даного методу до текстів природною мовою застосовується впорядкований набір універсальних процедур, що дозволяє виділити основні одиниці мови (фонеми, морфеми, слова, словоспо-лучення), провести їх класифікацію та встановити відносини семантичної схожості між ними.

**2.1. Призначення та функції мережевого засобу UkrVectōrēs.** Мережевий засіб UkrVectōrēs (у вигляді веб-сервісу з API (англ. *Application programming interface*, *API*)) — це інструмент, який дозволяє досліджувати семантичні відношення між словами в рамках прогностичних моделей дистрибутивної семантики, з використанням програмної бібліотеки з відкритим вихідним кодом для обробки та математичного моделювання природної мови gensim [11; 19–21] (яка включає інтерфейс прикладного програмування для роботи з алгоритмами Word2vec, fastText та інші).

Можна образно назвати Мережевий засіб UkrVectōrēs «семантичним калькулятором». Користувач може вибрати одну або кілька з ретельно підготовлених прогностичних моделей дистрибутивної семантики (або використати свою модель векторного представлення для слів української мови), навчених на різних корпусах текстів, зокрема таких наборів даних (англ. *Dataset*):

— проблеми поетики творчого доробку Олеся Гончара;

— художня література;

— книга «Серце віддаю дітям» Василя Сухомлинського.

Мережевий засіб UkrVectōrēs охоплює такі елементи дистрибутивно-семантичного аналізу:

• обчислення семантичної схожості/близькості між парами слів у рамках обраної прогностичної моделі дистрибутивної семантики;

• знаходження слова, найближчого до заданого (з можливістю фільтрації за алфавітом і коефіцієнтом косинусної схожості/близькості) в рамках обраної прогностичної моделі дистрибутивної семантики (обчислення семантичних асоціатів). Коефіцієнт косинусної близькості слів може приймати значення в проміжку [−1...1]. Якщо коефіцієнт косинусної схожості/близькості сутностей — слів приймає значення в проміжку [−1...0,5], — це свідчить про відсутність схожих контекстів в наборі даних та найменшу семантичну близькість слів. Якщо коефіцієнт косинусної схожості/близькості сутностей — слів приймає значення в проміжку [0,5...1], — це свідчить про наявність схожих контекстів у наборі даних та більшу семантичну близькість слів. Чим більше коефіцієнт косинусної схожості/близькості наближається до 1, тим більша семантична близькість слів та більше схожих контекстів в наборі даних;

• виконання над векторами слів алгебраїчних операцій (додавання, віднімання, пошук центру лексичного кластера і відстаней до

цього центру) в рамках обраної прогностичної моделі дистрибутивної семантики;

- генерування семантичної карти (з використанням програмного інструментарію з відкритим початковим кодом TensorFlow [12], а саме — TensorBoard [22]) відношень між словами (це дозволяє виявляти семантичні кластери або тестувати гіпотези на таких кластерах);

- отримання вектора (у вигляді масиву чисел) та його візуалізація для заданого слова в рамках обраної прогностичної моделі дистрибутивної семантики;

- вибір зі списку та завантаження для подальшого використання прогностичної моделі дистрибутивної семантики;

- використання інших прогностичних моделей дистрибутивної семантики, які вільно поширюються, за допомоги налаштування конфігураційного файлу.

**2.2. Програмні залежності мережевого засобу UkrVectōrēs.**

— Python 3.8.6 [23] — інтерпретатор та стандартні бібліотеки;

— gensim [11; 19—21] — програмна бібліотека з відкритим вихідним кодом для передової обробки та математичного моделювання природної мови;

— Flask [24] — мікрофреймворк для веб-додатків;

— Flask-CORS [25] — розширення Flask для обробки спільного використання ресурсів з різних джерел (англ. *Cross-Origin Resource Sharing, CORS*);

— uWSGI [26] — веб-сервер і сервер веб-додатків, спочатку реалізований для запуску додатків Python через протокол WSGI (і його бінарний варіант uwsgi);

— Pandas [27] — програмна бібліотека, написана для мови програмування Python для маніпулювання даними та їхнього аналізу. Вона зокрема пропонує структури даних та операції для маніпулювання чисельними таблицями та часовими рядами;

— nginx [28] — вільний веб-сервер і проксі-сервер;

— Angular [29] — написаний на TypeScript front-end фреймворк з відкритим кодом для розробки односторінкових застосунків (англ. *Single-page application, SPA*). В програмній інженерії терміни front-end та back-end розрізняють за принципом розділення відповідальності між рівнем представлення та рівнем доступу до даних відповідно. Front-end — це інтерфейс для взаємодії між користувачем і back-end. Front-end та back-end можуть бути розподілені між однією або кількома системами. В програмній архітектурі може бути багато рівнів між апаратним забезпеченням та кінцевим користувачем. Кожен з цих рівнів може мати як front-end, так і back-end. Front — це абстракція, спрощення базового компонента через надання користувачу зручного інтерфейсу взаємодію з SPA.

**2.3. Архітектурна організація мережевого засобу UkrVectōrēs**

**2.3.1. Опис прикладного програмного інтерфейсу веб-сервісів (back-end API) мережевого засобу UkrVectōrēs.** Розробнику доступні сервіси через кінцеві точки API (англ. *API endpoints*), наведені в таблиці 4.

Таблиця 4

**Кінцеві точки API мережевого засобу UkrVectōrēs**

| | Кінцева точка API | Сервіс | Метод http-запиту |
|---|---|---|---|
| S1 | host [:port] /api/word2vec/ similarity | обчислення семантичної схожості/близькості між парами слів у рамках обраної дистрибутивно-семантичної моделі | POST |
| S2 | host [:port] /api/word2vec/similar | обчислення семантичних асоціатів для заданого слова у рамках обраної дистрибутивно-семантичної моделі | POST |
| S3 | host [:port] /api/word2vec/center | обчислення центру лексичного кластера слів у рамках обраної дистрибутивно-семантичної моделі | POST |
| S4 | host [:port] /api/models | визначення списку доступних для використання дистрибутивно-семантичних моделей | GET |
| S5 | host [:port]/ | графічного інтерфейсу користувача односторінкового застосунку UkrVectōrēs | GET |

S1 — сервіс обчислення семантичної схожості/близькості між парами слів у рамках обраної дистрибутивно-семантичної моделі.

Опис вхідних даних.

Вхідними даними має бути JSON-структура, яка містить текстові поля *word_1*, *word_2* та числове поле *model*. Повна JSON-схема вхідних даних сервісу обчислення семантичної схожості/близькості між парами слів у рамках обраної дистрибутивно-семантичної моделі на-ведена на рисунку 4. Використовується метод http-запиту POST. Приклад POST запиту до кінцевої точки сервісу S1 на мові програмування JavaScript з використанням Fetch API [30] наведено на рисунку 5.

Опис вихідних даних.

Вихідними даними є спеціалізована JSON-структура, яка містить числове поле *similarity*, що відповідає коефіцієнту косинусної схожості/близькості. Повна JSON-схема вихідних даних сервісу обчислення семантичної схожості/близькості між парами слів у рамках обраної дистрибутивно-семантичної моделі наведена на рисунку 6.

S2 — сервіс обчислення семантичних асоціатів для заданого слова в рамках обраної дистрибутивно-семантичної моделі.

Опис вхідних даних.

Вхідними даними має бути JSON-структура, яка містить текстове поле *word* та числове поле *model*. Повна JSON-схема вхідних даних сервісу обчислення семантичних асоціатів для заданого слова в рамках обраної дистрибутивно-семантичної моделі наведена на рисунку 7. Використовується метод http-запиту POST. Приклад POST запиту до кінцевої точки сервісу S2 на мові програмування JavaScript з використанням Fetch API [30] наведено на рисунку 8.

Опис вихідних даних.

Вихідними даними є спеціалізована JSON-структура, яка містить поле *similar*, котре представляє собою масив масивів (елементи масиву містять текстове та числове поле) семантичних асоціатів та коефіцієнта косинусної схожості/близькості для заданого слова в рамках обраної дистрибутивно-семантичної моделі. Повна JSON-схема вихідних даних сервісу обчислення семантичних асоціатів для заданого слова в рамках обраної дистрибутивно-семантичної моделі наведена на рисунку 9.

S3 — сервіс обчислення центру лексичного кластера слів у рамках обраної дистрибутивно-семантичної моделі.

Опис вхідних даних.

Вхідними даними має бути JSON-структура, яка містить поле *words* (представляє собою масив текстових елементів) та числове поле *model*. Повна JSON-схема вхідних даних сервісу обчислення центру лексичного кластера слів у рамках обраної дистрибутивно-семантичної моделі наведена на рисунку 10. Використовується метод http-запиту POST. Приклад POST запиту до кінцевої точки сервісу S3 на мові програмування JavaScript з використанням Fetch API [30] наведено на рисунку 11.

```
{
    "$schema": "http://json-schema.org/draft-07/schema",
    "$id": "http://example.com/example.json",
    "type": "object",
    "title": "The root schema",
    "description": "Схема вхідних даних сервісу обчислення семантичної схожості/близькості між парами
слів
в рамках обраної дистрибутивно-семантичної моделі.",
    "default": {},
    "examples": [
        {
            "word_1": "Гончар",
            "word_2": "письменик",
            "model": 0
        }
    ],
    "required": [
        "word_1",
        "word_2",
        "model"
    ],
    "properties": {
        "word_1": {
            "$id": "#/properties/word_1",
            "type": "string",
            "title": "The word_1 schema",
            "description": "Лема слова для порівняння в рамках обраної дистрибутивно-семантичної моделі.",
            "default": "",
            "examples": [
                "Гончар"
            ]
        },
        "word_2": {
            "$id": "#/properties/word_2",
            "type": "string",
            "title": "The word_2 schema",
            "description": "Лема слова для порівняння в рамках обраної дистрибутивно-семантичної моделі.",
            "default": "",
            "examples": [
                "письменик"
            ]
        },
        "model": {
            "$id": "#/properties/model",
            "type": "integer",
            "title": "The model schema",
            "description": "Індекс обраної дистрибутивно-семантичної моделі. Індекс моделей міститься
в конфігураційному файлі config.models.simple.json",
            "default": 0,
            "examples": [
                0
            ]
        }
    },
    "additionalProperties": true
}
```

Рис. 4. JSON-схема вхідних даних сервісу обчислення семантичної схожості/
близькості між парами слів

```
var myHeaders = new Headers();
myHeaders.append("Content-Type", "application/json");

var raw = JSON.stringify({"word_1": "Гончар", "word_2": "письменик", "model": 0});

var requestOptions = {
 method: 'POST',
 headers: myHeaders,
 body: raw,
 redirect: 'follow'
};

fetch("host[:port]/api/word2vec/similarity", requestOptions)
 .then(response => response.text())
 .then(result => console.log(result))
 .catch(error => console.log('error', error));
var formData = new FormData();
var fileField = document.querySelector('input[type="file"]');
```

Рис. 5. Приклад POST-запиту до кінцевої точки сервісу S1

```
{
   "$schema": "http://json-schema.org/draft-07/schema",
   "$id": "http://example.com/example.json",
   "type": "object",
   "title": "The root schema",
   "description": "Схема вихідних даних сервісу обчислення семантичної схожості/близькості між парами слів
в рамках обраної дистрибутивно-семантичної моделі.",
   "default": {},
   "examples": [
      {
         "similarity": 0.939421488228
      }
   ],
   "required": [
      "similarity"
   ],
   "properties": {
      "similarity": {
         "$id": "#/properties/similarity",
         "type": "number",
         "title": "The similarity schema",
         "description": "Коефіцієнт косинусної схожості/близькості між парами слів
в рамках обраної дистрибутивно-семантичної моделі.",
         "default": 0.0,
         "examples": [
            0.939421488228
         ]
      }
   },
   "additionalProperties": true
}
```

Рис. 6. JSON-схема вихідних даних сервісу обчислення семантичної схожос-
ті/близькості між парами слів

**515**

```
{
    "$schema": "http://json-schema.org/draft-07/schema",
    "$id": "http://example.com/example.json",
    "type": "object",
    "title": "The root schema",
    "description": "Схема вхідних сервісу обчислення семантичних асоціатів для заданого слова
в рамках обраної дистрибутивно-семантичної моделі.",
    "default": {},
    "examples": [
        {
            "word": "Гончар",
            "model": 0
        }
    ],
    "required": [
        "word",
        "model"
    ],
    "properties": {
        "word": {
            "$id": "#/properties/word",
            "type": "string",
            "title": "The word schema",
            "description": "Лема слова для пошуку семантичних асоціатів
в рамках обраної дистрибутивно-семантичної моделі.",
            "default": "",
            "examples": [
                "Гончар"
            ]
        },
        "model": {
            "$id": "#/properties/model",
            "type": "integer",
            "title": "The model schema",
            "description": "Індекс обраної дистрибутивно-семантичної моделі. Індекс моделей міститься
в конфігураційному файлі config.models.simple.json.",
            "default": 0,
            "examples": [
                0
            ]
        }
    },
    "additionalProperties": true
}
```

Рис. 7. JSON-схема вхідних даних сервісу обчислення семантичних асоціатів
для заданого слова

Опис вихідних даних.

Вихідними даними є спеціалізована JSON-структура, яка містить
поле *center*, котре представляє собою масив масивів (елементи масиву
містять текстове та числове поле) семантичних асоціатів центру лек-
сичного кластера слів та коефіцієнта косинусної схожості/близькості
до центру лексичного кластера слів в рамках обраної дистрибутивно-
семантичної моделі. Повна JSON-схема вихідних даних сервісу об-
числення центру лексичного кластера слів в рамках обраної дистри-
бутивно-семантичної моделі наведена на рисунку 12.

```
var myHeaders = new Headers();
myHeaders.append("Content-Type", "application/json");

var raw = JSON.stringify({"word": "Гончар", "model": 0});

var requestOptions = {
 method: 'POST',
 headers: myHeaders,
 body: raw,
 redirect: 'follow'
};

fetch("host[:port]/api/word2vec/similar", requestOptions)
 .then(response => response.text())
 .then(result => console.log(result))
 .catch(error => console.log('error', error));
var formData = new FormData();
var fileField = document.querySelector('input[type="file"]');
```

Рис. 8. Приклад POST-запиту до кінцевої точки сервісу S2

S4 — сервіс визначення списку доступних для використання дистрибутивно-семантичних моделей.

Опис вхідних даних.

Вхідними даними має бути звичайний пустий метод http-запиту GET. Приклад GET запиту до кінцевої точки сервісу S4 на мові програмування JavaScript з використанням Fetch API [30] наведено на рисунку 13.

Опис вихідних даних.

Вихідними даними є спеціалізована JSON-структура. Повна JSON-схема вихідних даних сервісу визначення списку доступних для використання дистрибутивно-семантичних моделей наведена в додатку d1.

S5 — сервіс графічного інтерфейсу користувача односторінкового застосунку UkrVectōrēs. За замовчуванням сервіс графічного інтерфейсу користувача мережевого засобу UkrVectōrēs доступний через метод http-запиту GET на відповідний порт за відповідною адресою хоста. Графічний інтерфейс користувача односторінкового застосунку UkrVectōrēs розробляється окремо, використовуючи розроблені сервіси через кінцеві точки API та в залежності від предметної області і вирішуваних задач.

**2.3.2. Графічний інтерфейс користувача односторінкового застосунку UkrVectōrēs.** Розглянемо методику роботи користувача з графічним інтерфейсом односторінкового застосунку UkrVectōrēs, зокрема використання таких функцій (елементів) дистрибутивно-семантичного аналізу текстів природної мови.

```
{
    "$schema": "http://json-schema.org/draft-07/schema", "$id": "http://example.com/example.json", "type":
"object",  "title": "The root schema",  "description": "Схема вихідних даних сервісу обчислення семантичних
асоціатів для заданого слова в рамках обраної дистрибутивно-семантичної моделі.",
    "examples": [
        {
            "similar": [          [
                "український", 0.9999048709869385
            ]          ]       }
    ],
    "required": [
        "similar"
    ],
    "properties": {
        "similar": {
            "$id": "#/properties/similar", "type": "array", "title": "The similar schema",  "description": "Масив масивів
(елементи масиву містять текстове та числове поле) семантичних асоціатів та коефіцієнту косинусної
схожості/близькості для заданого слова в рамках обраної дистрибутивно-семантичної моделі.",
            "examples": [
                [             [
                    "український", 0.9999048709869385
                ]             ],
            "items": {
                "$id": "#/properties/similar/items",
                "anyOf": [
                    {
                        "$id": "#/properties/similar/items/anyOf/0",
                        "type": "array",
                        "title": "The first anyOf schema",
                        "examples": [
                            [
                                "український", 0.9999048709869385
                            ]
                        ],
                        "items": {
                            "$id": "#/properties/similar/items/anyOf/0/items",
                            "anyOf": [
                                {
                                    "$id": "#/properties/similar/items/anyOf/0/items/anyOf/0",
                                    "type": "string", "title": "The first anyOf schema",
                                    "description": "Лема семантичного асоціату для заданого слова в рамках обраної
дистрибутивно-семантичної моделі.",
                                    "examples": [
                                        "український"
                                    ]
                                },
                                {
                                    "$id": "#/properties/similar/items/anyOf/0/items/anyOf/1",
                                    "type": "number", "title": "The second anyOf schema",  "description": "Коефіцієнт
косинусної схожості/близькості семантичного асоціату до заданого слова в рамках обраної дистрибутивно-
семантичної моделі.",
                                    "examples": [
                                        0.9999048709869385
                                    ]  }  ]  }  }  }  ]  }  }  },
}
```

Рис. 9. JSON-схема вихідних даних сервісу обчислення семантичних асоціа-
тів для заданого слова

```
{
    "$schema": "http://json-schema.org/draft-07/schema",
    "$id": "http://example.com/example.json",
    "type": "object",
    "title": "The root schema",
    "description": "Схема вхідних даних обчислення центру лексичного кластера лем слів в рамках обраної
дистрибутивно-семантичної моделі.",
    "default": {},
    "examples": [
        {
            "words": [
                "Олесь",
                "Гончар",
                "письменник"
            ],
            "model": 0
        }
    ],
    "required": [
        "words",
        "model"
    ],
    "properties": {
        "words": {
            "$id": "#/properties/words",
            "type": "array",
            "title": "The words schema",
            "description": "Масив текстових елементів (лем слів) для обчислення центру їх лексичного кластера
в рамках обраної дистрибутивно-семантичної моделі.",
            "default": [],
            "examples": [
                [
                    "Олесь",
                    "Гончар"
                ]
            ],
            "model": {
            "$id": "#/properties/model",
            "type": "integer",
            "title": "The model schema",
            "description": "Індекс обраної дистрибутивно-семантичної моделі. Індекс моделей міститься в
конфігураційному файлі config.models.simple.json.",
            "default": 0,
            "examples": [
                0
            ]    }    }
}
```

Рис. 10. JSON-схема вхідних даних обчислення центру лексичного кластера слів

*Обчислення семантичних асоціатів для заданого слова* в рамках обраної дистрибутивно-семантичної моделі. Для використання цієї функції необхідно:

Запустити графічний інтерфейс односторінкового застосунку UkrVectōrēs з використанням актуальної версії веб-браузера Google Chrome, Mozilla Firefox або Microsoft Edge. Для цього в адрес-ному рядку веб-браузера потрібно вписати посилання: https://ukrvectores.ai-service.ml/ (посилання може відрізнятися, це залежить від особливостей розгортання мережевого засобу UkrVectōrēs) та в головному меню обрати режим роботи «Семантичні асоціати» (рисунок 14);

```
var myHeaders = new Headers();
myHeaders.append("Content-Type", "application/json");

var raw = JSON.stringify({"words": "Олесь", "Гончар", "письменник", "model": 0});

var requestOptions = {
 method: 'POST',
 headers: myHeaders,
 body: raw,
 redirect: 'follow'
};

fetch("host[:port]/api/word2vec/center", requestOptions)
 .then(response => response.text())
 .then(result => console.log(result))
 .catch(error => console.log('error', error));
var formData = new FormData();
var fileField = document.querySelector('input[type="file"]');
```

Рис. 11. Приклад POST-запиту до кінцевої точки сервісу S3

```
var requestOptions = {
 method: 'GET',
 redirect: 'follow'
};

fetch("host[:port]/api/word2vec/models", requestOptions)
 .then(response => response.text())
 .then(result => console.log(result))
 .catch(error => console.log('error', error));
```

Рис. 13. Приклад GET-запиту до кінцевої точки сервісу S4

За допомогою випадаючого списку компонента *select* під назвою «Моделі» (рисунок 15) обрати бажану дистрибутивно-семантичну модель, в рамках якої буде проводитися обчислення семантичних асоціатів (в загальному випадку, за замовчуванням, використовується нейронна векторна модель представлення слів «Олесь Гончар» (з використанням набору даних — проблеми поетики творчого доробку Олеся Гончара), алгоритм word2vec word embeddings розмірністю 500d. Сутність — слово, лематизовано, приведено до нижнього регістру. Параметри word2vec: -size 500 -negative 5 -window 5 -threads 24 -min_count 10 -iter 20);

```
{
  "$schema": "http://json-schema.org/draft-07/schema",
  "$id": "http://example.com/example.json",
  "type": "object",
  "title": "The root schema",
  "description": "Схема вихідних даних сервісу обчислення центру лексичного кластера слів в рамках
обраної дистрибутивно-семантичної моделі.",
  "default": {},
  "examples": [
    {
      "center": [
        [
          "український",
          0.8037014603614807
        ]
      ]
    }
  ],
  "required": [
    "center"
  ],
  "properties": {
    "center": {
      "$id": "#/properties/center",
      "type": "array",
      "title": "The center schema",
      "description": "Масив масивів (елементи масиву містять текстове та числове поле) семантичних
асоціатів центру лексичного кластера слів та коефіцієнту косинусної схожості/близькості до центру
лексичного кластера слів в рамках обраної дистрибутивно-семантичної моделі.",
      "default": [],
      "examples": [
        [
          [
            "український",
            0.8037014603614807
          ]
        ],
              ]
            }
          }
        ]
      }
    }
  }
}
```

Рис. 12. JSON-схема вихідних даних сервісу обчислення центру лексичного кластера слів

В полі компонента *input* під назвою «Введіть лему слова» вкажіть бажану лему слова, до якого треба обчислити семантичні асоціати (наприклад — Гончар, як наведено на рисунку 14), та натисніть клавішу «Enter» або кнопку «Обчислити»;

**521**

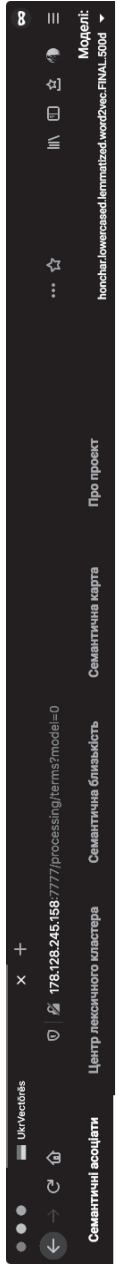

522Рис. 14. Графічний інтерфейс односторінкового застосунку UkrVectōrēs (режим роботи «Семантичні асоціати»)

Рис. 15. Графічний інтерфейс односторінкового застосунку UkrVectōrēs (режим роботи «Семантичні асоціати», компонент *select* під назвою «Моделі»)

1. На екрані (рисунок 14) відобразяться семантичні асоціати (за замовчуванням відображаються перші 100 асоціатів за зменшенням коефіцієнта косинусної схожості/близькості) для заданої леми слова «Гончар» в рамках обраної дистрибутивно-семантичної моделі «Олесь Гончар»;

2. Використовуючи елемент «Сутність», користувач може обрати відображення семантичних асоціатів за абеткою (рисунок 16);

3. Використовуючи елемент «Косинусна близькість», користувач може обрати відображення семантичних асоціатів за коефіцієнтом косинусної схожості/близькості (за збільшенням або за зменшенням) (рисунок 17).

*Обчислення семантичної схожості/близькості між парами слів* у рамках обраної дистрибутивно-семантичної моделі. Для використання цієї функції необхідно:

1. Запустити графічний інтерфейс односторінкового застосунку UkrVectōrēs з використанням актуальної версії веб-браузера Google Chrome, Mozilla Firefox або Microsoft Edge. Для цього в адресному рядку веб-браузера потрібно вписати посилання: https://ukrvectores.ai-service.ml/ (посилання може відрізнятися, це залежить від особливостей розгортання мережевого засобу UkrVectōrēs) та в головному меню обрати режим роботи «Семантична близькість» (рисунок 18);

2. За допомогою випадаючого списку компонента *select* під назвою «Моделі» (рисунок 19) обрати бажану дистрибутивно-семантичну модель, в рамках якої буде проводитися обчислення семантичної схожості/близькості між парами слів — коефіцієнта косинусної схожості/близькості (в загальному випадку, за замовчуванням, використовується нейронна векторна модель представлення слів «Олесь Гончар» (з використанням набору даних — проблеми поетики творчого доробку Олеся Гончара), алгоритм word2vec word embeddings розмірністю 500d. Сутність — слово, лематизовано, приведено до нижнього регістру. Параметри word2vec: -size 500 -negative 5 -window 5 -threads 24 -min_count 10 -iter 20);

3. В полях компонентів *intput* під назвами «Введіть лему слова для порівняння» вкажіть бажані леми слів, між якими треба обчислити семантичну схожість/близькість — коефіцієнта косинусної схожості/близькості (наприклад — Гончар та письменник, як наведено на рисунку 18) та натисніть клавішу «Enter» або кнопку «Обчислити»;

4. На екрані (рисунок 18) відобразиться коефіцієнт косинусної схожості/близькості для заданих лем слів «Гончар» та «письменник» в рамках обраної дистрибутивно-семантичної моделі «Олесь Гончар»;

*Обчислення центру лексичного кластера лем слів* в рамках обраної дистрибутивно-семантичної моделі. Для використання цієї функції необхідно:

1. Запустити графічний інтерфейс односторінкового застосунку UkrVectōrēs з використанням актуальної версії веб-браузера Google Chrome, Mozilla Firefox або Microsoft Edge. Для цього в адресному рядку веб-браузера потрібно вписати посилання: https://ukrvectores.ai-service.ml/ (посилання може відрізнятися, це залежить від особливостей розгортання мережевого засобу UkrVectōrēs) та в головному меню обрати режим роботи «Центр лексичного кластера» (рисунок 20);

2. За допомогою випадаючого списку компонента *select* під назвою «Моделі» (рисунок 21) обрати бажану дистрибутивно-семантичну модель, в рамках якої буде проводитися обчислення центра лексичного кластера (в загальному випадку, за замовчуванням, використовується нейронна векторна модель представлення слів «Олесь Гончар» (з використанням набору даних — проблеми поетики творчого доробку Олеся Гончара), алгоритм word2vec word embeddings розмірністю 500d. Сутність — слово, лематизовано, приведено до нижнього регістру. Параметри word2vec: -size 500 -negative 5 -window 5 -threads 24 -min_count 10 -iter 20);

3. В полі компонента *intput* під назвою «Введіть леми слів через пробіл» вкажіть бажані леми слів через пробіл, центр лексичного кластера яких необхідно обчислити (наприклад — «Гончар письменник герой», як наведено на рисунку 20) та натисніть клавішу «Enter» або кнопку «Обчислити»;

4. На екрані (рисунок 20) відобразяться леми слів, що формують центр лексичного кластера заданих лем слів (за замовчуванням відображаються перші 100 асоціатів за зменшенням коефіцієнта косинусної схожості/близькості до центру лексичного кластера) для заданої леми слова «Гончар» в рамках обраної дистрибутивно-семантичної моделі «Олесь Гончар»;

5. Використовуючи елемент «Сутність», користувач може обрати відображення семантичних асоціатів, що формують центр лексичного кластера заданих лем слів, за абеткою (рисунок 22);

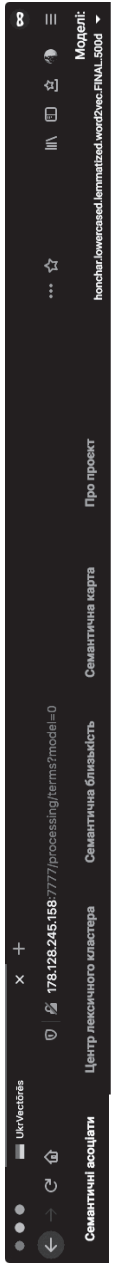

Рис. 16. Графічний інтерфейс односторінкового застосунку UkrVectōrēs (режим роботи «Семантичні асоціати», елемент «Сутність»)

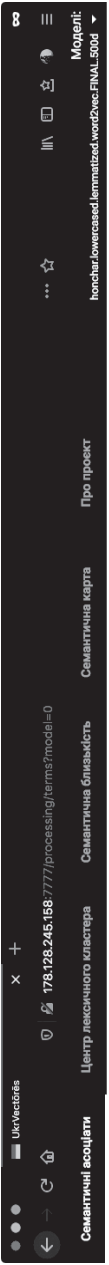

Рис. 17. Графічний інтерфейс односторінкового застосунку UkrVectōrēs (режим роботи «Семантичні асоціати», елемент «Косинусна близькість»)

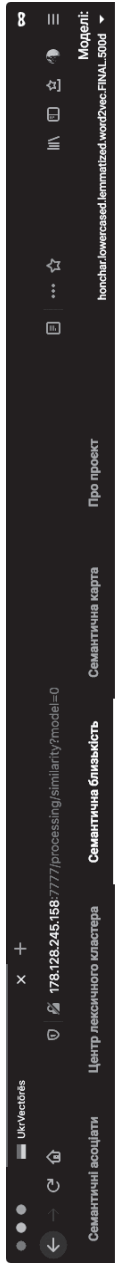

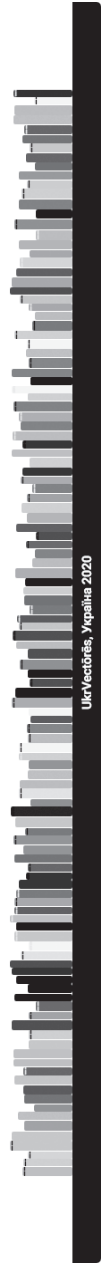

UkrVectōrēs, Україна 2020

## Обчислення семантичної близькості однослівних сутностей

Сервіс обчислює семантичні відношення між словами української мовою в рамках обраної моделі.

В дистрибутивній семантиці слова зазвичай представляються у вигляді векторів в багатовимірному просторі їх контекстів. Семантична схожість обчислюється як косинусна близькість між векторами двох слів і може приймати значення в проміжку [-1 ... 1] (на практиці часто використовуються тільки значення вище 0). Значення 0 приблизно означає, що у цих слів немає схожих контекстів і їх значення не пов'язані один з одним. Значення 1, навпаки, свідчить про повну ідентичність їх контекстів і, отже, про близькість значення.

Використовується нейронна векторна модель представлення слів «Олесь Гончар» (з використанням набору даних – проблеми поетики творчого доробку Олеся Гончара), алгоритм word2vec word embeddings розмірністю 500d. Сутність - слово, лематизовано, приведено до нижнього регістру. Параметри word2vec: -size 500 -negative 5 -window 5 -threads 24 -min_count 10 -iter 20.

Введіть лему слова для порівняння
Гончар

Введіть лему слова для порівняння
письменник

**Обчислити**

Косинусна близькість: **0.99990111589431776**

Рис. 18. Графічний інтерфейс односторінкового застосунку UkrVectōrēs (режим роботи «Семантична близькість»)

Рис. 19. Графічний інтерфейс односторінкового застосунку UkrVectōrēs (режим роботи «Семантична близькість», компонент *select* під назвою «Моделі»)

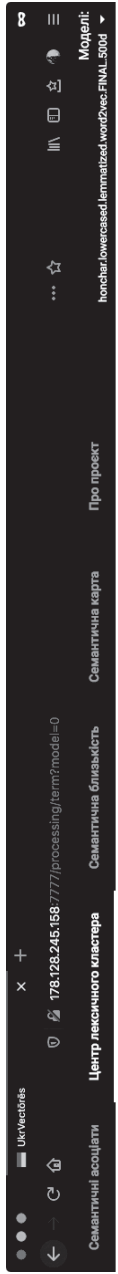

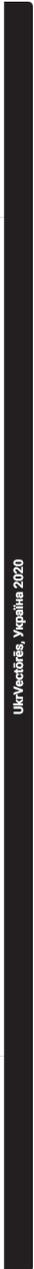

**530**

Browser tabs: UkrVectōrēs ×

178.128.245.158:7777/processing/term?model=0

Семантичні асоціації    Центр лексичного кластера    Семантична близькість    Семантична карта    Про проект

Model:
honchar.lowercased.lemmatized.word2vec.FINAL.500d ▼

## Обчислення центру лексичного кластера однослівних сутностей

Сервіс обчислює центр лексичного кластера слів українською мовою в рамках обраної моделі.

Використовується нейронна векторна модель представлення слів «Олесь Гончар» (з використанням набору даних – проблеми поетики творчого доробку Олеся Гончара), алгоритм word2vec word embeddings розмірністю 500d. Сутність – слово, лематизовано, приведено до нижнього регістру. Параметри word2vec: -size 500 -negative 5 -window 5 -threads 24 -min_count 10 -iter 20.

Введіть леми слів через пробіл
Гончар письменник герой

Обчислити

Filter

### Центр лексичного кластера

| Сутність | Косинусна близькість ↓ |
|---|---|
| автор | 0.9999430179595947 |
| художній | 0.9999415874481201 |
| творчий | 0.9999398589134216 |
| український | 0.9999394416809082 |
| контекст | 0.9999378919960144 |
| ставати | 0.9999368190765381 |

UkrVectōrēs, Україна 2020

Рис. 20. Графічний інтерфейс односторінкового застосунку UkrVectōrēs (режим роботи «Центр лексичного кластера»)

Рис. 21. Графічний інтерфейс односторінкового застосунку UkrVectōrēs (режим роботи «Центр лексичного кластера», компонент *select* під назвою «Моделі»)

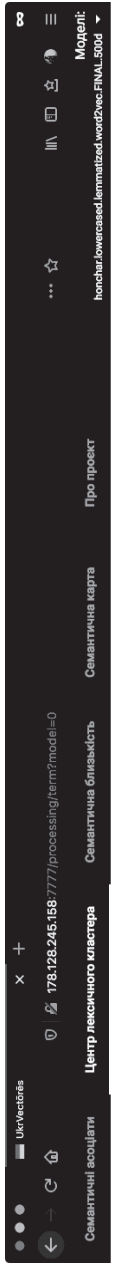

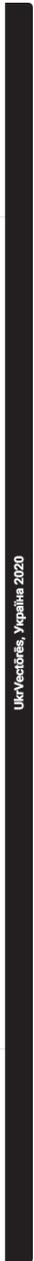

Рис. 22. Графічний інтерфейс односторінкового застосунку UkrVectōrēs (режим роботи «Центр лексичного кластера» елемент «Сутність»)

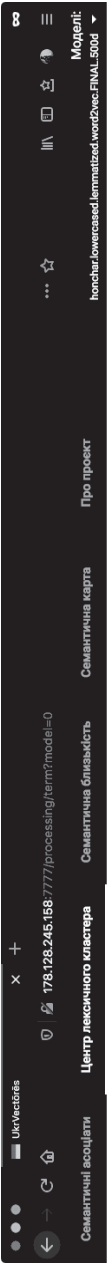

Рис. 23. Графічний інтерфейс односторінкового застосунку UkrVectōrēs (режим роботи «Центр лексичного класте-ра», елемент «Косинусна близькість»)

6. Використовуючи елемент «Косинусна близькість», користувач може обрати відображення семантичних асоціатів, що формують центр лексичного кластера заданих лем слів, за коефіцієнтом косинусної схожості/близькості (за збільшенням або за зменшенням) (рисунок 23).

*Генерування семантичних карт (з використанням програмного інструментарію з відкритим початковим кодом TensorFlow [12], а саме — TensorBoard [22]) відношень між словами* в рамках обраної дистрибутивно-семантичної моделі. Для використання цієї функції необхідно:

1. Запустити графічний інтерфейс односторінкового застосунку UkrVectōrēs з використанням актуальної версії веб-браузера Google Chrome, Mozilla Firefox або Microsoft Edge. Для цього в адресному рядку веб-браузера потрібно вписати посилання: https://ukrvectores.ai-service.ml/ (посилання може відрізнятися, це залежить від особливостей розгортання мережевого засобу UkrVectōrēs) та в головному меню обрати режим роботи «Семантична карта» (рисунок 24);

2. За допомогою випадаючого списку компонента *select* під назвою «Моделі» (рисунок 24) обрати бажану дистрибутивно-семантичну модель, в рамках якої буде проводитися обчислення центра лексичного кластера (в загальному випадку, за замовчуванням, використовується нейронна векторна модель представлення слів «Олесь Гончар» (з використанням набору даних — проблеми поетики творчого доробку Олеся Гончара), алгоритм word2vec word embeddings розмірністю 500d. Сутність — слово, лематизовано, приведено до нижнього регістру. Параметри word2vec: -size 500 -negative 5 -window 5 -threads 24 -min_count 10 -iter 20);

3. Документація користувача використання інструментарію візуалізації *TensorBoard* доступна за посиланням [22]. Приклад візуалізації семантичних асоціатів леми слова «Гончар» наведено на рисунку 25.

**2.4. Методика тренування дистрибутивно-семантичної моделі векторного представлення сутностей (з використанням набору даних — проблеми поетики творчого доробку Олеся Гончара).** Загальна методика тренування дистрибутивно-семантичних моделей векторного представлення сутностей (слів) дана у вигляді структури конвеєра (англ. Pipeline) обробки наборів даних (електронних текстових документів), що містять тексти природною мовою (рисунок 26), та складається з таких етапів та компонентів:

1. Технологія формування електронного корпусу текстів;

Рис. 24. Графічний інтерфейс односторінкового застосунку UkrVectōrēs (режим роботи «Семантична карта»)

Рис. 25. Графічний інтерфейс односторінкового застосунку UkrVectōrēs (режим роботи «Семантична карта», візуалі-
зація семантичних асоціатів леми слова «Гончар»)

2. Технологія попередньої лінгвістичної обробки електронного корпусу текстів;

3. Технологія тренування/навчання прогностичних моделей дистрибутивної семантики;

4. Технологія опрацювання прогностичних моделей дистрибутивної семантики;

5. Джерела текстових документів та корпусів текстів (аналогові тексти, мережа Інтернет, корпуси текстів Wikipedia, електронні колекції ТД, БД та ін.);

6. Електронний корпус текстів;

7. Анотований корпус текстів — лінгвістичний корпус текстів;

8. Прогностична модель дистрибутивної семантики — модель векторного представлення сутностей (слів).

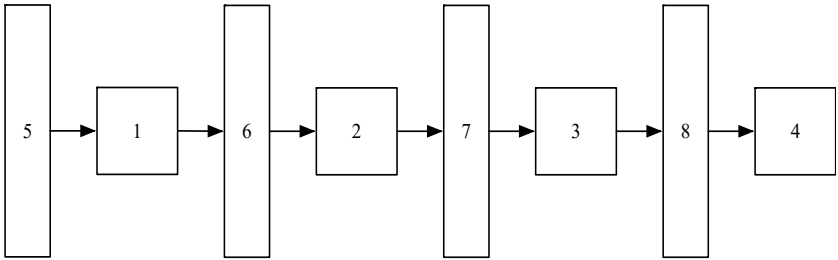

Рис. 26. Структура конвеєра узагальненої методики навчання прогностичних моделей дистрибутивної семантики

*Технологія формування електронного корпусу текстів* (рисунок 26, елемент 1). Електронний корпус текстів в загальному випадку зазвичай формується на основі (рисунок 26, елемент 5) загальнодоступного і достатньо об'ємного корпусу текстів статей Wikipedia (актуальні версії корпусу текстів Wikipedia доступні на сервері Wikimedia, що зберігає повні архіви — дампи різних проектів Wikipedia foundation [31]), з використанням програмної бібліотеки gensim [21] — програмне забезпечення з відкритим початковим кодом для векторного моделювання та тематичного моделювання, а саме її модулю gensim.corpora. wikicorpus (дозволяє побудувати електронний корпус текстів з дампів бази даних Wikipedia або інших MediaWiki ресурсів). Результатом застосування технології формування електронного корпусу текстів є — електронний корпус текстів (рисунок 26, елемент 6) представлений в текстовому форматі.

*Технологія попередньої лінгвістичної обробки електронного корпусу текстів* (рисунок 26, елемент 2). Конвеєр попередньої лінгвістичної обробки електронного корпусу текстів складається з таких компонентів:

— Лексичний аналіз (елементи лексичного аналізу) — токенізація (англ. *Tokenizing*) — в загальному випадку це процес перетворення послідовності символів у послідовність токенів: декомпозиція тексту в послідовність речень та декомпозиція речень в послідовність токенів (заздалегідь визначеної категорії) — слів (категорія — слово, словосполучення, термін), знаків пунктуації, пробілів тощо.

— Граматичний аналіз (елементи граматичного аналізу) — розмітка тексту частинами мови (англ. *POS tagging, Part-of-speech tagging*) — в загальному випадку це процес присвоєння синтаксичної категорії кожному з токенів.

— Морфологічний аналіз (елементи морфологічного аналізу) — в загальному випадку це процес лематизації — представлення слів у початковій словниковій формі. Лематизація значно скоротить словник і збільшить зв'язність текстів при побудові векторного представлення слів.

— Синтаксичний аналіз (елементи синтаксичного аналізу) — парсинг та чанкінг (англ. *Shallow parsing,* англ. *Chunking*) — в загальному випадку це процес аналізу речень, який спочатку визначає їх складні частини (іменники, дієслова, прикметники тощо), а потім зв'язує їх з одиницями вищого порядку, які мають дискретні граматичні значення (іменні групи або фрази, групи дієслів тощо).

Всі етапи конвеєра попередньої лінгвістичної обробки електронного корпусу текстів, зокрема українською мовою, виконуються за допомоги вільного програмного інструментарію обробки природної мови — LanguageTool API NLP UK [32] та проєктів відкритої спільноти фахівців у галузі комп'ютерної обробки текстів (програмістів, лінгвістів, дослідників) — lang-uk [33].

Результатом застосування технології попередньої лінгвістичної обробки електронного корпусу текстів є анотований корпус текстів — лінгвістичний корпус текстів (рисунок 26, елемент 7), представлений у текстовому форматі.

*Технологія тренування/навчання прогностичних моделей дистрибутивної семантики.* Найпоширенішим програмним інструментом/ технологією, що використовується для навчання прогностичних мо-делей дистрибутивної семантики — векторного представлення слів, є — *word2vec*. Бібліотека gensim надає досить багато засобів для ефективного використання та дослідження векторних представлень слів, в тому числі і *word2vec*.

Результати тренування моделей визначають якість векторних представлень і залежать від параметрів тренування, які може задавати користувач. До таких основних параметрів належать: власне корпус текстів; алгоритм тренування *skip-gram* чи *CBOW*; розмір векторів; максимальний розмір контекстного вікна; мінімальна частота слова, яке буде враховано; параметр для зменшення впливу високочастотних слів; модель для тренування — ієрархічний softmax (англ. *Hierarchical softmax*) або негативні вибірки (англ. *Negative sampling*); кількість негативних прикладів; кількість ітерацій (англ. *Epochs*) навчання; кількість виділених потоків для навчання.

Результатом застосування технології тренування/навчання прогностичних моделей дистрибутивної семантики є — дистрибутивно-семантична модель представлення сутностей (рисунок 26, елемент 8).

Для мережевого засобу UkrVectōrēs було розроблено (навчено) декілька прогностичних моделей дистрибутивної семантики, зокрема модель — «Олесь Гончар», з використанням набору даних невеликого (обмеженого) обсягу — «проблеми поетики творчого доробку Олеся Гончара». Початковий код програмного інструментарію навчання прогностичних моделей дистрибутивної семантики (програмна бібліотека gensim, алгоритм word2vec), зокрема моделі «Олесь Гончар» наведено в додатку d2. Відмінною рисою навчання дистрибутивно-семантичних моделей на текстових наборах даних невеликого обсягу є спеціалізований набір параметрів, застосування яких дозволяє отримати якісні векторні представлення сутностей. Нижче наведено набір параметрів для навчання моделі «Олесь Гончар»:

▪ алгоритм тренування — skip-gram;

▪ розмір векторів — 500;

▪ максимальний розмір контекстного вікна — 5;

▪ мінімальна частота слова, яке буде враховано — 10;

▪ модель для тренування — негативна вибірка (англ. *Negative sampling*);

▪ кількість ітерацій (англ. *Epochs*) навчання — 20;

▪ кількість виділених потоків для навчання — максимальна;

▪ параметр для зменшення впливу високочастотних слів — 1e-5;

▪ інші параметри — за замовчуванням.

*Оцінка векторних представлень.* Дослідження векторних представлень дають можливість зрозуміти характер результатів за виявленою подібністю між словами, але питання оцінки векторних представлень з позиції їхньої якості для розв'язання практичних завдань залишається не тільки актуальним, але і гостро дискусійним. Загалом розрізняють два підходи до оцінки векторних представлень слів: на основі внутрішніх оцінок та на основі зовнішніх оцінок. Використання зовнішніх оцінок передбачає оцінювання якості векторних представлень на основі результатів їхнього застосування для розв'язання реальних завдань. Наприклад, якщо вдалий добір параметрів під час побудови векторного представлення спричинив збільшення точності аналізу тональності тексту, то якість цього векторного представлення вважається вищою. Внутрішнє оцінювання базується на використанні спеціальних тестових наборів (тести аналогій — вирази виду «A до B як C до D») та вручну промаркованих даних.

*Технологія опрацювання прогностичних моделей дистрибутивної семантики.* В загальному випадку для розробки програмних засобів дистрибутивного аналізу (або опрацювання прогностичних моделей дистрибутивної семантики) використовують API бібліотеки gensim та реалізують такі функції: знаходження схожих векторів слів для вектора заданого слова на основі обчисленої косинусної подібності (англ. *Cosine similarity*) між вектором указаного слова та векторами всіх інших слів моделі; знаходження схожих векторів слів на основі множини позитивних та неґативних слів, які будуть мати відповідний вплив на виявлення схожості між векторами слів; знайдені схожі вектори слів на основі множини позитивних та неґативних слів можуть демонструвати наявність упереджень та стереотипів, які зберігаються в моделях; знайдені схожі вектори слів на основі множини позитивних та неґативних слів можуть ілюструвати асоціативні зв'язки між словами; знайдені схожі вектори слів на основі множини позитивних та неґативних слів можуть демонструвати також і граматичні зв'язки; знаходження косинусної відстані (англ. *Cosine distance*) між вектором слова та вектором іншого слова; знаходження слова, яке не трапляється разом з іншими словами із вказаного переліку; знаходження слова в переліку, яке найбільш подібне до вказаного слова; знаходження всіх слів у моделі, які ближчі до вказаного слова, ніж інше вказане слово; обчислення відстані між словами (англ. *Word mover's distance*) двох документів.

### 3. Розробка моделі розгортання мережевого засобу UkrVectōrēs

### 3.1. Інструментарій для управління ізольованими Linux-контейнерами Docker

Docker [34] — інструмент з відкритим сирцевим кодом, який автоматизує розгортання застосунку у середовищах, що підтримують контейнеризацію. Docker допомагає викладати код швидше, швидше тестувати, швидше викладати додатки і зменшити час між написанням і запуском коду. Docker робить це за допомогою легкої платформи контейнерної віртуалізації, використовуючи процеси й утиліти, які допомагають керувати і викладати програми. У своєму ядрі Docker дозволяє запускати практично будь-який додаток, безпечно ізольований в контейнері. Безпечна ізоляція дозволяє запускати на одному хості багато контейнерів одночасно.

*Переваги Docker:*

▪ пришвидшення процесу розробки;

▪ зручна інкапсуляція застосунку;

▪ однакова поведінка на локальній машині, а також dev/staging/production серверах;

▪ простий і чіткий моніторинг;

▪ зручність масштабування.

Термінологія Docker-інструментарію:

▪ Контейнер (Container) — запущений екземпляр, що інкапсулює необхідне ПЗ. Контейнери завжди створюються з образу і можуть надавати порти та дисковий простір для взаємодії з іншими контейнерами чи/та зовнішнім ПЗ. Контейнери можна з легкістю знищити/видалити та створити знову. Контейнери не зберігають стан.

▪ Образ (Image) — базовий елемент кожного контейнера. При створенні образу кожен крок кешується і може бути використаний повторно (копіювання під час запису). Час на збірку залежить від самого образу. З іншого боку, контейнери можна одразу запустити з образу.

▪ Порт (Port) — TCP/UDP порт у звичному розумінні. Для спрощення припустимо, що порти можуть бути відкриті для зовнішнього ПЗ (доступні з хостової ОС) або підключатися до інших контейнерів (тобто доступні лише з цих контейнерів та невидимі для іншого ПЗ).

▪ Volume можна вважати спільною текою. Volume ініціалізується при створенні контейнера і призначений для збереження даних, незалежно від життєвого циклу контейнера.

▪ Registry (Сховище) — сервер, що зберігає образи Docker. Ми можемо порівняти його з Github: витягуєте образ зі сховища, щоб розгорнути його локально, а потім відправляєте локально зібрані образи до віддаленого сховища.

▪ Docker Hub [35] — сховище з веб-інтерфейсом від Docker Inc. Зберігає багато Docker-образів з різним ПЗ. Docker Hub — джерело «офіційних» образів Docker, створених його командою або у співпраці з іншими компаніями (але це не обов'язково образи від офіційних виробників ПЗ). Якщо ви зареєстровані, можна переглянути перелік потенційних вразливостей таких образів. Доступні платні та безкоштовні облікові записи. Для безкоштовного облікового запису можна створювати один приватний образ та безліч публічних.

▪ Docker Store [36] — сервіс дуже подібний до Docker Hub. Це майданчик з рейтингами, відгуками тощо.

Архітектура Docker.

Docker складається з двох головних компонентів:

▪ Docker: платформа віртуалізації з відкритим кодом;

▪ Docker Hub: платформа-як-сервіс для поширення і управління Docker контейнерами.

Docker використовує архітектуру клієнт — сервер. Docker-клієнт спілкується з доменом Docker, який бере на себе створення, запуск, розподіл контейнерів. Обидва, клієнт і сервер, можуть працювати на одній системі, також можна підключити клієнт до віддаленого домена docker. Клієнт і сервер спілкуються через сокет або через RESTful API.

Користувач не взаємодіє з сервером напряму, а використовує для цього клієнт. Docker-клієнт — головний інтерфейс до Docker системи. Він отримує команди від користувача і взаємодіє з docker-доменом.

Щоб розуміти, з чого складається Docker, потрібно знати про три його компоненти:

▪ образи (images);

▪ реєстр (registries);

▪ контейнери.

Docker-образ — це read-only шаблон. Наприклад, образ може містити операційну систему Ubuntu з Apache і додатком на ній. Образи використовуються для створення контейнерів. Docker дозволяє легко створювати нові образи, оновлювати існуючі, або можна завантажити образи, створені іншими людьми. Образи — це компонента збірки Docker-а. Docker-реєстр зберігає образи. Є публічні і приватні реєстри, з яких можна скачати або завантажити образи. Публічний

Docker-реєстр — це Docker Hub. Там зберігається величезна колекція образів. Образи можуть бути створені вами або можна використовувати образи, створені іншими користувачами. Реєстри — це компонента поширення.

Контейнери схожі на директорії. У контейнерах міститься все, що потрібно для роботи програми. Кожен контейнер створюється з образу. Контейнери можуть бути створені, запущені, зупинені, перенесені або видалені. Кожен контейнер ізольований і є безпечною платформою для додатка. Контейнери — це компонента роботи. Виходячи з цих трьох компонентів в Docker можна:

- створювати образи, в яких знаходяться додатки;
- створювати контейнери з образів, для запуску додатків;
- розповсюджувати через Docker Hub або інший реєстр образів.

*Принцип роботи Docker.*

Отже образ — це read-only шаблон, з якого створюється контейнер. Кожен образ складається з набору рівнів. Docker використовує union file system для поєднання цих рівнів в один образ. Union file system дозволяє файлам і директоріям з різних файлових систем (різних гілок) прозоро накладатися, створюючи когерентну файлову систему.

Одна з причин, з якої Docker легкий — це використання таких рівнів. Коли змінюється образ, наприклад, проходить оновлення додатку, створюється новий рівень. Так, без заміни всього образу або його перезібрання, як вам можливо доведеться зробити з віртуальною машиною, тільки рівень додається або оновлюється. І вам не потрібно роздавати весь новий образ, публікується тільки оновлення, що дозволяє поширювати образи простіше і швидше.

В основі кожного образу знаходиться базовий образ. Наприклад, ubuntu, базовий образ Ubuntu, або Fedora, базовий образ дистрибутива Fedora. Так само можна використовувати готові образи як базу для створення нових образів. Наприклад, образ apache можна використовувати як базовий образ для веб-додатків. Docker зазвичай бере образи з реєстру Docker Hub.

Docker образи можуть створитися з цих базових образів, кроки опису для створення цих образів називаються інструкціями. Кожна інструкція створює новий образ або рівень. Інструкціями будуть такі дії:

- запуск команди;
- додавання файлу або директорії;
- створення змінної оточення;
- вказівки, що запускати, коли запускається контейнер цього способу.

Ці інструкції зберігаються в файлі Dockerfile. Docker зчитує цей Dockerfile, коли збирається образ, виконує ці інструкції і повертає кінцевий образ.

Реєстр — це сховище Docker образів. Після створення образу ви можете опублікувати його на публічному реєстрі Docker Hub або на вашому особистому реєстрі. За допомогою Docker-клієнта ви можете шукати вже опубліковані образи і завантажувати їх на машину з Docker для створення контейнерів.

Docker Hub надає публічні і приватні сховища образів. Пошук і скачування образів з публічних сховищ доступні для всіх. Вміст приватних сховищ не попадає в результат пошуку. І тільки ви і ваші користувачі можуть отримувати ці образи і створювати з них контейнери.

*Принцип роботи контейнера.*

Контейнер складається з операційної системи, призначених для користувача файлів і метаданих. Відомо, що кожен контейнер створюється з образу. Цей образ говорить Docker'у, що знаходиться в контейнері, який процес запустити, коли запускається контейнер та інші конфігураційні дані. Docker-образ доступний тільки для читання. Коли Docker запускає контейнер, він створює рівень для читання / запису зверху образу (використовуючи union file system, як було зазначено раніше), в якому може бути запущено додаток.

Або за допомогою програми Docker, або за допомогою RESTful API, Docker-клієнт говорить Docker-домену запустити контейнер.

$ sudo docker run -i -t ubuntu /bin/bash

Давайте розберемося з цією командою. Клієнт запускається за допомогою команди Docker, з опцією run, яка говорить, що буде запущений новий контейнер. Мінімальними вимогами для запуску контейнера є такі атрибути:

▪ який образ використовувати для створення контейнера. У нашому випадку ubuntu;

▪ команду яку ви хочете запустити, коли контейнер буде запущений. У нашому випадку /bin/bash.

Після запуску цієї команди Docker по порядку робить наступне:

▪ завантажує образ ubuntu: Docker перевіряє наявність образу ubuntu на локальній машині, і якщо його немає — то викачує його з Docker Hub. Якщо ж образ є, то використовує його для створення контейнера;

▪ створює контейнер: коли образ отриманий, Docker використовує його для створення контейнера;

▪ ініціалізує файлову систему і монтує read-only рівень: контейнер створений у файлової системі і read-only рівень доданий образ;

▪ ініціалізує мережу/міст: створює мережевий інтерфейс, який дозволяє Docker'у спілкуватися хост машиною;

▪ установка IP адреси: знаходить і задає адресу;

▪ запускає вказаний процес: запускає програму;

▪ обробляє та видає вихід додатку: підключається і залоговує стандартний вхід, вихід і потік помилок додатку, щоб можна було відслідковувати, як працює програма.

Тепер у вас є робочий контейнер. Ви можете управляти своїм контейнером, взаємодіяти з вашим додатком. Коли вирішите зупинити додаток, видаліть контейнер.

*Технології, використані у Docker.*

Докер написаний на мові Go і використовує деякі можливості ядра Linux, щоб реалізувати наведений вище функціонал.

Docker використовує технологію namespaces для організації ізольованих робочих просторів, які називаються контейнерами. Коли запускається контейнер, Docker створює набір просторів імен для даного контейнера. Це створює ізольований рівень, кожен контейнер запущений у своєму просторі імен, і не має доступу до зовнішньої системи.

Список деяких просторів імен, які використовує Docker:

▪ pid: для ізоляції процесу;

▪ net: для управління мережевими інтерфейсами;

▪ ipc: для управління IPC ресурсами (ICP: InterProccess Communication);

▪ mnt: для управління точками монтування;

▪ utc: для ізолювання ядра і контролю генерації версій (UTC: Unix timesharing system).

Control groups (контрольні групи). Docker також використовує технологію cgroups або контрольні групи. Ключ до роботи додатка в ізоляції, надання додатку тільки тих ресурсів, які йому потрібні. Це гарантує, що контейнери будуть добре співіснувати. Контрольні групи дозволяють розділяти доступні ресурси заліза і, якщо необхідно, встановлювати межі і обмеження. Наприклад, обмежити можливу кількість пам'яті, що використовується контейнером.

Union File Sysem або UnionFS — це файлова система, яка працює, створюючи рівні, що робить її дуже легкою і швидкою. Docker використовує UnionFS для створення блоків, з яких будується контейнер.

Docker може використовувати кілька варіантів UnionFS включаючи: AUFS, btrfs, vfs і DeviceMapper.

Docker поєднує ці компоненти в обгортку, яку ми називаємо форматом контейнера. Формат, який використовується за умовчанням, називається libcontainer. Так само Docker підтримує традиційний формат контейнерів в Linux з допомогою LXC. В майбутньому Docker можливо буде підтримувати інші формати контейнерів. Наприклад, інтегруючись з BSD Jails або Solaris Zones.

### 3.2. Компіляція, збірка та розгортання мережевого засобу UkrVectōrēs (з GitHub репозиторію) в середовищі UNIX-подібних операційних систем Linux

Системні вимоги:

▪ мінімальні апаратні ресурси: x86−64 сумісний процесор з тактовою частотою 1 ГГц; оперативна пам'ять: 4 Гб; місце на жорсткому диску: 20 Гб;

UNIX-подібна операційна система Linux (при тестуванні, компіляції, збірці та розгортанні мережевого засобу UkrVectōrēs використовувались дистрибутиви Ubuntu Server 18.04 LTS x86−64 та Alpine Linux 3.9.4 x86−64);

Git розподілена система керування версіями файлів та спільної роботи;

Docker CE інструментарій для управління ізольованими Linux/Windows-контейнерами;

швидкісне підключення до мережі Інтернет.

Компіляція, збірка та розгортання мережевого засобу UkrVectōrēs в середовищі UNIX-подібних операційних систем Linux складається з таких етапів:

Клонування початкового коду програми UkrVectōrēs з git-репозиторію [37] сервісу GitHub за посиланням https://github.com/malakhovks/docsim. Цей етап можна виконати, використовуючи таку команду:

$ git clone https://github.com/malakhovks/docsim.git.

Або клонувати початковий код програми UkrVectōrēs з git-репозиторію [37] сервісу GitHub з конкретної гілки/тега використовуючи таку команду:

$ git clone --depth=1 --branch=<tag_name> <repo_url>,

де tag_name — ім'я гілки/тега;

repo_url — https-адреса приватного репозиторію з параметрами авторизації.

Приклад:

$ git clone --depth=1 --branch=develop https://Velychko-Vitalii:ae9c2fa2 d73fbbb0bd0a5ffa746f1df59036815c@github.com/malakhovks/docsim.git.

Або отримати реліз у вигляді архіву (початковий код програми UkrVectōrēs) у розробника, розпакувати його та перейти до наступного етапу.

1. Перехід в діректорію docsim:

$ cd docsim.

2. Перехід в гілку, яку потрібно використовувати для компіляції/збірки, командою git checkout:

$ git checkout <branch_name>,

де branch_name — ім'я гілки;

git-репозиторій програми docsim має дві основні гілки: develop та master.

Гілка master містить стабільний початковий код програми docsim.

Гілка develop містить робочий початковий код програми docsim.

Приклад:

$ git checkout master.

Створення ізольованого застосунку Docker, так званого Docker image з файлу Dockerfile:

$ docker build. — t <image name>,

де: image name — ім'я ізольованого застосунку Docker image.

Приклад:

$ docker build. — t docsim_image.

Створення ізольованого застосунку docsim_image може зайняти тривалий час в залежності від потужностей апаратного забезпечення. Повна документація по командах Docker доступна за посиланням [38].

3. Запуск створеного ізольованого застосунку docsim_image в контейнері docsim:

$ docker run --restart always --name docsim -d -p 80:80 docsim_image.

Команда Docker run з параметром --restart always дозволяє автоматично перезапускати при перезавантаженні операційної системи, що дозволяє досягти безперебійної роботи сервісу.

Основні команди керування Docker-контейнером:

▪ docker attach docsim — побачити вихід консолі контейнера docsim;

▪ docker stop docsim — зупинити контейнер docsim;

▪ docker start docsim — відновити роботу (старт) контейнера docsim;

▪ docker rm docsim — видалення контейнера docsim (перед видаленням контейнера його потрібно зупинити);

Деякі корисні параметри для запуску Docker-контейнера:

▪ --name — дає контейнеру ім'я, яке можна знайти у виводі команди docker ps;

▪ -p 80:80 — публікує порт 80. Другий номер 80 після двокрапки повідомляє, який порт сервер nginx слухає всередині контейнера;

−d — запускає контейнер, від'єднаний від терміналу. Потім журнали можна переглядати за допомогою команди журналів Docker docker logs;

−t — щоб бачити консольний вихід Docker-контейнера;

−-restart on-failure — автоматичний перезапуск невдалих контейнерів. Перезапускає контейнер, якщо він вийде з ладу через помилку, яка виявляється як ненульовий код виходу;

−-restart always — завжди перезапускає контейнер, якщо він зупиняється. Якщо контейнер зупинено вручну, він перезапускається лише тоді, коли служба Docker перезапускається або сам контейнер перезапускається вручну.

### 3.3. Компіляція, збірка та розгортання мережевого засобу UkrVectōrēs (з GitHub репозиторію) в середовищі програми віртуалізації для операційних систем VirtualBox.

Системні вимоги:

мінімальні апаратні ресурси: x86−64 сумісний процесор з тактовою частотою 2 ГГц; оперативна пам'ять: 4 Гб; місце на жорсткому диску: 20 Гб;

x86−64 сумісна UNIX-подібна операційна система Linux; x86−64 сумісна операційна система Microsoft Windows 7 Service Pack 1 або новіша;

VirtualBox [39] програма віртуалізації для операційних систем версії VirtualBox 6.0.8 або новіша;

віртуальна машина з UNIX-подібною операційною системою Linux (при тестуванні, компіляції, збірці та розгортанні мережевого засобу UkrVectōrēs використовувались дистрибутиви Ubuntu Server 18.04 LTS x86−64 та Alpine Linux 3.9.4 x86−64), яка включає таке встановлене програмне забезпечення: Git-розподілена система керування версіями файлів та спільної роботи; Docker CE інструментарій для управління ізольованими Linux/Windows-контейнерами;

швидкісне підключення до мережі Інтернет;

віртуальна машина — модель обчислювальної машини, створена шляхом віртуалізації обчислювальних ресурсів: процесора, оперативної пам'яті, пристроїв зберігання та вводу і виводу інформації. Віртуальна машина на відміну від програми емуляції конкретного пристрою забезпечує повну емуляцію фізичної машини чи середовища виконання (для програми).

**VirtualBox** — програма для створення віртуальних машин, що належить Oracle Corporation. Ця програма є в вільному доступі та підтримується основними операційними системами *Linux, FreeBSD, Mac OS X, OS/2 Warp, Microsoft Windows*, які підтримують роботу гостьових операційних систем FreeBSD, Linux, OpenBSD, OS/2 Warp, Windows і Solaris.

Ключові можливості.

- кросплатформовість;
- модульність;
- жива міграція;
- підтримка USB 2.0, коли пристрої хост-машини стають доступними для гостьових ОС (лише в пропрієтарній версії);
- підтримка 64-бітних гостьових систем (починаючи з версії 2.0), навіть на 32-бітних хост-системах (починаючи з версії 2.1, для цього потрібна підтримка технології віртуалізації процесором);
- підтримка SMP на стороні гостьової системи (починаючи з версії 3.0, для цього потрібна підтримка технології віртуалізації процесором);
- вбудований RDP-сервер, а також підтримка клієнтських USB-пристроїв поверх протоколу RDP (лише в пропрієтарній версії);
- експериментальна підтримка апаратного 3D-прискорення (OpenGL, DirectX 8/9 (з використанням коду wine) (лише в 32-бітних Windows XP и Vista), для гостьових DOS/Windows 3.x/95/98/ME підтримка апаратного 3D-прискорення не передбачена;
- підтримка образів жорстких дисків VMDK (VMware) и VHD (Microsoft Virtual PC), включаючи snapshots (починаючи з версії 2.1);
- підтримка iSCSI (лише в пропрієтарній версії);
- підтримка віртуалізації аудіопристроїв (емуляція AC97 або SoundBlaster 16 на вибір);
- підтримка різноманітних видів мережевої взаємодії (NAT, Host Networking via Bridged, Internal);
- підтримка ланцюжка збережених станів віртуальної машини (snapshots), до яких можна повернутися з будь-якого стану гостьової системи;

▪ підтримка Shared Folders для простого обміну файлами між хостовою та гостьовою системами (для гостьових систем Windows 2000 і новіше, Linux та Solaris);

▪ підтримка інтеграції робочих столів (seamless mode) хостової та гостьової ОС;

▪ є можливість вибору мови інтерфейса (підтримується й україномовний інтерфейс).

Компіляція, збірка та розгортання мережевого засобу UkrVectōrēs (з git-репозиторію [37]) в середовищі програми віртуалізації для операційних систем VirtualBox складається з таких етапів.

Створення віртуальної машини з операційною системою Alpine Linux 3.9.4 x86−64 або новішою, згідно з настановами користувача, наведеними на офіційному сайті VirtualBox [39] nf, використовуючи відео-туторіали. Встановити апаратні ресурси для віртуальної машини згідно з прогнозованим навантаженням на сервіс UkrVectōrēs.

Встановлення Git та Docker CE в середовиші віртуальної машини з операційною системою Alpine Linux 3.9.4 x86−64 згідно з настановами користувача, наведеними на офіційному сайті wiki-документації wiki.alpinelinux.org.

Клонування початкового коду програми UkrVectōrēs з git-репозиторію [37] сервісу GitHub за посиланням https://github.com/malakhovks/docsim. Цей етап можна виконати, використовуючи команду:

$ git clone https://github.com/malakhovks/docsim.git.

Або клонувати початковий код програми UkrVectōrēs з git-репозиторію [37] сервісу GitHub з конкретної гілки/тега використовуючи команду:

$ git clone --depth=1 --branch=<tag_name> <repo_url>,

де tag_name — ім'я гілки/тега;

repo_url — https-адреса приватного репозиторію з параметрами авторизації.

Приклад:

$ git clone --depth=1 --branch=develop https://Velychko-Vitalii:ae9c2fa2 d73fbbb0bd0a5ffa746f1df59036815c@github.com/malakhovks/docsim.git.

Або отримати реліз у вигляді архіву (початковий код програми UkrVectōrēs) у розробника, розпакувати його та перейти до наступного етапу.

1. Перехід в директорію docsim:

$ cd docsim.

2. Перехід в гілку, яку потрібно використовувати для компіляції/збірки, командою git checkout:

$ git checkout <branch_name>,

де branch_name — ім'я гілки;

git-репозиторій програми UkrVectōrēs має дві основні гілки: develop та master.

Гілка master містить стабільний початковий код програми UkrVectōrēs.

Гілка develop містить робочий початковий код програми UkrVectōrēs.

Приклад:

$ git checkout master.

Створення ізольованого застосунку Docker, так званого docker image з файлу Dockerfile:

$ docker build. — t <image name>,

де image name — ім'я ізольованого застосунку docker image.

Приклад:

$ docker build. — t docsim_image.

Створення ізольованого застосунку docsim_image може зайняти тривалий час в залежності від потужностей апаратного забезпечення. Повна документація по командах Docker доступна за посиланням [38].

3. Запуск створеного ізольованого застосунку docsim_image в контейнері docsim:

$ docker run --restart always --name docsim -d -p 80:80 docsim_image.

Команда docker run з параметром --restart always дозволяє автоматично перезапускати при перезавантаженні операційної системи, що дозволяє досягти безперебійної роботи сервісу.

**4. Електронна бібліотека медіа-файлів підсистеми телереабілітації TISP — сервіс vHealth.**

Практична реалізація і впровадження алгоритмів та технологій, що входять в застосунок UkrVectōrēs, інтегрована у веб-сервіс *vHealth*, зокрема інтелектуальна інформаційно-пошукова підсистема vHealth повністю працює з використанням алгоритмів та технологій дистрибутивно-семантичного аналізу UkrVectōrēs.

Першочергові виклики постали перед системою медичної реабілітації в Україні. До особливо важливих завдань відноситься, у першу чергу, реабілітація хворих, які одужали від COVID-19. Цей факт добре усвідомлюється як суспільством, так і керівництвом МОЗ України, яке наразі створює спеціальну робочу групу з цієї проблеми.

Україна має систему лікувально-профілактичних закладів, призначених для психологічної та фізичної реабілітації військовослужбовців, в яких використовуються сучасні технології реабілітації. Однак довготривала реабілітація в таких центрах доступна далеко не всім. Тому застосування технології телереабілітації хворих з посттравматичним стресовим розладом та подібними розладами, в поєднанні з засобами об'єктивного контролю функціонального стану є вкрай важливим.

Роботи виконуються на перетині робіт за проектами Національного фонду досліджень України, які мають назву «Трансдисциплінарна інтелектуальна інформаційно-аналітична система супроводження процесів реабілітації при пандемії (TISP)» 2020—2020 рр. та «Розробка хмарної платформи пацієнтцентричної телереабілітації онкологічних хворих на основі математичного моделювання» 2022—2023 рр. Особливістю систем є те, що вони базується на знанняорієнтованій технології, онтологічному інжинірингу і трансдисциплінарній парадигмі. Когнітивні сервіси системи реалізують структуризацію і класифікацію інформації, синтезують необхідні документи на основі семантичного аналізу, виявляють характерні властивості інформаційних процесів і забезпечують підтримку прийняття рішень на всіх етапах їх життєвого циклу.

Методологія реабілітаційних заходів в умовах пандемії має ряд суттєвих особливостей, пов'язаних з непередбачуваністю і високою швидкістю виникнення (на відміну від звичайної ситуації) проблем високої складності, обмеженістю спілкування між реабілітологом і пацієнтом, необхідністю високої реактивності прийняття рішень і їх відповідністю, масштабністю процесу, і пов'язаною з нею необхідністю використання масштабованих операційних засобів тощо. Одним з ефективних рішень в наданні медичної реабілітаційної допомоги є дистанційна пацієнтцентрична реабілітація, яка потребує online-засобів теледіагностики, телеметрії і втручання з орієнтацією на можливості пацієнта, розвинутої Internet-взаємодії, інтелектуальних інформаційних технологій і сервісів, ефективних методів когнітивної підтримки в системі *«Реабілітолог — пацієнт — мультидисциплінарна команда»*, статистичної обробки великих об'ємів інформації (зокрема даних анкетування та телеметрії) з виділенням достовірних знань тощо. Звідси поряд з традиційними засобами реабілітації в процесі реалізації проєкту *«Трансдисциплінарна інтелектуальна інформаційно-аналітична система супроводження процесів реабілітації при пандемії*

*(TISP)»* [40], що переміг у конкурсі «Наука для безпеки людини та суспільства» Національного фонду досліджень України (НФДУ) [41] й отримав грантове фінансування, у складі системи TISP, з'явилася *Smart-система телемедичного супроводження реабілітаційних заходів* [42]. В поєднанні з інтелектуальними дистанційними засобами біологічного зворотного зв'язку [43] і ефективними мініатюрними приладами (англ. *Embedded systems, Wearable devices*) теледіагностики [40; 44], телеметрії і відновлення такі системи мають великі перспективи, про що свідчить також і світовий досвід.

Smart-система телемедичного супроводження реабілітаційних заходів — це комплексна, інтегрована, пацієнтцентрична інформаційна підсистема TISP надання медичної допомоги, вирішення різних клініко-організаційних та науково-дослідних задач у галузі реабілітаційної медицини (консультації; дистанційний нагляд і супроводження реабілітаційних процесів та заходів; виявлення, класифікація, прогнозування та вивчення знань; дослідження та огляд нових предметних галузей), з використанням засобів дистанційного зв'язку, елементів технологій штучного інтелекту, зокрема онтологічного інжинірингу [45 — 48] та машинного навчання [49].

Електронна бібліотека медіа-файлів підсистеми телереабілітації TISP — сервіс vHealth — це розподілена інформаційна система, що дозволяє зберігати, використовувати та розповсюджувати (функція шерингу) різнорідні колекції електронних документів (відео- та аудіоконтент) довільних предметних галузей, для дистанційного навчання пацієнтів і їх родичів, зокрема реабілітаційному комплексу вправ та заходів.

Реабілітація людей, що одужали від COVID-19, є безумовно вкрай важливою соціальною проблемою. Цей факт вже добре розуміється міжнародною лікарською спільнотою. У той же час доступність реабілітаційних заходів у різних країнах дуже різна. В країнах з розвиненою страховою медициною процес реабілітації в відповідних спеціалізованих центрах доступний багатьом. На жаль, в Україні ситуація інша — потреби пацієнтів в амбулаторних реабілітаційних послугах набагато перевищують наявні ресурси, що вимагає пошуку альтернативних рішень і підключення сучасних і передових технологій для підтримки пацієнтів. Тому є велика потреба у використанні нового сучасного напрямку відновлювальної медицини — телереабілітації.

Світове сучасне та загальноприйняте визначення поняття *Телереабілітація* або *E-реабілітація* (англ. *E-rehabilitation*) [44] — це комплекс реабілітаційних вправ і навчальних програм, які надаються пацієнту дистанційно за допомогою телекомунікаційних комп'ютерних технологій переважно на амбулаторному етапі лікування. Сенс цього сучасного напряму у тому, що пацієнт самостійно, як правило, у домашніх умовах виконує програми відновлювального лікування на амбулаторному етапі під дистанційним контролем і керівництвом лікаря-спеціаліста. Телереабілітація має супроводжуватися відповідним програмним забезпеченням, яке дозволяє спеціалісту з реабілітації, що спостерігає пацієнта в стаціонарі, швидко скласти індивідуальний комплекс вправ для самостійних занять у відеоформаті. Цей комплекс має коригуватися в залежності від динаміки відновлення. Таким чином пацієнти мають змогу продовжувати структуровану програму домашньої реабілітації, розроблену на стаціонарному етапі. Пацієнти отримують зворотний зв'язок від вже знайомих фахівців, відзначаються результати, досягнуті за минулий період, ставляться нові, актуальні для пацієнта завдання, здійснюється об'єктивний контроль відповідних функцій. Все це, безумовно, покращує і підтримує мотивацію пацієнта та забезпечує значно більшу ефективність амбулаторного етапу реабілітації.

Телереабілітація є втіленням відразу декількох сучасних технологічних трендів. По-перше, телереабілітація неможлива без власне телекомунікаційних технологій. По-друге, вона потребує застосування спеціалізованих для реабілітації медичних інформаційних систем. Ці системи потрібні як для адміністрування пацієнтів, так, і це є найголовніше, для створення низки реабілітаційних документів відповідно до структури реабілітаційного циклу, наприклад, індивідуального реабілітаційного плану, категоріального профілю, реабілітаційного прогнозу тощо. По-третє, важливою частиною телереабілітаційних технологій є функціональне оцінювання стану пацієнтів у домашніх умовах з використанням мініатюрних приладів та сучасних алгоритмів оцінювання даних у відповідності з тенденцією, яка англійською називається *point-of-care testing*, що у вільному перекладі означає медичний тест, який здійснюється безпосередньо в місці знаходження пацієнта, поза офісом лікаря. Нарешті, телереабілітація неможлива без активного залучення пацієнта до процесу прийняття рішень щодо його діагностики та лікування. Це є однією з основних тенденцій сучасної медицини, яка була підтримана в тому числі і на рівні законодавчих ініціатив у системі охорони здоров'я України.

Бурхливий розвиток телереабілітації у світі та набуття цим напрямком медицини трансдисциплінарних зв'язків з різноманітними предметними галузями, що виходять за рамки сучасної парадигми Е-здоров'я (англ. *E-health*), призвів до появи найсучаснішого різновиду реабілітації — **гібридна Е-реабілітація** (англ. *Hybrid E-rehabilitation*) [50].

Гібридна Е-реабілітація складається з ряду фундаментальних методів, підходів та технологій:

— *Телекомунікаційні технології* — надання реабілітаційних послуг через телекомунікаційні мережі та Інтернет. Крім того, це дозволяє пацієнтам дистанційно взаємодіяти з провайдерами та може використовуватися як для оцінки пацієнтів, так і для проведення терапії.

— *Насичені Інтернет-застосунки* (англ. *Rich Internet application*). Перш за все це медичні інформаційні системи (МІС). МІС — це всебічна, інтегрована інформаційна система, призначена для управління всіма аспектами роботи медичної установи (зокрема поліклініки, лікарні або реабілітаційного центру), такими, як медичні, адміністративні, фінансові та юридичні питання з відповідною обробкою/обміном даних (зокрема з реєстрами центральної бази даних електронної системи охорони здоров'я України [51]).

— *Телеметрія* — сукупність методів, підходів та технологій, що дають змогу проводити дистанційні вимірювання, збір, передачу та обробку інформації про показники діяльності (фізіологічні параметри) організму пацієнта (зокрема при виконанні реабілітаційного комплексу вправ та заходів в реальному часі).

— *Вбудовані системи* (англ. *Embedded systems)* та мініатюрні «розумні» прилади для носіння (англ. *Wearable devices, Wearables*). Такі пристрої для носіння можуть бути для загального використання, і в цьому випадку вони є лише особливим прикладом мобільних комп'ютерів. В якості альтернативи вони можуть бути призначені для спеціальних цілей, таких як фітнес-трекери або медичні пристрої. Вони можуть включати спеціальні датчики та сенсори, такі як акселерометри, монітори серцевого ритму, або більш просунуті, електрокардіограма, монітори насичення крові киснем та датчики контролю артеріального тиску;

— *Біологічний зворотний зв'язок* (БЗЗ, англ. *Biofeedback*). Технології БЗЗ [43] включають в себе комплекс дослідницьких, немедичних, фізіологічних, профілактичних і лікувальних процедур, в ході яких людині за допомогою зовнішньої ланцюга зворотного зв'язку, організо-ваного переважно за допомогою мікропроцесорної або комп'ютерної техніки, пред'являється інформація про стан і зміни тих чи інших власних фізіологічних процесів. БЗЗ-процедура полягає в безперервному моніторингу в режимі реального часу певних фізіологічних показників і свідомому управлінні ними за допомогою мультимедійних, ігрових та інших прийомів у заданій області значень. Таким чином протягом курсу БЗЗ-сеансів можливо посилити або послабити даний фізіологічний показник, а отже, рівень тонічної активації тієї регуляторної системи, чию активність показник відображає.

— *Інтелектуальні/віртуальні особисті помічники* (англ. *Intelligent/ Virtual personal assistant*). Це програмні агенти, що можуть надавати персональну інформацію, виконувати завдання та послуги для окремої особи. Сучасні програмні агенти класу інтелектуальних/віртуальних особистих помічників можуть взаємодіяти між собою задля виконання певного класу завдань. Взаємодія з такими помічниками з боку людини зазвичай відбувається за допомогою голосу або тексту. Іноді стосовно віртуальних помічників з текстовим інтерфейсом застосовують термін «чат-бот». Зокрема в рамках проєкту та системи TISP розроблено універсальну діалогову підсистему [52] (імплементовано в рамках предметної галузі «Фізична і реабілітаційна медицина» — ФРМ) у формах веб-застосунку та віртуального співрозмовника у сервісі «Telegram». Розроблена діалогова підсистема TISP використовує елементи онтологічного інжинірингу, зокрема онтологічне представлення «Білої книги з фізичної та реабілітаційної медицини в Європі» (БК, англ. *The White Book of Physical and Rehabilitation Medicine in Europe*) [53] та Міжнародну класифікацію функціонування, обмеження життєдіяльності та здоров'я» (МКФ, англ. *International Classification of Functioning, Disability and Health, ICF*) [54].

— Методи, технології та програмні застосунки на основі штучного інтелекту для обробки великих даних (англ. *Big data*) з метою отримання знань та вирішення аналітичних завдань [55]. Для виявлення та отримання знань та вирішення основних аналітичних завдань, таких як класифікація, діагностика чи прогнозування, ми використовуємо метод під назвою «Зростаючі пірамідальні мережі» [55] (ЗПМ, англ. *Growing pyramidal networks, GPN*). Результатом навчання є закономірність у вигляді логічної функції (функції алгебри логік). ЗПМ належить до класу статистичних методів. Інтелектуальний інформаційний пошук також базується на використанні прогностичних моделей дистрибутивної семантики [49].

### 4.1. Призначення та функції мережевого засобу vHealth.

Одним із основних завдань та призначень сервісу vHealth є інтеграція інформаційних ресурсів і ефективна навігація в них. Інтеграція інформаційних ресурсів — це їхнє об'єднання з метою використання різної інформації зі збереженням її властивостей, особливостей представлення і можливостей її обробляти. Об'єднання ресурсів може відбуватися як фізично, так і віртуально. Але при цьому таке об'єднання повинно забезпечувати користувачу сприйняття необхідної інформації як єдиного інформаційного простору: електронна бібліотека повинна забезпечити роботу з базами даних і високу ефективність інформаційних пошуків. Ефективна навігація в електронній бібліотеці — це можливість користувача знаходити інформацію, яка його цікавить, в усьому доступному інформаційному просторі з найбільшою повнотою і точністю при найменших затратах зусиль. Для вирішення цієї задачі сервіс vHealth використовує інтелектуальний пошук на основі прогностичних моделей дистрибутивної семантики.

Сервіс «Електронна бібліотека vHealth» має такі функціональні можливості:

• повний контроль над медіаконтентом та даними користувачів;

• підтримка декількох робочих процесів публікації контенту (режиму доступу): загальнодоступний контент (public), приватний контент (private), контент, що не входить до жодного списку та доступний тільки за посиланням (unlisted);

• підтримка декількох медіаформатів (медіатипів) даних: аудіоконтент, відеоконтент та в майбутньому планується підтримка текстових документів pdf, docx;

• можливість каталогізації об'єктів контенту і різних їхніх об'єднань (категорії та теги);

• обмін медіаконтентом з використанням спільного доступу до окремих медіаресурсів, списків відтворення (плейлистів), категорій, тегів. Автоматична генерація коду для вставки медіаконтенту на зовнішній Web-ресурс;

• інтелектуальний інформаційний пошук в реальному часі на основі прогностичних моделей дистрибутивної семантики (лексичний, символьний та атрибутний пошук);

• функція формування списків відтворення (плейлистів) медіаконтенту із налаштуванням робочих процесів публікації контенту (режиму доступу);

- сучасний дизайн графічного інтерфейсу користувача з підтримкою світлої та темної тем оформлення;
- функція розширеного адміністрування користувачами з використанням окремої панелі адміністратора сервісу;
- функції соціальної мережі: можливість додавати коментарі, вподобання (лайки та дизлайки) та завантаження медіаконтенту на локальний диск;
- наявність профілів кодування медіаконтенту: для кількох розширень (240p, 360p, 480p, 720p, 1080p) та декількох профілів кодування (h264, h265, vp9);
- функція адаптивної потокової передачі медіаконтенту: можливо використання протоколу HLS;
- підтримка багатомовних файлів субтитрів для відеоконтенту;
- функція поступового завантаження медіаконтенту (так званий chunked file upload);
- протоколювання сеансу роботи користувача із системою з можливістю переходу в кожний з раніше існуючих станів системи;
- маніпулювання зі структурою опису об'єкта медіаконтенту;
- підтримка апарату гіпертекстових і гіпермедійних зв'язків, що забезпечує користувачу оперативний перехід від об'єкта чи деякого його елемента до іншого взаємопов'язаного з ним об'єкта чи елемента.

**4.2. Програмні залежності мережевого засобу vHealth.**

— Python 3.8.6 — інтерпретатор та стандартні бібліотеки;

— gensim — програмна бібліотека з відкритим вихідним кодом для передової обробки та математичного моделювання природної мови;

— Django 3.1.6 — високорівневий відкритий Python-фреймворк (програмний каркас) для розробки веб-систем;

— PostgreSQL 12 — об'єктно-реляційна система керування базами даних;

— React 16.13.1 — відкрита JavaScript бібліотека для створення графічних інтерфейсів користувача, зокрема для Web-застосунків;

— Video.js — відкрита JavaScript бібліотека, веб-відеопрогравач;

— uWSGI — веб-сервер і сервер веб-додатків, спочатку реалізований для запуску додатків Python через протокол WSGI (і його бінарний варіант uwsgi);

— Celery 5.0.2 — є асинхронною чергою з відкритим кодом або чергою задач, яка базується на розподіленому передаванні повідомлень;

— Redis 6.2.1 — розподілене сховище пар ключ — значення, які зберігаються в оперативній пам'яті, з можливістю забезпечувати довговічність зберігання за бажанням користувача;

— nginx — вільний веб-сервер і проксі-сервер;

— Fine Uploader — відкрита JavaScript бібліотека для завантаження файлів.

**4.3. Графічний інтерфейс користувача мережевого засобу vHealth.**

За основу побудови графічного інтерфейсу користувача мережевого засобу vHealth було взято інтерфейс популярної медіаплатформи Youtube, який є еталоном для систем розповсюдження медіаконтенту. Під час розробки графічного інтерфейсу користувача мережевого засобу vHealth було дотримано одну з найважливіших вимог до сучасного графічного інтерфейсу програмної системи — концепцію «роби те, що я маю на увазі» або DWIM (англ. *Do What I Mean*). Тому система працює передбачувано, щоб користувач заздалегідь інтуїтивно розумів, яку дію виконає програма після отримання його команди. Це значно полегшує взаємодію користувача з системою та не потребує розробки додаткових методик та настанов користувачу для взаємодії з графічним інтерфейсом програмної системи.

Розглянемо основні елементи графічного інтерфейсу користувача мережевого засобу vHealth (рисунок 27):

— центральна частина інтерфейсу містить весь медіаконтент, доступний для користувача відповідно до його профілю. Медіаконтент (за замовчуванням) розподілено за категоріями:

— частина інтерфейсу зліва містить пункти головного меню мережевого засобу vHealth, зокрема пункти: «Завантажити» (функція завантаження нового медіаконтенту); «Мої файли» (функція перегляду завантаженого медіаконтенту в профілі користувача); «Мої плейлисти» (формування списків відтворення — плейлистів медіа-контенту із налаштуванням робочих процесів публікації контенту); «Історія» (функція перегляду списку медіаконтенту, що було вже переглянуто); «Персонал» (функція перегляду всіх користувачів сервісу vHealth); «Категорії» та «Теги» (функція каталогізації об'єктів медіа-контенту та різних їхніх об'єднань);

— графічний інтерфейс профілю (облікового запису) користувача наведено на рисунку 28. Користувач має доступ до персонального медіаконтенту та має змогу редагувати графічний інтерфейс свого профілю та кожен об'єкт медіаконтенту. Також є доступ до сторінки зі списками відтворення (плейлистами) користувача (рисунок 29).

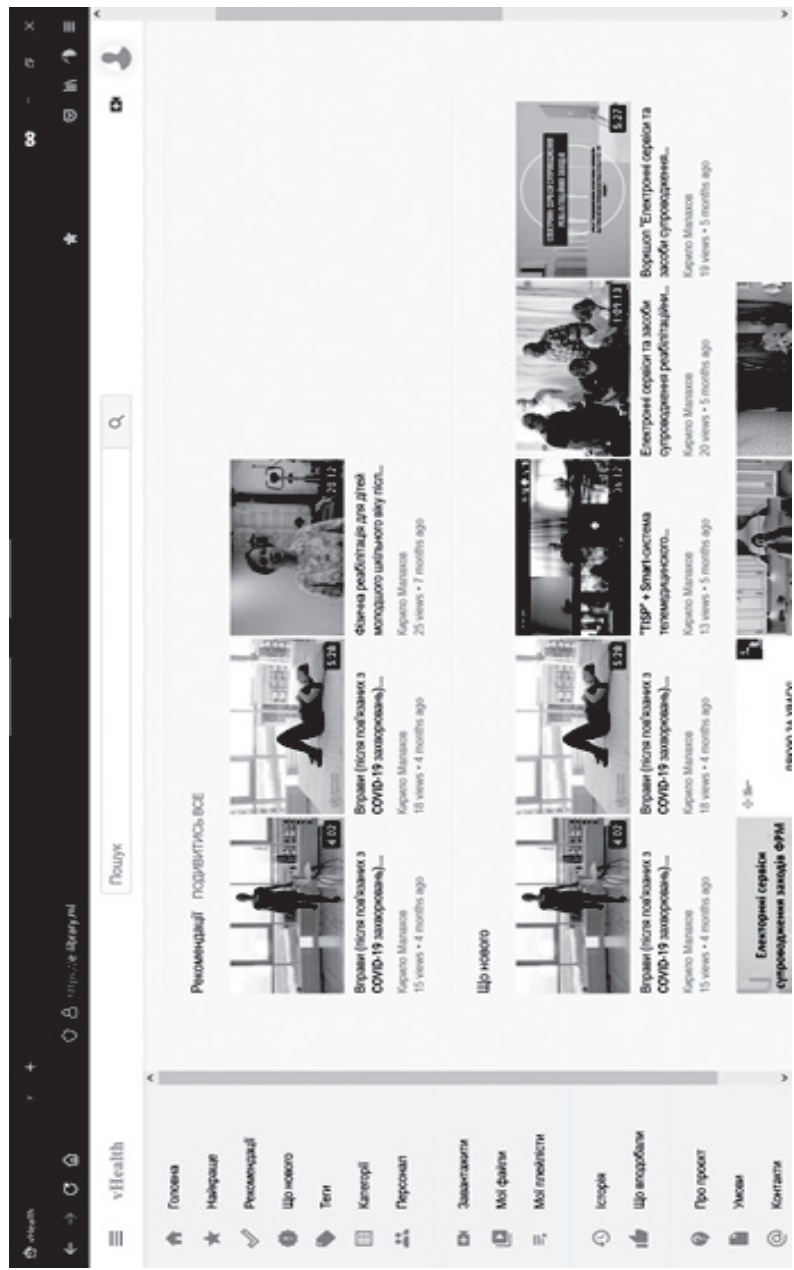

Рис. 27. Графічний інтерфейс користувача мережевого засобу vHealth (головна сторінка)

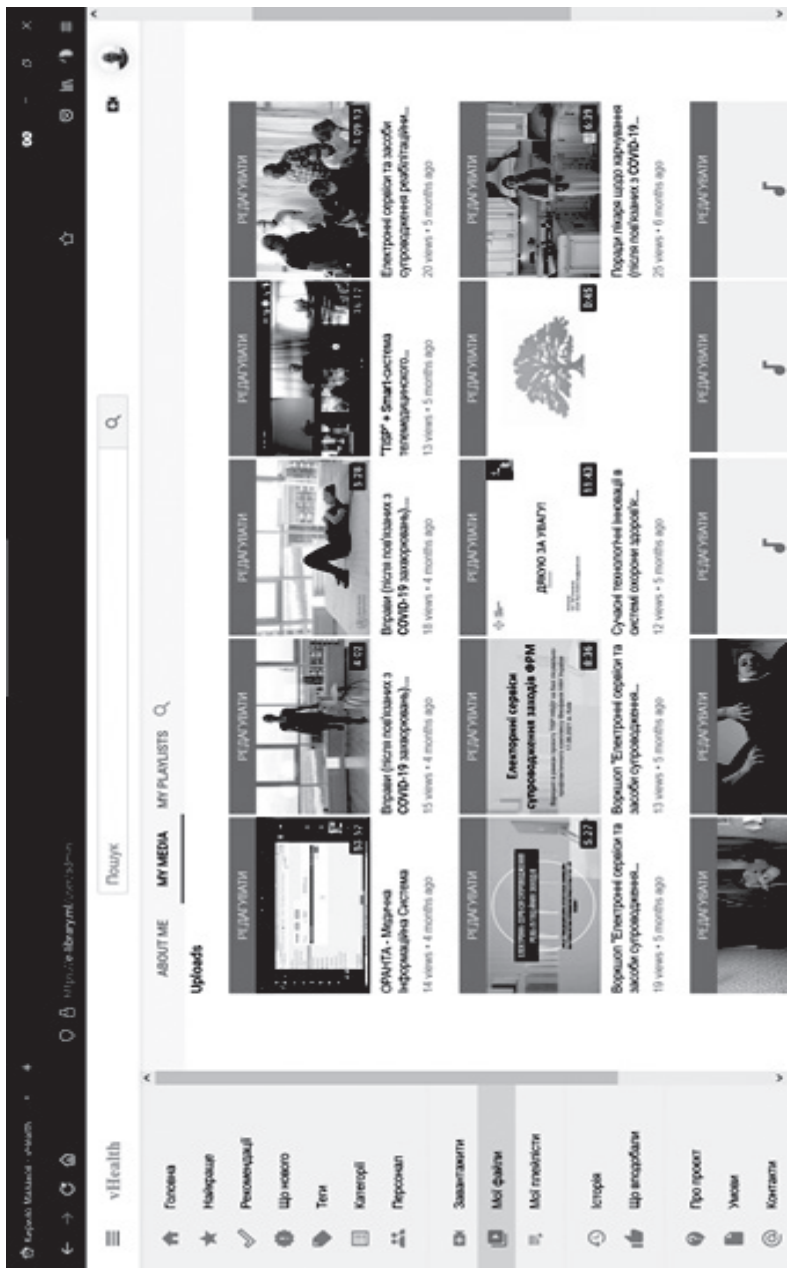

Рис. 28. Графічний інтерфейс користувача мережевого засобу vHealth (сторінка профілю користувача)

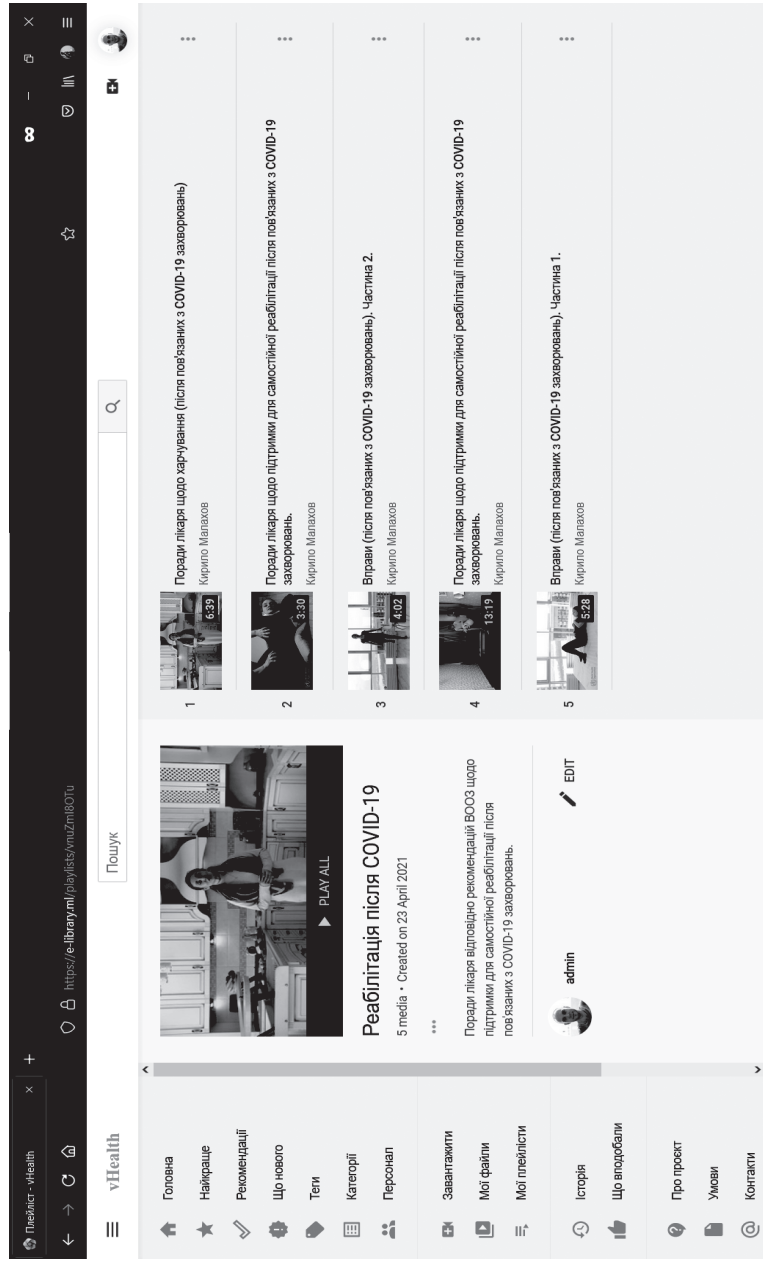

Рис. 29. Графічний інтерфейс користувача мережевого засобу vHealth (сторінка профілю користувача зі списками відтворення (плейлистами)

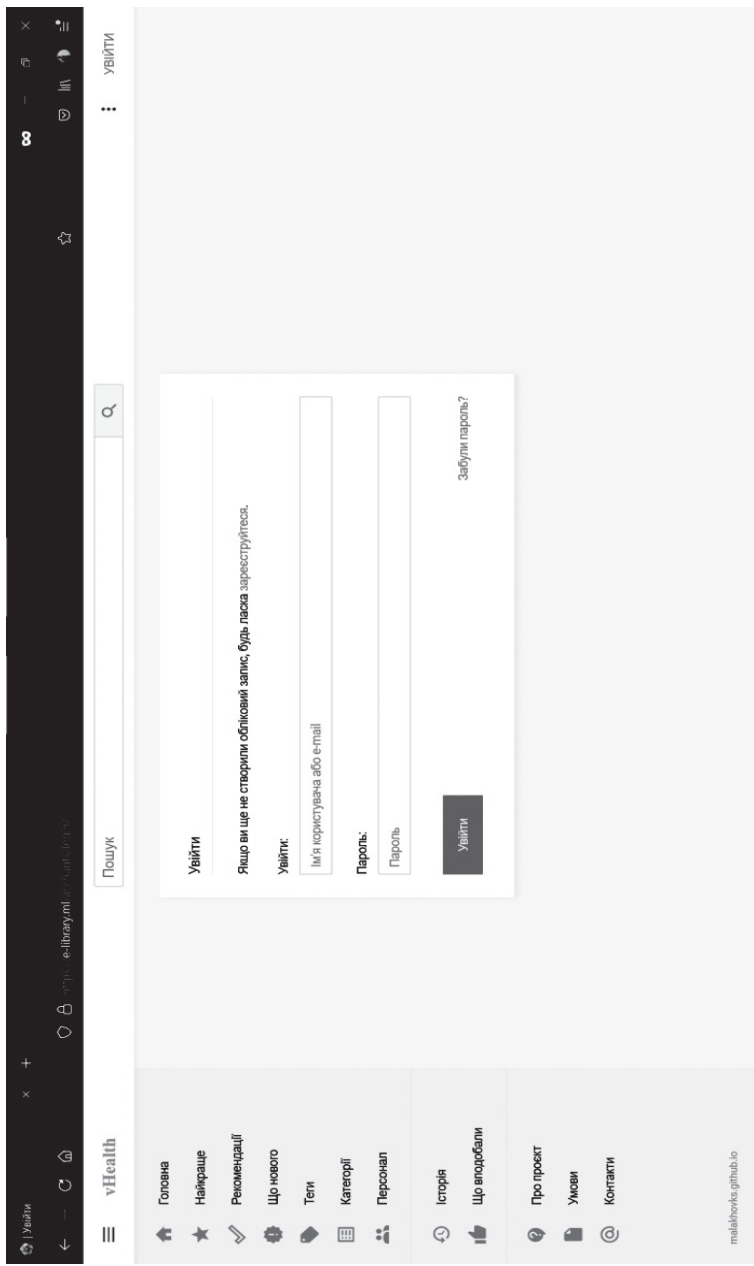

Рис. 30. Графічний інтерфейс користувача мережевого засобу vHealth (сторінка авторизації облікового запису користувача)

Технологія компіляції, збірки, розгортання та більш детальний опис початкового коду мережевого засобу vHealth, а також методика роботи користувача з графічним інтерфейсом застосунку vHealth буде наведено у фінальному звіті по проєкту «Трансдисциплінарна інтелектуальна інформаційно-аналітична система супроводження процесів реабілітації при пандемії (TISP)».

Актуальна версія сервісу vHealth доступна за посиланням: https://vhealth.ai-service.ml/

Для початку роботи з сервісом vHealth необхідно бути авторизованим користувачем (увійти), тому було створено обліковий запис для демонстрації роботи сервісу. Авторизуватися можна, використавши логін та пароль облікового запису демонстраційного профілю (рисунок 30):

Ім'я користувача (логін): demouser

Пароль: JyMyuC6nMdD494T

За посиланням: https://vhealth.ai-service.ml/accounts/login/

**Результати і висновки**

1. Систематизовано теоретичні засади методик мовного (дистрибутивно-семантичного) моделювання в математичній лінгвістиці.

2. Розроблено методику тренування дистрибутивно-семантичної моделі векторного представлення сутностей.

3. Розроблено програмну реалізацію сервісів Smart-системи телемедичного супроводження реабілітаційних заходів, зокрема електронної бібліотеки медіафайлів підсистеми телереабілітації TISP — сервіс vHealth та електронного засобу для дослідження, моделювання та вивчення довільних предметних галузей — сервіс UkrVectōrēs.

4. Розроблено моделі розгортання мережевих засобів UkrVectōrēs та vHealth.

5. Розроблено опис прикладного програмного інтерфейсу вебсервісів (back-end API) мережевого засобу UkrVectōrēs.

6. Визначено поняття *гібридна Е-реабілітація* та його фундаментальні основи.

**Перспективи подальших досліджень.**

У подальшій роботі планується впровадження в практику комп'ютерних програм для оптимізації затрат часу фахівцями мультидисциплінарної команди при застосуванні МКФ в реабілітації онкологічних хворих (зокрема на рак молочної залози). Також плануються подальші дослідження у визначенні та застосуванні ефективних математичних методів аналізу великих об'ємів даних, моделювання і побудови сценаріїв прогнозування й оптимізації усього комплексу реабілітаційних процедур і їх маршрутизації з використанням вже апробованих у колективі системних засобів, технологій та досвіду розробки реабілітаційних комплексів.

**Подяки**

**Науково-практичний вклад авторів**

*Величко Віталій Юрійович.* Розділи: Теоретичні засади методик мовного (дистрибутивно-семантичного) моделювання в математичній лінгвістиці; Розробка моделі розгортання мережевого засобу UkrVectōrēs.

*Малахов Кирило Сергійович.* Розділи: Розробка мережевого засобу (веб-сервісу) використання дистрибутивно-семантичних моделей векторного представлення сутностей природної мови — UkrVectōrēs; Електронна бібліотека медіафайлів підсистеми телереабілітації TISP — сервіс vHealth.

<div align="center">

**Ця книга присвячується моєму синові,
Марку Кириловичу Малахову.**

</div>

**JSON-схема вихідних даних сервісу визначення списку доступних для викорис-
тання дистрибутивно-семантичних моделей**

{

    «$schema»: «http://json-schema.org/draft-07/schema»,

    «$id»: «http://example.com/example.json»,

    «type»: «object»,

    «title»: «The root schema»,

    "description": «Схема вихідних даних сервісу визначення списку доступ-
них для використання дистрибутивно-семантичних моделей.»,

    «default»: {},

    «examples»: [

        {

            «models»: {

                «word2vec»: [

                    {

                        «description»: «Використовується нейронна вектор-
на модель представлення слів «Олесь Гончар» (з використанням набору
даних — проблеми поетики творчого доробку Олеся Гончара), алгоритм
word2vec word embeddings розмірністю 500d. Сутність — слово, лематизова-
но, приведено до нижнього регістру. Параметри word2vec: -size 500 -negative
5 -window 5 -threads 24 -min_count 10 -iter 20.»,

                        «name»: «honchar.lowercased.lemmatized.word2vec.
FINAL. 500d»,

                        «link»: «»,

                        «language»: «ua»,

                        «index»: 0

                    }

                ]

            }

        }

    ],

    «required»: [

        «models»

    ],

    «properties»: {

        «models»: {

            «$id»: «#/properties/models»,

            «type»: «object»,

            «title»: «The models schema»,

            «description»: «An explanation about the purpose of this instance.»,

            «default»: {},

«examples»: [
    {
        «word2vec»: [
            {
                «description»: «Використовується нейронна векторна модель представлення слів «Олесь Гончар» (з використанням набору даних — проблеми поетики творчого доробку Олеся Гончара), алгоритм word2vec word embeddings розмірністю 500d. Сутність — слово, лематизовано, приведено до нижнього регістру. Параметри word2vec: -size 500 -negative 5 -window 5 -threads 24 -min_count 10 -iter 20.»,
                «name»: «honchar.lowercased.lemmatized.word2vec.FINAL. 500d»,
                «link»: «»,
                «language»: «ua»,
                «index»: 0
            }
        ]
    }
],
«required»: [
    «word2vec»
],
«properties»: {
    «word2vec»: {
        «$id»: «#/properties/models/properties/word2vec»,
        «type»: «array»,
        «title»: «The word2vec schema»,
        «description»: «An explanation about the purpose of this instance.»,
        «default»: [],
        «examples»: [
            [
                {
                    «description»: «Використовується нейронна векторна модель представлення слів «Олесь Гончар» (з використанням набору даних — проблеми поетики творчого доробку Олеся Гончара), алгоритм word2vec word embeddings розмірністю 500d. Сутність — слово, лематизовано, приведено до нижнього регістру. Параметри word2vec: -size 500 -negative 5 -window 5 -threads 24 -min_count 10 -iter 20.»,
                    «name»: «honchar.lowercased.lemmatized.word2vec.FINAL. 500d»,
                    «link»: «»,
                    «language»: «ua»,

```
                                    «index»: 0
                                }
                            ]
                        ],
                        «additionalItems»: true,
                        «items»: {
                            «$id»: «#/properties/models/properties/word2vec/items»,
                            «anyOf»: [
                                {
                                    «$id»: «#/properties/models/properties/word2vec/
items/anyOf/0»,
                                    «type»: «object»,
                                    «title»: «The first anyOf schema»,
                                    «description»: «An explanation about the purpose of
this instance.»,
                                    «default»: {},
                                    «examples»: [
                                        {
                                            «description»: «Використовується не-
йронна векторна модель представлення слів «Олесь Гончар» (з використан-
ням набору даних — проблеми поетики творчого доробку Олеся Гончара),
алгоритм word2vec word embeddings розмірністю 500d. Сутність — слово,
лематизовано, приведено до нижнього регістру. Параметри word2vec: -size
500 -negative 5 -window 5 -threads 24 -min_count 10 -iter 20.»,
                                            «name»: «honchar.lowercased.lemmatized.
word2vec.FINAL. 500d»,
                                            «link»: «»,
                                            «language»: «ua»,
                                            «index»: 0
                                        }
                                    ],
                                    «required»: [
                                        «description»,
                                        «name»,
                                        «link»,
                                        «language»,
                                        «index»
                                    ],
                                    «properties»: {
                                        «description»: {
                                            «$id»: «#/properties/models/properties/
word2vec/items/anyOf/0/properties/description»,
                                            «type»: «string»,
```

«title»: «The description schema»,
«description»: «An explanation about the purpose of this instance.»,
«default»: «»,
«examples»: [
«Використовується нейронна векторна модель представлення слів «Олесь Гончар» (з використанням набору даних — проблеми поетики творчого доробку Олеся Гончара), алгоритм word2vec word embeddings розмірністю 500d. Сутність — слово, лематизовано, приведено до нижнього регистру. Параметри word2vec: -size 500 -negative 5 -window 5 -threads 24 -min_count 10 -iter 20.»
]
},
«name»: {
«$id»: «#/properties/models/properties/word2vec/items/anyOf/0/properties/name»,
«type»: «string»,
«title»: «The name schema»,
«description»: «An explanation about the purpose of this instance.»,
«default»: «»,
«examples»: [
«honchar.lowercased.lemmatized.word2vec.FINAL. 500d»
]
},
«link»: {
«$id»: «#/properties/models/properties/word2vec/items/anyOf/0/properties/link»,
«type»: «string»,
«title»: «The link schema»,
«description»: «An explanation about the purpose of this instance.»,
«default»: «»,
«examples»: [
«»
]
},
«language»: {
«$id»: «#/properties/models/properties/word2vec/items/anyOf/0/properties/language»,
«type»: «string»,
«title»: «The language schema»,

«description»: «An explanation about the purpose of this instance.»,
                                    «default»: «»,
                                    «examples»: [
                                        «ua»
                                    ]
                                },
                                «index»: {
                                    «$id»: «#/properties/models/properties/word2vec/items/anyOf/0/properties/index»,
                                    «type»: «integer»,
                                    «title»: «The index schema»,
                                    «description»: «An explanation about the purpose of this instance.»,
                                    «default»: 0,
                                    «examples»: [
                                        0
                                    ]
                                }
                            },
                            «additionalProperties»: true
                        }
                    ]
                }
            }
        },
        «additionalProperties»: true
    }
},
«additionalProperties»: true
}

**Початковий код програмного інструментарію навчання
прогностичних моделей дистрибутивної семантики
(програмна бібліотека gensim, алгоритм word2vec)**

```python
from __future__ import unicode_literals

import multiprocessing, time
from gensim.models import Word2Vec
from gensim.models import Word2Vec as WV_model
from gensim.models.word2vec import LineSentence
from gensim import utils

class MyCorpus(object):
    """An interator that yields sentences (lists of str)."""

    def __iter__(self):
        corpus_path = './dataset/honchar/extracted_lemmatized.txt'
        for line in open(corpus_path):
            # assume there's one document per line, tokens separated by whitespace
            yield utils.simple_preprocess(line)

# inp = "extracted.txt"

sentences = MyCorpus()

out_model = "../models/honchar.lowercased.lemmatized.word2vec.500d"

size = 500 # size is the dimensionality of the feature vectors.

window = 5 # window is the maximum distance between the current and predicted

word within a sentence.

sg = 1 # By default (sg=0), CBOW is used. Otherwise (sg=1), skip-gram is employed.

# cbow_mean = 1 # cbow_mean = if 0, use the sum of the context word vectors. If 1
(default), use the mean. Only applies when cbow is used.

sample = 1e-5 # sample = threshold for configuring which higher-frequency words are
randomly downsampled; default is 1e-3, useful range is (0, 1e-5).
```

negativeSampling = 5 # *negative = if > 0, negative sampling will be used, the int for negative specifies how many "noise words" should be drawn (usually between 5—20). Default is 5. If set to 0, no negative samping is used.*

hs = 0 # *hs = if 1, hierarchical softmax will be used for model training. If set to 0 (default), and negative is non-zero, negative sampling will be used.*

**iter** = 20

min_count = 10

workers = multiprocessing.cpu_count()

start = time.time()

*# model = Word2Vec(LineSentence(inp), sg = sg, size = size, window = window, workers = workers, negative = negativeSampling, iter = iter, min_count = min_count, hs = hs, sample = sample)*

model = Word2Vec(sentences = sentences, sg = sg, size = size, window = window, workers = workers, negative = negativeSampling, iter = **iter**, min_count = min_count, hs = hs, sample = sample)

*# trim unneeded model memory = use (much) less RAM*
model.init_sims(replace=True)

**print**(time.time()-start)

sim = model.wv.similarity('гончар', 'лист')
**print**(sim)
s = model.wv.most_similar('гончар')
**print**(s)
m = model.wv.most_similar('герой')
**print**(m)
lc = model.wv.most_similar(positive= ['гончар', 'герой'])
**print**(lc)

model.save(out_model)

# FUNDAMENTALS OF COMPUTER ECHOLOCATION
# IN DISTRIBUTER STRUCTURES

*Khoshaba O. M.*

*У роботі запропоновано основи комп'ютерної ехолокації в розподілених структурах на основі генерації та аналізу затримок повернутих сигналів різної частоти для виявлення структурно-функціональних порушень.*

*Показано класифікацію моделей впливу навантаження та обробки даних на основі детермінованих та ймовірнісних методів дослідження. Описано приклади використання комп'ютерної ехолокації в розподілених структурах. Представлено етапи дослідження розподілених структур за допомогою комп'ютерної ехолокації.*

*The paper proposes the basics of computer echolocation in distributed structures based on the generation and analysis of delays of reflected signals of different frequencies to detect structural and functional disorders.*

*The classification of models of loading impact and data processing based on deterministic and probabilistic research methods is shown. Examples of the use of computer echolocation in distributed structures are described. The stages of research of distributed structures using computer echolocation are presented.*

**1. Basic concepts and definitions.** Computer echolocation is a method for generating and analyzing delays of reflected signals of different frequencies to detect structural and functional disorders in computer systems or distributed structures.

The use of computer echolocation makes it possible to move to a higher, qualitative level in obtaining information about the structural and functional features of distributed systems and diagnosing their violations, which can manifest themselves both at earlier and later stages of the operation of corporate software and hardware systems.

Computer echolocation can diagnose structural and functional disorders in distributed systems by assessing the amplitude and frequency spectra of signals obtained due to loading effects on the objects of study by computer echolocation.

The importance of conducting echolocation studies is increasing since the developed infrastructure of an organization can consist of heterogeneous software and hardware complexes and devices where it is possible to determine abnormal areas of their functioning.

Computer echolocation is based on using the sent and reflected signals from the research object.

The use of computer echolocation is based on repeating sequences of stages that are aimed at studying the process of loading effects and data processing, selecting the parameters of the reference (required) trajectory of the loading effect on the research object, forming, cleaning (eliminating noise and distortion) and analyzing the signal.

The main stages of research of distributed structures using computer echolocation include the following.

1. Formation of a reference model of the trajectory of the load action.

2. Determination of the main parameters of the trajectory of the load action.

3. Preparation for signal shaping, sampling definition.

4. We are working on generating a signal, sending a signal, receiving, and shaping a reflected signal.

The stages of carrying out echolocation works are closely related to the use of subjects and objects of research.

As shown in Table 1, there are operations on the subjects of the study of computer echolocation. Operations used at the top level concerning the trajectory of loading actions are abbreviated as CRUD and mean the following.

C (create) — the operation of creating a trajectory with specific parameters in the form of an analog or discrete signal.

R (read) — operation of reading the parameters of the trajectory, which it can perform by displaying the required parameters. Also, such an operation can carry out the output of several types of trajectories.

Table 1

**Comparative characteristics of the levels of subjects of echolocation research**

| Specifications | Upper level | Lower level |
|---|---|---|
| Research subject | Load trajectory | Signal |
| Level feature | Abstract | Physical |
| The nature of the use of methods | Deterministic research methods | Probabilistic research methods |
| Operations on research subjects | CRUD operations | Sampling operations |

U (update) — the operation of adding a trajectory to the scenario of load action, since it is possible to combine several trajectories.

D (delete) — operation of deleting the added trajectory from the scenario of the load action. Such an operation can be carried out based on the input of the trajectory parameters or the identification number of the created or added trajectory.

It is convenient to perform the above operations at the stage of formalizing the task of the load action or when working with a database, where the parameters of the load action are processed.

**2. Classification of models of the process of exposure to load and data processing.** Load impact and data processing models describe the processes occurring in distributed structures. This can describe these processes using deterministic (Fig. 1) and probabilistic models (Fig. 2). In turn, each of the listed groups of deterministic and probabilistic models is divided into several categories that differ in the nature and characteristics of their use. Let's consider each group and category of models separately.

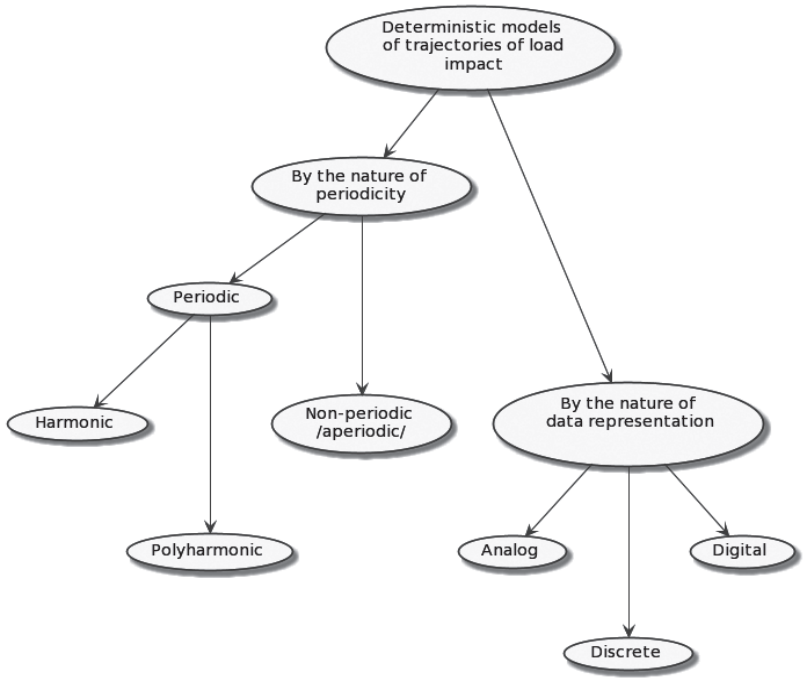

Fig. 1. Classification of deterministic models of trajectories of the loading action of the process of loading action and data processing

**2.1. Features of deterministic models of loading effects.** Deterministic models directly describe the processes of loading actions themselves. The model of the process of loading influence is deterministic (or non-random) when it is possible to describe its exact prediction (behavior) over any period of time.

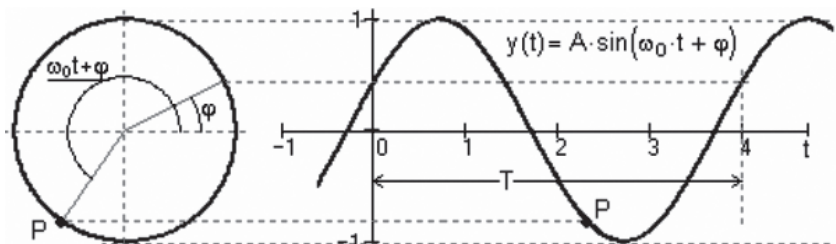

Fig. 2. Harmonic (or sinusoidal) functions that are used in the group of periodic trajectories of loading effects

The deterministic model of the load action and data processing ($M_{LI}$) process is expressed by the following relationship:

$$M_{LI} = F(t,z,\omega,...,A,B,C,...),$$

where $t$, $z$, $w$,... are independent arguments (time, spatial coordinate, frequency, etc.);

$A$, $B$, $C$... — trajectory parameters.

Using a deterministic model, the values of the trajectory of the load action are a priori known since they can be quite accurately determined (calculated) at an arbitrary moment in time on the numerical axis or at a point in space.

The physical meaning of the trajectory of the load action is that it is a reference function that contains information about the number of requests created at certain points in time to the object of study.

**2.2. Category of models of the nature of periodicity in trajectories of loading action.** In the category of models of periodicity nature, two groups in the load action trajectories can be distinguished, which correspond to periodic and non-periodic functions (Fig. 1).

In the periodic trajectories of the loading effects group, we include harmonic and polyharmonic functions. For periodic functions, the general condition is satisfied, under which the trajectory of the load impact ($Tr_{LI}$, the trajectory of load impact) takes the form:

$$Tr_{LI}(t) = Tr_{LI}(t + kT),$$

where $k = 1, 2, 3,...$ — any integer;

$T$ — period, which is a finite length of time.

**2.2.1. Harmonic paths of loading action.** Consider the harmonic trajectories of the loading action. Let us define the parameters of the trajectory of the loading action as amplitude, frequency, and phase shift, which will determine the general characteristics of the process of loading actions and data processing.

Then, let us designate the following information parameters of the load acts as a signal as follows:

A — signal amplitude (in units of measurement);

$f_o$ — cyclic frequency (in hertz);

$\omega_o = 2\pi f o$ — angular frequency (in radians);

$\varphi$ and $\phi$ are the initial phase angles (in radians).

In this case, the period of one swing will be:

$$T = 1/f_o = 2\pi/\omega_o.$$

This relationship also shows the relationship between cyclic and angular frequency.

If we take the group of harmonic (or sinusoidal) functions of the trajectory of loading actions as signals, then we describe the following relationships (Fig. 2):

$$Tr_{Lf}(t) = A\cdot sin\ (2\pi f_o\ t+\phi) = A\cdot sin\ (\omega_o\ t+\phi),$$

or

$$Tr_{Lf}(t) = A\cdot cos\ (\omega_o\ t+\varphi),$$

where A, $f_o$, $\omega_o$, $\phi$, $\varphi$ are constants that can play the role of information parameters of the signal.

We should also note that for $\varphi = \phi-\pi / 2$, the sine and cosine functions will describe the same signal. The signal's frequency spectrum can be represented (Fig. 3) by its amplitude and initial phase value of the frequency fo (at t = 0).

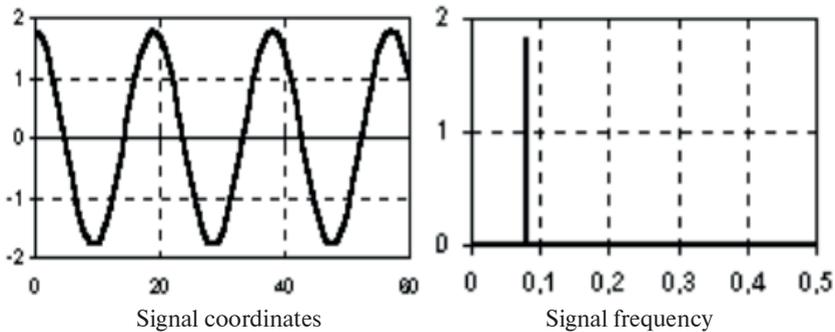

Signal coordinates          Signal frequency

Fig. 3. The trajectory of the loading effect in the form of a harmonic signal and the spectrum of its amplitude

**2.2.2. Polyharmonic paths of loading action.** The polyharmonic trajectories of the loading action are a set of trajectories and are described as follows:

$$Tr_{LI}(t) = \sum_{n=0}^{N-1} A_n sin(2\pi f_n t + \varphi_n),$$

or directly by function

$$Tr_{LI}(t) = y(t \pm kT_p),$$

where $k = 1,2,3,...,$

Tr is the period of one complete signal oscillation for the function $y(t)$, set for one period.

The value $fp = 1 / T_p$ is defined as the fundamental oscillation frequency.

At the same time, polyharmonic trajectories of loading actions, which are represented by signals, consisting of the sum of a certain constant component ($f_o = 0$) and an arbitrary (in the limit — infinite) number of harmonic components with arbitrary values of the amplitudes An and phases $\phi_n$, with periods that are multiples of the period of the fundamental frequency $f_p$.

Thus, on the period of the fundamental frequency fp, which is equal to or multiples of the minimum frequency of harmonics, multiple periods of all harmonics can fit. This creates a repetition rate for the signal.

The frequency spectrum of polyharmonic signals is discrete. In this regard, there is another mathematical representation of the signal, which is described in the form of spectra (or Fourier series).

For example, consider the representation of a periodic function (Fig. 4), obtained by summing the constant component (the frequency of the constant component is 0) and three harmonic oscillations with different values of the frequency and the initial phase of the oscillations.

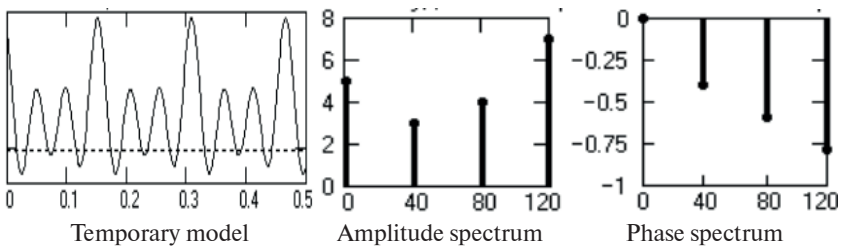

Temporary model   Amplitude spectrum   Phase spectrum

Fig. 4. Model of the trajectory of the loading action

The formula gives the mathematical description of the signal:

$$r_{LI}(t) = \sum_{n=0}^{3} A_n cos(2\pi f_n t + \phi_n),$$

where $A_k = \{5, 3, 4, 7\}$ — amplitude of harmonics;

$f_k = \{0, 40, 80, 120\}$ — frequency in hertz;

$\varphi_k = \{0, -0.4, -0.6, -0.8\}$ is the initial phase angle of the harmonic in radians;

$k = 0,1,2,3$ is the number of periods.

The fundamental frequency of the signal is 40 Hz.

The frequency representation of the trajectory of the load action (in the form of a signal spectrum) is shown in Fig. 4 (b). Also, the frequency representation of the periodic signal $Tr_{Lt}(t)$ is determined, limited by the number of spectrum harmonics.

It should note that a periodic signal of any arbitrary shape can be represented as a sum of harmonic oscillations with frequencies that are multiples of the fundamental frequency:

$$f_p = 1 \, / \, Tr.$$

To do this, it is enough to expand one period of the signal in a Fourier series in terms of trigonometric functions of sine and cosine with a frequency step equal to the fundamental frequency of oscillations $\Delta f = f_p$:

$$Tr(t) = \sum_{k=0}^{K} \left( a_k \cos 2\pi k \Delta f t + b_k \sin 2\pi k \Delta f t \right),$$

$$a_0 = (1/T) \int_0^T Tr(t) dt, a_k = (2/T) \int_0^T Tr(t) \cos 2\pi k \Delta f t dt,$$

$$b_k = (2/T) \int_0^T Tr(t) \sin 2\pi k \Delta f t dt.$$

The number of terms of the Fourier series $K = kmax$ is usually limited by the maximum frequencies fmax of harmonic components in the signals so that $f_{max} < K \cdot f_p$.

However, for signals with discontinuities and jumps, $f_{max} \to \infty$ exists. In this case, the number of members of the series is limited by the permissible error of approximation of the function $Tr_{Lt}(t)$.

Single frequency cosine and sine harmonics can be combined and decomposed in a more compact form as follows:

$$Tr(t) = \sum_{k=0}^{K} Tr_k \cos \left( 2\pi k \Delta f t - \varphi_k \right), ,$$

$$Tr_k = \sqrt{\phantom{x}} \; , \; \varphi_k = arctg \left( b_k \, / \, a_k \right).$$

There is also a representation of a rectangular periodic signal (meander) in an amplitude Fourier series in the frequency domain. For example, in fig. 5, the trajectory of the loading action in the form of a rectangular periodic signal, where the depicted signal is even concerning $t = 0$, does not have sinus harmonics, and all values of $\varphi_k$ for this signal model are equal to zero.

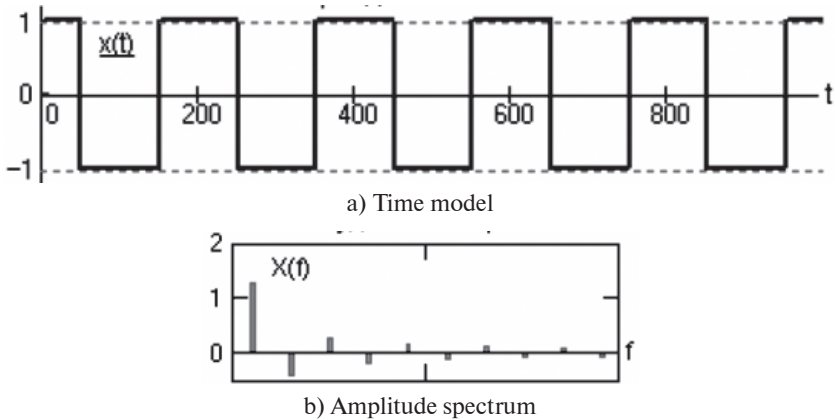

a) Time model

b) Amplitude spectrum

Fig. 5. Model of the trajectory of the load acting in the form of a rectangular periodic signal

The input parameters of the polyharmonic trajectory of the loading action can be:

– specific features of the waveform (swing from minimum to maximum, abnormal deviation from the mean, etc.);

– some parameters of signal harmonics.

For example, for rectangular pulses, the input parameters can be pulse repetition period, pulse duration, pulse duty ratio (i.e., the ratio of the period to duration).

When analyzing complex periodic signals, the input parameters can also be:

– the current average value ($A_C$) for a specific time, for example, for some time:

$$A_C = 1 / T \int_{t}^{t+T} Tr(t)dt \, ,$$

– constant component ($C_C$) of one period:

$$C_C = 1 / T \int_{0}^{T} Tr(t)dt \, ,$$

**584**

— average rectified ($A_R$) value:

$$A_R = 1/T \int_0^T |Tr(t)| dt \ .$$

**2.3. Models of the category of representation of the trajectory data of loading actions.** Models of the data presentation category describe processes based on trajectories in cases where they can represent the values of load actions as functions or signals:

— analog, in which there is no quantization;

— discrete, in which quantizations are used only on the time scale;

— digital, in which quantizations are used on all scales.

In this case, a signal can be understood as a physical or abstract process that contains information.

To the analog representation of the trajectory of the load, action belongs to such a continuous line, which has a set of values determined at each moment relative to the time axis (Fig. 6).

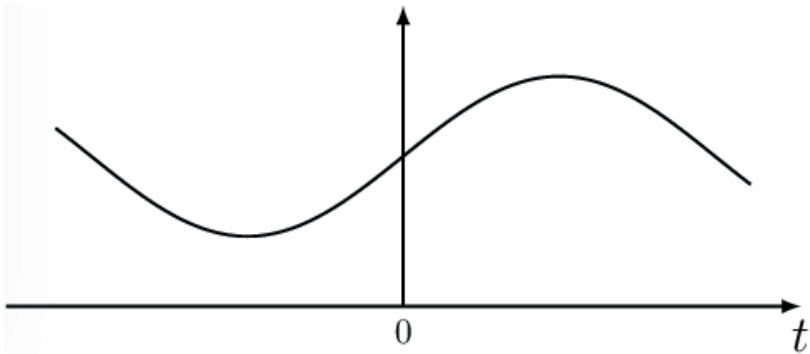

Fig. 6. Analog representation of the trajectory of the loading action

The values of the analog trajectory of the loading action (Fig. 6) are arbitrary at each moment. Therefore, such a trajectory can be represented as a continuous function (depending on time or on a variable) or as a piecewise continuous function of time.

The analog values of the trajectory of the loading action can have an infinite number of values within certain limits. They are continuous, and their values cannot change abruptly.

The analog form of the trajectory of the load action is written as *x(t)*, where t is a specific moment in time.

Such an imaginary line belongs to the discrete representation of the trajectory of the load action, at which there are many values only at specific points in time (Fig. 7).

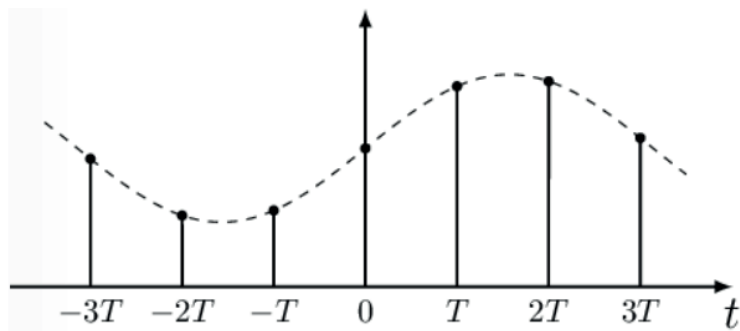

Fig. 7. Discrete representation of the trajectory of the loading action

Thus, the discrete representation of the trajectory of the load action takes on specific values only at certain moments of sampling. That is, the imaginary line is not continuous, in contrast to the analog representation of the trajectory of the load action. So, in fig. 7 shows an example of the formation of such a discrete representation of the trajectory of the load action with an interval (or step) of sampling $T$.

When using a discrete representation of the trajectory of the loading action, it is necessary to pay attention to the following provisions:

– for this type of representation, quantization is performed only at sampling intervals along the time scale, but not the values of the trajectory of the load action themselves;

– for this type of presentation, it is possible to use uniform and non-uniform intervals, which are located on the timeline.

The digital representation of the trajectory of the load action takes only fixed values (Fig. 8), which are located in the time axis with a certain interval (step).

Fig. 8 shows an example of the formation of a digital representation of the trajectory of the load acting on the basis of an analog one. In this case, it is necessary to pay attention to the fact that the values of digital representation cannot take intermediate values.

Thus, the digital representation of the trajectory of the load action is such a discrete representation in which quantization occurs not only in the time scale, but also in the level of values.

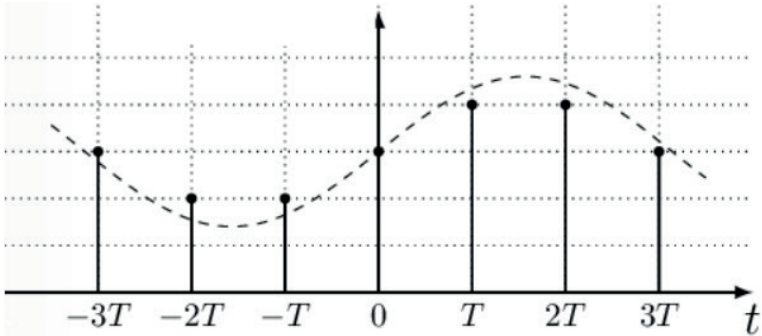

Fig. 8. Digital representation of the trajectory of the loading action

These can widely use such a representation of the trajectory of the load action in solving several problems. This representation of the load impact trajectory is especially in demand when using controllers or other computing devices to solve several modeling problems based on Boolean algebra.

**3. Probabilistic models of the process of loading effects and data processing**

**3.1. Features of using probabilistic models.** Random (probabilistic) models describe data processing processes due to loading effects on the research object. A data processing model describes the statistical characteristics of a random process by specifying various laws of a probability distribution, correlation function, spectral density, etc.

**4. An example of using the trajectory of the loading effect based on the frequency of sounding musical notes in octaves.** There is a table of the frequency of sounding musical notes in octaves. It is sometimes convenient to use this table to design the load path.

For example, the A note for the first octave (A) has a frequency of 440.00 hertz, the C# for the second octave is 554.37 hertz, and the E for the second octave (E) is 659.26 hertz. Let's round off the values of the proposed notes and get the following values:

$$A = 440 \text{ Hz};$$
$$C\# = 550 \text{ Hz};$$
$$E = 660 \text{ Hz}.$$

Let's write the program code for Matlab:
where the cyclic frequency will be:

$$fc = 440;$$

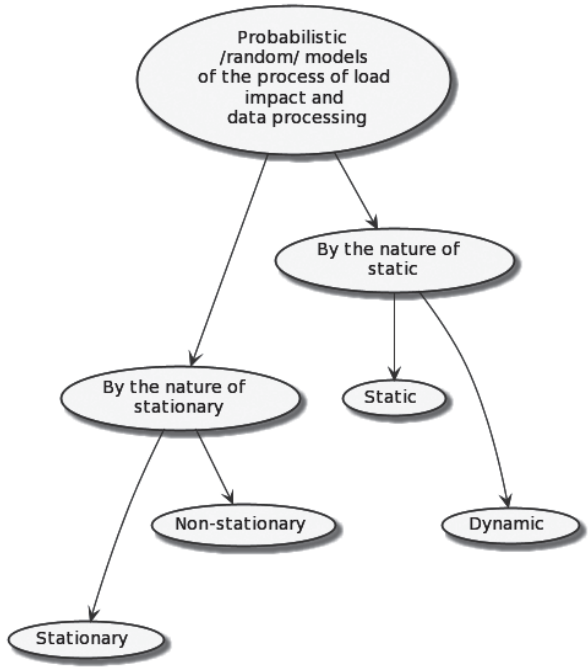

Fig. 9. Classification of probabilistic models of the process of loading impact and data processing

the sampling rate will be:

$$fs = 8000;$$

the time interval within the period will be:

$$dt = 1/fs;$$

the constant component ($A_k$) will be equal to:

$$Ak = 2.5.$$

Next, we create a vector of time t, which will have values from 0 to 0.1 seconds with a step of 1.25 * 10−4:

$$t = 0: dt: 0.1.$$

Next, we create a cosine wave as follows:

$$Tr = Ak + \cos(2 * pi * fc * t).$$

Consider the graph of the trajectory of the load action (Fig. 10):

plot (t, Tr)

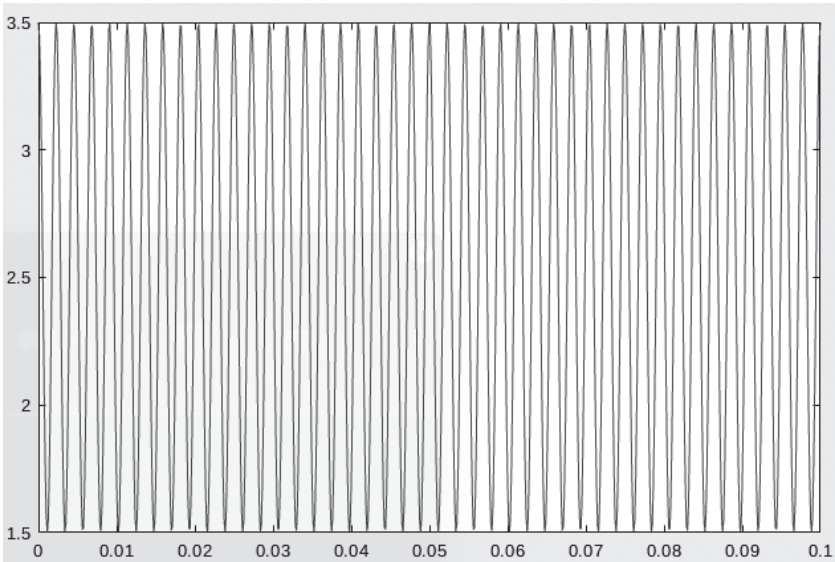

Fig. 10. Graph of the trajectory of the loading effect for the signal of the note A of the first octave

**5. Research structure of the distributed system using computer echolocation.** The study of distributed structures using computer echolocation has several functional components (Fig. 12): a load generator (or benchmark), a load balancer, a monitoring system, modules, and nodes.

Let's consider the main components of a distributed structure. Benchmark is a structural component of the research structure of the load balancer, which is necessary for the generation of request flows that are executed according to a specific trajectory.

A module is a structural component of the research structure of a load balancer. Its main functions include:

– processing and exchange of synchronous request flows;

– synchronization of requests between the pool and the load balancer.

The pool is a structural component of the research structure of the load balancer, which is necessary for:

– storage of request flows;

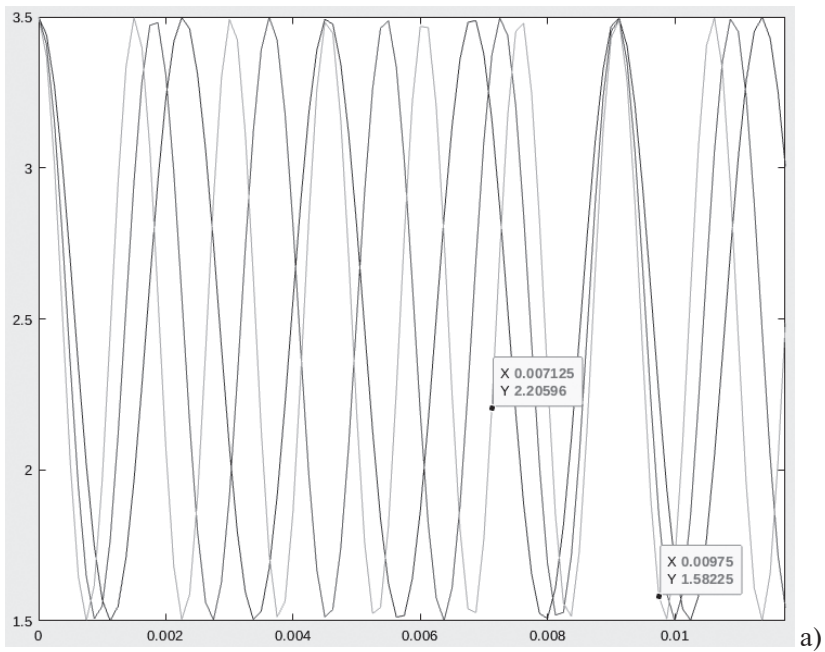

a)

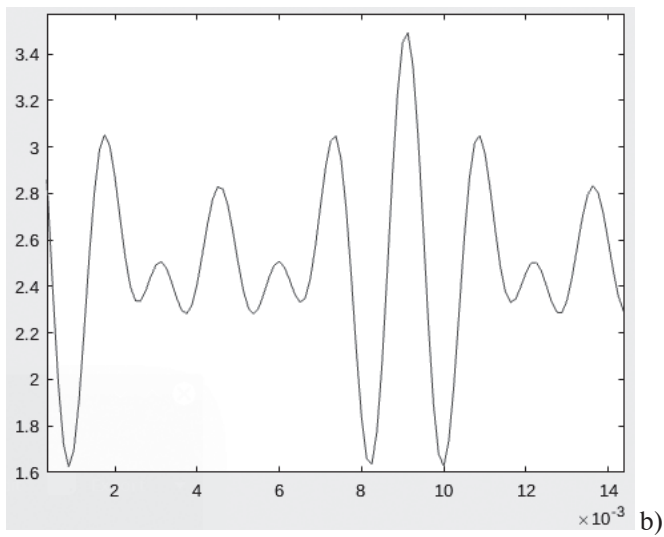

b)

Fig. 11. Graph of the trajectory of the loading effect for a signal of three notes:
a) presentation of individual signals b) presentation of the sum of three signals

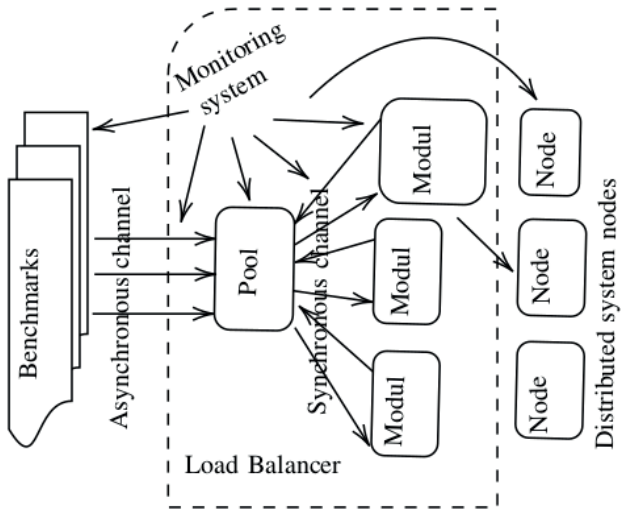

Fig. 12. Research structure of the distributed system.

— synchronization of request flows between the benchmark and the load balancer module.

In this load balancer research system, the pool is a dynamic bottleneck.

A distributed system node is a structural component of a load balancer. Its primary function is to implement request flows created by the benchmark in a specific scenario. Request flows are transferred to a distributed system node using a module.

The monitoring system of the research system is a structural component of the load balancer, which is necessary for measuring the performance of benchmarks, the pool of the load balancer, modules, and communication channels.

In this research structure of the load balancer, measuring the dynamics of changes in the bottleneck is based on the observation points. The structure can set observation points of the monitoring system of the research structure for a specific time interval of the process or the occurrence of some event.

Also, communication channels play an essential role in the work of the research structure of the load balancer. Communication channels are structural formations necessary to transfer requests to the main components of the load balancer.

The communication channels can transfer synchronous and asynchronous request flows in communication channels.

**6. Comparative characteristics of evaluations of the work of distributed systems.** Comparative characteristics of the assessments of the work of distributed systems are based on the assumption of the static and dynamic nature of the processes carried out in such structures. To solve the problems of determining the estimates of the operation of nodes in the literature, models of parallel and distributed structures are proposed based on reference algorithms for solving problems [1—4].

In this regard, the determination of the estimates of the work of nodes, which is based on the model of a parallel system, uses a reference sequential algorithm for solving some problem A by an application in time T [2, 4]. In this case, the acceleration estimate is used, which is defined as:

$$S = T_0 / T ,$$

where $T_0$ is the time to solve reference problem A by the application on one device (node) using the fastest sequential algorithm.

Acceleration S shows how many times the node can reduce the time for solving a problem by an application by using a parallel structure [3].

The next assessment of the work of applications on nodes in parallel structures is efficiency, which is usually defined [1, 3] as:

$$E = S / nT_0 / nT .$$

The described model of application operation on a parallel structure is simple for calculating scores. However, in this model, the values of their estimates are determined only after the solution of the problem, when the total time T is known. In addition, the model requires knowledge of the time of solving the problem by the best of the set of sequential algorithms T0 on one node on a parallel structure.

A different approach to using similar models for job appraisals has distributed structures due to two aspects. First, due to their heterogeneity, the nodes in distributed structures can have different values for evaluating the performance of applications since these nodes have different computing resources. Secondly, nodes in distributed structures may be unavailable at certain intervals throughout the solution of the entire problem.

Consider one of the models for distributed structures applications, which is characterized by the presence of a schedule and is defined as:

$$h_i(t): \mathfrak{R} \to \{0,1\} .$$

Moreover, hi (t) = 1 if the application on the node at time t is available for solving problem A. Otherwise, if hi (t) = 0, then the application on the node at time t is unavailable.

The performance estimates for applications on a distributed structure assume different relationships. So, for a model with a distributed structure schedule, the efficiency assessment takes the form [5]:

$$E_t = \frac{\underline{T}(A)}{T(A)},$$

where $T(A)$ is the reference time for solving problem $A$.

The reference time $T(A)$ means the time for solving problem $A$ by the $i$-th application at the node using the fastest sequential algorithm, where $T(A) > 0$.

Also, for applications on a distributed structure, such additional estimates are introduced as [5,6]: the performance of the reference system and the complexity of the task. Calculation of the reference system performance $(A, t)$ [5−7] is necessary to calculate $T(A)$.

The reference performance of the structure $i(A, t)$ is called [7] the performance of the $i$-th node of a distributed system when solving problem $A$:

$$\pi_i(A) = \frac{L(A)}{\underline{T_i}(A)},$$

where $L(A)$ is a function of the complexity of the task execution by an application on a distributed structure.

The function complexity of task execution $L(A)$ is defined on a certain set of tasks and expresses a priori knowledge of its complexity:

$$L(A): \Lambda \to \Re^+.$$

For example, an estimate of the number of elementary operations, calculated using complexity theory [8−10], can be referred to as the function of the complexity of task execution L (A) by an application on a distributed structure.

To solve the problems of determining the estimates of the operation of nodes in parallel and distributed structures, a model with a schedule is introduced, which describes the concept of the reference performance of the structure as the sum of the reference performance of the nodes [7]:

$$\pi(A,t) = \sum_{i=1}^{n} \pi_i(A) h_i(t) = (\vec{\pi}(A), \vec{h}(t)),$$

where with full availability of applications on nodes $h_i(t) \equiv 1$ during the solution of the problem and with the same reference performance, where $\pi_i(A) = \pi_0$, the reference performance of the distributed structure will coincide with the reference performance of the parallel structure since the condition $n \cdot \pi_0$ will be satisfied.

At the same time, a model with a schedule for solving problems from a set is called a set:

$$\Re = < \vec{\pi}(A), \vec{h}(t) > .$$

For a model with a schedule, the reference time for solving problem $A$ is called the value $\underline{T}(A)$ [5–7], which is determined by the following relation:

$$\underline{T}(A) = mint : \int_{t=0}^{t} \pi(A, \tau) d\tau .$$

In general, the acceleration rates for a parallel structure are defined as the ratio of the time it takes to solve a problem by an application at one node to the time it takes to solve a problem on the entire system. In a distributed structure with a schedule, the resources of the nodes can be different. Therefore, it is incorrect to introduce the concept of acceleration to applications in a distributed structure in the same way since it is unclear which node to calculate the acceleration parameter itself. Based on this, a more general concept of relative acceleration is introduced, which is defined as follows:

$$S(R_1, R_2) = \frac{T_1}{T_2} .$$

This relation shows the estimate of the acceleration $S$ for a model with a schedule $R_1$ relative to another system $R_2$ as the ratio of their time to solve problem $A$ on applications in a distributed structure.

Also, for a model with a schedule, it is customary to estimate the acceleration S for each node of the distributed structure as follows:

$$S(R_1, R_2) = \frac{T_1}{T_2} .$$

This relation shows the estimate of the acceleration Si for the $i$-th application on the node as the ratio of the reference time for solving problem A to the time for solving the same problem in a distributed structure based on a model with a schedule.

**Conclusions.** The paper proposes the basics of computer echolocation in distributed structures based on the generation and analysis of delays of reflected different frequency signals to detect structural and functional disorders.

The classification of models of loading and data processing based on determined and probabilistic research methods is shown. Examples of the use of computer echolocation in distributed structures are described.

The comparative characteristic of levels of subjects of echolocation research is described.

The stages of research of distributed structures with the help of computer echolocation are given.

# ЗАСТОСУВАННЯ МАТЕМАТИЧНИХ МОДЕЛЕЙ ТА ПРОГРАМНОГО ЗАБЕЗПЕЧЕННЯ ДЛЯ ПРОЕКТУВАННЯ НОВИХ ХАРЧОВИХ ПРОДУКТІВ

*Котлик С. В., Соколова О. П.*

*У роботі показано необхідність вживання людиною кисломолочних продуктів (сирів), показано їх переваги та користь, що вони приносять. До складу сирів входять різні тваринні жири, однак сучасні технології дозволяють замінювати тваринні жири рослинними. При цьому виникає необхідність застосовувати рослинні олії в такій пропорції, щоб одержання в суміші жирів НЖК, МНЖК, ПНЖК було близьким до співвідношень, які рекомендовані теорією раціонального харчування. Для цього були проведені відповідні експерименти, результати їх оброблені за допомогою методів регресійно-кореляційного аналізу, отримані адекватні математичні моделі. Для цього складено комп'ютерну програму, яка використовує власну базу даних та метод випадкового пошуку для оптимізації функцій. Програма була перевірена для деякої вибірки експериментальних даних, результати добре співвідносяться з реальними цифрами. Також були отримані математичні моделі гомогенізації емульсій при виробництві кисломолочного сиру з використанням купажів соняшникової та оливкової олій. Для визначення оптимальних режимів гомогенізації були побудовані математичні моделі залежності стійкості емульсії та відстою жирової фази від температури та тиску. Для їх використання також розроблено програмний додаток, проведено його перевірку, надано рекомендації щодо використання. В цілому розроблена програма дає в руки користувача-технолога інструмент, яким він може користуватися для розрахунку рецептури нових сортів кисломолочних продуктів з додаванням рослинних олій, не проводячи фізичних експериментів, досліджуючи властивості продукту на комп'ютері на підставі розроблених математичних моделей.*

*The paper shows the need for a person to consume fermented milk products (cheeses), shows their advantages and the benefits that they bring. The composition of cheeses includes various animal fats, but modern technologies make it possible to replace animal fats with vegetable ones. In this case, it becomes necessary to use vegetable oils in such a proportion that the production of SFA, MUFA, PUFA in a mixture of fats is close to the ratios recommended by the theory of rational nutrition. For this, appropriate experiments were carried out, their results were processed using regression-correlation analysis, and adequate mathematical models were obtained. To do this, a computer application has been compiled that uses its own database and a random search method to optimize functions. The program has been tested for some sample of experimental data, the results are*

*in good agreement with real numbers. Mathematical models of emulsion homogenization in the production of cheese were also obtained using blends of sunflower and olive oils. To determine the optimal modes of homogenization, mathematical models of the dependence of the stability of the emulsion and the settling of the fat phase on temperature and pressure were built. For their use, a software application was also developed, it was tested, and recommendations for use were given. In general, the developed program gives the user-technologist a tool that he can use to calculate the formulation of new varieties of fermented milk products with the addition of vegetable oils, without conducting physical experiments, examining the properties of the product on a computer based on the developed mathematical models.*

**Постановка проблеми.** Одне з найважливіших завдань із покращення структури харчування населення — збільшення продуктів масового споживання з високою харчовою та біологічною цінністю. Сучасне харчування має не лише задовольняти фізіологічні потреби організму людини в харчових речовинах та енергії, а й виконувати профілактичні та лікувальні функції та, звичайно, бути абсолютно безпечними.

Розробка продуктів харчування із заданими якісними характеристиками можлива за допомогою математичного моделювання їхнього рецептурного складу. Завдання моделювання полягає в обгрунтованому кількісному підборі основної сировини та збагачувальних добавок, що у сукупності забезпечує формування необхідних органолептичних та фізико-хімічних властивостей готового продукту із заданим рівнем споживчої та енергетичної цінності. При комп'ютерному моделюванні з'являється можливість оптимізації певних властивостей продукту, що розробляється, за встановленим критерієм (або критеріями) без використання дорогих експериментальних досліджень. Ця методологія дозволяє створювати продукти з певним вмістом білка, жиру, вуглеводів, вітамінів, харчових волокон, амінокислот, мінеральних та інших речовин [8; 9; 16; 21; 47].

Сьогодні поняття «проектування» продуктів включає в себе розробку моделей, що представляють математичні залежності, які відображають всі зміни одного або декількох ключових параметрів. При цьому необхідно проводити оптимізацію вибору і співвідношення вихідних компонентів для отримання рецептури, яка за кількісним вмістом і якісним складом максимально відповідає заданій формулі збалансованого харчування, відповідає заданим вимогам і володіє високими споживчими властивостями [22; 40; 46].

Розробка і виробництво нових продуктів заданої якості і складу в умовах сучасного розвитку науки і техніки (в першу чергу комп'ютерів і програмного забезпечення) вимагають застосування відповідного математичного апарату і високопродуктивного комп'ютерного обладнання.

Пошук і розробка ефективних чисельних методів, математичних моделей, алгоритмів і реалізація новітніх інформаційних технологій у вигляді комплексів проблемно-орієнтованих програм для вирішення задач оптимізації та проведення обчислювальних експериментів є актуальними для різних сфер виробничої діяльності, в тому числі при створенні нових харчових продуктів.

Створення таких ефективних рецептур в даний час базується на проведенні необхідних натурних експериментів, обробки результатів за допомогою методів регресійно-кореляційного аналізу, побудови адекватної математичної моделі, розробки відповідного програмного забезпечення і проведення комплексних розрахунків. Такий підхід дозволяє заощадити матеріальні засоби і отримати інструмент для розрахунку рецептури створення нових продуктів із заданими властивостями [21].

Виходячи зі сказаного, саме розширення можливостей оптимізаційних програмних засобів дозволить вийти на якісно новий рівень в розробці нових видів харчових продуктів із заданим хімічним складом, споживчими і технологічними характеристиками. Проектування харчових продуктів оптимального складу методами математичного моделювання дозволить знизити фінансові та часові витрати на розробку продуктів харчування, до яких відносяться і різні кисломолочні продукти, в тому числі сири, своєчасно реагувати на зміну потреб людського організму в умовах техногенного суспільства і суттєво розширити асортимент продукції функціонального, лікувально-профілактичного та лікувально-терапевтичного призначення, спрямованих на харчування окремих груп населення [23].

Майже у всіх лікувальних меню, що пропонуються лікарями, одним з перших значаться усі молочні та кисломолочні продукти, насамперед сир. Але він корисний і здоровим людям будь-якого віку. Сир є концентратом молочного білка і деяких інших складових частин молока. Важливість білка в нашому житті загальновідома: це той матеріал, з якого будуються всі клітини організму, ферменти, а також іммунні тіла, завдяки яким організм отримує стійкість до захворювань. Організм людини отримує білки разом з їжею, розщеплює їх до амінокислот і з цих своєрідних цеглинок будує молекули нових білків, властивих тільки нашому організму. Для цього йому необхідний набір з 20 амінокислот. Основним постачальником саме цих амінокислот і служить сир. Поряд з білками для нормальної життєдіяльності організму необхідні і мінеральні речовини, найважливіші з яких — з'єднання кальцію і фосфору. Саме останні складають основу кісткової тканини і зубів. До речі, цим і пояснюється той факт, що в період формування, росту організму діти та підлітки потребують додаткові кількості кальцію. Слід додати, що через насиченість кальцієм молочні продукти є незамінним при туберкульозі, переламах кісток, захворюваннях кровотворного апарату, рахіті.

Сирні продукти — це продукти, що виробляються сквашуванням молока або вершків чистими культурами молочнокислих бактерій. Деякі сирні продукти отримують в результаті тільки молочнокислого бродіння; при цьому утворюється досить щільний, однорідний згусток з вираженим кисломолочним смаком. Сирні продукти мають велике значення в харчуванні людини завдяки лікувальним і дієтичним властивостям, приємному смаку, легкій засвоюваності [42].

До складу сиру входить 14—17 % білків, до 18 % жиру, 2,4—2,8 % молочного цукру. Він багатий на кальцій, фосфор, залізо, магній — речовини, необхідні для росту і правильного розвитку молодого організму. Сир і вироби з нього дуже поживні, оскільки містять багато білків і жиру. Білки сиру частково пов'язані з солями фосфору і кальцію. Це сприяє кращому їх перетравлюванню в шлунку і кишечнику. Тому сир добре засвоюється організмом.

Вживання сиру і сирних виробів сприяє правильному обміну речовин в організмі, підтримці на певному рівні осмотичного тиску. Мінеральні речовини беруть участь в кісткоутворенні, харчуванні нирвової системи та сприяють підвищенню рівня гемоглобіну в крові. Сир містить різноманітні вітаміни груп А, В, С, D і багато інших. До складу сирів входять також різні тваринні жири, однак сучасні технології дозволяють замінювати тваринні жири рослинними [38; 42—44].

Це збільшує термін зберігання продукту, знижує вартість, покращує споживчі якості, рятує від шкідливого холестерину. Але продукт виходить лише за умови використання якісних фракцій, а не їх дешевих замінників.

Заміна тваринних жирів рослинними найчастіше зустрічається у молочній промисловості. Сучасні жирові системи, в яких заміна до 30–50 % молочного жиру на рослинні жири дозволяє виробити комбіновану олію, сметану, сир, морозиво, кефір, сирні вироби, які за смаковими якостями та консистенцією практично не відрізняються від традиційних продуктів зі 100 %-м молочним жиром.

Відповідно до концепції збалансованого харчування для нормальної життєдіяльності людини необхідно надходження до організму адекватної кількості енергії та основних харчових речовин, а також дотримання строго певних співвідношень між багатьма факторами харчування — білками, жирами, вуглеводами та іншими компонентами [9; 10; 38].

При проектуванні складу кисломолочних продуктів, що мають комплексний сировинний вид, слід врахувати, що застосування рослинної сировини, що має підвищену біологічну цінність, дозволяє отримувати композиції, які характеризуються поліпшеним вітамінним, мінеральним, вуглеводним і амінокислотним складом у порівнянні з окремо взятими компонентами, при цьому можливо більш тонке керування процесом формування продуктів.

При виробництві продуктів на молочній основі, які відповідають вимогам раціонального харчування, необхідним етапом є обгрунтування молочно-жирової основи та підбір інгредієнтів, які б сприяли корегуванню її складу, обгрунтування жирнокислотного складу обраних фізіологічних добавок [22; 43].

Робота присвячена розробці програмного забезпечення для створення та оптимізації рецептур сирних виробів з використанням купажів рослинних олій шляхом математичного моделювання різних складових та їх співвідношень.

**Аналіз розробки аналогічних комп'ютерних програм**. В основному запропоновані постановки моделей оптимізації рецептур зводяться до задач лінійного програмування, в яких в якості цільової функції виступають вимоги мінімальної вартості суміші, максимального виходу якогось одного компонента, необхідності утримання компонентів не менше певної величини, деякий адитивний критерій, який об'єднує кілька критеріїв з різними ваговими коефіцієнтами. У більшості публікацій дослідження завершується на етапі побудови математичної моделі з поясненням очікуваного результату їх застосування. Можна перерахувати поодинокі спроби реалізувати запропоновані постановки розробки рецептур харчових продуктів на практиці за допомогою комп'ютерних програм, хоча такий підхід є логічним і раціональним [8; 16; 25; 26; 34—37].

Дослідження з розробки комп'ютерних програм для розрахунку рецептур нових продуктів в основному ведуться в контексті реалізації математичних моделей, створених за результатами експериментів.

Існують різні програмні продукти для автоматизованого розрахунку рецептур. Однією з найбільш поширених програм для розрахунку рецептур є MS Excel. При використанні цього програмного продукту необхідні для обчислення дані, а також розрахункові формули заносяться у відповідні комірки електронної таблиці. Недоліком використання MS Excel є відсутність можливості автоматизованого введення вхідних даних і розрахункових залежностей [6; 9; 25].

Існуючі спеціалізовані пакети програм для проектування рецептур продуктів харчування діляться на два класи: програми в складі автоматизованих систем управління виробництвом і спеціалізовані програми, призначені для виконання разових розрахунків стосовно певних видів продовольчих продуктів. Для спеціалізованих пакетів програм, що працюють у складі математичного забезпечення автоматизованих систем управління виробництвом, характерна надмірно висока вартість, їх впровадження висуває підвищені вимоги до рівня комп'ютерної підготовки персоналу харчових підприємств. До недоліків спеціалізованих програм для проектування рецептур можна віднести обмеженість відомостей з альтернативних сировинних інгредієнтів, прив'язку до офісних програм загального призначення і конкретних видів продовольчих продуктів, а також недостатньо високий рівень захисту інтелектуальної власності. Загальним недоліком існуючих програмних продуктів, які застосовуються для проектування рецептур, є відсутність підсистеми (модуля) оптимізації рецептури за сукупністю критеріїв харчової, біологічної та енергетичної цінності [9; 16; 22; 25].

Спеціалізований програмний комплекс «Etalon» [37] призначений для проектування багатокомпонентних рецептур продуктів загального призначення, а також спеціалізованих продуктів, відповідних за складом фізіологічним потребам організму з урахуванням віку, патології, фізичних станів і навантажень, навколишнього середовища, призначених для дитячого, дієтичного, функціонального харчування, вагітних і жінок, що годують, спецконтингенту. Застосування програми в значній мірі дозволяє забезпечити впорядкова-ну роботу з даними і розробити продукт із заданими властивостями. Програмний комплекс складається з таких частин: 1) інформаційна база даних, в якій зберігається інформація про нутрієнтний склад харчової сировини і фізіологічні норми харчування різних соціальних груп населення; 2) спеціалізована база даних, розроблена для підвищення ефективності функціонування алгоритму моделювання рецептур харчових продуктів; 3) система управління інформаційною базою даних. Інформаційна база даних розроблена в середовищі Microsoft SQL Server 2000 на моделі «клієнт / сервер» і включає кілька взаємопов'язаних таблиць. Спеціалізована база даних побудована в програмному середовищі Microsoft Access 2002. Програма призначена для розрахунку й оптимізації рецептур м'ясних виробів. До недоліків цього програмного продукту слід віднести обов'язкову наявність на робочому комп'ютері Microsoft SQL Server і Microsoft Access.

Програма «Розробка рецептур композицій з рослинної сировини» [36] дозволяє відповідно до сучасних принципів створення здорових продуктів харчування розробити рецептури харчових концентратів підвищеної біологічної цінності на плодоовочевій основі. Завдання необхідних параметрів еталонного продукту дозволяє отримати рецептури зі збалансованим співвідношенням макроелементів і отримати максимально повне забезпечення добової потреби людини у вітамінах і мінеральних речовинах.

Generic 2.0 — програма Кубанського державного технологічного університету — призначена для автоматизованого проектування і розрахунку багатокомпонентних рецептур продуктів функціонального харчування. Цей програмний продукт призначений для розрахунку й оптимізації рецептур м'ясних, рослинних і молочних виробів [34].

Вищевказані програмні продукти не враховують специфіку розрахунку багатофазних рецептур кондитерських виробів. На сайті http://ttk.telenet.ru/index.htm представлений програмний комплекс «Система розрахунків для громадського харчування», що включає розробку промислових рецептур на кондитерські вироби. Недоліками комплексу є відсутність автономності програмного забезпечення і, як наслідок, недостатній захист права інтелектуальної власності користувача.

На сайті http://www.es-nsk.ru/programmi.html представлені розроблені компанією «Експерт Софт» комп'ютерні програми для техноло-

гів підприємств харчової промисловості та громадського харчування. Найбільший інтерес представляють програми: «Технолог-кулінар», «Технолог-кондитер», «Технолог-хлібопекар». Програма «Технолог-кулінар» розроблена для впровадження елементів системи якості і безпеки на підприємствах індустрії харчування. Функціональні можливості програми дозволяють повністю автоматизувати розробку технологічної документації на всіх основних етапах виробництва кулінарної продукції: при вхідному контролі якості сировини, при виробництві кулінарної продукції і при зберіганні та реалізації кулінарної продукції.

Основним недоліком перерахованих програмних комплексів є відсутність специфіки моделей тієї чи іншої якості для розрахунків рецептур. Програм розрахунків режимів виробництв сирних виробів у відкритому доступі дуже мало.

**Опис експериментальної частини досліджень жирнокислотного складу емульсії**. При виробництві продуктів на молочній основі, які відповідають вимогам раціонального харчування, необхідним етапом є обгрунтування молочно-жирової основи та підбір інгредієнтів, які б сприяли корегуванню її складу, обгрунтування жирнокислотного складу обраних фізіологічних добавок [23; 41; 42; 46; 47].

Відповідно вимогам раціонального харчування співвідношення між білком : жиром : вуглеводами повинно складати 1,0 : 1,2 : 4,6, а співвідношення НЖК : МНЖК : ПНЖК має певні особливості і повинно становити 0,3 : 0,6 : 0,1. Всі природні жири, в тому числі і жир молока, не задовольняють усім цим вимогам, тому одним із завдань розробки нових молочних продуктів є правильна оцінка (з точки зору збалансованості) жирнокислотного складу сировини з метою наступного його корегування і забезпечення оптимального жирнокислотного складу готового продукту. Для цього необхідно збільшити кількість рослинного жиру по відношенню до тваринного, щоб досягти необхідного співвідношення жирних кислот [22; 23].

У зв'язку з цим виникає необхідність вибору жирової добавки у вигляді рослинної олії, для чого було розглянуто жирнокислотний склад олій, які традиційно використовуються у молочній промисловості. Такими оліями є соняшникова, соєва та оливкова. Жирнокислотний склад перерахованих олій наведено у таблиці 1.

Для наближення складу основи для виробництва продуктів, що відповідають вимогам раціонального харчування, необхідно значно підвищити вміст ПНЖК і МНЖК. Кількість НЖК повинна залишатись майже такою ж. Як видно із даних, наведених в табл. 1, для корегування співвідношення між жирними кислотами доцільно використовувати оливкову олію, яка є основним постачальником МНЖК і соняшникову як джерело ПНЖК.

**Жирнокислотний склад рослинних олій**

| Показники | Соняшникова олія | Соєва олія | Оливкова олія |
|---|---|---|---|
| Сумарний вміст ліпідів, % | 99,9 | 99,9 | 99,8 |
| Тригліцериди | 99,2 | 99,2 | 99,0 |
| Фосфоліпіди | 0 | 0 | 0 |
| В-ситостерин | 0,57 | 0,30 | 0,30 |
| Холестерин | 0 | 0 | 0 |
| Жирні кислоти, % | 94,9 | 94,4 | 94,7 |
| **Насичені:** | 13,3 | 13,9 | 15,75 |
| Пальмітинова | 11,10 | 10,3 | 12,9 |
| Стеаринова | 2,20 | 3,5 | 2,5 |
| Арахідонова | 0 | 0 | 0,35 |
| **Мононенасичені:** | 24,0 | 19,8 | 66,9 |
| Олеїнова | 24,0 | 19,8 | 64,7 |
| Пальміолеїнова | – | – | 1,55 |
| Гадолеїнова | – | – | 0,50 |
| **Поліненасичені:** | 57,6 | 61,2 | 12,10 |
| Лінолева | 57,0 | 50,60 | 12,0 |
| Ліноленова | 0,6 | 10,30 | 0 |
| НЖК:МНЖК:ПНЖК | 1,0:1,8:4,3 | 1,0:1,4:4,4 | 1,3:5,5:1,0 |

Внесення у продукти рослинних олій дозволить збагатити їх не лише цінними МНЖК та ПНЖК, а й важливими вітамінами-антиоксидантами, зокрема жиророзчинним вітаміном Е та токоферолами. Ці біоантиоксиданти, які присутні в оліях, проявляють в організмі людини протиракову дію, стимулюють функцію серцевого м'яза, є стабілізаторами біологічних мембран.

З точки зору харчової і біологічної цінності, а також антиоксидантного статусу доцільним є використання суміші оливкової та соняшникової олій для нормалізації молочної суміші за масовою часткою жиру.

***Завдання дослідження полягає в тому, щоб підібрати таке співвідношення оливкової та соняшникової олій у суміші, щоб склад кислот НЖК :***

***МНЖК : ПНЖК якомога ближче підходив до співвідношення 0,3 : 0,6 : 0,1, визначеного теорією раціонального харчування.***

Для отримання математичної моделі кисломолочного продукту (жирнокислотного модуля молочно-жирової основи) на кафедрі ХХтаЕ Одеського національного технологічного університету, під керівництвом доцента Шарахматової Т. Є. були проведені відповідні експерименти, в яких вміст оливкової та соняшникової олій змінювали від 5 до 95 % (з інтервалом у 5 %) від загальної масової частки жиру у суміші, яка становить 1,6 %. Результати моделювання жирнокислотного модуля молочно-жирової основи наведено у табл. 2.

За результатами експериментів, наведених у таблиці 2, зроблено висновок про те, що при співвідношенні між соняшниковою та оливковою олією 0,4:0,6, досягається максимальне наближення у співвідношенні між НЖК:МНЖК:ПНЖК. Крім того, оливкова олія зовсім не має холестерину, що дуже суттєво для продуктів харчування, оскільки надлишок холестерину у продуктах харчування недопустимий. Отже часткова заміна молочного жиру сумішшю рослинних олій покращує збалансованість жирнокислотного складу молочно-жирової суміші.

Проте створення математичної моделі емульсії з соняшникової та оливкової олій, відповідного програмного забезпечення та проведення розрахунків з урахуванням процедур оптимізації функцій дозволяє покращити ці результати.

Для більш детальних розрахунків ефективності такої суміші необхідно побудувати таку математичну модель залежності співвідношення НЖК:МНЖК:ПНЖК від вмісту оливкової та соняшникової олій, яка б допомогла наблизити отримане співвідношення до 0,3 : 0,6 : 0,1.

**Опис експериментальної частини досліджень режимів гомогенізації емульсій різного хімічного складу.** Для промислового застосування емульсій необхідно піддавати їх подальшому ефективному диспергуванню за допомогою спеціального обладнання — гомогенізаторів клапанного типу, які широко застосовуються у молочній промисловості і дають змогу одержувати дрібнодисперсні стійкі жирові емульсії прямого типу. На сьогоднішній день в молочній промисловості гомогенізація є єдиним способом утворення стійкої емульсії, в тому числі і з рослинними оліями. Тому слід уточнити технологічні режими процесу гомогенізації емульсій визначеного хімічного складу [30; 38].

Таблиця 2

**Жирнокислотний склад емульсії з соняшниковою та оливковою оліями**

| Соняшникова олія | Оливкова олія | Вміст жирних кислот в 100 г олії | | | Сумарний вміст жирних кислот | Вміст жирних кислот в 100 г продукту | | |
|---|---|---|---|---|---|---|---|---|
| | | НЖК | МНЖК | ПНЖК | | НЖК | МНЖК | ПНЖК |
| 0,05 | 0,95 | 18,95195 | 68,66605 | 12,44605 | 100,0641 | 1,52272809 | 5,517095785 | 0,27600175 |
| 0,1 | 0,9 | 21,2729 | 66,6881 | 12,1151 | 100,0761 | 1,75589966 | 5,504543916 | 0,3189094 |
| 0,15 | 0,85 | 23,59385 | 64,71015 | 11,78415 | 100,0882 | 2,00216817 | 5,491287025 | 0,3646818 |
| 0,20 | 0,8 | 25,9148 | 62,7322 | 11,4532 | 100,1002 | 2,26266895 | 5,477263996 | 0,41310204 |
| 0,25 | 0,75 | 28,23575 | 60,75425 | 11,12225 | 100,1123 | 2,53867248 | 5,462406438 | 0,46475349 |
| 0,3 | 0,7 | 30,5567 | 58,7763 | 10,7913 | 100,1243 | 2,83160509 | 5,446637569 | 0,51988131 |
| 0,35 | 0,65 | 32,87765 | 56,79835 | 10,46035 | 100,1364 | 3,14307361 | 5,429870893 | 0,57884868 |
| 0,4 | 0,6 | 35,1986 | 54,8204 | 10,1294 | 100,1484 | 3,47489486 | 5,41208609 | 0,6420712 |
| 0,45 | 0,55 | 37,51955 | 52,84245 | 9,79845 | 100,1605 | 3,82913114 | 5,3929397 | 0,71002669 |
| 0,5 | 0,5 | 39,8405 | 50,8645 | 9,4675 | 100,1725 | 4,208133 | 5,3725376 | 0,78326731 |
| 0,55 | 0,45 | 42,16145 | 48,88655 | 9,13655 | 100,1846 | 4,61459194 | 5,350657524 | 0,86243456 |
| 0,6 | 0,4 | 44,4824 | 46,9086 | 8,8056 | 100,1966 | 5,05160353 | 5,327132734 | 0,94827814 |
| 0,65 | 0,35 | 46,80335 | 44,93065 | 8,47465 | 100,2087 | 5,52274725 | 5,301770575 | 1,04167979 |
| 0,7 | 0,3 | 49,1243 | 42,9527 | 8,1437 | 100,2207 | 6,03218439 | 5,274347041 | 1,14368363 |
| 0,75 | 0,25 | 51,44525 | 40,97475 | 7,81275 | 100,2328 | 6,58478129 | 5,244600173 | 1,25553542 |
| 0,8 | 0,2 | 53,7662 | 38,9968 | 7,4818 | 100,2448 | 7,18626534 | 5,212221658 | 1,37873364 |
| 0,85 | 0,15 | 56,08715 | 37,01885 | 7,15085 | 100,2569 | 7,84342421 | 5,176846109 | 1,51509704 |
| 0,9 | 0,1 | 58,4081 | 35,0409 | 6,8199 | 100,2689 | 8,56436311 | 5,138037215 | 1,66685502 |
| 0,95 | 0,05 | 60,72905 | 33,06295 | 6,48895 | 100,281 | 9,3588408 | 5,095269651 | 1,83677046 |

Для обгрунтування режимів механічного оброблення емульсій різного хімічного складу їх піддавали гомогенізації при тиску у межах від 7 до 15 МПа та температурі від 55 до 70 °C, що є загальноприйнятими режимами для молочної промисловості. Раціональні режими гомогенізації визначали за стійкістю емульсії (Y, %) та відстоєм жирової фази (v, %). При цьому стійкість емульсії повинна бути максимальною (100 %), відстій жирової фази повинний бути мінімальним.

Результати експериментальних досліджень наведено в табл. 3. Аналізуючи наведені дані, бачимо, що з підвищенням гомогенізації відстій жирової фази зменшується. Це пов'язано з тим, що при радіусі жирових кульок менше 0,5 мкм в гомогенізованій суміші при тиску 12—15 МПа таких кульок абсолютна більшість, електричні сили відштовхування перевищують ван-дер-вальсові сили притягування, такі кульки не утворюють скупчення. Саме тому при високому тиску гомогенізації спостерігається менший відстій жирової фази. Так при тиску 15 МПа відстій складає 0,6 % — 4,2 % від загальної кількості жиру, при 12 МПа — 0,9—5,2 %, а при 7 МПа — 3,5—9,2 %. Таким чином найкращий тиск, обчислений емпіричним шляхом, 12—15 МПа.

Таблиця 3

**Фізичні характеристики гомогенізованих емульсій**

| Тиск, МПа | Температура, °C | | | | | | | |
|---|---|---|---|---|---|---|---|---|
| | 55 | | 60 | | 65 | | 70 | |
| | Y, % | v, % | Y, % | v, % | Y, % | v, % | Y, % | v, % |
| Олія соняшникова | | | | | | | | |
| 7 | 98,5 | 8,3 | 98,8 | 7,6 | 99,1 | 4,9 | 99,3 | 3,4 |
| 10 | | 6,1 | | 5,2 | | 3,8 | | 2,0 |
| 12 | 100,0 | 4,4 | 100,0 | 3,1 | 100,0 | 1,5 | 100,0 | 0,9 |
| 15 | | 3,8 | | 2,6 | | 1,0 | | 0,6 |
| Олія оливкова | | | | | | | | |
| 7 | 98,6 | 8,6 | 98,7 | 7,8 | 99,2 | 4,9 | 99,6 | 3,6 |
| 10 | | 6,4 | | 5,5 | | 3,7 | | 2,4 |
| 12 | 100,0 | 4,8 | 100,0 | 3,4 | 100,0 | 1,6 | 100,0 | 0,9 |
| 15 | | 3,7 | | 2,8 | | 0,9 | | 0,8 |
| Купаж (олія соняшникова+олія оливкова) | | | | | | | | |
| 7 | 98,1 | 9,2 | 98,4 | 7,7 | 99,0 | 5,0 | 99,3 | 3,5 |
| 10 | 98,3 | 6,9 | | 5,3 | | 3,6 | | 2,1 |
| 12 | 98,6 | 5,2 | 100,0 | 3,2 | 100,0 | 1,5 | 100,0 | 0,9 |
| 15 | 98,8 | 4,2 | | 2,8 | | 1,0 | | 0,6 |

Емпіричним шляхом встановлюють гранично допустимі коефіцієнти відстою, при значеннях вище яких якість продукту недопустима.

Результати розрахунків коефіцієнта відстою відповідно до суміші з масовою часткою жиру 1,6 % наведені в табл. 4.

Таблиця 4

**Коефіцієнт відстою для суміші з масовою часткою жиру 1,6 %**

| Тиск гомогенізації, МПа | Коефіцієнт відстою |
|---|---|
| 7 | 0,224 |
| 10 | 0,194 |
| 12 | 0,176 |
| 15 | 0,170 |

Аналізуючи дані таблиці, помітно, що зі збільшенням кількості рослинного жиру в суміші коефіцієнт відстою збільшується. Найбільш оптимальний тиск гомогенізації суміші 12 МПа та 15 МПа. При такому тиску коефіцієнт відстою жирової фази найменший і становить при 12 МПа — 0,176, при 15 МПа — 0,170 відповідно.

Отже, за результатами експерименту, при гомогенізації соняшникової та оливкової олій окремо достатнім режимом гомогенізації є температура 55 °C при тиску 15 МПа. При гомогенізації купажу (соняшникова олія + оливкова олія) рекомендованим режимом гомогенізації є температура 60 °C при тиску 12 МПа. При цьому стійкість емульсії максимальна і становить 100 % і відстій жирової фази становить 2,8 %, що є достатнім для виробництва продукції з кисломолочного сиру. Подальше підвищення тиску та температури недоцільне, тому що збільшуються енерговитрати на виробництво, за рахунок чого значно зростає собівартість продукції.

Порівнюючи отримані дані по зміні стійкості емульсії, відстою жирової фази, можна зробити висновок, що експериментальні дані для тиску 12 МПа та 15 МПа майже не мають суттєвих відмінностей. Стійкість емульсії складає 100 %, відстій жирової фази — 2,8–3,2 %. Тому оптимальним тиском за результатами експерименту при виробництві кисломолочного сиру з використанням купажів рослинних олій є 12 ± 0,5 МПа.

***Завданням дослідження є побудова математичної моделі та побудова комп'ютерної програми для розрахунку режимів гомогенізації різного хімічного складу емульсій при виробництві кисломолочного сиру з вико-***

***ристанням купажів рослинних олій (соняшникова олія + оливкова олія), тобто знаходження оптимальних значень тиску та температури.***

**Методика складання математичних моделей.** Для розробки нової рецептури необхідно застосувати сучасний математичний апарат, побудувати математичну модель процесу, оптимізувати її та отримати найкращі параметри. Математична модель має складатися за допомогою методів регресійно-кореляційного аналізу на основі натурних експериментів [22; 33; 40; 44; 46].

Сьогодні одним з важливих чинників підвищення ефективності та якості досліджень є застосування ймовірнісно-статистичних методів і комп'ютерів для створення математичних моделей різних процесів і об'єктів. Їх застосування дозволяє від простих розрахунків і оцінок перейти до нової стадії роботи — детального математичного моделювання і дослідження складних реальних процесів і об'єктів. Статистичні методи планування й обробки експерименту з використанням комп'ютера дозволяють значно інтенсифікувати працю дослідника, скоротити терміни і витрати на реальний фізичний експеримент, підвищити достовірність і якість висновків за результатами експериментів. Метою застосування таких методів є отримання математичної моделі досліджуваного процесу, яка має описувати його досить повно. Після отримання такої моделі з'являється можливість замінити подальше експериментальне дослідження реального процесу аналізом його математичної моделі, що, природно, різко знижує витрати часу і матеріальних вкладень. Це дозволяє визначити оптимальні режими та інші характеристики технологічного процесу, конструктивні параметри машини або апарату [15; 22; 23].

У загальному випадку модель є тією або іншою формою відображення реальної дійсності. Надалі під моделлю будемо розуміти таку формалізовану систему, яка, відображаючи і відтворюючи об'єкт дослідження, здатна заміщати його в розрахунках.

Критерії відповідності моделі об'єкта можуть бути різними. Найчастіше таким критерієм є ступінь відхилення показника якості процесу (вихідні характеристики, витрата енергії робочого агента, витрати та ін.), виміряного безпосередньо на об'єкті, та отриманого шляхом розрахунків знайденої математичної моделі z.

Іноді при складанні математичної моделі досліджуваного процесу деякі параметри її можуть бути відомі заздалегідь, однак в загальному випадку об'єкт дослідження уявляється у вигляді «чорного ящика» (кібернетичний термін), внутрішня структура якого невідома; на вхід його діють вхідні впливу $x_i$, $i = 1,2,..., k$ (так звані «фактори», k — кількість факторів), а вихідні впливу y («відгуки») можуть вимірюватися реєструючими приладами (рис.1). Якщо об'єкт має кілька відгуків, вони можуть розглядатися незалежно один від одного, або входити в розрахункову формулу з ваговими коефіцієнтами, тому в подальшому будемо розглядати системи з одним виходом.

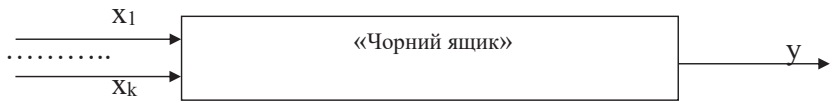

Рис. 1. Представлення досліджуваного об'єкта у вигляді «чорного ящика»

Для знаходження математичної моделі зазвичай проводять експериментальні дослідження, які встановлюють зв'язок між вхідними факторами, що впливають на перебіг процесу, і вихідними параметрами процесу, які характеризують його властивості. Перші з них є незалежними і можуть приймати довільні значення, другі — залежними [12; 22].

Експеримент проводять пасивними і активними методамі. Під пасивними експериментами розуміють отримання будь-яких даних без планування умов проведення дослідів при випадковій зміні вхідних факторів. Пасивні методи, хоча і прості в застосуванні, але не досить точні, а також відрізняються складними методами обробки отриманих даних. Активний, заздалегідь спланований експеримент, ставлять в тому випадку, якщо досліджуваний об'єкт допускає можливість зміни вхідних факторів в необхідних межах. Методика зміни вхідних факторів у цьому випадку має передбачати мінімізацію числа експериментів, застосування простих і найменш трудомістких методів обробки отриманих результатів. Принципами створення ефективних планів проведення дослідів займається спеціальний розділ математики — «планування експерименту» [12].

Під математичною моделлю досліджуваного об'єкта на рис.1 будемо розуміти рівняння, що пов'язує відгук і фактори

$$y = f(x_1, x_2,..., x_k). \tag{1}$$

Потрібно виразити аналітично (тобто у вигляді формули) залежність між значеннями x і y, в результаті чого замість функції $y = f(x_1, x_2,..., x_k)$ повинна вийти інша, апроксимуюча (тобто приблизно описуюча) її функція $z = \varphi(x_1, x_2,..., x_k)$. Залежно від вимог, що пред'являються до апроксимуючої функції, розрізняють два типи апроксимації [6; 12].

Апроксимація першого типу. Допускається, що результати експерименту, зафіксовані в таблиці, є наближеними, мають деякі погрішності. Це часто відбувається в реальних експериментах в силу недосконалості застосовуваних при вимірах приладів. Завдання полягає в тому, щоб вибрати таку апроксимуючу функцію з деякого класу функцій, яка б з мінімальними відхиленнями відповідала результатам експериментів.

Апроксимація другого типу. Передбачається, що результати експерименту, зафіксовані в таблиці, є точними. Потрібно із заданого класу вибрати таку функцію $\varphi(x_1, x_2,..., x_k)$, щоб для вузлів сітки $x_1$, $x_2,..., x_n$ виконувалася строга рівність $\varphi(x_1, x_2,..., x_k) = y(x_1, x_2,..., x_k)$. Такий вид апроксимації називають інтерполяцією. Для вирішення такого завдання часто використовують інтерполяційний многочлен Лагранжа, поліноми Ньютона. Однак точність такого наближення гарантована лише в невеликому інтервалі, для іншого проміжку потрібно заново обчислювати коефіцієнти інтерполяційної формули.

Надалі будемо вирішувати завдання апроксимації першого типу як найбільш поширеної в інженерних задачах; іноді таке завдання називають також завданням знаходження емпіричної залежності.

Для складання математичної моделі за результатами експериментальних даних необхідно вирішення таких завдань: вибір виду аналітичної формули; визначення її найкращих параметрів; доказ адекватності отриманої моделі досліджуваного об'єкта.

Загальної теорії вибору виду емпіричної залежності не існує, функцію вибирає дослідник виходячи з прогностичних здібностей, специфіки завдання і характеру розташування на координатній площині точок, відповідних експериментальним даним. У деяких випадках вибір виду емпіричної формули може бути проведений на основі теоретичних уявлень про характер досліджуваної залежності або про зміну вимірюваних величин. В інших випадках доводиться підбирати формулу, порівнюючи графік, побудований за даними спостережень, з типовими графіками залежностей, наведеними в довідковій літературі. Деякі рекомендації щодо вибору виду емпіричної формули наведені в [12].

На практиці для знаходження параметрів емпіричної залежності (після висунення гіпотези про її вигляд) використовують метод вибраних точок, метод середніх, метод найменших квадратів. Для чи-сельних розрахунків коефіцієнтів регресійної формули нами буде використовуватися метод випадкового пошуку.

Для знаходження кривої, яка приблизно відповідає вихідної інформації, необхідно виробити критерій. Назвемо відхиленням експериментальної точки різницю між експериментальною ординатою $y_i$ і тією, яка обчислена з теоретично знайденою функціональною залежністю $z_i = \varphi(x_i)$ (на рис. 2 розрахунковий графік залежності $z_i = \varphi(x_i)$ і відхилення показані суцільними лініями).

В якості сумарного критерію, що визначає отримане загальне відхилення, можна прийняти формулу квадратів відхилень:

$$\sum_{i=1}^{n} (y_i - z_i)^2 \ .$$

Метод апроксимації, в якому критерієм якості обрана показана формула, називають методом найменших квадратів [5; 12; 16; 31]. Основна мета регресійного аналізу полягає у визначенні аналітичної форми зв'язку, в якій зміна результативної ознаки зумовлена впливом однієї або декількох факторних ознак.

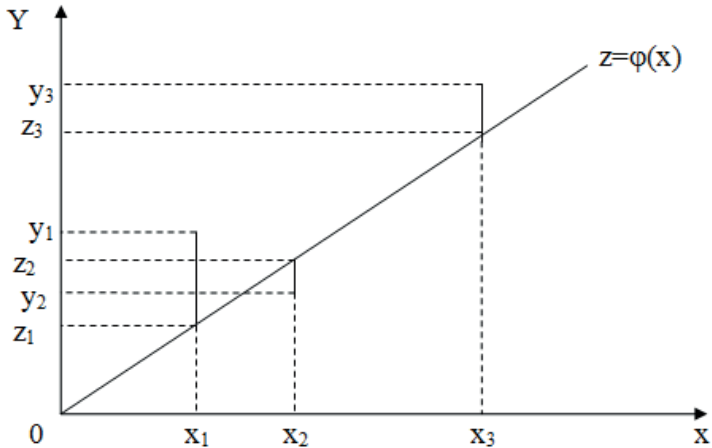

Рис. 2. Ілюстрація відхилень між $y_i$ і $z_i = \varphi\,(x_i)$

Завдання регресійного аналізу [12; 44]:

а) встановлення форми залежності. Щодо характеру і форми залежності між явищами розрізняють позитивну лінійну і нелінійну, а також негативну лінійну і нелінійну регресію;

б) визначення функції регресії у вигляді математичного рівняння того або іншого типу і встановлення впливу вхідних змінних на залежну змінну;

в) оцінка невідомих значень залежної змінної. За допомогою функції регресії можна відтворити значення залежної змінної всередині інтервалу заданих значень (тобто вирішити задачу інтерполяції) або оцінити перебіг процесу поза заданим інтервалом (тобто вирішити задачу екстраполяції). Результат являє собою оцінку значення залежної змінної.

Нехай функціональна залежність між x і y має вигляд:

$$z = \varphi\,(x,\ a,\ b,\ c,...),$$

де a, b, c, ... — невідомі параметри у формулі емпіричної залежності, які необхідно підібрати (якщо функція $z_i = \varphi\,(x_i)$ має вигляд полінома, то шукану математичну модель об'єкта іноді називають рівнянням регресії). Вираз

$$S\,(a,\ b,\ c,...) = \sum_{i=1}^{n}\ [y_i - \varphi(x_i,\ a,\ b,\ c,...)]^2 \qquad (2)$$

називають квадратичним відхиленням емпіричної формули від експериментальних даних.

У методі найменших квадратів параметри a, b, c,... підбирають таким чином, щоб мінімізувати функцію (2), тобто знаходячи найменше сумарне відхилення розрахункових даних від експериментальних. Для цього відповідно до правил класичної математики можна прирівняти нулю окремі похідні від функції S (2) в невідомих параметрах a, b, c,**.**.:

$$\frac{\partial S}{\partial a} = 0;\ \ \frac{\partial S}{\partial b} = 0;\ \ \frac{\partial S}{\partial c} = 0;\ ........... \qquad (3)$$

У розгорнутому вигляді систему (3) записують так:

$$........ \sum_{i=1}^{n}\ 2\,[y_i\text{-}\varphi(x_i,a,b,c,...)] \cdot (-\,(\partial\varphi(x_i,a,b,c,...))\,/\partial a) = 0,$$
$$\sum_{i=1}^{n}\ 2\,[y_i\text{-}\varphi(x_i,a,b,c,...)] \cdot (-\,(\partial\varphi(x_i,a,b,c,...))/\partial b) = 0, \qquad (4)$$
$$...................$$

В отриманій системі число рівнянь дорівнює числу невідомих параметрів. Вирішивши систему будь-яким відомим методом, знайдемо значення параметрів a, b, c,.... Однак не можна забувати про труднощі при реалізації методу найменших квадратів, які можуть бути викликані такими причинами [2; 19; 31]:

1. Система (4) може бути несумісна.

2. Система (4) може вийти надзвичайно важкою для вирішення.

3. Система (4) може мати безліч рішень, в цьому випадку необхідно додатково з'ясовувати, які з них відповідають мінімуму функції (2).

У загальному випадку методом найменших квадратів можна знайти формули для обчислення параметрів будь-якої емпіричної формули, проте рішення системи (4) і пов'язані з її отриманням розрахунки можуть виявитися досить складними.

Для оцінки точності отриманої апроксимації можна застосовувати середньоквадратичну помилку (чим менше ця величина, тим точніше проведена апроксимація)

$$\delta^2 = \frac{\sum\limits_{i=1}^{n}(y_i - z_i)^2}{n-k-1}. \tag{5}$$

Це значення характеризує міру розсіювання фактичних значень щодо розрахункових, отриманих за емпіричною формулою. Чим менше помилка $\Delta$, тим краще отримана математична модель описує існуючий зв'язок між відгуком і факторами. Варіюючи види функції (1) і оцінюючи результати за допомогою середньоквадратичної помилки (5), можна серед розглянутих емпіричних формул вибрати найкращу. Приклади використання декількох емпіричних формул і розрахунків (5) наведені в [5; 12; 31]. Однак слід ще перевірити, чи значимо отримане рівняння регресії; при відсутності паралельних дослідів це проводиться за розрахунковим критерієм Фішера

$$F_r = \frac{\delta_y^2}{\delta^2},$$

де $\delta_y^2 = \dfrac{\sum\limits_{i=1}^{n}(y_i - y_s)^2}{n-1}$ — дисперсія у щодо середнього $y_s$;

$y_s$ — середнє значення у за експериментальними даними.

Розраховане рівняння регресії вважають надійним і адекватним експериментальним даним, якщо виконується умова

$$F_r < F_t(\alpha, \nu_1, \nu_2),$$

де $\nu_1 = n - 1$, $\nu_2 = n - k - 1$ — ступені свободи;

α — рівень значущості (у технічних розрахунках найчастіше використовують рівень значущості α = 0,05);

$F_t(\alpha, \nu_1, \nu_2)$ — табличний критерій Фішера.

Якщо виявиться, що умова $F_r < F_t$ не виконується, то це означає, що вид емпіричної залежності (1) обраний неправильно і необхідно повторювати всі розрахунки спочатку.

**Використання методів оптимізації**. У розроблюваному додатку для застосування методології найменших квадратів та оптимізації отриманих математичних моделей реалізовано метод випадкового пошуку.

Метод випадкового пошуку заснований на застосуванні послідовностей випадкових чисел, за допомогою яких у сфері зміни незалежних змінних проводиться вибірка випадкових точок або визначення випадкових напрямків. Цей метод є прямим розвитком відомого методу спроб і помилок, коли рішення шукається випадково і при удачі приймається, а при невдачі відкидається, щоб негайно знову звернутися до випадковості як до джерела можливостей. Така випадкова поведінка розумно спирається на впевненість, що випадковість містить у собі всі можливості, у тому числі й шукане рішення у всіх його випадках [4; 20; 31].

Метод випадкового пошуку при оптимальному проектуванні дозволяє порівняно невеликими витратами машинного часу визначити екстремум функції великої кількості змінних. Перевагою цього методу є те, що, крім необхідності існування в області єдиного локального екстремуму, він не пред'являє істотних вимог ні до виду безлічі параметрів, за якими відшукується оптимальне значення, ні до виду залежностей, що пов'язують параметри, які вибираються з оптимізуючим критерієм і обмеженнями. Він дозволяє знайти всі локальні мінімуми функції від 10—20 змінних зі складним рельєфом. Він корисний і для дослідження функції з єдиним мінімумом.

Цей метод має дві переваги. По-перше, він придатний для будь-якої цільової функції незалежно від того, є вона унімодальною чи ні. По-друге, ймовірність успіху при спробах залежить від розмірності аналізованого простору. Хоча цей метод не дозволяє безпосередньо знайти оптимальне рішення, він створює відповідні передумови для подальшого застосування інших методів пошуку. Тому його часто застосовують у поєднанні з одним чи декількома методами інших типів.

Недолік методу полягає в тому, що треба заздалегідь задати область, де вибираються випадкові точки. Якщо ми поставимо занадто широку область, то її важче детально досліджувати, а якщо виберемо занадто вузьку область, то багато локальних мінімумів можуть опинитися поза нею.

Існують кілька методів випадкового пошуку, які схожі один на одного і відрізняються лише кількома кроками чи умовами.

Найзагальніша ітераційна формула методів випадкового пошуку має вигляд

$$X_{k+1} = X_k + \xi_k \, ,$$

де $\varepsilon\_(k)$ — n-вимірна випадкова величина. Імовірнісні розподіли цієї випадкової величини, їх зміни у різних кроках методу визначають метод пошуку. Звичайно для випадкової величини $\varepsilon\_(k)$ ставляться специфічні вимоги.

Норма цієї векторної величини має бути обмеженою, щоб точка $X\_(k+1)$ залишалася поблизу точки $X\_k$.

Закони розподілу залежать від результатів подальших випробувань (адаптація випадкового пошуку).

До найпростіших алгоритмів належать такі:

*Алгоритм із парною пробою*. У випадковому напрямку по обидва боки вихідного стану $X\_k$ роблять пробні кроки $X\_(k)\pm L_k$, де L — довжина пробного кроку. Обчислюють значення цільової функції у цих точках. Робочий крок роблять у напрямі меншого значення цільової функції. Характерною особливістю алгоритму є висока тенденція до «блукання».

*Алгоритм із поверненням при невдалому кроці.* Роблять крок у випадковому напрямку. Якщо значення цільової функції у новій точці більше, ніж у вихідній, тобто крок виявився невдалим, то повертаються у вихідну точку. Після цього випадкові кроки повторюються.

*Алгоритм найкращої проби.* З вихідної точки роблять m випадкових кроків і запам'ятовують той крок, який призвів до найменшого значення цільової функції. Робочий крок роблять саме у цьому напрямку.

Оскільки у математичних моделях кількість змінних невелика, в програмному забезпеченні для знаходження коефіцієнтів рівнянь регресії та мінімізації отриманих функцій реалізований метод з парною пробою як найбільш універсальний і простий метод оптимізації.

**Обґрунтування вибору мови програмування.** Для реалізації обраних алгоритмів практично потрібно побудувати програму на комп'ютері, і тут центральне питання — вибір необхідної мови програмування. Мова програмування — формальна знакова система, призначена для запису комп'ютерних програм. Мова програмування визначає набір лексичних, синтаксичних та семантичних правил, що задають зовнішній вигляд програми та дії, які виконає комп'ютер під її керуванням [14; 32].

Високорівнева мова програмування — мова програмування, розроблена для швидкості та зручності використання програмістом. Основна риса високорівневих мов — це абстракція, тобто введення смислових конструкцій, що коротко описують такі структури даних та операції над ними, опис яких на машинному коді (або іншій низькорівневій мові програмування) дуже довгий і складний для розуміння.

Так, високорівневі мови прагнуть не лише полегшити вирішення складних програмних завдань, а й спростити портування програмного забезпечення. Використання різноманітних трансляторів та інтерпретаторів забезпечує зв'язок програм, написаних за допомогою мов високого рівня, з різними операційними системами та обладнанням, у той час як їхній вихідний код залишається, в ідеалі, незмінним.

Загалом мова називається безпечною, якщо програми на ній, які можуть бути прийняті компілятором як правильно побудовані, в динаміці ніколи не вийдуть за межі допустимої поведінки. Це не означає, що такі програми не містять помилок взагалі. Термін «хороша поведінка програми» (англ. well behavior) означає, що навіть якщо програма містить якийсь баг (зокрема логічну помилку), вона, проте, не здатна порушити цілісність даних і обрушитися. Хоча терміни неформальні, безпека деяких мов (наприклад, Standard ML) математично доведена. Безпека інших (наприклад, Ada) була забезпечена ad hoc-чином, без забезпечення концептуальної цілісності, що може призвести до катастроф, якщо покластися на них у відповідальних завданнях.

Мова C та її нащадок C++ є небезпечними. У програмах з ними широко зустрічаються ситуації ослаблення типізації (приведення типів) та її порушення (каламбур типізації), отже помилки доступу до пам'яті є у них статистичною нормою (але крах програми настає далеко не відразу, що утруднює пошук місця помилки у коді). Найпотужніші системи статичного аналізу здатні виявляти трохи більше 70—80 % помилок, та їх використання дуже дороге фінансове. Достовірно ж гарантувати безвідмовність програм цими мовами неможливо, не вдаючись до формальної верифікації, що ще дорожче та вимагає спеціальних знань [14].

Сі має і безпечні нащадки, такі як Cyclone, C# або Rust. Мова Forth не претендує на звання «безпечної», але, проте, на практиці існування програм, здатних пошкодити дані, майже виключено, оскільки програма, що містить потенційно небезпечну помилку, аварійно завершується на першому ж тестовому запуску, примушуючи до корекції вихідного коду. У співтоваристві Erlang прийнято підхід «let it crash», також націлений на раннє виявлення помилок.

Програма компільованою мовою за допомогою спеціальної програми компілятора перетворюється (компілюється) на набір інструкцій для даного типу процесора (машинний код) і далі записується в модуль, який може бути запущений на виконання окремих програм. Інакше кажучи, компілятор переводить вихідний текст програми з мови програмування високого рівня у двійкові коди інструкцій процесора.

Якщо програма написана мовою, що інтерпретується, то інтерпретатор безпосередньо виконує (інтерпретує) вихідний текст без попереднього перекладу. При цьому програма залишається вихідною мовою і може бути запущена без інтерпретатора. Можна сказати, що процесор комп'ютера це інтерпретатор машинного коду.

Коротко кажучи, компілятор перекладає вихідний текст програми на машинну мову відразу і повністю, створюючи при цьому окрему програму, що виконується, а інтерпретатор виконує вихідний текст прямо під час виконання програми [14; 28; 45].

Поділ на компільовані та інтерпретовані мови є дещо умовним. Так, для будь-якої традиційно компільованої мови, наприклад, Паскаль, можна написати інтерпретатор. Крім того, більшість сучасних «чистих» інтерпретаторів не виконують конструкції мови безпосередньо, а компілюють їх у деяке високорівневе проміжне уявлення (наприклад, з розіменуванням змінних та розкриттям макросів).

Для будь-якої інтерпретованої мови можна створити компілятор — наприклад, мова Лісп спочатку інтерпретується, може компілюватися без будь-яких обмежень. Створюваний під час виконання програми код може динамічно компілюватися.

Як правило, скомпільовані програми виконуються швидше і не вимагають виконання додаткових програм, оскільки вже перекладені на машинну мову. Про те при кожній зміні тексту програми потрібна її перекомпіляція, що створює труднощі розробки. Крім того, скомпільована програма може виконуватися тільки на тому ж типі комп'ютерів і, як правило, під тією самою операційною системою, на яку розрахували компілятор. Щоб створити файл для машини іншого типу, потрібна нова компіляція.

Інтерпретовані мови мають деякі специфічні додаткові можливості, крім того, програми на них можна запускати відразу ж після зміни, що полегшує розробку. Програма, що інтерпретується, може бути часто запущена на різних типах машин і операційних систем без додаткових зусиль.

Однак програми, що інтерпретуються, виконуються помітно повільніше, ніж компільовані, крім того, вони не можуть виконуватися без додаткової програми-інтерпретатора. Приклади компільованих мов: assembler, C++, Pascal. Приклади мов, що інтерпретуються: PHP, JavaScript, Python. Деякі мови, наприклад, Java і C#, знаходяться між компільованими та інтерпретованими [28; 32].

Після аналізу застосовуваних алгоритмічних мов та особливостей їх використання було прийнято рішення для програмування додатку використовувати мову C# в силу її поширеності, універсальності та відносної простоти.

Мова C#, розроблена компанією Microsoft (вона з'явилася в 2000 році), одна з найпопулярніших сучасних мов програмування. Вона затребувана на ринку розробки в різних країнах, C# застосовують під час роботи з програмами для ПК, створення складних веб-сервісів або мобільних додатків. Мова перетерпіла велику кількість оновлень та нововведень.

C# — вкрай гнучка, потужна та універсальна алгоритмічна мова. У сучасному вигляді C# здатна на дуже багато речей. Сьогодні вона не дарма займає лідируючі позиції в списках популярних мов, тому що на її основі можливо будувати практично будь-які проекти.

Крім того, після появи ігрового двигуна Unity мова набула додаткової сили на ринку. Тепер на її основі у зв'язці з мегапопулярним двигуном Unity можна легко та швидко створювати ігри будь-якого жанру та будь-якої складності.

Мова C# є об'єктно орієнтованою мовою програмування. Це означає, що кожен файл являє собою певний клас.

Мова C# практично універсальна. Можна використовувати її для створення будь-якого програмного забезпечення: просунутих бізнес-додатків, відеоігор, функціональних веб-додатків, програм для Windows, macOS, мобільних програм для iOS та Android [18; 27; 28].

Інструментарій C# дозволяє вирішувати широке коло завдань, мова справді дуже потужна й універсальна. На цій мові розробляють:

- Програми для WEB.
- Різні ігрові програми.
- Програми платформ Андроїд або iOS.
- Програми для Windows.

На ній пишуть практично все, від невеликих веб-додатків до потужних програмних систем, що поєднують у собі веб-структури, додатки для десктопів та мобільних пристроїв. Все це стало можливим завдяки зручному Сі-подібному синтаксису, строгому структуруванню, величезній кількості фреймворків та бібліотек (до кількох сотень).

Список можливостей розробки практично не має обмежень завдяки найширшому набору інструментів та засобів. Звичайно, все це можна реалізувати за допомогою інших мов, але деякі з них вузькоспеціалізовані, в інших доведеться використовувати додаткові інструменти сторонніх розробників. У C# вирішення широкого кола завдань можливо швидше, простіше і з меншими витратами часу та ресурсів.

Розвиток об'єктноорієнтованого програмування та мережі Internet сприяв появі нової технології програмування —.NET технології, що дозволяє на єдиній платформі розробляти компоненти програм різними мовами програмування та забезпечити їхнє спільне виконання. В рамках .NET технології запропоновано нову мову програмування C#, засновану на мові C++, що перейняла з мови Java риси, які забезпечують створення безпечних програм. З урахуванням .NET технології мова C++ розширена новими можливостями і отримала назву C++/CLI, з'явилася також мова J# — мова Java стосовно .NET технології [45].

Мова C# розроблена після мови Java. Вона не тільки успадкувала найкраще з мови Java, але модифікувала її, надавши стрункість та зручність використання, наприклад, таких конструкцій, як делегати та події. Але, будучи відкритою і легкодоступною із сайту фірми Sun Microsystems в Інтернеті, мова Java, мабуть, стала найпопулярнішою мовою програмування у світі. Сайт фірми Sun Microsystems доступний програмістам усього світу. Доступність сайту об'єднала професійних програмістів, небайдужих до долі мови Java, сприяючи просуванню компонентноорієнтованого програмування цією мовою. Запропонована фірмою Microsoft мова J#, що є варіантом мови Java для .NET платформи, може використовувати (імпортувати — import) як бібліотеку .NET Framework, і бібліотеку Java.

Коли говорять C#, часто мають на увазі технології платформи .NET (Windows Forms, WPF, ASP.NET, Xamarin). І навпаки, коли говорять .NET, нерідко мають на увазі C#. Проте, хоча ці поняття пов'язані, ототожнювати їх не слід. Мова C# була створена спеціально для роботи з фреймворком .NET, проте саме поняття .NET дещо ширше.

Фреймворк .NET є потужною платформою для створення додатків. Можна виділити такі основні риси [45]:

• Підтримка кількох мов. Основою платформи є загальномовне середовище виконання Common Language Runtime (CLR), завдяки чому .NET підтримує кілька мов: поряд з C# це також VB.NET, C++, F#, а також різні діалекти інших мов, прив'язані до .NET, наприклад, Delphi.NET. При компіляції код будь-якої з цих мов компілюється у складання загальною мовою CIL (Common Intermediate Language) — свого роду асемблер платформи .NET. Тому за певних умов ми можемо зробити окремі модулі однієї програми окремими мовами.

• Кросплатформність. .NET є платформою, що переноситься (з деякими обмеженнями). Наприклад, остання версія платформи на даний момент — .NET 6 підтримується на більшості сучасних ОС Windows, MacOS, Linux. Використовуючи різні технології на платформі .NET, можна розробляти програми мовою C# для різних платформ — Windows, MacOS, Linux, Android, iOS, Tizen.

• Потужна бібліотека класів. .NET представляє єдину бібліотеку класів, що підтримує всім мов. І яку б програму ми не збиралися писати на C# — текстовий редактор, чат або складний веб-сайт, — так чи інакше ми використовуємо бібліотеку класів .NET.

• Різноманітність технологій. Загальномовне середовище виконання CLR та базова бібліотека класів є основою цілого стеку технологій, які розробники можуть задіяти при побудові тих чи інших додатків. Наприклад, для роботи з базами даних у цьому стеку технологій призначено технології ADO.NET та Entity Framework Core. Для побудови графічних програм з багатим насиченим інтерфейсом — технологія WPF і WinUI, для створення більш простих графічних програм — Windows Forms. Для розробки кросплатформових мобільних та десктопних програм — Xamarin/MAUI. Для створення веб-сайтів та веб-додатків — ASP.NET і т. д. До цього варто додати активний Blazor — фреймворк, що розвивається і набирає популяність, який працює поверх .NET і який дозволяє створювати веб-додатки як на стороні сервера, так і на стороні клієнта. А в майбутньому підтримуватиме створення мобільних додатків і, можливо, десктоп-додатків.

• Продуктивність. Відповідно до ряду тестів веб-програми на .NET 6 у ряді категорій сильно випереджають веб-програми, побудовані за допомогою інших технологій. Програми на .NET 6 у принципі відрізняються високою продуктивністю.

Також слід відзначити таку особливість мови C# і фреймворку .NET, як автоматичне складання сміття. А це означає, що нам здебільшого не доведеться, на відміну від C++, дбати про звільнення пам'яті. Вищезазначене загальномовне середовище CLR само викличе збирач сміття та очистить пам'ять.

Варто відзначити, що .NET тривалий час розвивався переважно як платформа для Windows під назвою .NET Framework. У 2019 році вийшла остання версія цієї платформи — .NET Framework 4.8. Вона більше не розвивається.

З 2014 року Microsoft став розвивати альтернативну платформу — .NET Core, яка вже призначалася для різних платформ і повинна була увібрати в себе всі можливості застарілого .NET Framework і додати нову функціональність. Потім Microsoft послідовно випустив ряд версій цієї платформи: .NET Core 1, .NET Core 2, .NET Core 3, .NET 5. І поточною версією є платформа .NET 6. Тому слід розрізняти .NET Framework, який призначений переважно для Windows, і кросс­платформений .NET 6 (його будемо розглядати у зв'язці з C#).

Нерідко програму, створену на C#, називають керованим кодом (managed code). Це означає, що ця програма створена на основі платформи .NET і тому управляється загальномовним середовищем CLR, яке завантажує програму і при необхідності очищає пам'ять. Але є також програми, наприклад, створені мовою C++, які компілюються не у спільну мову CIL, як C#, VB.NET чи F#, а у звичайний машинний код. У цьому випадку .NET не керує програмою. У той же час платформа .NET надає можливості для взаємодії з некерованим кодом [27; 45].

Вихідний код C# компілюється у програми або окремі збірки на CIL з розширеннями dll, exe. У процесі запуску готової програми виконується JIT-компіляція — це скорочення від Just-In-Time (просто зараз). На виході буде машинний код, який передається на виконання.

Для роботи програм на C# необхідно встановити та налаштувати платформу .NET Framework. Вона поставляється повністю безкош-товно, застосовується дуже широко, а тому проблем із пристроями зазвичай не виникає. Платформа вбудована в інсталяційний пакет Windows, при необхідності її також можна завантажити та поставити окремо. Існують версії для Linux та MAC.

В рамках платформи до обробки коду, що виконується, підключається середовище CLR — єдиний об'єднаний набір бібліотек та класів, який був розроблений Microsoft і є реалізацією світового стандарту Common Language Infrastructure (CLI).

Після роботи компілятора текст програми перекладається проміжною мовою IL, яка «розуміє» CLI. IL і всі необхідні ресурси, включаючи рядки та малюнки формату BMP, зберігаються на жорсткий диск у вигляді файлу dll або exe. З таких файлів з проміжним кодом формується збірка програми, яка включає опис з повною інформацією про всі важливі параметри роботи [14; 18].

Безпосередньо при виконанні програми CLR звертається до складання та виконує дії залежно від отриманих відомостей. Якщо код написаний правильно і проходить перевірку безпеки системи, проводиться компіляція з IL в інструкції до машинних команд. Середовище CLR одночасно виконує ще багато побічних функцій:

• видалення «програмного» сміття;
• робота з винятками;
• розподіл ресурсів;
• контроль типізації;
• керування версіями;
• типізація.

В результаті код C# вважається керованим, тобто він компілюється у двійковий вигляд на власному пристрої з урахуванням особливостей встановленої системи.

C# популярний за рахунок своєї «простоти». Простоти для сучасних програмістів і великих команд розробників, щоб ті могли в стислий термін створювати функціональні та продуктивні програми. Цьому сприяють нетипові конструкції мови та специфічний синтаксис, що допомагає максимально органічно реалізувати намічені функції.

На додаток до всіх переваг мова C# полегшує розробку програмних компонентів за допомогою кількох інноваційних конструкцій мови:

• Інкапсульовані сигнатури методів, названі делегатами, включають оповіщення безпеки типів.

• Властивості служать акцесорами до змінних закритих елементів.

• Атрибути надають декларативні метадані щодо типів під час виконання.

• Рядкові документаційні коментарі XML.

• Інтегрована мова запитів (LINQ) надає вбудовані можливості запитів між джерелами даних.

Популярність мови — ще одна значима перевага. Багато шанувальників C# сприяють її розвитку. Також це позитивно впливає на зростання кількості вакансій, пов'язаних з розробкою мовою Microsoft. Програмісти, добре знайомі з C#, затребувані в індустрії, незважаючи на їх кількість, що постійно збільшується.

Важливим достоїнством програми є можливість компіляції лише необхідних нині частин програми. Якщо програма не звертається до якоїсь частини коду, її компіляція не відбувається. У момент звернення виконується моментальна компіляція із CIL у машинний код [32].

C# протягом тривалого часу впевнено лідирує у рейтингу найкращих та найбільш затребуваних на ринку розробки мов. Спочатку ним зацікавилися лише розробники, які пишуть програми під Windows. Але в процесі розвитку C# «навчився» працювати на Mac, Linux, IoS та Android. А після того, як код платформи відкрили для всіх бажаючих, було знято практично всі можливі обмеження до застосування C#. В результаті мова активно розвивається, застосовується все ширше. Рекомендована до вивчення як одна з базових мов для розробників будь-якого профілю.

Компанія Microsoft приділяє значну увагу підтримці мови розробки, тому регулярно з'являються оновлення і доповнення, виправляються виявлені баги в компіляторі, розширюються бібліотеки. Розробники зацікавлені у популяризації інструменту та докладають до цього багато зусиль.

**Методика розрахунку коефіцієнтів математичної моделі жирнокислотного складу емульсії.** Як зазначалося вище, для наближення складу сирного виробу до продуктів, що відповідають вимогам раціонального харчування, необхідно додати суміш оливкової та соняшникової олій. Однак при цьому також необхідно, щоб співвідношення кислот НЖК : МНЖК : ПНЖК було максимально наближено до співвідношення 0,3 : 0,6 : 0,1. Для моделювання залежності співвідношення кислот від суміші олій будемо використовувати таблицю експериментальних даних табл.1, перетворивши її у вигляді табл.5, відкинувши проміжні дані.

**Вхідні дані для моделі співвідношення кислот**

| Соняшникова олія | Оливкова олія | Вміст жирних кислот в 100 г продукту | | |
|---|---|---|---|---|
| | | НЖК | МНЖК | ПНЖК |
| 0,05 | 0,95 | 1,5227 | 5,5171 | 0,2760 |
| 0,1 | 0,9 | 1,7559 | 5,5045 | 0,3190 |
| 0,15 | 0,85 | 2,0022 | 5,4913 | 0,3646 |
| 0,2 | 0,8 | 2,2627 | 5,4773 | 0,4131 |
| 0,25 | 0,75 | 2,5387 | 5,4624 | 0,4648 |
| 0,3 | 0,7 | 2,8316 | 5,4466 | 0,5199 |
| 0,35 | 0,65 | 3,1431 | 5,4299 | 0,5788 |
| 0,4 | 0,6 | 3,4749 | 5,4120 | 0,6421 |
| 0,45 | 0,55 | 3,8291 | 5,3929 | 0,7100 |
| 0,5 | 0,5 | 4,2081 | 5,3725 | 0,7833 |
| 0,55 | 0,45 | 4,6146 | 5,3507 | 0,8624 |
| 0,6 | 0,4 | 5,0516 | 5,3271 | 0,9483 |
| 0,65 | 0,35 | 5,5227 | 5,3018 | 1,0417 |
| 0,7 | 0,3 | 6,0322 | 5,2743 | 1,1437 |
| 0,75 | 0,25 | 6,5848 | 5,2446 | 1,2555 |
| 0,8 | 0,2 | 7,1863 | 5,2122 | 1,3787 |
| 0,85 | 0,15 | 7,8434 | 5,1768 | 1,5151 |
| 0,9 | 0,1 | 8,5644 | 5,1380 | 1,6669 |
| 0,95 | 0,05 | 9,3588 | 5,0953 | 1,8368 |

Для побудови та оптимізації математичної моделі раціонально вибрати єдиний критерій, який характеризуватиме отриманий результат. Досить часто у цій якості вибирають адитивний критерій, до якого входять з деякими ваговими коефіцієнтами як складові інші математичні висловлювання. Однак в даному випадку це не можна використовувати, тому що ми повинні отримати критерій, який показує ступінь наближення відношення кислот до шуканого співвідношення $0,3 : 0,6 : 0,1$, визначеного теорією раціонального харчування. Для створення єдиного критерію пропонується визначити «еталонні співвідношення» між числами $0,3, 0,6, 0,1$ і порівнювати отримані пропорції з цім «еталоном».

Позначимо еталонні співвідношення (до яких будемо наближатися при побудові моделі) як:

$$Y_1 = 0,3 : 0,6 = 0,5;$$
$$Y_2 = 0,6 : 0,1 = 6; \qquad (6)$$
$$Y_3 = 0,3 : 0,1 = 3.$$

**Розрахункові дані для моделі співвідношення кислот**

| Соняшни-кова олія, $l_i$ | Оливкова олія, $m_i$ | Вміст жирних кислот в 100 г продукту | | | НЖК/ МНЖК, $T1_i$ | МНЖК/ ПНЖК, $T2_i$ | НЖК/ ПНЖК $T3_i$ | Сумарне відхилен-ня, $F_i$ |
|---|---|---|---|---|---|---|---|---|
| | | НЖК, $c_i$ | МНЖК, $d_i$ | ПНЖК, $e_i$ | | | | |
| 0,05 | 0,95 | 1,5227 | 5,5171 | 0,2760 | 0,2760 | 19,9894 | 5,5171 | 16,7304 |
| 0,1 | 0,9 | 1,7559 | 5,5045 | 0,3190 | 0,3190 | 17,2561 | 5,5045 | 13,9417 |
| 0,15 | 0,85 | 2,0022 | 5,4913 | 0,3646 | 0,3646 | 15,0608 | 5,4913 | 11,6875 |
| 0,2 | 0,8 | 2,2627 | 5,4773 | 0,4131 | 0,4131 | 13,2589 | 5,4773 | 9,8230 |
| 0,25 | 0,75 | 2,5387 | 5,4624 | 0,4648 | 0,4648 | 11,7533 | 5,4624 | 8,2510 |
| 0,3 | 0,7 | 2,8316 | 5,4466 | 0,5199 | 0,5199 | 10,4767 | 5,4466 | 6,9432 |
| 0,35 | 0,65 | 3,1431 | 5,4299 | 0,5788 | 0,5788 | 9,3805 | 5,4299 | 5,8892 |
| 0,4 | 0,6 | 3,4749 | 5,4120 | 0,6421 | 0,6421 | 8,4290 | 5,4120 | 4,9831 |
| 0,45 | 0,55 | 3,8291 | 5,3929 | 0,7100 | 0,7100 | 7,5954 | 5,3929 | 4,1984 |
| 0,5 | 0,5 | 4,2081 | 5,3725 | 0,7833 | 0,7833 | 6,8591 | 5,3725 | 3,5149 |
| 0,55 | 0,45 | 4,6146 | 5,3507 | 0,8624 | 0,8624 | 6,2041 | 5,3507 | 2,9172 |
| 0,6 | 0,4 | 5,0516 | 5,3271 | 0,9483 | 0,9483 | 5,6177 | 5,3271 | 3,1577 |
| 0,65 | 0,35 | 5,5227 | 5,3018 | 1,0417 | 1,0417 | 5,0896 | 5,3018 | 3,7538 |
| 0,7 | 0,3 | 6,0322 | 5,2743 | 1,1437 | 1,1437 | 4,6117 | 5,2743 | 4,3063 |
| 0,75 | 0,25 | 6,5848 | 5,2446 | 1,2555 | 1,2555 | 4,1772 | 5,2446 | 4,8230 |
| 0,8 | 0,2 | 7,1863 | 5,2122 | 1,3787 | 1,3787 | 3,7804 | 5,2122 | 5,3105 |
| 0,85 | 0,15 | 7,8434 | 5,1768 | 1,5151 | 1,5151 | 3,4168 | 5,1768 | 5,7751 |
| 0,9 | 0,1 | 8,5644 | 5,1380 | 1,6669 | 1,6669 | 3,0825 | 5,1380 | 6,2224 |
| 0,95 | 0,05 | 9,3588 | 5,0953 | 1,8368 | 1,8368 | 2,7740 | 5,0953 | 6,6580 |

Позначимо також (табл. 6):

$l_i$ — вміст соняшникової олії в $i$-му експерименті;

$m_i$ — вміст оливкової олії в $i$-му експерименті;

$c_i$ — вміст кислоти НЖК в $i$-му експерименті;

$d_i$ — вміст кислоти МНЖК в $i$-му експерименті;

$e_i$ — вміст кислоти ПНЖК в $i$-му експерименті.

Визначимо також співвідношення кислот експериментальних даних НЖК/МНЖК, МНЖК/ПНЖК, НЖК/ПНЖК в $i$-му експерименті, як $T1_i$, $T2_i$, $T3_i$:

$$T1_i = \frac{c_i}{d_i};$$

$$T2_i = \frac{d_i}{e_i}; \qquad (7)$$

$$T3_i = \frac{c_i}{e_i}.$$

Розрахуємо $T1_i$, $T2_i$, $T3_i$ в експериментальних даних та занесемо до таблиці 6. В якості критерію оптимальності відношення кислот будемо використовувати розрахункову величину, представлену в останньому стовпці таблиці 4:

$$F_i = |T1_i - Y_1| + |T2_i - Y_2| + |T3_i - Y_3|. \qquad (8)$$

Очевидно, що чим менше показник F, тим ближче співвідношення кислот НЖК : МНЖК : ПНЖК до найбільш раціонального відношення 0,3 : 0,6 : 0,1 (гіпотетично F може дорівнювати 0, тоді співвідношення рівні).

Для моделювання необхідно побудувати математичну модель залежності F від l і m, шукатимемо її у вигляді полінома, як найбільш універсальної функції. Обмежимося 4-м ступенем і шукатимемо модель у такому вигляді:

$$G(l, m) = a_1 l^2 + a_2 m^2 + a_3 l + a_4 m + a_5 lm + a_6 + a_7 l^3 + a_8 m^3 + a_9 l^4 + a_{10} m^4. \qquad (9)$$

Розраховуємо невідомі коефіцієнти $a_i$ за методикою найменших квадратів з використанням методу випадкового пошуку в розробленій програмі, отримуємо таку модель:

$$G = 1{,}552162 + 0{,}589001 l + 0{,}955900 m +$$
$$+ 5{,}549843 l^3 + 17{,}483025 m^4. \qquad (10)$$

Таким чином, розраховані коефіцієнти регресійної моделі такі:

| | |
|---|---|
| $a_1=$ | 0,000000 |
| $a_2=$ | 0,000000 |
| $a_3=$ | 0,589001 |
| $a_4=$ | 0,955900 |
| $a_5=$ | 0,000000 |
| $a_6=$ | 1,552162 |
| $a_7=$ | 5,549843 |
| $a_8=$ | 0,000000 |
| $a_9=$ | 0,000000 |
| $a_{10}=$ | 17,483025 |

Мінімізуючи отриману функцію, маємо такі значення:

$$l = 0,579062;$$

$$m = 0,420938.$$

При цьому значення критерію оптимальності дорівнюватиме $F = 2,8949$, що добре корелює з експериментальними даними з табл. 2.

**Методика розрахунку коефіцієнтів математичної моделі** гомогенізації емульсій при виробництві кисломолочного сиру з використанням купажів соняшникової та оливкової олій.

Як зазначалося вище, для визначення оптимальних режимів гомогенізації необхідно побудувати математичні моделі залежності стійкості емульсії та відстою жирової фази від температурі та тиску. Розглянемо окремі залежність стійкості емульсії від температури та тиску та залежність жирової фази від температури та тиску. Для зручності обробки вхідних даних під час створення математичної моделі стійкості емульсії перетворимо таблицю 3, відкинувши проміжні дані:

Для знаходження залежності стійкості емульсії від тиску та температури введемо такі позначення:

$D_i$ — значення стійкості емульсії в i-му експерименті;

$x_i$ — значення тиску в i-му експерименті;

$y_i$ — значення температури в i-му експерименті.

Шукатимемо математичну залежність у вигляді поліноміальної функції

$$S = a_1{*}x + a_2{*}y + a_3{*}x^2 + a_4{*}y^2 + a_5xy + a_6 + a_7{*}x^3 + a_8{*}y^3 + a_9{*}x^4 + a_{10}{*}y^4. \quad (11)$$

**Вибрані експериментальні дані для моделювання стійкості емульсії**

| Тиск, Мпа | Температура, С | Стійкість емульсії Y, % |
|---|---|---|
| 7 | 55 | 98,1 |
| 10 | 55 | 98,3 |
| 12 | 55 | 98,6 |
| 15 | 55 | 98,8 |
| 7 | 60 | 98,4 |
| 10 | 60 | 100 |
| 12 | 60 | 100 |
| 15 | 60 | 100 |
| 7 | 65 | 99 |
| 10 | 65 | 100 |
| 12 | 65 | 100 |
| 15 | 65 | 100 |
| 7 | 70 | 99,3 |
| 10 | 70 | 100 |
| 12 | 70 | 100 |
| 15 | 70 | 100 |

Розраховуємо невідомі коефіцієнти $a_i$ за методикою найменших квадратів: шукаємо мінімум функції

$$\sum_{i=1}^{n} \left( S(x_i, y_i) - D(x_i, y_i) \right)^2 \to \min .$$ (12)

Таблиця 8

**Розрахунок стійкості емульсії за математичною моделлю**

| Тиск, $x_i$ | Температура, $y_i$ | Стійкість емульсії $D_i$ | Розрахунок S | Відхилення |
|---|---|---|---|---|
| 7 | 55 | 98,1 | 91,357 | 6,743 |
| 10 | 55 | 98,3 | 101,393 | 3,093 |
| 12 | 55 | 98,6 | 103,777 | 5,177 |
| 15 | 55 | 98,8 | 94,798 | 4,002 |
| 7 | 60 | 98,4 | 93,491 | 4,909 |
| 10 | 60 | 100 | 103,526 | 3,526 |
| 12 | 60 | 100 | 101,910 | 1,910 |
| 15 | 60 | 100 | 96,931 | 3,069 |
| 7 | 65 | 99 | 93,971 | 5,029 |
| 10 | 65 | 100 | 104,006 | 4,006 |
| 12 | 65 | 100 | 106,390 | 6,390 |

| Тиск, $x_i$ | Температура, $y_i$ | Стійкість емульсії $D_i$ | Розрахунок S | Відхилення |
|---|---|---|---|---|
| 15 | 65 | 100 | 97,411 | 2,589 |
| 7 | 70 | 99,3 | 92,862 | 6,438 |
| 10 | 70 | 100 | 102,897 | 2,897 |
| 12 | 70 | 100 | 105,280 | 5,280 |
| 15 | 70 | 100 | 96,301 | 3,699 |

Використовуючи метод випадкового пошуку в розробленій програмі, отримуємо таку модель:

$$S = -0,004666x - 0,026530y + 0,577439x^2 + 0,076405y^2 - 0,000010xy +$$

$$+ 1,550705 - 0,021941x^3 - 0,001249y^3 - 0,000656x^4\ 0,000005y^4. \quad (13)$$

Таким чином, розраховані коефіцієнти регресійної моделі такі:

| $a_1 =$ | -0,004666 |
|---|---|
| $a_2 =$ | -0,026530 |
| $a_3 =$ | 0,577439 |
| $a_4 =$ | 0,076405 |
| $a_5 =$ | -0,000010 |
| $a_6 =$ | 1,550705 |
| $a_7 =$ | -0,021941 |
| $a_8 =$ | -0,001249 |
| $a_9 =$ | -0,000656 |
| $a_{10} =$ | 0,000005 |

Для експериментальних даних розрахунок за цією математичною моделлю призводить до таких результатів (табл. 8). В останньому стовпці показано відхилення експериментальних даних від розрахункових, отриманих за побудованою математичною моделлю.

Очевидно, що відхилення невеликі, тому побудованою моделлю можна користуватися для розрахунку реальних значень стійкості емульсії сирних виробів.

Також для визначення оптимальних режимів гомогенізації необхідно побудувати математичну модель залежності відстою жирової фази від температури та тиску. Для цього перетворимо таблицю 3 на таку, відкинувши проміжні дані:

**Вибрані експериментальні дані до моделювання відстою жирової фази**

| Тиск, МПа | Температура, С | Відстій жирової фази v, % |
|-----------|----------------|---------------------------|
| 7 | 55 | 9,2 |
| 10 | 55 | 6,9 |
| 12 | 55 | 5,2 |
| 15 | 55 | 4,2 |
| 7 | 60 | 7,7 |
| 10 | 60 | 5,3 |
| 12 | 60 | 3,4 |
| 15 | 60 | 2,6 |
| 7 | 65 | 5 |
| 10 | 65 | 3,6 |
| 12 | 65 | 1,5 |
| 15 | 65 | 1 |
| 7 | 70 | 3,5 |
| 10 | 70 | 2,1 |
| 12 | 70 | 0,9 |
| 15 | 70 | 0,6 |

За аналогією з попередніми розрахунками введемо позначення відстою жирової фази $V_i$ (яке залежить від експериментальних даних $x_i$ — значення тиску в i-му експерименті, $y_i$ — значення температури в i-му експерименті) та будемо шукати математичну модель як функцію

$$G = a_1*x + a_2*y + a_3*x^2 + a_4*y^2 + a_5xy + a_6 + a_7*x^3 + a_8*y^3 + a_9*x^4 + a_{10}*y^4. \quad (14)$$

Застосування розрахунків за методикою найменших квадратів, як зазначено вище, дає таку модель:

$$G = -0,004837x - 0,027562y + 0,005950x^2 + 0,034753y^2 - 0,000011xy +$$

$$+ 1,599765 - 0,008708x^3 - 0,000901y^3 + 0,000438x^4 + 0,000006y^4. \quad (15)$$

Розраховані коефіцієнти регресійної моделі такі:

| $a_1=$ | -0,004837 |
|--------|-----------|
| $a_2=$ | -0,027562 |
| $a_3=$ | 0,005950 |
| $a_4=$ | 0,034753 |
| $a_5=$ | -0,000011 |
| $a_6=$ | 1,599765 |
| $a_7=$ | -0,008708 |
| $a_8 =$ | -0,000901 |
| $a_9 =$ | 0,000438 |
| $a_{10} =$ | 0,000006 |

Для експериментальних даних розрахунок за цією математичною моделлю відстою жирової фази призводить до таких результатів (табл. 10). В останньому стовпці показано відхилення експериментальних даних від розрахункових, отриманих за побудованою математичною моделлю.

Відхилення експериментальних даних від розрахункових (останній стовпець таблиці 10) невеликі, тому знайдену математичну модель можна використовувати у роботах дослідників.

Використовуючи ці дві математичні моделі гомогенізації емульсій різного хімічного складу, можна знайти оптимальні параметри процесу гомогенізації таким чином: стійкість емульсії має бути максимальною (100 %), відстій жирової фази має бути мінімальний. Розрахунки за цими критеріями дають такі результати: тиск дорівнює 12 МПа, температура 60°C, при цьому стійкість емульсії дорівнює 100 %, відстій жирової фази 3,167 %. Ці дані добре корелюють з даними, отриманими іншим експериментальним шляхом.

Таблиця 10

**Розрахунок відстою жирової фази за математичною моделлю**

| Тиск, $x_i$ | Температура, $y_i$ | Відстій жирової фази $v_i$ | Розрахунок $G_i$ | Відхилення |
|---|---|---|---|---|
| 7 | 55 | 9,2 | 8,915 | 0,285 |
| 10 | 55 | 6,9 | 6,809 | 0,091 |
| 12 | 55 | 5,2 | 5,424 | 0,224 |
| 15 | 55 | 4,2 | 4,639 | 0,439 |
| 7 | 60 | 7,7 | 7,058 | 0,642 |
| 10 | 60 | 5,3 | 4,953 | 0,347 |
| 12 | 60 | 3,4 | 3,167 | 0,233 |
| 15 | 60 | 2,6 | 2,782 | 0,182 |
| 7 | 65 | 5 | 5,360 | 0,360 |
| 10 | 65 | 3,6 | 3,254 | 0,346 |
| 12 | 65 | 1,5 | 1,868 | 0,368 |
| 15 | 65 | 1 | 1,084 | 0,084 |
| 7 | 70 | 3,5 | 4,277 | 0,777 |
| 10 | 70 | 2,1 | 2,171 | 0,071 |
| 12 | 70 | 0,9 | 0,785 | 0,115 |
| 15 | 70 | 0,6 | 0,000 | 0,600 |

**Програмна підтримка дослідження.** З метою розрахування оптимальної рецептури сирного виробу були проведені натурні експерименти, на основі отриманих даних з яких будується математична модель. Всі отримані дані потрібно структуровано зберігати для зручного доступу до них, саме тому використання бази даних є необхідним. Приклад створено універсальною алгоритмічною мовою C# [18; 27; 28].

Оскільки обсяг даних є невеликим, а доступ до них буде виключно локальним, використання таких великих та потужних СУБД як PostgreSQL або Oracle Database не є раціональним, тому враховуючи вищезазначене найбільш оптимальним вибором стає SQLite, це легка вбудована СУБД що повністю задовольняє поставленим умовам, а також є безкоштовною, тому її використання не потребує залучення додаткових коштів.

SQLite — компактна СУБД, що вбудовується. Вихідний код бібліотеки передано до загального використання. Слово «вбудований» (embedded) означає, що SQLite не використовує парадигму клієнт — сервер, тобто двигун SQLite не є окремо працюючим процесом, з яким взаємодіє програма, а являє собою бібліотеку, з якою програма компонується, і двигун стає складовою програми. Таким чином, як протокол обміну використовуються виклики функцій (API) бібліотеки SQLite. Такий підхід зменшує накладні витрати, час відгуку та спрощує програму. SQLite зберігає всю базу даних (включаючи визначення, таблиці, індекси та дані) в єдиному стандартному файлі на тому комп'ютері, на якому виконується програма. Простота реалізації досягається за рахунок того, що перед початком виконання транзакції запису весь файл, що зберігає базу даних, блокується; ACID-функції досягаються у тому числі для створення файлу журналу [14; 17].

Декілька процесів або потоків можуть одночасно без будь-яких проблем читати дані з однієї бази. Запис до бази можна здійснити лише у тому випадку, якщо жодних інших запитів на даний момент не обслуговується; інакше спроба запису закінчується невдачею, й у програму повертається код помилки. Іншим варіантом розвитку подій є автоматичне повторення спроб запису протягом заданого інтервалу часу.

SQLite — це вбудована бібліотека, яка реалізує автономний, безсерверний, нульової конфігурації механізм транзакції СУБД SQL. Це база даних, яка налаштована на нуль. Це означає, що вам не потрібно налаштовувати її у вашій системі.

SQLite не є автономним процесом, як інші бази даних, ви можете пов'язати його статично або динамічно відповідно до вашої вимоги з вашою програмою. SQLite безпосередньо звертається до своїх файлів зберігання [8].

Особливості SQLite:

• SQLite не вимагає окремого процесу сервера або системи для роботи (без сервера).

• SQLite поставляється з нульовою конфігурацією, що означає відсутність необхідності в налаштуванні або адмініструванні.

• Повна база даних SQLite зберігається в одному кросплатформному диску.

• SQLite дуже маленька і легка, менше 400KiB повністю налаштована або менше 250KiB з додатковими функціями.

• SQLite є автономною, що означає відсутність зовнішніх залежностей.

• Транзакції SQLite повністю сумісні з ACID, забезпечуючи безпечний доступ до кількох процесів або потоків.

• SQLite підтримує більшість функцій мови запитів, знайдених у стандарті SQL92 (SQL2).

• SQLite написана на ANSI-C і надає простий спосіб у використанні API.

• SQLite доступна у UNIX (Linux, Mac OS-X, Android, iOS) та Windows (Win32, WinCE, WinRT).

СУБД SQLite багато в чому підтримує стандартний SQL. Діалект мови SQL, що використовується в SQLite, за синтаксисом схожий на той, який використовується в PostgreSQL. Однак SQLite накладає низку специфічних особливостей на SQL.

Слід розрізняти саму SQLite як бібліотеку, що містить СУБД, і базу даних як таку. За допомогою SQLite створюються бази даних, що являють собою один кросплатформений текстовий файл. Файл бази даних, на відміну SQLite, не вбудовується в додаток, не стає його частиною, він існує окремо. Так можна створити базу даних, користуючись консольним sqlite3, після чого використовувати її у програмі за допомогою бібліотеки SQLite мови програмування. При цьому файл бази даних зберігається на локальній машині.

Програма, що включає SQLite, використовує її функціональність за допомогою простих викликів функцій. Оскільки функції викликаються в тому самому процесі, що працює програма, виклики працюють швидше, ніж це було б у випадку міжпроцесної взаємодії.

Відхід від клієнт-серверної моделі зовсім не означає, що SQLite це навчальна або урізана СУБД. Це означає лише специфіку її застосування в ролі компонента, що вбудовується. Існує безліч типів додатків, від «записних книжок» до браузерів та операційних систем, що потребують невеликих локальних баз даних.

Оскільки SQLite працює в рамках іншої програми, під час запису файл бази даних блокується. Таким чином, записувати дані можна лише послідовно. У той же час, читати базу можуть відразу кілька процесів. SQLite — не найкращий вибір, якщо передбачаються часті звернення до БД на запис. Перелічені особливості СУБД SQLite сприяють тому, що вона була використана для розробки додатка.

На рис. 3 представлена схема бази даних SQLite, що використовується для роботи додатку.

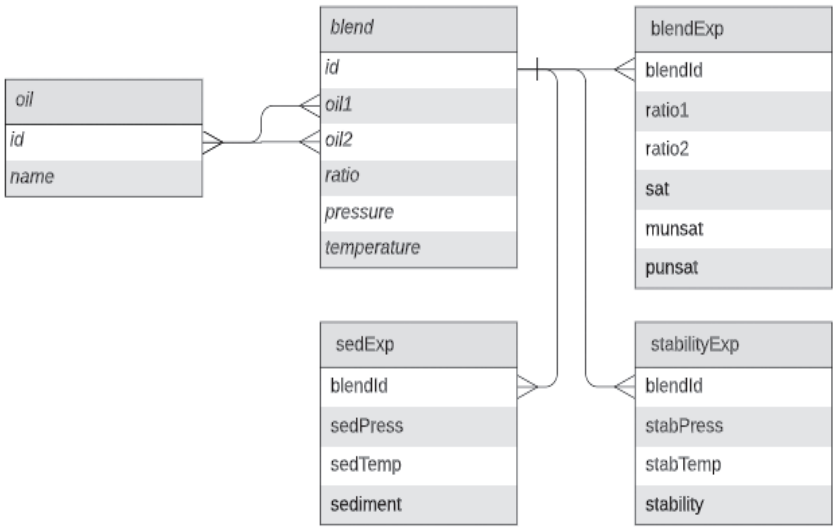

Рис. 3. Схема бази даних додатку

Також нижче наводяться фрагменти опису таблиць бази даних:
Таблиця олій:

```
CREATE TABLE "oil" (
    "id"    INTEGER NOT NULL UNIQUE,
    "name"  TEXT NOT NULL,
    PRIMARY KEY("id" AUTOINCREMENT)
```

**635**

Таблиця сумішей:

```
CREATE TABLE "blend" (
    "id"      INTEGER NOT NULL UNIQUE,
    "oil1"    INTEGER,
    "oil2"    INTEGER,
    "ratio"   INTEGER,
    "pressure"   INTEGER,
    "temperature"   INTEGER,
    FOREIGN KEY("oil1") REFERENCES "oil"("id"),
    FOREIGN KEY("oil2") REFERENCES "oil"("id"),
    PRIMARY KEY("id")
```

Таблиця експериментів жирнокислотного складу сумішей:

```
CREATE TABLE "blendExp" (
    "blendId"   INTEGER NOT NULL,
    "ratio1"    INTEGER NOT NULL,
    "ratio2"    INTEGER NOT NULL,
    "sat"    REAL NOT NULL,
    "munsat"    REAL NOT NULL,
    "punsat"    REAL NOT NULL,
    FOREIGN KEY("blendId") REFERENCES "blend"("id")
```

Таблиця експериментів стійкості суміші:

```
CREATE TABLE "stabilityExp" (
    "blendId"   INTEGER NOT NULL,
    "stabPress" INTEGER NOT NULL,
    "stabTemp"  INTEGER NOT NULL,
    "stability" REAL,
    FOREIGN KEY("blendId") REFERENCES "blend"("id")
```

Таблиця експериментів відстою жирової фази:

```
CREATE TABLE "sedimentExp" (
    "blendId"   INTEGER NOT NULL,
    "sedPress"  INTEGER NOT NULL,
    "sedTemp"   INTEGER NOT NULL,
    "sediment"  REAL,
    FOREIGN KEY("blendId") REFERENCES "blend"("id")
```

**Опис основних процедур та функцій**. Першочергово при запуску додатку перевіряється актуальність посилання на файл бази даних.

```
var db = Properties.Settings.Default.pathToDB;
    if (!File.Exists(db))
        MessageBox.Show («Файл бази даних відсутній, вкажіть нове
розташування файлу»);
```

**636**

```
        using (OpenFileDialog dialog = new OpenFileDialog() {Filter = «Файл
бази даних (*.db)|*.db"}»)
            dialog.ShowDialog();
            if (string.IsNullOrEmpty(dialog.FileName) || !File.Exists(dialog.File-
Name))
                return;
            db = dialog.FileName;
            Properties.Settings.Default.Save();
    else if(Path.GetExtension(db) != ".db")
        MessageBox.Show («Файл не є базою даних. Оберіть файл з
розширенням.db»);
        using (OpenFileDialog dialog = new OpenFileDialog() {Filter = «Файл
бази даних (*.db)|*.db}»)
            dialog.ShowDialog();
        if (string.IsNullOrEmpty(dialog.FileName) || !File.Exists(dialog.File-
Name))
            return;
            db = dialog.FileName;
            Properties.Settings.Default.Save();
```

У випадку недоступності файлу з тих чи інших причин буде від-
крито діалогове вікно системного файлового менеджеру з метою вка-
зування шляху до нового файлу, сам файл може знаходитись у будь-
якому місці.

```
public static DataTable select(string columns, string from)
    CheckConnection();
    using (SQLiteDataAdapter adapter = new SQLiteDataAdapter(String.For-
mat(«select {0} from {1}», columns, from), connection))
    using (var dataSet = new DataSet())
        adapter.Fill(dataSet);
        return dataSet.Tables [0];
```

Процедура запиту даних з таблиці приймає рядок назв потрібних
стовбців та назву таблиці, також є однойменна процедура з додатко-
вим параметром для вказування параметрів фільтрування даних.

```
public static void update(string table, string upd, string where)
    CheckConnection();
    using (var cmd = new SQLiteCommand(String.Format(«update {0} set {1}
where {2}», table, upd, where), connection))
        cmd.ExecuteNonQuery();
```

Процедура оновлення існуючих записів.

```
public static void delete(string from, string where)
    CheckConnection();
    using (var cmd = new SQLiteCommand(String.Format(«delete from {0} where
{1}», from, where), connection))
        cmd.ExecuteNonQuery();
```

Процедура видалення. Для відображення та роботи з даними БД віконний додаток використовує елемент DataGridView. Далі наведено процедури роботи з ними.

```
private void loadBlendExp()
    if (blendExp.Rows.Count != 0)
        blendExp.Rows.Clear();
    var rows = select("rowid, ratio1,ratio2,sat,munsat,punsat", "blendExp",
$"blendId={selectedBlend.ItemArray [0]}").Rows;
    if (rows.Count != 0)
        foreach (DataRow row in rows)
            blendExp.Rows.Add(row.ItemArray);
```

Заповнення даних з таблиці експериментів складу суміші.

```
private void DataGridUserDeletingRow(object sender, DataGridViewRowCan-
celEventArgs e)
    var dg = (sender as DataGridView);
    string id = dg.Columns [0].Name.Contains(«row») ? «id» : «rowid»;
    delete(dg.Name, $»{id}={dg.Rows [e.Row.Index].Cells [0].Value}»);
```

Процедура видалення, оскільки всі таблиці посилаються на спільні процедури, в першу чергу, визначається, до якої таблиці адресовано запит, після чого вже виконується робота над вказаними запитами

```
private string cellValue;
private void DataGridCellBeginEdit(object sender, DataGridViewCellCancelEv-
entArgs e)
    var dg = (sender as DataGridView);
    cellValue = dg.Rows [e.RowIndex].Cells [e.ColumnIndex].Value.ToString();
private void DataGridCellEndEdit(object sender, DataGridViewCellEventArgs e)
    var dg = (sender as DataGridView);
    string id = dg.Columns [0].Name.Contains(«row») ? «id» : «rowid»;
    if (dg.Rows [e.RowIndex].Cells [e.ColumnIndex].Value.ToString() != cellValue)
```

```
    update(dg.Name,
        $»{dg.Columns [e.ColumnIndex].Name}='{dg.Rows [e.RowIndex].
Cells [e.ColumnIndex].Value}'»,
        $»{id}={dg.Rows [e.RowIndex].Cells [0].Value}»);
    if (dg.Name == «oil»)
        FillCalcBlendCombo();
```

Процедура редагування розбита на два етапи, спочатку зберігається поточне значення у редагованій комірці, яке потім перевіряється на збіг з новим записом, це зроблено для уникнення перезапису, якщо виклик редагування був випадковим.

Вибір суміші робиться шляхом обрання першого та другого компонентів з двох випадаючих списків, черговість вибору неважлива, вона збережена після додавання запису про нову суміш. Випадаючі списки розміщені на трьох сторінках додатку та є синхронізованими, немає потреби обирати компоненти на кожній сторінці.

```
private void FillCalcBlendCombo()
    var CB = blendCB1.ComboBox;
    var selected = CB. SelectedIndex == -1 ? CB. Items.Count : CB. SelectedIn-
dex;
    var DataSource = GetIdNamePairs(«oil»);
    CB. DataSource = DataSource;
    CB. ValueMember = «id»;
    CB. DisplayMember = «name»;
    CB. SelectedIndex = selected;
    CB = blendCB2.ComboBox;
    selected = CB. SelectedIndex == -1 ? CB. Items.Count : CB. SelectedIndex;
    CB. BindingContext = new BindingContext();
    CB. DataSource = DataSource;
    CB. ValueMember = «id»;
    CB. DisplayMember = «name»;
    CB. SelectedIndex = selected;
```

Процедура заповнення випадаючих списків.

```
private void SynchronizeCB(ComboBox CB1, ComboBox CB2)
    CB1.DataSource = blendCB1.ComboBox.DataSource;
    CB1.ValueMember = «id»;
    CB1.DisplayMember = «name»;
    CB2.BindingContext = blendCB2.ComboBox.BindingContext;
    CB2.DataSource = blendCB2.ComboBox.DataSource;
```

```csharp
    CB2.ValueMember = «id»;
    CB2.DisplayMember = «name»;
private void BindCB()
    SynchronizeCB(sedimentCB1.ComboBox, sedimentCB2.ComboBox);
    SynchronizeCB(stabilityCB1.ComboBox, stabilityCB2.ComboBox);
    (blend.Columns [1] as DataGridViewComboBoxColumn).DataSource =
blendCB1.ComboBox.DataSource;
    (blend.Columns [1] as DataGridViewComboBoxColumn).ValueMember =
«id»;
    (blend.Columns [1] as DataGridViewComboBoxColumn).DisplayMember =
«name»;
    (blend.Columns [2] as DataGridViewComboBoxColumn).DataSource =
blendCB2.ComboBox.DataSource;
    (blend.Columns [2] as DataGridViewComboBoxColumn).ValueMember =
«id»;
    (blend.Columns [2] as DataGridViewComboBoxColumn).DisplayMember =
«name»;
```

Процедура синхронізації випадаючих списків.

```csharp
private void selectBlend()
    var CB1 = blendCB1.ComboBox;
    var CB2 = blendCB2.ComboBox;
    var s = select(«*», «blend», $»oil1={CB1.SelectedValue} and oil2={CB2.
SelectedValue} or oil1={CB2.SelectedValue} and oil2={CB1.SelectedValue}»);
    if (s.Rows.Count == 0)
        using (var cmd = new SQLiteCommand($»insert into blend (oil1, oil2)
values ('{CB1.SelectedValue}', '{CB2.SelectedValue}'); SELECT last_insert_row-
id()», connection))
            selectedBlend = select(«*», «blend», $»id={cmd.ExecuteScalar()}»).
Rows [0];
            loadExp();
    else
selectedBlend = s.Rows [0];
    loadExp();
    if (selectedBlend != null)
        blendExp.Columns [1].HeaderText = selectedBlend.ItemArray [1].
ToString() == CB1.SelectedValue.ToString() ? StringExtensions.FirstCharToUp-
per(CB1.Text) : StringExtensions.FirstCharToUpper(CB2.Text);
        blendExp.Columns [2].HeaderText = selectedBlend.ItemArray [2].
ToString() == CB2.SelectedValue.ToString() ? StringExtensions.FirstCharToUp-
per(CB2.Text) : StringExtensions.FirstCharToUpper(CB1.Text);
```

Процедура обрання суміші. Оскільки списки синхронізовані, процедура завжди посилається на ті, що розташовані на першій сторінці. Коди компонентів зчитуються та перевіряється наявність суміші з такою комбінацією, якщо такого запису немає — буде створено новий та отримано посилання на нього.

```
StringBuilder newValues = new StringBuilder();
    var CB1 = blendCB1.ComboBox;
    var CB2 = blendCB2.ComboBox;
    if (CB1.SelectedValue == null)
        if (CB1.Text != «»)
            newValues.Append($»('{CB1.Text}'),»);
        if (CB2.SelectedValue == null)
if (CB2.Text != «»)
newValues.Append($»('{CB2.Text}'),»);
if (newValues.Length != 0)
insert(«oil», «name», newValues.Remove(newValues.Length — 1, 1).ToString());
FillCalcBlendCombo();
selectBlend();
```

Списки передбачають можливість введення нових компонентів не переходячи до вікна всіх записів. Не знайшовши ідентифікатора компонентів, додаток автоматично збереже їх та суміш на їх основі.

Для додавання нових записів експериментів на кожній сторінці є відповідні поля та кнопка підтвердження.

```
if (selectedBlend != null)
    var error = false;
    for (int i = 0; i < TS. Items.Count — 1; i += 2)
        if (TS. Items [i].Text == «»)
        TS. Items [i].BackColor = Color.Red;
        error = true;
    if (!error)
        StringBuilder newValues = new StringBuilder();
        newValues.Append($»({selectedBlend.ItemArray [0]},»);
        for (int i = 0; i < TS. Items.Count — 1; i += 2)
        newValues.Append($»'{TS. Items [i].Text.Replace(',', '.')}',»);
        TS. Items [i].Text = «»;
        newValues [newValues.Length — 1] = ')';
        return (newValues.ToString());
    else return null;
else return null;
```

```
var newValues = addExp((sender as ToolStripButton).Owner);
    if (newValues != null)
        insert(«blendExp», «blendid, ratio1,ratio2,sat,munsat,punsat», newValues.
ToString());
        loadBlendExp();
```

Процедура додавання результату експерименту спільна для всіх вікон, спочатку отримується посилання на панель з елементами введення, потім перевіряється, чи обрана суміш, після чого процедура зчитує наявні поля, перевіряє їх заповнення та додає новий запис.

**Опис програмного додатка.** Програмне забезпечення являє собою віконний додаток з п'ятьма основними сторінками та вбудованою системою управління базою даних. Розроблена програма може застосовуватися не тільки для створення та оптимізації математичних моделей сумішей при виробництві кисломолочних продуктів, але й для будь-яких завдань, у яких необхідно знайти коефіцієнти рівняння регресії за еспериментальними даними. Особливістю додатка є можливість визначення близькості одержуваних трьох величин до заданого співвідношення трьох еталонних чисел.

Основне вікно програми містить кілька вкладок, на які користувач може переходити у процесі роботи над створенням математичних моделей та їх оптимізацією, — «розрахунок суміші», «стійкість емульсії», «відстій фази», «суміші», «олії» (рис. 4). Дві останні вкладки є допоміжними, в них заносяться і вибираються вихідні дані — назви олій та ті олії, які будуть у суміші для подальших розрахунків. Хоча зараз додаток використовується для оптимізації сумішей з соняшниковою та оливковою оліями, в принципі він носить універсальний характер і може застосовуватися і для інших інгредієнтів.

На рис. 4 представлений екран програми з відкритою вкладкою для занесення та редагування інгредієнтів, що використовуються в задачі, — олій.

На рис. 5 представлений фрагмент роботи додатка при виборі двох олій, що беруть участь у подальших розрахунках, у цьому вікні також можна заносити деякі вхідні характеристики суміші. Для початку роботи необхідно обрати з випадаючих списків перший та другий компонент суміші та підвердити вибір, натиснувши на кнопку.

При виборі олій можна скористатися табл.1, але в реальних розрахунках будуть використовуватися соняшникова та оливкова олії. У табл.1 наведено жирнокислотний склад різних рослинних олій од-нак наступні міркування дозволяють вибрати з усього списку дві — соняшникову та оливкову олії.

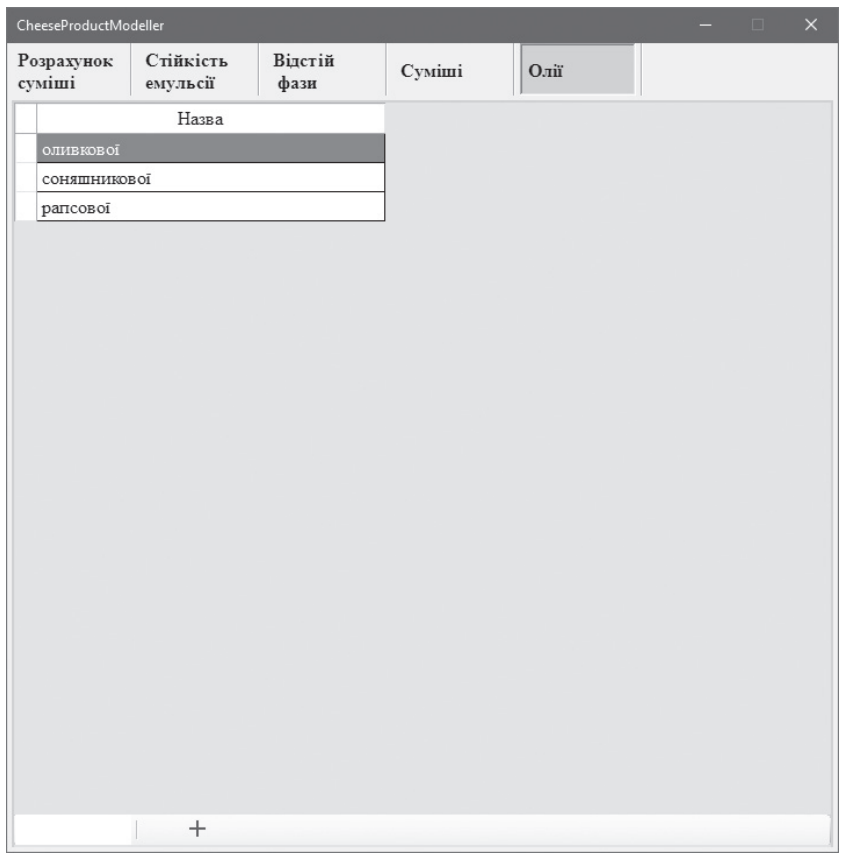

Рис. 4. Екран роботи з вкладкою «олії»

Для наближення складу основи для виробництва продуктів, що відповідають вимогам раціонального харчування необхідно значно підвищити вміст ПНЖК і МНЖК. Кількість НЖК повинна залишатись майже такою ж. Як видно із даних, наведених в табл. 1, для корегування співвідношення між жирними кислотами доцільно використовувати оливкову олію, яка є основним постачальником МНЖК, і соняшникову як джерело ПНЖК.

Рис. 5. Екран роботи з вкладкою «суміші»

На рис. 6 показаний процес занесення експериментальних вхідних даних для створення математичної моделі емульсії з соняшниковою та оливковою оліями. На цьому етапі необхідно ввести відсоток участі в суміші кожної олії (соняшникова та оливкова), а також отримані значення кислот НЖК, МНЖК, ПНЖК у 100 г кисломолочного продукту (тобто занести дані з табл. 5). Результат занесення експериментальних даних в «розрахунок суміші», представлений на рис. 7. Внесені дані повністю відповідають табл. 5. Вкладка «розрахунок суміші» відповідає за перегляд, додавання, видалення та редагування записів експериментальних даних й розрахунок на їх основі оптимального співвідношення олій в суміші для максимального наближення їх жирнокислотного складу до еталонного.

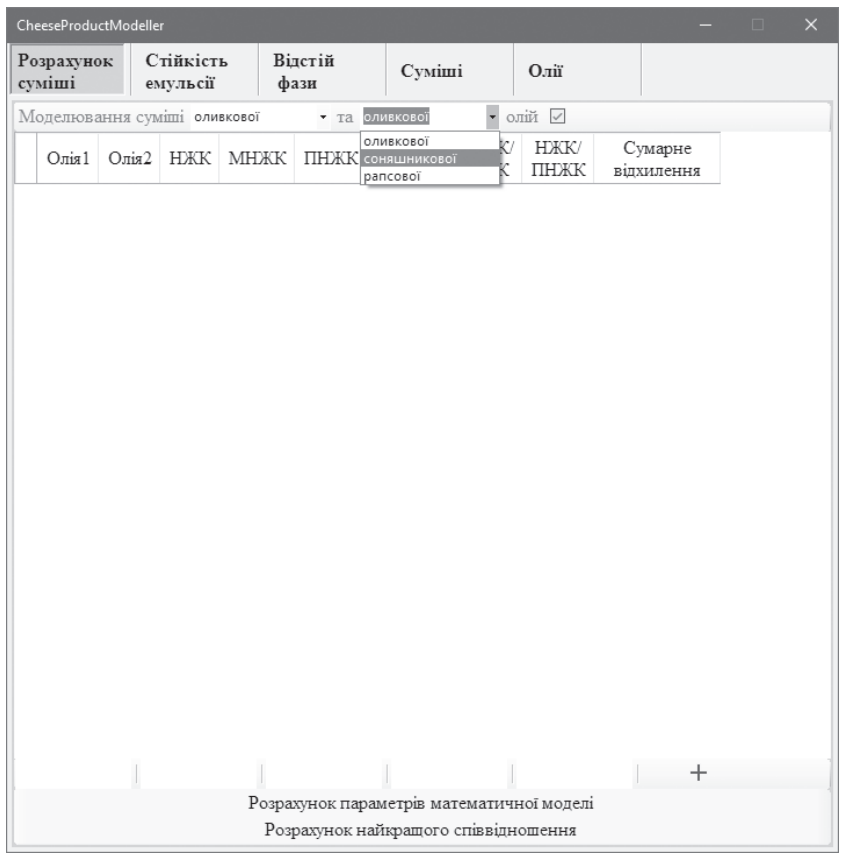

Рис. 6. Екран демонстрації занесення даних у вкладку «розрахунок суміші»

При цьому програма автоматично розраховує співвідношення НЖК/МНЖК, МНЖК/ПНЖК, НЖК/ПНЖК за формулами (7), а також інтегральний показник сумарного відхилення від заданого співвідношення 0,3:0,6:0,1 за формулою (8) — рис. 8. Результати такого розрахунку наведено у табл.6. Для додавання нових записів слід заповнити поля у нижній частині екрану та підвередити додаваня, після чого таблиця автоматично оновиться і нові дані будуть виведені.

| CheeseProductModeller | | | | | | | | | |
|---|---|---|---|---|---|---|---|---|---|

**Розрахунок суміші** | **Стійкість емульсії** | **Відстій фази** | **Суміші** | **Олії**

Моделювання суміші оливкової ▼ та соняшникової ▼ олій ☑

| | Соняшникової | Оливкової | НЖК | МНЖК | ПНЖК | НЖК/МНЖК | МНЖК/ПНЖК | НЖК/ПНЖК | Сумарне відхилення |
|---|---|---|---|---|---|---|---|---|---|
| | 5 | 95 | 1,523 | 5,517 | 0,276 | | | | |
| | 10 | 90 | 1,756 | 5,505 | 0,319 | | | | |
| | 15 | 85 | 2 | 5,491 | 0,365 | | | | |
| | 20 | 80 | 2,263 | 5,426 | 0,413 | | | | |
| | 25 | 75 | 2,539 | 5,477 | 0,465 | | | | |
| | 30 | 70 | 2,831 | 5,447 | 0,5199 | | | | |
| | 35 | 65 | 3,1431 | 5,42999 | 0,5788 | | | | |
| | 40 | 60 | 3,4749 | 5,412 | 0,6421 | | | | |
| | 45 | 55 | 3,8291 | 5,3929 | 0,71 | | | | |
| | 50 | 50 | 4,2081 | 5,3725 | 0,7833 | | | | |
| | 55 | 45 | 4,6146 | 5,3507 | 0,8624 | | | | |
| | 60 | 40 | 5,0516 | 5,3271 | 0,9483 | | | | |
| | 65 | 35 | 5,5227 | 5,3018 | 1,0417 | | | | |
| | 70 | 30 | 6,0322 | 5,2743 | 1,1437 | | | | |
| | 75 | 25 | 6,5848 | 5,2446 | 1,2555 | | | | |
| | 80 | 20 | 7,1863 | 5,2122 | 1,3787 | | | | |
| | 85 | 15 | 7,8434 | 5,1768 | 1,5151 | | | | |
| | 90 | 10 | 8,5644 | 5,138 | 1,6669 | | | | |
| | 95 | 5 | 9,3588 | 5,0953 | 1,8368 | | | | |

+

Розрахунок параметрів математичної моделі

Розрахунок найкращого співвідношення

Рис. 7. Результат занесення експериментальних даних у вкладку «розрахунок суміші»

Для розрахунку параметрів математичної моделі необхідно натиснути відповідну кнопку, після чого розрахункові величини будуть відображені у таблиці, а також з'явиться вікно зі збереженою математичною моделлю та її коефіцієнтами. Розрахунок ведеться за алгоритмом випадкового пошуку (Монте-Карло), викладеним вище. При цьому методом найменших квадратів вирішується завдання

$$\sum_{i=1}^{n} \left( G(l_i, m_i) - F(l_i, m_i) \right)^2 \to \min, \qquad (16)$$

Рис. 8. Автоматичний розрахунок інтегрального показника сумарного відхилення де значення G та F знаходяться за формулами (8) та (9). Для рівняння регресії використовується поліноміальна функція як найбільш універсальна функція (при цьому для зменшення кількості коефіцієнтів обмежуємося 4-м ступенем). Результати розрахунку коефіцієнтів математичної моделі (9) за допомогою розробленої програми наведено на рис. 9 та в моделі (10). Алгоритм випадкового пошуку (Монте-Карло) в програмі використовує вбудовану функцію псевдовипадкових чисел із дуже великим періодом повторення (можна вважати для нашого завдання, що це нескінченний ряд випадкових нормованих у діапазоні [0, 1] чисел). Для кожного кроку алгоритму застосовується 10 тисяч невдалих кроків, якщо не знаходиться точка з меншим значенням функції, крок зменшується вдвічі. Мінімальне знання кроку, після якого алгоритм припиняє свою роботу, — 0,0001.

Рис. 9. Результати розрахунку коефіцієнтів математичної моделі суміші олій

Для розрахунку найкращого співвідношення, використовуючи отриману математичну модель, потрібно натиснути відповідну кнопку, після чого вікно з ним з'явиться на екрані, а також буде збережено для обраної суміші. Це досягається мінімізацією функції двох змінних (10) за допомогою описаного алгоритму випадкового по-шуку. Результати подано на рис.10, при цьому змінні дорівнюють l = 0,579062, m = 0,420938, а мінімум функції дорівнює F = 2,8949. Ці результати добре корелюють з експериментальними даними з табл.6, де мінімум, який знаходиться шляхом вибору найкращого експерименту, знаходиться у точці l = 0,55, m = 0,45 зі значенням функції F = 2,9172. Аналіз показує, що оптимізація співвідношення оливкової та соняшникової олій у суміші з використанням математичної моделі дає краще, більш точне значення інтегральної функції, що дозволяє використовувати розроблені модель і програму при реальному розрахунку параметрів кисломолочної суміші.

| | | | | | | | | CheeseProductModeller − □ × |
|---|---|---|---|---|---|---|---|---|
| Розрахунок суміші | Стійкість емульсії | Відстій фази | | Суміші | | Олії | | |

Моделювання суміші оливкової ▾ та соняшникової ▾ олій ☑

| | Соняшникової | Оливкової | НЖК | МНЖК | ПНЖК | НЖК/ МНЖК | МНЖК/ ПНЖК | НЖК/ ПНЖК | Сумарне відхилення |
|---|---|---|---|---|---|---|---|---|---|
| | 5 | 95 | 1,523 | 5,517 | 0,276 | 0,276 | 19,989 | 5,518 | 16,731 |
| | 10 | 90 | 1,756 | 5,505 | 0,319 | 0,319 | 17,257 | 5,505 | 13,943 |
| | 15 | 85 | 2 | 5,491 | 0,365 | 0,364 | 15,044 | 5,479 | 11,659 |
| | 20 | 80 | 2,263 | 5,426 | 0,413 | 0,417 | 13,138 | 5,479 | 9,7 |
| | 25 | 75 | 2,539 | 5,477 | 0,465 | 0,464 | 11,778 | 5,46 | 8,274 |
| | 30 | 70 | 2,831 | 5,447 | 0,5199 | 0,52 | 10,477 | 5,445 | 6,942 |
| | 35 | 65 | 3,1431 | 5,42990 | 0,5788 | 0,579 | 9,381 | 5,43 | 5,89 |
| | 40 | 60 | 3,4749 | | | × | | 8,429 | 5,412 | 4,983 |
| | 45 | 55 | 3,8291 | | | | | 7,596 | 5,393 | 4,199 |
| | 50 | 50 | 4,2081 | Найкраще співвідношення: 58 на 42 Сумарне відхилення 2.895 | | | | 6,859 | 5,372 | 3,514 |
| | 55 | 45 | 4,6146 | | | | | 6,204 | 5,351 | 2,917 |
| | 60 | 40 | 5,0516 | | | OK | | 5,618 | 5,327 | 3,157 |
| | 65 | 35 | 5,5227 | | | | | 5,09 | 5,302 | 3,754 |
| | 70 | 30 | 6,0322 | 5,2743 | 1,1437 | 1,144 | 4,612 | 5,274 | 4,306 |
| | 75 | 25 | 6,5848 | 5,2446 | 1,2555 | 1,256 | 4,177 | 5,245 | 4,824 |
| | 80 | 20 | 7,1863 | 5,2122 | 1,3787 | 1,379 | 3,781 | 5,212 | 5,31 |
| | 85 | 15 | 7,8434 | 5,1768 | 1,5151 | 1,515 | 3,417 | 5,177 | 5,775 |
| | 90 | 10 | 8,5644 | 5,138 | 1,6669 | 1,667 | 3,082 | 5,138 | 6,223 |
| | 95 | 5 | 9,3588 | 5,0953 | 1,8368 | 1,837 | 2,774 | 5,095 | 6,658 |

+

Розрахунок параметрів математичної моделі
Розрахунок найкращого співвідношення

Рис. 10. Результати розрахунку найкращого співвідношення олій у суміші

Цю ж програму можна використати для розрахунку оптимальних параметрів гомогенізації різного хімічного складу емульсій. Розглянемо спочатку модель залежності стійкості емульсії від температури та тиску. Експериментальні дані наведено в табл. 7, відповідне вікно програми у вкладці «стійкість емульсії» наведено на рис. 11.

Рис. 11. Введення експериментальних даних у вкладці «стійкість емульсії»

Вкладка «стійкість емульсії» відповідає за перегляд, додавання, видалення та редагування записів експериментальних даних й розрахунок на їх основі оптимального режиму гомогенізації для макси-мальної стійкості емульсії отриманої суміші. Для опису процесу гомогонезації у вигляді залежності стійкості емульсії від температурі та тиску будемо використовувати функцію (11) двох змінних S(x,y) як поліном 4-го ступеня, де x — значення тиску, y — значення температури, S — значення стійкості емульсії. Для знаходження коефіцієнтів математичної моделі за методом найменших квадратів шукати будемо мінімум функції (12) за допомогою викладеного раніше методу випадкового пошуку::

$$\sum_{i=1}^{n} \left( S(x_i, y_i) - D(x_i, y_i) \right)^2 \to \min.$$

Результати розрахунків у вигляді коефіцієнтів наведено на рис.12 та у формулі (13).

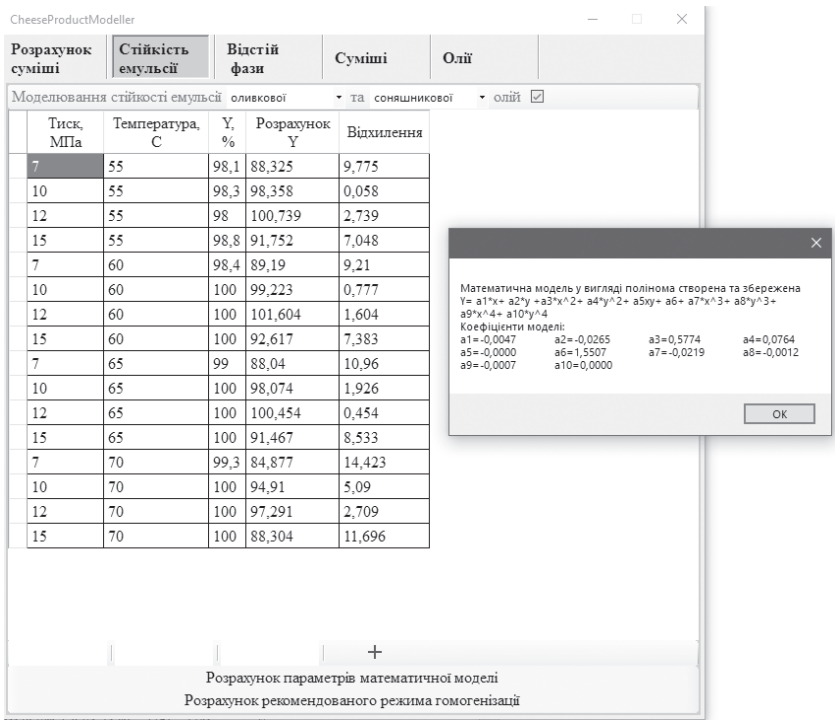

Рис. 12. Результати розрахунку коефіцієнтів математичної моделі стійкості емульсії суміші

Таким чином, отримана математична модель у вигляді

S(x, y) = -0,004666x-0,026530y+0,577439x²+0,076405y²−

−0,000010xy+1,550705−0,021941x³−0,001249y³−0,000656x⁴+0,000005y⁴.

Її можна використовувати при розрахунку режиму гомоненізації емульсій різного хімічного складу при виробництві кисломолочного сиру для знаходження значень стійкості в залежності від температури та тиску.

Розглянемо математичну модель залежності відстою жирової фази також від температури та тиску. Вхідні експериментальні дані наведено у табл.9. та на рис.13. Вкладка «відстой фази» використовує ті ж прийоми для заповнення даними, що і раніше описані процедури, тому на рис.12 наведено лише результат розрахунків коефіцієнтів моделі. Після знаходження коефіцієнтів шуканої функції (14) математична модель набуде вигляду

G(x,y) = -0,004837x-0,027562y+0,005950x²+0,034753y²−

−0,000011xy+1,599765−0,008708x³−0,000901y³+0,000438x⁴+0,000006y⁴.

Цю модель можна використовувати при розрахунках та оптимізації залежності відстою жирової фази також від температури та тиску.

Однак у більшій мірі використовують дві отримані моделі разом, для розрахунків режимів гомогенізації, оскільки на сьогоднішній день в молочній промисловості гомогенізація є єдиним способом утворення стійкої емульсії, в тому числі і з рослинними оліями. При цьому стійкість емульсії повинна бути максимальною, відстій жирової фази — мінімальним. Якщо використовувати дві останні отримані математичні моделі для оптимізації процесу гомогенізації, отримаємо рішення, яке представлене на рис. 12. Оптимальні параметри процесу гомогенізації, моделі виглядають таким чином: тиск дорівнює 12 Мпа, температура 60°С, при цьому стійкість емульсії дорівнює 100 %, відстій жирової фази 3,167 %. Ці дані добре корелюють з даними, отриманими іншим експериментальним шляхом в табл. 3, 4.

**Висновки.** У роботі представлений підхід до побудови математичних моделей отримання кисломолочних продуктів для покращення розрахунків параметрів одержуваної суміші з рослинними оліями.

В ході виконання досліджень були вирішені такі завдання: проаналізовано результати натурного фізичного експерименту і вибрано клас математичної моделі; оброблено експериментальні дані

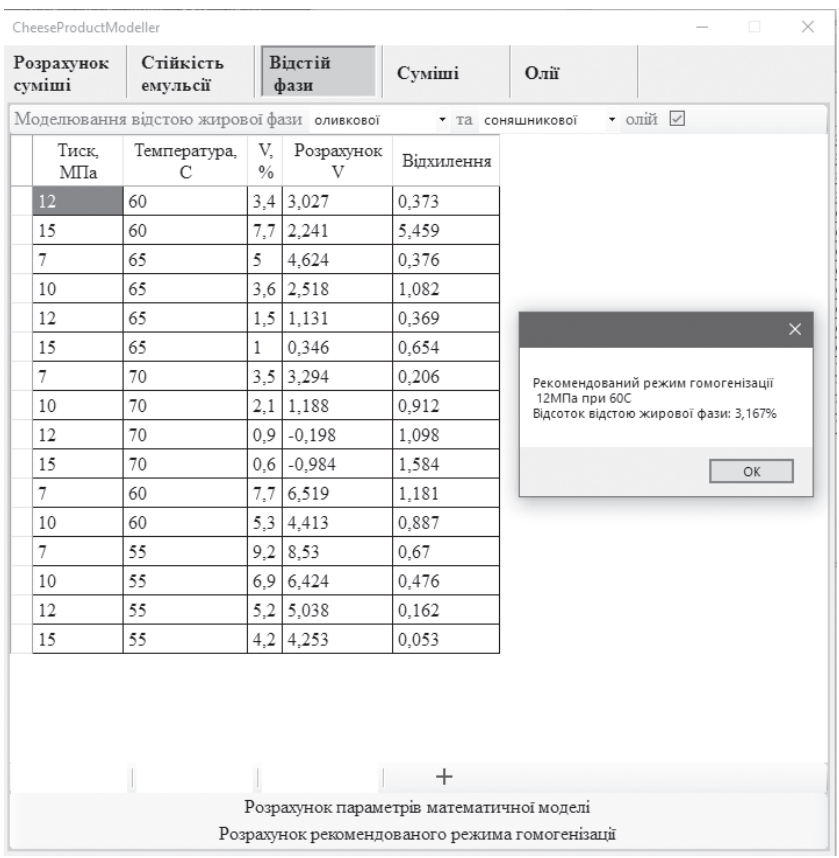

Рис. 13. Результати розрахунку оптимальних параметрів і процесу гомогенізації за допомогою методів регресійно-кореляційно аналізу в розробленій програмі, знайдено числові коефіцієнти математичних моделей; проаналізовані отримані коефіцієнти на предмет адекватності моделі вхідним даним; проаналізовано переваги та недоліки застосовуваних алгоритмічних мов та обрано для програмування мову C#; побудовано комп'ютерну програму для розрахунку співвідношення вихідних інгредієнтів для отримання заданих властивостей кисломолочного продукту; в якості алгоритму оптимізації запропонована модифікація методу випадкового пошуку Монте-Карло, який, хоча і вимагає значних часових комп'ютерних ресурсів, проте є досить універсальним і стійким до вхідних даних; в якості керуючого параметра алгоритму використовується кількість невдалих спроб генерації чисел, тому в процесі декількох експериментальних прогонів програми знайдено оптимальне значення кількості невдалих спроб; розроблено практичні рекомендації користувачеві для використання програми при розробці рецептури нового типу кисломолочного продукту.

Практичні результати дослідження полягають у тому, що розроблена програма дає в руки користувача-технолога інструмент, яким він може користуватися для розрахунку рецептури нових сортів кисломолочних продуктів з додаванням рослинних олій, не проводячи фізичних експериментів, досліджуючи властивості продукту на комп'ютері на підставі розроблених математичних моделей.

## Розділ V

# КОМП'ЮТЕРНІ ТЕЛЕКОМУНІКАЦІЙНІ МЕРЕЖІ ТА ТЕХНОЛОГІЇ

## МЕТОДИКА РІВНОМІРНОГО РОЗПОДІЛУ ЗАВДАНЬ МІЖ ОБЧИСЛЮВАЛЬНИМИ КОМПЛЕКСАМИ

*Завертайло К. С.*

*В роботі розглядається проблема балансування навантаження і рівномірний розподіл завдань між обчислювальними комплексами. Ця проблема розглядається в загальному вигляді, щоб запропоновані методи могли бути базовими для розробки програмних засобів для всіх сфер інформатики, де виникає проблема нерівномірного навантаження і розподілу завдань. Описані основні причини виникнення нерівномірного навантаження. Зроблені висновки і запропоновано методи усунення вказаної проблеми.*

*The paper considers the problem of load balancing and uniform distribution of tasks between computer systems. This problem is considered in general, the reason is that the proposed methods could be the basis for the development of software for all areas of computer science, where there is a problem of uneven workload and distribution of tasks. The article will describe the main causes of uneven load. Conclusions are made and methods of elimination of the specified problem are offered.*

Проблема нерівномірного розподілу завдань виникає там, де застосовують обчислювальні комплекси. У випадках, коли в обчислювальних елементів різна продуктивність роботи, ризик виникнення нерівномірного навантаження між ними підвищується. Оскільки при різних можливостях обчислювальних елементів значно складніше планувати розподіл завдань і навантаження між ними. Також причиною неправильного балансу може бути висока завантаженість системи в цілому, тому що в таких випадках потрібно розраховувати не тільки теперішнє навантаження між компонентами, а й їх продуктивність. В такій складній ситуації, коли інтенсивність надходження завдань на обробку до комплексу зростає, ризик виникнення дисбалансного навантаження також зростає.

Також до причин нерівномірного навантаження можна віднести різний обсяг завдань, що надходять до обчислювальних комплексів.

Зрозуміло, що в такій ситуації простим кількісним шляхом розподілу завдань між обчислювальними елементами досить складно рівномірно їх розподілити.

Основним негативним наслідком від некоректного балансування навантаження в комплексних системах є те, що одні обчислювальні елементи використовуються всією системою не на повні свої можливості. Тому таким елементам, можна сказати, притаманно часткове простоювання впусту. Також при нерівномірному балансуванні навантаження практично завжди є елементи, які отримують занадто багато завдань на відміну від інших. Це сильно впливає на продуктивність всієї системи, особливо це відчутно, коли інтенсивність надходження задач значно зростає.

Також з попередньо описаного негативного наслідку від нерівномірного навантаження випливає ще один. Він полягає в тому, що значне навантаження на певні елементи системи підвищує ризик відмови в роботі обчислювального елемента. Ще можна додати, що надмірне навантаження знижує зносостійкість апаратної частини обчислювальних компонентів.

Звісно, що неправильне балансування навантаження негативно впливає на пропускну спроможність в мережах і час відгуку. Коли запити розподіляються між серверами не рівномірно, значна частина запитів простоює, очікуючи своєї черги, а це знижує ефективність роботи системи серверів в цілому.

З наведених вище негативних наслідків можна зробити висновок, що балансування навантаження є дуже важливим елементом в тих інформаційних галузях, де застосовуються комплекси обчислювальних систем. Тим більше застосування ефективного балансувальника навантаження важливо, якщо продуктивність обчислювальних елементів різна, що на практиці виникає досить часто.

**Теоретичні відомості.** Питання планування навантаження потрібно вирішувати на ранній стадії розвитку будь-якого проекта. Вихід з ладу системи загрожує серйозними наслідками — як моральними, так і матеріальними. Спочатку проблеми недостатньої продуктивності системи у зв'язку зі зростанням навантажень можна вирішувати шляхом нарощування потужності або оптимізацією використовуваних алгоритмів, програмних кодів і тому подібне. Але рано чи пізно настає момент, коли ці заходи виявляються недостатніми [1; 7].

Доводиться вдаватися до кластеризації, наприклад, кілька серверів поєднуються в кластер; навантаження між ними розподіляєть-ся за допомогою комплексу спеціальних методів, які називаються балансуванням. Крім вирішення проблеми високих навантажень, кластеризація допомагає також забезпечити взаємне резервування серверів. Ефективність кластеризації безпосередньо залежить від того, як розподіляється навантаження між елементами кластера. Загальна схема роботи балансувальника навантаження наведена на рисунку 1.

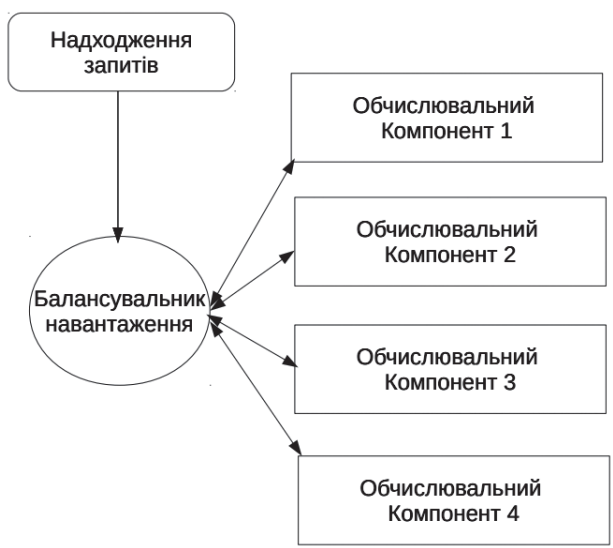

Рис. 1. Узагальнена схема роботи балансувальника навантаження

В подальшому буде проведено короткий огляд теорії балансування навантаження у двох важливих в сфері комп'ютерних наук напрямках:

– Балансування навантаження для багатопроцесорних комп'ютерів;

– Балансування навантаження мережі.

**Балансування навантаження в багатопроцесорних системах.** Комунікаційний дисбаланс навантаження обумовлений різністю в продуктивності комунікаційних зв'язків між багатоядерними процесорами або іншого кластера в групі багатоядерних процесорів. З іншого боку, дисбаланс в обмінах може бути пов'язаний і з паралельним додатком і визначатися алгоритмом розв'язання задачі. Важливим прикладом для балансування є гетерогенні обчислювальні системи з розділеною пам'яттю. Тут необхідні ресурси можна визначити лише під час роботи програми, і балансування має проводитися динамічно, можливо між різними операційними системами [2; 12].

Потрібно виділити кілька рівнів і дати характеристику алгоритмів, що їх реалізують, для досягнення балансування навантаження процесорів для окремого паралельного застосування.

Балансування лише на рівні операційної системи, механізми — кластеризація, поділ навантаження та міграція процесів на етапі виконання; балансування на рівні проміжного програмного забезпечення, механізм — високорівневе балансування навантаження в контексті сесії або запиту, розподіл команд у стадії трансляції; балансування на рівні користувацького додатку, реалізовується механізм балансування за допомогою прикладних паралельних програм. Існуючі операційні системи покладаються на однозначний статичний розподіл завдань користувачем, який може призводити до значного розбалансування процесорів. Кластеризація допоможе подолати труднощі, що виникають. Технологія кластеризації має дві основні переваги. Вона підвищує масштабованість та обчислювальну потужність. Більшість кластерних рішень забезпечують збалансованість навантаження та автоматичне перемикання з одного вузла на інший у разі перевантаження або відмови. Одним з таких кластерних рішень для Linux є система MOSIX.

OpenMOSIX — системне програмне забезпечення для ядер, таких як Linux, що складається з адаптивних алгоритмів розподілу ресурсів. Алгоритми розділу ресурсів OpenMOSIX розроблені відповідно до використання ресурсів вузлів у режимі реального часу.

Усі ці алгоритми реалізуються завдяки механізму міграції процесів. З кожним процесом асоціюється ідентифікатор унікального домашнього вузла, з якого він був запущений. Для міграції процес розбивається на дві частини: контекст користувача і системний контекст. Частина користувача складається з коду програми, даних, стека, карт пам'яті та регістрів процесу. Системна частина включає опис ресурсів, що належать даному процесу, та визначає машиннозалежну частину, яка завжди залишається на унікальному домашньому вузлі процесу. Мігруючий процес використовує ресурси нового вузла, наскільки це можливо, але взаємодіє з ОС через домашній вузол.

Міграція базується на інформації, що забезпечується одним із алгоритмів поділу ресурсів. Стратегія призначення завдань, заснована також на економічних засадах та конкурентному аналізі. Ця стратегія дає можливість керувати гетерогенними ресурсами способом, близьким до оптимального.

Прикладне програмне забезпечення, інтерфейс для передачі обміну повідомленнями забезпечують вихідне фіксоване розміщення процесів по вузлах кластера, тоді як openMOSIX виконує це динамічно залежно від конфігурації доступних ресурсів. Прикладне програмне забезпечення й інтерфейс для передачі повідомлень працюють на рівні користувача, на якому діють звичайні програми. Рішення openMOSIX функціонує як модуль ядра операційної системи і є повністю прозорим для додатків. Немає необхідності модифікувати програми під openMOSIX або пов'язувати їх з бібліотеками. Операційна система openMOSIX — це, з одного боку, альтернатива технології інтерфейсу передачі повідомлень, а з іншого боку, їхній розвиток. Тут необхідно зазначити, що openMOSIX і інтерфейс передачі повідомлень можуть працювати одночасно на одному і тому ж кластері.

Користувачі та програми можуть безпосередньо взаємодіяти з openMOSIX через API інтерфейс, який забезпечує інформацію про стан локальних процесорів та процесів. Очевидним недоліком openMOSIX є великі накладні витрати при виконанні системних викликів. Додаткові витрати виникають під час операцій мережного доступу. Наприклад, всі сокети створюються в ідентифікаторі домашнього вузла, це призводить до великих комунікаційних витрат, якщо процес мігрує з ідентифікатора домашнього вузла.

Загальні недоліки такого балансування у гнучкості та адаптивності. Перші виникають через неможливість у режимі виконання приймати рішення про балансування навантаження у додатку; другі — через відсутність зворотного зв'язку з репліками, об'єктами, які працюють на стороні сервера балансування, і під його керуванням всі разом представляють процеси, що піддаються балансуванню. Ще один недолік — стандартне проміжне програмне забезпечення обмежено використовує ці можливості операційної системи, вони реалізовані по-різному в різних операційних системах.

Основне проміжне програмне забезпечення для розподілу процесів у паралельних системах передбачає середовище виконання, що потребує адаптованих додатків та поінформованості користувача про це. Воно включає утиліти для ініціалізації прив'язки процесу до вузла, ігноруючи доступні ресурси, наприклад, вільна пам'ять і процеси введення-виводу. Це прикладне програмне забезпечення запускаєть-ся на рівні користувача, як звичайна програма, таким чином, стає нездатним реагувати на нестійке завантаження і адаптивно перерозподіляти обчислювальне навантаження.

Балансування навантаження на проміжному рівні виявляється кращим, ніж балансування на нижчих рівнях мережі або операційної системи, які відрізняються відсутністю гнучкості та адаптованості. Проміжне програмне забезпечення може забезпечити багатий набір метрик для балансування, у тому числі користувацьких, залежних від додатків (гнучкість); тоді як мережеві або операційної системи балансувальники працюють лише з фіксованими описами навантаження. Проміжне програмне забезпечення може бути використане спільно як зі стандартними, так і зі спеціалізованими мережами, операційними системами, а також із системами балансування навантаження, тоді як низькорівневі балансувальники тісно пов'язані з апаратно-програмним середовищем, для якого вони призначені [3; 10].

Механізми балансування навантаження можуть реалізуватися в прикладному програмному забезпеченні, наприклад, MIST, Dynamite, mpC. У системах обміну повідомленнями процеси, що містять код усієї прикладної програми, запускаються на різних обчислювальних вузлах відповідно до заданої конфігурації. При розбалансуваннях, що виникають під час виконання, їх необхідно перервати, перемістити між вузлами та знову запустити в колишньому контексті. У системах обміну повідомленнями балансування навантаження формулюється через ефективний перерозподіл між обчислювальним вузлами, при цьому міграція процесів — це основний механізм балансування.

Також механізми балансування навантаження можуть реалізовуватися шляхом планування ресурсів. Розподілені системи спочатку складаються з окремих модулів, які, взаємодіючи один з одним, призводять до розбалансування системи. Тому необхідно ефективно пов'язувати модулі під час роботи розподіленої системи для вирівнювання навантаження. У розподілених системах балансування може бути описано за допомогою структур даних (або об'єктно-орієнтованих інтерфейсів), які дозволяють побудувати відповідність між постачальниками та споживачами ресурсів.

Проміжне програмне забезпечення динамічного балансування паралельних і розподілених багатопроцесорних обчислювальних систем має забезпечувати оптимальне розподілення паралельних додатків при динамічно змінних ресурсах, за рахунок міграції таких процесів між обчислювальними вузлами багатопроцесорної системи.

До складу проміжного програмного забезпечення для динамічного балансування навантаження входять такі компоненти: монітор навантаження, планування процесу виконання, міграції завдань та ін. Менеджер балансування навантаження розраховує балансування навантаження. Якщо система розбалансована, починає працювати менеджер міграції, який визначає нові місця виконання процесорів та міграції. Окремий сервіс створює копію процесу, включаючи в нього інформацію про встановлені комунікаційні з'єднання і відкриті файли. Для переміщення процесу в новий вузол або процесор дані передаються міграторові завдань. Процес за допомогою системи перезапуску завдань відновлює вихідний стан процесу на новому вузлі, і паралельна програма продовжує виконуватись.

Прикладне програмне забезпечення обміну повідомленнями складається з сервісу та бібліотеки, яка включається до коду прикладної програми. Шляхом розширення, тобто модифікації вихідного коду можна реалізувати додаткову функцію міграції процесів.

Основна ідея полягає в такому:

— Мігруючий процес припиняється. Приймаючий сервіс визначає відповідний цьому процесу виконуваний файл;

— Стан мігруючого процесу, завантажений у пам'ять код і дані, стек, відкриті канали (файли, мережеві з'єднання тощо), передається на сторону, що приймає;

— Усі накопичені за перші два кроки повідомлення, адресовані мігруючому процесу, також передаються на сторону, що приймає;

— Вихідний екземпляр мігруючого процесу повністю зупиняється, новий — запускається з того місця (стану), на якому було призупинено вихідний.

Для реалізації міграції процесів потрібна наявність відповідних можливостей (API — прикладний програмний інтерфейс) операційних систем, на яких використовуватиметься розширене прикладне програмне забезпечення. API має включати функції призупинення процесів, визначення адресного простору процесів, запуск процесів у контексті і т. д. Міграція процесів може бути реалізована в однорідній обчислювальній мережі, оскільки у процедурі міграції задіяні такі параметри як адресний простір процесу (переміщення процесу між вузлами з різними розмірами оперативної пам'яті реалізувати важко).

Рішення, мігрувати завданню чи ні, залежить від низки параметрів, які мають бути оцінені керуючою програмою переміщення (демон UNIX). У разі паралельного завдання, що виконується на клас-тері, програма, що управляє, бере до уваги: навантаження кожного вузла; середнє навантаження у МВС, продуктивність мережі; час, необхідний для запровадження контрольних точок, переміщення і перезапуску завдання; прогнозоване подальше навантаження процесора.

**Балансування навантаження мережі.** Для будь-якої хмарної мікросервісної архітектури однією з основних характеристик є масштабованість. У сучасному світі будь-який сервіс, призначений для надання послуг широкому числу споживачів, повинен відрізнятися гарною масштабованістю і швидкістю відповіді. Потенційні користувачі ні за що не захочуть терпіти постійні помилки, втрати запитів, відмову в обслуговуванні та інші проблеми, пов'язані з нестачею продуктивності сервісу. Вони просто звернуться до конкурентів, навіть якщо їх пропозиція буде менш вигідною, але працюватиме стабільно.

Найочевиднішим вирішенням проблеми може бути збільшення потужності. Чим потужнішим буде центральний сервер і чим більше у нього буде ресурсів, тим менша ймовірність виходу з ладу сервісу або недоступність послуг. Так, це буде працювати до певного часу. Доки серверу не буде потрібне обслуговування [4; 11].

Другим рішенням буде зробити кілька серверів, які зможуть підтримувати працездатність сервісу надання послуг. Але при цьому відразу виникає кілька проблем:

— Яким чином клієнтська програма зможе визначити сервер, який обслуговуватиме клієнта в даний момент часу;

— Як саме визначити доступність сервера та доступність сервісу в цілому;

— Якщо один із серверів вже перебуває під навантаженням численних запитів і не має змоги дозволити собі прийняти додаткове навантаження, то як визначити, до якого доступного сервера необхідно звернутися, щоб не створити другий вкрай завантажений сервер.

Вирішенням цієї проблеми є балансування навантаження. Це сукупність методів розподілу задач між пристроями в мережі, з метою оптимізації використовуваних як апаратних та обчислювальних, так і мережевих ресурсів, скорочення часу обслуговування запитів та збільшення максимального обсягу завдань, які може виконати система.

Для вирішення цього завдання в комп'ютерній мережі необхідна розробка додаткового програмного забезпечення, яке взяло б на себе функцію балансувальника. Балансувальник може бути не обов'язково програмним засобом, а й окремим мережевим пристроєм.

Введення ще одного елемента в систему допоможе уникнути помилок на стороні клієнта в системі, і зробити його роботу більш легкою і продуктивною. Клієнт може безпосередньо звертатися до балансувальника через заздалегідь визначений механізм, згідно з алгоритмами та стратегією, закладеними в балансувальник.

Балансувальник зможе у фоновому режимі проводити моніторинг доступності серверів і маршрутів до них, щоб у момент запиту від клієнта маршрутизувати запити навколо перевантаженого елемента мережі найкоротшим шляхом.

І найважливіша функція балансувальника — це усунення неоднорідності в мережі, тому що часто створити великий сервіс в одній зоні або регіоні недостатньо, і тоді на плечі балансувальника лягає подолання подібних дрібних системних нестиковок, які для стандартної клієнт-серверної моделі суттєво ускладнили б завдання.

Процедура балансування здійснюється за допомогою комплексу алгоритмів та методів, що відповідають рівням моделі OSI: мережевому, транспортному і прикладному.

Балансування навантаження на мережевому рівні передбачає вирішення такого завдання: потрібно зробити так, щоб за одну конкретну IP-адресу сервера відповідали різні фізичні машини. Таке балансування може здійснюватися за допомогою багатьох різноманітних способів:

— DNS-балансування. На одне доменне ім'я виділяється кілька IP-адрес. Сервер, на який буде направлений запит клієнта, зазвичай визначається за допомогою алгоритму Round Robin.

— Побудова класу NLB. При використанні цього способу сервери поєднуються в кластер, що складається з вхідних та обчислювальних вузлів.

— Розподіл навантаження здійснюється за допомогою спеціального алгоритму. Використовується в рішеннях компанії Microsoft.

— Балансування IP за допомогою додаткового маршрутизатора.

— Балансування за територіальною ознакою здійснюється шляхом розміщення однакових сервісів з однаковими адресами у територіально різних регіонах Інтернету.

Балансування навантаження на транспортному рівні. Цей вид балансування є найпростішим: клієнт звертається до балансувальника, той у свою чергу перенаправляє запит на один з серверів, який його оброблятиме. Вибір сервера, на якому оброблятиметься запит, може здійснюватися відповідно до найрізноманітніших алгоритмів: шляхом простого кругового перебору, шляхом вибору найменш завантаженого сервера з пулу. Іноді балансування на транспортному рівні важко відрізнити від балансування на мережному рівні [5; 9].

Відмінність між рівнями балансування можна пояснити так. До мережного рівня відносяться рішення, які не термінують на собі сесії користувача. Вони просто перенаправляють трафік і не працюють у режимі проксі.

На мережному рівні балансувальник просто вирішує, на який сервер передавати пакети. Сесію з клієнтом здійснює сервер.

На транспортному рівні спілкування з клієнтом замикається на балансувальник, який працює як проксі-сервер. Він взаємодіє із серверами від свого імені, передаючи інформацію про клієнта у додаткових даних та заголовках. Таким чином працює, наприклад, популярний програмний балансувальник HAProxy

Балансування навантаження на прикладному рівні балансувальник працює в режимі «розумного проксі». Він аналізує клієнтські запити і перенаправляє їх у різні сервери залежно від характеру запитуваного контенту.

Так працює, наприклад, веб-сервер Nginx, розподіляючи запити між фронтендом та бекендом. За балансування у Nginx відповідає модуль Upstream.

Як ще один приклад інструменту балансування на прикладному рівні можна навести pgpool — проміжний шар між клієнтом та сервером СУБД PostgreSQL. З його допомогою можна розподіляти запити по серверах баз даних в залежності від їх змісту. Наприклад, запити на читання будуть передаватися на один сервер, а запити на запис — на інший.

**Класифікація стратегій балансування навантаження.** Для розробки методу балансування навантаження необхідно розглянути існуючі методи, алгоритми та стратегії, що застосовуються в цій галузі. Буде розглянуто основні класи стратегій балансування навантаження. Причиною того, чому будуть розглянуті саме класифікації підходів балансування навантаження є те, що саме виокремлені класифікації та їх ознаки наглядно демонструють теоретичну основу і розуміння самих підходів рівномірного розподілу завдань [6].

Перший клас балансування навантаження — коли принципи балансування навантаження розподіляються на статичні, напівдинамічні, динамічні. При використані статичної стратегії балансування навантаження план розподілу задач між обчислювальними комплексами відомий заздалегідь, оскільки розподіл навантаження виконується заздалегідь. Статична стратегія більш проста в реалізації, але менш ефективна, ніж динамічна чи напівдинамічна.

При використані напівдиноміної стратегії розподілу навантаження план визначається на етапі ініціалізації, до того як починаються основні обчислення. Тобто попередньо оцінюються всі ресурси, що є в наявності. На відміну від статистичної стратегії напівдинамічна є більш ефективною в роботі, оскільки оцінка на початку ситуації в системі дозволяє краще в подальшому балансувати навантаження між обчислювальними елементами. Статистична стратегія балансування є більш простою в реалізації.

Найбільш ефективною в роботі є динамічна стратегія розподілу навантаження. Вона змінюється протягом роботи всього обчислювального комплексу. Ця стратегія змінюються під впливом певних факторів і регулярно оцінює стан середовища, у якому працює. Зазвичай динамічна стратегія базується на тому, що проводить оцінку стану середовища через певний встановлений інтервал часу. Хоча також можна не фіксувати цей інтервал чітко, а він буде визначатись після кожної оцінки середовища, в залежності від того, в якому стані буде це середовище. Також динамічна стратегія може змінюватися не тільки між інтервалами. В складних системах, де є підвищений ризик виходу з-під контролю нормального балансу навантаження, динамічна система може в будь-який момент отримувати сигнали, коли навантаження різко змінюється. Ці сигнали можуть свідчити про те, що необхідно виконувати певні зміни до закінчення встановленого інтервалу. Описана стратегія є найбільш складною в реалізації, проте вона є найбільш ефективною на практиці. Слід зауважити, що використання динамічної стратегії є доцільним лише в складних обчислювальних системах. Тому що в незначних за складністю обчислювальних комплексах з завданням балансування навантаження може впоратись і напівдинамічна, і в навіть в зовсім простих статична. Тому дуже важливо розуміти доцільність використання динамічної стратегії балансування навантаження.

Наступним класом поділу стратегій балансування є принцип поділу на залежний і незалежний розподіл. Залежним розподілом завдань, між обчислювальними елементами можна вважати розподіл, який змінює принцип розподілу завдань між вузлами в залежності від певних подій, що виникають при роботі цілої комплексної систе-ми. Конкретно умови, при яких відбуваються ці зміни в плануванні, визначаються заздалегідь, і вони залежать конкретно від системи, під яку працює планувальник. Зазвичай цими параметрами є підвищення чи зменшення навантаження, як і в усій системі, так і на одному, конкретному обчислювальному елементі. Також до цих параметрів можна віднести зменшення під час роботи продуктивності одного з обчислювальних елементів. Враховуючи те, що в залежності від певних факторів змінюється стратегія планування, то залежний розподіл планування співвідноситься з динамічною стратегією планування.

При розгляді незалежного принципу балансування навантаження мається на увазі те, що немає ніякого впливу якихось факторів на роботу балансувальника навантаження чи ці впливи зведені до мінімуму. Зазвичай незалежний можна порівняти і зі статичним чи напівдинамічним підходом, оскільки робота балансувальника навантаження планується заздалегідь, перед роботою самої системи. Тому зазвичай незалежний розподіл задач між обчислювальними елементами застосовується до простих систем.

Також необхідно виокремити класифікацію, що поділяє способи балансування навантаження на спеціалізовані та універсальні. В принципі з самої назви можна визначити, що спеціалізована стратегія має на увазі розробку алгоритму під конкретну топологію мережевого середовища чи конкретну архітектуру розподіленої системи. Також зрозуміло, що ефективність від спеціалізованих підходів розробки балансувальника навантаження вища від універсальних. До недоліків можна віднести те, що ціна розробки таких балансувальників вища і вузька направленість не дає змоги застосовувати до багатьох напрямів, де потрібне балансування навантаження.

Про універсальні можна сказати, що, як видно з їх назви, вони є багатонаправленими. І фактично це і є їх основною перевагою над спеціалізованими. Зазвичай універсальними є методи, які є базою для алгоритмів, а спеціалізовані підходи — це вже безпосередньо алгоритми, що розроблені для конкретних архітектур.

Наступна класифікація полягає у врахуванні і неврахуванні причин зміни балансу навантаження в системі. Необхідно визначити при розробці методу і алгоритму, чи потрібно враховувати причини зміни балансу навантаження. Розглядаючи ті алгоритми, що не враховують причини переміни балансу розподілу задач в системі, можна зазначи-ти, що перевагою цього є те, що це спрощує розробку балансувальника навантаження. Недоліки цього підходу очевидні, вони полягають в тому, що навіть ефективний алгоритм балансування навантаження, який не враховує причини перепаду навантаження, будуть поступатися алгоритмам, які враховують принципи зміни навантаження, та як саме на майбутнє змінювати стратегію роботи відносно обчислювального комплексу.

На відміну від вищеописаного ті методи, що враховують причину зміни навантаження в майбутній роботі системи, визначають більш точно, яким чином необхідно перерозподілити саме завдання між обчислювальними елементами. З переваги оберненого до цього підходу випливає і обернений недолік. Визначається він тим, що розробка способів балансування навантаження між обчислювальним елементами є більш складною і затратною. Саме такі обернені один до одного переваги і недоліки ставлять основні питання з приводу того, який саме підхід необхідно застосувати.

Системи, де враховуються причини перепаду балансу між компонентами системи, застосовують у складних системах, де просто необхідна ефективність і точність виконання рівномірного розподілу між вузлами. Підходи, що не враховують виникнення причин зміни балансу в системах, застосовують до простих у своїй структурі системах.

Потрібно відмітити те, що клас балансувальників навантаження, що враховують причини зміни балансу, можна поділити на два підкласи. Першими є фактори, що впливають на баланс навантаження на систему ззовні. Другими є ті, що впливають на баланс навантаження всередині самої системи. До прикладу, збільшення запитів до комплексу серверів у час пік — є зовнішніми факторами. А, наприклад, зменшення продуктивності з причини поганої зносостійкості, одного з обчислювальних елементів є фактором, що впливає на систему зсередини.

Стратегії балансування навантаження також можна поділити на прогностичні, тобто ті, що моделюють навантаження на обчислювальні компоненти в системі в майбутньому, і на ті, що не здатні прогнозувати поведінку навантаження в подальшій роботі обчислювального комплексу. Зрозуміло, що прогностичні підходи є більш ефективними. Оскільки завдяки коректному прогнозу і розподілу навантаження в подальшій роботі системи продуктивність її роботи значно підвищується. Завдяки прогностичним підходам підвищу-ється робота не частини певних факторів, а всіх разом. Це дозволяє підвищити як пропускну спроможність, так і швидкість обробки обчислювальними елементами поставлених перед ними завдань. Недоліком врахування прогностичності в підходах балансування навантаження є значні витрати на розробку безпосередньо способу прогнозу роботи системи в майбутньому. Саме з цієї причини такий спосіб застосовують лише з урахуванням того, що розроблені алгоритми балансування будуть застосовуватися до тих систем, які є значними за своїм обсягом, дуже чутливі до точності визначення розподілу задач між вузлами і потребують значної ефективності у виконанні поставлених перед ними завдань.

При огляді тих систем балансування навантаження, що не здатні прогнозувати подальшу поведінку навантаження в обчислювальному комплексі, слід відразу відмітити, що такі системи значно поступаються у своїй ефективності прогностичним. Проте все ж потрібно в черговий раз зауважити, що існують обчислювальні комплекси, які, хоч і можуть бути значними за своїм обсягом, але не є складними за своєю суттю і не такі вибагливі щодо високої точності розподілу завдань між обчислювальним елементами. Саме для таких систем дуже часто застосування не прогностичних стратегій балансування є більш доцільним, аніж прогностичних.

Однією з найбільш важливих класифікацій поділу способів балансування є розподіл на централізовані і децентралізовані. Такий поділ підходів до стратегій балансування навантаження є дуже важливим при розробці методу балансування навантаження, тому його потрібно розглянути дуже детально. Для початку розглянемо централізований підхід до розробки методу балансування навантаження. Цей спосіб пропонує єдиний балансувальник навантаження для всіх обчислювальних елементів у системі. Саме балансувальник повинен оцінювати інтенсивність надходження завдань на обробку, оцінювати можливості і ступінь завантаженості в конкретний момент часу обчислювальних елементів, аналізувати ситуацію в системі в цілому і проводити розподіл завдань. Перевага такого способу очевидна: єдиний балансувальник навантаження значно краще координує роботу між обчислювальним елементами, швидше оцінює стан системи в цілому і може вплинути на неї при її подальшій роботі чи подати сигнал щодо її критичного стану при необхідності. Звісно такий підхід має свої недоліки. Одним з таких недоліків є те, що при відмові в роботі балансувальника навантаження чи початку його некоректної роботи балансування навантаження у всій системі починає проводитися некоректно, що швидко призведе до значних негативних наслідків. Також потрібно відмітити, що централізовані балансувальники навантаження зазвичай більш негативно впливають на масштабованість, ніж децентралізовані.

При розгляді децентралізованих слід відмітити, що в кожного обчислювального компонента є свій балансувальник навантаження, завданням якого є оцінка навантаження свого компонента, визначення інтенсивності надходження завдань чи запитів до його компонента і обчислення для визначення подальшої продуктивності роботи обчислювального елемента, за який він відповідає. Всі балансувальники навантаження обмінюються між собою інформацією з приводу навантаження на своєму обчислювальному вузлі. В підсумку, кожен балансувальник навантаження визначає ступінь навантаження його елемента і робить висновок з приводу того, чи його обчислювальний елемент є занадто навантаженим, чи його обчислювальний елемент використовується не на повну потужність. У всій системі всі балансувальники навантаження підтримують між собою зв'язок, обмінюються інформацією. Завдяки цьому обміну даними всі балансувальники навантаження містять інформацію про завантаженість кожного балансувальника навантаження в системі. Тому, коли є необхідність, менш завантажений балансувальник може прийняти частину запитів з черги більш завантаженого. Перевага децентралізованого балансування навантаження перед централізованим походить від недоліку централізованого підходу. Ця перевага полягає в тому, що при відмові в роботі чи початку некоректного планування виходить з ладу лише один балансувальник навантаження зі всієї системи, на відміну від того, що в централізованих алгоритмах це відразу негативно впливає на баланс навантаження всієї системи, оскільки там балансувальник один. Недоліками децентралізованого підходу балансування навантаження є те, що, на відміну від централізованого, така стратегія балансування не може загалом оцінювати ситуацію в усій системі, оскільки в кожного обчислювального компонента свій балансувальник. Тому при критичній ситуації, яка впливає на систему і потрібно приймати рішення, яке повинно глобально впливати на весь обчислювальний комплекс, централізований спосіб балансування явно переважає децентралізований. На рисунку 2 зображено схему з п'яти обчислювальних компонентів, що обмінюються даними.

Причиною того, чому саме на класифікацію поділу на централі-
зовані і децентралізовані стратегії балансування було звернено ува-
гу, полягає в тому, що вибір одного з цих підходів часто змінює весь
принцип підходу в моделюванні і створені методу розподілу завдань
між обчислювальним елементами. Звертаючи увагу на те, для чого
саме будуть застосовуватися розроблені методи, потрібно вибирати
на початковому етапі одну з цих стратегій: централізовану чи децен-
тралізовану.

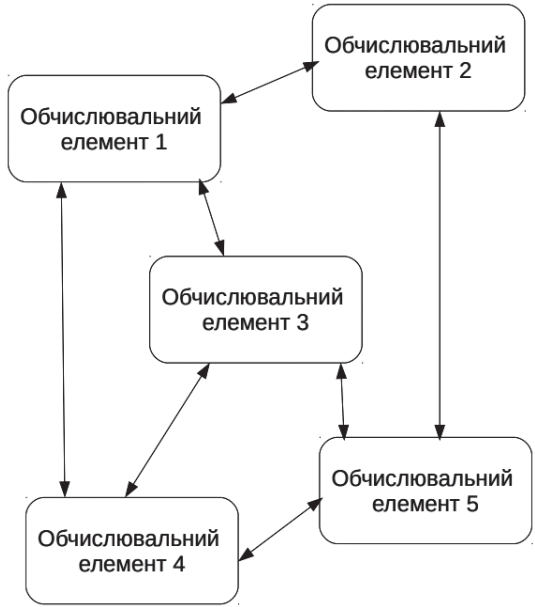

Рис. 2. Децентралізоване балансування навантаження

Останнім поділом на класи стратегій балансування є розподіл
на адаптивні і неадаптивні способи балансування навантаження.
Адаптивна стратегія балансування навантаження передбачає при-
стосування до кардинальних змін в системі. Цими змінами можуть
бути, наприклад, додавання нових обчислювальних вузлів у систему
чи виключення з неї існуючих. Звісно, тим, що така стратегія може
пристосовуватися до кардинальних змін, вона має перевагу над не-
адаптивним підходом. Проте недоліком такого способу є те, що при
її розробці виникає додаткова робота і необхідно враховувати, чи взагалі потрібно буде її реалізовувати. Адже, якщо в системі взагалі не передбачається ні включення, ні виключення нових обчислювальних елементів, не буде змінюватися ресурсна конфігурація розподіленої системи й інші значні зміни, то застосування такого підходу просто не має сенсу. Адаптивна стратегія балансування навантаження дуже сильно подібно до динамічної, її навіть можна вважати динамічною, тільки в більш вузькому розумінні цього визначення. На рисунку 3 зображено централізований адаптивний балансувальник навантаження, до якого додаються ще два обчислювальних елемента, вони відмічені штрихом. На рисунку 4 зображено децентралізований адаптивний балансувальник навантаження, до трьох взаємодіючих між собою елементів додаються ще два, які також відмічені штрихом.

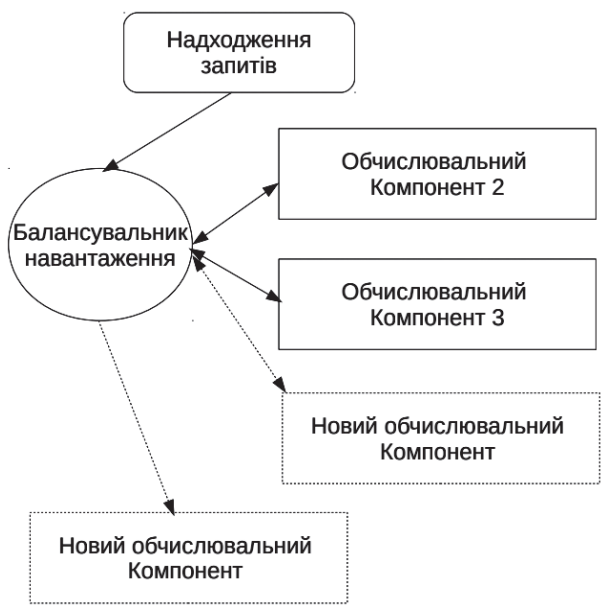

Рис. 3. Централізований адаптивний балансувальник навантаження

Неадаптивна стратегія балансування навантаження не враховує і не передбачає кардинальних змін у роботі. Це є її недоліком перед адаптивною. Проте для систем, що не передбачають кардинальних змін у роботі, цей спосіб розподілу завдань підходить ідеально.

**Опис методу балансування навантаження.** Було проведено огляд того, за якими показниками можна класифікувати стратегії балансування навантаження, щоб застосувати їх для методу балансування навантаження, що пропонується в нашій роботі. Причиною того, чому саме через класифікації буде описано теоретичну базу методу є те, що саме така характеристика чітко показує можливості і його властивості.

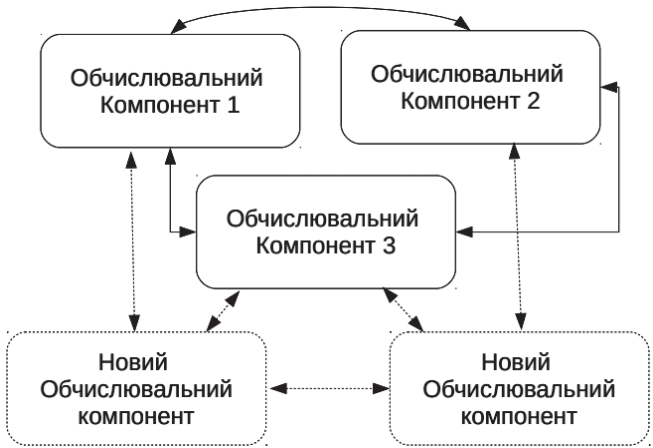

Рис. 4. Децентралізований адаптивний балансувальник навантаження

При короткому огляді методу балансування навантаження, перш за все, слід відзначити, що його роботу відразу потрібно поділити на дві складові. Функції першої складової передбачають глобальну оцінку всієї системи, відслідковування інтенсивності надходження завдань на обчислення до системи, рівномірний розподіл між обчислювальними елементами задач відносно можливостей продуктивності обчислювальних елементів і завдяки визначенню інтенсивності надходження запитів до обчислювального комплексу визначати спосіб розподілу завдань між обчислювальними вузлами в майбутньому.

Перерозподіл надходження запитів до обчислювальних елементів буде здійснюватися щоразу через встановлений проміжок часу. Який саме проміжок часу буде встановлюватися, залежать від конкретної системи. Також допускається, що під час роботи обчислювального комплексу цей інтервал часу буде змінюватися в залежності від роботи самої системи.

Розподіляючи завдання між вузлами, оперувати балансувальник навантаження буде обсягом надходжень завдань, зіставляючи його з продуктивними можливостями кожного обчислювального компонента в системі. Тому для роботи балансування навантаження потрібно оцінювати продуктивність кожного обчислювального вузла за допомогою базових характеристик цього вузла. Передбачається, що характеристики потужності роботи кожного елемента потрібні балансувальнику навантаження лише на початку роботи системи. В подальшому продуктивність буде визначатися за допомогою оцінки ефективності роботи кожного обчислювального елемента, тобто на практиці.

Оцінка інтенсивності надходження завдань до системи є ключовою складовою першої частини балансувальника навантаження. За допомогою однієї з основних і базових формул теорії вірогідності і оцінки інтенсивності надходження запитів у попередніх ітераціях буде визначатися прогноз на роботу в подальших ітераціях, що будуть виникати через встановлені проміжки часу. Сенс цього прогнозу полягає в тому, щоб порівнювати при кожному перерозподілі завдань між обчислювальним компонентами системи точність прогнозу. Порівняння буде відбуватися за допомогою співставлення кількості задач, що надішли і спрогнозовані, також оцінка хиби. Якщо прогноз дуже близький до реально отриманих завдань, то перерозподіл не відбувається, а все працює за способом, визначеним в попередній ітерації. Перевага такого підходу полягає в тому, що просте порівняння даних між собою займає дуже мало часу і обчислювальних можливостей і фактично не впливає на роботу як системи, так і балансувальника навантаження. А ось якщо щоразу перерозподіляти, то це вимагає певних витрат. Тому навіть якщо прогноз не буде співпадати з реальними даними, що надійшли значну кількість разів, це буде вимагати перерозподіляти стратегію поділу навантаження, що і так потрібно б було виконувати. Коли прогноз співпадає з даними, що надійшли, то перерозподіляти нічого не потрібно, і так буде зрозуміло, яким чином надалі проводити розподіл навантаження. А це вже буде давати певну економію як часу, так і обчислювальних можливостей, як балансувальника навантаження, так і всієї системи.

Також у функції першої частини балансувальника навантаження буде входити слідкування за навантаженням і роботою системи в цілому. Наприклад, при значному збільшенні навантаження на всю систему такий балансувальник зможе швидко визначати це і оперативно реагувати на таку проблему, чи визначати проблеми в роботі одного з обчислювальних елементів. Це є основною перевагою централізованої стратегії балансування навантаження.

Підбиваючи підсумки щодо першої складової балансувальника навантаження, необхідно визначити основні її характеристики з приводу описаних класифікацій. Безумовно, такий підхід є централізованим, що дає можливість краще контролювати розподіл навантаження в системі. Також передбачається, що перша частина методу балансування навантаження буде адаптивною, тобто вона може пристосуватися до роботи системи в тій ситуації, коли, наприклад, якийсь обчислювальний компонент вийде з ладу чи будуть додані нові. Звісно, враховуючи перерозподіл навантаження через кожний встановлений проміжок часу, першу складову методу розподілу завдань можна назвати динамічною стратегією. Характеризується залежний розподіл, оскільки, спираючись на зміни в системі і інтенсивність надходження запитів, балансування навантаження безумовно буде враховувати вказані фактори. І, звичайно, ця стратегія є прогностичною, оскільки прогноз того, яким саме чином буде розподілено навантаження в подальшому, є її основною суттю.

В другій частині методу розподілу задач між обчислювальними елементами передбачається, що балансувальник навантаження буде працювати безпосередньо з елементами системи. Причому оцінювати можливість роботи кожного елемента індивідуально. Також робота буде здійснюватися циклічно, через встановлені проміжки часу, і ітерації будуть проходити після того, як перша частина виконала свою роботу.

Можна зробити висновки, що друга складова балансувальника навантаження виглядає як децентралізована, оскільки, працюючи з кожним обчислювальним компонентом, балансувальник оцінює його спроможності з приводу того, яке конкретно навантаження він зможе витримати. При цьому потрібно врахувати, що навантаження не зменшить продуктивність роботи компонента. Звісно, якщо навантаження зростає у всій системі з більш інтенсивним надходженням нових задач, то через збільшення обсягу роботи всьому обчислювальному комплексу потрібно буде більше часу для того, щоб обробити всю інформацію, що надійшла. Проте навіть у ситуації, коли обсяг роботи зростає в цілому, то все одно балансувальник зобов'язаний розділити його між обчислювальними вузлами рівномірно.

Таким чином, передбачається, що коли розпочне свій перероз-
поділ навантаження друга частина балансувальника навантаження,
всі обчислювальні елементи будуть обмінюватися інформацією між
собою. Конкретна інформація про стан розрахункових вузлів перед-
бачає в собі швидкість обробки запитів за той інтервал часу, який об-
межується перерахунками навантаження в системі, що виконуються
другою складовою балансувальника навантаження. Оцінюється про-
дуктивність роботи кожного обчислювального елемента в системі
протягом всієї її роботи. Також, оцінюючи роботу кожного вузла,
буде зроблена опора на практичний підхід, тобто буде оцінюватися
швидкість, з якою обробляє запити кожен компонент протягом всієї
роботи системи, і під час росту інтенсивності надходження нових за-
вдань, і під час її зменшення.

Друга частина буде виконувати функцію порівняння продуктив-
ності роботи між обчислювальними елементами і ступінь заванта-
женості. Також буде створена схема, що буде динамічно змінюватися
після кожного перерахунку навантаження. В такій схемі буде оціню-
ватися вплив обчислювальних елементів один на одного. Під впливом
розуміється те, яку частину завдань якийсь конкретний елемент, що
сильно перевантажений, може перекласти на інший, навантаження
якого незначне. Така схема є основною для другої частини балан-
сувальника навантаження. Адже таким чином створюється цілий
алгоритм, що дозволяє динамічно обмінюватися обчислювальним
елементам запитам із власної черги. При динамічній оцінці ступеня
навантаження кожного елемента буде враховуватися зміна його на-
вантаження відносно останньої і передостанньої ітерації, динаміка
зміни навантаження під час зростання чи зменшення інтенсивності
надходження нових завдань до системи і продуктивність компонента
від самого початку роботи всієї системи. Таким чином, за обрахунок
цих трьох складових будуть оцінюватися продуктивність і наванта-
ження в конкретний момент часу кожного обчислювального компо-
нента системи.

Слід зауважити, що визначення ефективності роботи кожного об-
числювального вузла в комплексі є важливим, оскільки це надасть
достовірну оцінку при перерахунку навантаження. В тому випадку,
коли компонент системи, що виконує розрахункові функції, не має
занадто великого навантаження, але при цьому його продуктивність
протягом роботи всієї системи значно поступається іншим елемен-
там, то перенаправляти частину роботи від більш навантажених ком-понентів системи потрібно в останню чергу. Мається на увазі, що спочатку навантаження перерозподіляється на менш навантаженні обчислювальні елементи, що показують гарну продуктивність в цілому, а лише потім частку завдань потрібно розподіляти на менш навантаженні елементи, що поступаються у своїй продуктивності попереднім обчислювальним елементам.

При розгляді другої складової балансувальника навантаження видно, що вона відрізняється від першої складової тим, що є децентралізованою. Незважаючи на те, що функції цих складових різні, алгоритми, що будуть розробляти на базі цих методик будуть значно відрізнятися за своєю суттю. Також причиною різниці складових є те, що перша полягає в прогностичності подальшого навантаження і його подальшого розподілу, а друга складова займається оцінкою навантаження на кожному обчислювальному елементі його перерозподілом.

Потрібно звернути увагу на те, що від самого початку роботи обчислювального комплексу необхідно оцінити продуктивні можливості системи [8]. Враховуючи те, що практичних даних на самому початку не буде, то потрібно оцінити обчислювальну продуктивність компонентів, що будуть займатися обрахунком завдань, що надходять до них. Для цього потрібно розглянути спрощену модель обчислювальної машини, що складається із набору інструкцій I та доступної пам'яті M. Відмітимо, що під інструкцією $x \in I$ мається на увазі ще і значення всіх її операндів, а не лише назва цієї інструкції. Таким чином, при наявності двох інструкцій з одним ім'ям, але з різними значеннями операндів, вони вважаються різними і обидві включаються в набір I.

Позначимо завдання перед процесором символом A, як деяку послідовність інструкцій $x(P) = x_1, x_2, ..., x_i \in I$. При цьому, якщо задача A, наприклад, містить цикл, який повторюється декілька разів, то послідовність $x(A)$ міститиме в собі тіло цього циклу, що повторюється вказану кількість разів. Очевидно, що не всі послідовності інструкцій можуть бути допустимі. Наприклад, можливе існування пари інструкцій, при послідовному виконанні яких може статися помилка під час роботи процесора. Для цього потрібно визначити послідовність лише тих інструкцій, що є допустимими.

Припустимо, що час виконання інструкції $x$ дорівнює $\tau(x)$. Для простоти будемо вважати, що для всіх інструкцій час їх виконання є цілим числом та найбільший загальний дільник всіх τ (x), x ∈ I дорівнює 1 (це уточнення правильне для більшості процесорів, тому що як час виконання завжди можна розглядати кількість тактів процесора), завжди існує найпростіша інструкція, час виконання якої дорівнює одиниці, тобто τ (x) = 1). У цій роботі це припущення при описі обчислювальної здатності дозволить використовувати lim замість lim sup. Таким чином, час виконання τ (y) послідовності інструкцій y = x 1 x 2 x 3..., x t описується формулою:

$$\tau(y) = \sum_{i=1}^{t < \tau} (x_i). \tag{1}$$

Виведена формула (1) є простою і базовою в обчисленнях, які оцінюють продуктивність процесорів і обчислювальних систем. Вона виводить час, за який виконується задача, яка, в свою чергу, розбивається на окремі завдання. Оскільки для початку нас цікавить тільки час виконання обчислювальними елементами поставлених задач, то в простих випадках такої формули буде достатньо, щоб надати базу для методу балансування в подальших розрахунках, опираючись на те, з якою швидкістю обробляються запити, що надійшли. Якщо таких простих формул не буде достатньо, то потрібно буде застосовувати більш складні формули та підходи, спираючись на роботи, що присвячені продуктивності комп'ютерів.

Підводячи підсумки теоретичного огляду, потрібно зауважити, що при такому поділі на два етапи роботи балансування навантаження основною перевагою є так звана додаткова перевірка і перерозподіл. Зрозуміло, що це виконується другою частиною балансувальника навантаження. Після виконання прогностичних обчислень на наступний перерозподіл навантаження алгоритми першої частини балансувальника навантаження розподіляють навантаження між обчислювальними компонентами згідно з оцінкою навантаженості на цих же компонентах і їх продуктивних можливостей. Після того, як робота другої частини була виконана, знову через встановлений інтервал викликається на роботу друга частина і, оцінюючи навантаженість на обчислювальних компонентах, при потребі виконує новий перерозподіл навантаження. Це підвищує точність роботи балансувальника. А прогностичність першої частини може заощадити час при точному прогнозі.

Також потрібно зауважити, що дві частини, які різними способами будуть перераховувати навантаження між обчислювальними компонентами, не будуть між собою конфліктувати, незважаючи на те, що стратегії балансування в них відрізняються. Причиною відсутності конфліктів є те, що обидві частини опираються на одні і ті ж дані — показники швидкості розрахунку завдань обчислювальними елементами та інтенсивність надходження запитів. Загальна схема роботи балансувальника навантаження зображена на рисунку 5.

При більш детальному розгляді запропонованого методу спочатку потрібно звернути увагу на першу частину балансувальника навантаження. Перш за все, будуть проводитися прогностичні обрахунки, які спираються на базову формулу в теорії вірогідності, що має назву формула повної ймовірності:

$$P(A) = \sum_{i=1}^{n} P(B_i) P(B_i \vee A). \tag{2}$$

Формула (2) застосовується при розрахунку прогнозу того, яка кількість завдань буде поставлена перед обчислювальним комплексом в наступній ітерації. Для цього застосовують дані з попередньої ітерації. Цими даними є кількість запитів, що було надіслано за попередній проміжок часу. Дані, що відображають кількість запитів з попередньої ітерації, вважаються подією, що відбулася, тобто P(B), оскільки вони вже були розподілені по всіх обчислювальних вузлах і обчислювальний комплекс працював з тим навантаженням, яке йому розподілили. Дані, які відображають кількість запитів, що надійшли під час діючої ітерації, вважаються подією P(A), що залежна від даних попередньої ітерації. Такі параметри, як i та n, співвідносяться кількістю останніх ітерацій, що будуть задіяні в цих обрахунках. За рахунок обчислень інформації про кількість надісланих завдань за дві поспіль ітерації отримуються дані, які вважаються прогностичними та наближеними до тієї ітерації, що буде наступною.

Звісно, можна сперечатись з приводу точності такого прогнозу. Тому що інтенсивність надходження завдань до обчислювального комплексу не є стабільною. Проте, як зазначалось, для ефективності роботи прогнозу інтенсивність підвищення навантаження чи його зменшення повна бути рівномірною. При нерівномірній інтенсивності буде проходити щоразу перерахунок прогнозу на наступну ітерацію. Але сенс застосування такого підходу полягає в простоті описаної формули та її простій програмній реалізації. Оскільки вона не є громіздкою, обчислення будуть проходити швидко і це ніяк не вплине на ефективність роботи балансувальника навантаження. Тим більше, якщо зростання або зменшення інтенсивності надходження завдань відбувається поступово, то потрібно зауважити, що достатньо простого порівняння і, якщо хиба незначна, скажімо, в 10—20 відсотків, то немає потреби в тому, щоб виконувати перерахунки стану навантаження на всіх елементах, а можна розподіляти навантаження між ними згідно зі стратегією, що застосовувалась у попередній ітерації. Тим більше, слід зауважити, що, якщо хиба все-таки відчутно вплине на продуктивність роботи і навантаження на окремих обчислювальних компонентах, то друга частина балансувальника, оцінюючи роботу кожного компонента, перерозподілить навантаження між ними більш доцільно.

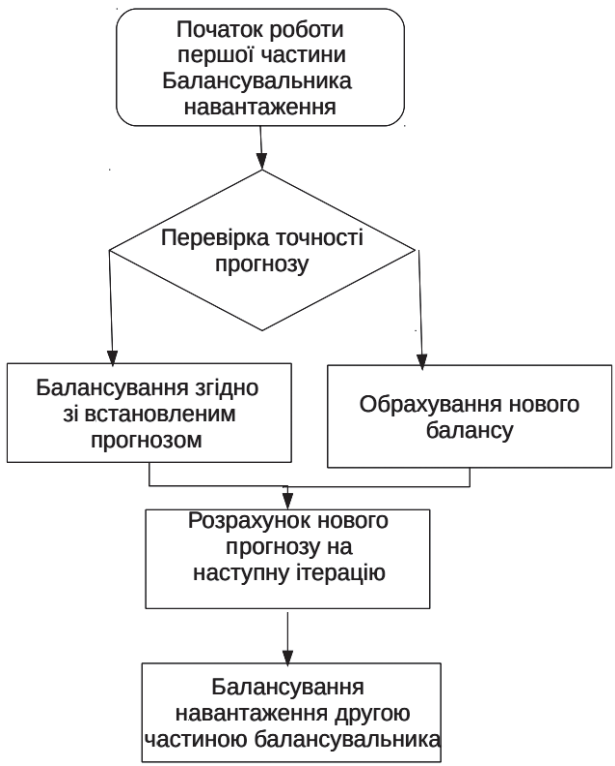

Рис. 5. Загальна схема роботи балансувальника навантаження

Тому можна підвести підсумки відносно прогностичної частини балансувальника навантаження: прогноз відбувається в тому випадку, якщо інтенсивність надходження запитів до обчислювального комплексу є рівномірною і поступовою. У випадках, коли між надходженнями запитів різниця в їх обсягу є значною, то прогностична частина не буде корисною для балансування навантаження. Враховуючи вже зауважену простоту в реалізації, застосування прогностичної частини ніяк не вплине на ефективність роботи всієї системи, оскільки вона не буде вимагати багато часу та ресурсних витрат на своє обчислення.

Безпосередньо саме балансування навантаження буде вираховуватися таким відношенням:

$$\frac{k_{i(n-1)}}{k_{i(n)}}. \tag{3}$$

В наведеній формулі (3) значеннями є відношення кількості поставлених запитів відносно діючої і попередньої ітерації. Символ $i$ позначає обчислювальний компонент, до якого застосовують це відношення, $k$ відображає кількість завдань, що надійшли за вказану ітерацію, а $n$ відображає порядковий номер ітерації. Спочатку запити розділяються рівномірно, потім порівнюються між собою отримані коефіцієнти, якщо знаходяться ті обчислювальні елементи, значення яких суттєво відрізняється, то частина запитів розподіляється від більш навантаженого до менш навантаженого. Таким чином сортується до тих пір, поки відношення не стануть рівномірними в порівнянні між собою. Далі проводяться обрахунки продуктивності кожного обчислювального елемента, в такому випадку робиться відношення швидкості роботи за попередній інтервал часу і той, що щойно завершився. Таким чином оцінюється продуктивність. Потім проводяться обрахунки у відношенні між рівномірно розподіленими запитами на елементах відносно продуктивності елементів. Це потрібно для того, щоб визначати більш продуктивні обчислювальні вузли і розподіляти більше навантаження саме на них. Таким чином навантаження між компонентами, що роблять обрахунки в системі, розподілиться рівномірно не тільки відносно обсягу надісланих запитів, а й відносно продуктивності роботи кожного компонента.

Точно визначити інтервал між ітераціями неможливо, тому що це залежить конкретно від системи, для якої буде застосовуватись ме-тод рівномірного розподілу завдань. Адже при реалізації є різниця, наприклад, між комплексом серверів, що обраховують запити, що надійшли, і багатопроцесорним комп'ютером. Тому конкретні реалізації можуть відрізнятись. Проте можна встановити, що проміжок між ітераціями не може бути значним. Причиною цього є невелика складність запропонованих формул для програмної реалізації. Так само складно знайти точну похибку, яка дозволяється в прогнозуванні, це залежить від системи, де балансувальник реалізується. Можна відмітити, що вимагати точного прогнозу і відкидати можливість похибки не є доцільним тому, що точно спрогнозувати інтенсивність надходження запитів неможливо. Так само не потрібно завищувати допустимість похибки, тому що це буде призводити до некоректної роботи першої частини балансувальника навантаження. Як було вказано, краще звертати увагу на похибку в діапазоні до 15 відсотків, хоча все ж потрібно перш за все враховувати середовище, де саме цей метод балансування навантаження застосовується.

Друга частина балансувальника навантаження працює, оцінюючи і порівнюючи навантаження і продуктивність обчислювальних елементів один з одним. Це виконується відразу після роботи першої частини балансувальника навантаження. Друга частина оцінює, наскільки дисбалансним є розподіл, встановлений першою частиною, і робить перерозподіл. Тільки це робиться, не оцінюючи ситуацію в цілому, а сортуючи дані навантаження всіх обчислювальних елементів, порівнюючи між собою. Спираються розрахунки на формулу (3), тільки для двох елементів. Таким чином друга частина роботи балансувальника навантаження є так званою корекцією першої. Вона вирівнює навантаження між обчислювальними вузлами, яке встановила перша частина.

Інтервал часу, який необхідно вказувати між закінченням роботи першої частини і початком роботи другої, також встановити складно, це залежить конкретно від середовища, де описаний метод балансування навантаження буде застосовуватись. Необхідно зауважити, що інтервал часу між закінченням ітерації першої частини балансувальника і початком роботи ітерації другої повинен бути не більшим, ніж інтервал часу між закінченням роботи ітерації другої частини балансувальника і початком роботи наступної ітерації першої частини балансувальника навантаження. У випадку, коли друга частина почне свою роботу пізно, втрачається сенс того, щоб вирівняти баланс навантаження, встановлений першою частиною.

Незважаючи на те, що визначення похибки в прогнозі буде більш ефективним для конкретної системи, рекомендується використовувати формулу:

$$k = \frac{\sum_{i=1}^{n} \frac{i_{n-1}}{i_n}}{n}. \tag{4}$$

Формула (4) передбачає визначення середньої арифметичної похибки за весь час роботи системи. Параметри $i$ та $n$ відображають числове значення запитів і номер ітерації відповідно. Рекомендується ввести константні значення, що зменшують чи збільшують інтервал між ітераціями. Наприклад, можна визначити, що константні числа інтервалу і середньої похибки можуть бути пропорційним один до одного. Наприклад, при похибці в 50 % випадків інтервал між ітераціями зменшується вдвічі. Все ж при застосуванні наведеної вище формули і визначені констант, при яких будуть проводитися зміни, рекомендується враховувати, для яких саме сфер комп'ютерних наук буде розроблятися алгоритм, що буде базуватися на запропонованому методі. Таким чином можна вираховувати динаміку зміни інтервалу, базуючись на двох складових — похибки в прогнозуванні й у відношенні кількості перерахованого навантаження, мається на увазі кількості запитів діючої ітерації до попередньої. Важливо, щоб взаємозв'язок перерахованого відношення було визначено, спираючись на дані другої частина роботи балансувальника навантаження. Тому що саме друга частина корегує навантаження після роботи першої. Так само можна застосувати формулу до двох вищевказаних факторів і вираховувати середнє значення і порівнювати його зі встановленими константами. Саме це вкаже на ефективність роботи методу планування розподілу навантаження. Якщо значення похибки і коректності навантаження будуть значними, це є сигналом того, що інтенсивність не є рівномірною. Зрозуміло, що при зменшенні часового інтервалу між ітераціями зменшиться кількість відчутних перепадів інтенсивності надходження завдань в самому інтервалі. Інтервали між ітераціями можна робити відносно короткими, оскільки, як зазначалося, програмна реалізація та виконання цього методу нескладні.

Підсумовуючи роботу балансувальника навантаження, пропонуємо ввести схему всіх обчислювальних елементів в системі. За весь час роботи в системі динамічно визначається швидкість роботи кожного обчислювального компонента, що вказує на продуктивність кожного. Суть схеми полягає в тому, щоб відзначити, які компоненти за час роботи в системі працюють більш ефективно, а які менш. Такі дані нададуть можливість визначати заздалегідь, з яких саме елементів переносити частину задач на інші, менш продуктивні. На рисунку 6 зображено схему, на якій найменш продуктивний обчислювальний елемент перенаправляє частину завдань на інші обчислювальні елементи. Рисунок 7 відображає схему, на якій частина завдань перенаправляється від двох найменш продуктивних елементів до найбільш продуктивного в системі.

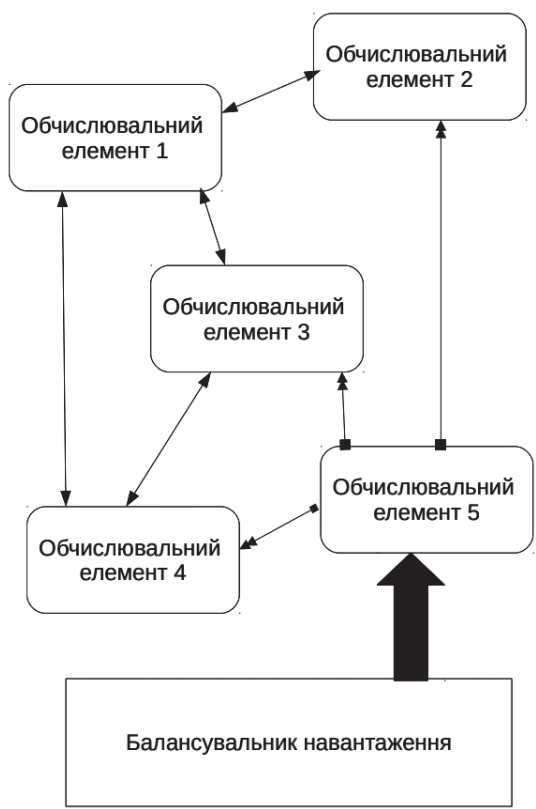

Рис. 6. Перенаправлення завдань до найбільш продуктивних обчислювальних компонентів

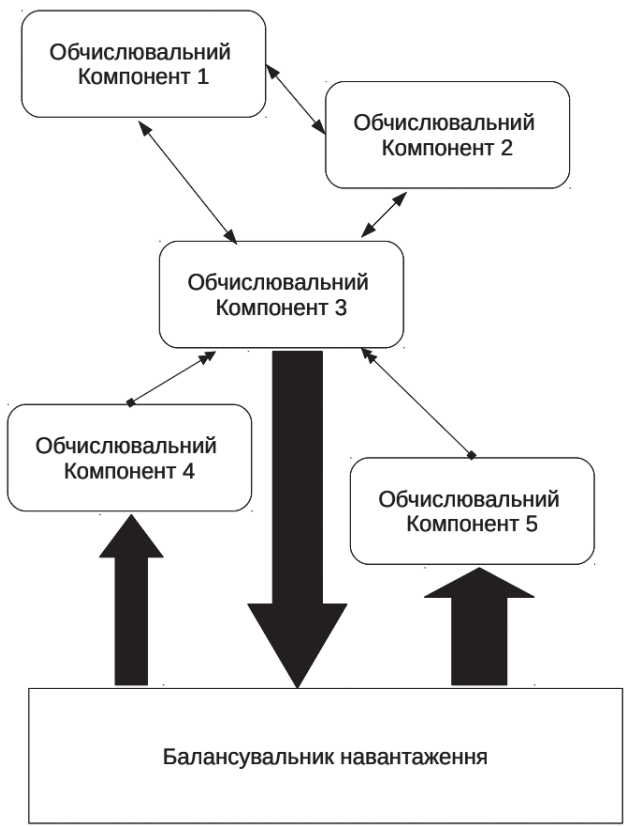

Рис. 7. Перенаправлення завдань від найменш продуктивного компонента

**Висновки.** Спираючись на описані стратегії балансування ми запропонували нову теорію балансування навантаження. Враховувалось, що запропонований метод буде універсальним і буде основою для розробки алгоритмів для різних інформаційних систем, які потребують балансування навантаження. Також описаний метод прогнозує розподіл навантаження на подальшу роботу системи при рівномірній інтенсивності надходження завдань. Описаний метод балансування навантаження є динамічним, він змінює розподіл навантаження між обчислювальними компонентами системи через встановлені інтервали часу. Метод спирається не лише на оцінку інтенсивності надходження завдань, а і продуктивність роботи обчис-лювальних компонентів. Продуктивність визначається через оцінку швидкості роботи кожного компонента за весь час роботи всієї системи.

## АКТУАЛЬНІСТЬ РОЗВИТКУ МЕРЕЖІ NGN

*Кунуп Т. В.*

*Сучасний етап розвитку людства характеризується перебудовую технічного та економічного базису суспільства, де основою стають інформатика, зв'язок, енергетика і транспорт. Оптимальна побудова інфокомунікаційних систем та мереж, що забезпечують рух інформації, матеріальних цінностей, суттєво зменшує витрати суспільства на нормальне функціонуваннях. Розглянуто етапи розвитку телекомунікаційних мереж, актуальність розвитку сучасних мереж та зростання попиту на використання таких систем зв'язку та мереж NGN. Одним із основних аспектів NGN є забезпечення відповідності якості сервісу, що надається, це пов'язано з ефективністю функціонування системи та надання сервісів у мережах.*

*The current stage of human development is characterized by a restructuring of the technical and economic basis of society, where computer science, communications, energy, and transport become the basis. Optimal construction of infocommunication systems and networks that ensure the movement of information and material values significantly reduces the costs of society for its normal functioning. The article considers the stages of development of telecommunications networks, the relevance of the development of modern networks and the growing demand for the use of such communication systems and NGN networks. One of the main aspects of NGN is ensuring compliance with the quality of the service provided.this is related to the efficiency of the system and the provision of services in networks.*

В історичному розвитку мереж та послуг зв'язку можна виділити такі етапи: PSTN (Public Switched Telephone Network), IDN (Integrated Digital Network), ISDN (Integrated Service Digital Network), IN (Intelligent Network), NGN, FN (Future Network).

Перший етап — побудова телефонної мережі загального користування (PSTN). Телефонний зв'язок ототожнювався з єдиною послугою — передачею мовних повідомлень. Надалі телефонними мережами за допомогою модемів стала здійснюватися передача даних.

Другий етап — цифровизація телефонної мережі: були створені інтегральні цифрові мережі IDN, які також надавали в основному послуги телефонного зв'язку на базі цифрових систем комутації та передачі.

Третій етап — інтеграція послуг: з'явилася концепція цифрової мережі з інтеграцією служб ISDN. У процесі розвитку мереж зв'язку особлива увага стала приділятися інтелектуальним послугам. Саме тому інтеграція служб починає замінюватися концепцією IN.

Четвертий етап — інтелектуальна мережа (IN). *Інтелектуальна мережа — це архітектура, для якої характерні такі принципи* [14; 15]:

— широке використання сучасних засобів обслуговування інформації;

— ефективне використання мережних ресурсів;

— модульність мережних функцій з можливістю багаторазового їх використання;

— одночасне створення і впровадження сервісів завдяки модульним, повторно використовуваним мережним функціям;

— інваріантність засобів розміщення мережних функцій у різних фізичних об'єктах;

— взаємодія мережних функцій через незалежні від сервісів стандартизовані інтерфейси;

— можливість керування деякими атрибутами сервісів з боку абонентів і користувачів;

— стандартизоване керування логікою сервісів.

Функціональну архітектуру IN можна представити у вигляді формули: інтелектуальна мережа = комутатор + комп'ютер [22].

Ця мережа призначена для швидкого, ефективного та економічного надання інтелектуальних сервісів масовому користувачеві. Принципова відмінність IN від попередніх мереж — у гнучкості та економічності надання сервісів [14].

Інтелектуальна мережа = комутатор + комп'ютер [22], до цієї формули протягом багатьох років прагнули як виробники комутаційного обладнання, так і виробники засобів обчислювальної техніки. При цьому перші отримували можливість гнучкого й оперативного створення і впровадження нових послуг зв'язку без істотних змін в комутаційному обладнанні, а другі — вихід на один з найбільших сегментів ринку нових інформаційних технологій.

Сучасні мережі зв'язку отримали масу нових можливостей завдяки розвитку концепції «інтелектуальних мереж» (IN, Intelligent

Networks). Основна ідея IN полягає в тому, щоб комутатор (MSC) займався виключно комутацією. А різні послуги організовувались за допомогою сторонніх платформ, які по мережі сигналізації «спілкуються» з комутатором, вказуючи, що йому робити з викликом. При такому підході вся логіка реалізується на окремому керуючому вузлі (SCP, Service Control Point), а на комутаторі описуються тільки умови, при яких необхідно перенаправити запит на потрібний SCP. Оскільки всі зміни логіки відбуваються поза комутатором — теоретично на одному вузлі — то це дозволяє скоротити час на розгортання послуг і зробити їх дуже гнучкими. IN-архітектура спочатку розроблялася для PSTN. Потім протоколи INAP (IN Application Part) поширилися і в мобільному зв'язку. Однак у INAP був істотний недолік — закритість архітектури і, як наслідок, несумісність рішень різних виробників одне з одним. У підсумку, INAP-сервіси можуть бути надані тільки всередині домашньої мережі. У роумінгу абонент їх втрачає. Для вирішення цієї проблеми був розроблений протокол CAP (CAMEL Application Part). CAP — це протокол для роботи додатків (послуг) з комутаційним обладнанням. Представляє собою розширений протокол INAP з додаванням інформації про місцезнаходження абонента. Розроблений ETSI (TS GSM 02.78 від липня 1996 року) [25]. Ключовою його відмінністю, крім функціональної відмінності з INAP, можна вважати великий ступінь відповідності стандартам і уніфікації його реалізацій у різних виробників базового комутаційного обладнання. Точково протокол впроваджується в Україні та інших країнах з кінця 2002 року, повсюдно з середини 2003 року.

CAMEL, або CAP дозволяє забезпечити повний пакет інтелектуальних додаткових послуг (перш за все роумінг) своїм абонентам (включаючи абонентів препейд) не тільки в домашній мережі, але в роумінгу, в мережах, що підтримують стандарт CAMEL за рахунок можливості контролю рахунку і тарифікації в домашній мережі в режимі реального часу. На відміну від, наприклад, USSD, забезпечує мінімальний час з'єднання. Завдяки його відкритості обладнання різних виробників може працювати між собою без проблем. Тобто, навіть в роумінгу, в мережі іншого оператора є можливість користуватися своїми «розумними» сервісами.

Система надання інтелектуальних послуг — це платформа, що працює на основі протоколу CAMEL і дозволяє вирішувати найрізноманітніший спектр завдань у сфері надання додаткових послуг зв'язку [25]. Особливістю організації роумінгу за протоколом CAMEL, на відміну від USSD, є те, що вона може бути надана тільки в мережі того оператора, який також підтримує CAMEL. Розрізняють реалізації в'їзного роумінгу, внутрішньомережевого роумінгу та виїзного роумінгу. Існують версії протоколу: CAP2, CAP3 (CAMEL Application Protocol phase 2 / phase 3) [25].

Приклади послуг на базі системи:

1. Система надання передплачених послуг. Будь-які послуги зв'язку (вихідні дзвінки, SMS, інтернет-трафік, надання контенту) можуть тарифікуватися в реальному часі — перед початком або в процесі надання послуги.

2. Віртуальний телефонний номер. Єдина точка доступу для корпоративного абонента. Будь-які правила маршрутизації вхідних дзвінків, інтеграція з голосовою поштою або службою автоінформатора і багато іншого.

3. Мелодії замість гудків (RBT). Той, хто подзвонить вам, почує замість гудків виклику вашу улюблену мелодію. Або корпоративний гімн, а може рекламу — і це вже не просто розважальна послуга.

4. Call Back («Передзвоніть мені!»). Якщо раптом у вас немає можливості виконати вихідний дзвінок (не дозволяє баланс особового рахунку або умови роумінгу) — ця послуга допоможе передати вашому адресату SMS або USSD повідомлення з проханням передзвонити вам.

5. Чорний / білий список. Хочете позбутися небажаних дзвінків або SMS? Або хочете мати номер, на який додзвоняться тільки потрібні люди? Ця послуга блокує небажані дзвінки на рівні комутатора, і ці дзвінки більше не турбують вас і не займають вашу лінію зв'язку.

Інтелектуальні послуги зв'язку та можливості комп'ютерної телефонії (Computer — Telephony Integration, CTI), що об'єднує два різнорідних інформаційних простора, цій темі присвячено немало публікацій. Набагато менше уваги приділяється її засобам — системам інтелектуальної комутації і маршрутизації в звичайній телефонній мережі, які забезпечують формування мовного і факсимільного трафіку та створюють технічні можливості для надання нетрадиційних (інтелектуальних) послуг зв'язку. Телекомунікаційні послуги, пов'язані з нетрадиційною процедурою обробки дзвінків, встановлення з'єднання або нарахування оплати набувають все більшої популярності. До подібних сервісів належать Freephone (дзвінки за рахунок сторони, що викликається), Premium Rate Service (дзвінки з нарахуванням додаткової оплати, наприклад, за доступ до інформаційних ресурсів або за участь у телефонних лотереях, голосуваннях і т. п.), Prepaid Calling (дзвінки за передоплатою з доступом абонентів по паролях), Least Cost Routing (маршрутизація за найбільш вигідним маршрутом) і ряд інших.

Розглянемо, як реалізуються такі послуги. Наприклад, якийсь оператор має намір організувати для певної категорії абонентів один з доданих додаткових сервісів. Для цього необхідно забезпечити спеціальну процедуру обробки виклику і встановлення з'єднання (наприклад, для послуги Prepaid Calling потрібна можливість динамічного списування грошей з рахунку абонента в ході розмови). Однак базова АТС телефонної мережі оператора не підтримує подібну процедуру. Вийти з цієї ситуації можна «традиційними» способами, вклавши «ну дуже великі гроші» в модернізацію обладнання мережі або зовсім відмовившись від ідеї надання абонентам інтелектуальних послуг.

Ще один спосіб (настільки ж безрадісний для оператора) — вдатися до допомоги «телефонних панянок». Невже становище безнадійне? Зовсім ні, якщо скористатися системою комп'ютерної телефонії Call Routing, заснованої на технології CTI.

Ідея застосування спеціалізованого ПЗ і апаратного забезпечення досить очевидна, хоча і незвична для зв'язківців. Якщо АТС не дозволяє реалізувати інтелектуальні послуги, то нехай вона вирішує традиційні завдання, а відсутній інтелект додасть система Call Routing. Така система підключається до АТС аналоговими або цифровими каналами (двопровідними абонентськими, T1/E1/ISDN) і «спілкується» останньою зрозумілою їй мовою (тобто підтримує практично будь-які протоколи сигналізації — loop start, CAS, CCS і навіть ОКС 7). Єдине, що повинна робити АТС, — обслуговувати стандартні дзвінки звичайним чином і направляти дзвінки, що вимагають спеціальної процедури встановлення з'єднання, в систему Call Routing.

Як кажуть, відчуйте різницю: заміна базового обладнання мережі або установка Call Routing без будь-якої модернізації. Оператору зв'язку потрібно лише запрограмувати свою АТС для комутації певного типу викликів в систему комп'ютерної телефонії і виділити необхідну кількість портів для її підключення. Своєю чергою, Call Routing комплектується і програмується для забезпечення конкретного набору послуг. Тепер про економічний аспект проблеми.

Рішення на базі Call Routing коштують значно дешевше, ніж установка спеціалізованих АТС, що включають в себе ПЗ і апаратні мо-дулі для організації інтелектуальних послуг. Справа в тому, що всі процедури, пов'язані з інтелектуальними послугами (такі як голосове спілкування з абонентом декількома мовами «за вибором», перевірка паролів, встановлення з'єднання на замовлення абонента, динамічне списування грошей з рахунку абонента), є для CTI «рідними» і зазвичай підтримуються програмними засобами навіть найпростіших систем Call Routing. Більш витончене ПЗ забезпечує індивідуальні тарифікацію і маршрутизацію, елементи IP-телефонії, функцію зворотного виклику (Call Back) і безліч інших можливостей [24].

Перелік основних переваг систем Call Routing виглядає солідно:

1. Підтримка декількох типів ліній зв'язку. Системи підключаються до АТС по каналах будь-якого типу (аналогових, цифрових — T1 / E1 / ISDN);

2. Гнучкість. Підтримуються будь-які протоколи взаємодії з базовим комутаційним вузлом мережі (loop start, CAS, CCS);

3. Масштабованість. Можна нарощувати потужність системи як збільшуючи кількість обслуговуваних каналів (від декількох десятків абонентських ліній дрібних операторів до десятків цифрових потоків, що обслуговуються великими провайдерами), так і розширюючи набір можливостей;

4. Простота розгортання. Не потрібно модернізації вже існуючої мережі, досить виділити на АТС оператора зв'язку необхідну кількість аналогових або цифрових портів для підключення системи;

5. Економічна ефективність. Нерідко для впровадження нових послуг зв'язку або підвищення ефективності використання застарілої мережі зв'язку потрібні величезні капіталовкладення (наприклад, при організації нових каналів зв'язку, заміні комутаційного обладнання і т. ін.).

Call Routing дозволяє організувати інтелектуальні послуги, властиві сучасним цифровим мережам, навіть на застарілих аналогових каналах і декадно-крокових АТС.

До середини 80-х років основним завданням під час проектування систем зв'язку було забезпечення високої пропускної спроможності за прийнятною ціною. Оскільки ця мета була частково досягнута з розгортанням волоконно-оптичних систем і впровадженням технологій SDH, B-ISDN, ATM, на телекомунікаційному ринку набуває значущості інший фактор — можливість швидкого розвитку комплексних телекомунікаційних послуг, що задовольняють зростаючі потреби абонентів. Загалом в історичному розвитку телекомуні-каційних мереж і послуг можна умовно виділити такі основні етапи (рисунок 1):

1. Побудова телефонної мережі загального користування ТМЗК (Public Switched Telephone Network, PSTN). На цьому етапі створювалася національна аналогова телефонна мережа загального користування, орієнтована на передачу мовних повідомлень. Надалі в ТМЗК за допомогою модемів стала здійснюватися передача даних.

2. Цифровізація телефонної мережі. Для підвищення якості послуг зв'язку, підвищення автоматизації управління та технологічності обладнання на цьому етапі починають створюватися інтегральні цифрові мережі (Integrated Digital Network, IDN), які надають також в основному послуги телефонного зв'язку на базі цифрових систем комутації та передачі інформації.

3. Інтеграція послуг. На цьому етапі розширюється спектр послуг, що надаються абонентам мережі, і з'являється концепція цифрової мережі з інтеграцією служб (Integrated Service Digital Network, ISDN). Однак ця концепція не мала значного поширення через високу вартість обладнання.

4. Створення інтелектуальної мережі (Intelligent Network, IN). Концепція IN була розроблена для більш швидкого впровадження нових послуг при максимально ефективному використанні існуючої інфраструктури телекомунікаційної мережі [20; 22].

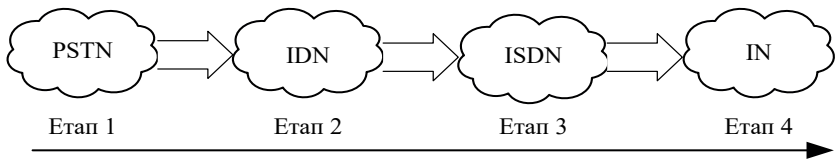

Рис. 1. Етапи розвитку телекомунікаційних послуг і мереж

Подальшим розвитком стала поява мереж зв'язку наступного покоління.

До 50-х років XX століття телекомунікаційні мережі обмежувалися тільки передачею аналогової телефонії. Основне завдання, яке в цей період ставили розробники мереж, полягало в якісній передачі мови на великі відстані з високою надійністю та мінімальною вартістю.

До середини XX століття в розвинених країнах оператори опинилися перед фактом значного зниження темпів зростання доходів від надання традиційних послуг. Попит на такі послуги був повністю за-доволений і в міру розвитку всіх галузей людської діяльності на ринку зв'язку стали з'являтися користувачі, які вимагали нових типів послуг окрім традиційного двоточкового мовного з'єднання. Звичайні мовні з'єднання між двома абонентами більше не могли задовольнити потреби клієнтів ділового сектора, і вже з другої половини 60-х років оператори мереж зв'язку почали пропонувати ряд послуг, щоб залучити нових користувачів: послуги, зроблені на замовлення для простих абонентів, і послуги Centrex для ділових абонентів.

Перелік послуг, зроблених на замовлення, визначався можливостями комутаційних станцій, наприклад, абонент міг попросити розбудити його дзвінком у певний час тощо.

Під Centrex спочатку розумівся спосіб надання послуг зв'язку абонентам декількох компаній на основі спільно використовуваної відомчої комутаційної станції. З появою комутаційних станцій із програмним управлінням термін «Centrex» став означати спосіб надання додаткових послуг діловим абонентам мережі загального користування, аналогічних послугам відомчої АТС. Для цього АТС доустатковувалися спеціальним блоком, а багато компаній-користувачів заощаджували засоби на закупівлю, монтаж і експлуатацію власних відомчих АТС, оскільки могли за допомогою Centrex створювати свої корпоративні мережі, використовуючи ресурси телефонної мережі загального користування. Centrex дозволяв користуватися скороченим набором номера, тристороннім конференц-зв'язком, переадресацією виклику, переведенням з'єднання на інший номер, постановкою виклику на очікування, встановленням з'єднання із зайнятим у цей момент абонентом після його звільнення тощо.

Centrex — це складене слово з двох слів: Central Exchange, і відноситься до виду телефонного зв'язку. Centrex — це назва централізованих послуг АТС / УАТС, які забезпечує провайдер телефонії і які зазвичай забезпечуються офісними АТС (Private Branch Exchange PBX). Пакет послуги Centrex дозволяє організації або компанії відмовитися від придбання та обслуговування офісної PBX власними силами, тому що всі необхідні кошти забезпечує централізований телефонний вузол оператора фіксованого зв'язку. Схожий пакет послуг може надаватися й оператором мобільного зв'язку. У мобільних мережах це зазвичай називається FMC (Fixed Mobile Convergence). Мінімально необхідний набір послуг Centrex: використання внутрішніх 2- або 3-значних номерів для дзвінків всередині офісу; пе-реведення вхідного з міста дзвінка; переведення вихідного в місто дзвінка (класичний приклад «Наталія Петрівна! Наберіть мені Миколая!»); перехоплення дзвінка [4].

У випадку Centrex надання нових послуг вимагало узгодження оператором із замовником ряду специфічних вимог, виконувати які повинен був працівник АТС. Внаслідок цього стандартизація для Centrex практично відсутня, а його послугами користується досить незначна кількість абонентів з тих, кому ці послуги доступні.

З початком чергового бурхливого етапу інформаційної революції з'явилася потреба передачі даних у величезних обсягах з високою швидкістю, у тому числі й комутованими каналами. При цьому швидкість комутації аналогової телефонії перестала бути задовільною. Подальший розвиток розвинених аналогових мереж із досить високою якістю передачі мови в країнах Заходу до кінця 60-х років став недоцільним. У 70-х роках з'явилася концепція цифрової телефонії на базі можливості здійснення з'єднання зі швидкістю 64 кбіт/с, а також розвитку цифрових сполучних ліній (наприклад, ІКМ). З середини 80-х років основна мета забезпечення високої пропускної спроможності за прийнятною ціною була почасти досягнута з розгортанням волоконно-оптичних систем і впровадженням таких технологій як SDH і ATM.

Цифровізація мереж не тільки дозволила підвищити якість послуг, але й сприяла зростанню їх кількості. Внаслідок цього були сформульовані основні принципи створення цифрової мережі з інтеграцією послуг, або ISDN. Абонент ISDN одержує два інформаційних канали по 64 кбіт/с і один канал сигналізації 16 кбіт/с для управління з'єднанням (2b+d). Загальна пропускна спроможність каналу становить 144 кбіт/с, що дозволяє передавати навіть один відеоканал у реальному часі.

Ці нові мережі складаються з ISDN-станцій, які комутують цифрові потоки, що містять будь-яку інформацію: мова, дані, стисле відео тощо. Перетворення в аналоговий сигнал відбувається безпосередньо в абонентському терміналі. Крім того, всі комутаційні станції мережі ISDN на відміну від аналогових можуть працювати як одна велика станція, дозволяючи здійснювати автоматичну маршрутизацію виклику, рівномірний розподіл навантаження, мати єдиний план номерів, створювати віртуальну мережу та надають ряд інших додаткових послуг. За час свого розвитку концепція ISDN пережила зльоти й падіння, пов'язані з коливанням потреб ринку та наявністю в абонентів комп'ютерів. Нині практично всі комутаційні станції на мережах розвинених країн мають функції ISDN.

Однак ISDN та інші методи надання абонентам додаткових послуг мають свої проблеми, що виражаються також у недостатній стандартизації, внаслідок чого у світі діє кілька несумісних стандартів. Крім того, для введення нових послуг необхідно замінити програмне забезпечення кожної ISDN-станції, що вимагає значних капіталовкладень і колосальної інтуїції від оператора мережі, оскільки в цьому випадку крок «не в ту сторону» коштує ще дорожче. Час життя комутаційного обладнання триває кілька десятків років, тому заміняти його щоразу для надання нової послуги недоцільно (і не заміняти не можна, оскільки існує різке зростання вимог до збільшення кількості функцій, які мають бути підтримані мережею). Крім того, при введенні в такий спосіб нових послуг ускладнюються структура мереж, а також процеси управління та експлуатації.

Необхідність модернізації обладнання та програмного забезпечення на всіх АТС мережі є слабким місцем усіх перерахованих технологій при розширенні набору послуг, які надає мережа. Для вирішення питання про послуги, які оператор мережі хотів би запропонувати своїм клієнтам, необхідно було погоджувати з виробником обладнання характеристики майбутніх послуг і можливості їхньої реалізації. Розгортання кожної нової послуги вимагало модифікації апаратних засобів у всіх комутаційних станціях. Цей процес ускладнювався ще й тим, що мережа оператора, як правило, складалася з обладнання декількох різних виробників, і іноді траплялося, що послуги в зоні обслуговування цього оператора виявлялися не повністю ідентичними. Крім того, після введення послуги в експлуатацію модифікувати її з урахуванням вимог нових груп клієнтів також було дуже непросто. Найчастіше для цього доводилося погоджувати з постачальником обладнання додаткові зміни апаратних засобів. Як наслідок, оператору були потрібні роки, щоб спланувати та реалізувати у своїй мережі нову послугу. Створення комутаційних станцій із програмним управлінням було суттєвим кроком уперед, який дозволив зробити логіку надання послуг програмованою, що значно полегшило реалізацію процесу надання послуг. Однак концепція надання послуг не була модульною. В міру зростання складності окремих послуг і залежності між ними додати нову послугу до вже існуючих ставало все складніше. Оператор не міг сам використати логіку, що підтримує одну послугу, для підтримки іншої.

Велика кількість, складність і частота введення нових функцій у мережі зв'язку вимагали нового підходу, який міг кардинально змінити всі аспекти створення та надання послуг, а також експлуатаційного управління ними. Виникла необхідність заміни консервативного підходу, який використовувався протягом тривалих років і передбачав надання невеликого переліку послуг, до створення нової платформи, що дозволяє вводити широкий спектр нетрадиційних послуг і надає можливість «налаштовувати» їх під індивідуальні вимоги клієнта. Таким новим підходом стала концепція інтелектуальної мережі.

Подальшим розвитком стала поява мереж зв'язку наступного покоління. Основу мережі NGN складає мультипротокольна мережа — транспортна мережа зв'язку, яка входить до складу мультисервісної мережі, що забезпечує перенос різних типів інформації з використанням різних протоколів передачі. NGN являє собою єдину транспортну платформу, на базі якої об'єднуються різні види сервісів.

Ключовими особливостями мережі NGN є:

— використання режиму комутації пакетів для передачі даних;

— поділ функцій управління на функції, пов'язані з управлінням транспортом, управлінням викликами/сесіями і додатками/сервісами;

— відділення процесу надання сервісів від процесу транспорту, використання відкритих інтерфейсів;

— підтримка великого набору сервісів, додатків і механізмів, заснованих на конструктивних блоках, включаючи потокові сервіси, сервіси в режимі реального та нереального часу, мультимедійні сервіси;

— підтримка широкосмугових технологій з наскрізним («з кінця в кінець», end-to-end) забезпеченням якості обслуговування;

— взаємодія з існуючими мережами через відкриті інтерфейси;

— мобільність в загальному сенсі (generalized mobility);

— необмежений доступ користувачів до різних постачальників сервісів;

— безліч схем ідентифікації абонента;

— одні й ті ж характеристики для однакових з погляду користувача сервісів;

— конвергенція сервісів мобільних і фіксованих мереж;

— незалежність сервіс-орієнтованих функцій від транспортних технологій;

— підтримка різних технологій для реалізації мережі доступу та ін.

На сьогоднішній день можна говорити про пост-NGN, а точніше, про використання підходів IMS.

Розглянемо більш детально основні етапи розвитку мультисервісних мереж.

Сьогодні існує концепція мереж наступного покоління, в яких ключове місце відведено поняттю «послуга» — NGS (New Generation Services).

Інтелектуальна мережа була першим кроком на шляху переходу до модульної архітектури мережі і дозволила відокремити шар комутації від шару надання сервісів. Завдяки успіху IN, розвитку пакетних технологій на сучасному етапі виявилося можливим створити NGN (Next Generation Network — мережа наступного покоління). Перехід до NGN можна вважати радикальною модернізацією телекомунікаційної системи. Міняються не тільки технологічні принципи передачі і комутації. Вельми істотні зміни відбудуться на ринку інфокомунікаційних сервісів, в системі технічної експлуатації і не тільки.

Огляду архітектури NGN присвячені роботи багатьох сучасних вчених. У працях Б. Гольдштейна і О. Гольдштейна та інших розкриваються питання переходу до мереж наступного покоління, аналізуються дві конкуруючі концепції NGN — IPCC і TISPAN, а також доповнюючі технології NGN — MPLS, Softswitch, Call-центри, протокол SIP. Основні положення зазначеного напрямку можна знайти в [10].

Для мережі NGN характерні істотні особливості, що виділяють її в новий клас телекомунікаційних систем [9]:

• передача з пакетною комутацією;

• розділення функцій управління між пропускною спроможністю каналу-носія викликом/сеансом, а також додатком/сервісами;

• розмежування між наданням сервісів і транспортуванням і надання відкритих інтерфейсів;

• підтримка широкого спектру сервісів, додатків і механізмів на основі уніфікованих блоків обслуговування (включаючи сервіси в реальному масштабі часу, в потоковому режимі, в автономному режимі і мультимедійні сервіси);

• можливості широкосмугової передачі наскрізною функцією QoS (якості обслуговування);

• взаємодія з існуючими мережами за допомогою відкритих інтерфейсів;

• універсальна мобільність;

• необмежений доступ користувачів до різних постачальників сервісів;

• різноманітність схем ідентифікації;

• єдині характеристики обслуговування для одного і того ж сервісу з точки зору користувача;

• зближення сервісів між фіксованим і рухомим зв'язком;

• незалежність пов'язаних з обслуговуванням функцій від використовуваних технологій транспортування;

• підтримка різних технологій «останньої милі»;

• виконання всіх регламентарних вимог, наприклад, для аварійного зв'язку, захисту інформації, конфіденційності, законного перехоплення і так далі.

З'ясувавши особливості мережі наступного покоління, слід дати визначення цього терміна.

Мережа зв'язку наступного покоління (NGN) — концепція побудови мереж зв'язку, що забезпечують надання необмеженого набору сервісів з гнучкими можливостями з управління, персоналізації і створення нових сервісів за рахунок уніфікації мережних рішень, що припускає реалізацію універсальної транспортної мережі з розподіленою комутацією, винесення функцій надання сервісів в крайові мережеві вузли й інтеграцію з традиційними мережами зв'язку [14].

В роботі [13] Б. С. Гольдштейн запропонував дещо інший варіант трактування терміна NGN, в якому за основу була взята можливість мережею надавати потрійний сервіс (Triple — play services — мова, відео, дані): NGN — це мережа, здатна забезпечити обслуговування виду Triple-play services за рахунок використання устаткування передачі і комутації, заснованого на пакетних технологіях.

Говорячи про NGN, маємо на увазі мультисервісну мережу на основі пакетів з відокремленням функцій надання сервісів від функцій комутації.

Архітектура мережі NGN представлена на рисунку 2 [13].

NGN запозичила в IN принцип відділення функції комутації від функції надання сервісів. Функції IN були розподілені між Softswitch та серверами. Відповідність IN та NGN представлена на рисунку 3.

В такому випадку Softswitch виконує функцію комутації інтелектуальних сервісів, а функцію управління сервісом здійснює сервер додатків.

Вводиться новий елемент мережі: програмний комутатор Softswitch, що виконує функцію управління викликами і сесіями

CSCF (Call Session Control Function), який, з одного боку, управляє з'єднанням, а з іншого — взаємодіє з серверами надання сервісів за SIP протоколом (Session Initiation Protocol) [11].

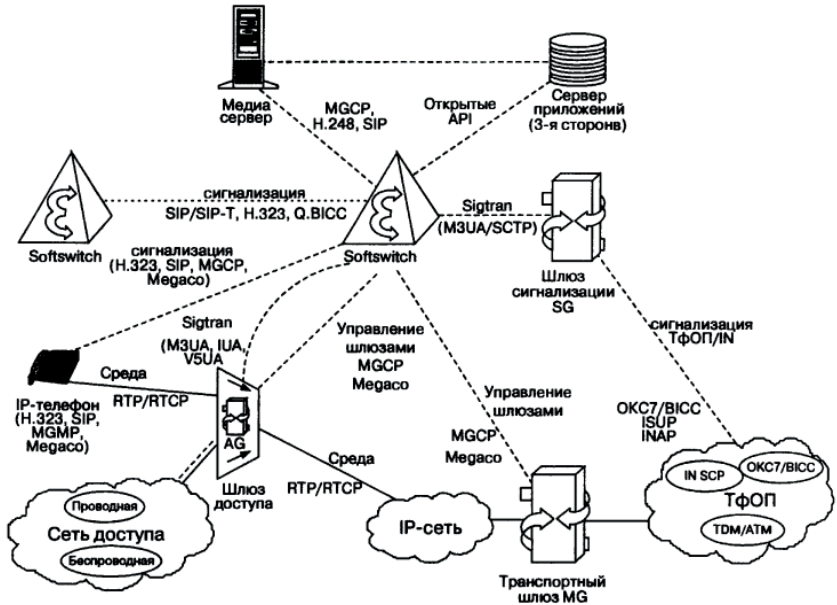

Рис. 2. Архітектура мережі NGN

У термінах NGN платформа надання інтелектуальних сервісів називається SDP (Service Delivery Platform). Основа ідеології NGN — це відкриті стандарти консорціуму 3GPP (3'rd Generation Partnership Project).

Ідеологія побудови NGN забезпечує можливість надання абонентам сервісів Triple-Play (передача мови, даних і відео) на базі мультисервісних мереж.

На сьогоднішній день мова йде про об'єднання стільникового та стаціонарного зв'язку — у відповідності з концепцією IMS.

Для IMS розроблена багаторівнева архітектура з поділом транспорту для перенесення трафіку і сигнальної мережі IMS для управління сеансами (рисунок 4) [11]. Таким чином, 3GPP при розробці IMS фактично переніс на мобільні мережі основну ідеологію Softswitch. Хоча деякі функції не завжди легко віднести до того чи іншого рів-ня, але такий підхід забезпечує мінімальну залежність між рівнями. У IMS можна виділити [11]:

— User Plane — рівень користувачів, або рівень передачі даних;
— Control Plane — рівень управління;
— Application Plane — рівень додатків.

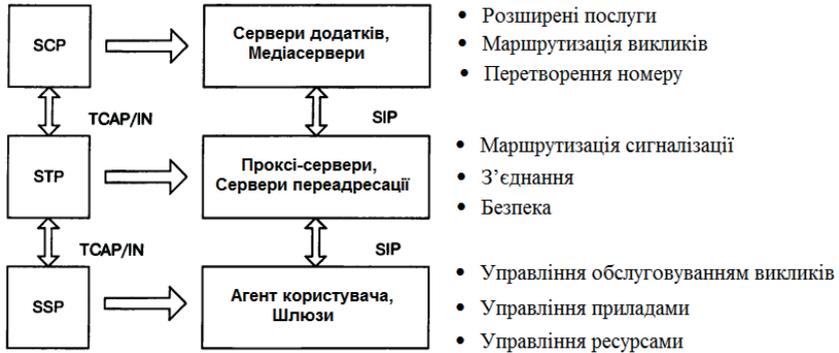

Рис. 3. Відповідність інтелектуальної мережі та NGN

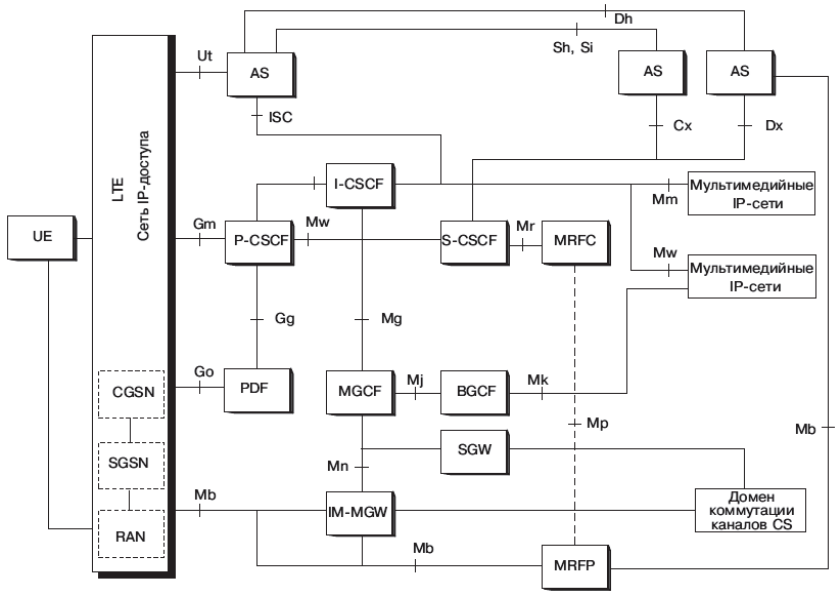

Рис. 4. Архітектура IMS

3GPP, за прикладом IN, а потім Softswitch в IPCC, специфікує не вузли мережі, а функції. Це означає, що IMS-архітектура, як і IN або Softswitch, теж являє собою набір функцій, з'єднаних стандартними інтерфейсами.

Розробники мають право комбінувати кілька функцій в одному фізичному об'єкті або, навпаки, реалізувати одну функцію розподілено, однак найчастіше фізичну архітектуру ставлять у відповідність до функціональної і реалізують кожну функцію в окремому вузлі.

Таким чином, в IMS інтелектуальна надбудова практично трансформувалася в рівень додатків, що включає в себе три типи серверів додатків:

— SIP AS (SIP Application Server — сервер додатків);

— OSA-SCS (Open Service Access-Service Capability Server — сервер сервісів, який забезпечує інтерфейс до сервісів, які базуються на відкритому доступі до сервісів);

— IM-SSF (IP Multimedia Service Switching Function — платформа для надання сервісів IN мережі).

В IMS поняття інтелектуальний сервіс замінене поняттям новий сервіс.

Наступницею NGN вважається мережа майбутнього (Future Network — FN).

Згідно з рекомендацією MCE-T Y. 3001, мережа майбутнього — це мережа, здатна надавати сервіси, можливості і засоби, які важко надати з використанням існуючих мережевих технологій [9].

Мережею майбутнього є:

— або нова компонентна мережа чи вдосконалений варіант існуючої компонентної мережі;

— або різнорідна група нових компонентних мереж чи група, що складається з нових та існуючих компонентних мереж, які працюють як єдина мережа.

Рекомендується, щоб FN надавали сервіси, функції яких спроектовані так, щоб відповідати потребам додатків і користувачів. Очікується, що в майбутньому кількість і вибір сервісів будуть стрімко зростати. Рекомендується, щоб FN забезпечувала можливість впровадження цих сервісів, не вимагаючи, наприклад, істотного додаткового розгортання і збільшення експлуатаційних витрат.

На рисунку 5 зображені взаємозв'язки між чотирма цільовими установками та дванадцятьма цілями проектування FN [9].

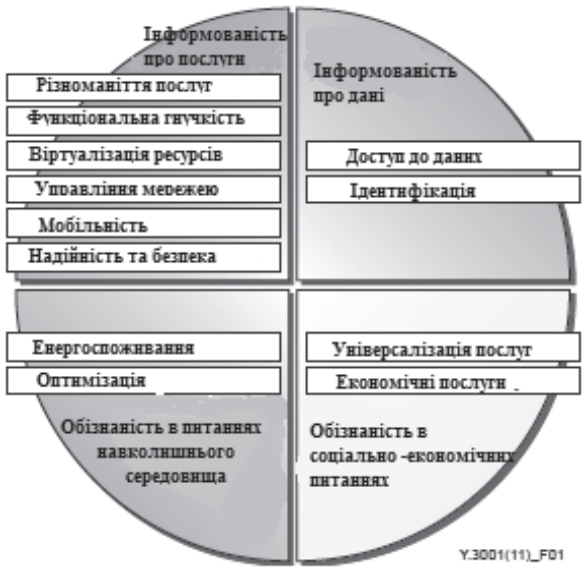

Рис. 5. Чотири цільові установки та 12 задач проектування мереж майбутнього

Безпосередньо стосуються задач проектування:

– різноманітність послуг — у майбутніх мережах мають підтримуватись різноманітні послуги, пристосовані для передачі трафіку з широким вибором характеристик і властивостей;

– функціональна гнучкість — майбутні мережі мають бути функціонально гнучкими для підтримки та забезпечення стійкості нових послуг, які стануть відповіддю на потреби користувачів;

– управління мережею — у майбутніх мережах необхідно ефективно експлуатувати, обслуговувати та надавати все більше послуг і додатків;

– оптимізація — майбутні мережі забезпечуватимуть достатню якість роботи шляхом оптимізації можливостей мережевого обладнання, виходячи з вимоги до послуги і потреб користувача;

– надійність і безпека — проектування, експлуатацію та розвиток майбутніх мереж необхідно здійснювати таким чином, щоб забезпечити надійність і здатність до відновлення, з урахуванням складних умов.

Прогнозували, що до 2020 року повинні були з'явитися мережі майбутнього — FN.

Архітектура NGN, розроблена IPCC, розподілена на такі рівні [15; 16].

На нижньому рівні архітектури знаходиться транспортний рівень (Transport Layer), що відповідає за перенесення по мережі сигнальних повідомлень і мультимедійної інформації. Крім того, він забезпечує взаємодію і обмін сигнальною і медіаінформацією з PSTN та іншими пакетними мережами.

Транспортний рівень, своєю чергою, підрозділяється на три під­рівні: IP-транспорту, міжмережної взаємодії і відмінного від IP (NON-IP) доступу.

Підрівень IP-транспорту надає магістральну мережу передачі і структуру комутації/маршрутизації для транспортування пакетів по VoIP-мережі. До цього рівня належать маршрутизатори і комутатори, а також пристрої, що відповідають за забезпечення якості обслугову­вання (Quality of Service, QoS) і політики передачі даних.

Підрівень міжмережної взаємодії відповідає за перетворення сигнальної і мультимедійної інформації, що отримується із зовніш­ніх мереж, у форму, придатну для передачі усередині VoIP-мережі, і навпаки. Тут функціонують такі пристрої, як шлюзи сигналіза­ції (Signaling Gateways), медіашлюзи (Media Gateways) і міжмережні шлюзи (Interworking Gateways).

Підрівень NON-IP доступу об'єднує несумісні термінали і безпро­водові радіомережі, що мають доступ до VoIP-мережі. До цього під­рівня відносяться шлюзи доступу або резидентні шлюзи для несуміс­них терміналів або телефонів, ISDN-термінали, кабельні модеми або MTA (Multimedia Terminal Adaptors) для HFC-мереж (Hybrid/Fiber Coaxial), медіашлюзи мереж GSM/3G і мереж радіодоступу.

Наступний рівень — управління викликами і сигналізації (Call Control & Signaling). Управляє основними елементами VoIP-мережі, що знаходяться на транспортному рівні. Пристрої і функції цього рівня управляють викликом, ґрунтуючись на сигнальній інформації, отриманій від транспортного рівня, зокрема здійснюють встановлен­ня і розрив медіазв'язку в VoIP-мережі, передаючи команди мереже­вим компонентам. Рівень управління викликами і сигналізації міс­тить такі пристрої, як контролери медіашлюзів (MGC, Call Agent, Call Controller), LDAP-сервери.

Третій рівень — сервісів і додатків (Service & Application) — забез­печує управління, логіку і виконання деякого числа сервісів або до­датків. Пристрої, що належать до цього рівня, управляють потоком викликів, ґрунтуючись на запрограмованій логіці виконання сервісів, за допомогою взаємодії з пристроями рівня управління викликами і сигналізації. До самого рівня сервісів і додатків належать такі пристрої, як сервери додатків і сервери сервісів.

Останній рівень — управління (Management) виконує функції призначеного для користувача забезпечення, підтримку операцій і надання сервісів, а також вирішує завдання білінга й інші завдання мережевого управління. Рівень управління може взаємодіяти з будь-яким з трьох перерахованих, використовуючи стандартні або внутрішньофірмові протоколи і програмні інтерфейси API.

Тепер звернемося до концепції ETSI TISPAN. В цьому проекті, на відміну від концепції IPCC, архітектура мереж описана не сукупністю вузлів, а як набір функціональних модулів, які можуть бути реалізовані в різних фізичних елементах. Взаємодія між модулями здійснюється за стандартизованим інтерфейсом. Найчастіше взаємодія відбувається за сигнальним протоколом SIP-I, інколи H.248 та ін.

Мережна архітектура, запропонована ETSI TISPAN, зображена на рисунку 6 [15; 16].

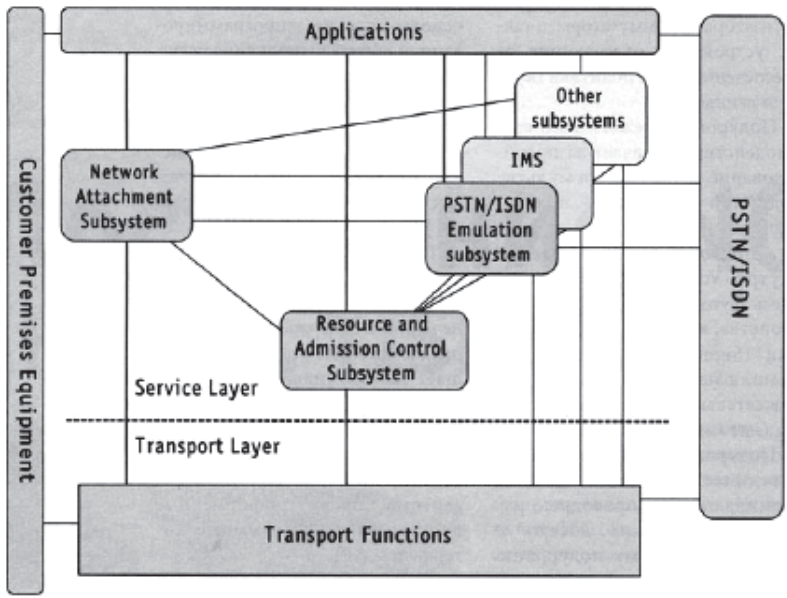

Рис. 6. Підсистеми TISPAN

Однією з найважливіших підсистем TISPAN вважається система управління викликами і сервісами IMS.

Серед важливих принципів IMS слід зазначити, що вона базується на відкритих Інтернет-стандартах і тому без додаткової адаптації може використовувати всі сервіси і додатки мережі Інтернет, проте усередині самої IMS передбачено застосування протоколу IPv6.

Другою особливістю архітектури IMS є інноваційний підхід до надання сервісів, що дозволяє операторові створювати різні сервіси й інтегрувати їх один з одним, що забезпечує широкі можливості для персоналізації і збільшення кількості сервісів.

Підхід IMS припускає горизонтальну архітектуру (рисунок 6) [15; 16], що дозволяє операторові просто й економічно упроваджувати нові сервіси, які персоналізуються, причому користувачі можуть стримати доступ до різних сервісів в рамках однієї сесії зв'язку.

Нова архітектура надання сервісів дозволила змінити традиційний погляд на їх створення і стандартизацію. Можливості, які привносить впровадження IMS, безумовно, додають плюсів до рішення TISPAN.

В МСЕ був створений форум з мереж майбутнього, який зараз продовжує активно функціонувати. Мережа наступного покоління спроможна надати найширший спектр сервісів. Дамо визначення терміна «сервіс» та спробуємо класифікувати існуючі сервіси. У рекомендації МСЕ I.112 термін сервіс визначається як: «те, що пропонується споживачам для задоволення певної комунікаційної потреби». У цій же рекомендації сервіс надання зв'язку визначений як «вид обслуговування, що повністю реалізує можливості (включаючи функції термінального устаткування) зв'язку між користувачами відповідно до протоколів, встановлених для відповідного виду зв'язку». Під сервісами користувача розуміється те, що пропонується користувачеві, здається йому в оренду або оплачується ним.

На сьогоднішній день оператори часто класифікують сервіси за одним з критеріїв. Це у свою чергу приводить до певних труднощів, наприклад, при розрахунку тарифів. Тому інколи доцільно класифікувати сервіси, використовуючи систему класифікаторів.

Найбільш розповсюджені види класифікацій такі:

1. Класифікація сервісів за типом інформації, котра передається (контенту).

2. Класифікація сервісів за способом забезпечення доступу клієнта та до сервісу.

3. Класифікація сервісів за типом клієнта.

4. Класифікація сервісів за типом обміну інформацією.

Для кожного типу сервісів можливий їх підрозділ за наступними ознаками:

1. За пріоритетом впровадження та важливості — базові (основні) сервіси та додаткові.

2. За маркетинговою функцією — сервіси, орієнтовані в основному на отримання доходу, та сервіси, направлені на залучення нових клієнтів.

Особливий інтерес викликає розподіл сервісів на основні *(basic services)* і додаткові *(supplementary services)*.

Основний сервіс визначається функціональним призначенням пристрою. Передача мови при з'єднанні двох користувачів телефонної служби є прикладом надання основного сервісу.

Додаткові сервіси (сервіси з додатковою вартістю) можуть бути постійними і разовими. Постійний додатковий сервіс надає на певний період часу власникові пристрою додаткові можливості, які для пристрою даного типу не є обов'язковими. Разовий додатковий сервіс надається клієнтові за його запитом.

Найбільш поширеними в даний час додатковими сервісами є:

1. **Безумовне перенаправлення виклику** *(CFU)* — можливість направляти всі вхідні виклики на інший номер.

2. **Перенаправлення виклику при зайнятості абонента** *(CFB)* — можливість направляти на інший номер всі вхідні виклики, що надходять під час зайнятості крайового пристрою користувача.

3. **Перенаправлення виклику при невідповіді абонента** *(CFNR)* —можливість всі вхідні виклики, на які немає відповіді впродовж певного проміжку часу, направляти на інший номер.

4. **Конференц**-зв'язок *(CONF)* — надає можливість брати участь і управляти одночасним зв'язком декількох користувачів.

5. **Утримання виклику** *(HOLD)* — дозволяє користувачеві переривати і відновлювати зв'язок на існуючому з'єднанні.

6. Інші.

Відповідно до Рекомендацій МСЕ-Т Y.1540 якість послуг оцінюється за трьома показниками [9]:

— швидкість — це один з найважливіших показників, який характеризує якість надання більшості ІС. Показник швидкості визначається контрольними термінами. Контрольні терміни — це максимальний час, протягом якого повинен бути наданий сервіс;

— точність і достовірність — це характеристики споживчих властивостей сервісу, тобто наскільки він придатний для використання.

— надійність — це властивість засобів зв'язку надавати якісні сервіси.

В останніх працях з цього напряму зазначається, що надання сервісу залежить від таких мережних показників: IPTD (затримки передачі пакету IP з інформацією управління), IPDV (зміни затримки пакета IP), IPLR (відсотка втрачених пакетів IP) і IPER (відсотка помилкових пакетів IP).

Згідно з рекомендаціями МСЕ-ТI.380/Y.1540, 2007 [9] визначення якості функціонування NGN має спиратися на формування таких показників:

• затримка перенесення пакетів;
• варіація затримки пакетів (джиттер);
• коефіцієнт втрати пакетів;
• коефіцієнт помилок по пакетах.

Згідно з рекомендацією МСЕ-Т [7—9] якість обслуговування (QoS) визначено як сукупність характеристик послуги електрозв'язку, які стосуються її можливості задовольняти встановлені і передбачувані потреби користувача сервісу.

Паралельно існує поняття якості сприйняття (QoSE) — рівня якості, який, за заявою абонентів / користувачів, вони відчували.

Показники роботи мережі (NP), зокрема показники якості роботи інтелектуальної надбудови, характеризують здатність останньої або її частини забезпечувати функції, пов'язані з наданням ІС користувачам та управлінням цим сервісом.

У зв'язку з ускладненням сервісів з'являється нове устаткування, яке окрім стандартних функцій комутації виконує функції управління сервісами. В такому випадку використовується так званий мережний інтелект.

Мережний інтелект — це програмне забезпечення, призначене для управління процесами з'єднання крайового устаткування і надання користувачам інфокомунікаційних сервісів [19]. Використання мережного інтелекту та розподіл сервісної логіки і логіки комутації передбачає створення так званої інтелектуальної надбудови.

Додаткові сервіси, що надаються за допомогою інтелектуальної надбудови, називають інтелектуальними сервісами. Інтелектуальні сервіси включають і персоніфіковані сервіси, що базуються на постійно оновлюваній інформації про місцеположення користувача, про його записи в органайзері, особистих перевагах і тому подібне, підказують цьому користувачеві найбільш доцільний напрям пере-сування, нагадують йому про покупку подарунка до дня народження, організують поїздки, бронювання квитків, отримання інформації про погоду у вказаному пункті, надають банківську інформацію, проводять фінансові операції і багато іншого.

У Рекомендаціях МСЕ з інтелектуальних сервісів специфіковано чотири набори ІС: CS-1, CS-2, CS-3, CS-4 [5—9]. Набір сервісів CS-1 (Capability Set 1) включає 25 сервісів і 38 властивостей (основоположних і допоміжних), які мають дві загальні характеристики, що стандартизовані МСЕ:

— сервіс замовлюється єдиним користувачем (single ended);

— виконання сервісу контролюється єдиною точкою контролю сервісу (single control).

Сьогодні перелік інтелектуальних сервісів значно збільшився. До найбільш популярних сервісів можна віднести Freephone (дзвінки за рахунок сторони, що викликається), Premium Rate Service (дзвінки з додатковою платою, наприклад, за участь в лотереях, голосуваннях тощо), VAS (Value Added Services, послуги з доданою вартістю) та інші.

Сервіси CS-1 відносяться до сервісів типу А (однокінцеві) з централізованою логікою управління. CS-1 включає 25 видів сервісів, підтримується мережами PSTN, ISDN.

Останніми роками телекомунікаційні оператори не тільки працюють над поліпшенням якості і поширенням традиційних послуг зв'язку, а й активно пропонують нові сервіси, які стають найважливішою точкою зростання обороту компаній в умовах гострої конкурентної боротьби на ринку. При цьому для реалізації різних сервісів потрібен відповідний розвиток мереж зв'язку і зокрема їх транспортної інфраструктури. Світове телекомунікаційне співтовариство прийшло до висновку про необхідність створення мереж наступного покоління, так званих (Next Generation Networks). Велика частина особливостей NGN схожі з характеристиками сучасного Інтернету. Однак NGN повинна підтримувати набагато більшу кількість протоколів виробників різного устаткування — як «старого», так і перспективного.

Поставлене запитання «NGN: мода чи необхідність?» сьогодні виглядає абсолютно недоречним — про жодну моду тепер годі й казати, провідні телекомунікаційні оператори не тільки успішно впроваджують фрагменти мереж наступного покоління, а й повністю формують свою інфраструктуру за принципами NGN. Багато компаній тепер повідомляють, що їхні міжміські та міжнародні мережі зв'язку побудовані на основі NGN.

За минулі роки була остаточно осмислена концепція NGN і стався помітний прогрес у випуску обладнання для IP-мереж. Визначилися можливості і вигоди створення інфраструктури мереж NGN, з'явилася комерційна складова проектів. Відбувся перехід від захопленого уявлення про нові технології до їх комерційного впровадження. При цьому NGN стає передовою основою для впровадження послуг Triple Play (голос, передача даних і відеосервіси по одній абонентської лінії).

У рекомендаціях Міжнародного союзу електрозв'язку (МСЕ/ITU) дано таке визначення Next Generation Network: NGN це мережа з комутацією пакетів, здатна надавати телекомунікаційні послуги за допомогою широкосмугових транспортних технологій, що підтримують якість обслуговування (QoS), в якій функції послуг не залежать від використовуваних транспортних технологій [1; 7].

Відмінною рисою моделі NGN, пропонованої сектором МСЕ-Т, є її функціональний розподіл на рівень послуг і транспортний рівень [1]. Останній забезпечує виконання функції обміну дискретною інформацією будь-якого типу між будь-якими двома географічно рознесеними точками [8].

Перший рівень реалізує прикладні функції, пов'язані з затребуваними послугами, наприклад, з організацією передачі мови і відеозображень окремо або в комбінації. Відповідно до рекомендацій МСЕ-Т, NGN повинна здійснювати конвергенцію послуг передачі даних, мови, відео-, аудіо- та візуальних даних в індивідуальному, груповому і широкомовному режимах [1].

Мережі NGN повинні забезпечувати надання необмеженого набору послуг з гнучкими можливостями щодо їх управління, персоналізації і створення нових послуг за рахунок уніфікації мережевих рішень.

Властивості NGN:

1. Мультисервісність — незалежність технологій надання послуг від транспортних технологій;

2. Широкополосність — можливість гнучкої і динамічної зміни швидкості передачі інформації в широкому діапазоні в залежності від поточних потреб користувача;

3. Мультимедійність — здатність мережі передавати багатокомпонентну інформацію (мова, дані, відео, аудіо) з необхідною синхронізацією цих компонентів в реальному часі і використанням складних конфігурацій з'єднань;

4. Інтелектуальність — можливість управління послугою, викликом і з'єднанням з боку користувача або постачальника послуг;

5. Інваріантність доступу (або можливість) організації доступу до послуг незалежно від використовуваної технології;

6. Багатооператорність — участь декількох операторів у процесі надання послуги і поділ їх відповідальності в залежності від області їх діяльності [2].

На основі аналізу існуючих сьогодні концептуальних документів та експертних оцінок можна зробити висновок про те, що NGN являє собою універсальну багатоцільову мережу, призначену для передачі мови, зображень і даних з використанням технології комутації пакетів.

Її фундаментом є мультипротокольна-мультисервісна транспортна мережа зв'язку, що забезпечує перенесення різнорідного трафіку по різних протоколах передачі.

Концепція NGN передбачає підтримку необмеженого набору послуг з гнучкими можливостями управління ними, реалізацію універсальної транспортної мультипротокольної мережі з розподіленою комутацією, інтеграцію з традиційними мережами зв'язку. Базовим принципом NGN є поділ функцій перенесення і комутації, управління викликом і управління послугами.

Замість прийнятої в традиційних мережах канальної парадигми, в рамках якої з'єднання між абонентами будуються за принципом «точка — точка», в NGN реалізується перехід до ідеології віртуальних приватних мереж (VPN), які організовують доставку сервісів кінцевому користувачеві поверх протоколу IP.

Технологія NGN відкриває масу можливостей побудови накладених сервісів поверх універсального транспортного середовища — від пакетної телефонії (VoIP) до інтерактивного телебачення і Web-служб. Вона характеризується доступністю сервісів незалежно від місця розташування клієнта і використовуваних ним інтерфейсів (Ethernet, xDSL, Wi-Fi і т. д.). Таким чином, будь-який сервіс, створений в будь-якій точці NGN, стає доступним кожному споживачеві [3].

Гетерогенність інфраструктури, зростаюча конкуренція і зниження продажів базових сервісів, вважають західні експерти, сьогодні можуть розглядатися як головна загроза телекомунікаційній індустрії. Мережеві оператори прагнуть переосмислити свої бізнес-моделі і перетворити їх інфраструктуру в платформу, повністю засновану на IP. Головна мета і основна мотивація переходу до NGN — знизити витрати і створити нові джерела доходів.

Останніми роками на ринку склалася ситуація, яка підготувала ґрунт для просування NGN. На ринку зв'язку сформувалися такі умови:

— відкрита конкуренція між операторами, що стала наслідком приватизації підприємств зв'язку і ослаблення державного регулювання ринку;

— конвергенція мереж електрозв'язку та інформаційно-обчислювальних мереж, розвиток інформаційно-комунікаційних мереж;

— бурхливе зростання цифрового трафіку, в основному за рахунок розширення використання Інтернету;

— високий рівень попиту на рухомий зв'язок і нові мультимедійні служби;

— конвергенція операторів, мереж, терміналів, служб/послуг електрозв'язку.

Зазначені фактори створюють передумови до впровадження операторами широкого спектру нових послуг. За статистикою операторів, дохід від одного користувача нових телекомунікаційних послуг в кілька разів вищий, ніж від абонента традиційної телефонії.

Зазначимо також, що оператори фіксованих мереж, впроваджуючи NGN, мають на меті — скорочення капітальних витрат і операційних витрат за рахунок створення єдиного мультисервісного транспортного середовища для пропуску різнорідного трафіку.

Підходи до побудови транспортних мереж NGN представляють однаковий інтерес як для операторів мереж зв'язку загального користування (стаціонарних і мобільних), так і для операторів технологічних мереж зв'язку — відомчих і корпоративних. Незважаючи на те, що технологічні мережі зв'язку, як правило, мають певну професійну орієнтацію і спеціалізацію, при їх розвитку також враховується ідеологія NGN.

Розвиток мереж NGN та корпоративних відеокомунікацій є взаємовигідними і взаємопосилюючими процесами. Мережа NGN може з високою якістю передавати відеотрафік, дозволяє споживачеві самому керувати пропускною здатністю й іншими параметрами мережі, домагаючись найефективнішого використання доступної смуги пропускаючими.

Якщо подивитися на динаміку розвитку відеозв'язку, то мережі NGN з'явилися вчасно. З одного боку, в сучасному відеообладнанні реалізовані новітні технології для управління сеансами (SIP), стиснення даних (Н.264), динамічного керування смугою пропуску, проходження міжмережевих екранів і ін. Все це «піднімає» якість, підвищує керованість, що особливо важливо у зв'язку з поступовим переходом на телебачення високої чіткості HD.

Останнім часом у всьому світі, особливо у період пандемії COVID-19 швидко виросла потреба в відеокомунікаціях. А у зв'язку з розвитком та удосконаленням корпоративного управління, а саме скороченням витрат на відрядження, зниженням навантаження на навколишнє середовище, розвитком телемедицини, оперативнішим реагуванням на надзвичайні ситуації, відеоконференцзв'язок став дуже привабливим для корпоративних і інших користувачів.

Будь-яку сучасну послугу, доступ до якої надається оператором, можна представити як сукупність трафіків даних, мови та відео. Дослідження останніх років показують, що динаміка розвитку мовного трафіку в загальносвітовому масштабі залишається стабільною. В протилежність йому трафік даних зростає від року в рік експоненціально. Сьогодні більшість інформаційних потоків, що проходять мережами зв'язку, належать саме до трафіку даних. Існуючі телекомунікаційні мережі, концепції яких розроблювалися більше десяти років і були орієнтовні, в першу чергу, на передачу мовного трафіку, скоро не в змозі будуть задовольнити всі потреби користувачів, що продовжують збільшуватися.

Концепція NGN — це технічні рішення, що з'явилися на етапі розвитку цифрового зв'язку, коли трафік даних став важливішим за мовний. Мережа NGN представляє собою мультисервісну систему зв'язку, ядром якої є транспортна IP-мережа, що підтримує інтеграцію послуг передачі даних, мови і відео та реалізує принцип конвергенції технології.

В мережі NGN нам може знадобитися захист таких ресурсів:

— послуги в галузі зв'язку та комп'ютерних операцій;

— інформація та дані, включаючи програмне забезпечення, і дані, пов'язані з послугою забезпечення безпеки;

— обладнання та засоби.

В мережі NGN необхідно гарантувати захист інформації від таких загроз.

DoS (Denial of service — відмова сервісу) — напад на комп'ютерну систему з наміром зробити комп'ютерні ресурси недоступними для користувачів, для яких комп'ютерна система була призначена.

Одним з поширених методів нападу є насичення атакованого комп'ютера або мережевого устаткування великою кількістю зовнішніх запитів (часто безглуздих або невірно сформованих) так, що ата-коване устаткування не може відповісти користувачам або відповідає так повільно, що стає фактично недоступним. Взагалі відмова сервісу здійснюється:

1) примушенням атакованого устаткування зупинити роботу програмного забезпечення або устаткування або витратити наявні ресурси так, що устаткування не може продовжувати роботу;

2) заняттям комунікаційних каналів між користувачами і атакованим устаткуванням так, що якість сполучення перестає відповідати вимогам.

Підслуховування — загроза конфіденційності, що виникає завдяки перехвату повідомлень між відправником та отримувачем інформації каналом зв'язку.

Неавторизований доступ — доступ до мережних об'єктів повинен бути обмеженим та відповідати політиці безпеки. У випадку якщо зловмисники його отримають, система буде не захищена від інших несанкціонованих дій, таких як DoS атаки, підслуховування, маскування та інше [3].

Зміна інформації — зловмисники пошкоджують дані, запобігаючи тим самим доступу до ресурсів з боку авторизованих користувачів.

Відмова — зловмисники запобігають доступу до ресурсів з боку активних учасників телекомунікаційного з'єднання. Можливі методи атаки включають відмову в передачі, отримані, модифікації даних під час розмови.

Оскільки всі шлюзи NGN будуть приєднані до Інтернету, то для цих мереж будуть актуальними всі загрози, які є в інших IP-мережах. Наприклад, шахрай може виконати атаку підміни IP-адреси.

У традиційних телекомунікаційних мережах використовуються, як правило, пропрієтарні алгоритми та протоколи. Це ускладнює порушнику досягнення його цілей, вимагаючи наявності певної інсайдерської інформації. На відміну від цієї ситуації протоколи IP-мереж добре відомі і задокументовані.

Телефонна мережа загального користування має централізовану архітектуру, «інтелект» мережі зосереджений в АТС, а телефони не володіють великою функціональністю. На противагу цьому IP-мережі децентралізовані за своєю природою, абонентськими терміналами є, по суті, комп'ютери, використовуючи які, шахраї можуть створювати численні загрози

Користувацькі термінали знаходяться в тому ж просторі IP-адрес, що й елементи. Порушники можуть одночасно використовувати для організації атак кілька різних способів доступу: мідний кабель, опто-волокно, радіоканал. Для виявлення шахрайства необхідний постійний обмін інформацією між всіма елементами NGN, що досить проблематично.

Білінгові моделі мереж NGN відрізнятимуться від нині прийнятих і в них будуть враховуватися не тільки обсяг трафіку або час з'єднання, але і тип трафіку, вибрана якість обслуговування і т. п. Відповідно можна очікувати появи нових типів шахрайства.

Вважається, що розвиток NGN дасть поштовх поширенню мобільної торгівлі. Тому й увага шахраїв буде звернена в основному в цьому напрямку: вартість контенту буде істотно перевищувати вартість самих сполук. Так що зловмисники перейдуть від махінацій з незаконними дзвінками ТфЗК / GSM до махінацій з контентом:

1) споживання неоплаченого трафіку;

2) незаконний перепродаж сервісів;

3) завищення плати за послуги [2; 3].

Головна тенденція, яка простежується при аналізі сучасних стандартів в області інформаційної безпеки, полягає у відмові від жорсткої універсальної шкали класів безпеки і гнучкому підході до оцінки безпеки.

**Висноки.** Мережа NGN представляє собою мультисервісну систему зв'язку, ядром якої є транспортна IP-мережа, що підтримує інтеграцію послуг передачі даних, мови і відео та реалізує принцип конвергенції технологій. Останнім часом, а особливо в період пандемії COVID-19 в усьому світі, швидко виросла потреба в відеокомунікаціях. Відеоз'єднання стало чи не єдиним засобом, спроможним вести справи в епоху, коли особисті контакти не бажані. А у зв'язку з розвитком та удосконаленням корпоративного управління, а саме скорочення витрат на відрядження, зниження навантаження на навколишнє середовище, розвиввиток телемедицини, оперативне реагування на надзвичайні ситуації, саме відеоконференцзв'язок став дуже привабливим для корпоративних і інших користувачів.

# Список авторів

**Величко Віталій Юрійович (Vitalii Velychko)**, д. т. н., доцент, с. н. с., Інститут кібернетики ім. В. М. Глушкова Національної академії наук України (Київ)

**Воінова Світлана Олександрівна (Svitlana Voinova)**, к. т. н., доцент, Одеський національний технологічний університет (Одеса)

**Граняк Валерій Федорович (Valery Granyak)**, к. т. н., доцент, Вінницький національний аграрний університет (Вінниця)

**Гурський Олександр Олександрович (Alexander Gurskiy)**, к. т. н., доцент, Одеський національний технологічний університет (Одеса)

**Завертайло Костянтин Сергійович (Kostiantyn Zavertailo)**, аспірант, Інститут проблем математичних машин і систем (Київ)

**Іванова Лілія Вікторівна (Liliia Ivanova)**, к. т. н., директор, Відокремлений структурний підрозділ «Одеський технічний фаховий коледж ОНТУ» (Одеса)

**Котлик Діана Олександрівна (Diana Kotlyk)**, викладач, Відокремлений структурний підрозділ «Одеський технічний фаховий коледж ОНТУ» (Одеса)

**Котлик Сергій Валентинович (Sergii Kotlyk)**, к. т. н., доцент, Одеський національний технологічний університет (Одеса)

**Кудряшова Альона Вадимівна (Alona Kudriashova)**, к. т. н., старший викладач, Українська академія друкарства (Львів)

**Кунуп Тетяна Василівна (Tetiana Kunup)**, к. т. н., викладач, Відокремлений структурний підрозділ «Одеській технічний фаховий коледж ОНТУ», (Одеса)

**Малахов Кирило Сергійович (Kyrylo Malakhov)**, магістр (Інформаційні технології), н. с., Інститут кібернетики ім. В. М. Глушкова Національної академії наук України (Київ)

**Піх Ірина Всеволодівна (Iryna Pikh)**, д. т. н., професор, Українська академія друкарства (Львів)

**Пунченко Наталія Олегівна (Nataliia Punchenko)**, к. т. н., доцент, Одеський державний екологічний університет (Одеса)

**Сеньківський Всеволод Миколайович (Vsevolod Senkivskyy)**, д. т. н., професор, Українська академія друкарства (Львів)

**Сергєєва Олександра Євгенівна (Olexandra Sergeeva)**, д. ф.-м. н., професор, Одеський національний технологічний університет (Одеса)

**Соколова Оксана Петрівна** (**Oksana Sokolova**), старший викладач, Одеський національний технологічний університет (Одеса)

**Федосов Сергій Никифорович** (**Sergiy Fedosov**), д. ф.-м. н., професор, Одеський національний технологічний університет (Одеса)

**Хошаба Олександр Мирославович** (**Oleksandr Khoshaba**), к. т. н., доцент, Вінницький національний технічний університет (Вінниця)

**Цира Олександра Василівна (Olexandra Tsyra)**, к. ф. н., доцент, Державний університет інтелектуальних технологій і зв'язку (Одеса)

**Чаплінський Юрій Петрович (Yuri Chaplinskyy)**, к. т. н., с. н. с., Інститут кібернетики імені В. М. Глушкова НАН України (Київ)

| | |
|---|---|
| Vitalii Velychko |  https://orcid.org/0000-0002-7155-9202 |
| Svitlana Voinova |  https://orcid.org/0000-0003-0203-0599 |
| Valery Granyak |  https://orcid.org/0000-0001-6604-6157 |
| Liliia Ivanova |  https://orcid.org/0000-0003-1738-7697 |
| Sergii Kotlyk |  https://orcid.org/0000-0001-5365-1200 |
| Alona Kudriashova |  https://orcid.org/0000-0002-0496-1381 |
| Tetiana Kunup |  https://orcid.org/0000-0003-0246-0951 |
| Kyrylo Malakhov |  https://orcid.org/0000-0003-3223-9844 |
| Iryna Pikh |  https://orcid.org/0000-0002-9909-8444 |
| Nataliia Punchenko |  https://orcid.org/0000-0003-1382-4490 |
| Vsevolod Senkivskyy |  https://orcid.org/0000-0002-4510-540X |
| Olexandra Sergeeva |  https://orcid.org/0000-0002-5534-9563 |
| Oksana Sokolova |  https://orcid.org/0000-0001-9224-6734 |
| Sergiy Fedosov |  https://orcid.org/0000-0002-5775-1468 |
| Oleksandr Khoshaba |  https://orcid.org/0000-0001-5375-6280 |
| Olexandra Tsyra |  https://orcid.org/0000-0003-3552-2039 |
| Yuri Chaplinskyy |  https://orcid.org/0000-0001-7697-3958 |
| Olexander Gurskiy | |
| Kostiantyn Zavertailo | |
| Diana Kotlyk | |

**NEW INFORMATION TECHNOLOGIES,**
**SIMULATION AND AUTOMATION**

MONOGRAPH

**IOWA STATE UNIVERSITY DIGITAL PRESS**
2022